\documentclass[a4paper]{article}
\usepackage[top=2.5cm,bottom=2.5cm,left=2.0cm,right=2.0cm]{geometry} 
\usepackage{lscape} 
\usepackage{authblk}
\usepackage{amsmath}
\usepackage{amssymb}
\usepackage{amsthm}
\usepackage{color}
\usepackage{bm}
\usepackage{cases}
\usepackage[colorlinks,linkcolor=blue,anchorcolor=green,citecolor=red]{hyperref}
\usepackage{appendix}
\usepackage{graphicx}
\usepackage{epstopdf}
\usepackage{float}                  
\usepackage{subfig}                 
\usepackage{overpic}                
\usepackage{subcaption} 
\usepackage{booktabs}
\usepackage{lscape}
\usepackage[figuresright]{rotating} 
\usepackage{algorithm}  
\usepackage{algpseudocode}  
\usepackage{multirow}
\usepackage{lscape}
\usepackage{tikz} 
\usetikzlibrary{patterns.meta}
\usepackage{enumerate}
\usetikzlibrary{calc}

\allowdisplaybreaks[4]
\numberwithin{equation}{section}

\newtheorem{remark}{Remark}[section] 

\title{ Pump-driven droplet electrohydrodynamics: deformation, pinch-off and recoalescence}
\author[a]{Yuzhe Qin}
\author[b,c,*]{Huaxiong Huang} 
\author[d]{Zilong Song} 
\author[c,*]{Shixin Xu}
\affil[a]{School of Mathematics and Statistics, 
Shanxi University, Taiyuan 030006, China}
\affil[b]{Department of Mathematics and Statistics, York University, Toronto, ON. M3J 1P3, Canada} 
\affil[c]{Zu Chongzhi Center, Duke Kunshan University, 8 Duke Ave, Kunshan, Jiangsu, China} 
\affil[d]{Math and Statistics Department, Utah State University, Old Main Hill, Logan, Utah 84322, USA}
\affil[*]{Corresponding authors: huaxiong.huang@dukekunshan.edu.cn;  shixin.xu@dukekunshan.edu.cn; }

\date{}
\begin{document}
\maketitle
\begin{abstract}
We investigate pump-driven droplet electrohydrodynamics with an emphasis on deformation, pinch-off, and recoalescence. A thermodynamically consistent phase-field framework is developed that couples Nernst–Planck–Poisson electrodiffusion with incompressible Navier–Stokes–Cahn–Hilliard flow, and incorporates interfacial ionic pumps as prescribed surface fluxes. In the pump-free baseline, applied fields merely polarise the droplet and deformation is negligible. By contrast, surface-localised pumping drives the accumulation of positive ions within the droplet, elevates the interior potential, and generates non-uniform electric fields. The resulting Lorentz stresses stretch and displace the droplet, thin interfacial necks, and trigger pinch-off; the daughter droplets subsequently recoalesce, often after wall contact, yielding flattened remnants. In multiple-droplet settings, pump-induced charging produces lateral electrostatic repulsion and asymmetric deformation; under geometric confinement, crescent bending and star-like morphologies emerge. Shear-flow tests further show that a pumped droplet can be immobilised and ruptured while an unpumped neighbour is advected downstream, suggesting a route to sorting. Taken together, the results establish interfacial pumping as an internal actuation mechanism that robustly controls droplet morphology and dynamics across configurations.
\end{abstract}

\paragraph{Keywords:} Active pumps; Droplet electrohyropdynamcs; Break-up and coalescence;
\section{Introduction}

Fluid–structure interaction and mass transport across interfaces are fundamental processes in many physical, biological, and engineered systems. 
In electrolyte solutions, these interactions are further complicated by the presence of mobile ions and external electric fields, 
which generate electrostatic forces and electrokinetic flows that couple tightly with hydrodynamics and interfacial motion. 
For example, in electroosmotic systems, electric fields drive ionic motion and induce fluid flows near interfaces, leading to deformation, migration, or instability of droplets, vesicles, and other soft structures. 
Accurately modeling these coupled phenomena requires accounting for electrohydrodynamic forces, charge redistribution, and interfacial dynamics simultaneously.

Traditionally, such systems are modeled using a combination of the Navier--Stokes (NS) equations for fluid flow, the Poisson--Nernst--Planck (PNP) equations for ion transport and electrostatics, and the Cahn--Hilliard (CH) equation for diffuse interface representation. 
The NS--PNP--CH system forms a robust theoretical framework for simulating multiphase electrohydrodynamics in the presence of ionic species and moving interfaces. 
Each component captures a distinct physical mechanism-viscous fluid motion, ion migration and diffusion, and interfacial energy minimization, respectively. 
This coupled system has been successfully applied to model electrokinetic flows in microfluidic devices~\cite{Chatterjee2005Modeling}, electrolyte transport in electrochemical cells~\cite{Karoline2023Nano}, and ionic behavior in biological membranes~\cite{Cucchi2020A}.

However, in most existing models, ion transport across interfaces is assumed to occur passively, driven solely by diffusion and electric field gradients. 
This assumption overlooks a critical feature of many biological and bioinspired systems: the presence of \emph{active ion pumps}, which use external energy to transport ions against their electrochemical gradients. These pumps play a central role in regulating ion distributions and maintaining membrane potential in cells. 
For example, the Na\textsuperscript{+}/K\textsuperscript{+} ATPase consumes ATP to maintain the ionic imbalance necessary for nerve signal transmission and osmotic homeostasis~\cite{Pivovarov2018Pump}. 
Inspired by such biological mechanisms, synthetic ion pumps are being developed for use in drug delivery, energy conversion, and selective separation~\cite{Mei2022Bioinspired,Zhang2013Bioinspired}. 
Yet, most mathematical models of electrohydrodynamics neglect the directional flux contributions from these active mechanisms.

In this work, we introduce a thermodynamically consistent phase-field framework that explicitly incorporates active ion pumps into the NS–PNP–CH system. 
The model considers a charged, deformable droplet suspended in an ionic solution under an external electric field. 
The novelty lies in the formulation of an interfacial pump flux that actively transports selected ions from the exterior to the interior of the drop. 
This is achieved by extending the classical energy variation approach to include an energy input term associated with active transport, which drives directional ionic fluxes localized near the interface.

The inclusion of ion pumps introduces new coupling pathways between ionic transport, fluid motion, and interfacial dynamics. 
When combined with electroosmotic effects, pump-induced fluxes can significantly alter charge distributions, flow fields, and droplet behavior. 
Our numerical framework allows us to simulate these effects and examine how the interplay of pump strength, applied field, and flow conditions influences the system. 
These insights are valuable for designing and optimizing artificial ion pump technologies and understanding natural systems that rely on active ionic regulation.

\medskip

The remainder of the paper is organized as follows. 
Section~\ref{sec:model} presents the mathematical model, derived via an energetic variational method framework, together with the nondimensional governing equations. 
Details of the numerical implementation and accuracy assessment are deferred to Appendices~\ref{sec:scheme}–\ref{Ap:constudy}. 
Section~\ref{subsec:1droplet} examines how interfacial pumping modulates the deformation and migration of a single droplet under grounded and biased electric boundaries, using pump–free baselines for comparison. 
Section~\ref{subsec:2dropsnoelectric} analyses two–droplet interactions, highlighting lateral electrostatic repulsion, asymmetric bending, and pump–induced pinch–off and recoalescence. 
Section~\ref{subsec:selection} demonstrates a microfluidic application by showing shear–assisted separation in which a pumped droplet is immobilised and ruptures while an unpumped neighbour is advected downstream. 
We conclude with a brief discussion of the model’s capabilities and avenues for extension.

\section{Mathematical model}\label{sec:model}
In this section, we focus on deriving the phase field model based on the Energy Variation method 
to describe the ion transportation in the Newtonian fluid within the moving interface.

Let $\Omega = \Omega^{+} \cup \Omega^{-}$ denote the entire domain, 
where $\Omega^{+}$ represents the droplet (interior) region and $\Omega^{-}$ corresponds to the surrounding (exterior) fluid. 
These two immiscible regions are separated by a sharp interface $\Gamma = \partial \Omega^{+} \cap \partial \Omega^{-}$.
To model the interface using a diffuse interface framework, 
we introduce an order parameter $\psi(\bm{x}, t)$ defined as:
\begin{align}\label{def: psi}
    \psi(\bm{x}, t) =
    \begin{cases}
        1, & \text{in } \Omega^{+}, \\
       -1, & \text{in } \Omega^{-},
    \end{cases}
\end{align}
with the interface $\Gamma$ corresponding to the zero level set of $\psi$, 
i.e., $\Gamma = \{\bm{x} : \psi(\bm{x}, t) = 0\}$. 
Then with the label function, $\bm{n}$ is the outer unit vector to the $\Omega$ and defined as follows
\begin{equation}
\bm{n} = \frac{\nabla \psi}{\left|\nabla \psi \right|}, 
\end{equation}
which is pointing from the exterior to the interior of the droplet. 

Based on the laws of conservation of mass and momentum, 
coupled ion transport and hydrodynamics in the two-phase system are governed by the following equations:
\begin{subequations}\label{def:main}
    \begin{align}
        & \frac{\partial \psi}{\partial t} 
        + \nabla \cdot (\bm{u} \psi) 
        + \nabla \cdot \bm{j}_{\psi} = 0,
        && \text{in } \Omega, 
        \label{def:psi} \\
        & \frac{\partial c_{i}}{\partial t} 
        + \nabla \cdot (\bm{u} c_{i}) 
        + \nabla \cdot \bm{j}_{i} +\nabla\cdot \bm{I}_{\rm{pump}}=0, 
        \qquad i = 1, \dots, N,
        && \text{in } \Omega,
        \label{def:ci} \\
        & \nabla \cdot \bm{D} 
        = \sum_{i=1}^{N} z_{i} e c_{i},
        && \text{in } \Omega,
        \label{def:D} \\
        & \bm{D} = \epsilon_{\mathrm{eff}} \bm{E} 
        = -\epsilon_{\mathrm{eff}} \nabla \phi,
        && \text{in } \Omega,
        \label{def:phi} \\
        & \rho \left( \frac{\partial \bm{u}}{\partial t} 
        + (\bm{u} \cdot \nabla) \bm{u} \right) 
        = \nabla \cdot \sigma_{\eta} 
        + \nabla \cdot \sigma_{e} 
        + \nabla \cdot \sigma_{\psi},
        && \text{in } \Omega,
        \label{def:u} \\
        & \nabla \cdot \bm{u} = 0,
        && \text{in } \Omega.
        \label{def:nabla_u}
    \end{align}
\end{subequations}

Equation~\eqref{def:psi} describes the conservation of the order parameter $\psi$, 
where $\bm{u}$ is the velocity field and $\bm{j}_{\psi}$ is the flux associated with $\psi$.  
Equation~\eqref{def:ci} represents the mass conservation of the $i$-th ionic species, 
with $\bm{j}_i$ denoting its flux; we consider $N$ ionic species in total.
Equation~\eqref{def:D} is Gauss's law in differential form, 
where $\bm{D}$ is the electric displacement field. 
From Equation~\eqref{def:phi}, it follows that 
$\bm{D} = \epsilon_{\mathrm{eff}} \bm{E} = -\epsilon_{\mathrm{eff}} \nabla \phi$, 
where $z_i$ is the valence of the $i$-th ion, $e$ is the elementary charge, 
$\epsilon_{\mathrm{eff}}$ is the effective dielectric permittivity, $\bm{E}$ is the electric field, 
and $\phi$ is the electrostatic potential.
Equations~\eqref{def:u} and \eqref{def:nabla_u} describe the momentum balance and incompressibility condition, respectively. 
Here, $\rho$ denotes the fluid density, $\sigma_{\eta}$ is the viscous stress tensor, 
$\sigma_{e}$ is the Maxwell (electric) stress tensor, 
and $\sigma_{\psi}$ captures the contribution of interfacial (capillary) stresses.
If the dielectric constant is different  insdie and outside of droplet, 
we define the effective dielectric constant over the whole domain as follows 
\begin{equation}
    \epsilon_{\mathrm{eff}}^{-1}
    = \frac{1-\psi}{2\epsilon^{-}}
    + \frac{1+\psi}{2\epsilon^{+}}.
\end{equation}

At the interface, two mechanisms contribute to transmembrane ion flux:  
(1) a passive leakage flux driven by chemical potential gradients $\bm{j}_i$, and  
(2) an active pump flux that selectively transports specific ions from the exterior to the interior $\bm{I}_{\rm{pump}}$. We define the directional pump flux as
\[
\bm{I}_{\mathrm{pump}} = \mathcal{I}_{\mathrm{pump}}\frac{\nabla \psi}{\left|\nabla \psi \right|},
\]
where $\mathcal{I}_{\mathrm{pump}}$ is the scalar pump strength. Away from the interface, the label function $\psi$ is constant and does not induce any spatial gradient, so the pump effect vanishes.

\paragraph{Modeling the Pump Strength.}  
A classical representation of ion pumping can be described by a reversible enzymatic reaction occurring at the interface:
\begin{equation}
E + \beta S_{o} + ATP \overset{k_1}{\underset{k_{-1}}{\rightleftharpoons}} ES\text{-}ATP \overset{k_2}{\underset{k_{-2}}{\rightleftharpoons}} E + \beta S_i + P_i,
\end{equation}
where $E$ denotes the free pump protein, $S_o$ and $S_i$ represent the extracellular and intracellular substrates, respectively, $ES$–$ATP$ is the enzyme-substrate complex, $ATP$ is the energy source, and $P_i$ denotes the phosphate product. 

When $\beta = 1$, the pump rate can be approximated by the standard Michaelis–Menten form:
\begin{align}
\mathcal{I}_{\mathrm{pump}} = I_0 \left( \frac{[S_o]}{K_0 + [S_o]} \right),
\end{align}
where $I_0$ is the maximum pump rate and $K_0$ is a saturation constant. A detailed derivation of this formula is provided in the Supporting Information (SI).

In biological systems, many pumps simultaneously transport multiple substrate molecules per cycle, i.e., $\beta > 1$. There are two main modeling approaches for $\mathcal{I}_{\mathrm{pump}}$~\cite{zaheri2020comprehensive}:

\begin{itemize}
\item \textbf{Cooperative Binding:}  
Cooperative binding describes a kinetic phenomenon where the binding of one substrate molecule to a site on an enzyme or transporter alters the likelihood of additional substrate molecules binding to other sites (positive or negative cooperativity). In this case, the pump flux generally increases as $\beta$ increases, and the Hill equation is employed:
\begin{align}\label{eq:Hillpump}
\mathcal{I}_{\mathrm{pump}} = I_0 \left( \frac{[S_o]^\beta}{K_0^\beta + [S_o]^\beta} \right),
\end{align}
where $I_0$ is the maximum pump rate and $K_0$ is an empirical constant. This formulation is commonly used to describe phenomena such as oxygen binding by hemoglobin or $\rm{Ca^{2+}}$ binding to the SERCA pump.

\item \textbf{Simultaneous (Stoichiometric) Transport:}  
Stoichiometric transport—sometimes referred to as structural cooperativity---occurs when a transporter or pump must bind exactly $\beta$ substrate molecules simultaneously to complete a transport cycle. The pump remains inactive unless all required substrates are bound. In this scenario, the pump flux decreases as $\beta$ increases, and a modified Michaelis–Menten (MM) model is adopted:
\begin{align}\label{eq:pumpmm}
\mathcal{I}_{\mathrm{pump}} = I_0 \left( \frac{[S_o]}{K_0 + [S_o]} \right)^\beta.
\end{align}
This form is suitable for modeling pumps such as $\rm{Na}^+/\rm{K}^+$-ATPase (which exchanges 3 $\rm{Na}^+$ and 2 $\rm{K}^+$ per cycle) and $\rm{Ca^{2+}}$-ATPase (which transports 2 $\rm{Ca^{2+}}$ per ATP hydrolyzed).
\end{itemize}

In this study, we approximate $[S_o]$ via the phase-field as $\frac{1 - \psi}{2} c_i$.
Future work will explore the direct incorporation of the full mass-action-based enzymatic kinetics to capture the intrinsic pump dynamics.

The following homogeneous boundary conditions are used 
\begin{equation}
\bm{j}_{\psi} \cdot \bm{n}|_{\partial \Omega} = 0, \quad 
\bm{j}_{i} \cdot \bm{n}|_{\partial \Omega} = 0, \quad 
\phi|_{\partial \Omega} = 0, \quad 
\bm{u}|_{\partial \Omega} = 0, \quad 
\left( \sigma_{\eta} + \sigma_{e} + \sigma_{\psi}\right)|_{\partial \Omega} = 0. 
\end{equation}

We derive the constitutive relations for the unknown quantities---$\bm{j}_{\psi}$, $\bm{j}_{c_{i}}$, $\sigma_{\eta}$, $\sigma_{e}$, and $\sigma_{\psi}$---using an energy variational approach. Following the framework in \cite{xu2023coupled}, for an isothermal system with energy input, the time rate of change of total energy $E_{\mathrm{total}}$ is balanced by the energy dissipation $\Delta$ and the power input $\mathcal{P}$ required to sustain active pumping:
\begin{equation}\label{def: energy dissipation}
    \frac{\mathrm{d}E_{\mathrm{total}}}{\mathrm{d}t} = -\Delta + \mathcal{P}.
\end{equation}

The total energy of the coupled system \eqref{def:main} consists of the kinetic energy, electrostatic energy, ionic entropy, and phase-field mixing energy:
\begin{equation}\label{def:energy}
\begin{aligned}
    E_{\mathrm{total}} 
    &= E_{\mathrm{kin}} + E_{\mathrm{es}} + E_{\mathrm{ion}} + E_{\mathrm{mix}} \\
    &= \int_{\Omega} \left[ 
        \frac{\rho}{2} |\bm{u}|^{2} 
        + \frac{1}{2} \bm{E} \cdot \bm{D} 
        + k_{B} T \sum_{i=1}^{N} c_{i} \left( \ln \frac{c_{i}}{\tilde{c}} - 1 \right) 
        + \lambda \left( \frac{\delta^{2}}{2} |\nabla\psi|^{2} + F(\psi) \right) 
    \right] \, \mathrm{d}\bm{x},
\end{aligned} 
\end{equation}
where $k_{B}$ is the Boltzmann constant, $T$ is temperature, and $\tilde{c}$ is a reference concentration. The parameter $\lambda$ characterizes the interfacial energy density, and $\delta$ represents the characteristic thickness of the diffuse interface. The double-well potential is given by $F(\psi) = \frac{1}{4}(1 - \psi^{2})^{2}$.

Then the chemical potentials are defined as 
\begin{subequations}
\begin{align}
    \mu_{i} = & z_{i} e \phi + k_{B}T \ln \frac{c_{i}}{\tilde{c}}, \\
    \mu_{\psi} = & \lambda \left( - \delta^{2} \nabla^{2} \psi + F^{\prime}(\psi) \right) - \frac{1}{2} \frac{\partial \epsilon_{\it eff}}{\partial \psi} |\bm{E}|^{2}.
\end{align}
\end{subequations} 

The total energy dissipation $\Delta$ accounts for three major sources: viscous dissipation from fluid deformation, irreversible ionic diffusion, and phase-field relaxation:
\begin{equation}\label{def: dissipation} 
\begin{aligned}
    \Delta ={} & \int_{\Omega} 2 \eta \left|\bm{D}_{\eta}\right|^{2} \, \mathrm{d}\bm{x} 
    + \int_{\Omega} \sum_{i=1}^{N} \frac{D_{i} c_{i}}{k_{B} T} \left| \nabla \mu_{i} \right|^{2} \, \mathrm{d}\bm{x} 
    + \int_{\Omega} \mathcal{M} \left| \nabla \mu_{\psi} \right|^{2} \, \mathrm{d}\bm{x},
\end{aligned}
\end{equation}
where $\eta$ is the dynamic viscosity, and $\bm{D}_{\eta} = \frac{1}{2}(\nabla\bm{u} + \nabla\bm{u}^{T})$ is the strain-rate tensor. For each ion species, $D_{i}$ denotes its diffusivity and $\mu_{i}$ its chemical potential. Similarly, $\mu_{\psi}$ and $\mathcal{M}$ are the chemical potential and mobility associated with the phase-field variable $\psi$.  Taking into account the differences in diffusion on both sides of the droplets, we model diffusion coefficients as follows \cite{Qin2022JCP},
\begin{equation}
    D_{i}^{-1}
    = \frac{1-\psi}{2D_{i}^{-}}
    +\frac{1+\psi}{2D_{i}^{+}}+\frac{(1-\psi^2)^2}{\delta gq(c_i)}, 
\end{equation}
where $D_{i}^{\pm}$ is the
diffusion coefficient of $i$ -th ion in the domain $\Omega^{\pm}$ and $g$ is the membrane conductance. 

The external energy input $\mathcal{P}$ accounts for active ionic pumping driven by non-conservative forces and is defined as:
\begin{equation}
    \mathcal{P} = -\int_{\Omega} \sum_{i} \mu_i  \nabla\cdot \bm{I}_{\mathrm{pump}} \, \mathrm{d}\bm{x}.
\end{equation}

Using the energy variation method \cite{qin2023droplet}, the governing equations can be derived as 
\begin{subequations}\label{main_eqn:pump}
    \begin{align}
        &\frac{\partial \psi}{\partial t} 
        + \nabla \cdot \left( \bm{u} \psi \right) 
        + \nabla \cdot \bm{j}_{\psi} = 0, 
        & \mbox{in} \quad \Omega, 
        \label{eqn:pump_psi}\\ 
        &\bm{j}_{\psi} = -\mathcal{M} \nabla \mu_{\psi},  
        & \mbox{in} \quad \Omega, 
        \label{eqn:pump_j}\\ 
        &\mu_{\psi} = \lambda \left( - \delta^{2} \nabla^{2} \psi 
        + F^{\prime}(\psi) \right) 
        - \frac{1}{2} \frac{\partial \epsilon_{\mathrm{eff}}}{\partial \psi} |\bm{E}|^{2}, 
        & \mbox{in} \quad \Omega, 
         \label{eqn:pump_mu_psi}\\
        &\frac{\partial c_{i}}{\partial t} 
        + \nabla \cdot \left( \bm{u} c_{i} \right) 
        + \nabla \cdot \bm{j}_{i}+\nabla\cdot \bm{I}_{\mathrm{pump}}= 0,  \qquad i=1,\cdots,N, 
        & \mbox{in} \quad \Omega, 
        \label{eqn:pump_ci}\\
        &\bm{j}_{i} = -\frac{D_{i}c_{i}}{k_{B}T} \nabla \mu_{i}, 
        & \mbox{in} \quad \Omega, 
        \label{eqn:pump_ji}\\ 
        & \bm{I}_{\mathrm{pump}} 
		= I_{0} \left(\frac{\frac{1-\psi}{2}c_{i}}{K_{0} 
		+ \frac{1-\psi}{2}c_{i}}\right)^{\beta}\bm{n}, 
		& \mbox{in} \quad \Omega,  
		\label{eqn:pump_pump}
		\\
        &\mu_{i} = z_{i} e \phi + k_{B}T \ln \frac{c_{i}}{\tilde{c}}, 
        & \mbox{in} \quad \Omega, 
        \label{eqn:pump_mui}\\
        &\nabla \cdot \bm{D} = \sum_{i=1}^{N}z_{i}ec_{i}, 
        & \mbox{in} \quad \Omega, 
        \label{eqn:pump_E}\\ 
        &\bm{D} = \epsilon_{\mathrm{eff}} \bm{E} = - \epsilon_{\mathrm{eff}} \nabla \phi, 
        & \mbox{in} \quad \Omega, 
        \label{eqn:pump_phi} \\
        &\rho \left( \frac{\partial \bm{u}}{\partial t} 
        + \left( \bm{u} \cdot \nabla \right) \bm{u} \right) 
        = \nabla \cdot \sigma_{\eta} 
        + \nabla \cdot \sigma_{\psi} 
        + \nabla \cdot \sigma_{e}, 
        & \mbox{in} \quad \Omega, 
        \label{eqn:pump_u}\\ 
        &\sigma_{\eta} = 2 \eta \bm{D}_{\eta}  - p \textbf{I},  
        & \mbox{in} \quad \Omega, 
        \label{eqn:pump_sigma_eta}\\
        &\sigma_{\psi} = - \lambda \delta^{2} \nabla \psi \otimes \nabla \psi,  
        & \mbox{in} \quad \Omega, 
        \label{eqn:pump_sigma_psi}\\
        &\sigma_{e} = \epsilon_{\mathrm{eff}} \left( \nabla \phi \otimes \nabla \phi 
        -\frac{1}{2} \left| \nabla \phi \right|^{2} \textbf{I} \right), 
        & \mbox{in} \quad \Omega, 
        \label{eqn:pump_sigma_e}\\
        & \nabla \cdot \bm{u} = 0, & \mbox{in} \quad \Omega, 
        \label{eqn:pump_nabla_u} \\
        &\epsilon_{\mathrm{eff}}^{-1} 
        = \frac{1-\psi}{2\epsilon^{-}} 
        + \frac{1+\psi}{2\epsilon^{+}}, 
        & \mbox{in} \quad \Omega, 
        \label{eqn:pump_epsilon}\\
        &D_{i}^{-1} 
        = \frac{1-\psi}{2D_{i}^{-}} 
        + \frac{1+\psi}{2D_{i} ^{+}}+\frac{(1-\psi^2)^2}{\delta g(c_i)}, 
        & \mbox{in} \quad \Omega,  
        \label{eqn:pump_D}
    \end{align}
\end{subequations}

The dimensionless form is given below, where the details of nondimensionalization are given in Appendix~\ref{sec: dimensionlization}
\begin{subequations}\label{eqn:simplified_system}
	\begin{align}
		& \frac{\partial \bm{u}}{\partial t} 
        + \left( \bm{u} \cdot \nabla \right) \bm{u} 
        + \nabla P 
        = \frac{1}{Re}\nabla^{2}\bm{u} 
        + \mu_{\psi} \nabla \psi 
        - \frac{Ca_{E}}{\zeta^{2}} \sum_{i=1}^{N} z_{i} c_{i} \nabla \phi,
        \label{eqn:u}
        \\ 
        & \nabla \cdot \bm{u} = 0, 
        \label{eqn:nabla_u}
        \\
		&\frac{\partial c_{i}}{\partial t} 
		+ \nabla \cdot \left( \bm{u} c_{i} \right) 
		= \frac{1}{Pe} 
		\nabla \cdot \left( D_{i} \nabla c_{i} \right) 
        + \alpha_{i} \frac{\zeta^{2}}{Pe_{E}} \nabla \cdot \left( D_{i} c_{i} \nabla \phi \right) - \nabla \cdot \bm{I}_{\mathrm{pump}},
		\label{eqn:ci}
        \\ 
        & \bm{I}_{\mathrm{pump}} 
		= I_{0} \left(\frac{\frac{1-\psi}{2}c_{i}}{K_{0} 
		+ \frac{1-\psi}{2}c_{i}}\right)^{\beta}\bm{n}, 
        \label{eqn:pump}
		\\
        &- \zeta^{2} 
        \nabla \cdot \left( \epsilon_{\mathrm{eff}} \nabla \phi \right) 
        = \sum_{i=1}^{N} z_{i} c_{i}, 
        \\ 
		&\frac{\partial \psi}{\partial t} 
		+ \nabla \cdot \left( \bm{u} \psi \right) 
		= M \nabla^{2} \mu_{\psi}, 
		\label{eqn:psi}
        \\
		&\mu_{\psi} = - \delta \nabla^{2} \psi 
        + \frac{1}{\delta} F^{\prime} \left( \psi \right)
        - \frac{Ca_{E}}{2} 
        \frac{\partial \epsilon_{\mathrm{eff}}}{\partial \psi} 
        \left|\nabla\phi\right|^{2}, 
        \label{eqn:mu_psi}\\
        &\epsilon_{\mathrm{eff}}^{-1} 
        = \frac{1-\psi}{2} 
        + \frac{1+\psi}{2\epsilon_{r}}, 
        \label{eqn:eps}\\
        & D_{i}^{-1} 
        = \frac{1-\psi}{2}
        + \frac{1+\psi}{2D_{i}^{r}}+\frac{(1-\psi^2)^2}{\sigma \delta g}, 
        \label{eqn:Deff}
	\end{align}
\end{subequations}
with $\epsilon_{r} = \epsilon^{+} / \epsilon^{-}$ and $D_{i}^{r} = D_{i}^{+} / D_{i}^{-}$.

Details of the numerical implementation and the mesh–refinement study are given in Appendices~\ref{sec:scheme}–\ref{Ap:constudy}. The one–dimensional tests in Appendix~\ref{App:1d} demonstrate how interfacial pumps sustain ionic asymmetry in the absence of flow.

\section{Pump-induced droplet deformation}\label{subsec:1droplet}

In this section, we investigate how active ion pumping affects the shape and electrohydrodynamic behavior of a two-dimensional droplet under various electric fields. 
Simulations are performed using the full coupled Navier--Stokes--Poisson--Nernst--Planck--Cahn--Hilliard (NS--PNP--CH) system, 
with the pump term localized near the droplet interface.

The computational domain is set as $\Omega = [-4, 4] \times [-4, 4]$, 
with the droplet initially centered at the origin. 
The initial conditions are given by:
\begin{subequations}
\begin{align}
& \psi(x, y, 0) = \tanh \left( \frac{\sqrt{1 - (x^2 + y^2)}}{\sqrt{2} \delta} \right), \\
& p(x, y, 0) = 1, \quad n(x, y, 0) = 1, \quad \phi(x, y, 0) = 0,
\end{align}
\end{subequations}
representing a circular droplet with uniform ion concentrations and zero initial potential.  

The following boundary conditions are applied:
\begin{equation}
\begin{aligned}
& \left. \nabla \psi \cdot \bm{n} \right|_{\partial \Omega} = 0, 
\quad \left. \nabla \mu_{\psi} \cdot \bm{n} \right|_{\partial \Omega} = 0, 
\quad \left. \bm{u} \right|_{\partial \Omega} = \bm{0}, \\
& \left. \nabla p \cdot \bm{n} \right|_{\partial \Omega} = 0, 
\quad \left. \nabla n \cdot \bm{n} \right|_{\partial \Omega} = 0,
\end{aligned}
\end{equation}
where $\bm{n}$ is the outward unit normal to the boundary.

Unless otherwise specified, the parameter values are:
\begin{equation}
\begin{aligned}
& Re = 1, \quad \zeta = 0.1, \quad Pe =1, \quad Pe_E = 1, \quad \alpha_1 = 0.5, \quad \alpha_2 = -0.5, \quad   \beta = 2, \\
&  K_0 = 3.5, \quad\delta = 0.1, \quad M = \delta^2, \quad \epsilon_r = 1, \quad D_i^r = 1, \quad Ca_E= 1.
\end{aligned}
\end{equation}

To mimic an applied electric field, Dirichlet boundary conditions are imposed on the electric potential. We examine how the interaction between the externally imposed electrostatic field and the pump-driven ionic flux influences droplet deformation and flow fields.

\paragraph{Vertical electric field}
We first consider the electric field to be added in the vertical direction using the following clamped boundary condition. 
  
\begin{equation}
  \left. \phi \right|_{y=4} = \phi_{0b}, 
\quad 
\left. \phi \right|_{y=-4} = \phi_{0u}, 
\quad 
\left. \nabla \phi \cdot \bm{n} \right|_{x=\pm 4} = 0.
\end{equation}
 
Initially, the positive and negative ions, along with the electric potential, are uniformly distributed throughout the domain, establishing a symmetric and electrically neutral configuration. The black contour indicates the initial droplet interface where $\psi = 0$. 

\begin{figure}[!ht] 
\vskip -0.4cm
\centering
	\subfloat[$p~(t=10)$]{
		\includegraphics[width=0.3\linewidth]{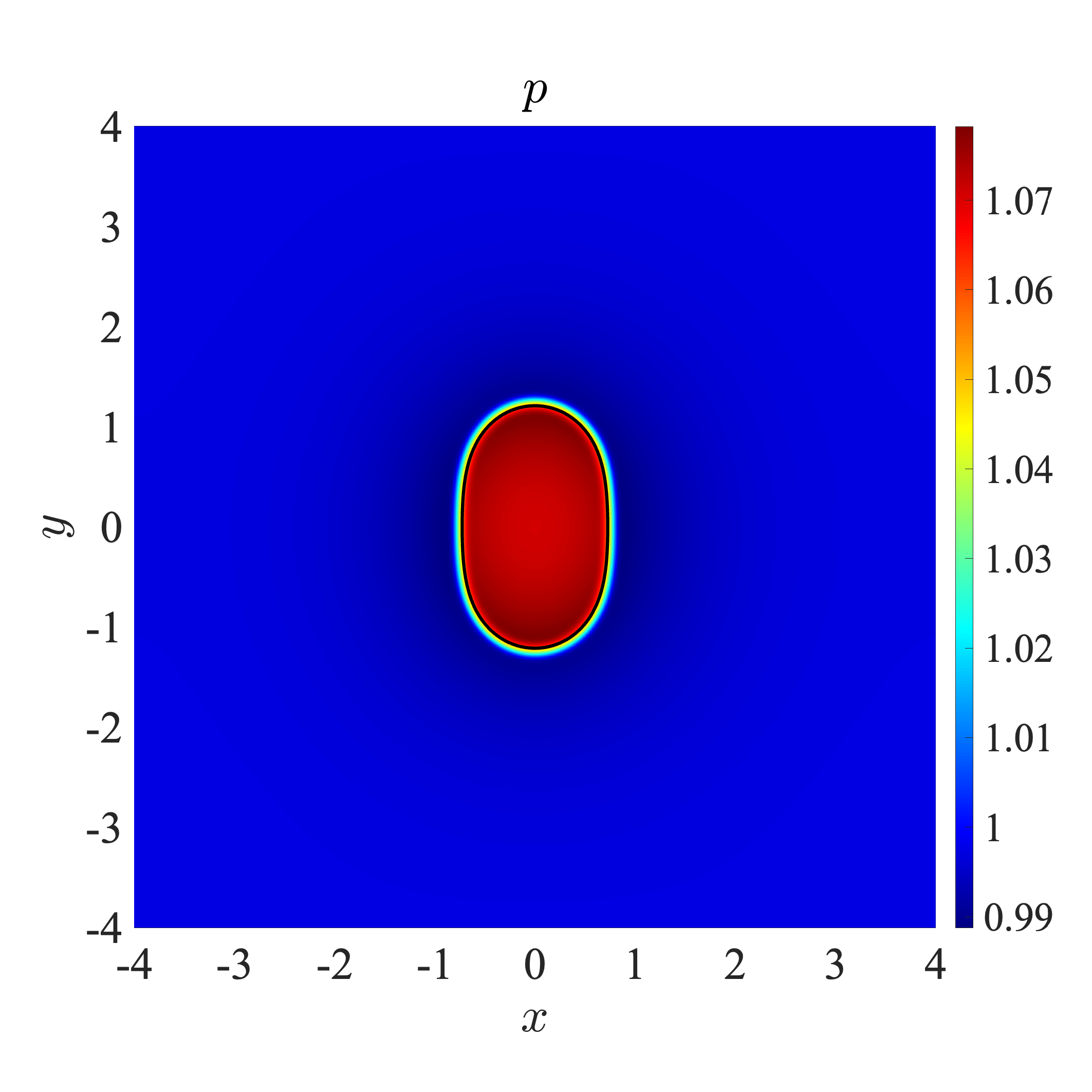}
        \label{subfig:1DropD0N0Pump25P10}
		}
    \hskip -0.3cm
	\subfloat[$n~(t=10)$]{
		\includegraphics[width=0.3\linewidth]{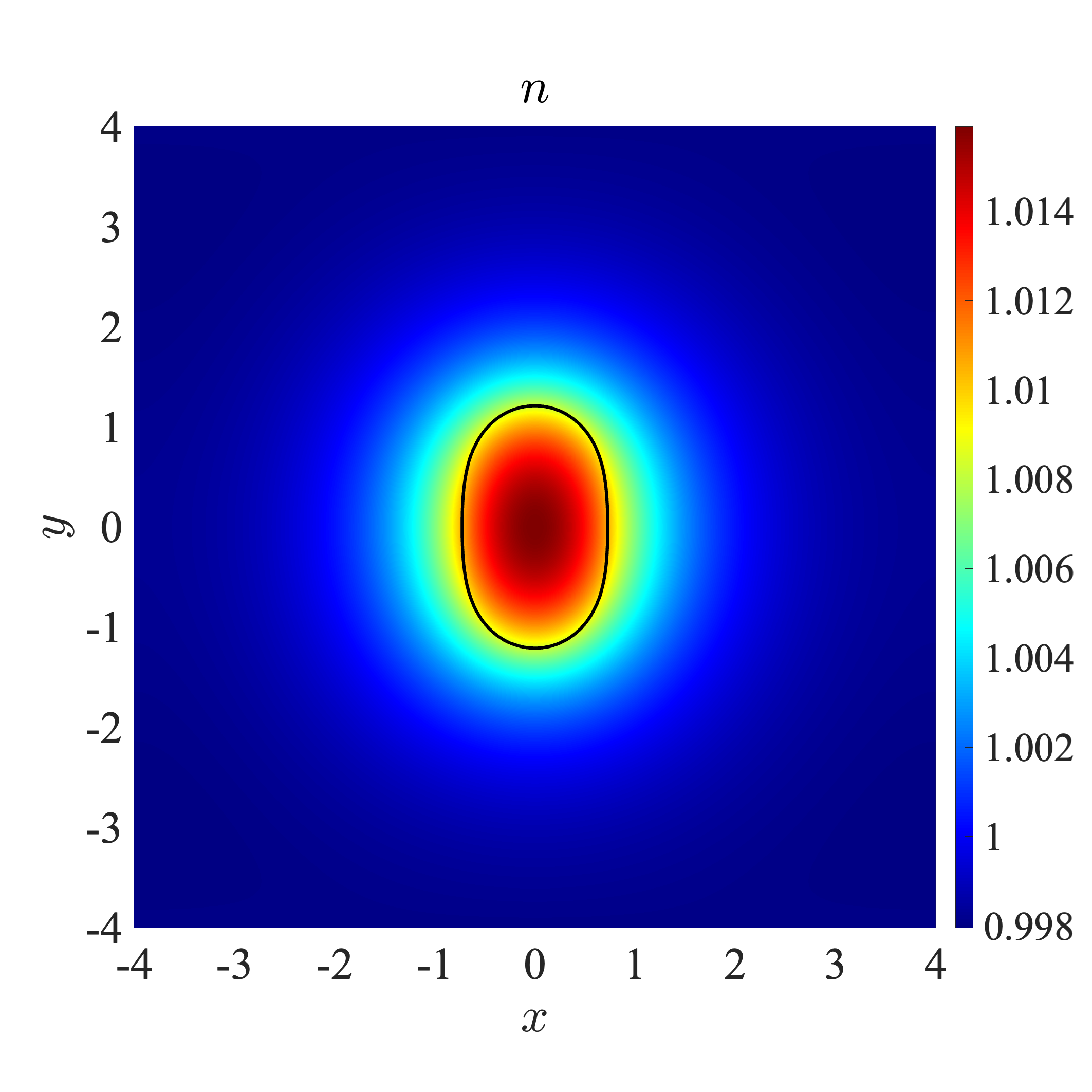}
        \label{subfig:1DropD0N0Pump25N10}
		}
    \hskip -0.3cm
	\subfloat[$\phi~(t=10)$]{
		\includegraphics[width=0.3\linewidth]{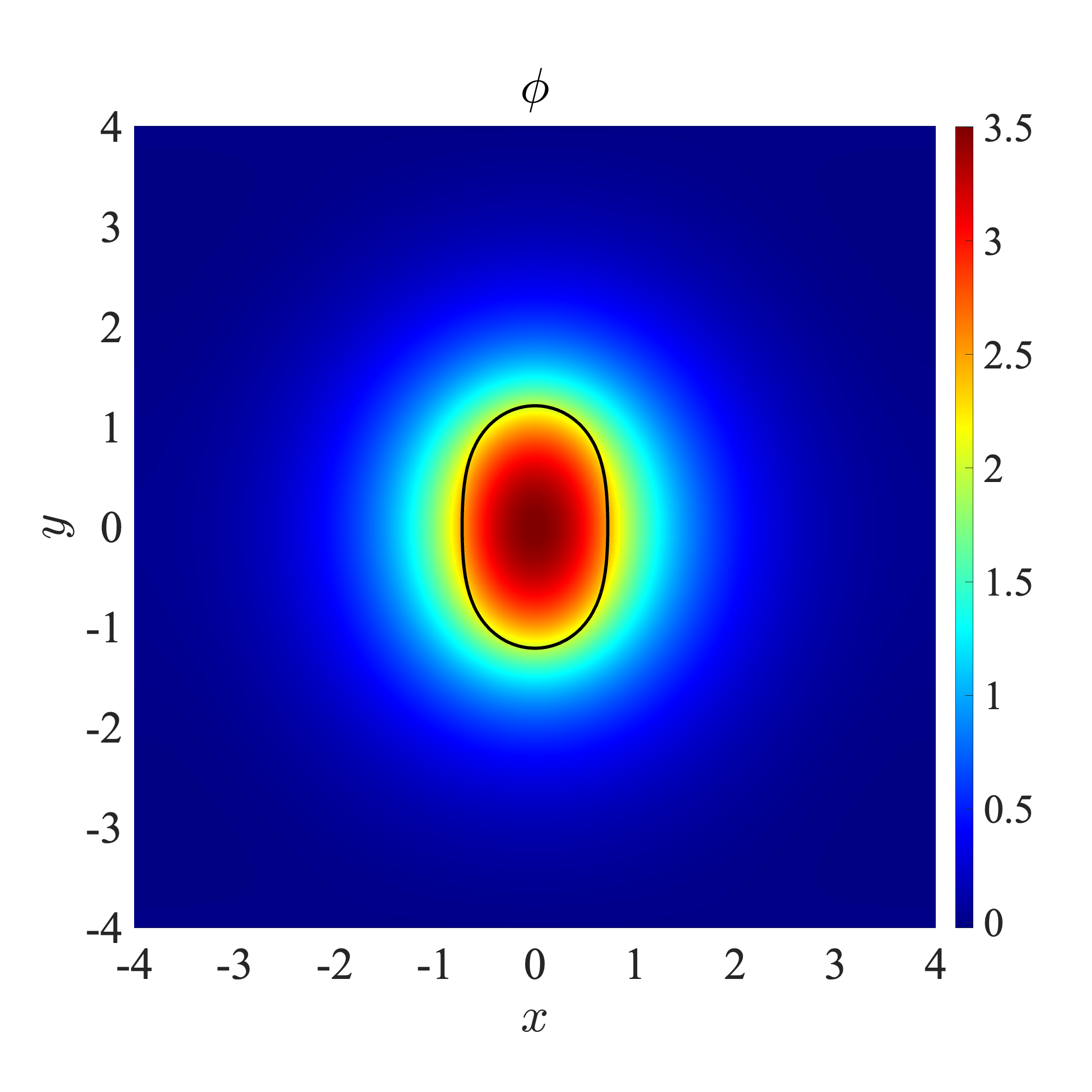}
        \label{subfig:1DropD0N0Pump25Phi10}
		}
    \\
    \vskip -0.3cm
	\subfloat[$p+n~(t=10)$]{
		\centering
		\includegraphics[width=0.3\linewidth]{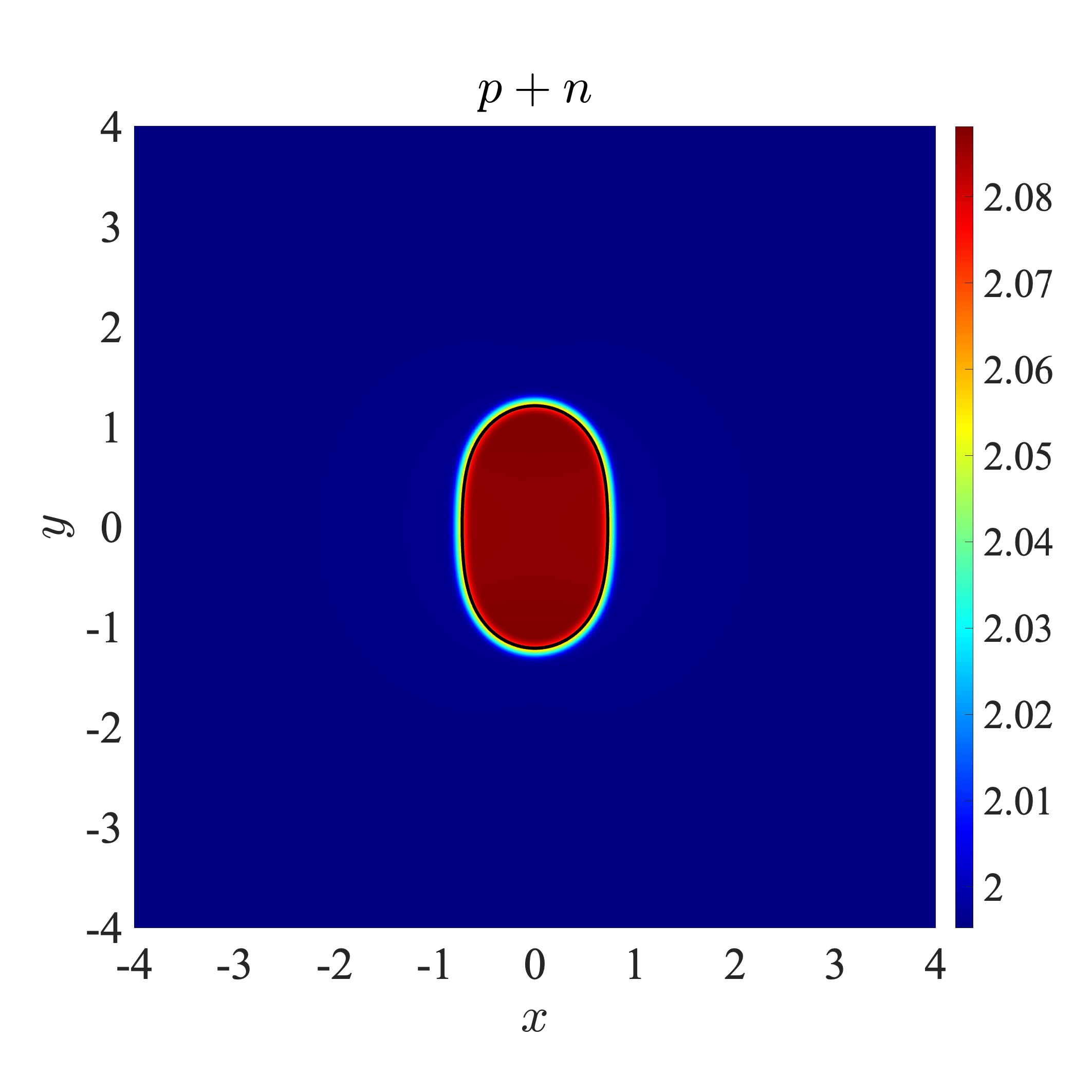}
        \label{subfig:1DropD0N0Pump25Sum10}
		} 
    \hskip -0.3cm
	\subfloat[$p-n~(t=10)$]{
		\centering
		\includegraphics[width=0.3\linewidth]{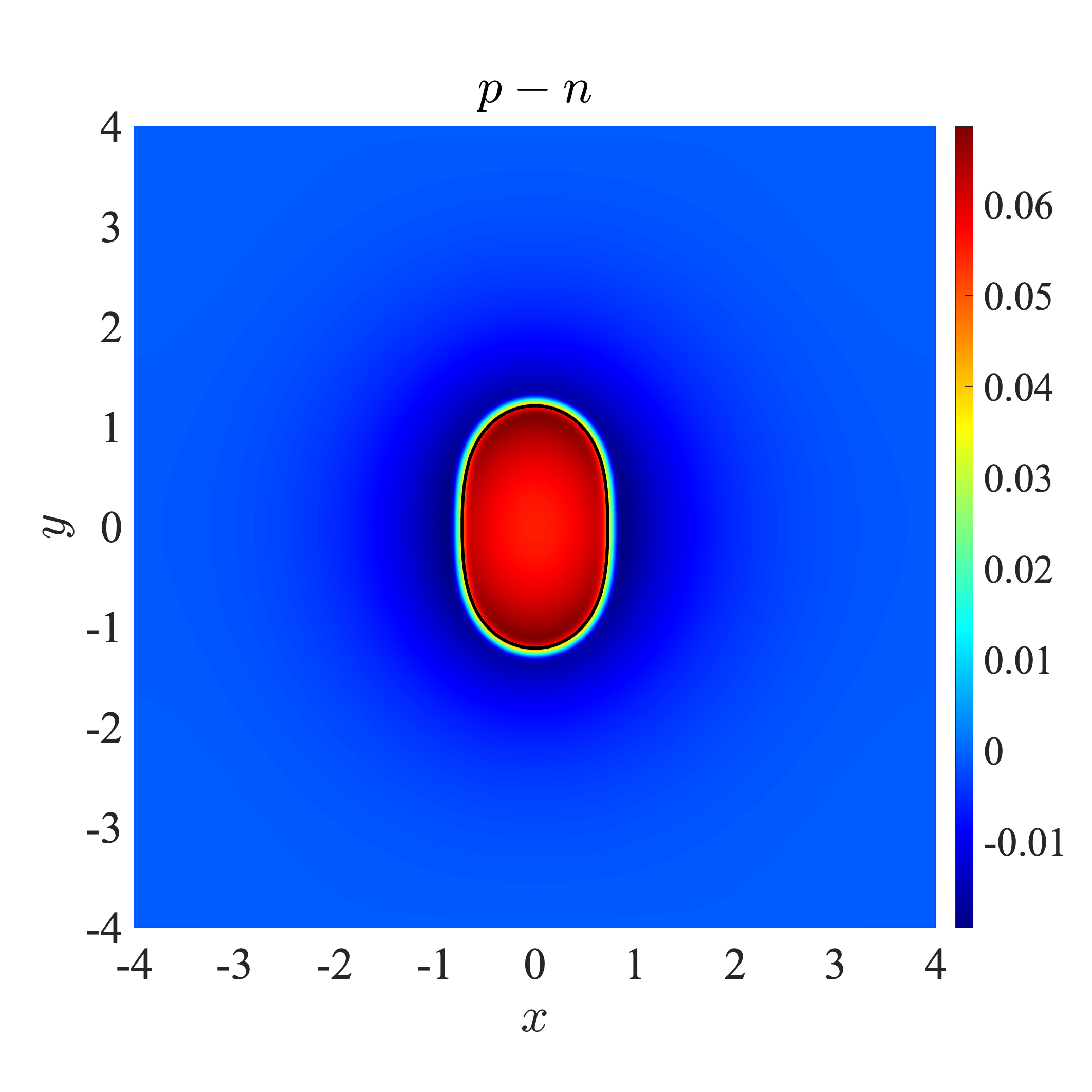}
        \label{subfig:1DropD0N0Pump25Dif10}
	}
    \hskip -0.3cm
	    \subfloat[$\bm{u}~(t=10)$]{
		\centering
		\includegraphics[width=0.3\linewidth]{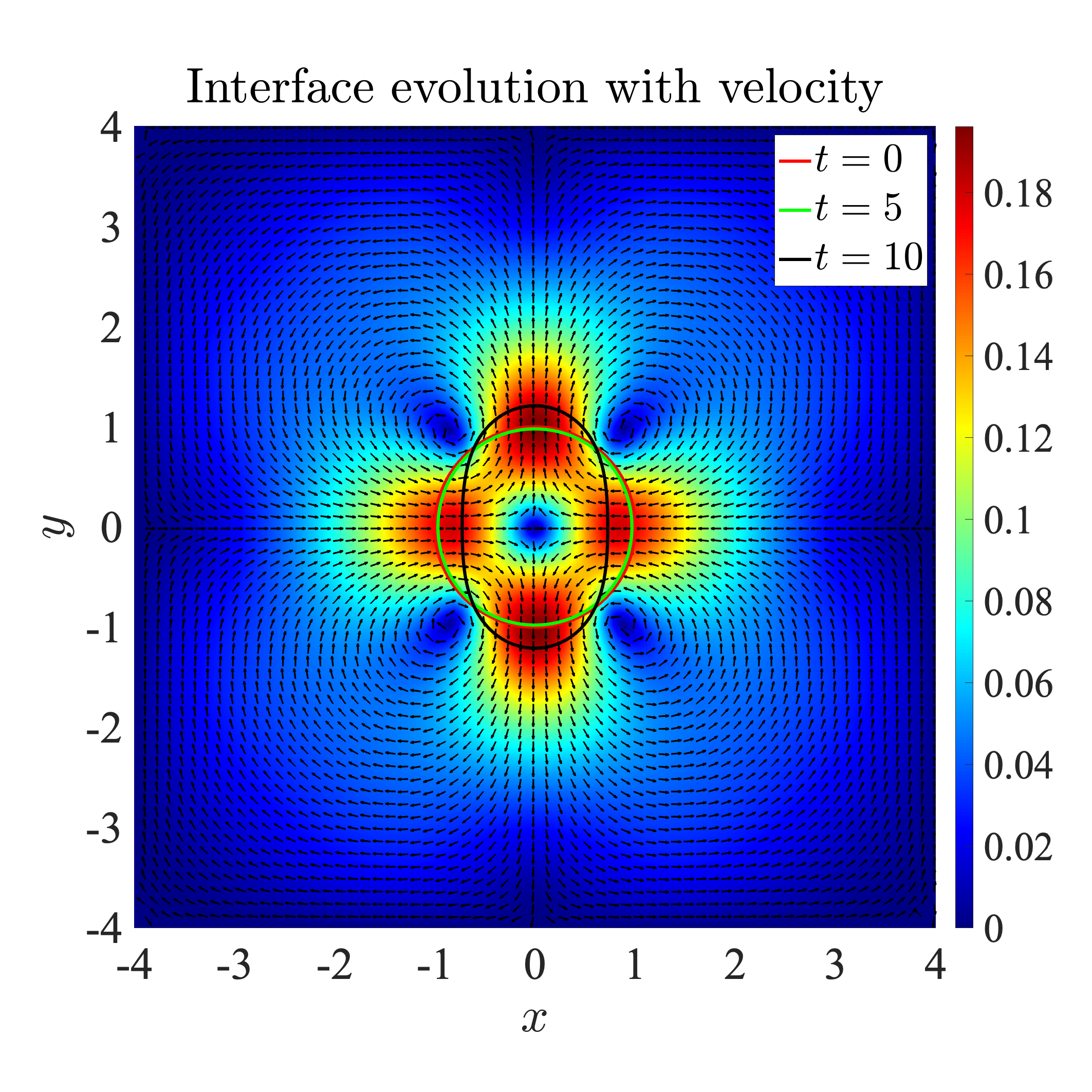}
        \label{subfig:1DropD0N0Pump25U10}
		}
    \hskip -0.2cm
	\caption{The snapshots for the drop deformation with positive ion pump at $t=10$. 
    Top row: (left) positive ion; (middle) negative ion; (right) potential; 
    Bottom row: the total charge (left), net charge (middle), and the velocity (right). 
    The black solid circle represents the location of the drop, 
    which is denoted by the level set $\psi=0$. 
    The concentration and electric potential distribution are shown on the color map.  
    The bottom and upper plates are grounded. 
    }\label{fig:1DropD0N0Pump25}
\end{figure}

We first consider the case in which both the upper and lower boundaries are grounded, i.e., $\phi_{0l} = \phi_{0u} = 0$. In the absence of active pumping, the ion and potential distributions remain unchanged. However, when pumps with strength $I_0 = 25$ are activated along the interface, additional positive ions are transported from the exterior into the droplet, which subsequently attracts negative ions via electrostatic forces, as shown in Fig.~\ref{fig:1DropD0N0Pump25}. This inward accumulation of charge elevates the electric potential within the droplet (Fig.~\ref{subfig:1DropD0N0Pump25Phi10}), establishing an electric field directed from the droplet interior toward the exterior.

Due to the clamped Dirichlet boundary conditions in the $y$-direction and Neumann boundary conditions in the $x$-direction, the induced electric field is stronger along the $y$-axis. This results in a larger Lorentz force in the vertical direction, as demonstrated in Appendix Fig.~\ref{fig:1DropD0N0bdPump25Eforce}. Consequently, the droplet undergoes vertical elongation, as shown in Fig.~\ref{subfig:1DropD0N0Pump25U10}.

\begin{figure}[!ht] 
\vskip -0.4cm
\centering
	\subfloat[$p~(t=10)$]{
		\includegraphics[width=0.33\linewidth]{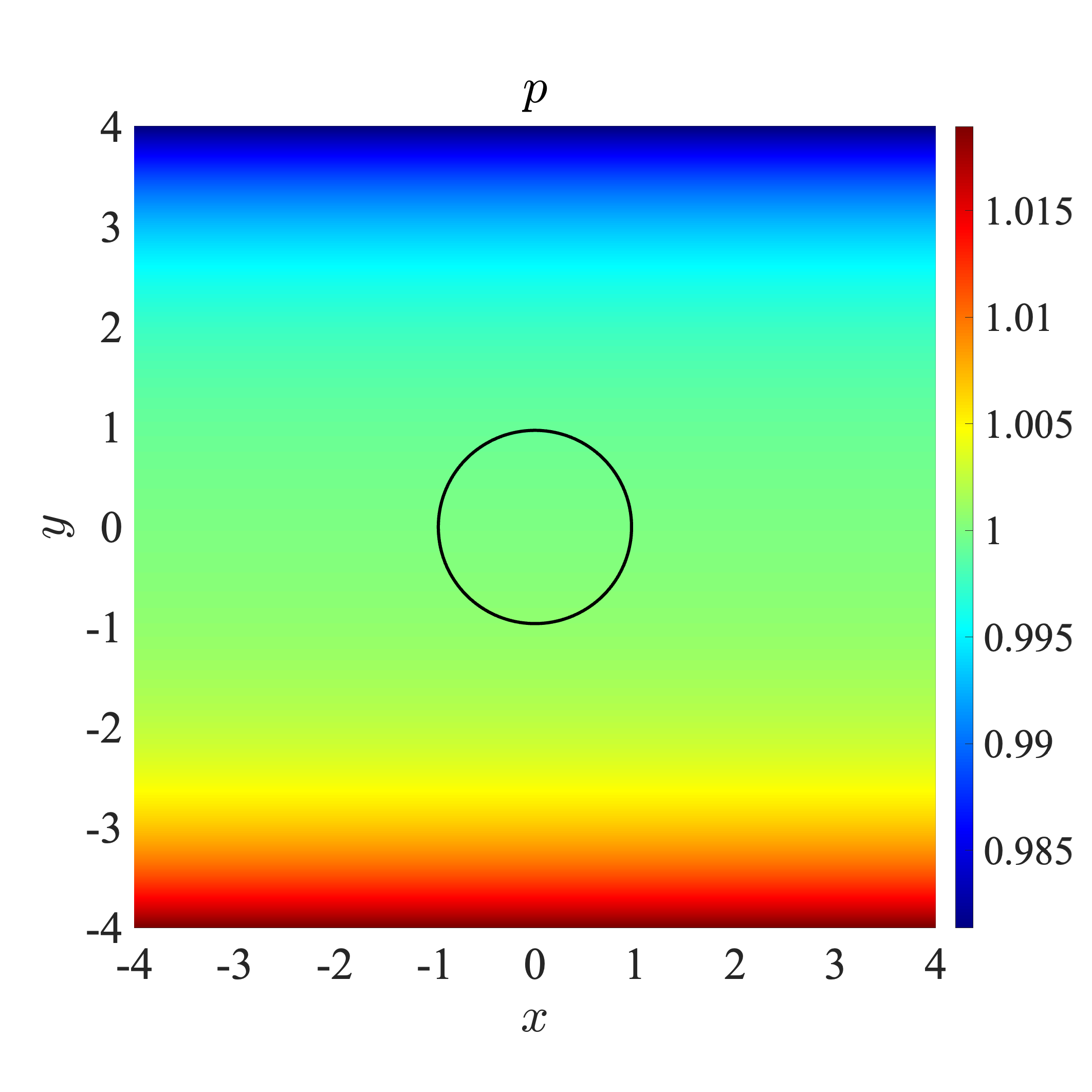}
        \label{subfig:1DropNoPumpP10}
		}
    \hskip -0.3cm
	\subfloat[$n~(t=10)$]{
		\includegraphics[width=0.33\linewidth]{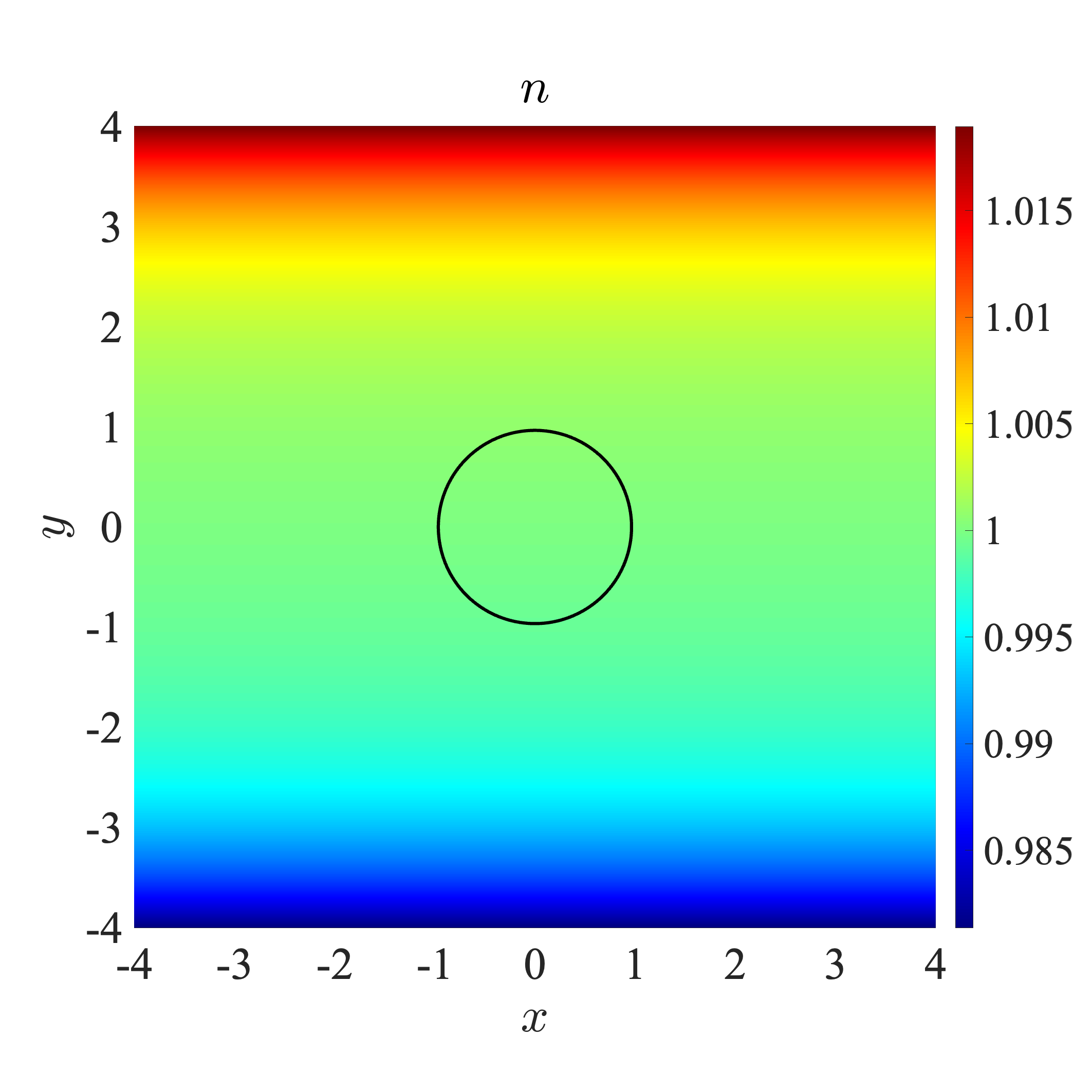}
        \label{subfig:1DropNoPumpN10}
		}
    \hskip -0.3cm
	\subfloat[$\phi~(t=10)$]{
		\includegraphics[width=0.33\linewidth]{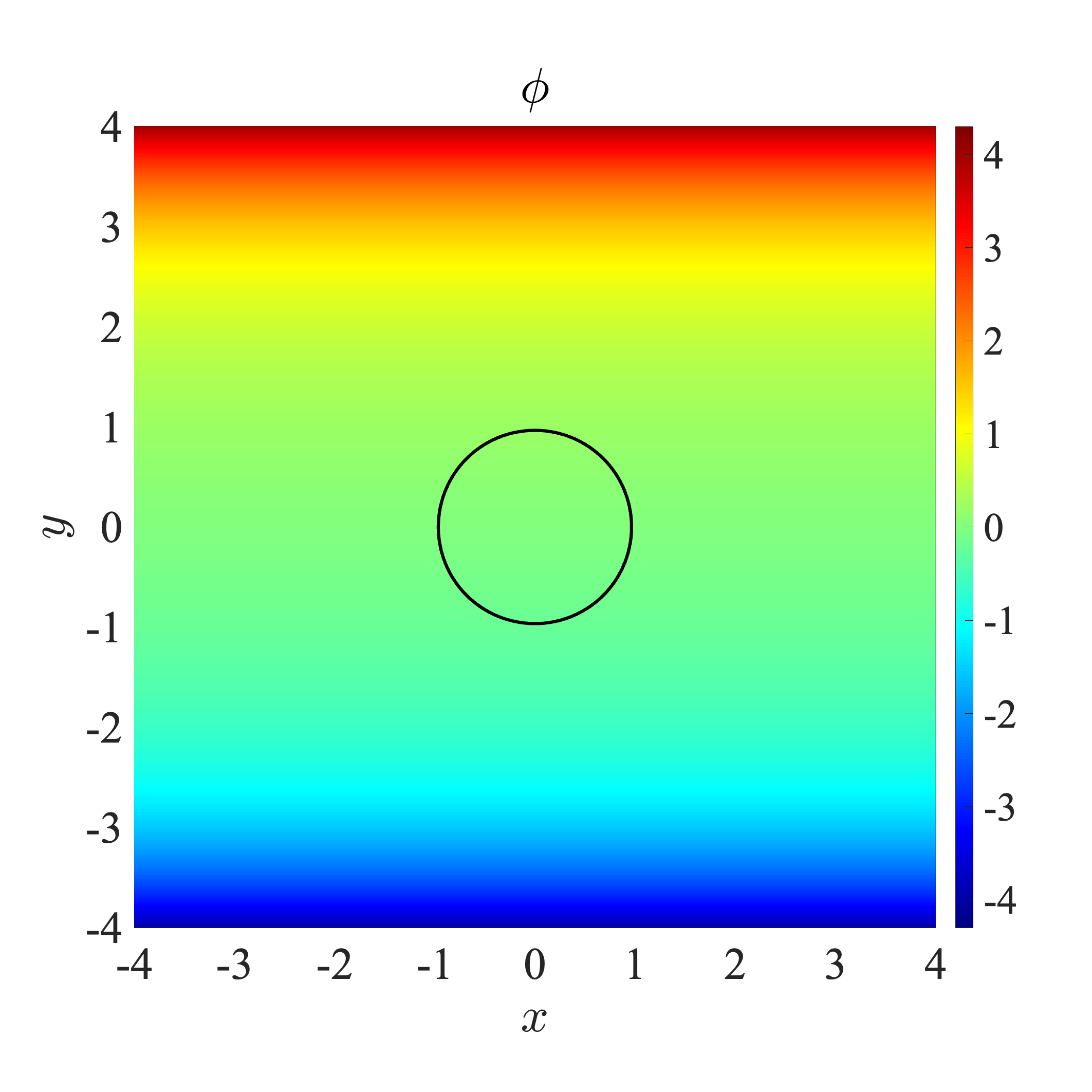}
        \label{subfig:1DropNoPumpPhi10}
		}
    \hskip -0.2cm
	\caption{The snapshots for the drop deformation without ion pump. 
    The black solid circle represents the location of the drop, 
    which is denoted by the level set $\psi=0$. 
    The concentration and electric potential distribution at the final time, chosen as $t=10$, 
    are shown on the color map.  Here $\phi_{0b} = -4, \quad \phi_{0u} = 4.$
    }\label{fig:1DropNoPump}
\end{figure}

Next, we apply an electric potential difference between the bottom and upper plates by setting
\[
\phi_{0b} = -4, \quad \phi_{0u} = 4.
\]
Simulation results for the case without active pumping are shown in Fig.~\ref{fig:1DropNoPump}. Under the imposed electric field, positive ions migrate toward the upper boundary, while negative ions move downward. The ionic distributions remain laterally uniform along the $x$-direction, as illustrated in Fig.~\ref{subfig:1DropNoPumpP10y0} and Fig.~\ref{subfig:1DropNoPumpN10y0} in Appendix~\ref{subsec:figs}. The net charge surrounding the droplet remains nearly zero, and the Lorentz force distribution (provided in the SI) is negligible. As a result, no appreciable deformation of the droplet is observed.

\begin{figure}[!ht]
    \vskip -0.4cm
    \centering
	\subfloat[$p~(t=3)$]{
		\includegraphics[width=0.16\linewidth]{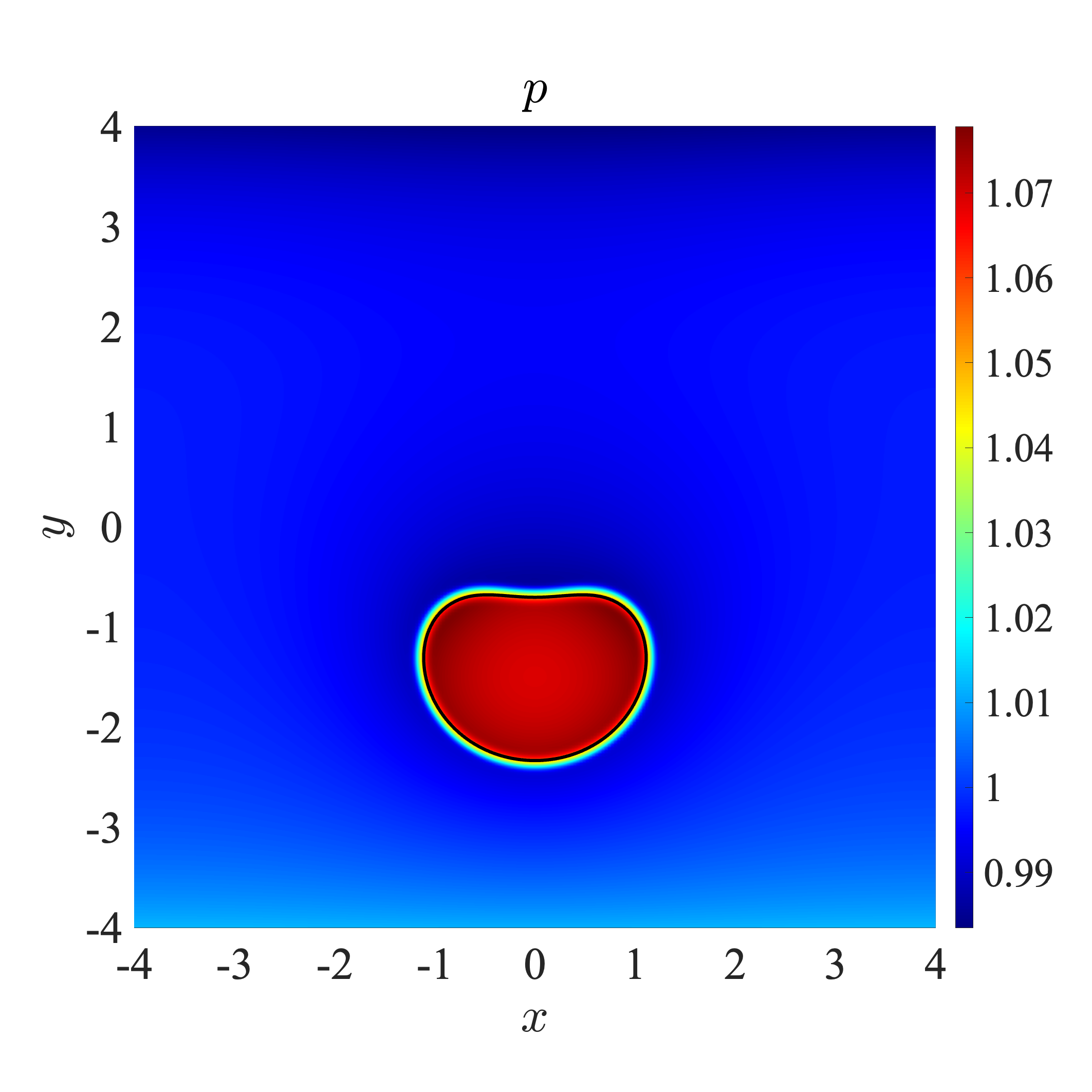}
        \label{subfig:1DropD4N0Pump25P3}
		}
    \hskip -0.3cm
    \subfloat[$p~(t=5)$]{
		\includegraphics[width=0.16\linewidth]{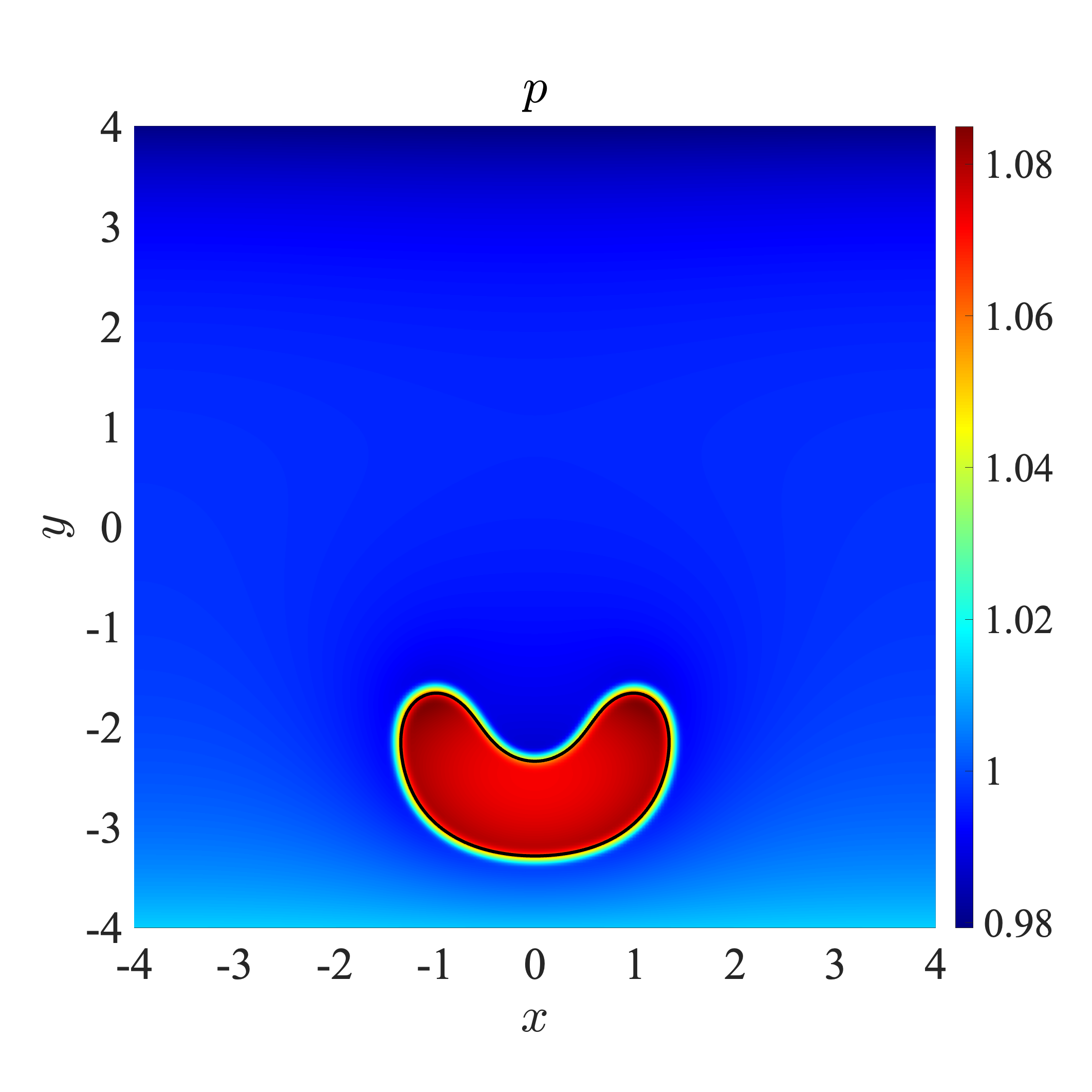}
        \label{subfig:1DropD4N0Pump25P5}
		}
    \hskip -0.3cm
	\subfloat[$p~(t=8)$]{
		\includegraphics[width=0.16\linewidth]{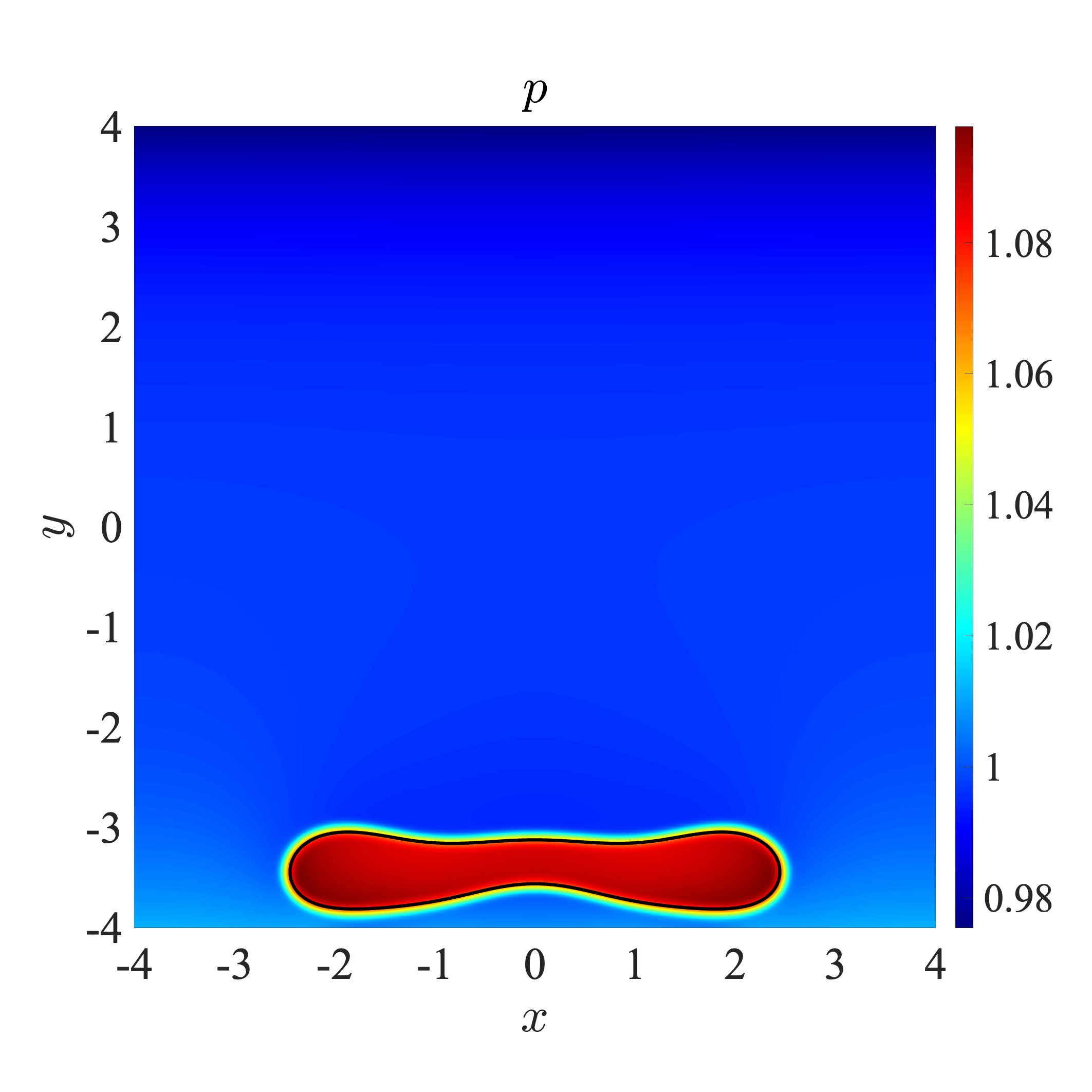}
        \label{subfig:1DropD4N0Pump25P8}
		}
    \hskip -0.3cm
        \subfloat[$p~(t=10)$]{
		\includegraphics[width=0.16\linewidth]{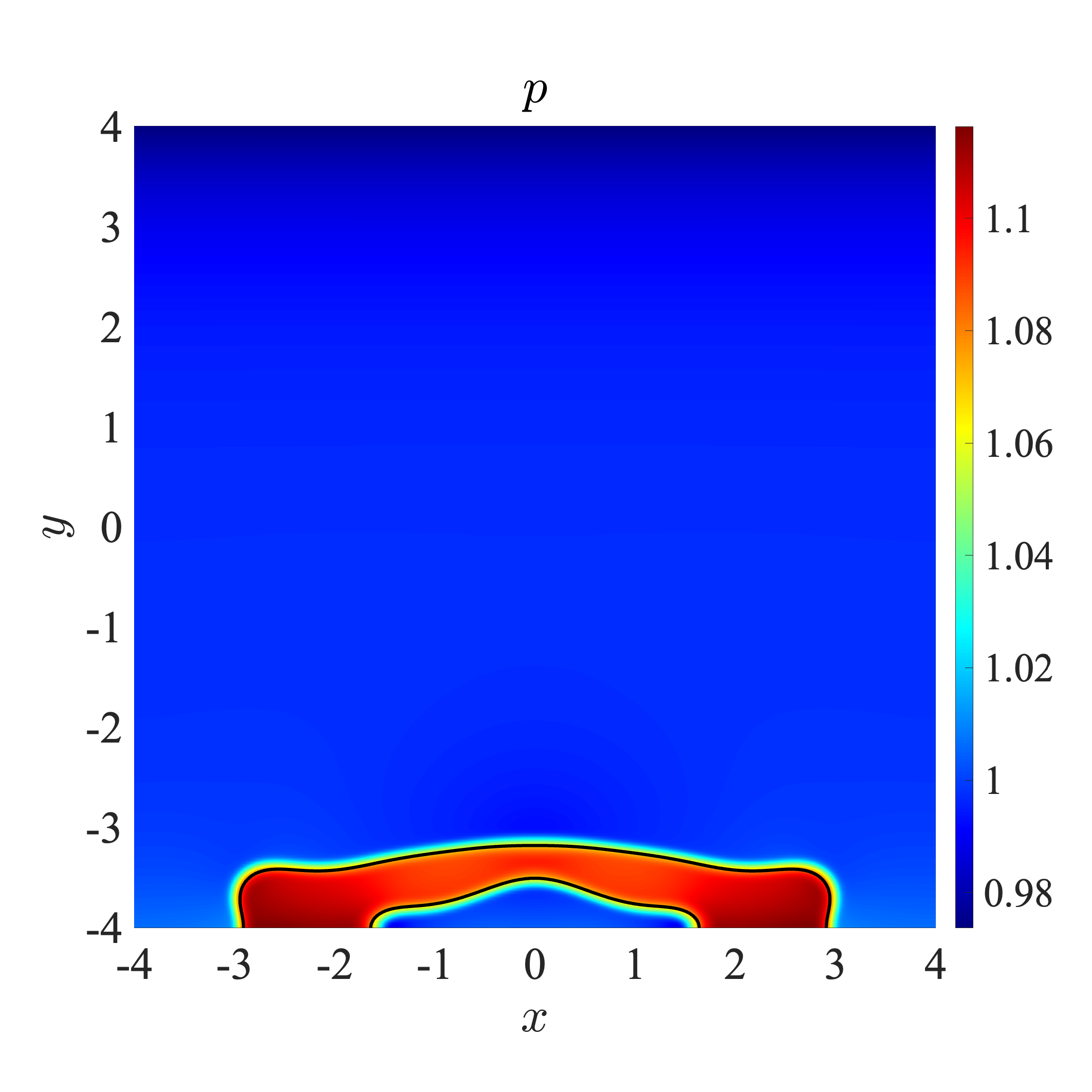}
        \label{subfig:1DropD4N0Pump25P10}
		}
    \hskip -0.3cm
        \subfloat[$p~(t=12)$]{
		\includegraphics[width=0.16\linewidth]{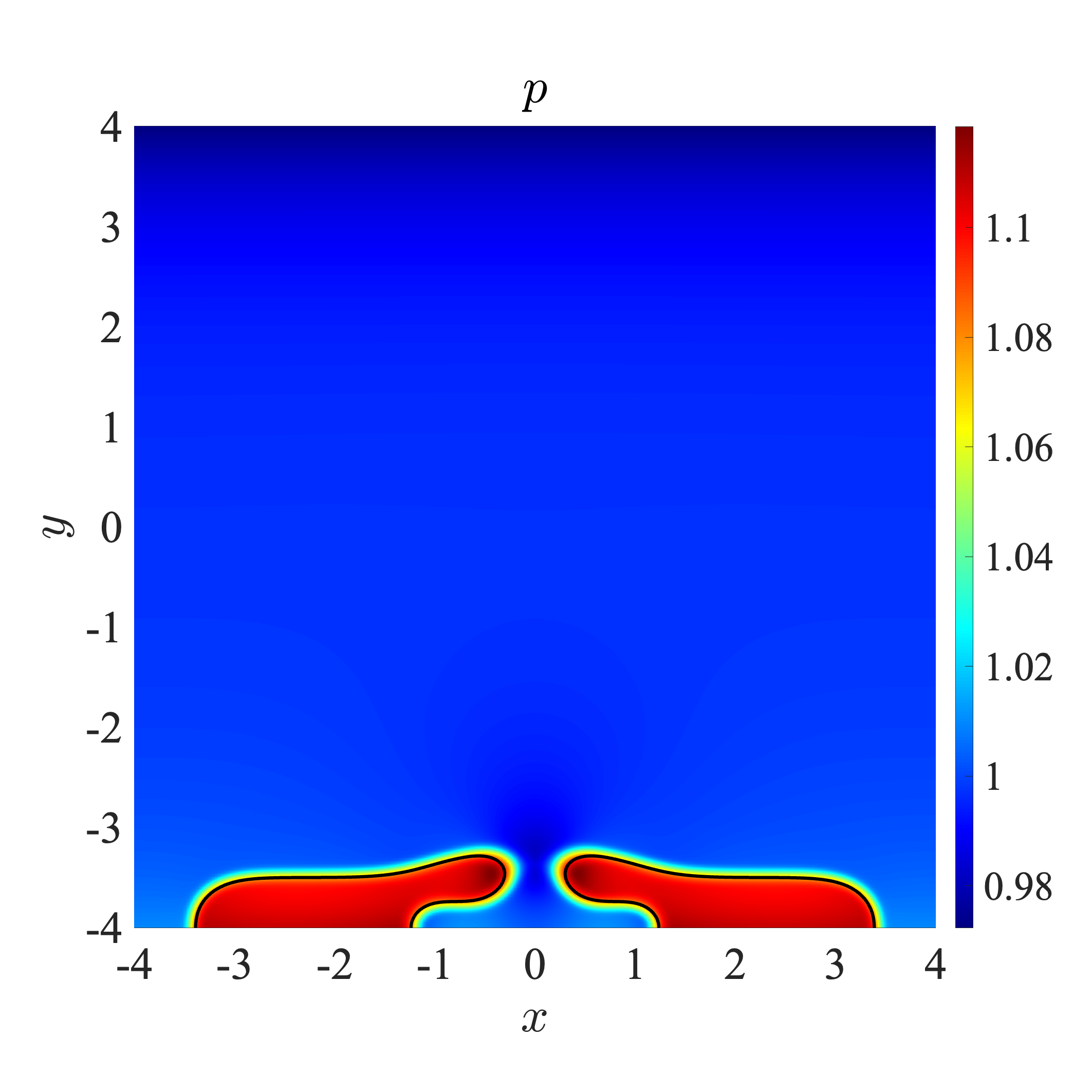}
        \label{subfig:1DropD4N0Pump25P12}
		}
    \hskip -0.3cm
        \subfloat[$p~(t=30)$]{
		\includegraphics[width=0.16\linewidth]{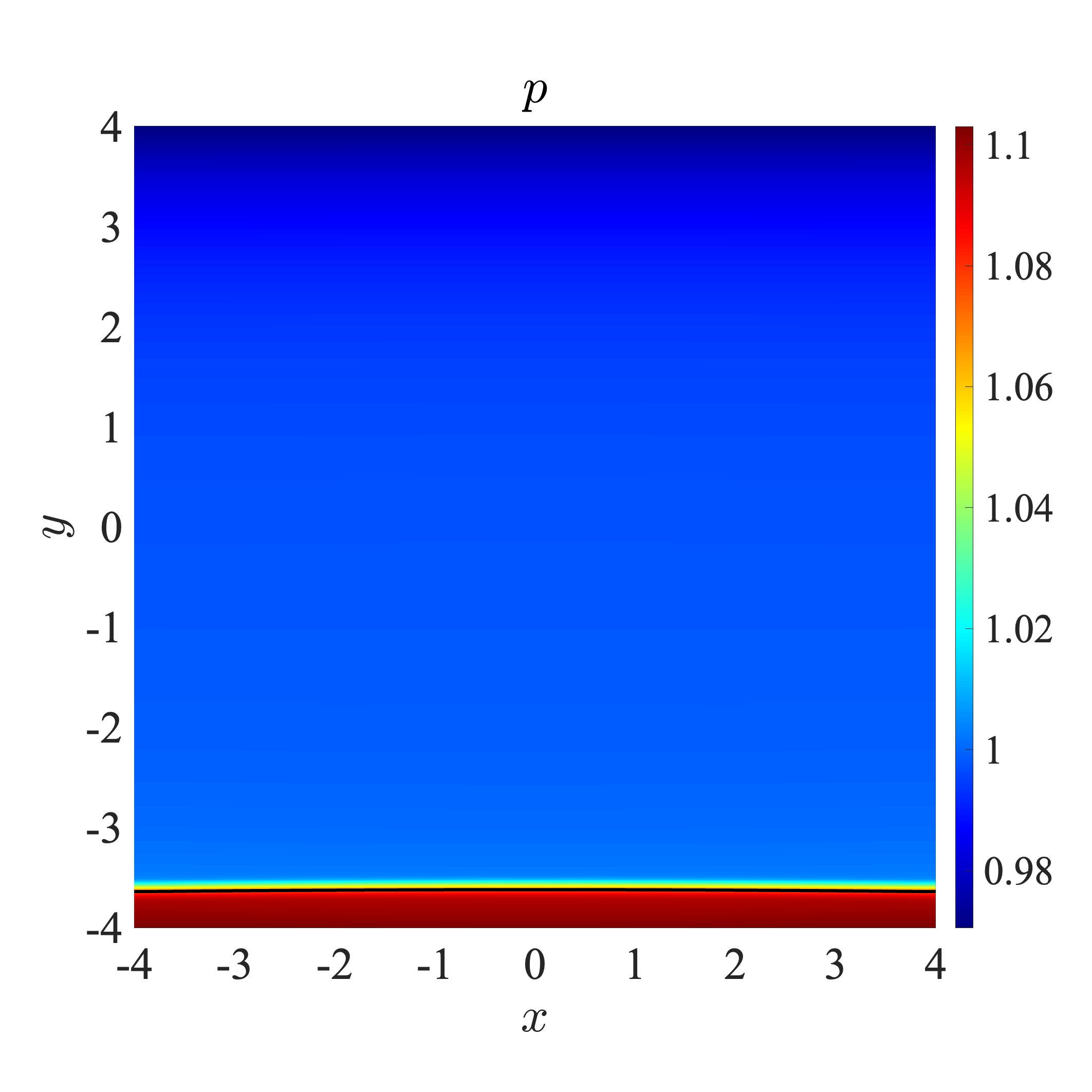}
        \label{subfig:1DropD4N0Pump25P30}
		}
        \\
        \vskip -0.3cm
    \subfloat[$n~(t=3)$]{
		\includegraphics[width=0.16\linewidth]{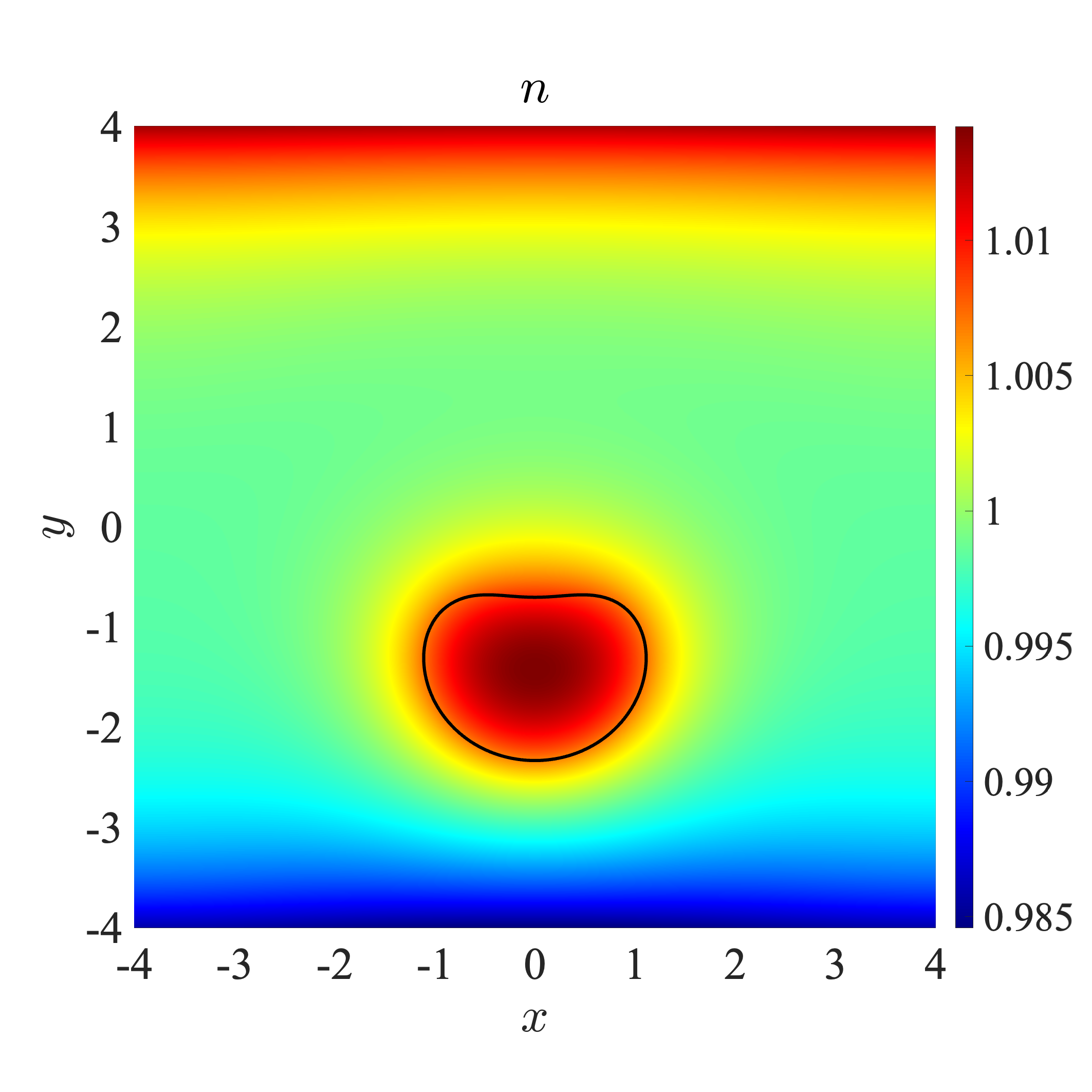}
        \label{subfig:1DropD4N0Pump25N3}
		}
    \hskip -0.3cm
	\subfloat[$n~(t=5)$]{
		\includegraphics[width=0.16\linewidth]{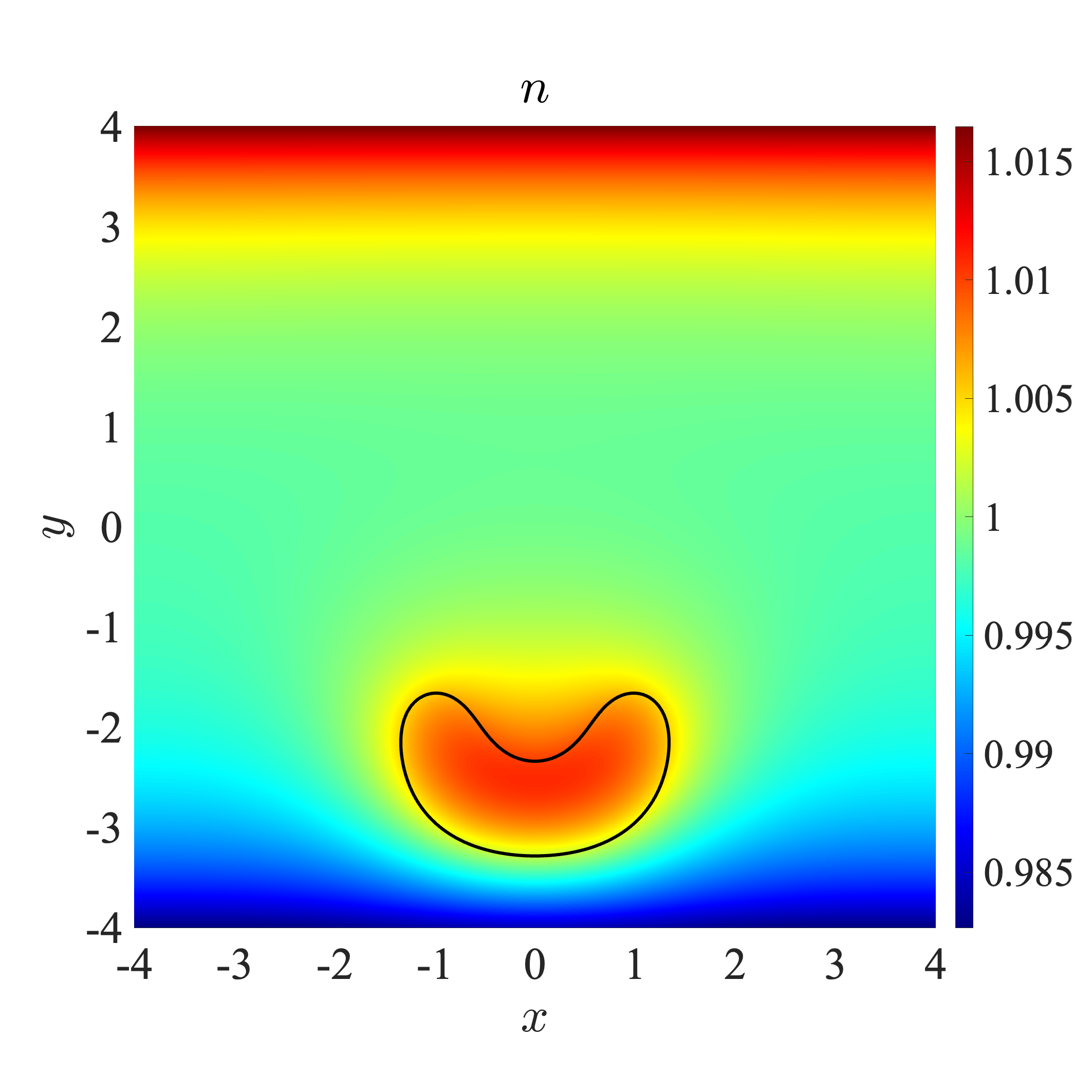}
        \label{subfig:1DropD4N0Pump25N5}
		}
    \hskip -0.3cm
	\subfloat[$n~(t=8)$]{
		\includegraphics[width=0.16\linewidth]{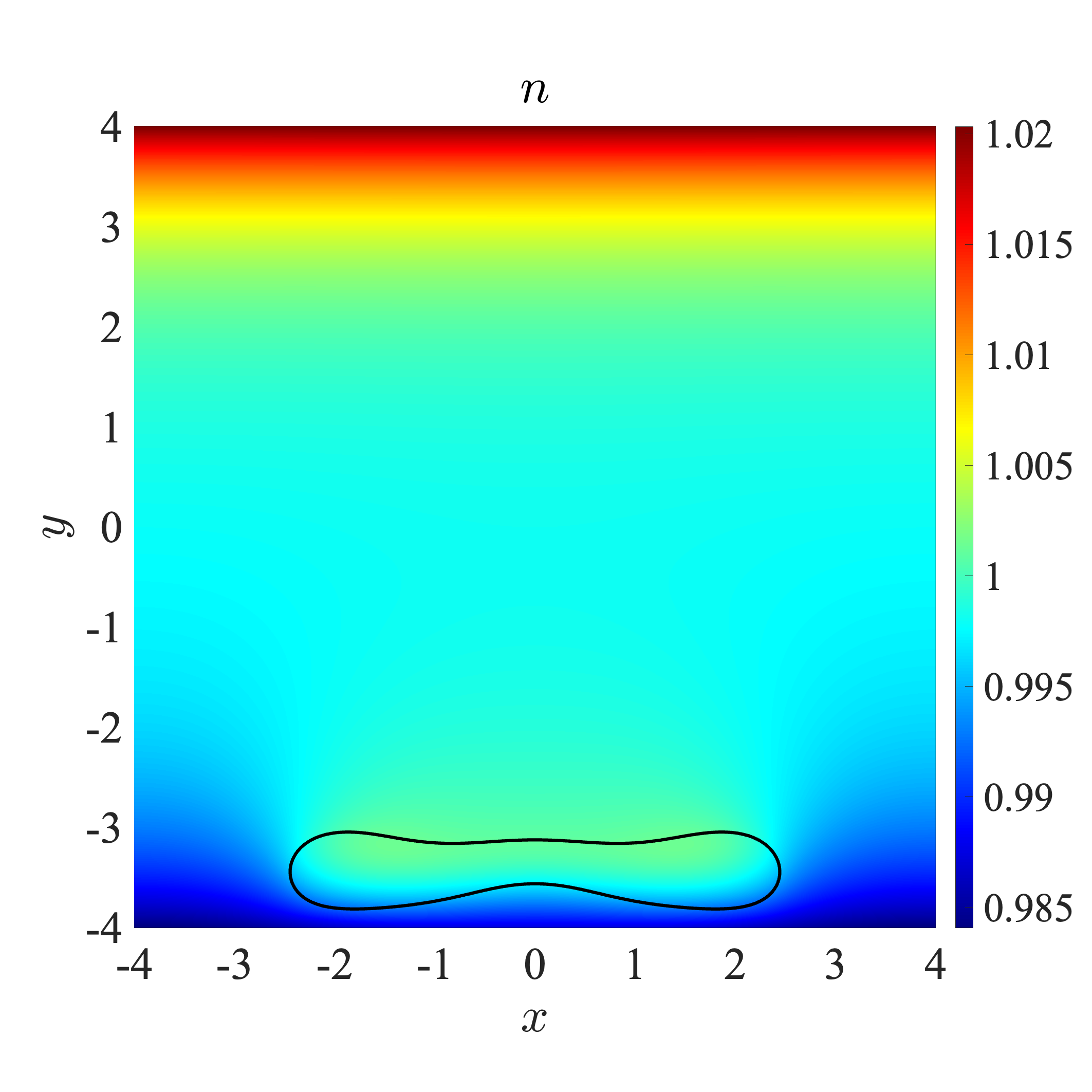}
        \label{subfig:1DropD4N0Pump25N8}
		}
    \hskip -0.3cm
	\subfloat[$n~(t=10)$]{
		\includegraphics[width=0.16\linewidth]{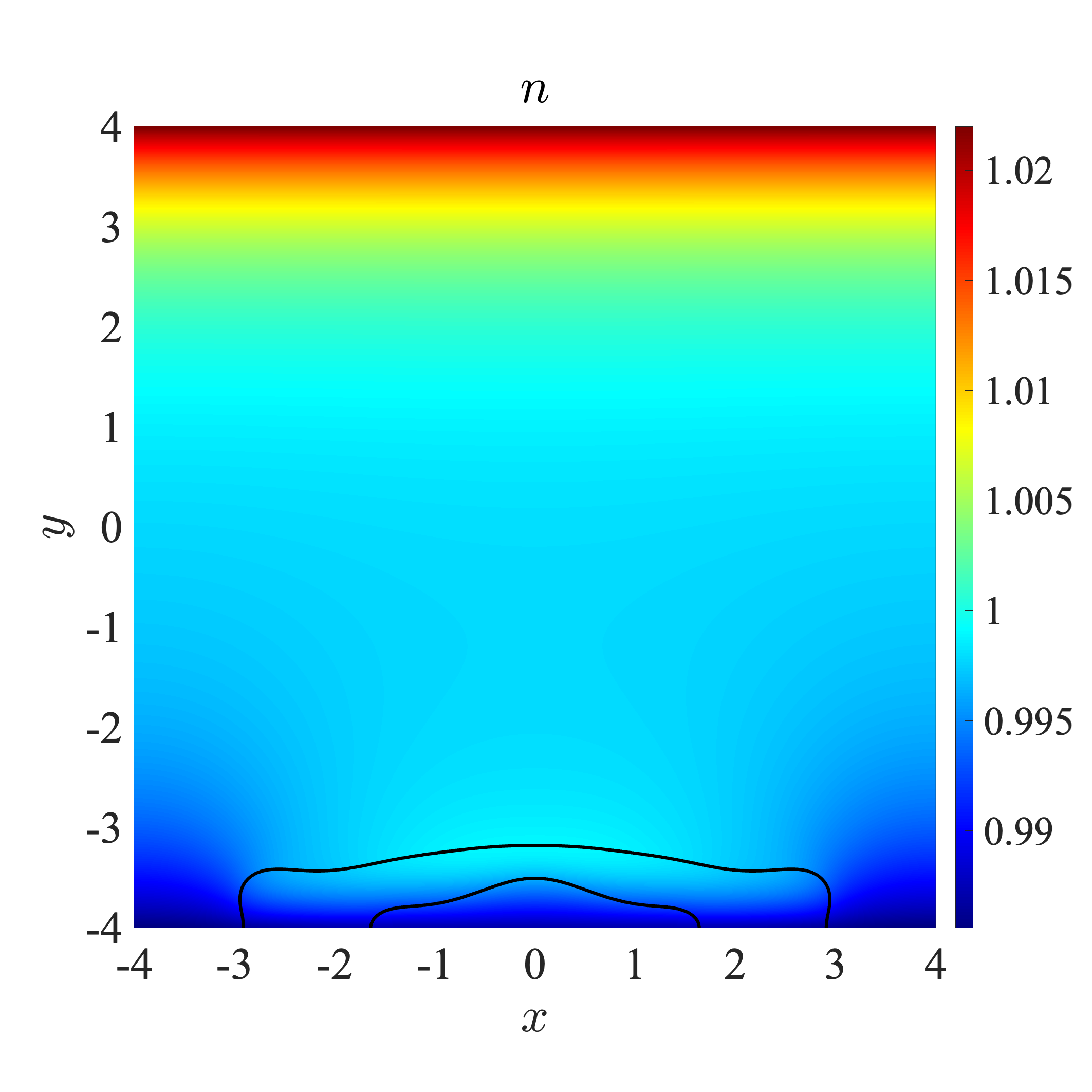}
        \label{subfig:1DropD4N0Pump25N10}
		}
    \hskip -0.3cm
	\subfloat[$n~(t=12)$]{
		\includegraphics[width=0.16\linewidth]{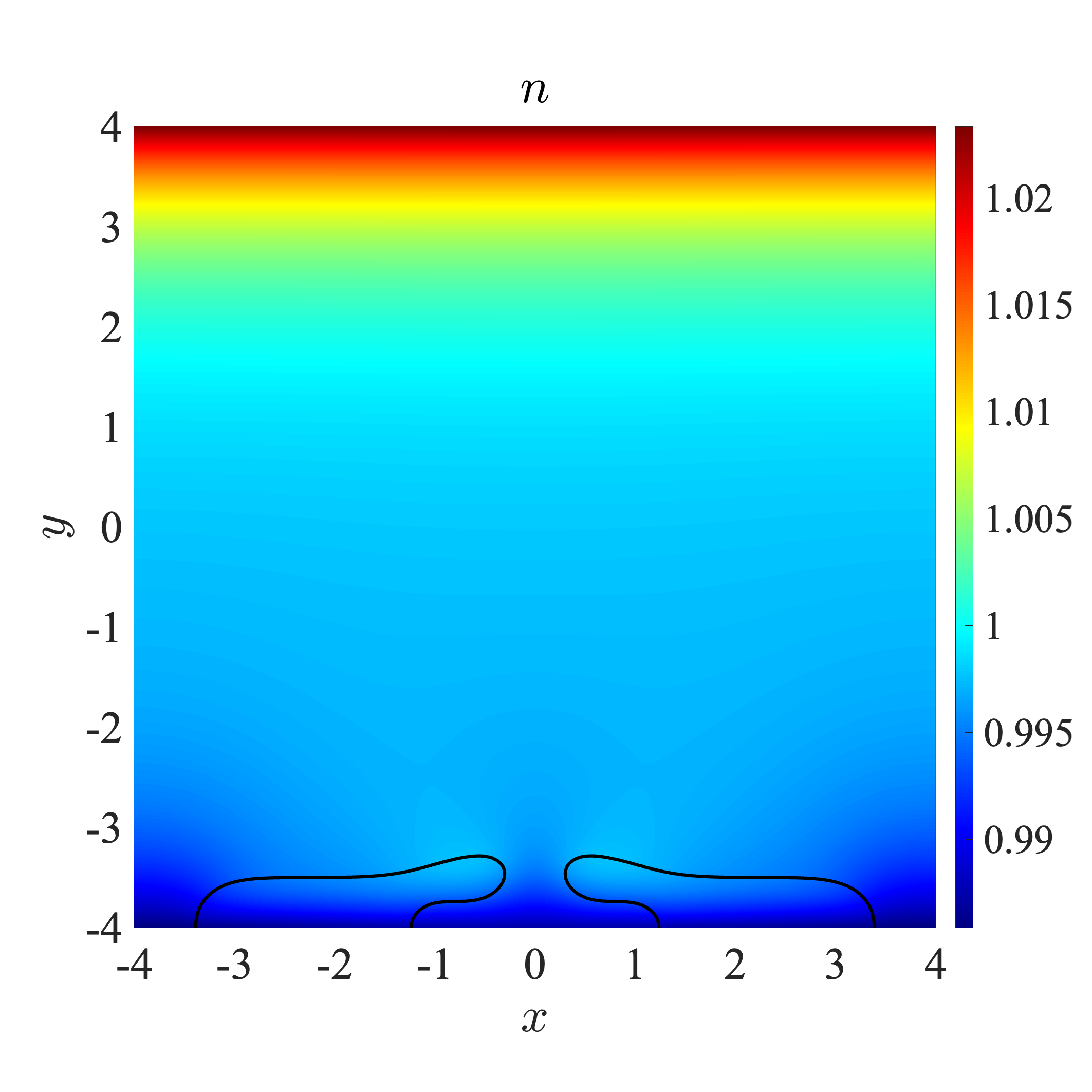}
        \label{subfig:1DropD4N0Pump25N12}
		}
    \hskip -0.3cm
	\subfloat[$n~(t=30)$]{
		\includegraphics[width=0.16\linewidth]{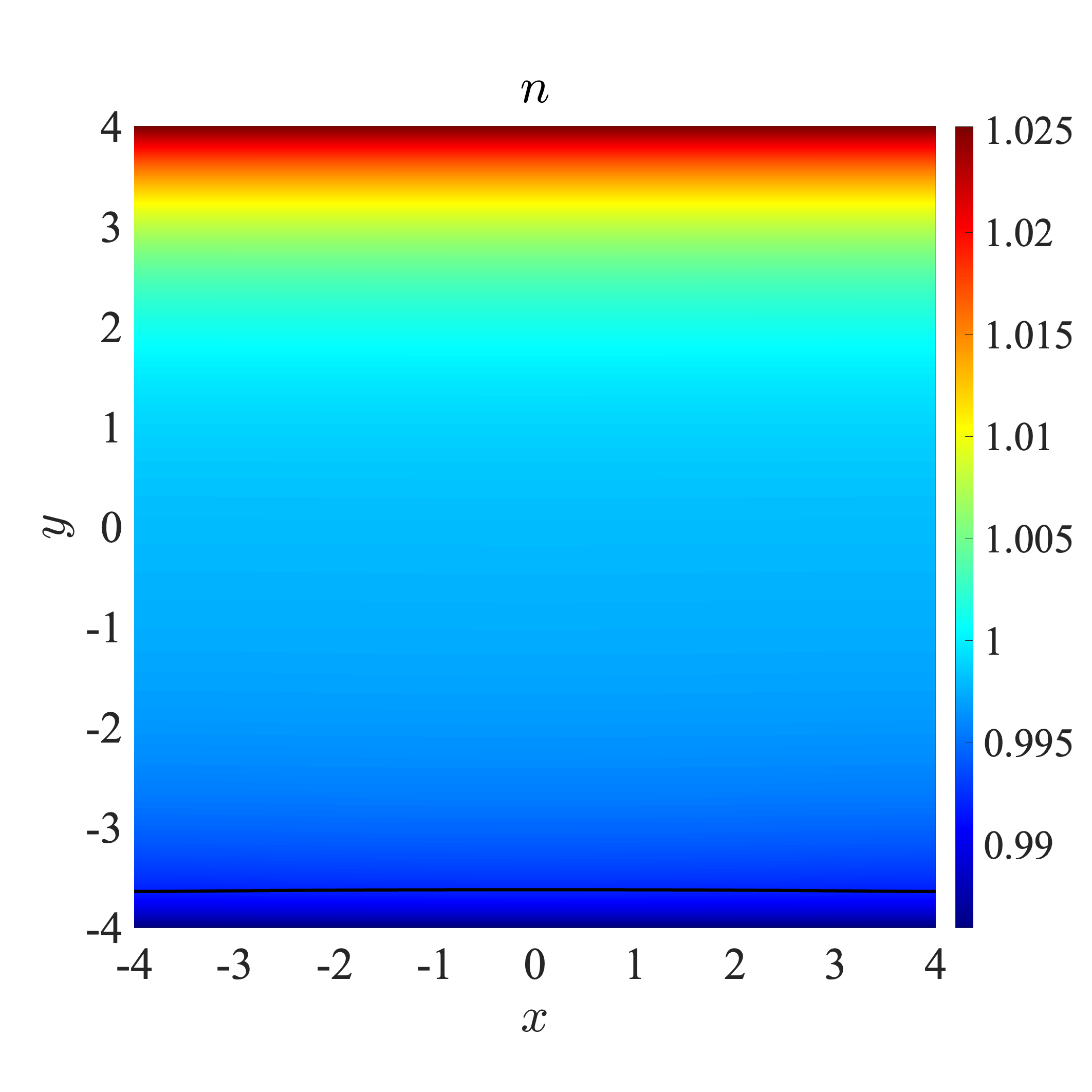}
        \label{subfig:1DropD4N0Pump25N30}
		}
        \\ 
        \vskip -0.3cm
	\subfloat[$\phi~(t=3)$]{
		\includegraphics[width=0.16\linewidth]{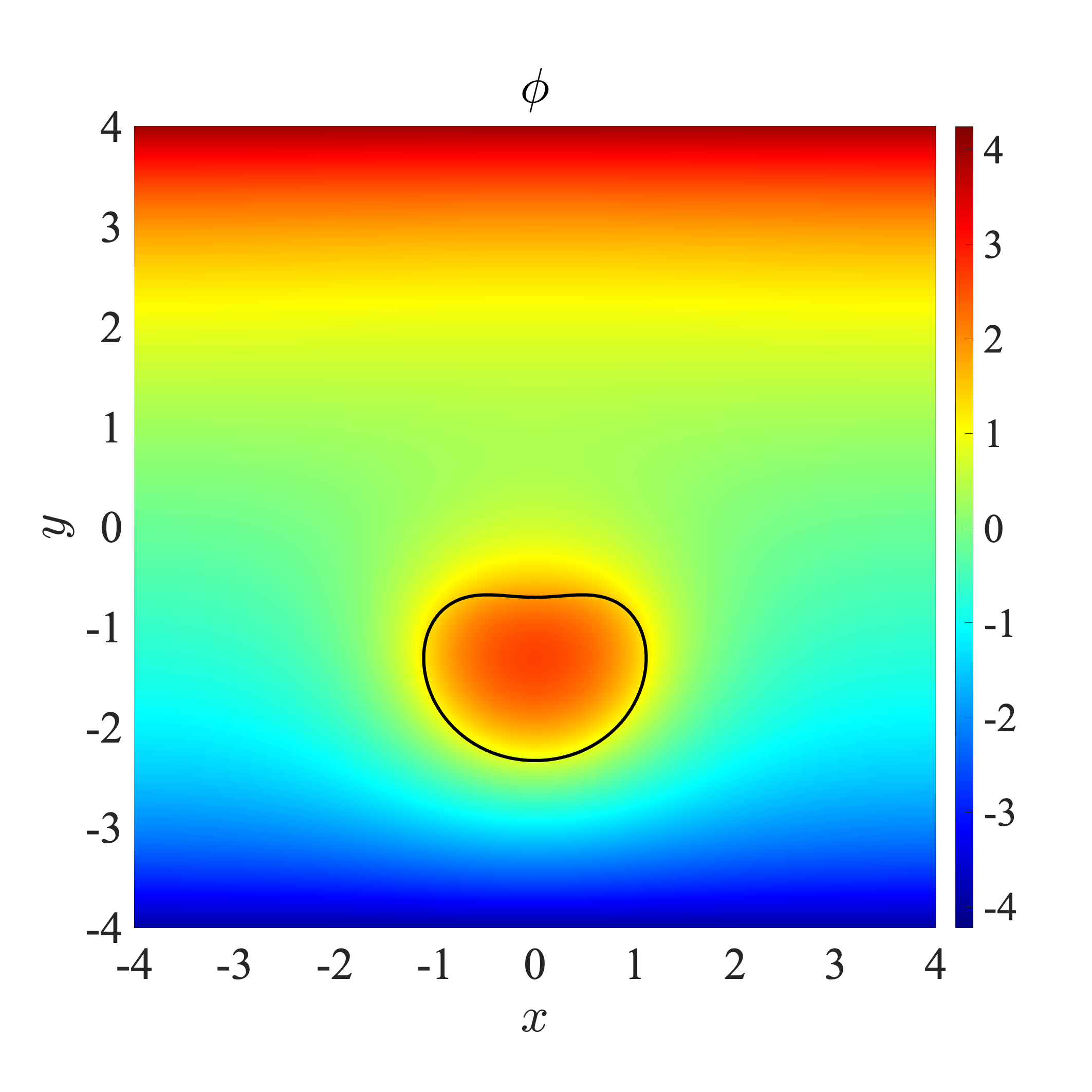}
        \label{subfig:1DropD4N0Pump25Phi3}
		}
    \hskip -0.3cm
    \subfloat[$\phi~(t=5)$]{
		\includegraphics[width=0.16\linewidth]{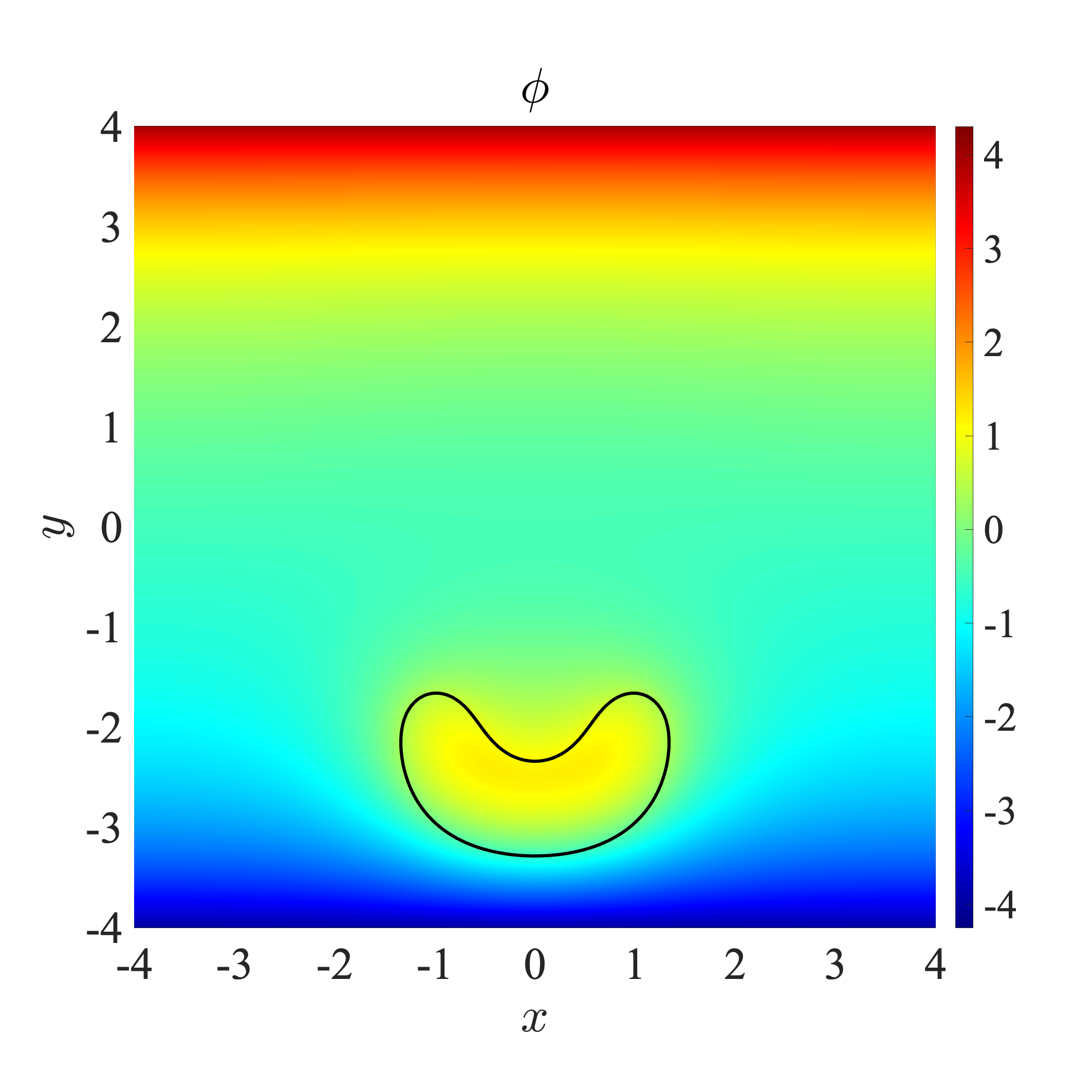}
        \label{subfig:1DropD4N0Pump25Phi5}
		}
    \hskip -0.3cm
	\subfloat[$\phi~(t=8)$]{
		\includegraphics[width=0.16\linewidth]{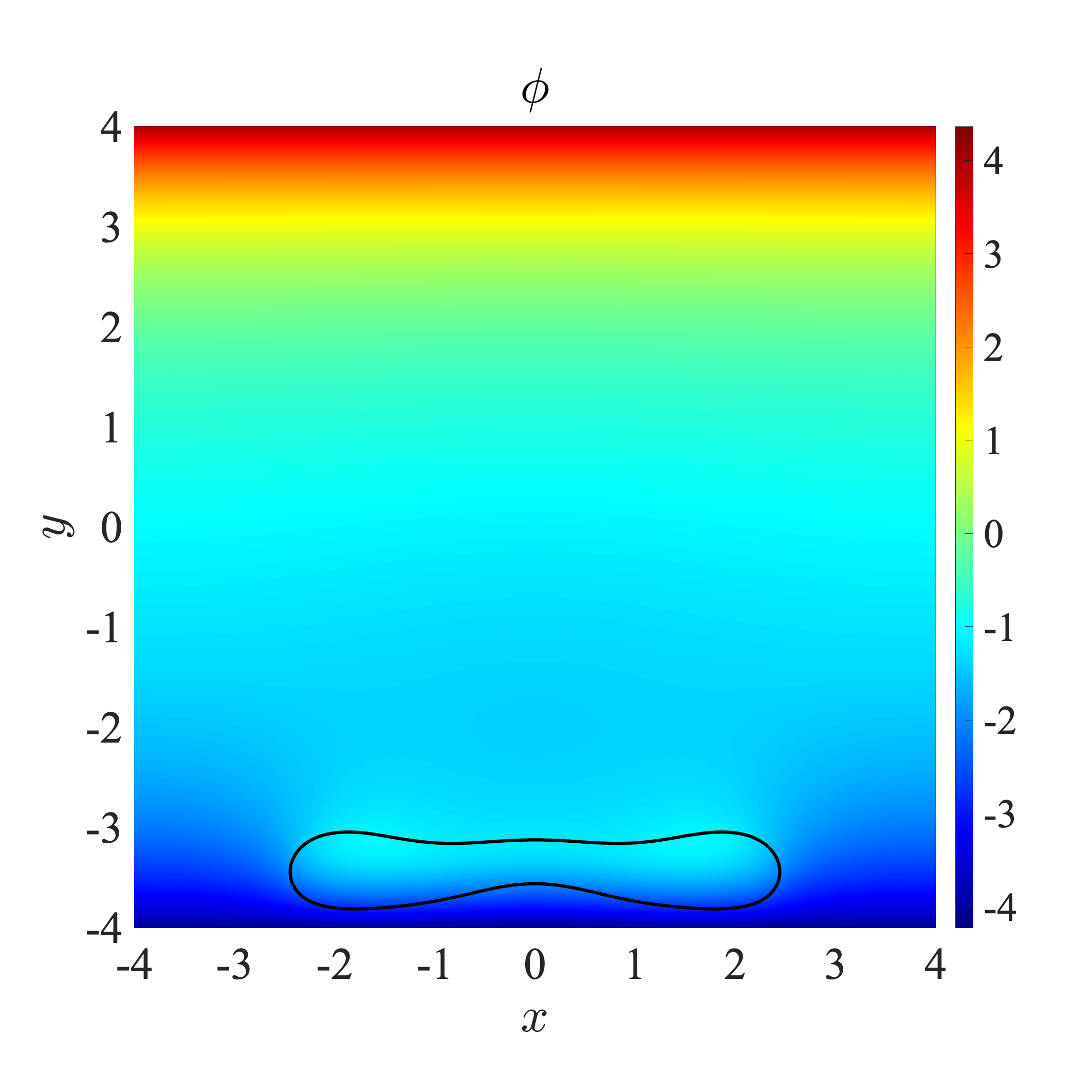}
        \label{subfig:1DropD4N0Pump25Phi8}
		}
    \hskip -0.3cm
	\subfloat[$\phi~(t=10)$]{
		\includegraphics[width=0.16\linewidth]{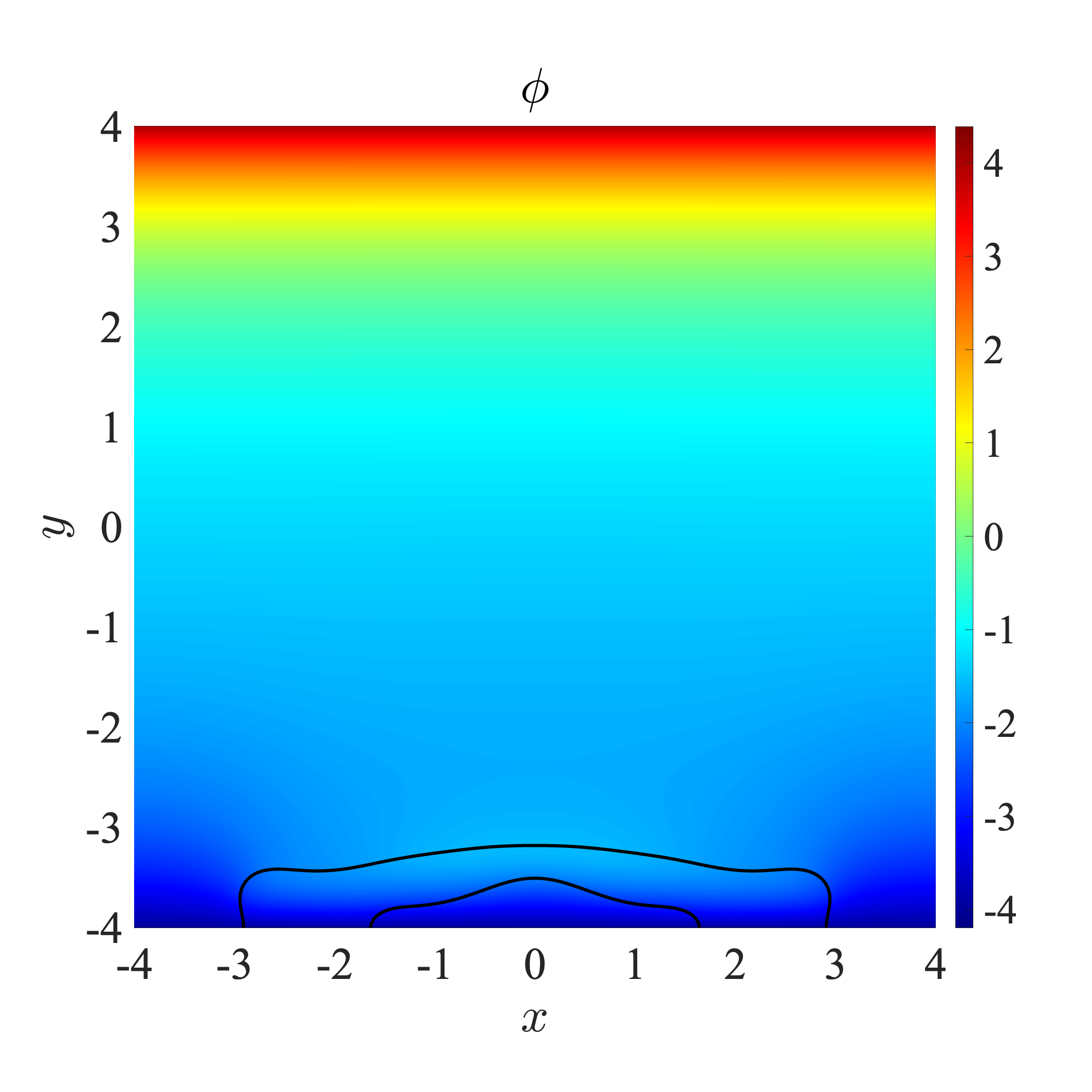}
        \label{subfig:1DropD4N0Pump25Phi10}
		}
    \hskip -0.3cm
    \subfloat[$\phi~(t=12)$]{
		\includegraphics[width=0.16\linewidth]{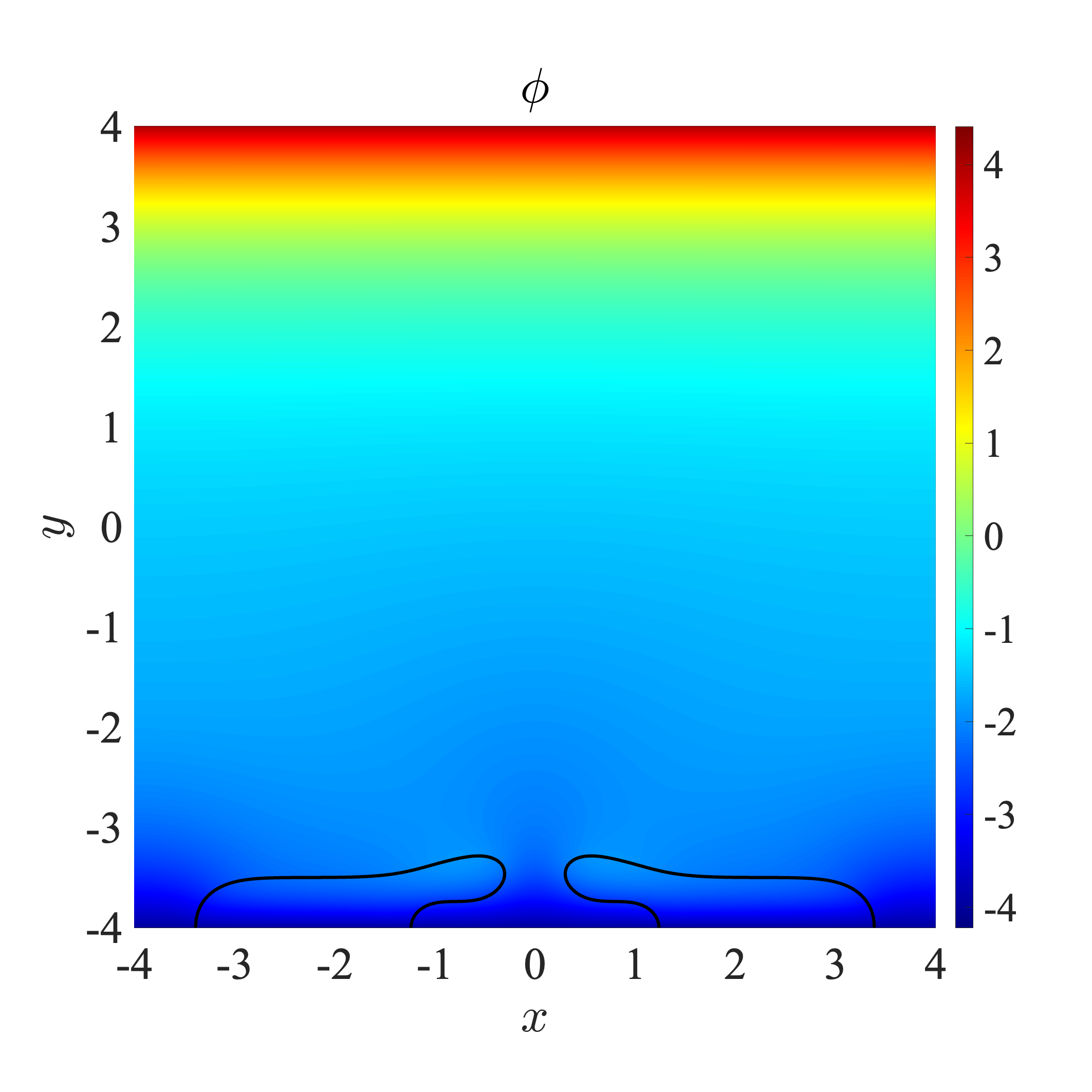}
        \label{subfig:1DropD4N0Pump25Phi12}
		}
    \hskip -0.3cm
    \subfloat[$\phi~(t=30)$]{
		\includegraphics[width=0.16\linewidth]{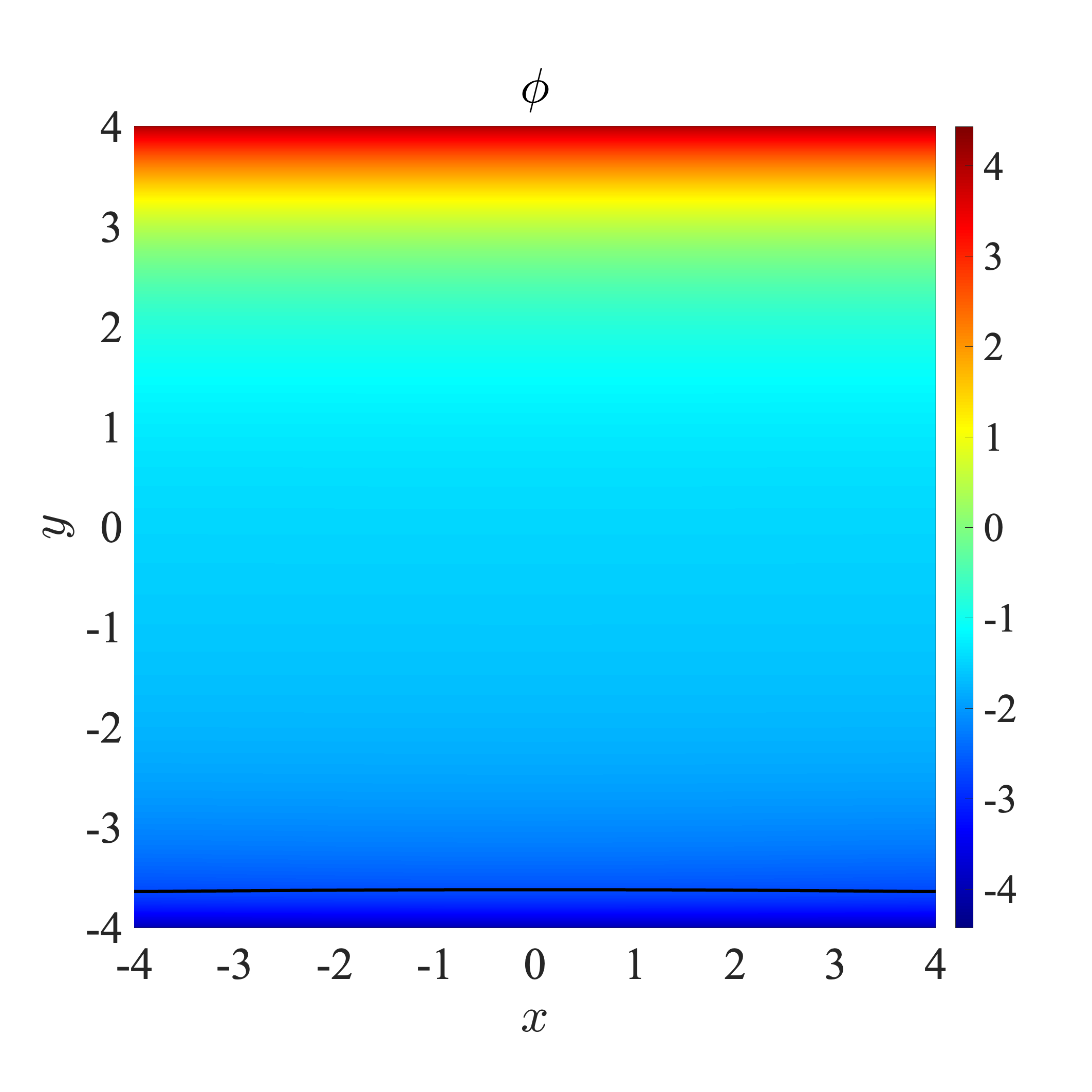}
        \label{subfig:1DropD4N0Pump25Phi30}
		}
        \vskip -0.2cm
	\caption{The snapshots for the drop deformation with positive ion pump when the vertical electric field is added. The black solid circle represents the location of the drop, 
    which is denoted by the level set $\psi=0$. 
    The concentration and electric potential distribution are shown on the color map. We choose the final time as $t = 30$.  Here $\phi_{0b} = -4, \quad \phi_{0u} = 4$.
    }\label{fig:1DropD4N0Pump25}
\end{figure}

When uniform pumps are applied on the droplet interface, the internal electric potential increases due to the influx of positive ions. However, as seen in Fig.~\ref{subfig:1DropD4N0Pump25Phi5} and Appendix Fig.~\ref{subfig:1DropD4N0Pump25Phi10x0}, the elevated internal potential still remains below that of the upper boundary, and hence no significant elongation is observed. Nonetheless, the droplet experiences a net downward motion due to the Lorentz force arising from its positive charge.

As the droplet descends, viscous shear exerted by the surrounding fluid introduces a drag force on its upper interface. This shear, combined with the incompressibility of the fluid, induces a concave (crescent-shaped) deformation at the droplet's rear. Such shapes are characteristic of droplets undergoing sustained viscous drag.

Upon nearing the solid lower boundary, the droplet encounters resistance due to the no-slip condition and fluid incompressibility. 
A sharp pressure gradient develops beneath the center of the droplet, preventing further downward motion in this region. In contrast, the pressure gradient near the lateral sides is less steep (see Appendix Figs.~\ref{fig:1DropD4N0Pump25eforceSection}), allowing the droplet to first make contact with the wall at its sides. This results in an arched interface, where the center of the droplet remains suspended above the wall, trapping a small fluid pocket beneath the arch.

\begin{figure}[!ht]
    \vskip -0.4cm
    \centering 
	\subfloat[$p+n~(t=3)$]{
		\includegraphics[width=0.16\linewidth]{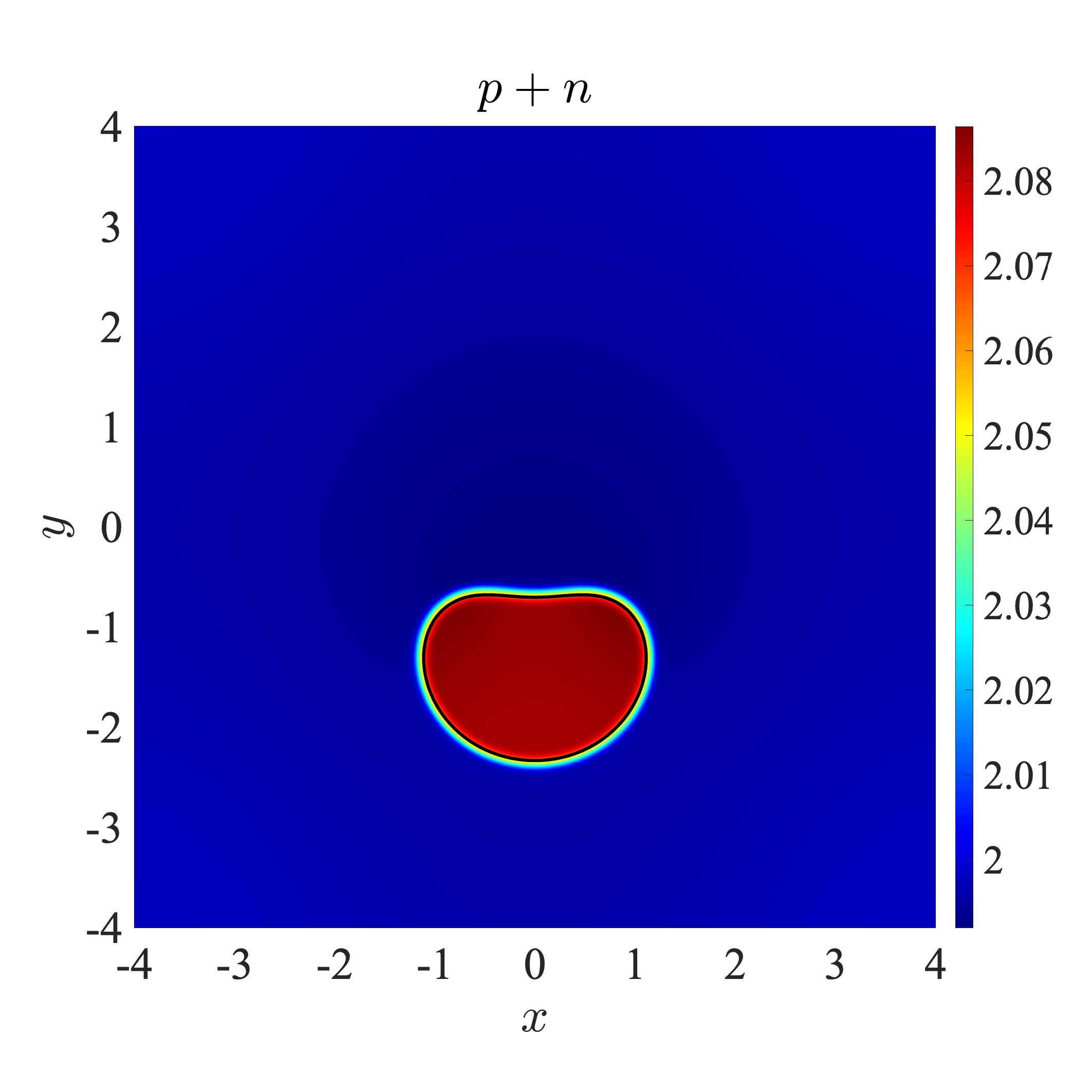}
        \label{subfig:1DropD4N0Pump25Sum3}
		} 
    \hskip -0.3cm
    \subfloat[$p+n~(t=5)$]{
		\includegraphics[width=0.16\linewidth]{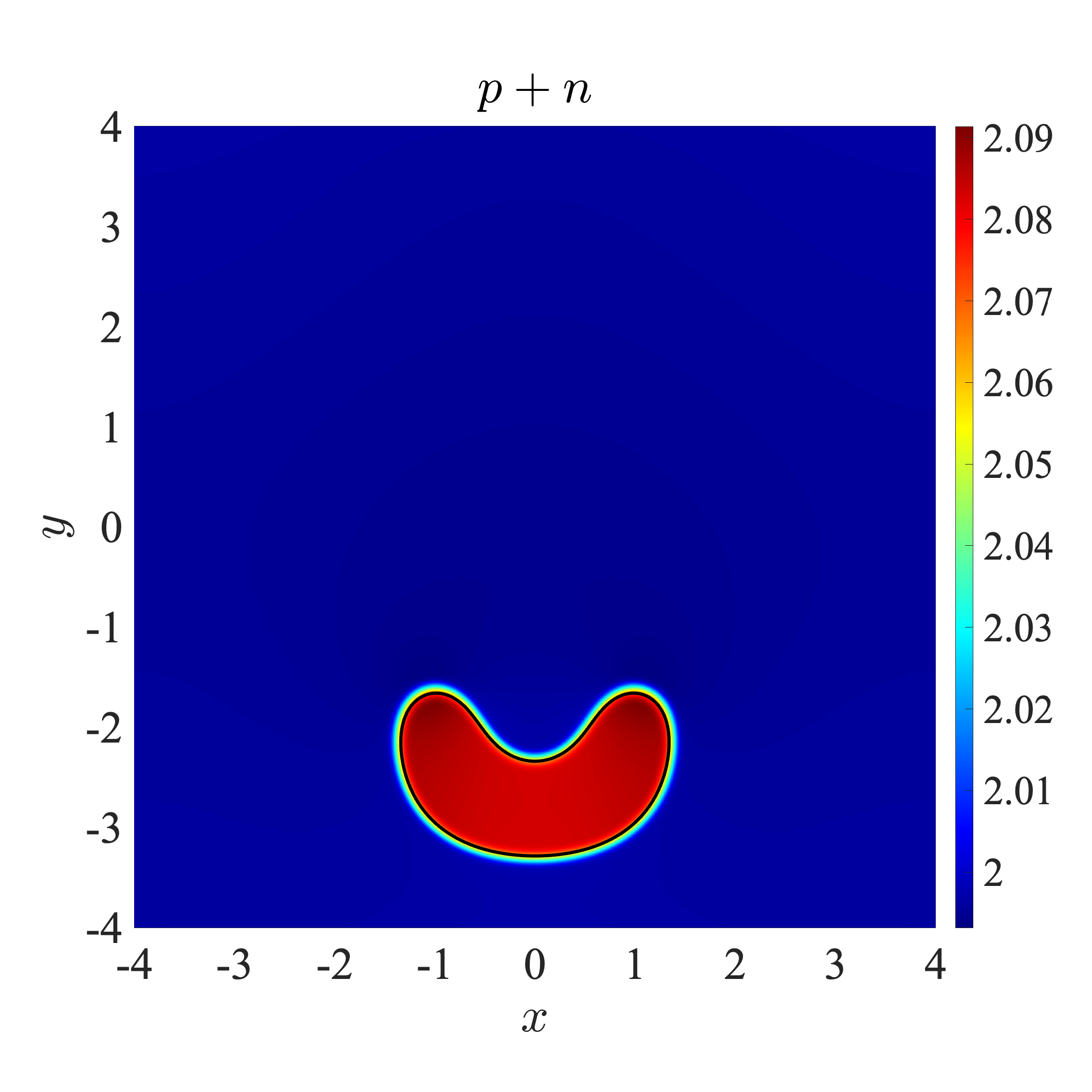}
        \label{subfig:1DropD4N0Pump25Sum5}
		} 
    \hskip -0.3cm
    \subfloat[$p+n~(t=8)$]{
		\includegraphics[width=0.16\linewidth]{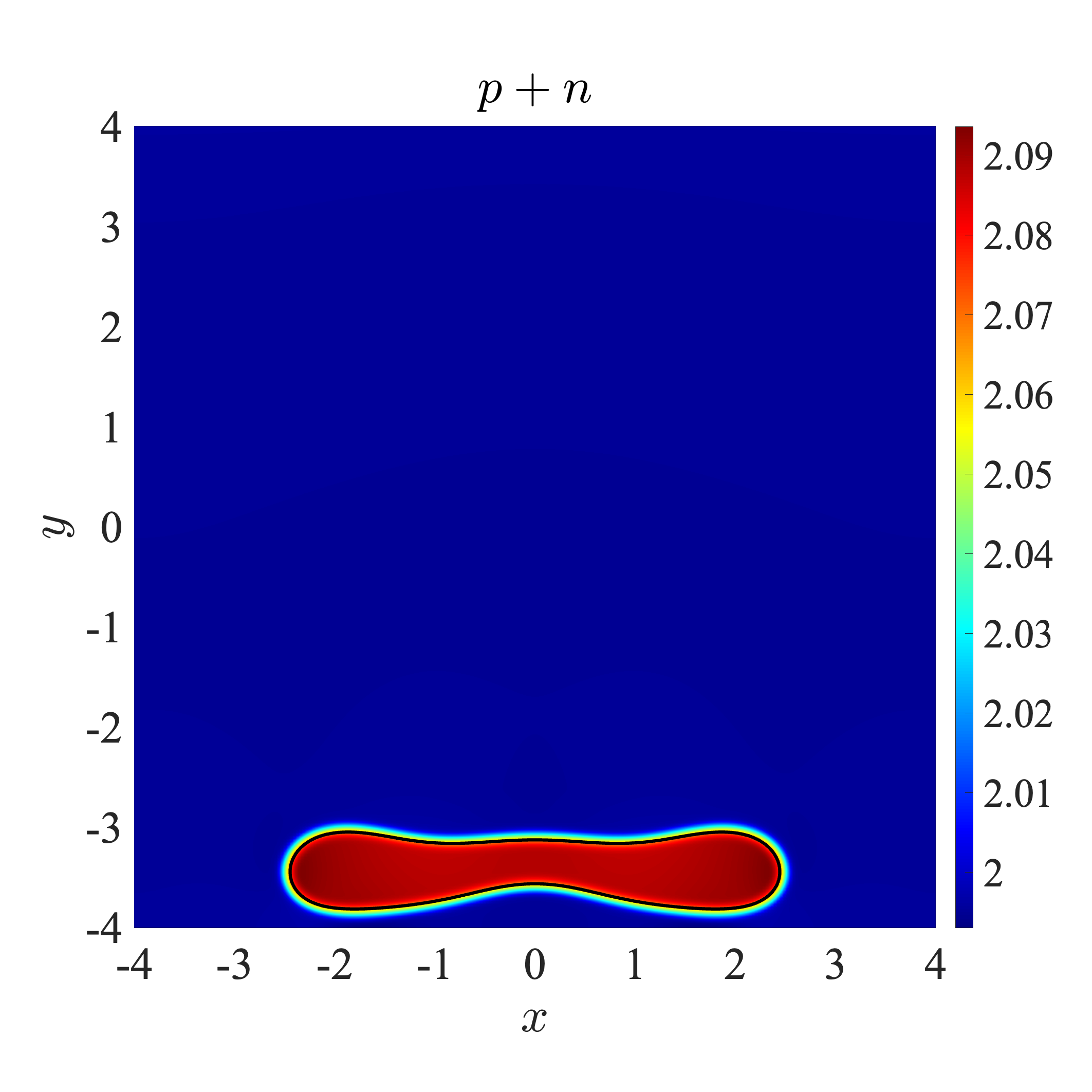}
        \label{subfig:1DropD4N0Pump25Sum8}
		} 
    \hskip -0.3cm
    \subfloat[$p+n~(t=10)$]{
		\includegraphics[width=0.16\linewidth]{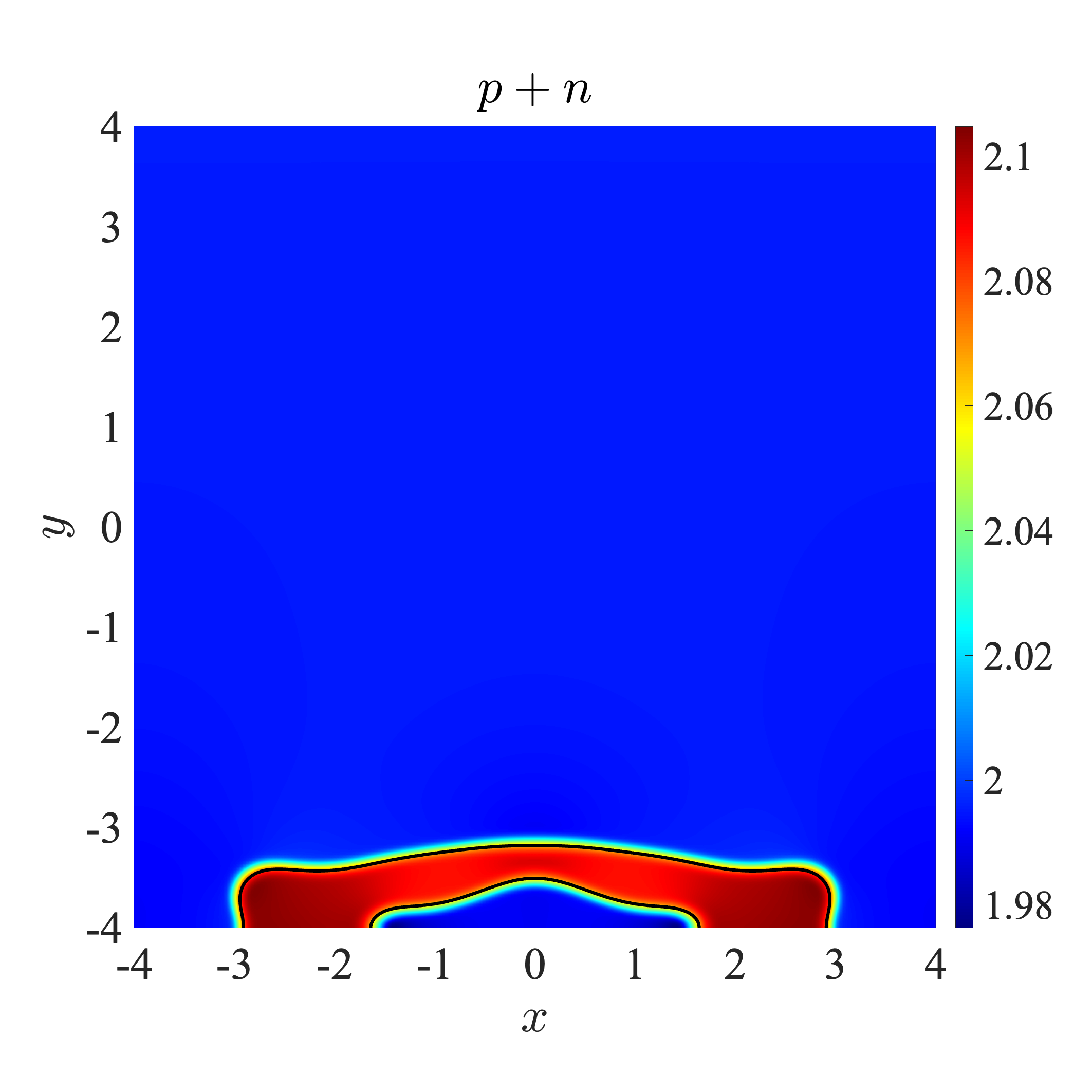}
        \label{subfig:1DropD4N0Pump25Sum10}
		} 
    \hskip -0.3cm
    \subfloat[$p+n~(t=12)$]{
		\includegraphics[width=0.16\linewidth]{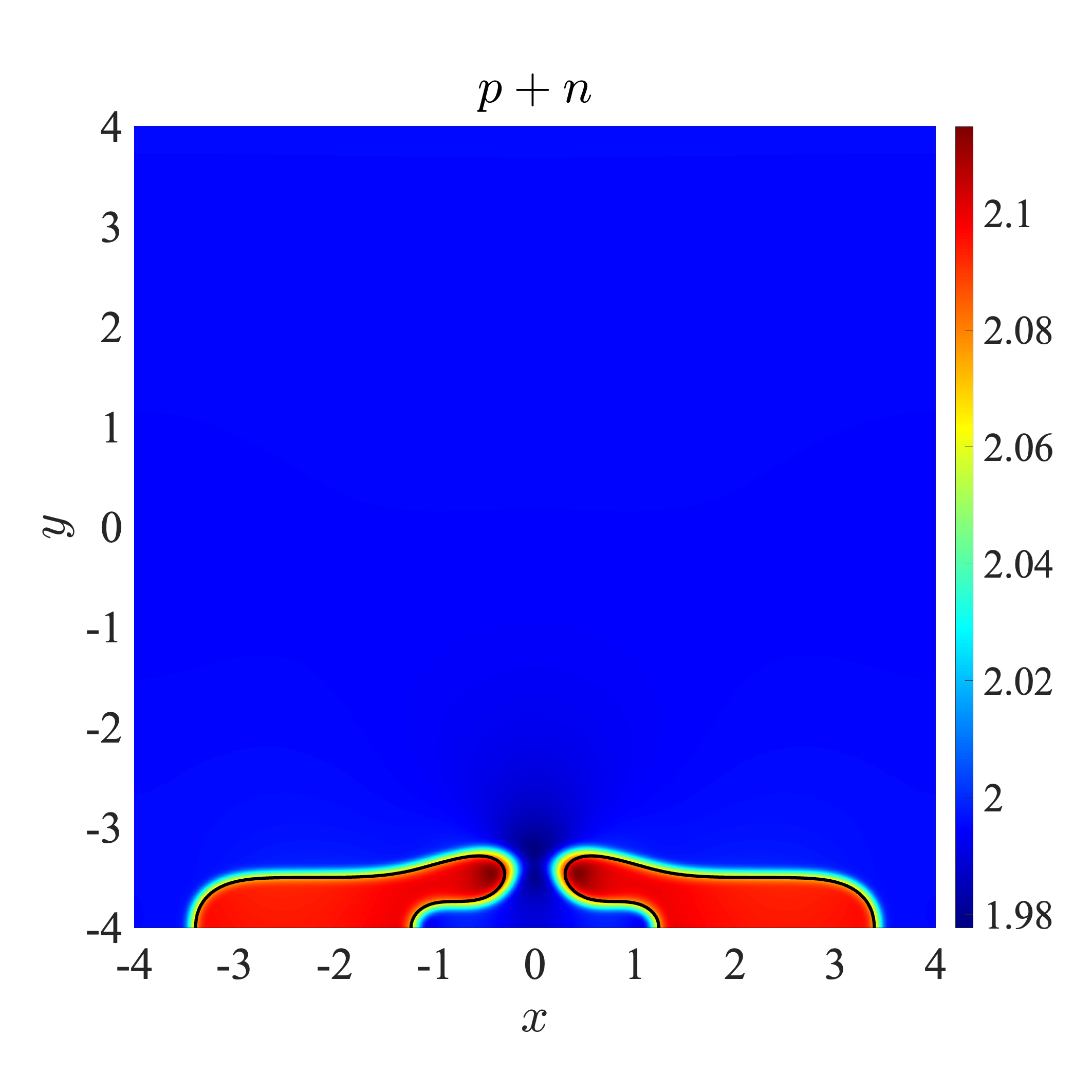}
        \label{subfig:1DropD4N0Pump25Sum12}
		} 
    \hskip -0.3cm
    \subfloat[$p+n~(t=30)$]{
		\includegraphics[width=0.16\linewidth]{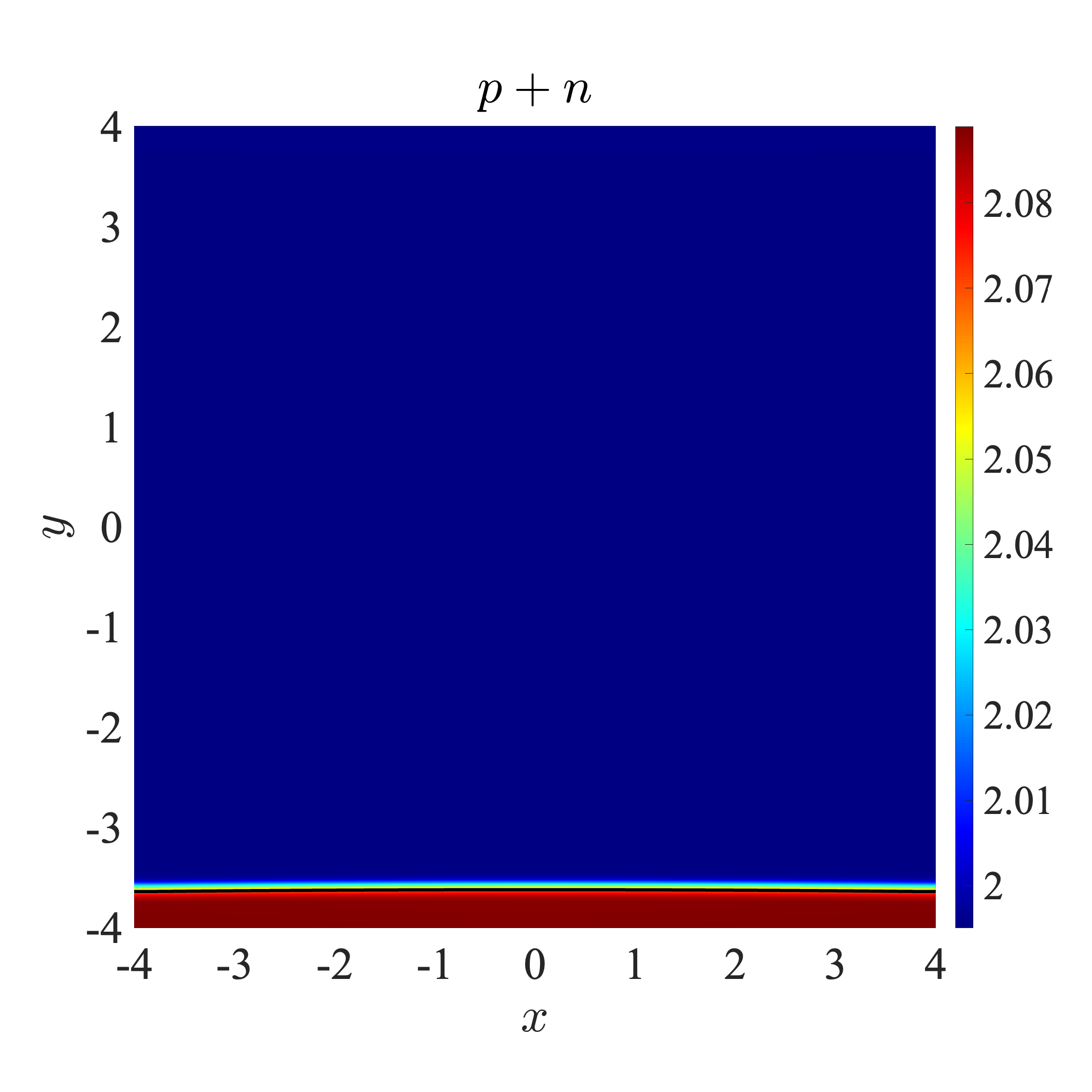}
        \label{subfig:1DropD4N0Pump25Sum30}
		}
    \\
    \vskip -0.3cm
	\subfloat[$p-n~(t=3)$.]{
		\centering
		\includegraphics[width=0.16\linewidth]{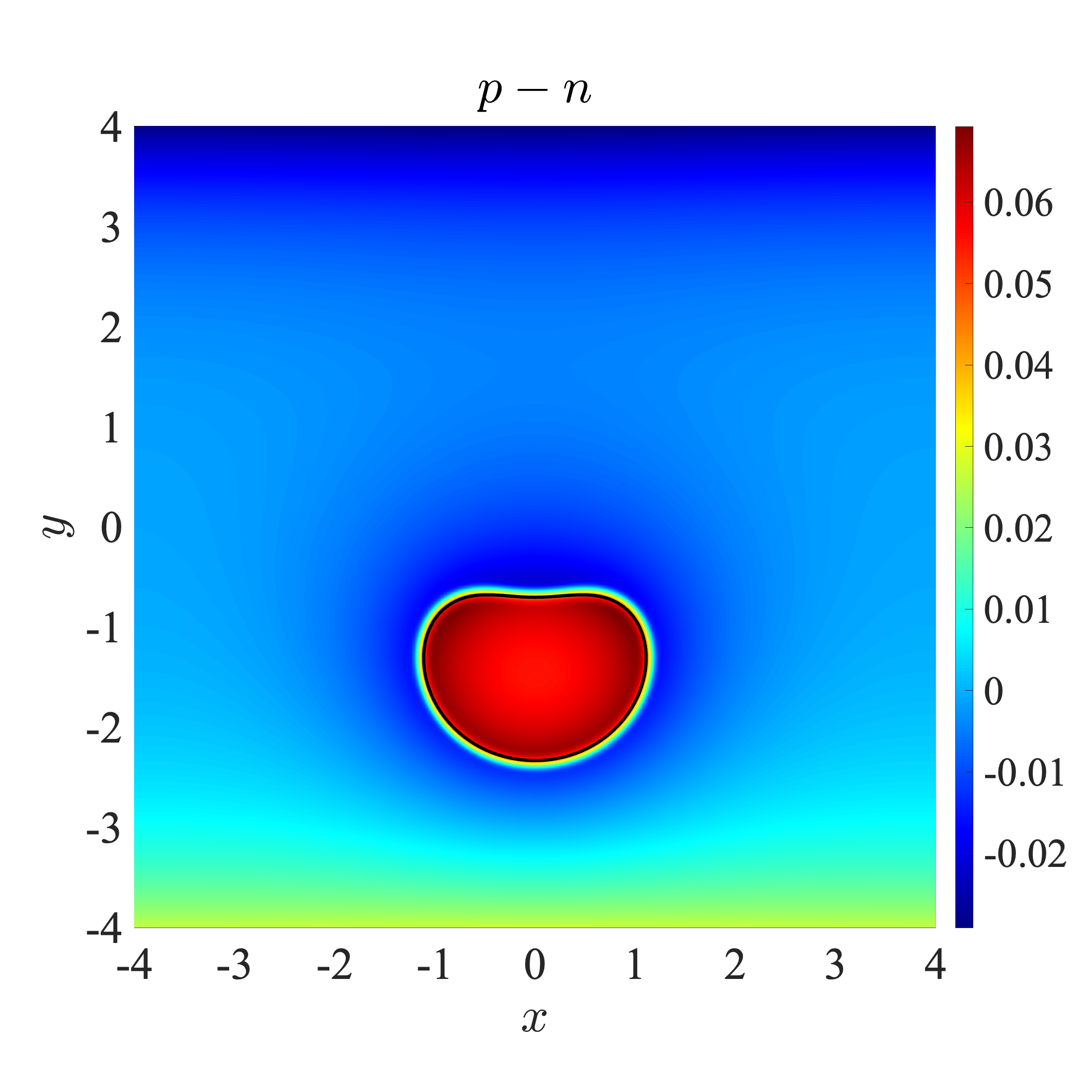}
        \label{subfig:1DropD4N0Pump25Dif3}
	}
    \hskip -0.3cm
    \subfloat[$p-n~(t=5)$.]{
		\centering
		\includegraphics[width=0.16\linewidth]{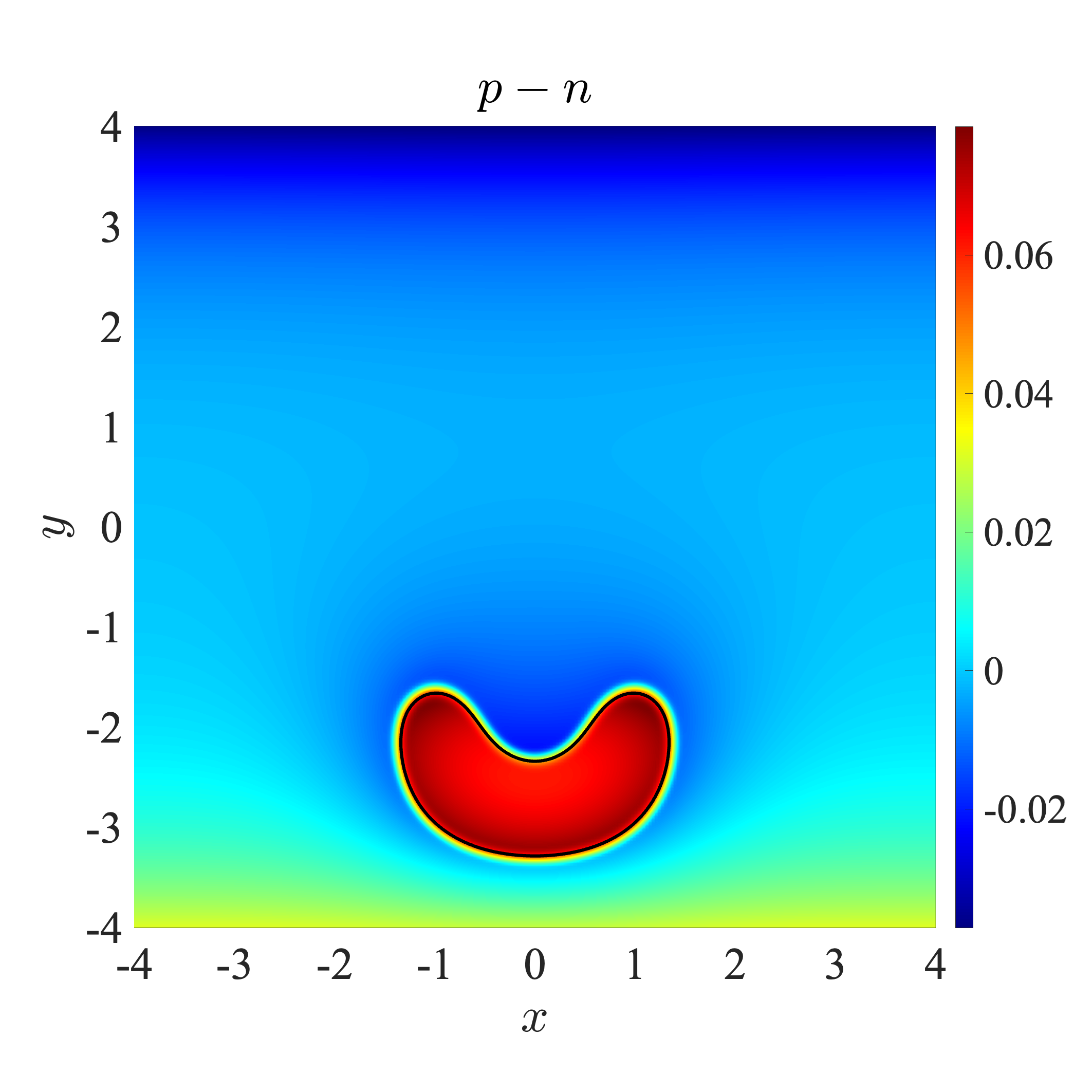}
        \label{subfig:1DropD4N0Pump25Dif5}
	}
    \hskip -0.3cm
    \subfloat[$p-n~(t=8)$.]{
		\centering
		\includegraphics[width=0.16\linewidth]{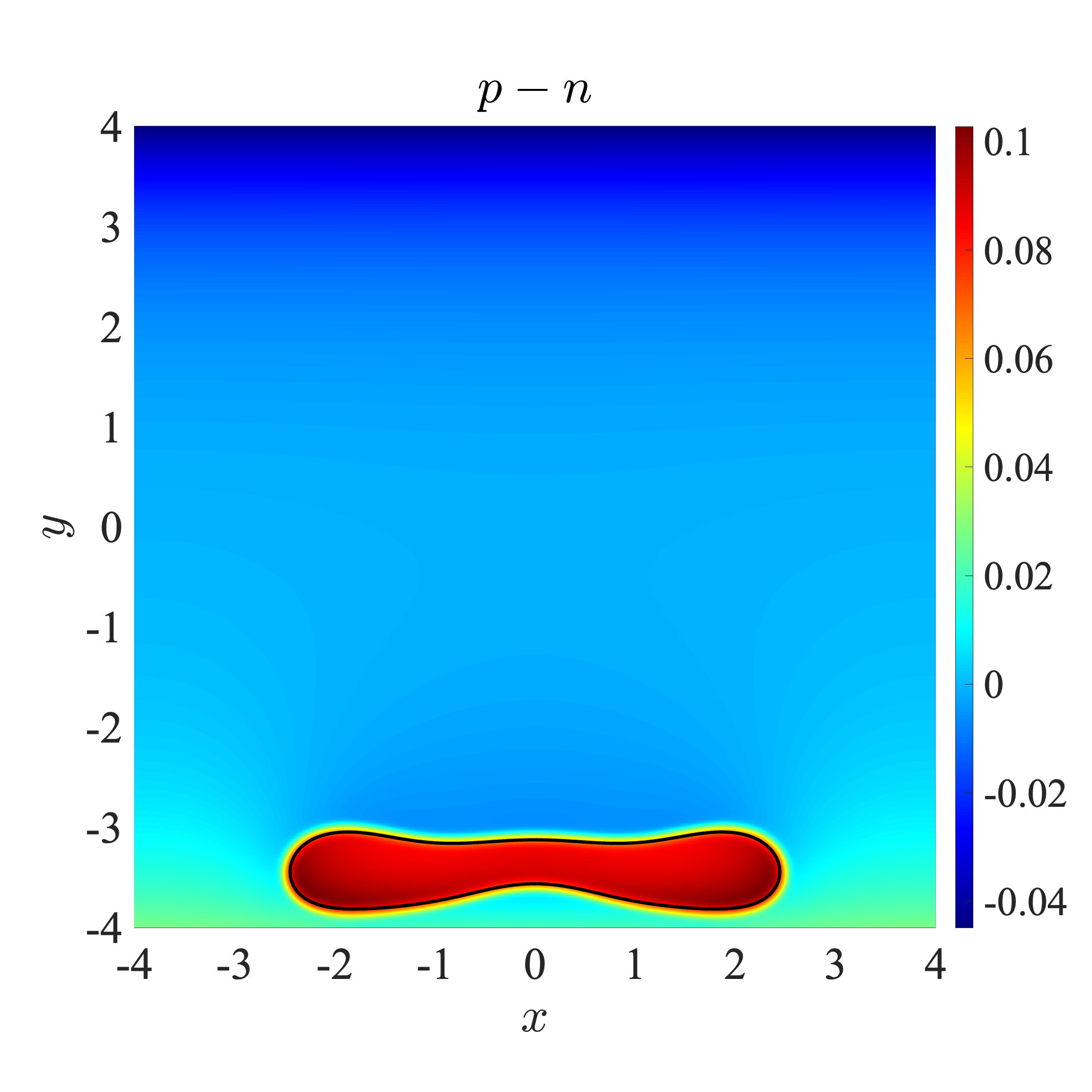}
        \label{subfig:1DropD4N0Pump25Dif8}
	}
    \hskip -0.3cm
    \subfloat[$p-n~(t=10)$.]{
		\centering
		\includegraphics[width=0.16\linewidth]{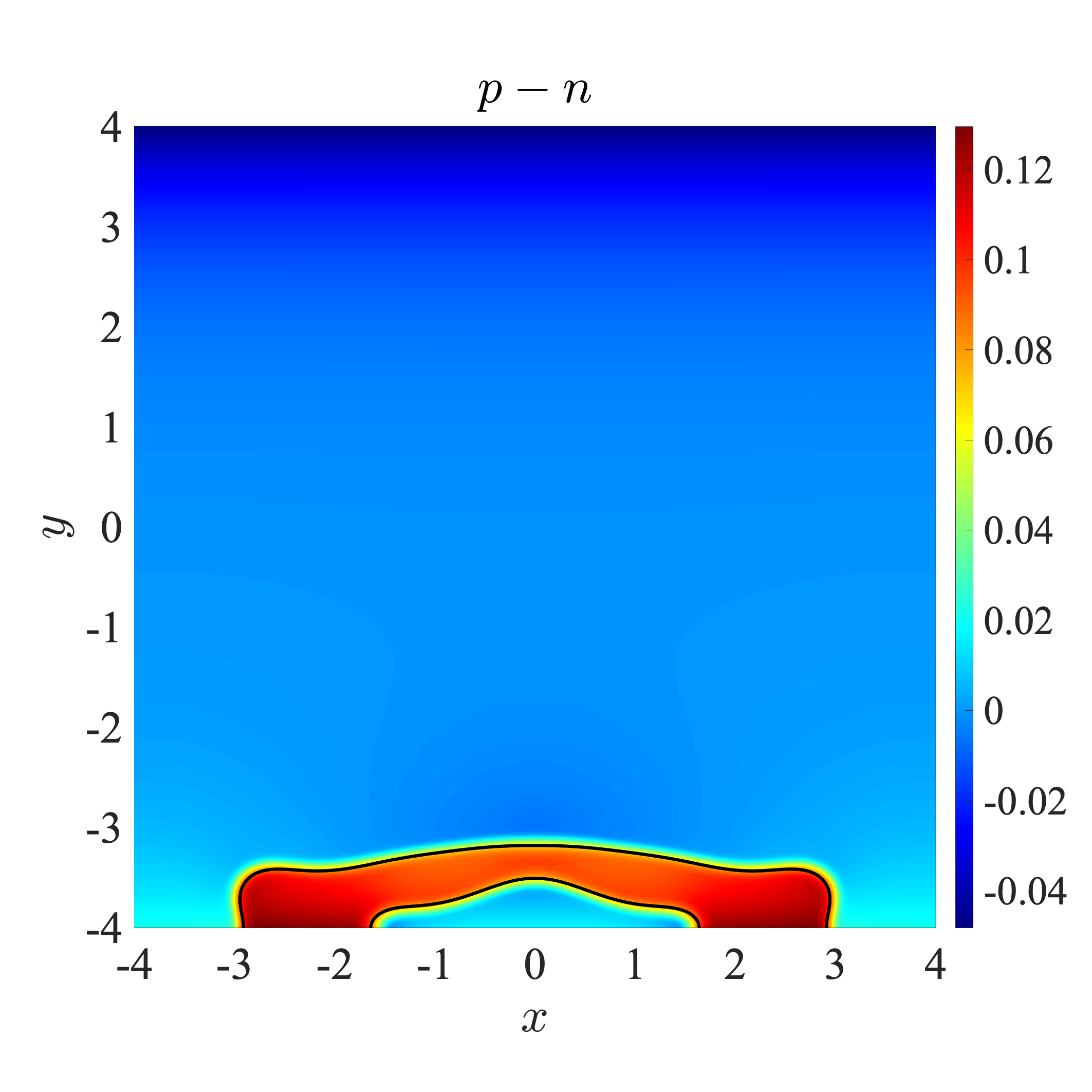}
        \label{subfig:1DropD4N0Pump25Dif10}
	}
    \hskip -0.3cm
    \subfloat[$p-n~(t=12)$.]{
		\centering
		\includegraphics[width=0.16\linewidth]{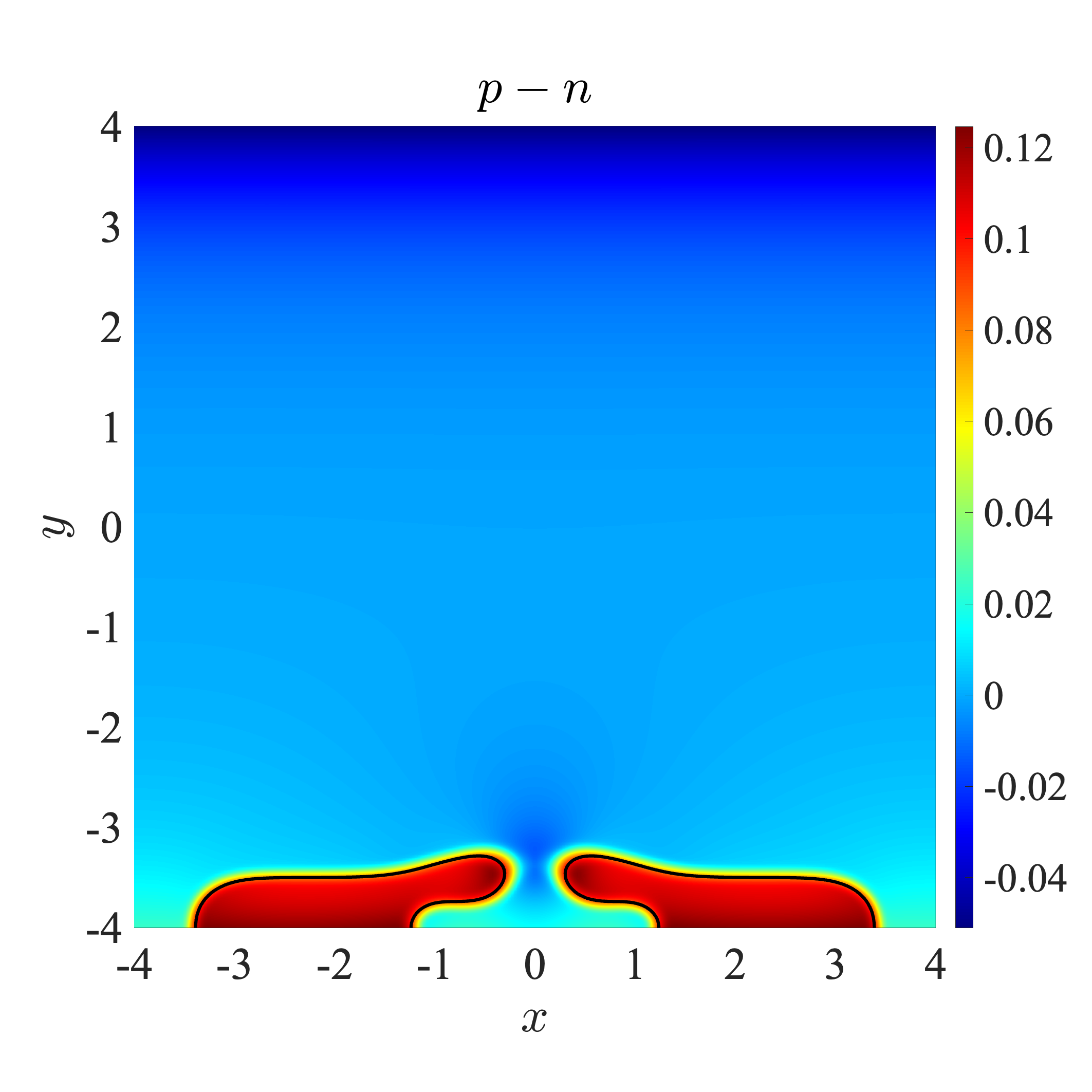}
        \label{subfig:1DropD4N0Pump25Dif12}
	}
    \hskip -0.3cm
    \subfloat[$p-n~(t=30)$.]{
		\centering
		\includegraphics[width=0.16\linewidth]{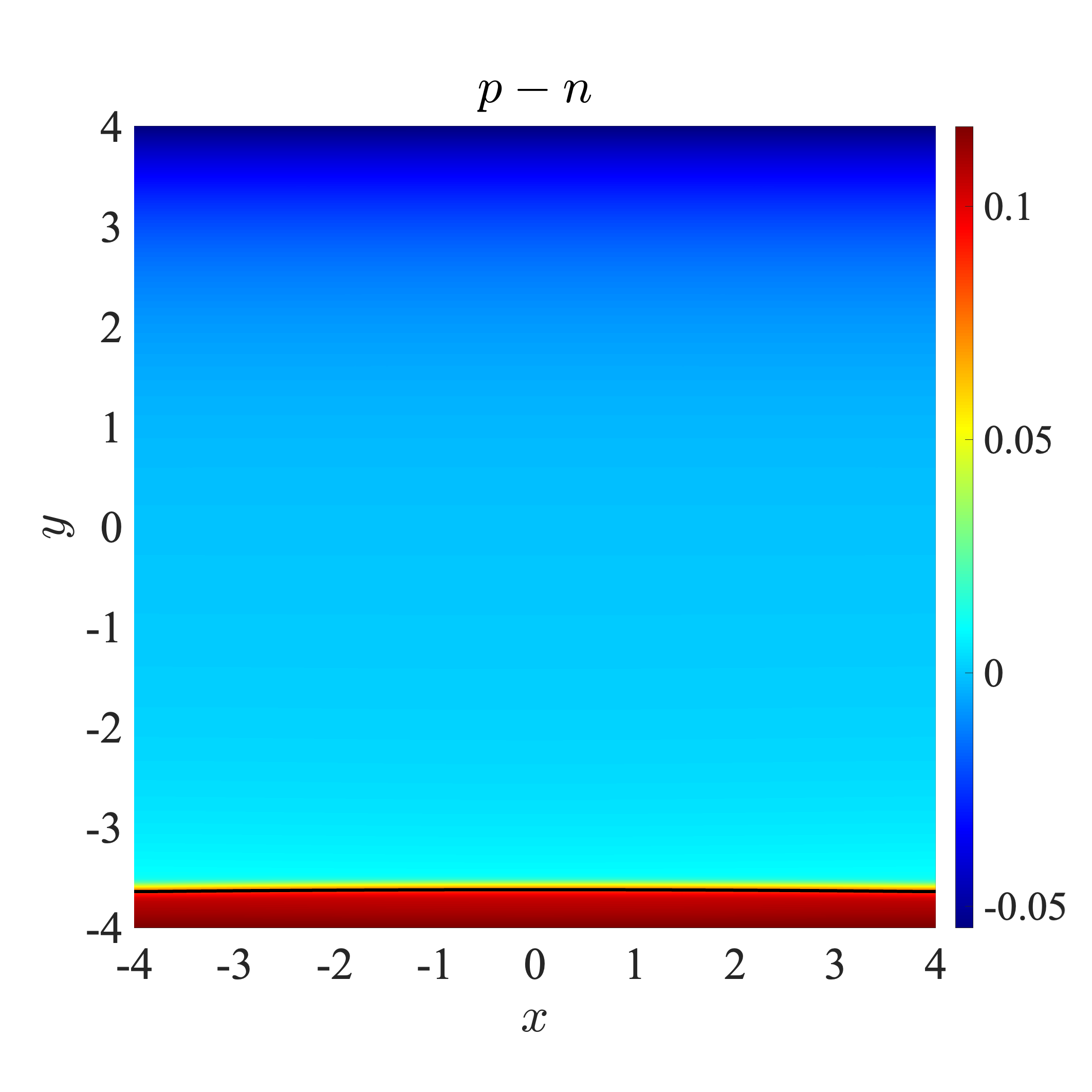}
        \label{subfig:1DropD4N0Pump25Dif30}
	}
        \vskip -0.2cm
	\caption{The snapshots of the total charge (top) and net charge (bottom). 
    The total charge and net charge both accumulate in the droplet due to the redistribution of positive and negative ions. Here $\phi_{0b} = -4, \quad \phi_{0u} = 4$.}\label{fig:1DropD4N0Pump25V}
\end{figure}

Following side contact, the droplet begins to spread laterally along the bottom surface. Due to the arched geometry, the central region becomes increasingly thin. Surface tension, which minimizes interfacial curvature and area, destabilizes this narrow neck, eventually triggering a pinch-off event. This leads to the formation of two daughter droplets.
Because these droplets remain spatially close and the surrounding flow is relatively quiescent, interfacial forces and short-range capillary attraction act to pull the droplets back together. Over time, this results in a recoalescence event, wherein the droplets merge and subsequently spread along the bottom boundary, forming a flattened configuration as shown in Figs. \ref{fig:1DropD4N0Pump25} and \ref{fig:1DropD4N0Pump25V}. 
 
\paragraph{Fully voltage clump boundary condition}

\begin{figure}[!ht]
\vskip -0.4cm
\centering
	\subfloat[$p~(t=10)$]{
		\includegraphics[width=0.33\linewidth]{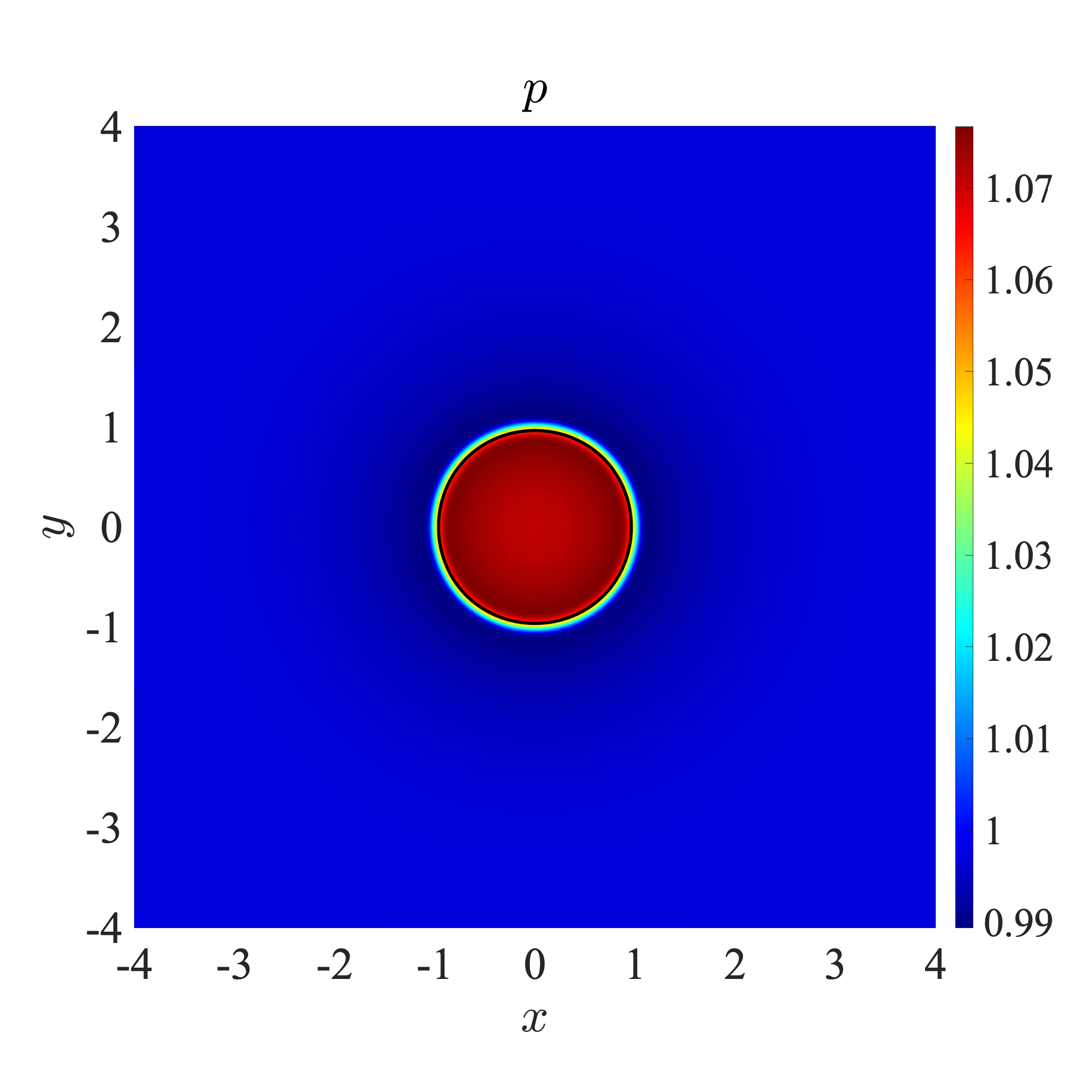}
        \label{subfig:1Drop0bdPump25P10}
		}
    \hskip -0.3cm
	\subfloat[$n~(t=10)$]{
		\includegraphics[width=0.33\linewidth]{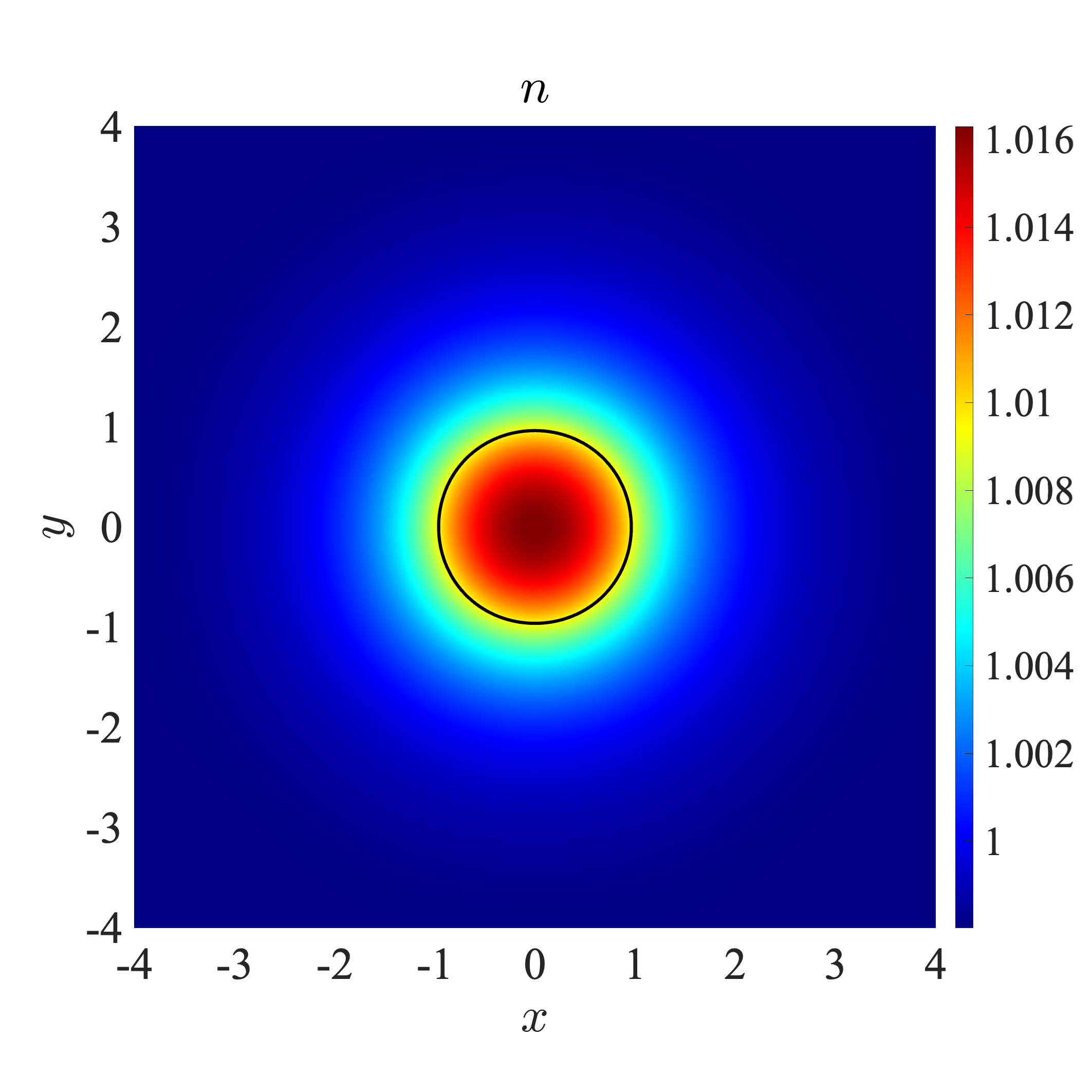}
        \label{subfig:1Drop0bdPump25N10}
		}
    \hskip -0.3cm
	\subfloat[$\phi~(t=10)$]{
		\includegraphics[width=0.33\linewidth]{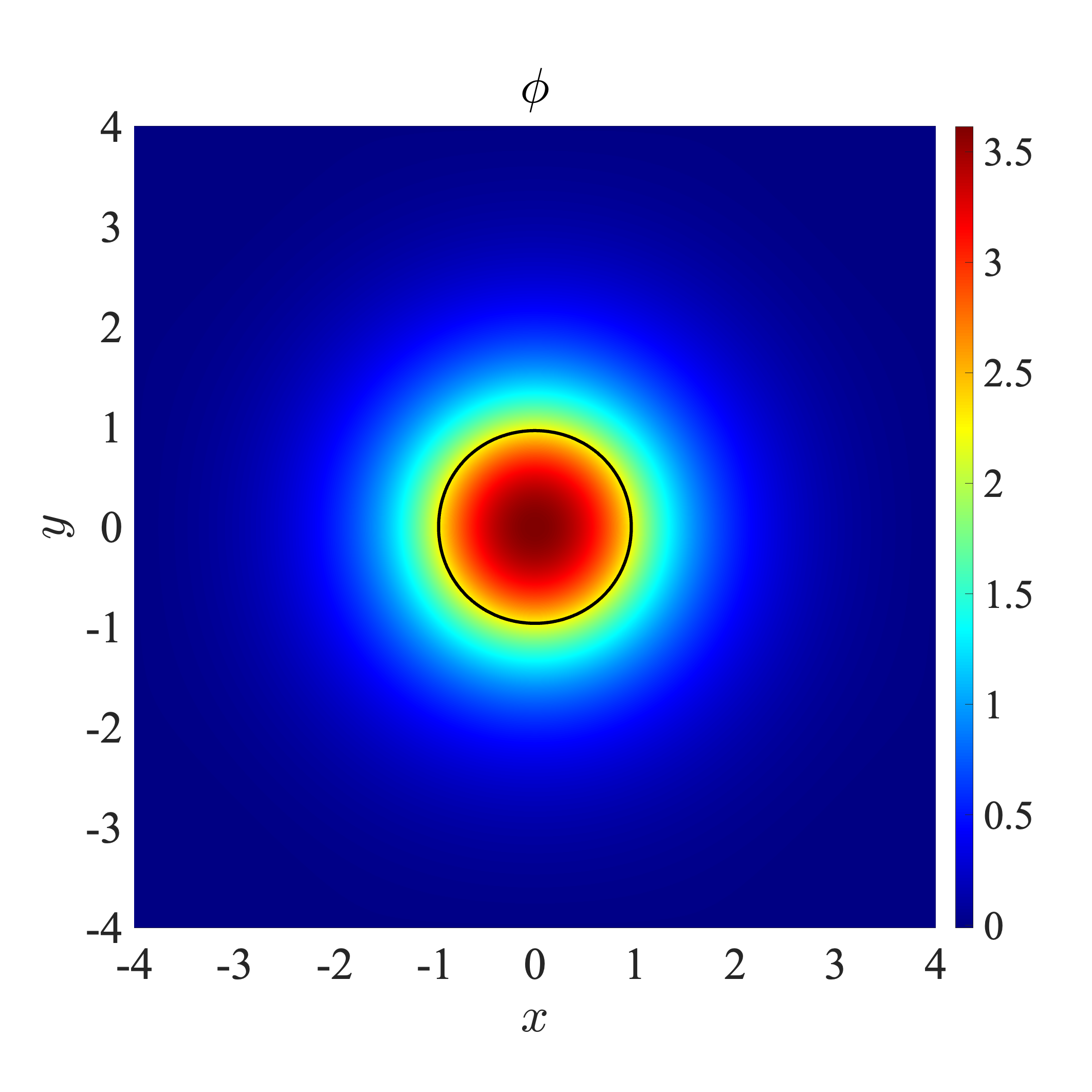}
        \label{subfig:1Drop0bdPump25Phi10}
		}
    \\
	\subfloat[$p+n~(t = 10)$]{
		\includegraphics[width=0.33\linewidth]{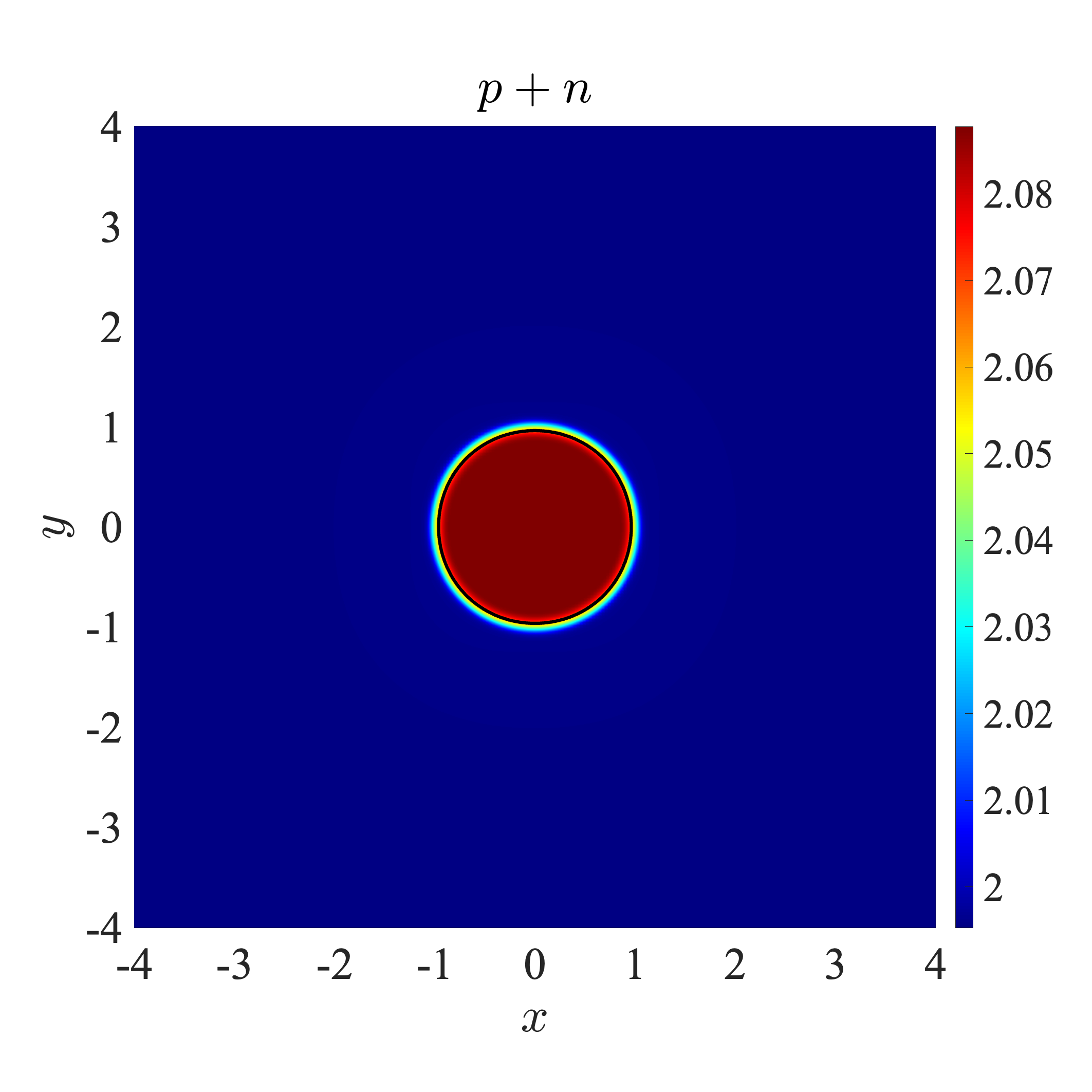}
        \label{subfig:1Drop0bdPump25Sum10}
		} 
    \hskip -0.3cm
	\subfloat[$p-n~(t = 10)$]{
		\includegraphics[width=0.33\linewidth]{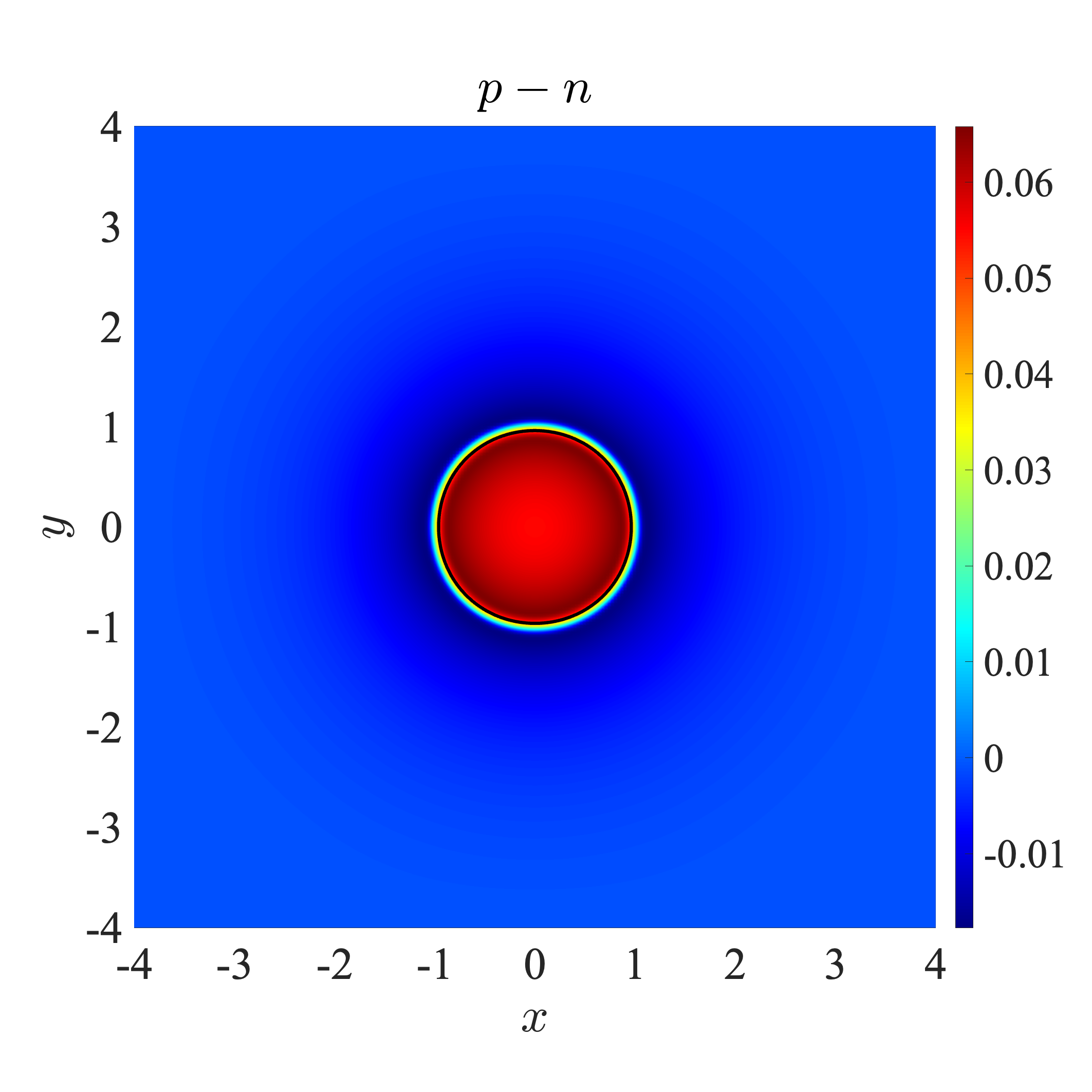}
        \label{subfig:1Drop0bdPump25Dif10}
	}
    \hskip -0.3cm
	    \subfloat[$\bm{u}~(t=10)$]{
		\includegraphics[width=0.33\linewidth]{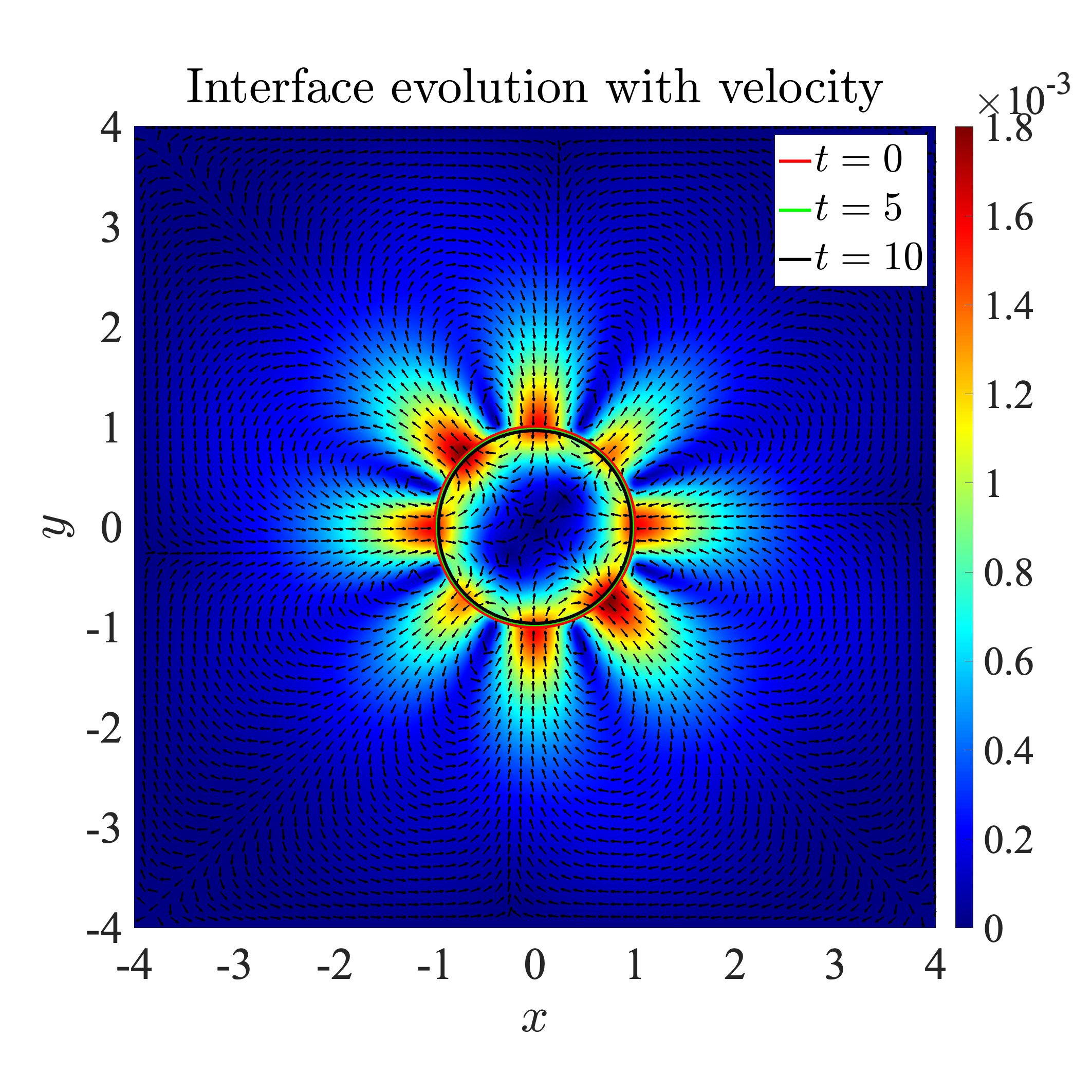}
        \label{subfig:1Drop0bdPump25U10}
		}
    \vskip -0.2cm
	\caption{The snapshots for the drop deformation with uniform positive ion pump and full grounded electric potential condition on the boundary at time $t=10$. 
    The black solid circle represents the location of the drop, 
    which is denoted by the level set $\psi=0$. 
    The concentration and electric potential distribution are shown on the color map.  
    The black solid lines show the positive and negative ions and the electric potential distribution along the $ x$- and $ y$-axes. 
    }\label{fig:1Drop0bdPump25}
\end{figure}

In this case, if no pumps are present on the interface, the ion concentrations remain unchanged and the droplet retains its original shape. However, with uniformly distributed pumps, as shown in Fig.~\ref{fig:1Drop0bdPump25}, positive ions are actively transported into the droplet, leading to an elevated electric potential inside. Despite this, no visible deformation occurs, as the resulting electrostatic forces remain symmetric along the interface and thus exert no net mechanical effect on the droplet shape.

Then we consider a nonuniformly distributed pump case, where the pump density depends on the angle. 
\begin{equation}\label{case:nonuniformpump}
    I = I_{0}(1 + 0.5\cos(4\theta)),~\theta = \arctan(n_y/n_x)
\end{equation}
where  the pump achieves maximum at $\theta = 0, \pi/2, \pi, 3\pi/2$. 

\begin{figure}[!ht] 
\vskip -0.4cm
\centering 
	\subfloat[Positive ion $p$ at $t=5$.]{
		\includegraphics[width=0.33\linewidth]{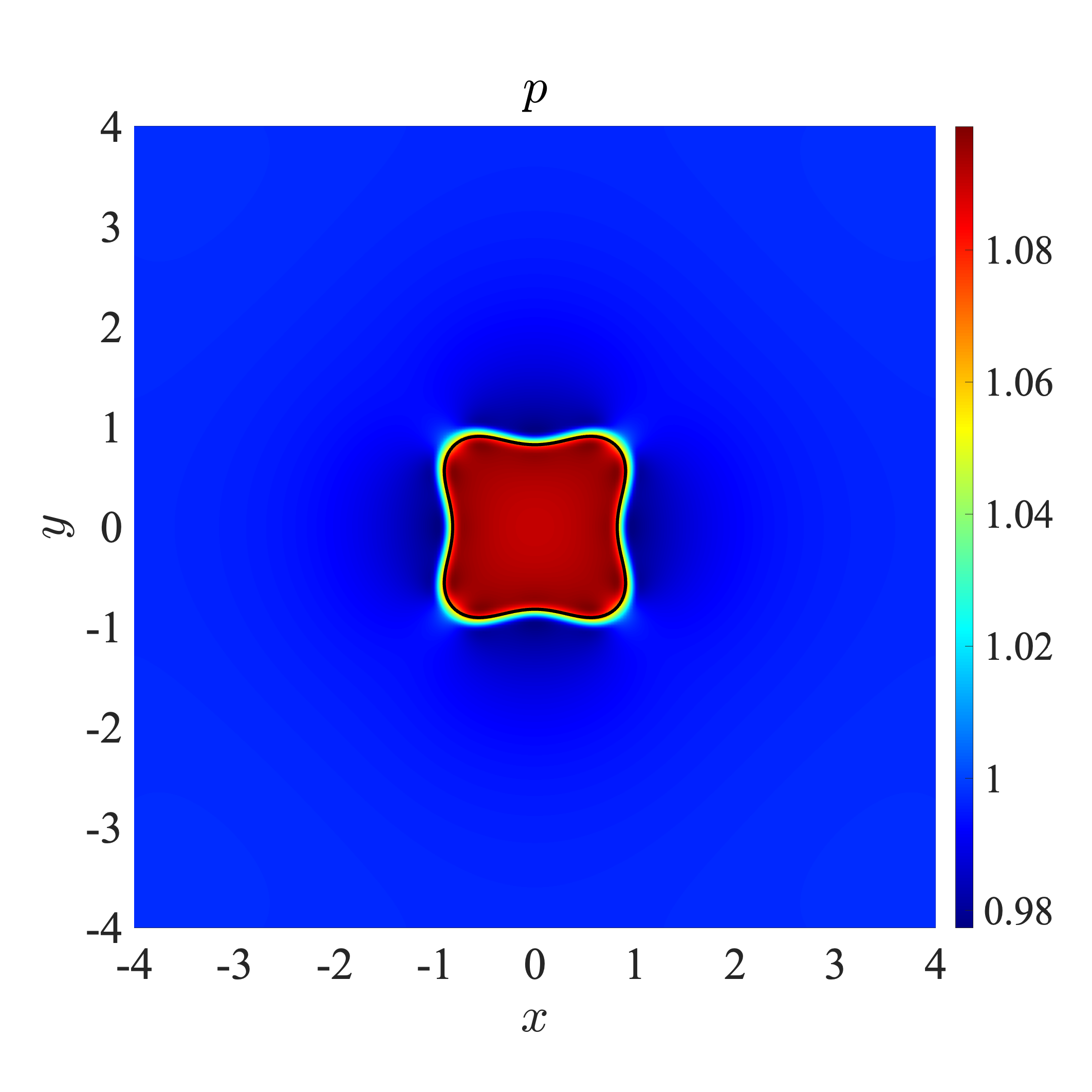}
        \label{subfig:1Drop0bdgroundPump25P5}
		}
    \hskip -0.3cm
	\subfloat[Negative ion $n$ at $t=5$.]{
		\includegraphics[width=0.33\linewidth]{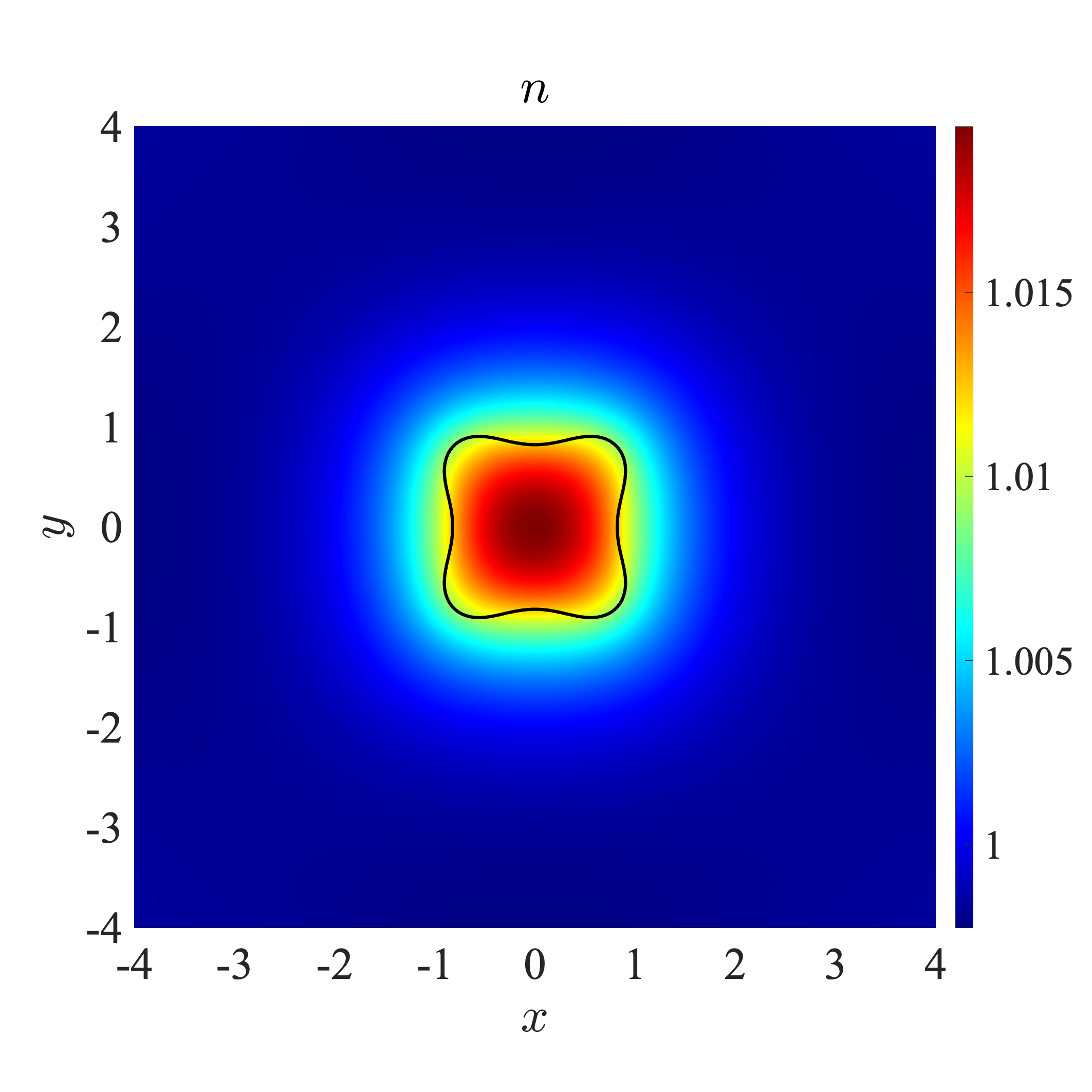}
        \label{subfig:1Drop0bdgroundPump25N5}
		}
    \hskip -0.3cm
	\subfloat[Electric potential $\phi$ at $t=5$.]{
		\includegraphics[width=0.33\linewidth]{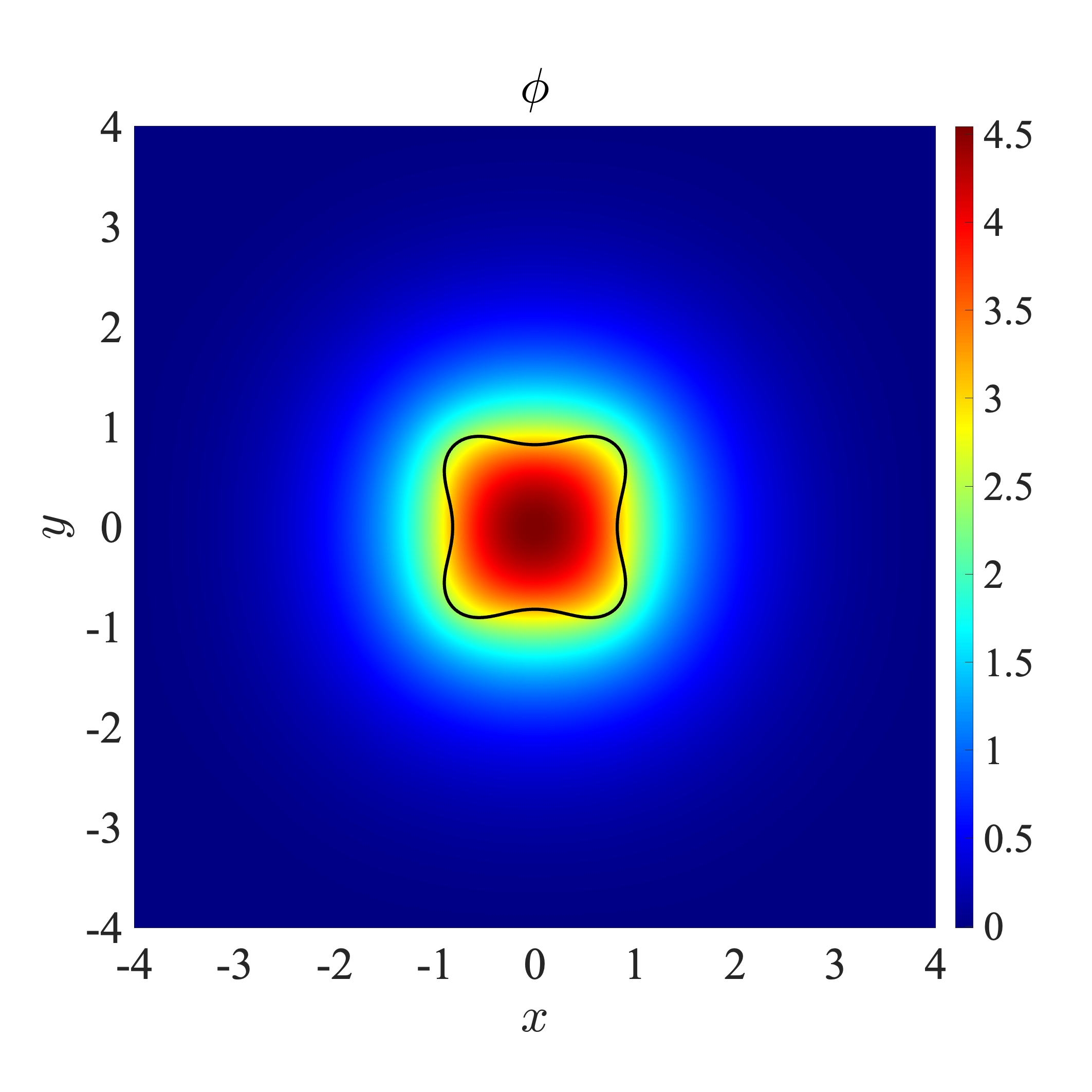}
        \label{subfig:1Drop0bdgroundPump25Phi5}
		}
        \\
    \vskip -0.3cm
	\subfloat[Positive ion $p$ at $t=10$.]{
		\includegraphics[width=0.33\linewidth]{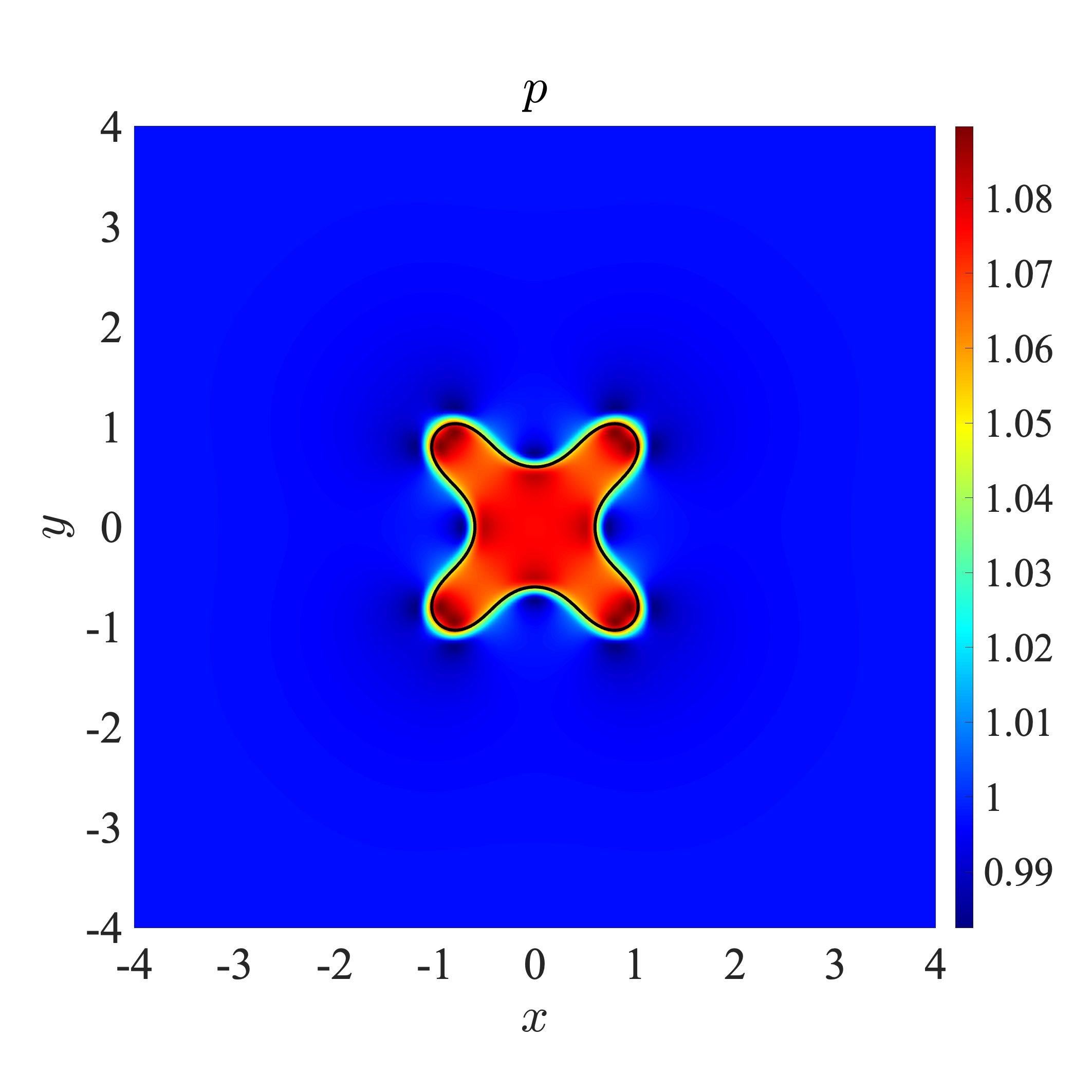}
        \label{subfig:1Drop0bdgroundPump25P10}
		}
    \hskip -0.3cm
	\subfloat[Negative ion $n$ at $t=10$.]{
		\includegraphics[width=0.33\linewidth]{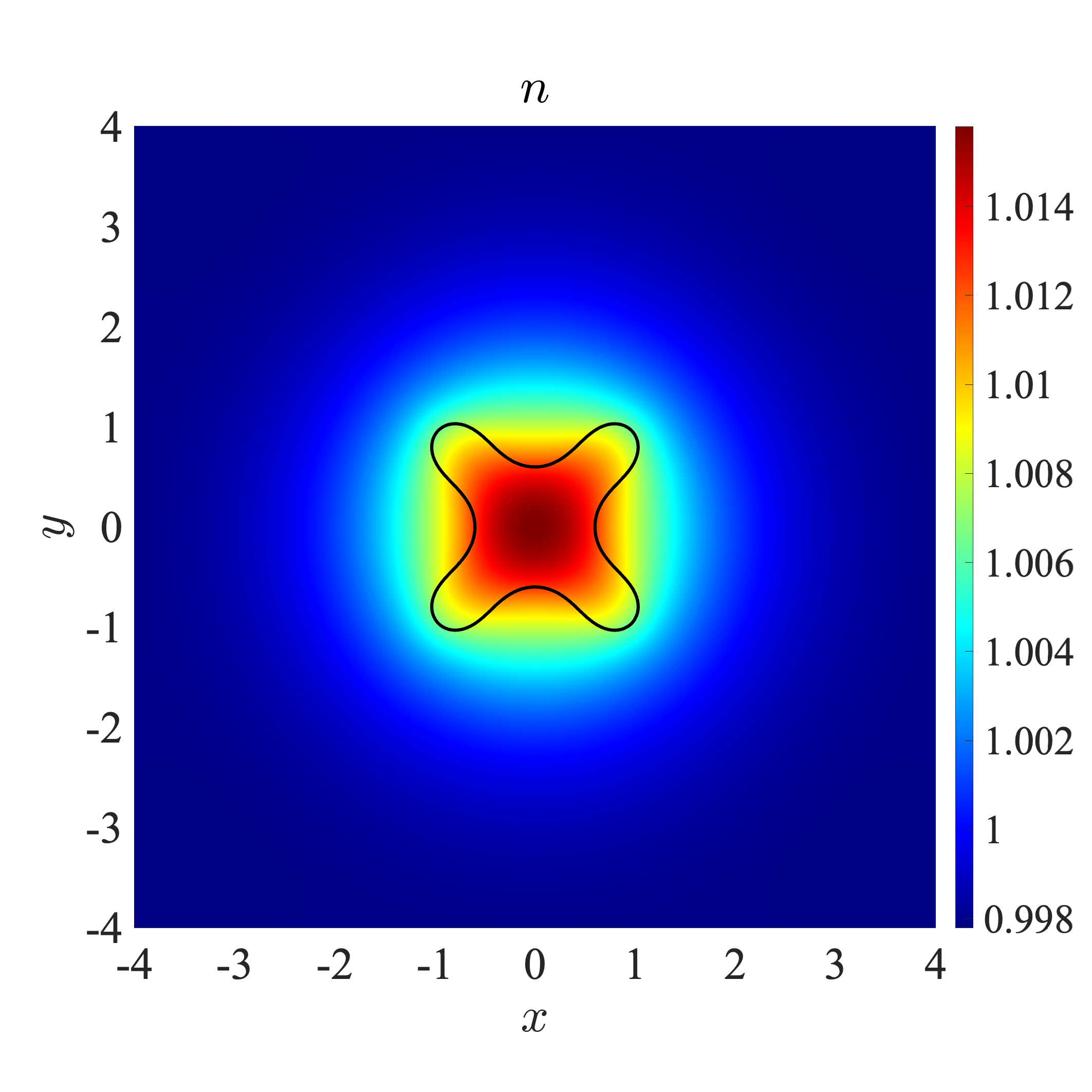}
        \label{subfig:1Drop0bdgroundPump25N10}
		}
    \hskip -0.3cm
	\subfloat[Electric potential $\phi$ at $t=10$.]{
		\includegraphics[width=0.33\linewidth]{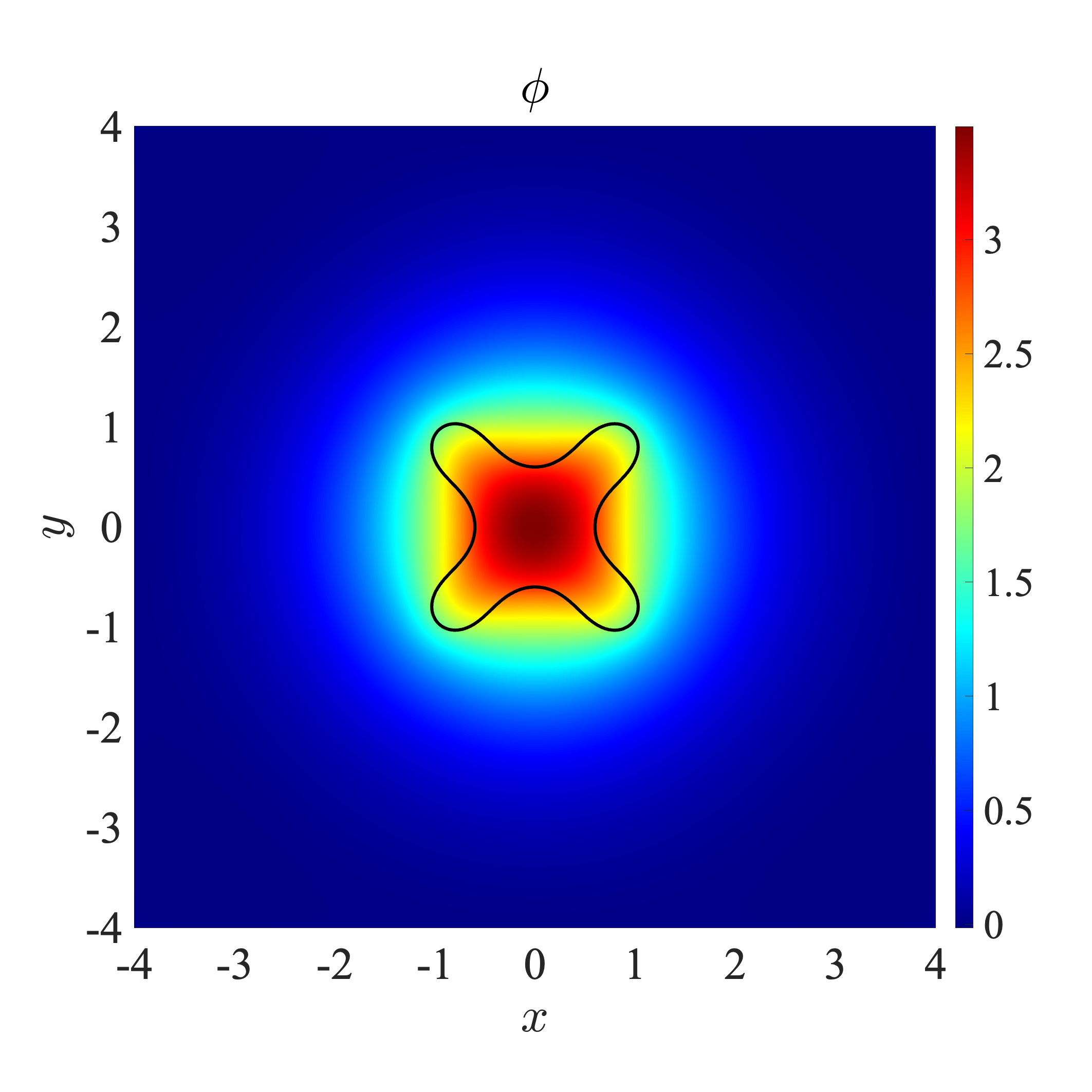}
        \label{subfig:1Drop0bdgroundPump25Phi10}
		}
    \vskip -0.2cm
	\caption{The snapshots for the drop deformation with a nonuniformly distributed positive ion pump and a fully grounded Dirichlet boundary condition.   First row: $t=5$  and Second row: $t=10$.
    The black solid circle represents the location of the drop, 
    which is denoted by the level set $\psi=0$. 
    The concentration and electric potential distribution are shown on the color map. 
    }\label{fig:1Drop0bdgroundPump25}
\end{figure}

\begin{figure}[!ht]
\vskip -0.4cm
\centering 
	\subfloat[Total charge $p+n$ at $t = 10$.]{
		\centering
		\includegraphics[width=0.33\linewidth]{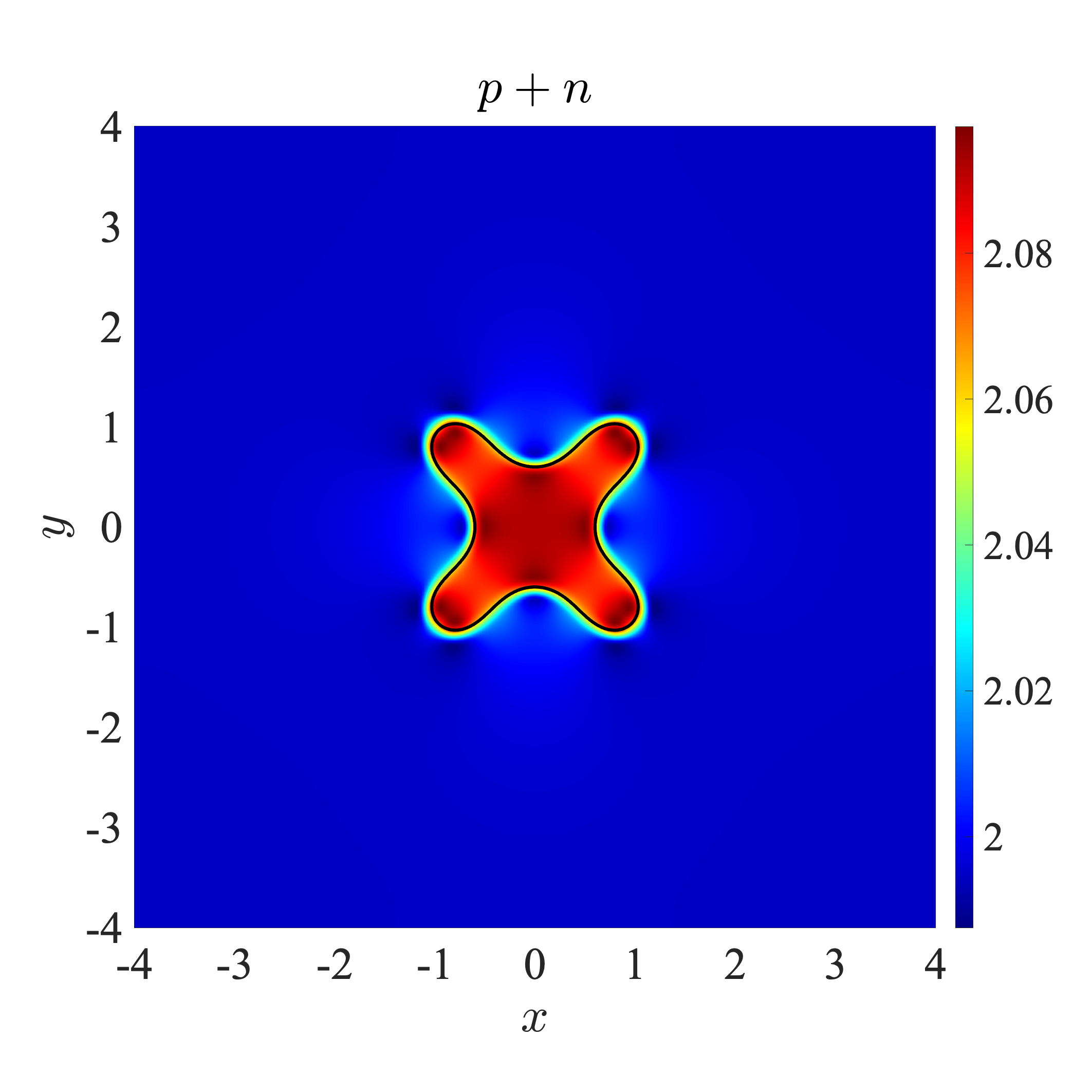}
        \label{subfig:1Drop0bdgroundPump25Sum10}
		} 
    \hskip -0.3cm
	\subfloat[Net charge $p-n$ at $t = 10$.]{
		\centering
		\includegraphics[width=0.33\linewidth]{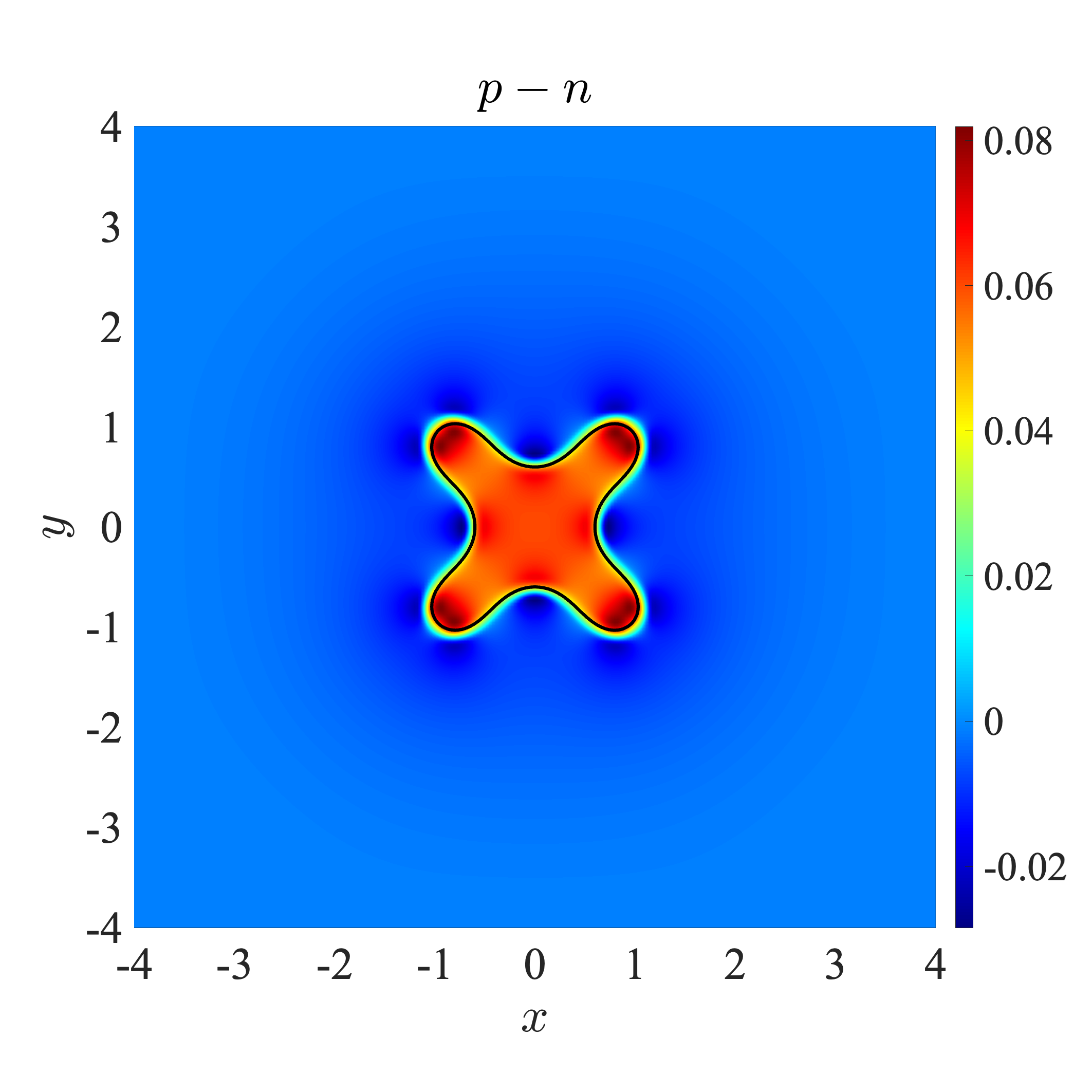}
        \label{subfig:1Drop0bdgroundPump25Dif10}
	}
    \hskip -0.3cm
	    \subfloat[Velocity and deformation at $t=10$.]{
		\centering
		\includegraphics[width=0.33\linewidth]{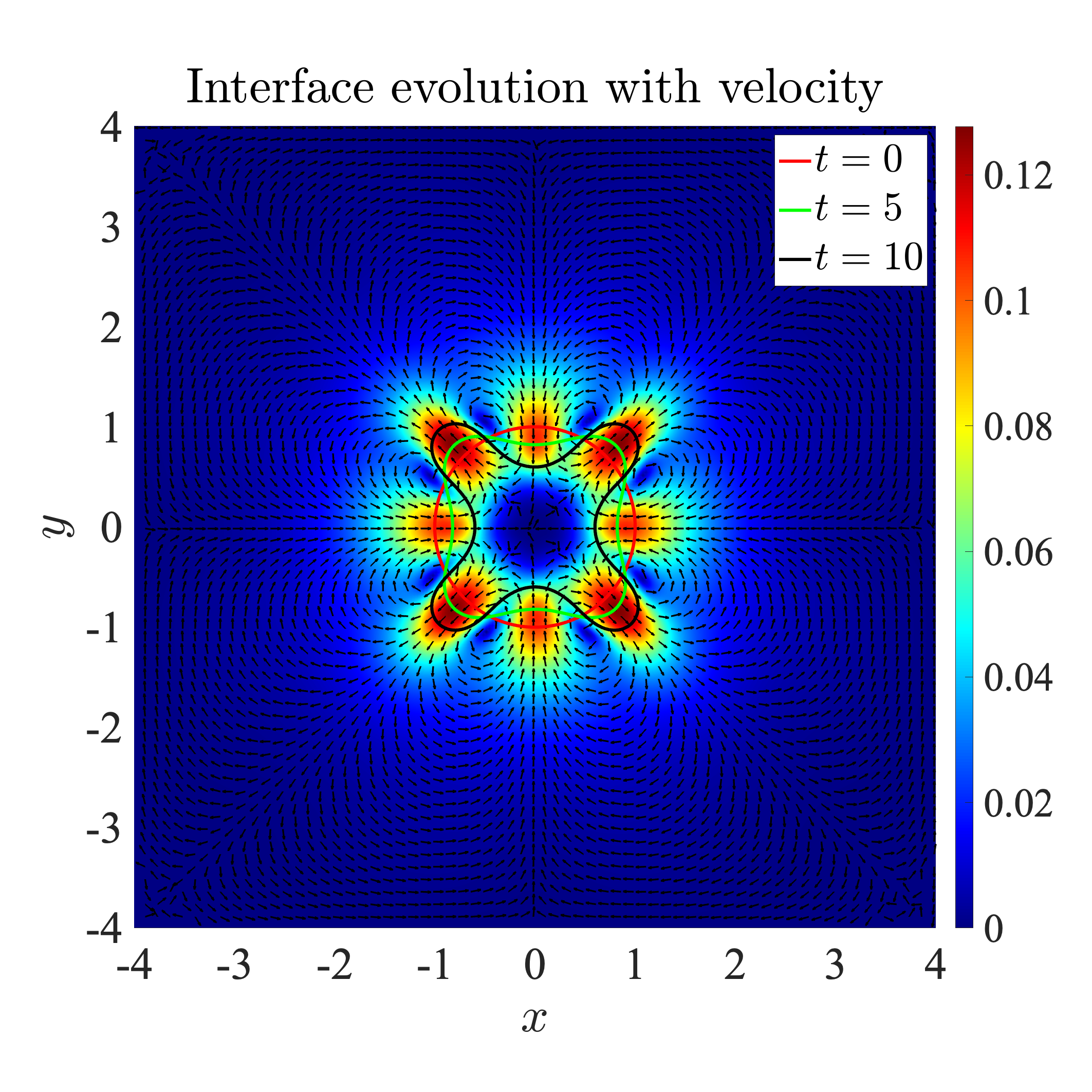}
        \label{subfig:1Drop0bdgroundPump25U10}
		}
    \vskip -0.2cm
	\caption{The snapshots of the total charge (left), net charge (middle), and the velocity (right). 
    The total charge and net charge both accumulate in the droplet due to the redistribution of positive and negative ions.}\label{fig:1Drop0bdNonuPump25V}
\end{figure}

Figs.~\ref{fig:1Drop0bdgroundPump25}-\ref{fig:1Drop0bdNonuPump25V} illustrate that, 
under a nonuniform pump distribution, more positive ions are transported into the droplet at angular positions $\theta = 0, \pi/2, \pi, 3\pi/2$. 
This localized influx leads to net negative charge accumulation at these sites outside of the droplet. 
The resulting elevated electric potential induces an outward-pointing electric field, 
generating inward Lorentz forces along the lines $x = 0$ and $y = 0$ outside the droplet. The total viscous stress induced force $\nabla\cdot\bm{\sigma}_{\eta}$ induces inwards forces that compress the droplet along the vertical and horizontal axes (see Fig. \ref{fig:1Drop0bdNonuPumpstess}).

Due to fluid incompressibility, the induced flow redirects along the diagonals, 
stretching the droplet in those directions. 
This flow further transports positive ions toward the diagonals, reinforcing charge accumulation there. 
The associated Lorentz forces act to elongate the droplet even more, 
progressively amplifying the deformation until a dynamic equilibrium is reached between the electrohydrodynamic forces and surface tension. 
The final configuration exhibits a characteristic star-like shape.  Moreover, when the pump strength is increased to $I_0 = 30$,  the deformation becomes more extreme. In this regime, the protrusions along the diagonal directions may detach from the main droplet body, as illustrated in Appendix Fig.~\ref{fig:1Drop0bdNonuPump}.

To investigate the influence of horizontal asymmetry on electrohydrodynamic behavior, we place a single droplet off-center---closer to the left boundary. The results are shown in the Appendix  Fig. \ref{fig:1LeftDropD4N0Pump25}.  The electric field intensity is higher beneath the droplet due to the shorter distance between the high-potential interior and the negatively charged lower boundary. This strong vertical field causes the droplet to be elongated along the $y$-axis.

Interestingly, the deformation is not strictly vertical. As the droplet descends, it develops a crescent-like profile with a rightward bend. This is attributed to the asymmetric confinement of the surrounding fluid. On the left side, the fluid domain is spatially constrained by the nearby vertical wall, while the right side offers a larger free space. As a result, vortical structures develop symmetrically on both sides of the droplet, but the vortex on the right becomes larger and stronger, inducing a net bias in the flow direction toward the right. This leads to an asymmetric drag distribution that deforms the rear interface into a right-bending crescent shape.

As the droplet continues to descend under the influence of the Lorentz force and the induced flow, it approaches the bottom boundary. The strong electric field in the narrow region below the droplet accelerates the downward motion of the lower interface. When the thin neck between the upper and lower parts of the droplet becomes unstable, a pinch-off event occurs, dividing the droplet into two separate parts. Subsequently, the lower part makes contact with the substrate and spreads as the symmetric case. The upper part also continues descending and eventually recoalesces with the lower portion. The final state is a flattened droplet adhered to the bottom plate.

\section{Pump-induced droplets interaction}\label{subsec:2dropsnoelectric}

In this section, we consider the motion and deformation of two droplets with the positive pump effect. 
We use the following two different label functions as the initial condition to describe the left and right droplets in the 2D computational domain $\Omega = \left[-4,4\right]^{2}$ and $\psi\left(x,y\right)=\psi_{L}+\psi_{R}+1$ where 
\begin{subequations}
\begin{align}
& \psi_{L}\left(x,y,0\right) = \tanh\frac{\sqrt{1-(\left(x+1.5\right)^{2}+y^{2})}}{\sqrt{2}\delta},
\\
& \psi_{R}\left(x,y,0\right)  = \tanh\frac{\sqrt{1-(\left(x-1.5\right)^{2}+y^{2})}}{\sqrt{2}\delta}.
\end{align}
\end{subequations}
The ion and potential initial conditions are chosen as follows  
\begin{subequations}
\begin{align}
& p\left(x,y,0\right) = 1, \quad n\left(x,y,0\right) = 1, \quad \phi\left( x,y,0 \right) = 0. 
\end{align}
\end{subequations}

The following boundary condition is set to make this system self-contained 
\begin{equation}
\begin{aligned}
\left. \nabla \psi \cdot \bm{n} \right|_{\partial \Omega} = 0, 
\quad 
\left. \nabla \mu_{\psi} \cdot \bm{n} \right|_{\partial \Omega} = 0, 
\quad 
\left. \bm{u} \right|_{\partial \Omega} = \bm{0}, 
\quad 
\left. \nabla p \cdot \bm{n} \right|_{\partial \Omega} = 0, 
\quad 
\left. \nabla n \cdot \bm{n} \right|_{\partial \Omega} = 0. 
\end{aligned}
\end{equation}
The parameters are chosen as 
\begin{equation}
\begin{aligned}
& Re = 1, \quad Ca_{E} = 0.5, \quad \zeta = 0.1, \quad Pe = 1, \quad Pe_{E} = 1, 
\\ 
& \alpha_{1} = 0.5, \quad \alpha_{2} = -0.5, \quad 
I_{0} = 25, \quad \beta = 2, \quad K_{0} = 3.5, 
\\ 
& \delta = 0.1, \quad M = \delta^{2}, \quad \epsilon_{r} = 1, \quad D_{ir} = 1.
\end{aligned}
\end{equation}

We first consider a configuration where the computational domain is bounded by grounded vertical boundaries and electrically insulating horizontal boundaries. The electric potential satisfies the following boundary conditions:
\begin{equation}\label{bd:2drops_b0d0}
\left. \phi \right|_{y=4} = 0, 
\quad 
\left. \phi \right|_{y=-4} = 0, 
\quad 
\left. \nabla \phi \cdot \bm{n} \right|_{x = \pm 4} = 0. 
\end{equation}

\begin{figure}[!ht]
\vskip -0.4cm 
\centering
	\subfloat[Positive ion $p$ at $t=10$.]{
		\includegraphics[width=0.33\linewidth]{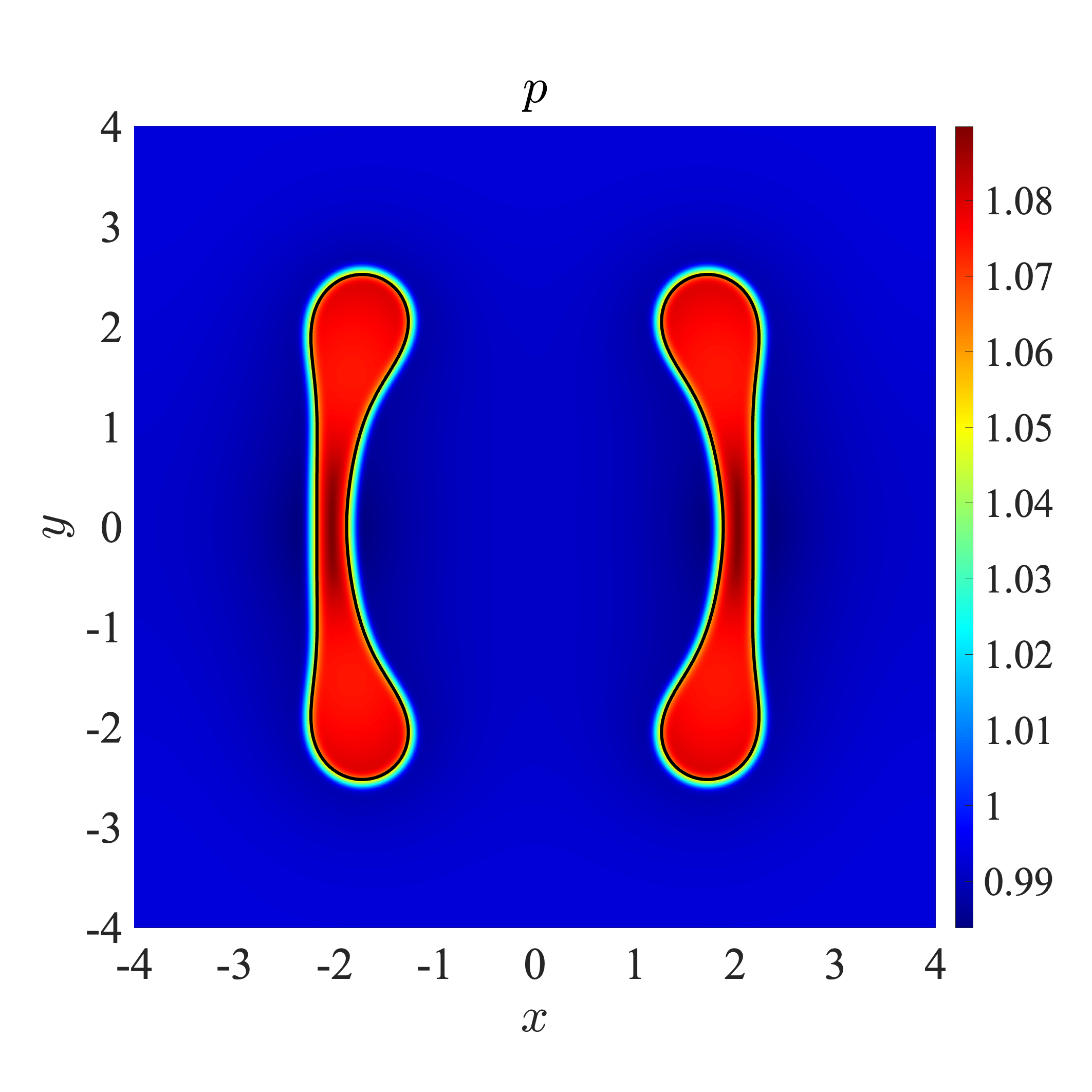}
        \label{subfig:2Drops0bdPump25P10}
		}
    \hskip -0.3cm
	\subfloat[Negative ion $n$ at $t=10$.]{
		\includegraphics[width=0.33\linewidth]{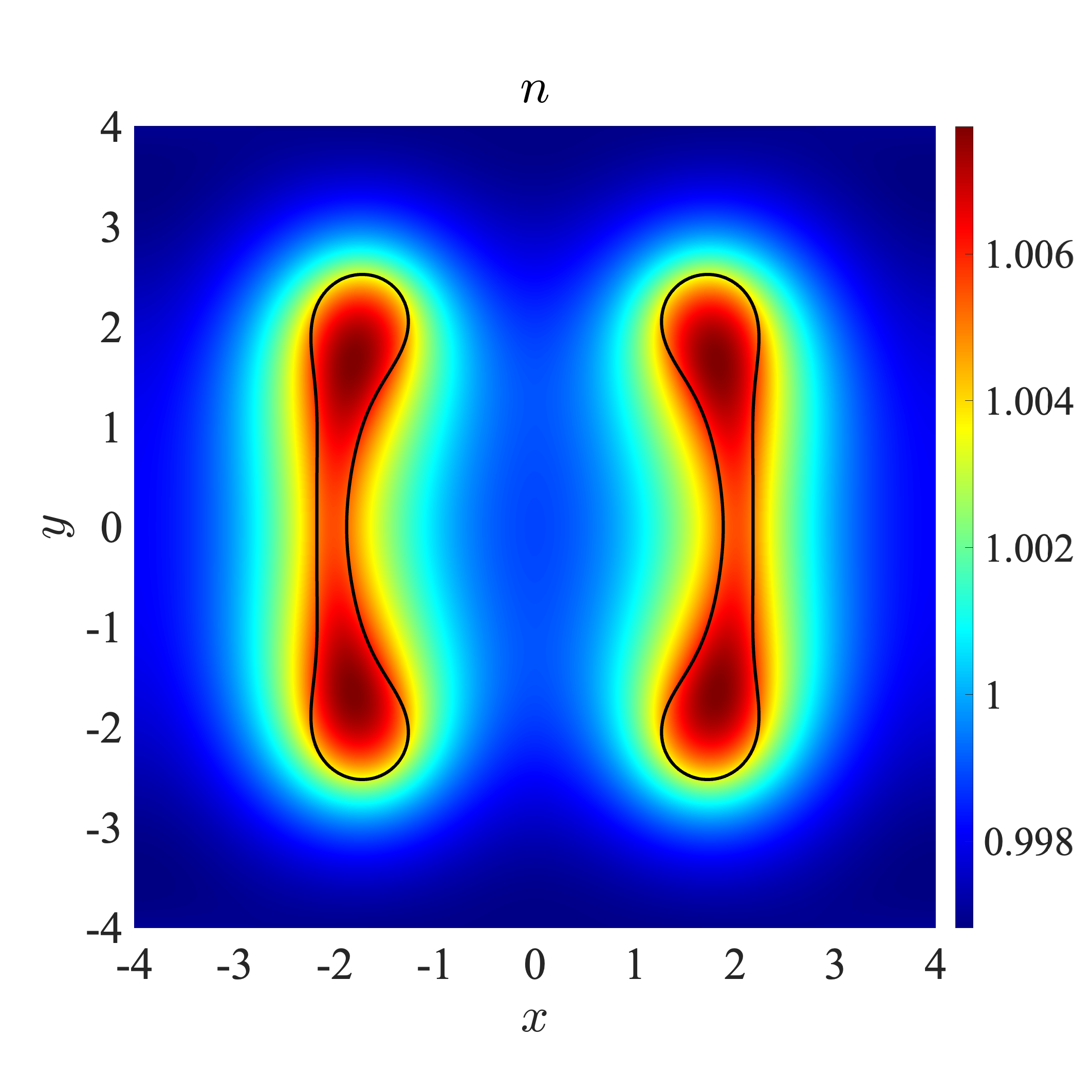}
        \label{subfig:2Drops0bdPump25N10}
		}
    \hskip -0.3cm
	\subfloat[Electric potential $\phi$ at $t=10$.]{
		\includegraphics[width=0.33\linewidth]{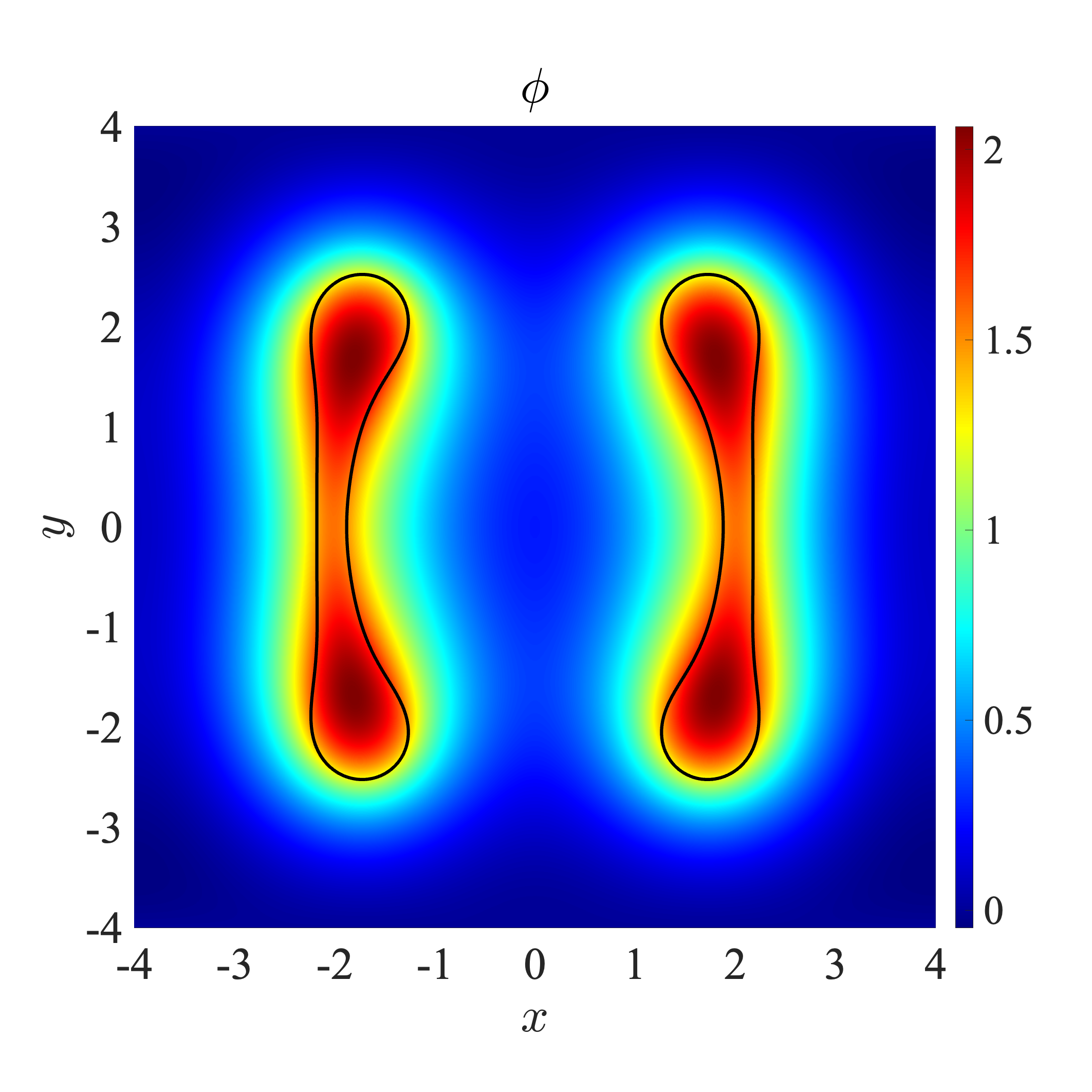}
        \label{subfig:2Drops0bdPump25Phi10}
		}
    \\
    \vskip -0.3cm
    \subfloat[Total charge $p+n$ at $t = 10$.]{ 
		\includegraphics[width=0.33\linewidth]{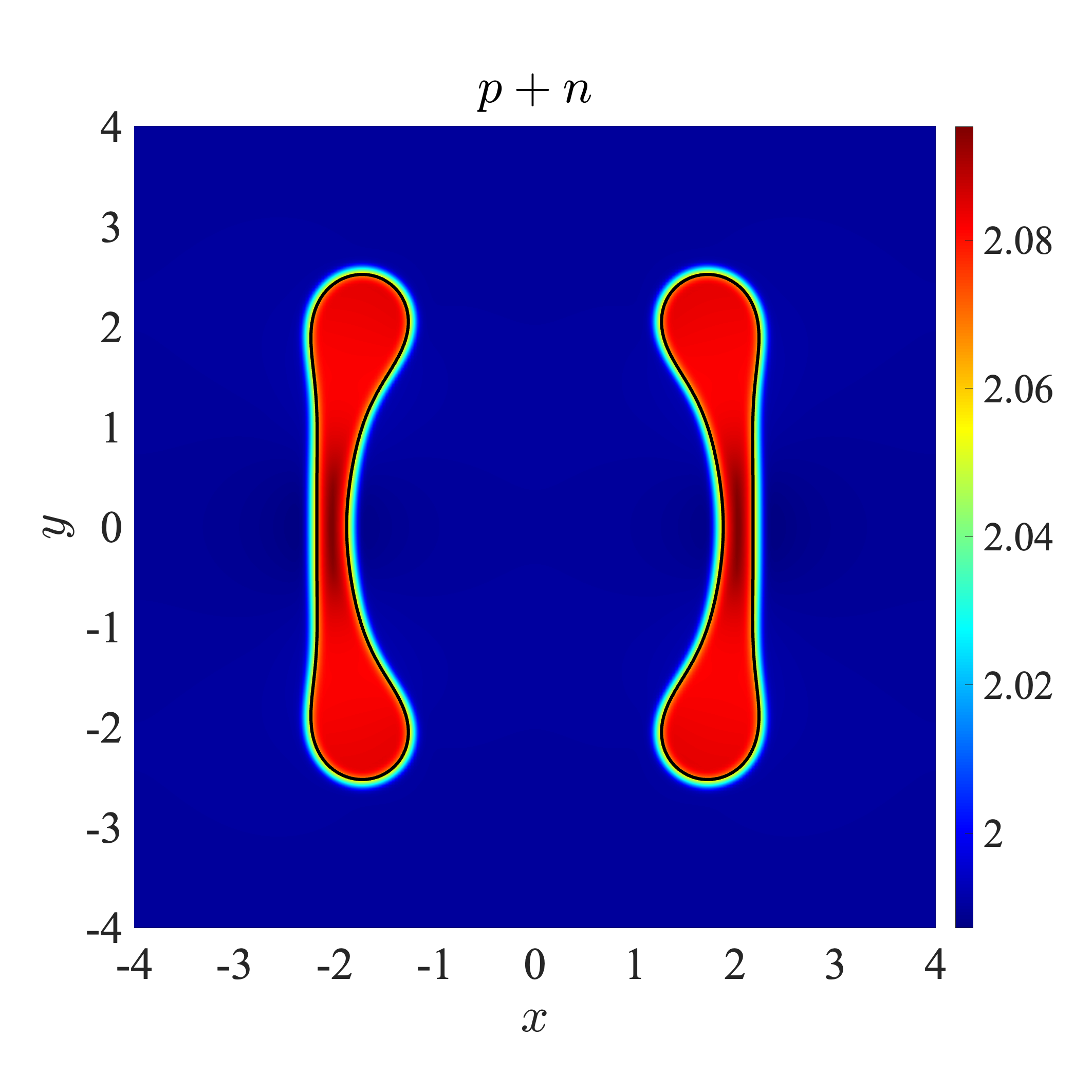}
        \label{subfig:2Drops0bdPump25Sum10}
	} 
    \hskip -0.3cm
	\subfloat[Net charge $p-n$ at $t = 10$.]{ 
		\includegraphics[width=0.33\linewidth]{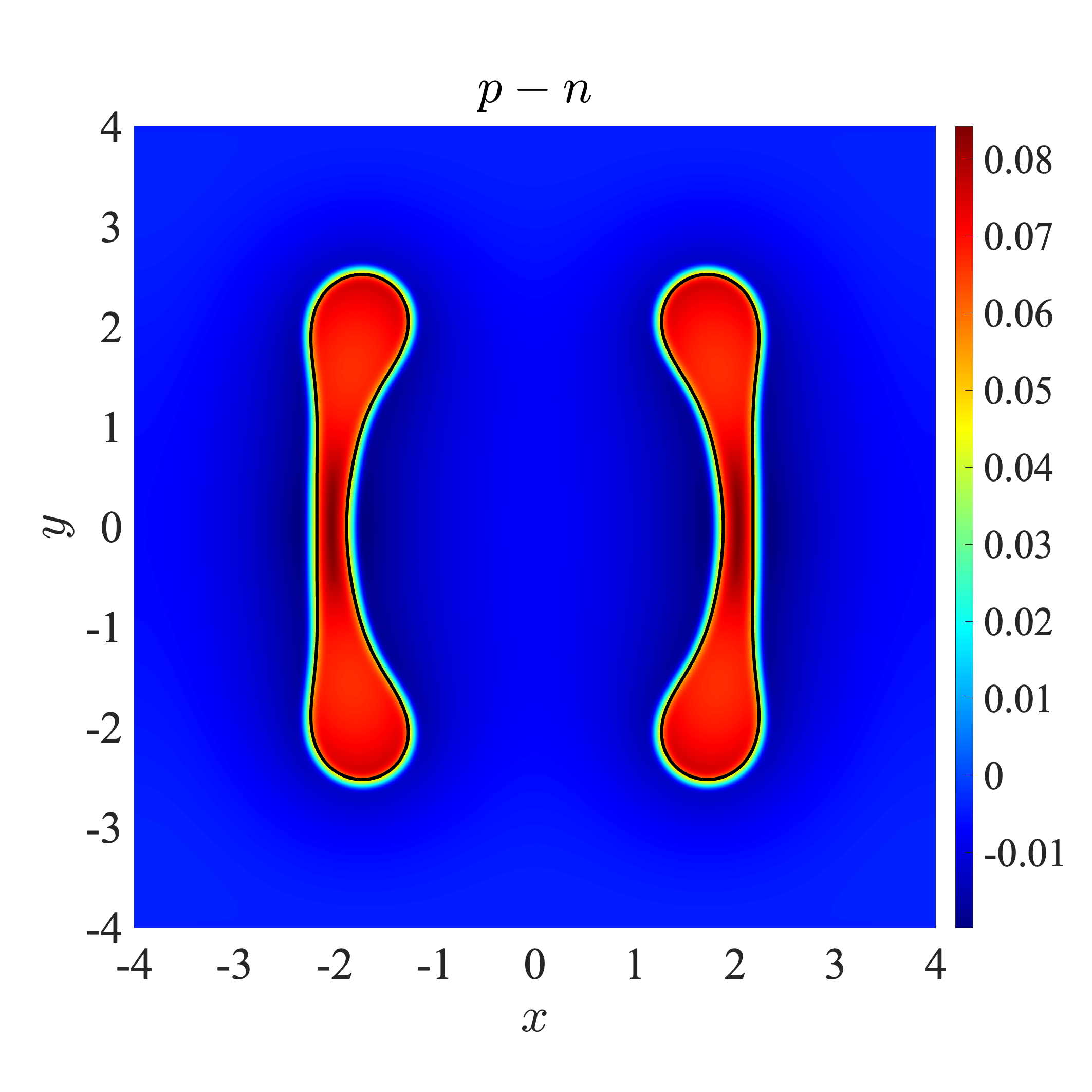}
        \label{subfig:2Drops0bdPump25Dif10}
	}
    \hskip -0.3cm
	\subfloat[Velocity and deformation at $t=10$.]{ 
		\includegraphics[width=0.33\linewidth]{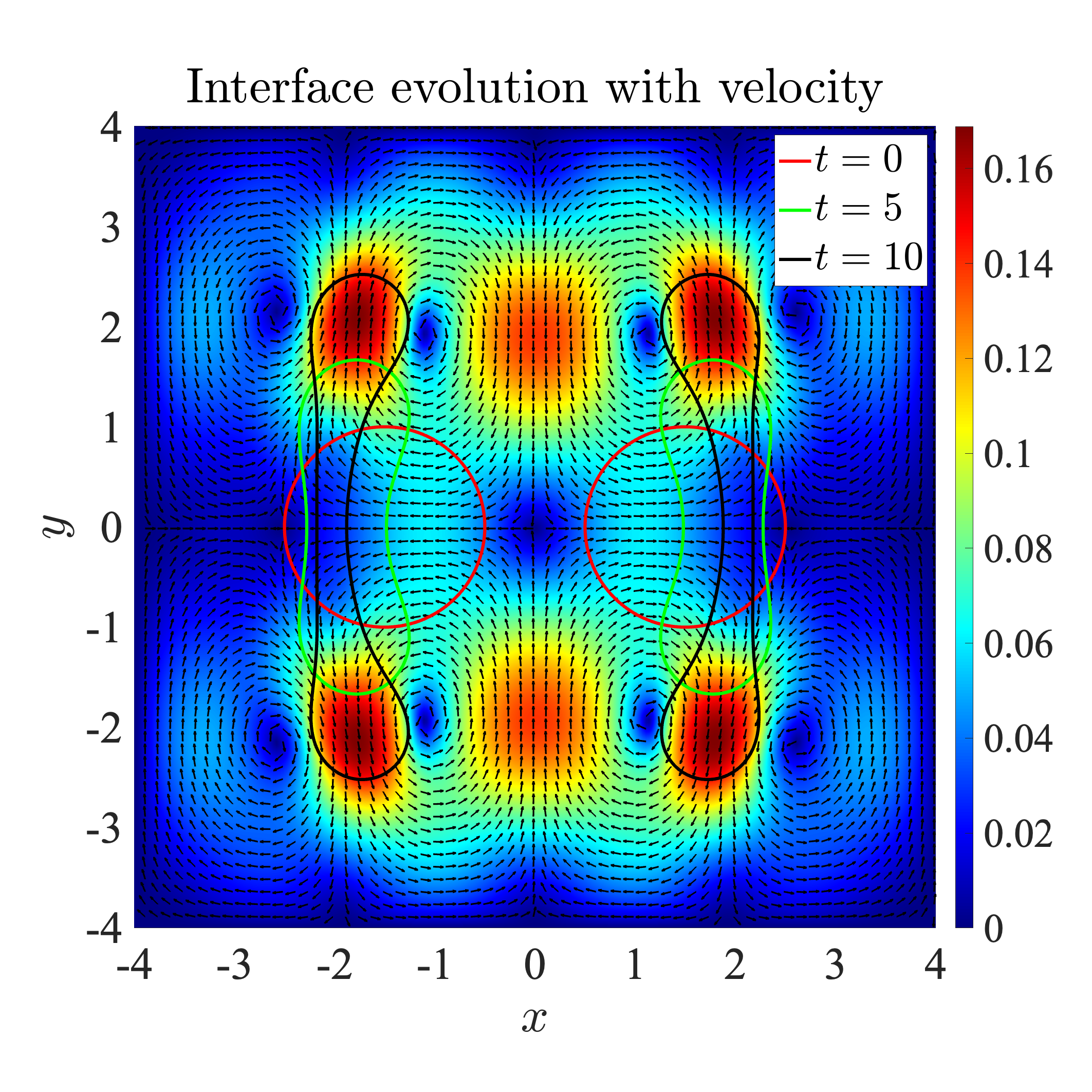}
        \label{subfig:2Drops0bdPump25U10}
	}
	\caption{The snapshots for the drop deformation with positive ion pump for the boundary condition \eqref{bd:2drops_b0d0}. 
    Top row: (left) positive ion; (middle) negative ion; (right) potential; 
    Bottom row: the total charge (left), net charge (middle), and the velocity (right). 
    The black solid circle represents the location of the drop, 
    which is denoted by the level set $\psi=0$. 
    The concentration and electric potential distribution are shown on the color map. 
    The bottom and upper plates are grounded. 
    We choose the final time as $t = 10$. 
    }\label{fig:2Drops0bdPump25}
\end{figure}

The corresponding simulation results are presented in Fig.~\ref{fig:2Drops0bdPump25}. 
The accumulation of positive charge within each droplet generates electrostatic repulsion in the horizontal direction. This causes the two droplets to repel one another laterally. At the same time, the elevated internal electric potential creates nonuniform electric fields around each droplet, which in turn generate Lorentz forces acting both in the vertical ($y$) and horizontal ($x$) directions. Due to the stronger electric field gradients along the vertical axis—arising from domain boundary conditions—the $y$-component of the Lorentz force dominates, leading to vertical elongation of each droplet as shown in the single droplet case.

During this deformation process, the induced flow fields exhibit symmetric counter-rotating vortices around each droplet, as visualized in the velocity field plots Fig. \ref{subfig:2Drops0bdPump25U10}. These vortical structures impose compressive shear stress near the mid-height of each droplet. The combined effects of electrohydrodynamic stretching and flow-induced compression lead to a characteristic biconcave deformation, where the droplets appear narrow at their centers and broaden toward the top and bottom edges.

Then, we add a vertical applied electric field, i.e., the boundary condition for the electric potential is chosen as 
\begin{equation}\label{bd:2drops_D4N0}
\left. \phi \right|_{y= 4} = 4, 
\quad 
\left. \phi \right|_{y= - 4} = - 4, 
\quad 
\left. \nabla \phi \cdot \bm{n} \right|_{x = \pm 4} = 0. 
\end{equation}

\begin{figure}[!ht]
\vskip -0.4cm
\centering
	\subfloat[$p(t=1)$]{
		\includegraphics[width=0.165\linewidth]{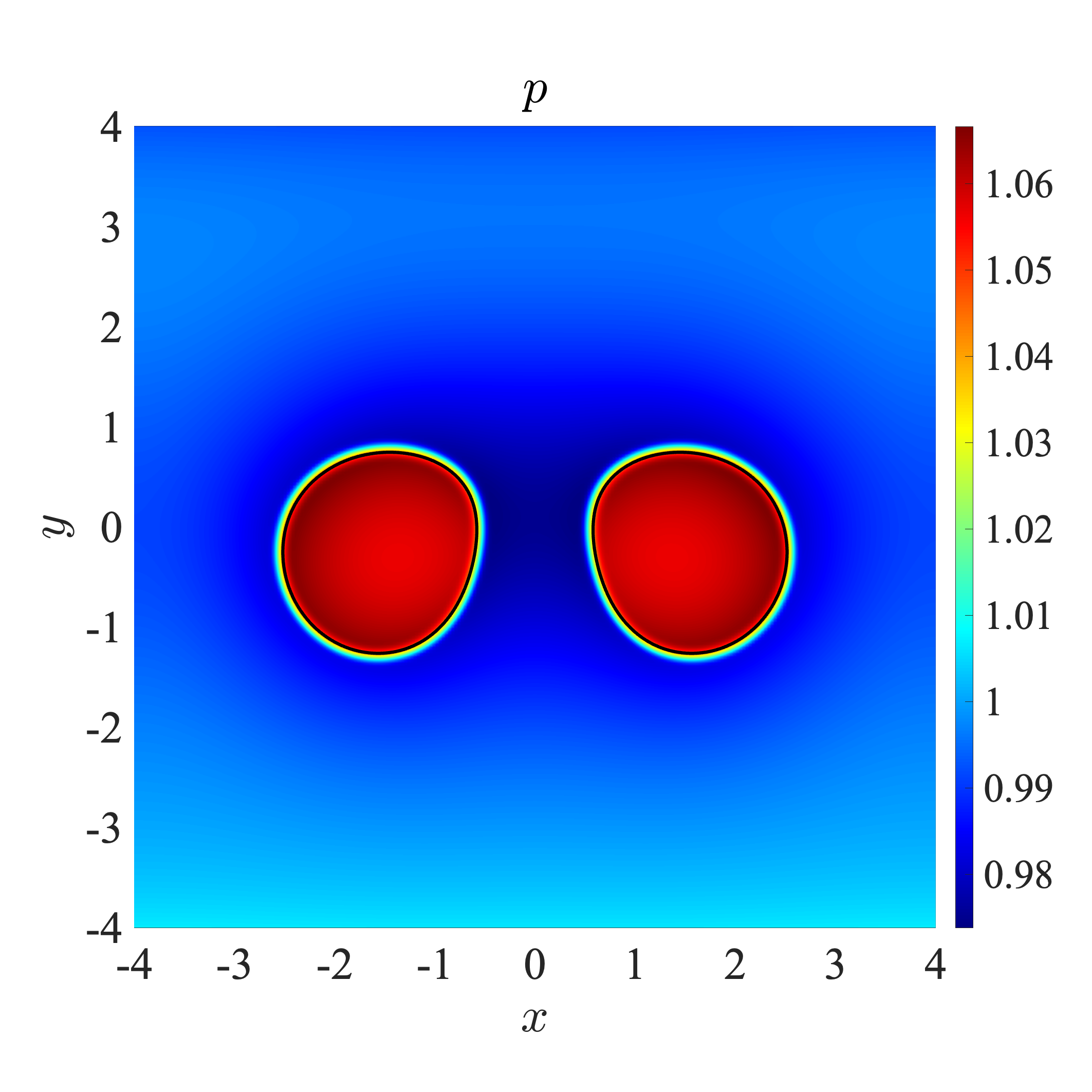}
        \label{subfig:2DropsD4N0Pump25P1}
		}
    \hskip -0.3cm
    \subfloat[$p(t=3)$]{
		\includegraphics[width=0.165\linewidth]{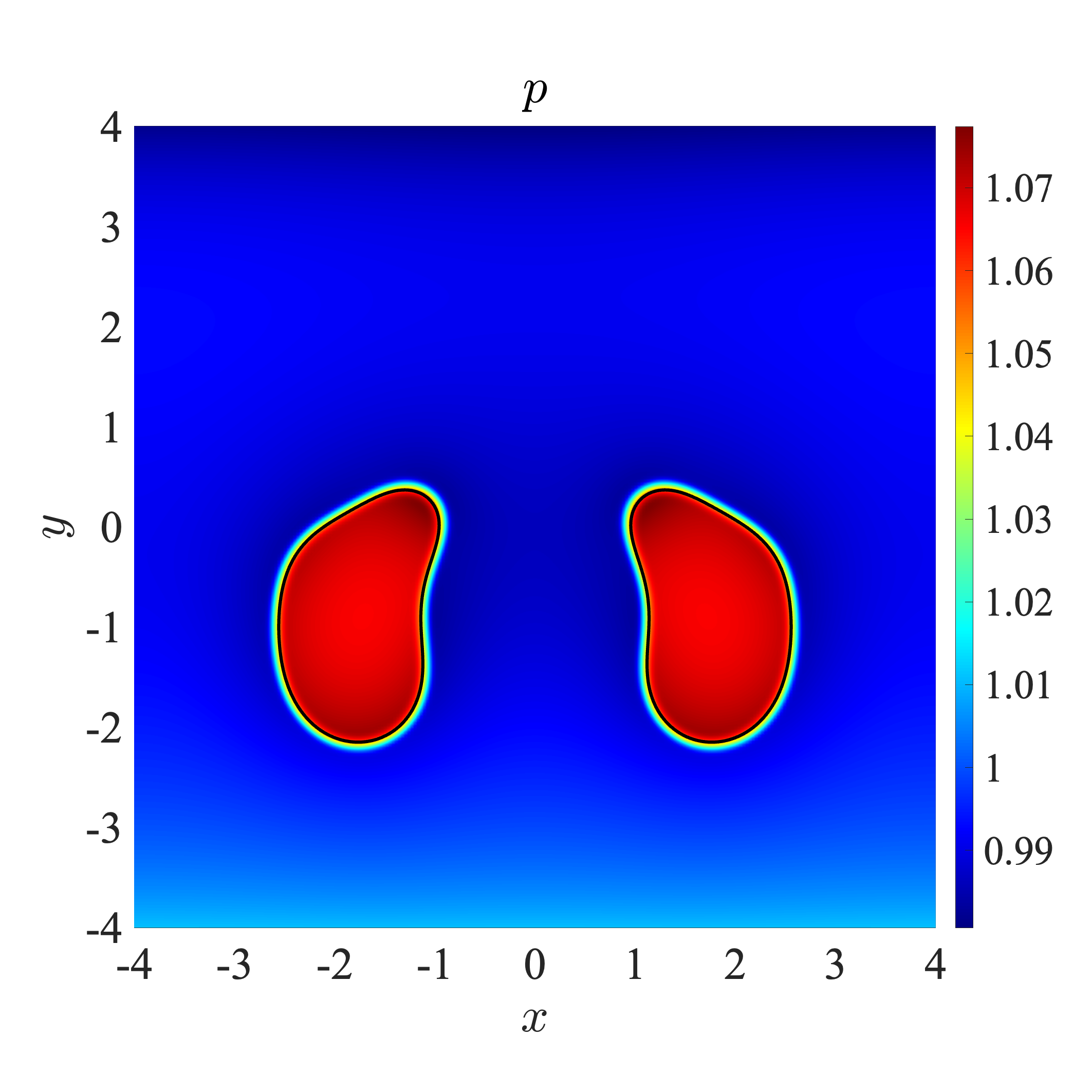}
        \label{subfig:2DropsD4N0Pump25P3}
		}
    \hskip -0.3cm
    \subfloat[$p(t=5)$]{
		\includegraphics[width=0.165\linewidth]{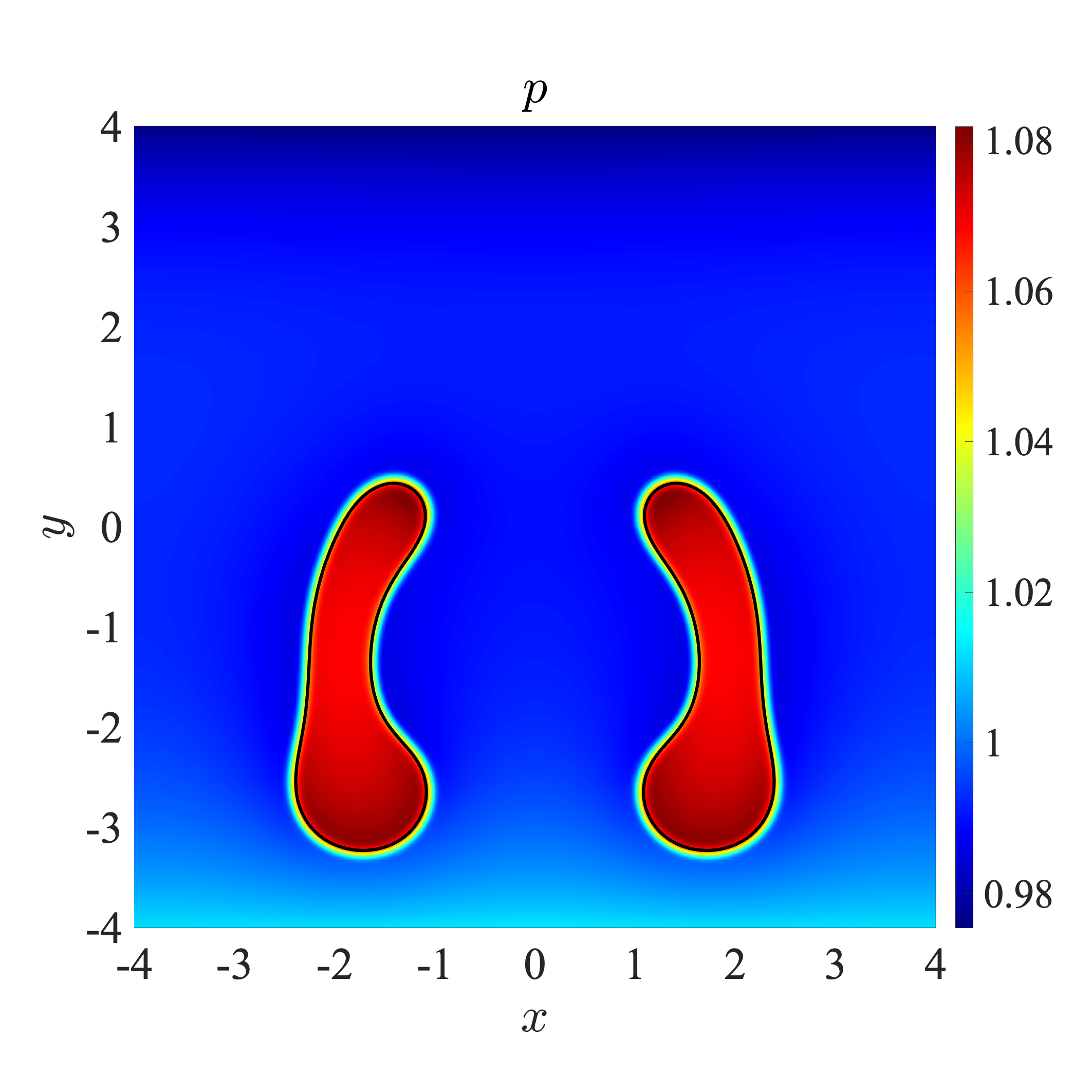}
        \label{subfig:2DropsD4N0Pump25P5}
		}
    \hskip -0.3cm
    \subfloat[$p(t=8)$]{
		\includegraphics[width=0.165\linewidth]{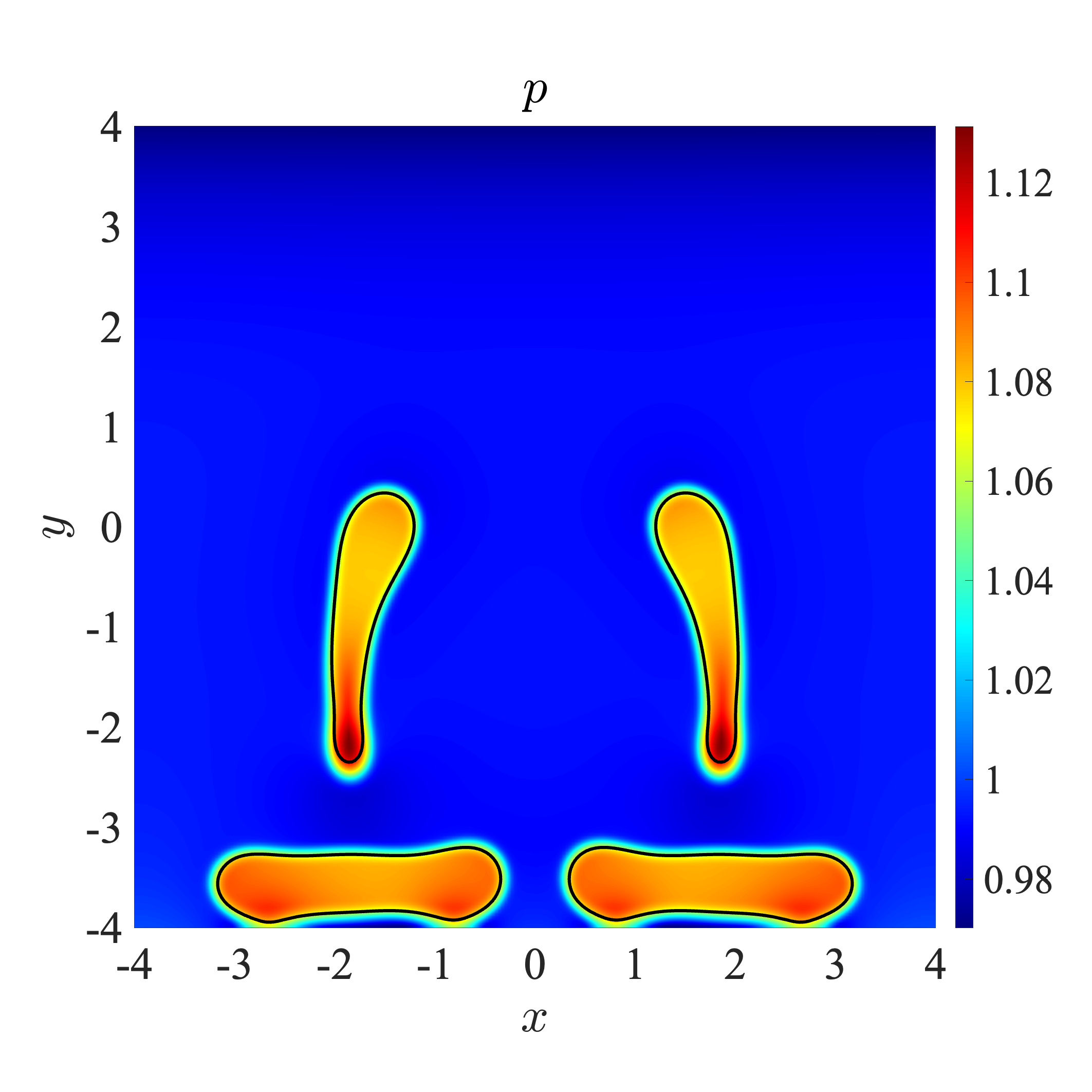}
        \label{subfig:2DropsD4N0Pump25P8}
		}
    \hskip -0.3cm
    \subfloat[$p(t=10)$]{
		\includegraphics[width=0.165\linewidth]{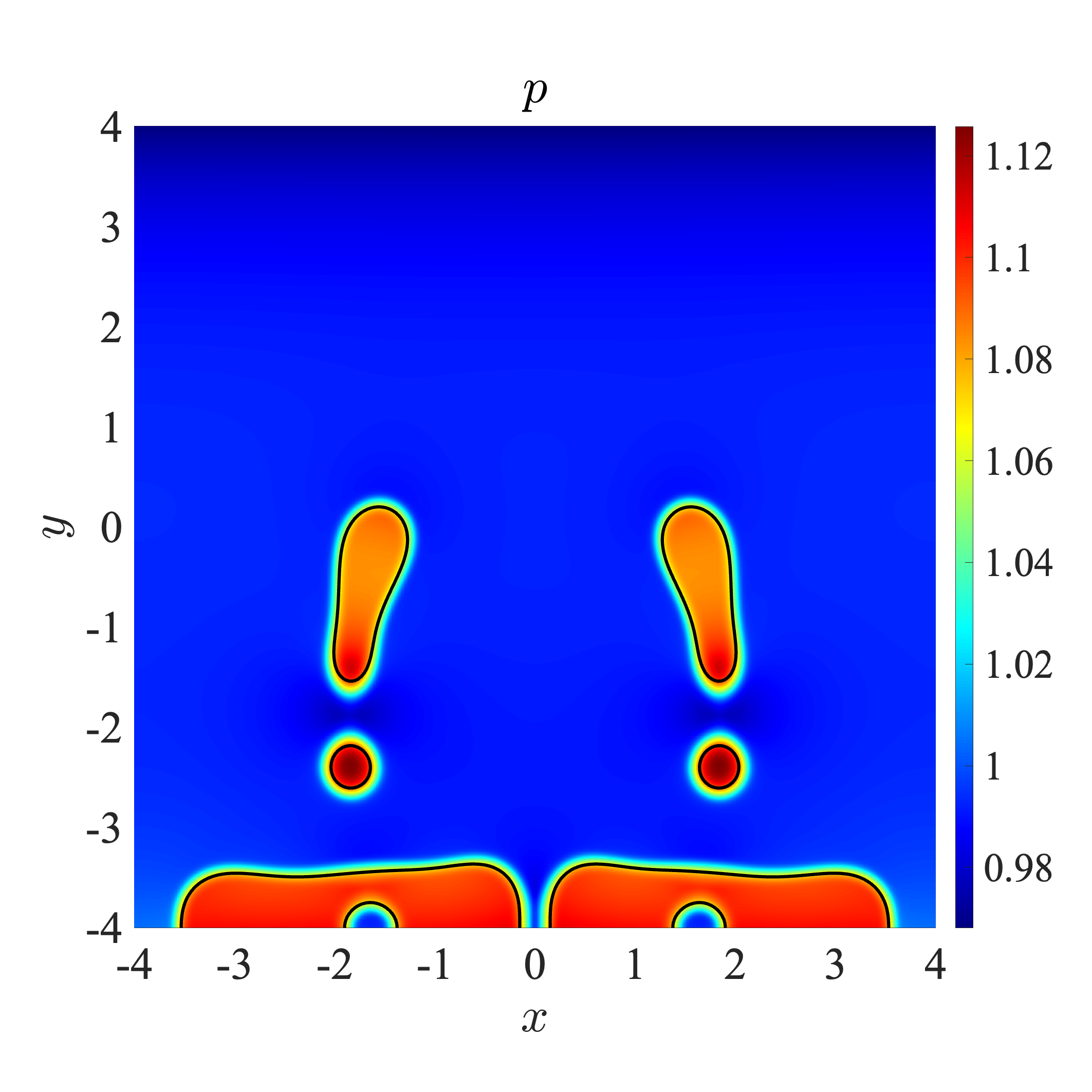}
        \label{subfig:2DropsD4N0Pump25P10}
		}
    \hskip -0.3cm
    \subfloat[$p(t=60)$]{
		\includegraphics[width=0.165\linewidth]{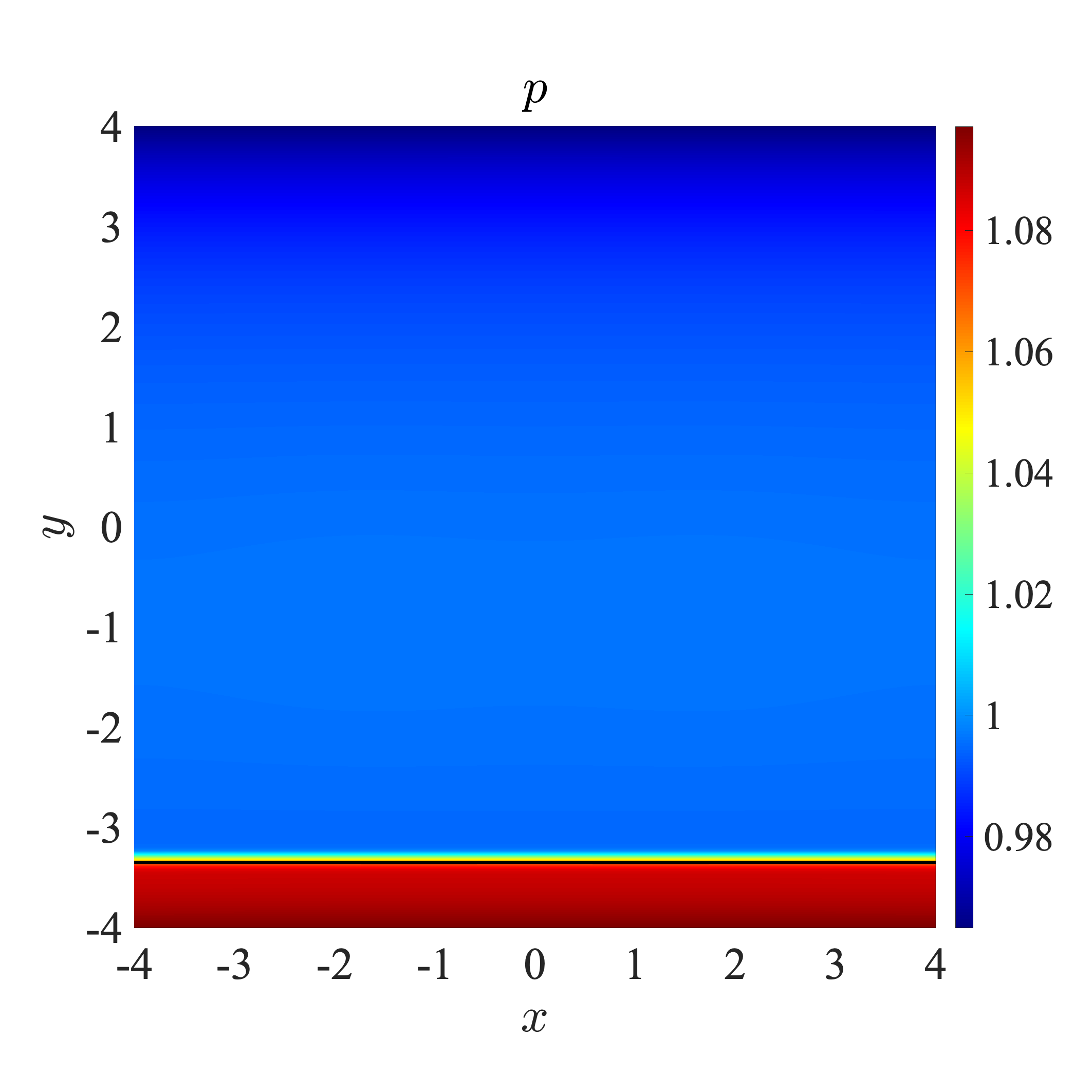}
        \label{subfig:2DropsD4N0Pump25P60}
        }
    \\
    \vskip -0.3cm
    \subfloat[$n(t=1)$]{
		\includegraphics[width=0.165\linewidth]{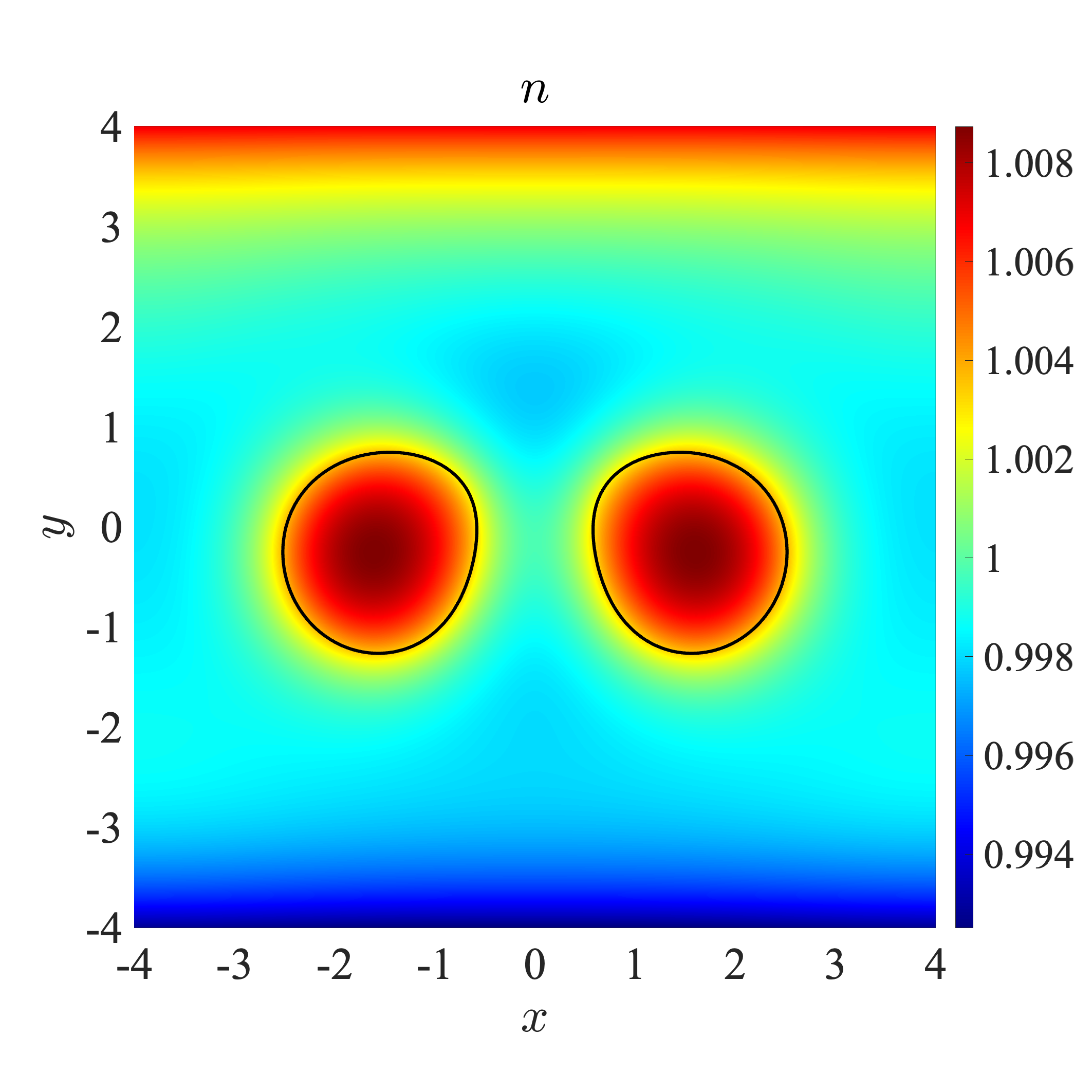}
        \label{subfig:2DropsD4N0Pump25N1}
		}
    \hskip -0.3cm
    \subfloat[$n(t=3)$]{
		\includegraphics[width=0.165\linewidth]{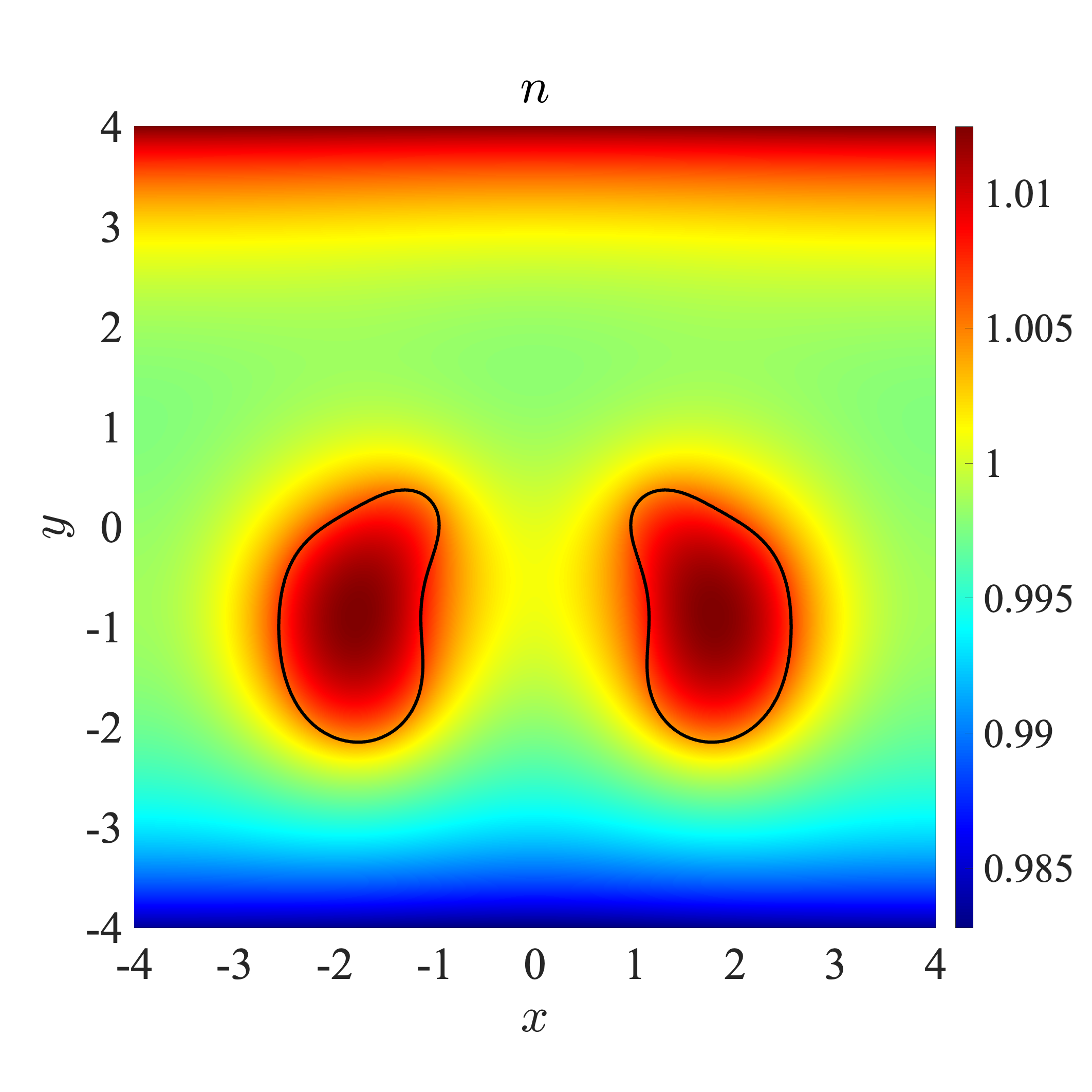}
        \label{subfig:2DropsD4N0Pump25N3}
		}
    \hskip -0.3cm
    \subfloat[$n(t=5)$]{
		\includegraphics[width=0.165\linewidth]{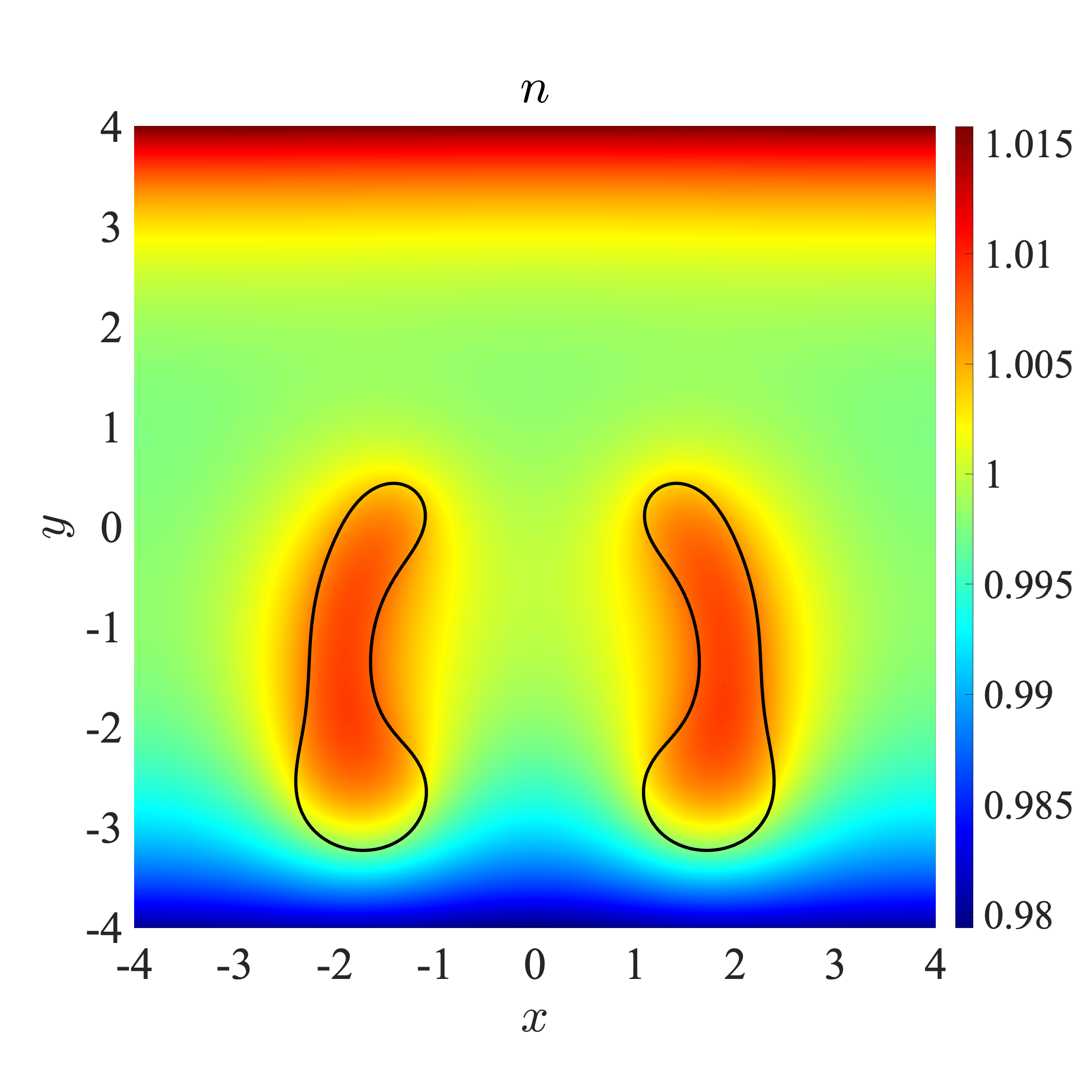}
        \label{subfig:2DropsD4N0Pump25N5}
		}
    \hskip -0.3cm
    \subfloat[$n(t=8)$]{
		\includegraphics[width=0.165\linewidth]{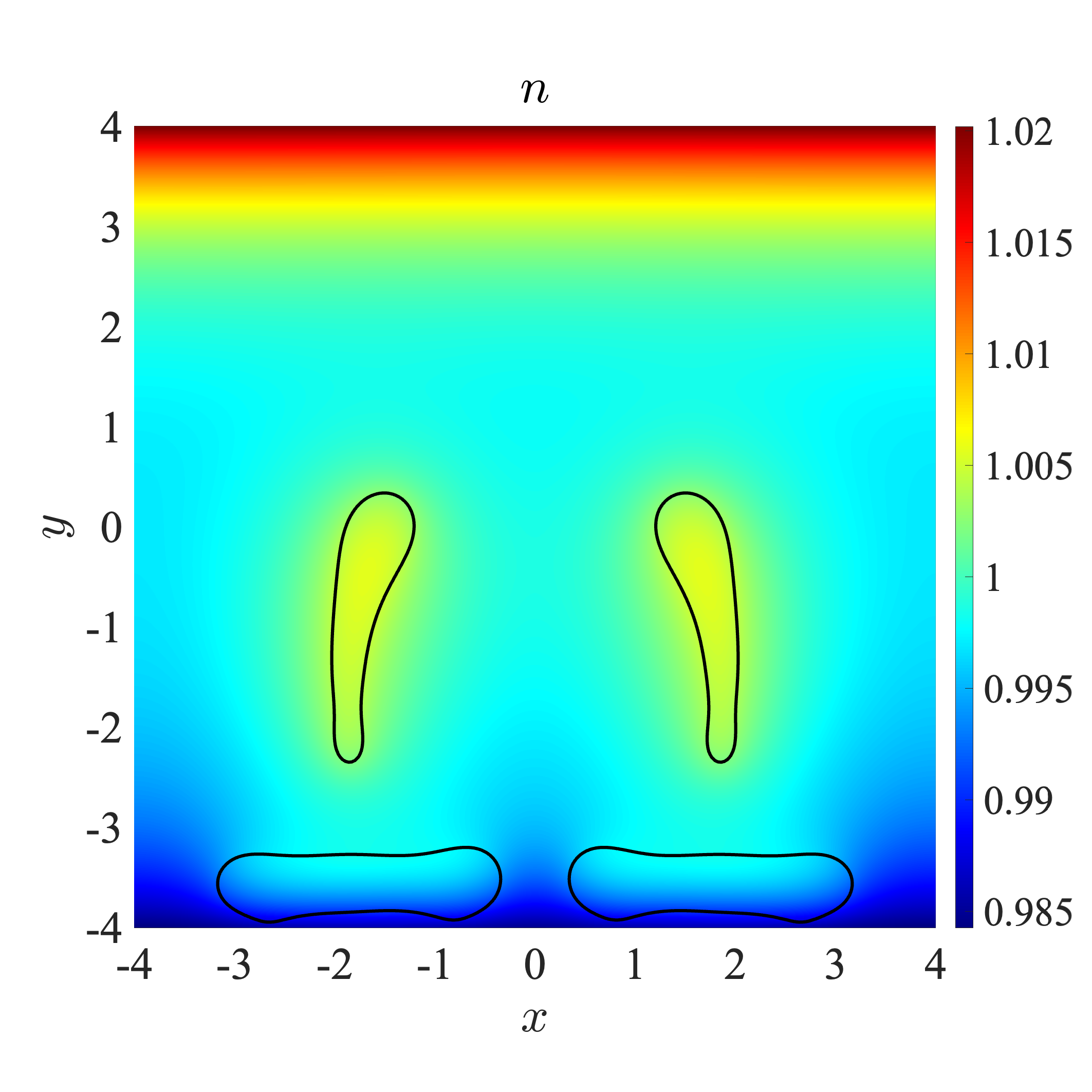}
        \label{subfig:2DropsD4N0Pump25N8}
		}
    \hskip -0.3cm
    \subfloat[$n(t=10)$]{
		\includegraphics[width=0.165\linewidth]{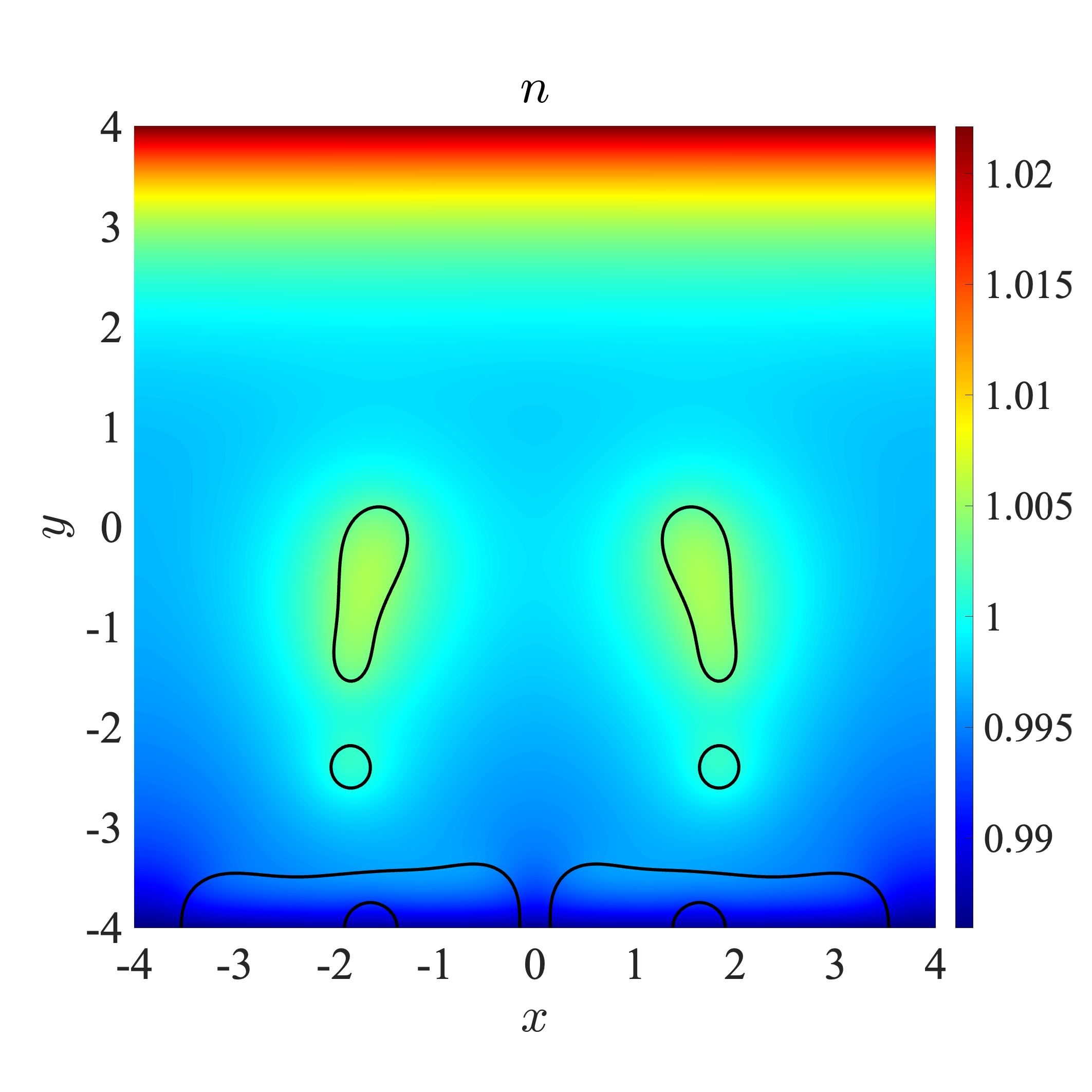}
        \label{subfig:2DropsD4N0Pump25N10}
		}
    \hskip -0.3cm
    \subfloat[$n(t=60)$]{
		\includegraphics[width=0.165\linewidth]{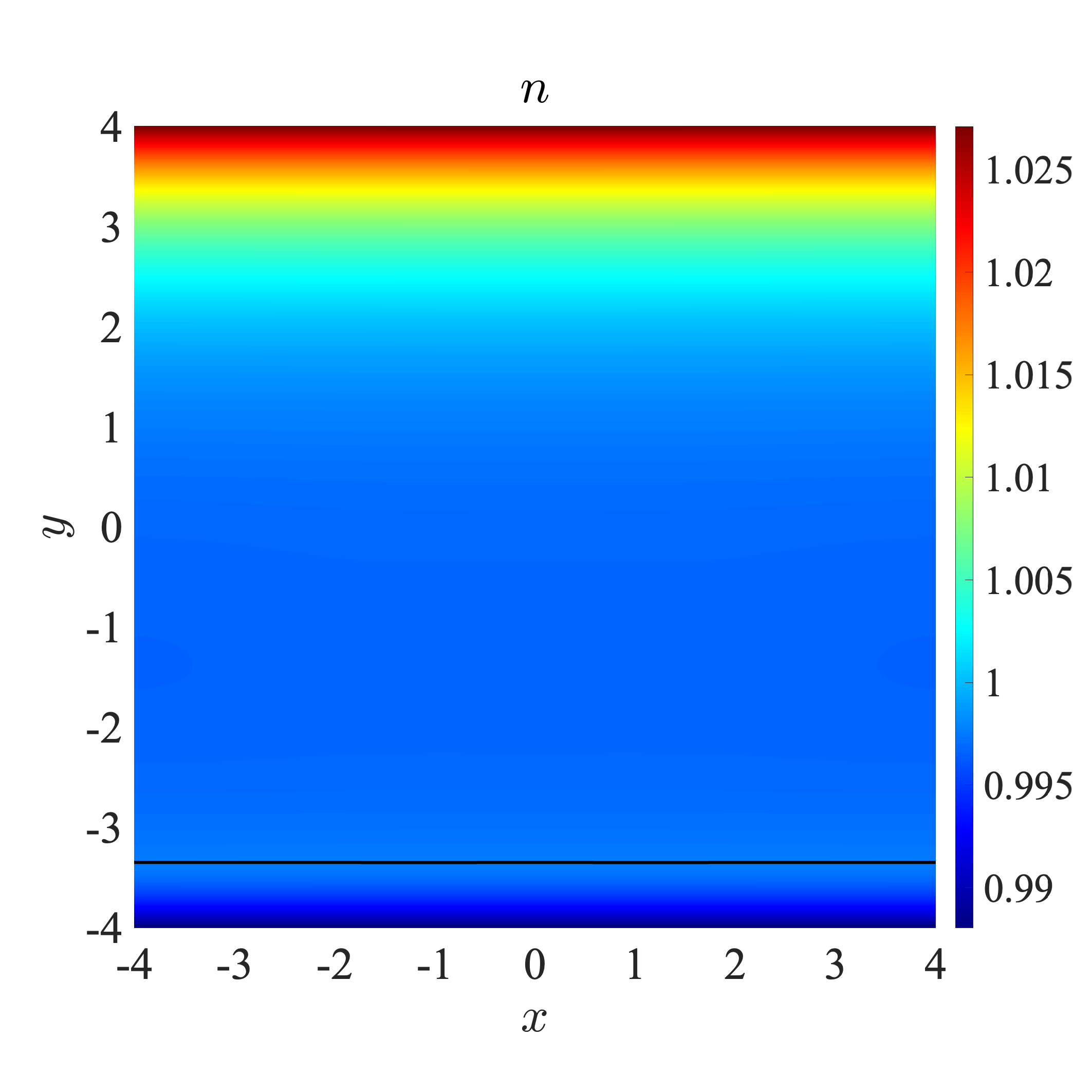}
        \label{subfig:2DropsD4N0Pump25N60}
		}
    \\
    \vskip -0.3cm
    \subfloat[$\phi(t=1)$]{
		\includegraphics[width=0.165\linewidth]{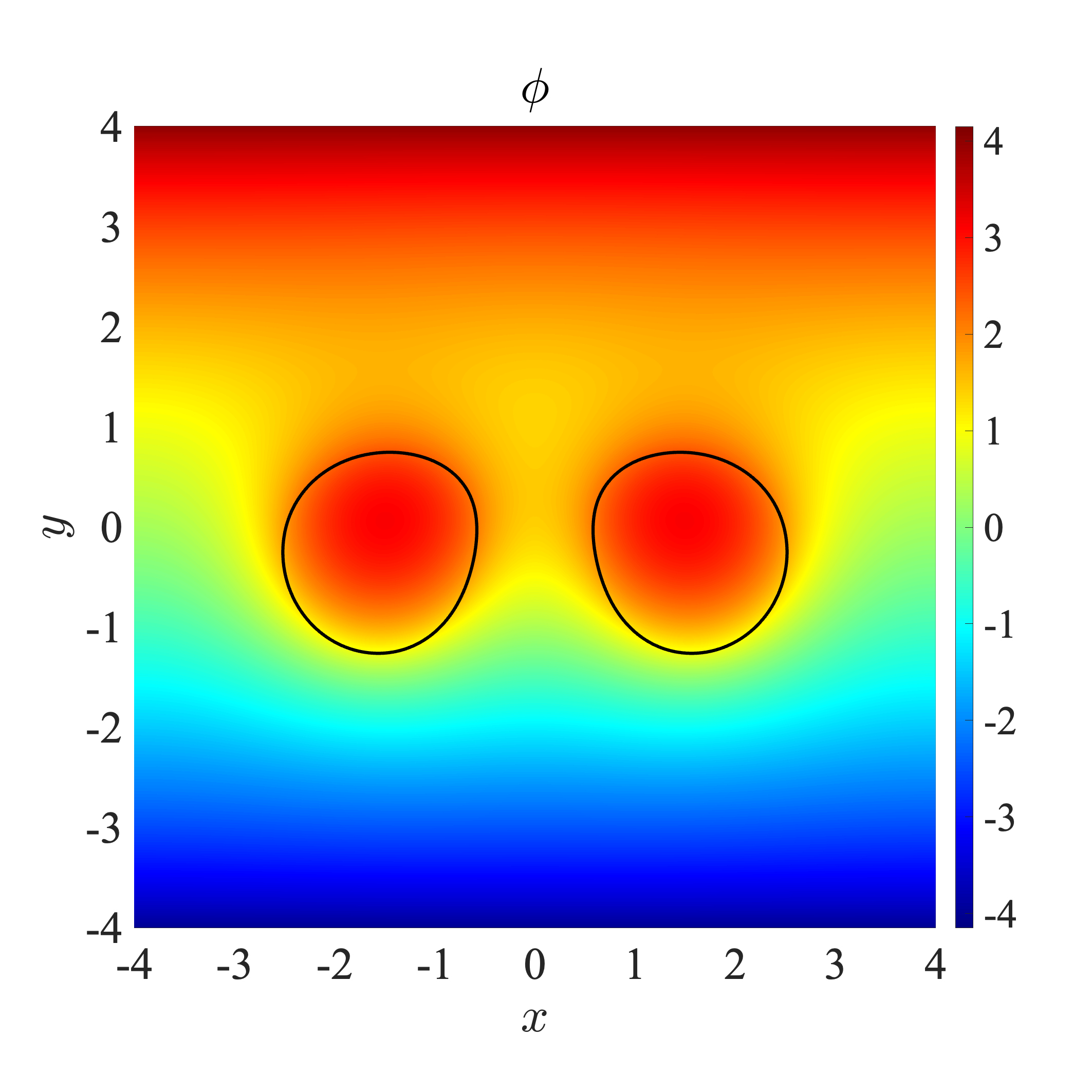}
        \label{subfig:2DropsD4N0Pump25Phi1}
		}
    \hskip -0.3cm
    \subfloat[$\phi(t=3)$]{
		\includegraphics[width=0.165\linewidth]{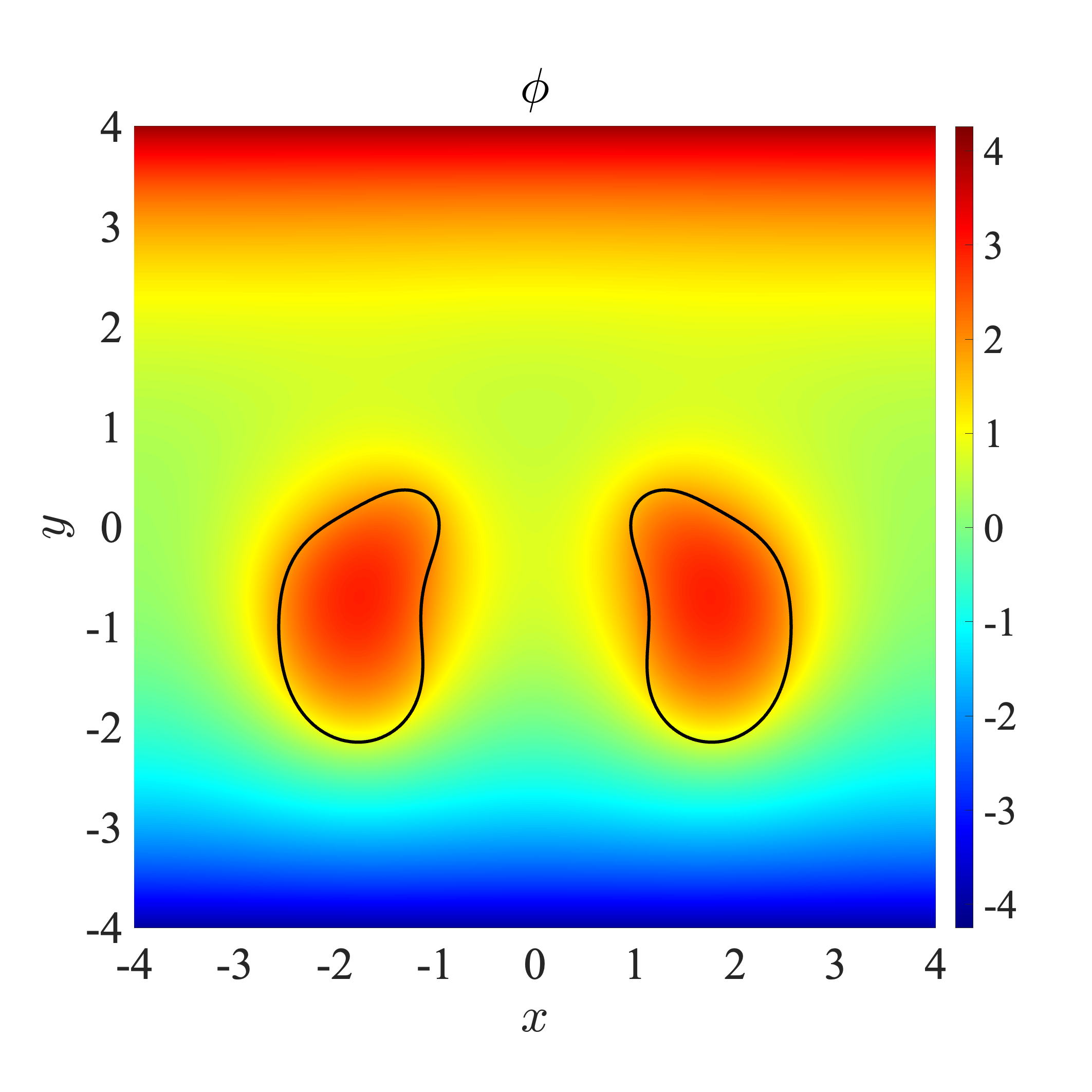}
        \label{subfig:2DropsD4N0Pump25Phi3}
		}
    \hskip -0.3cm
    \subfloat[$\phi(t=5)$]{
		\includegraphics[width=0.165\linewidth]{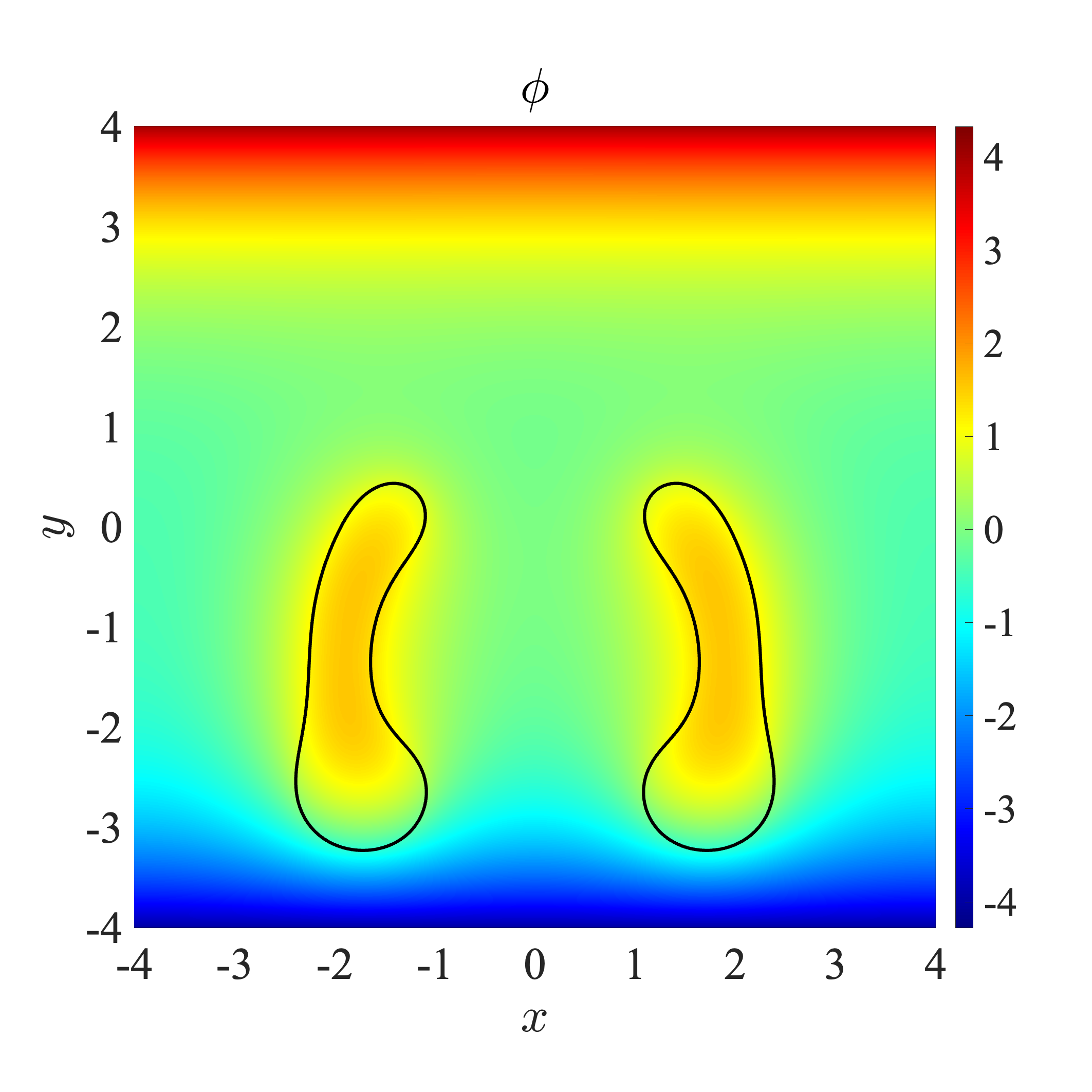}
        \label{subfig:2DropsD4N0Pump25Phi5}
		}
    \hskip -0.3cm
    \subfloat[$\phi(t=8)$]{
		\includegraphics[width=0.165\linewidth]{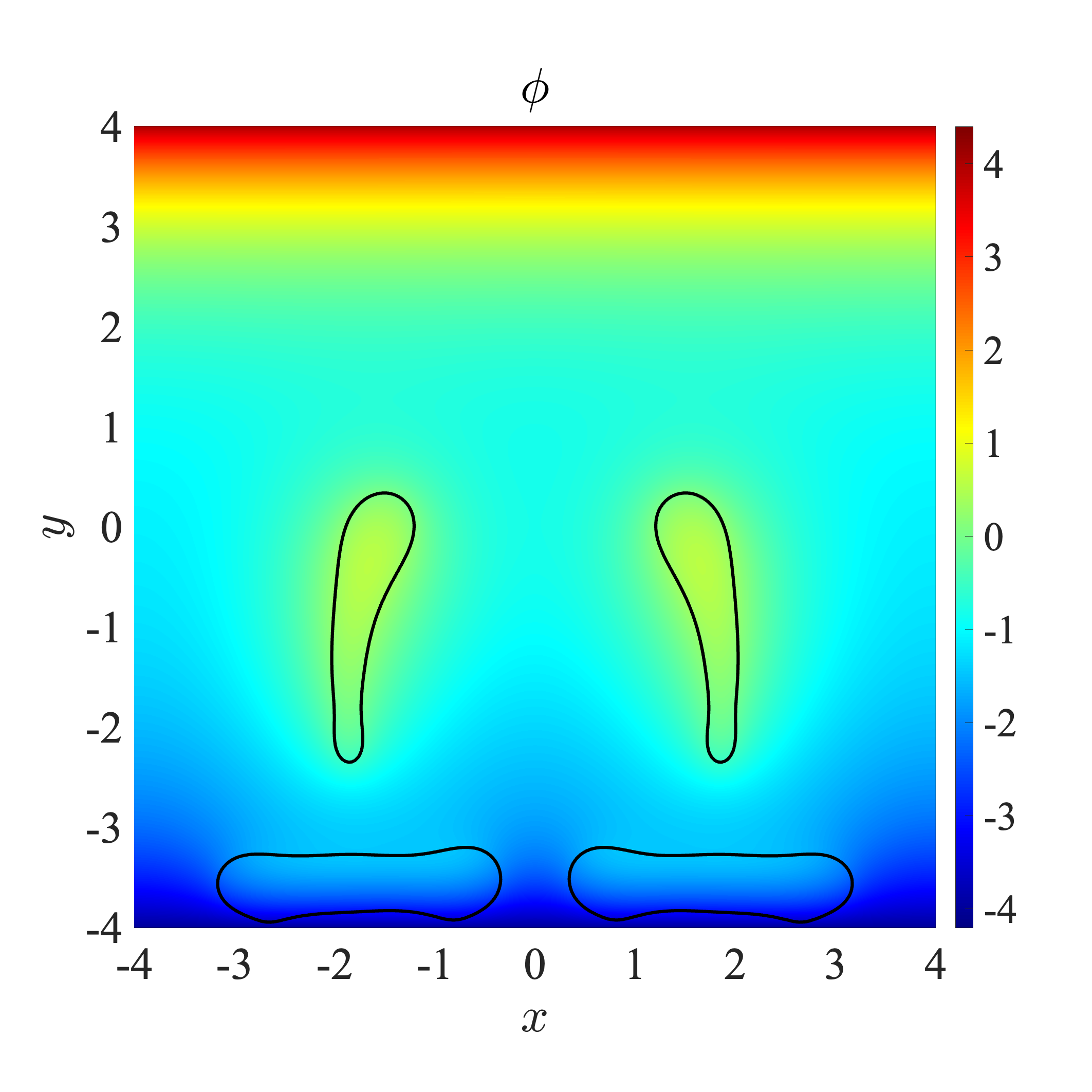}
        \label{subfig:2DropsD4N0Pump25Phi8}
		}
    \hskip -0.3cm
    \subfloat[$\phi(t=10)$]{
		\includegraphics[width=0.165\linewidth]{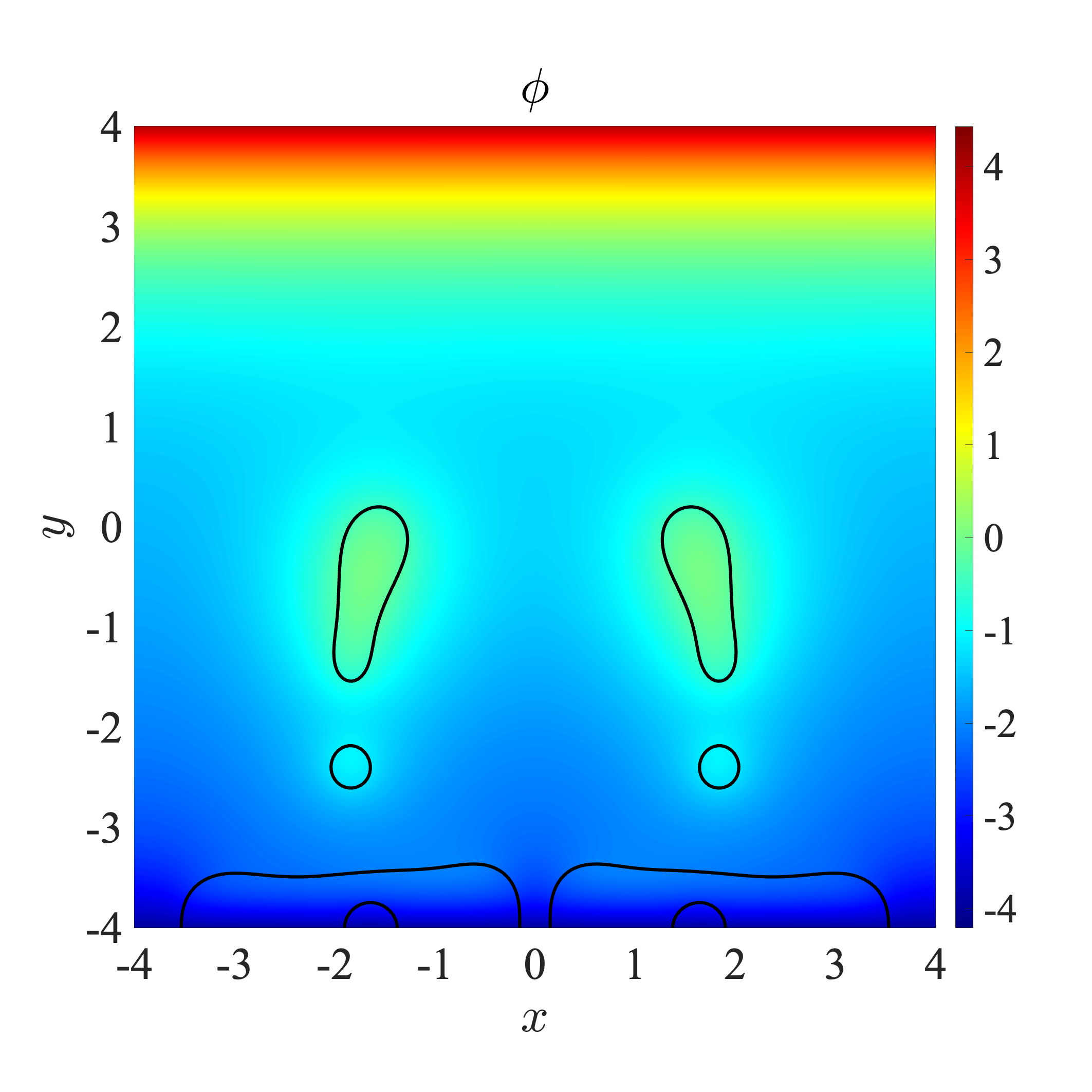}
        \label{subfig:2DropsD4N0Pump25Phi10}
		}
    \hskip -0.3cm
    \subfloat[$\phi(t=60)$]{
		\includegraphics[width=0.165\linewidth]{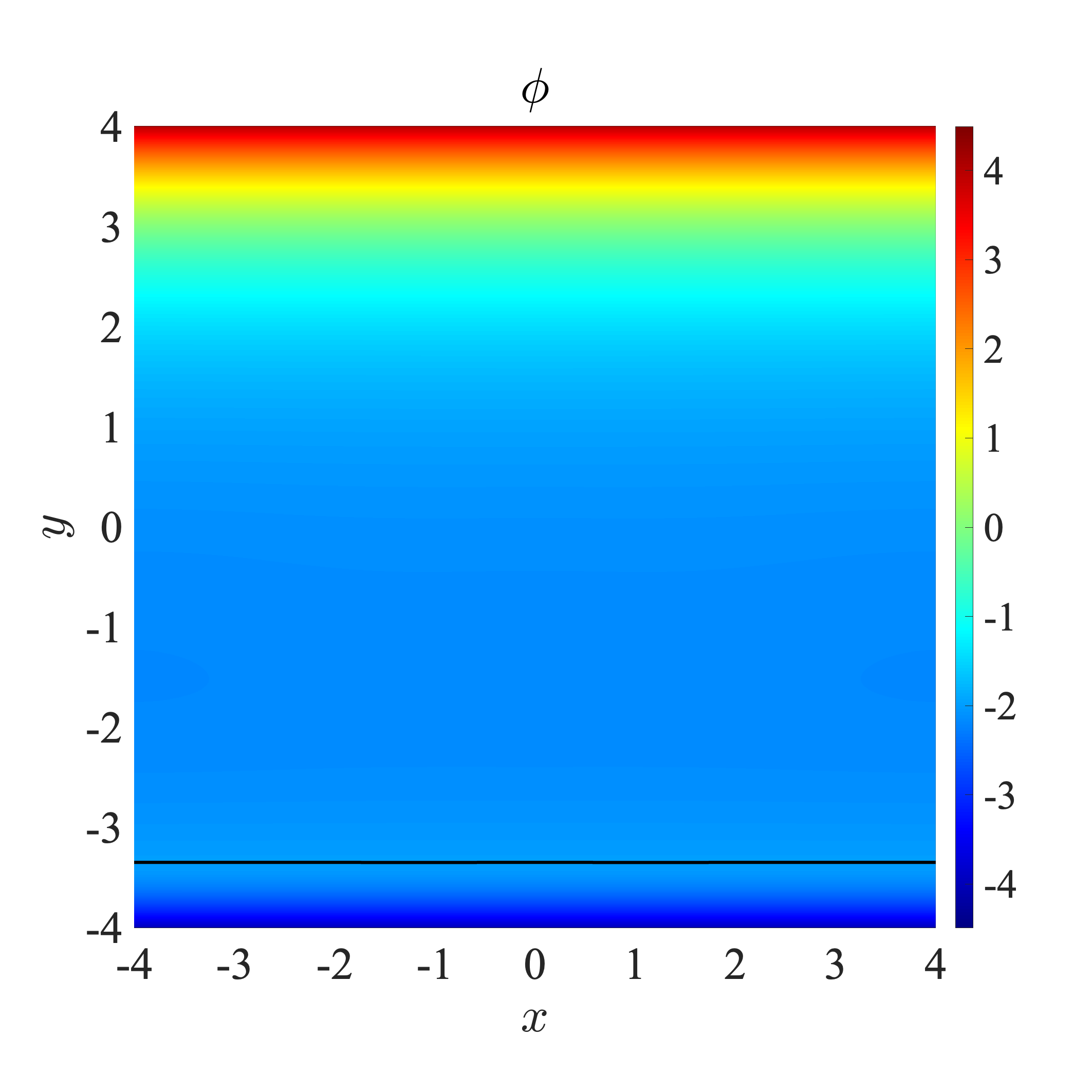}
        \label{subfig:2DropsD4N0Pump25Phi60}
        }
    \\
    \vskip -0.2cm
	\caption{The snapshots for the drop deformation and motion with positive ion pump for the boundary condition \eqref{bd:2drops_D4N0}. 
    Top: positive ion; middle: negative ion; bottom: electric potential. 
    The black solid circle represents the location of the drop, 
    which is denoted by the level set $\psi=0$. 
    The concentration and electric potential distribution are shown on the color map. 
    }\label{fig:2DropsverticalPump25}
\end{figure}

\begin{figure}[!ht]
    \vskip -0.0cm
    \centering 
	\subfloat[$-\nabla \phi(t=1)$]{
		\includegraphics[width=0.33\linewidth]{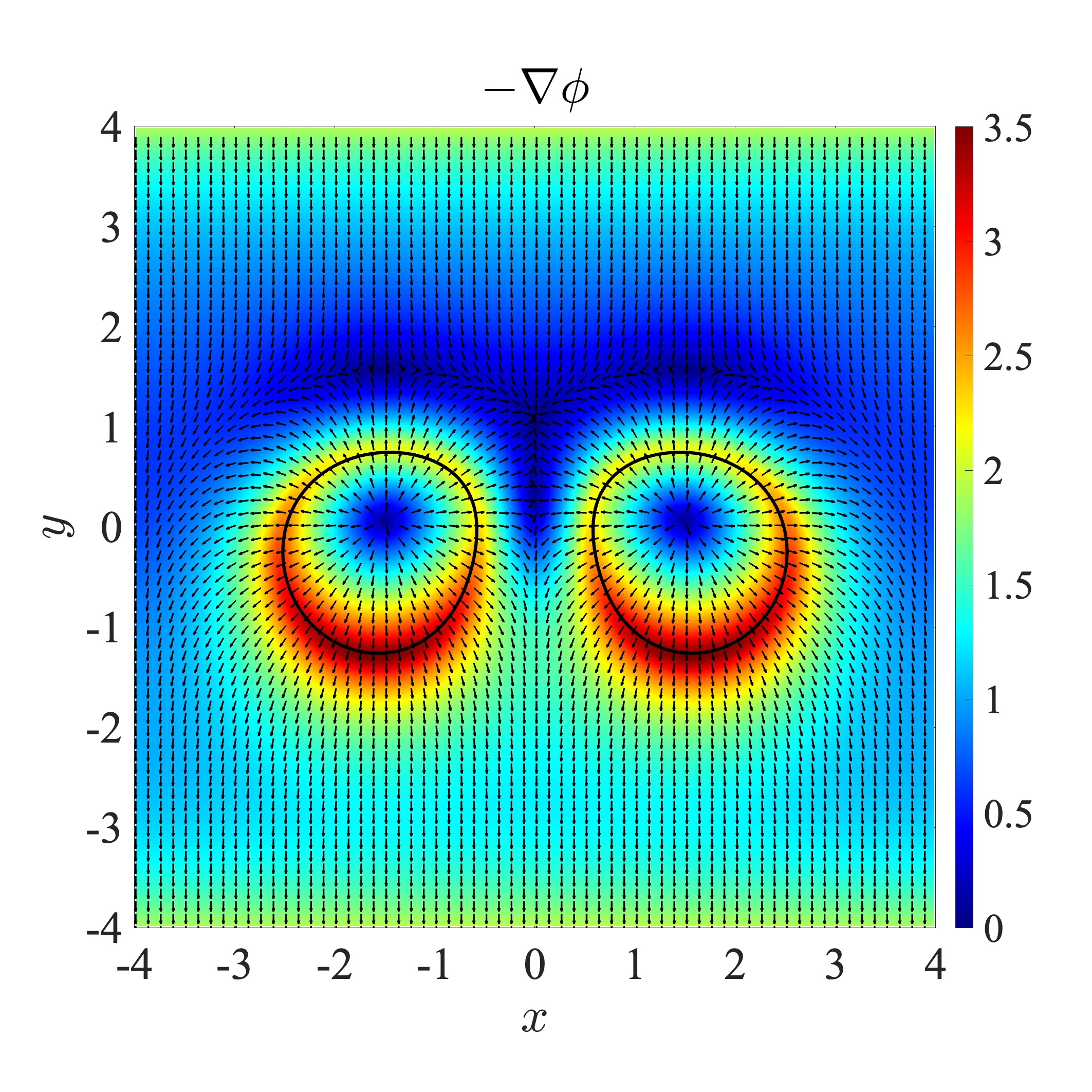}
        \label{subfig:2DropsD4N0Pump25Efield1}
		} 
    \subfloat[$\bm{u}(t=1)$]{
		\includegraphics[width=0.33\linewidth]{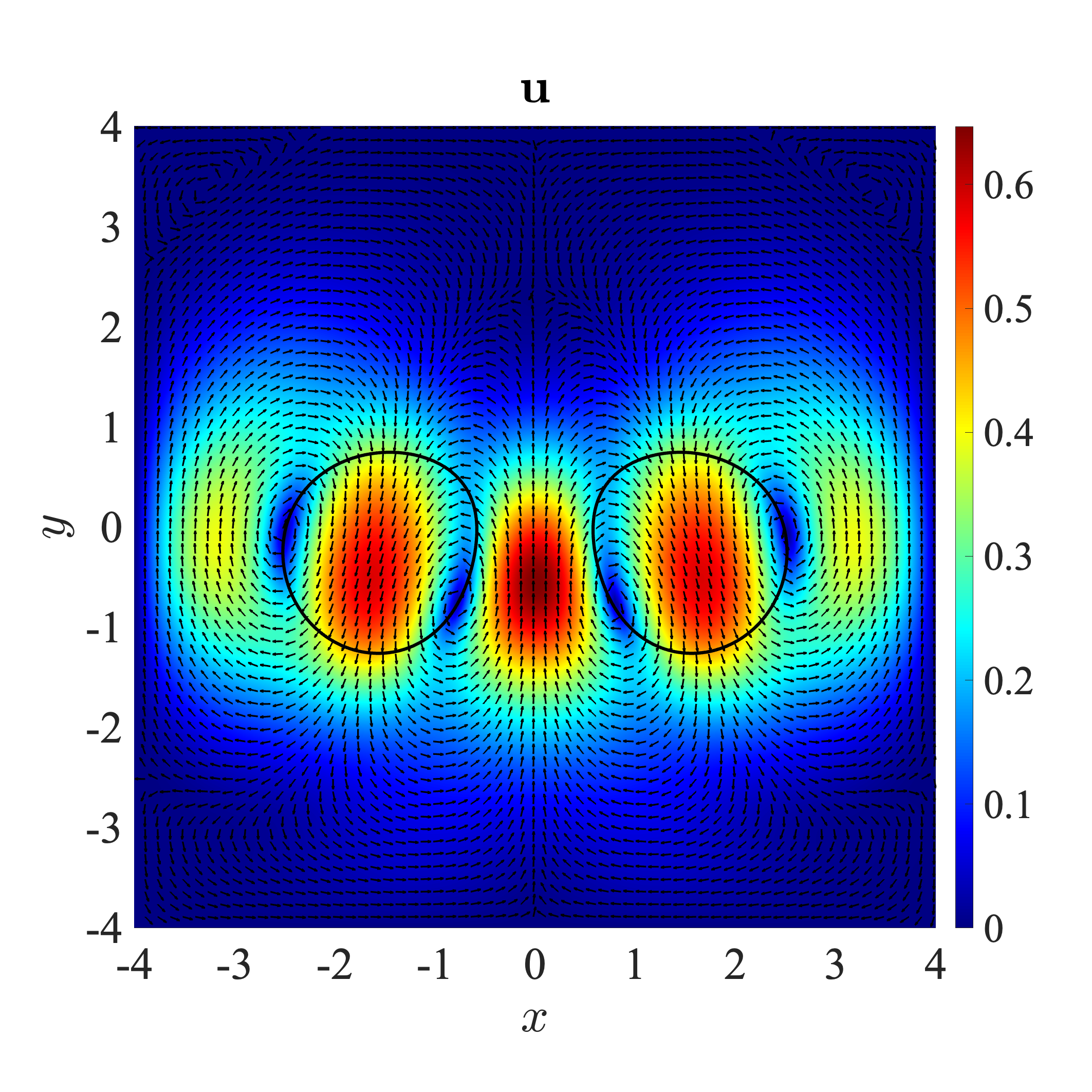}
        \label{subfig:2DropsD4N0Pump25Velocity5}
		} 
    \subfloat[$\frac{1}{Re}\Delta\bm{u}-\nabla P-\frac{Ca_{E}}{\zeta^{2}}\rho\nabla\phi~(t=1)$]{
    \includegraphics[width=0.33\linewidth]{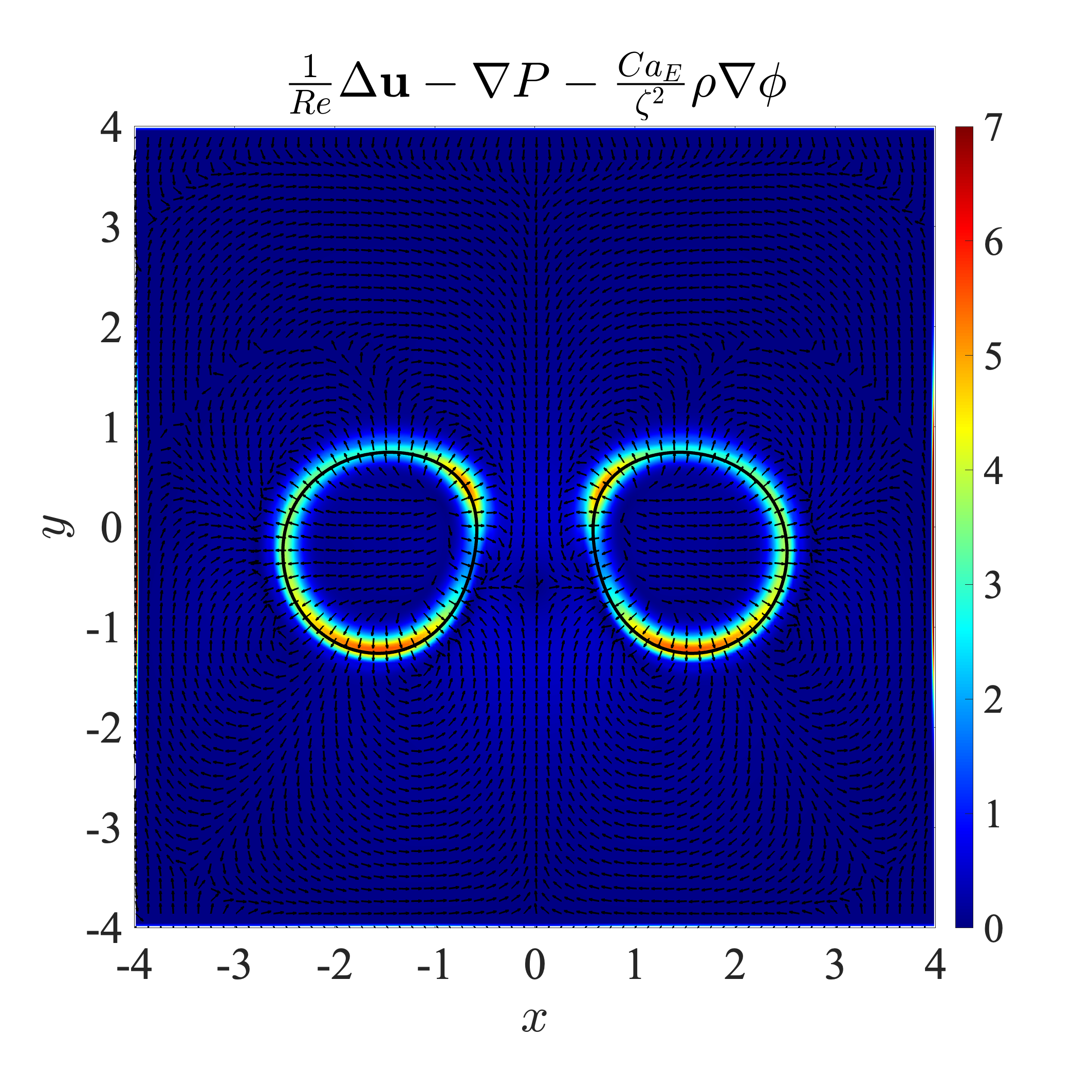}
        \label{subfig:2DropsD4N0Pump25Force5}
    }
        \vskip -0.2cm
	\caption{The snapshots of the electric field (left), velocity (middle) and fluid viscous stress induced force and Lorentz force (right) at time t=1. 
    Here $\phi_{0b} = -4, \quad \phi_{0u} = 4$.}\label{fig:2DropsD4N0Pump25EV}
\end{figure}

The simulation results are shown in Fig.~\ref{fig:2DropsverticalPump25}. Positive ions are actively pumped into each droplet, leading to significant charge accumulation and an increase in the local electric potential, as seen in Fig.~\ref{subfig:2DropsD4N0Pump25Phi1}. This charge buildup induces strong electrostatic repulsion between the droplets in the horizontal direction.
Due to the vertical electric field, which is stronger beneath the droplets, especially near the grounded lower boundary, the electric field distribution becomes vertically asymmetric (see Fig. \ref{subfig:2DropsD4N0Pump25Efield1}). As a result, the horizontal separation between the lower halves of the two droplets is noticeably larger than that between their upper halves (see Fig.~\ref{subfig:2DropsD4N0Pump25P1}). The stronger downward electric field also induces a larger Lorentz force in the vertical direction, causing the droplets to elongate along the $y$-axis.

However, lateral confinement from the no-slip walls on the left and right, along with the proximity of the neighboring droplet, creates an asymmetric flow environment (see Fig. \ref{subfig:2DropsD4N0Pump25Velocity5}). The \textit{inner lower sides} of the droplets are exposed to relatively open space, while the \textit{outer sides} are constrained. This asymmetry gives rise to complex flow patterns, including vortical structures near the lateral edges. As a result, the deformation of the droplets is not purely vertical: the left droplet bends slightly to the right, and the right droplet bends to the left, forming a pair of crescent-shaped interfaces curving away from each other. This behavior emerges from the interplay of vertical stretching, electrostatic repulsion, interfacial tension, and anisotropic flow confinement. Over time, as in the single-droplet case, the deformation progresses until a pinch-off event occurs. Eventually, the resulting daughter droplets move downward, contact the bottom boundary, and recoalesce into a flattened configuration.

\begin{figure}[!ht]
    \vskip -0.0cm
    \centering 
	\subfloat[$p+n(t=1)$]{
		\includegraphics[width=0.165\linewidth]{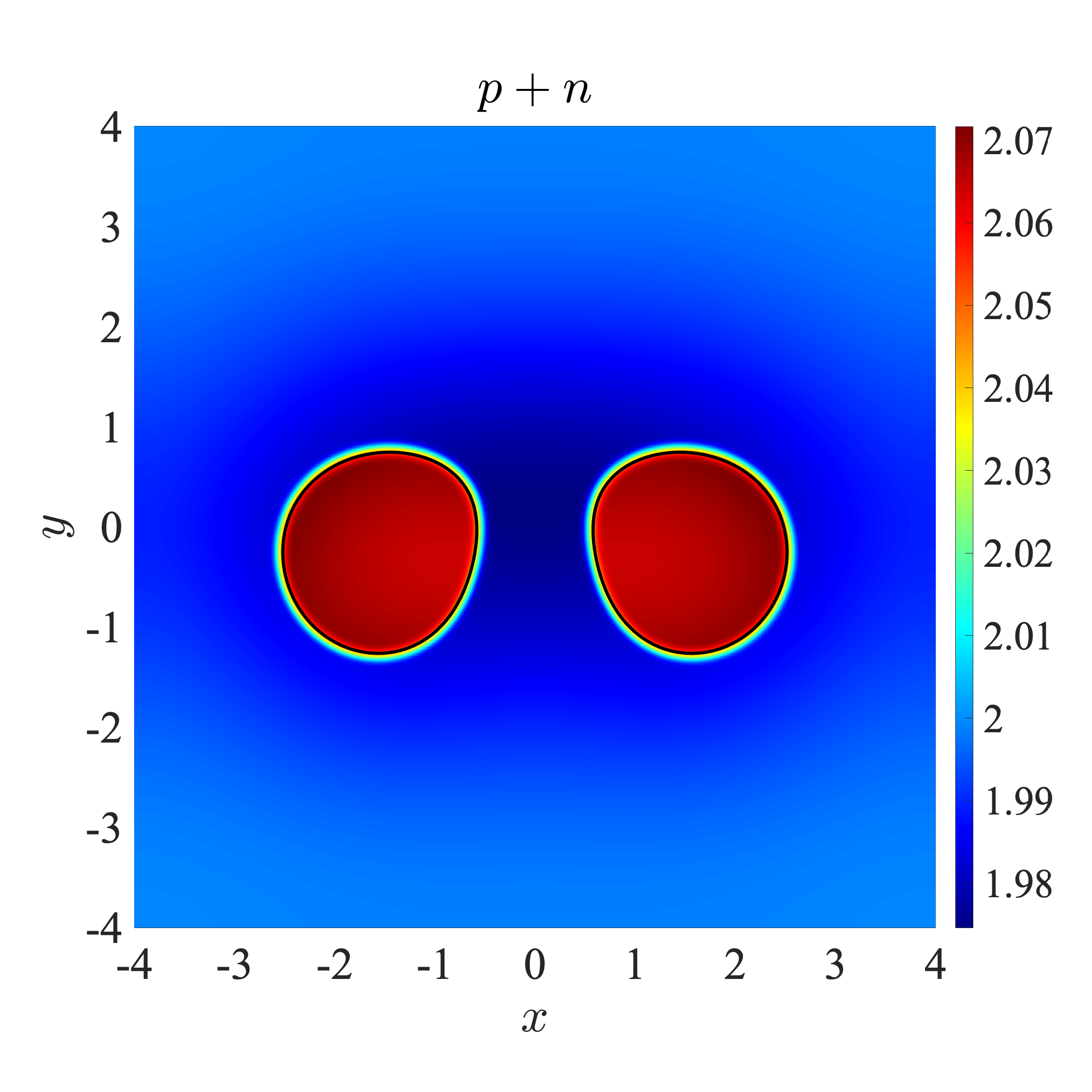}
        \label{subfig:2DropsD4N0Pump25Sum1}
		} 
    \hskip -0.3cm
    \subfloat[$p+n(t=3)$]{
		\includegraphics[width=0.165\linewidth]{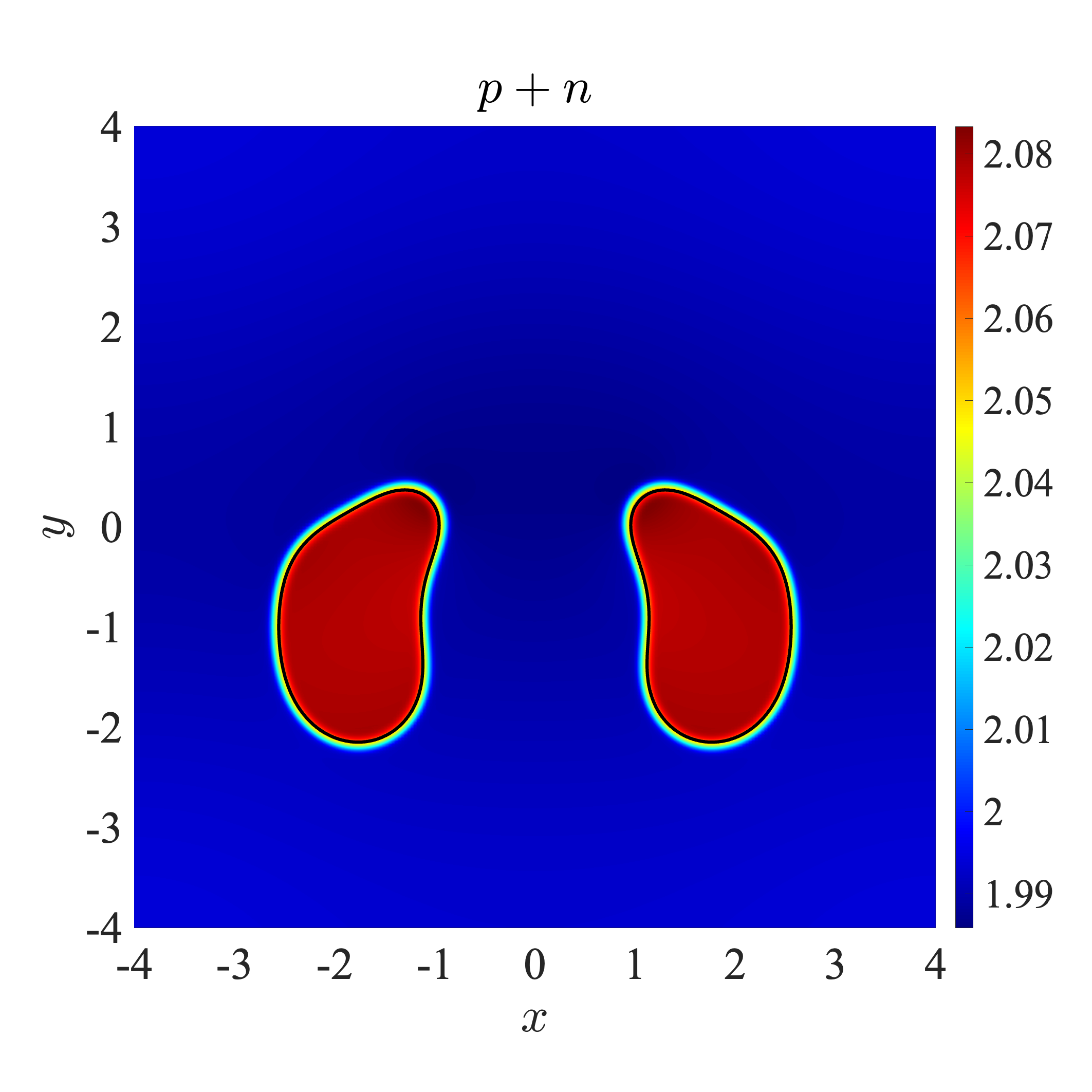}
        \label{subfig:2DropsD4N0Pump25Sum3}
		} 
    \hskip -0.3cm
    \subfloat[$p+n(t=5)$]{
		\includegraphics[width=0.165\linewidth]{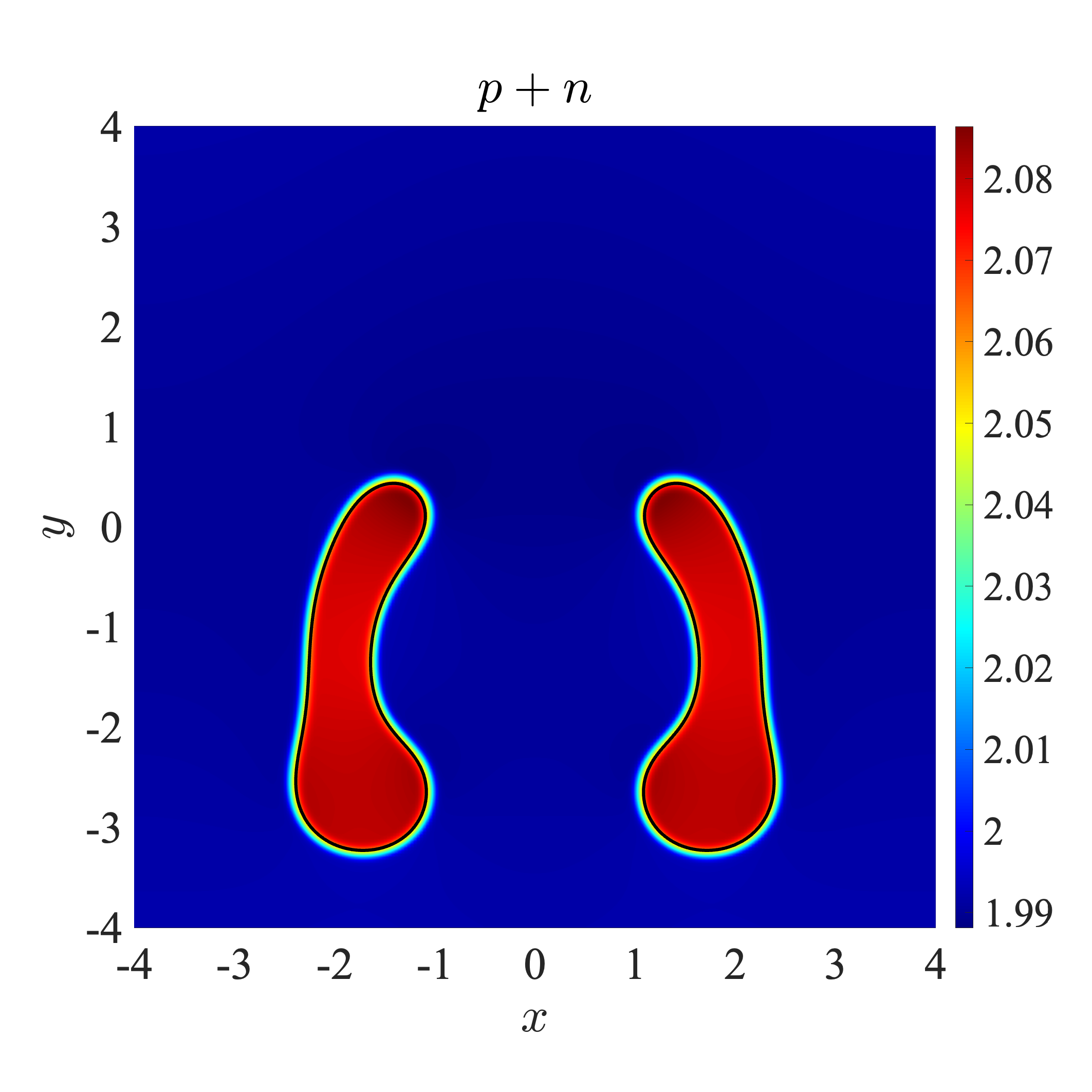}
        \label{subfig:2DropsD4N0Pump25Sum5}
		} 
    \hskip -0.3cm
    \subfloat[$p+n(t=8)$]{
		\includegraphics[width=0.165\linewidth]{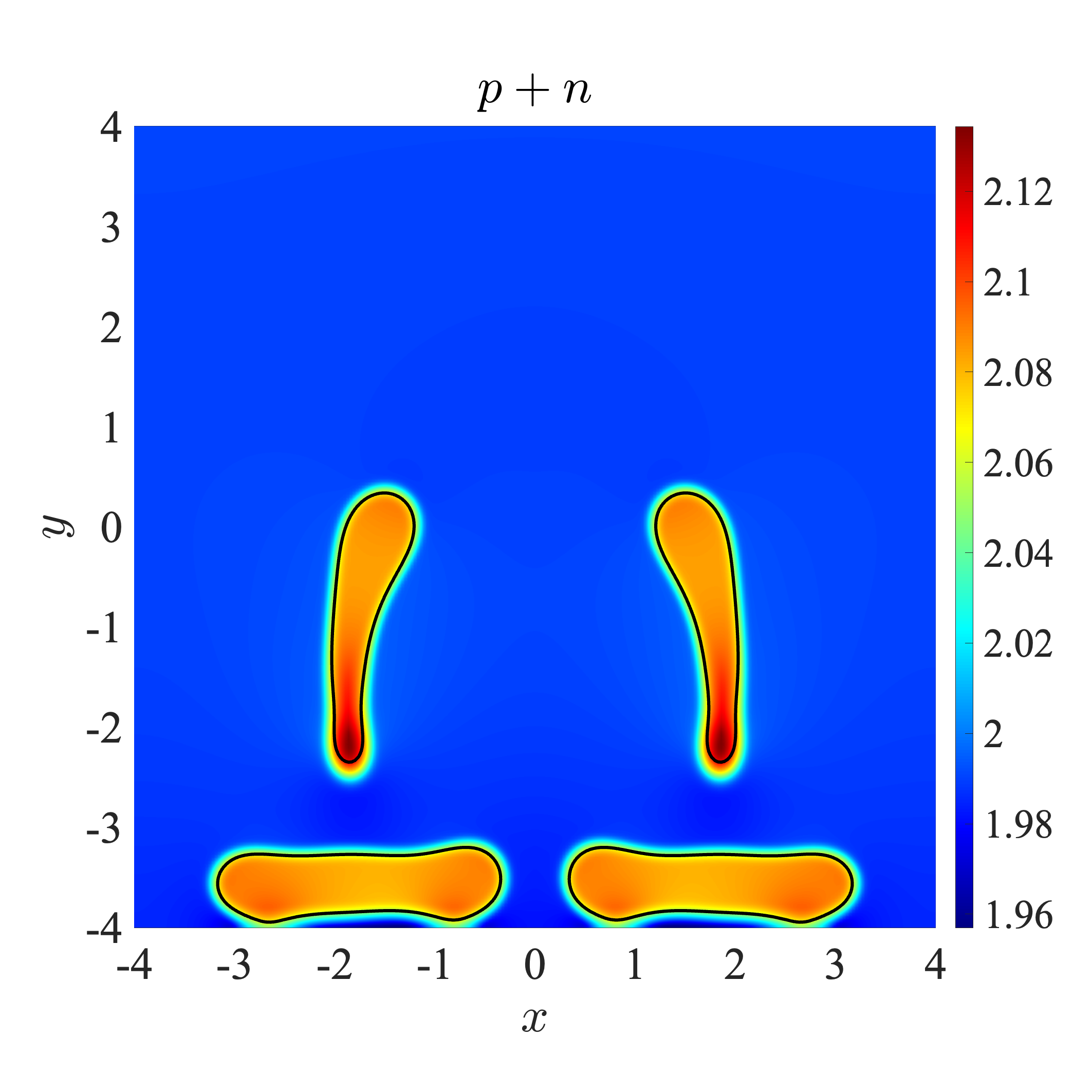}
        \label{subfig:2DropsD4N0Pump25Sum8}
		} 
    \hskip -0.3cm
    \subfloat[$p+n(t=10)$]{
		\includegraphics[width=0.165\linewidth]{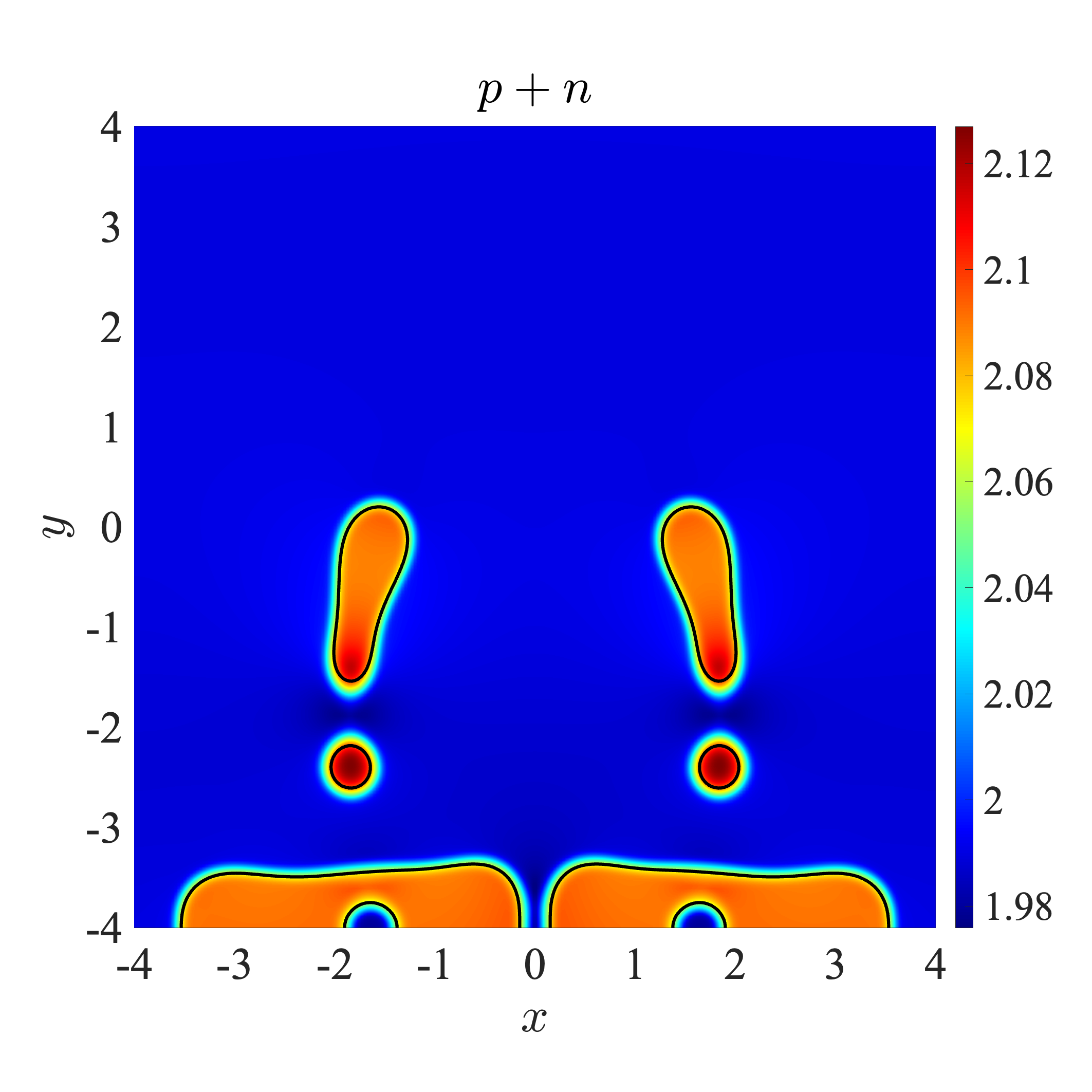}
        \label{subfig:2DropsD4N0Pump25Sum10}
		} 
    \hskip -0.3cm
    \subfloat[$p+n(t=60)$]{
		\includegraphics[width=0.165\linewidth]{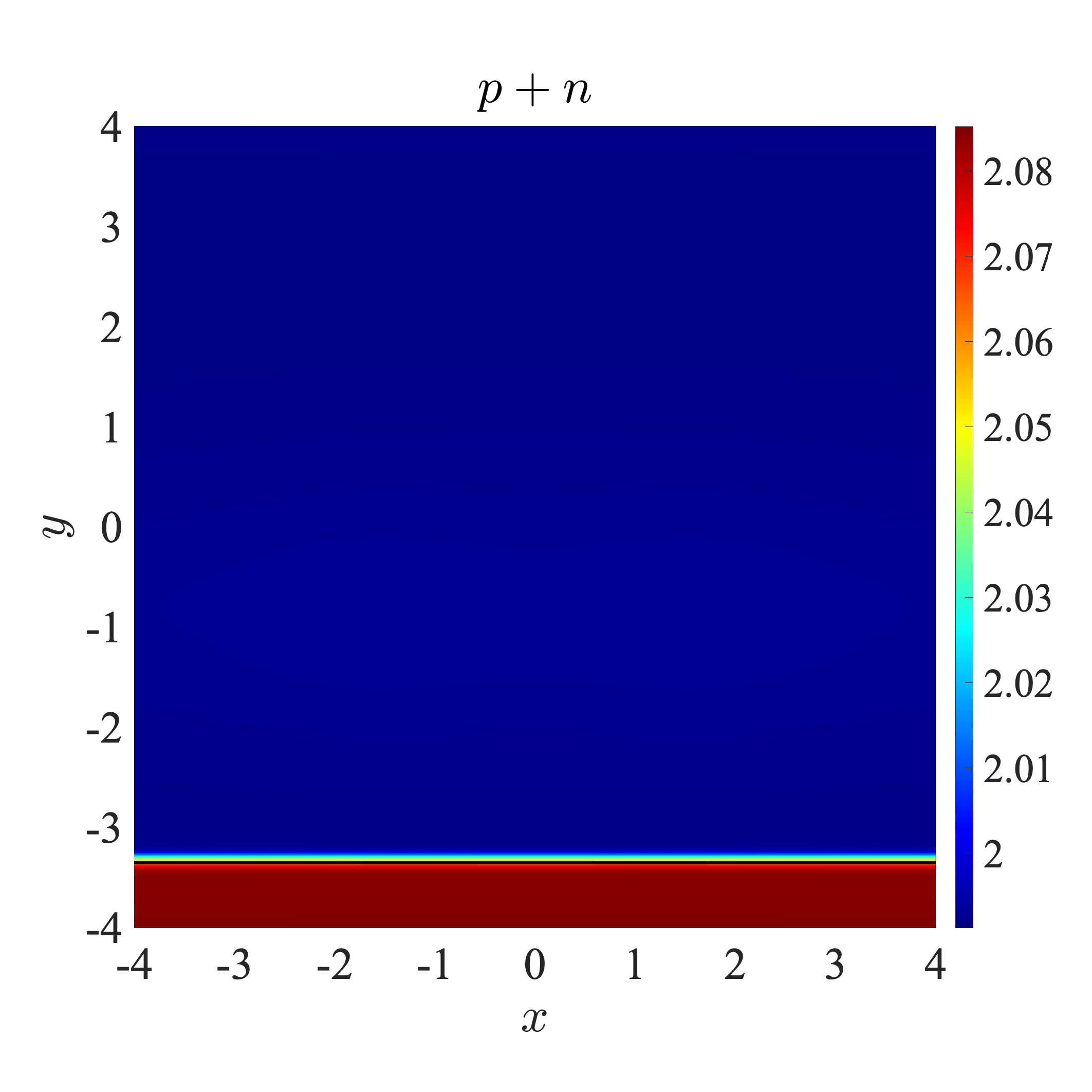}
        \label{subfig:2DropsD4N0Pump25Sum60}
		}
    \\
    \vskip -0.3cm
	\subfloat[$p-n(t=1)$.]{
		\centering
		\includegraphics[width=0.165\linewidth]{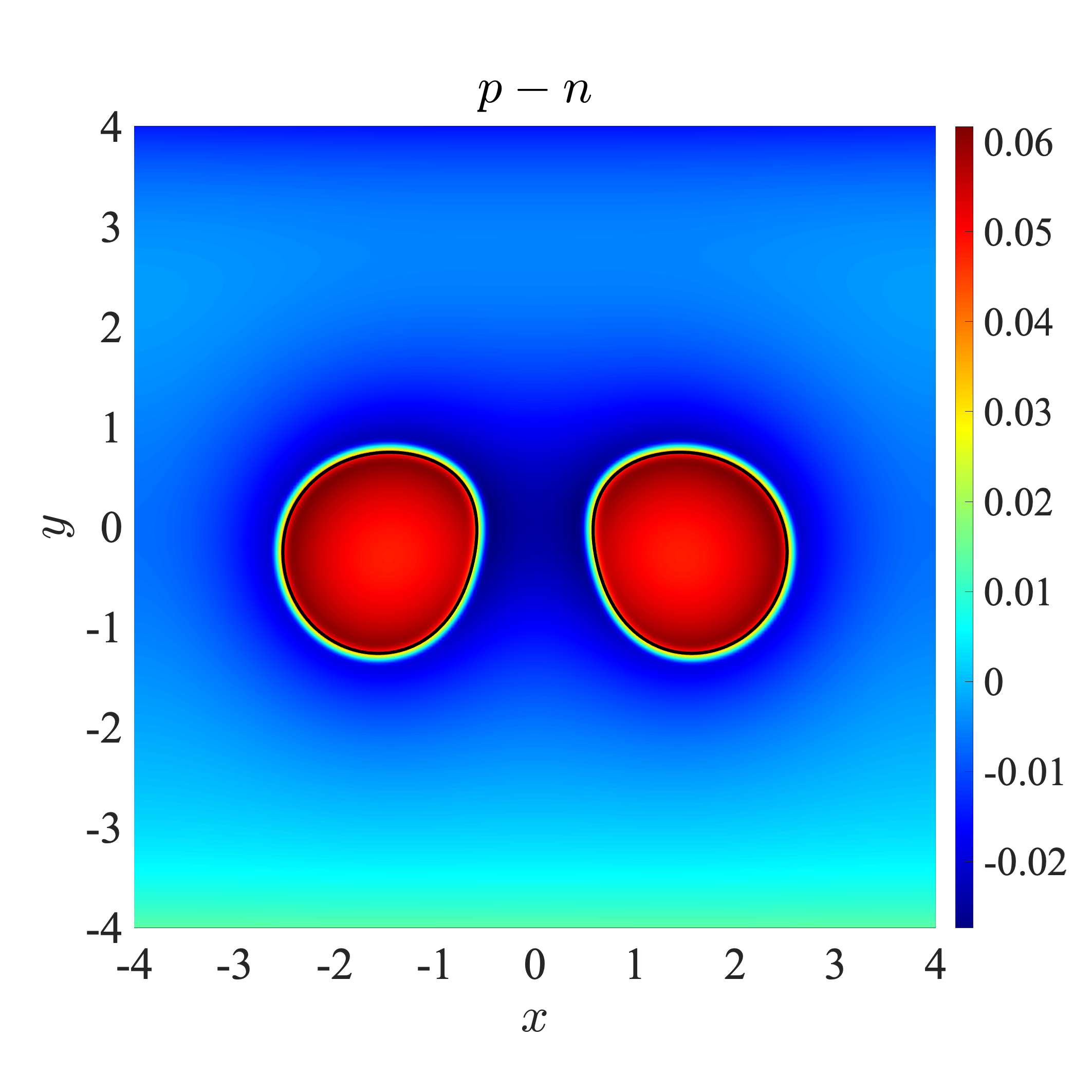}
        \label{subfig:2DropsD4N0Pump25Dif1}
	}
    \hskip -0.3cm
    \subfloat[$p-n(t=3)$.]{
		\centering
		\includegraphics[width=0.165\linewidth]{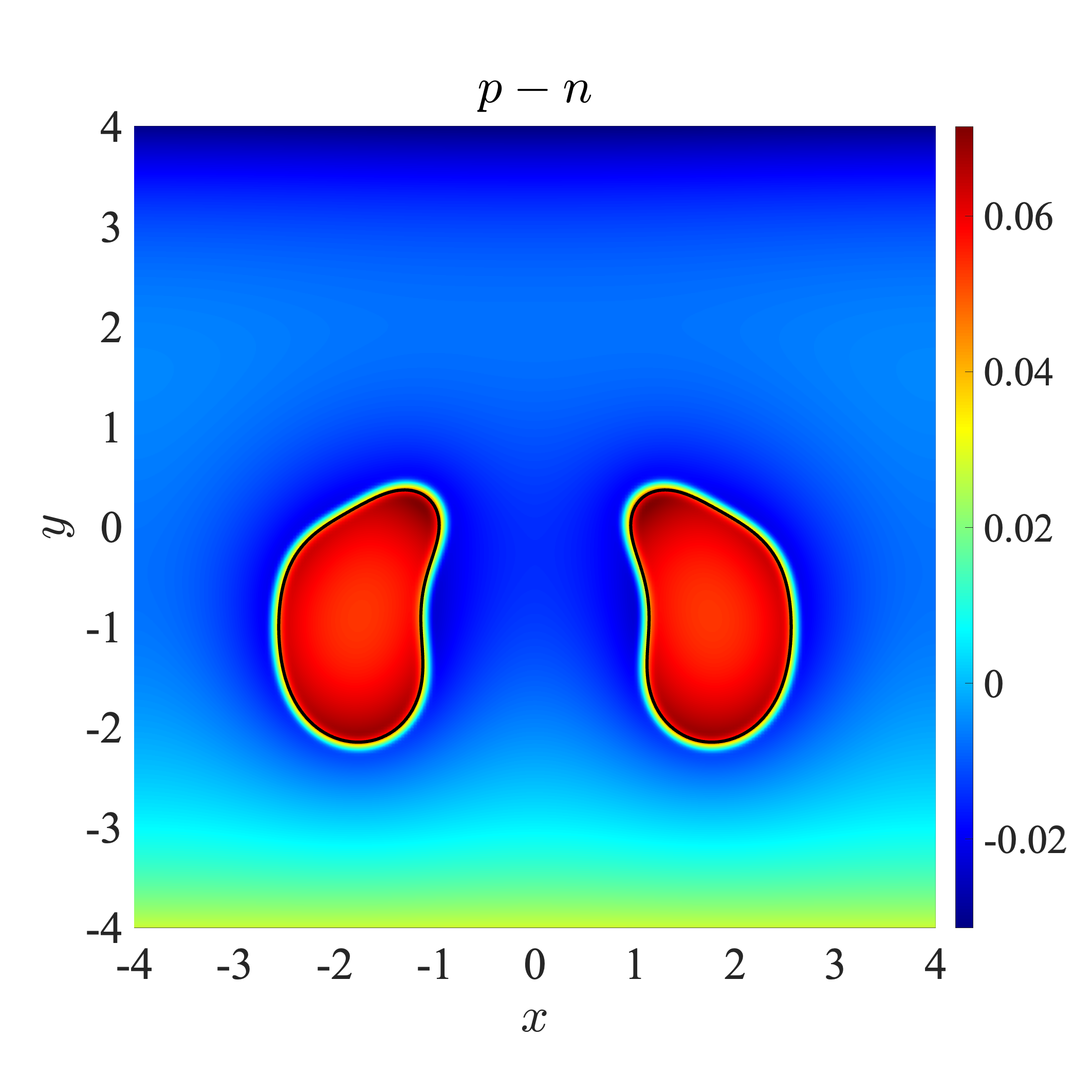}
        \label{subfig:2DropsD4N0Pump25Dif3}
	}
    \hskip -0.3cm
    \subfloat[$p-n(t=5)$.]{
		\centering
		\includegraphics[width=0.165\linewidth]{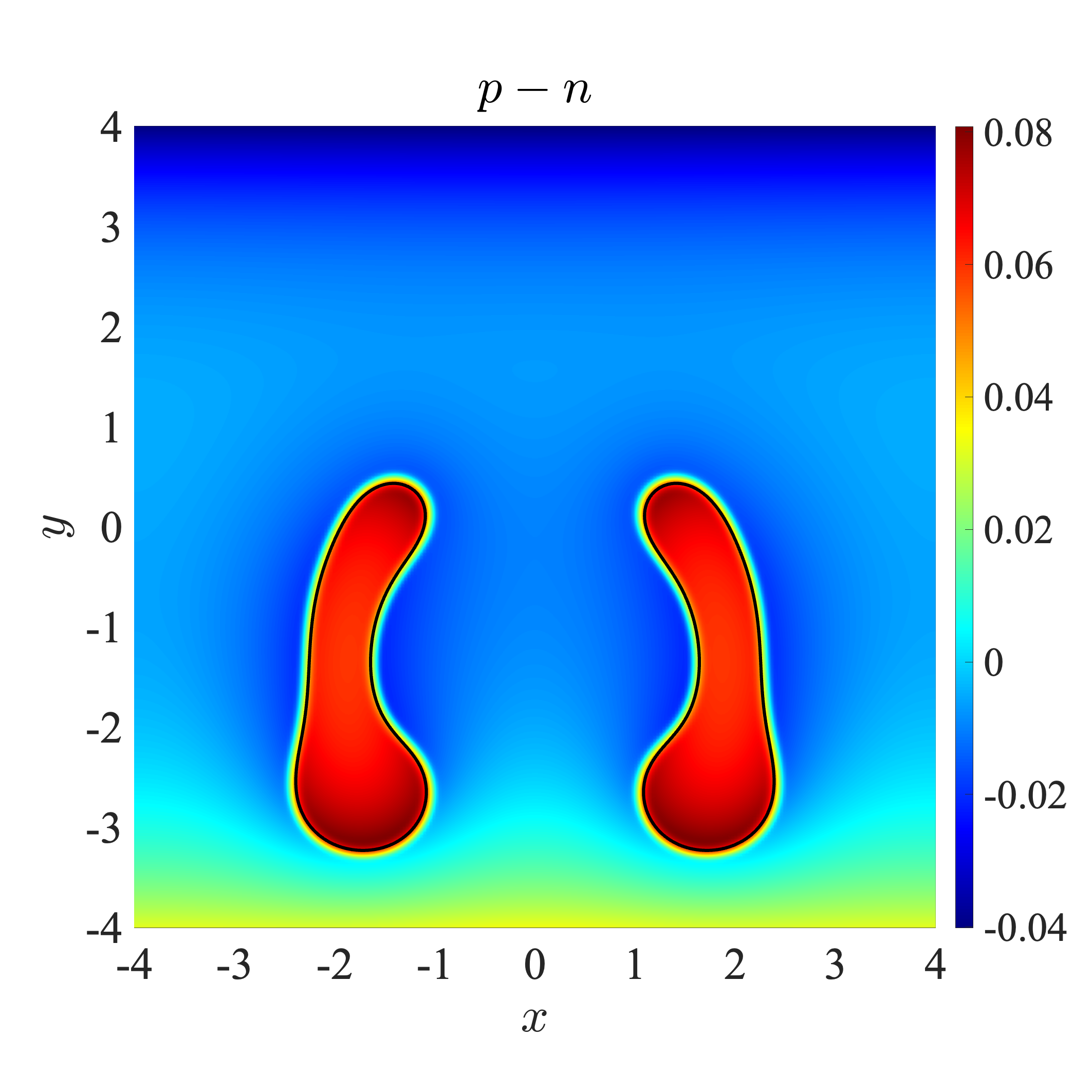}
        \label{subfig:2DropsD4N0Pump25Dif5}
	}
    \hskip -0.3cm
    \subfloat[$p-n(t=8)$.]{
		\centering
		\includegraphics[width=0.165\linewidth]{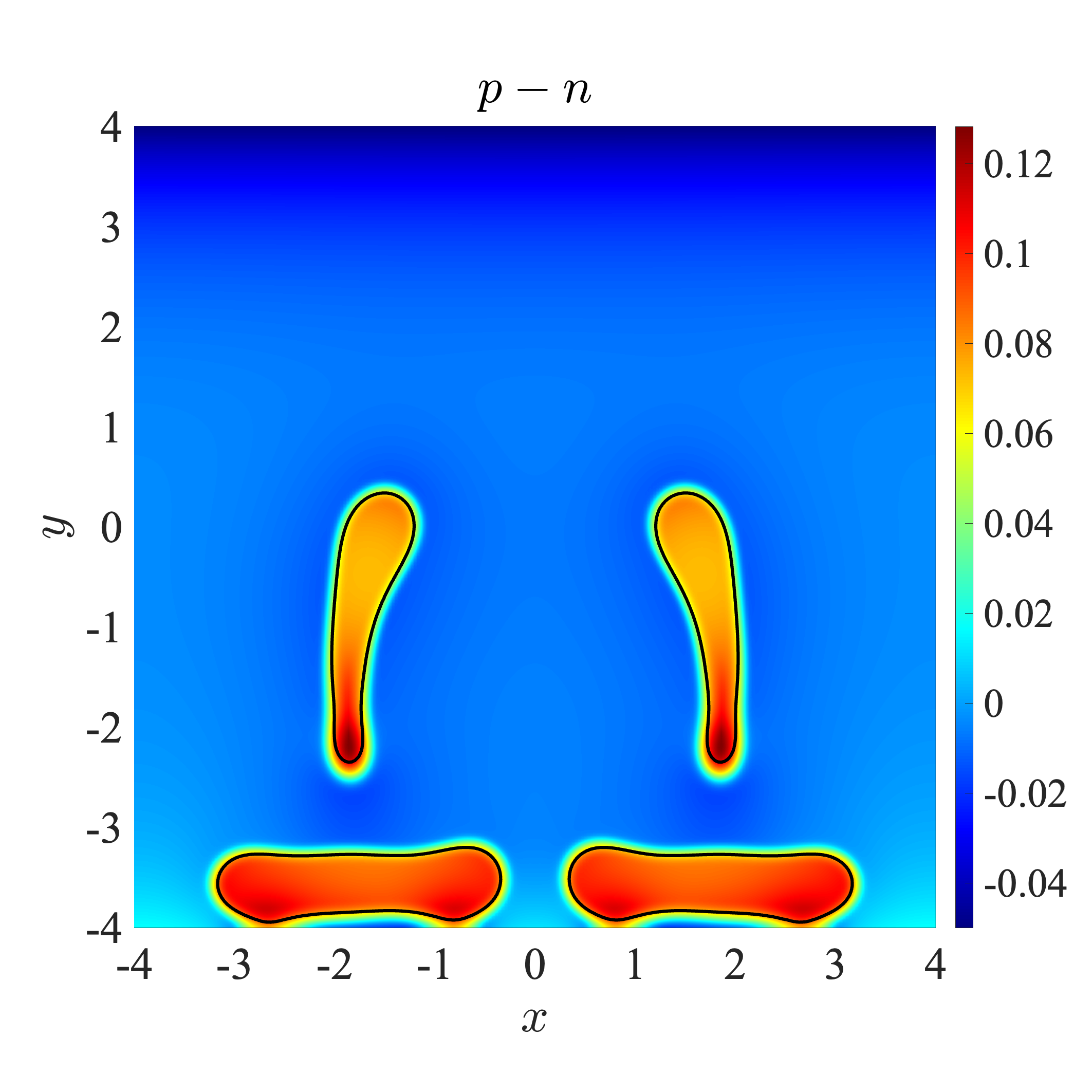}
        \label{subfig:2DropsD4N0Pump25Dif8}
	}
    \hskip -0.3cm
    \subfloat[$p-n(t=10)$.]{
		\centering
		\includegraphics[width=0.165\linewidth]{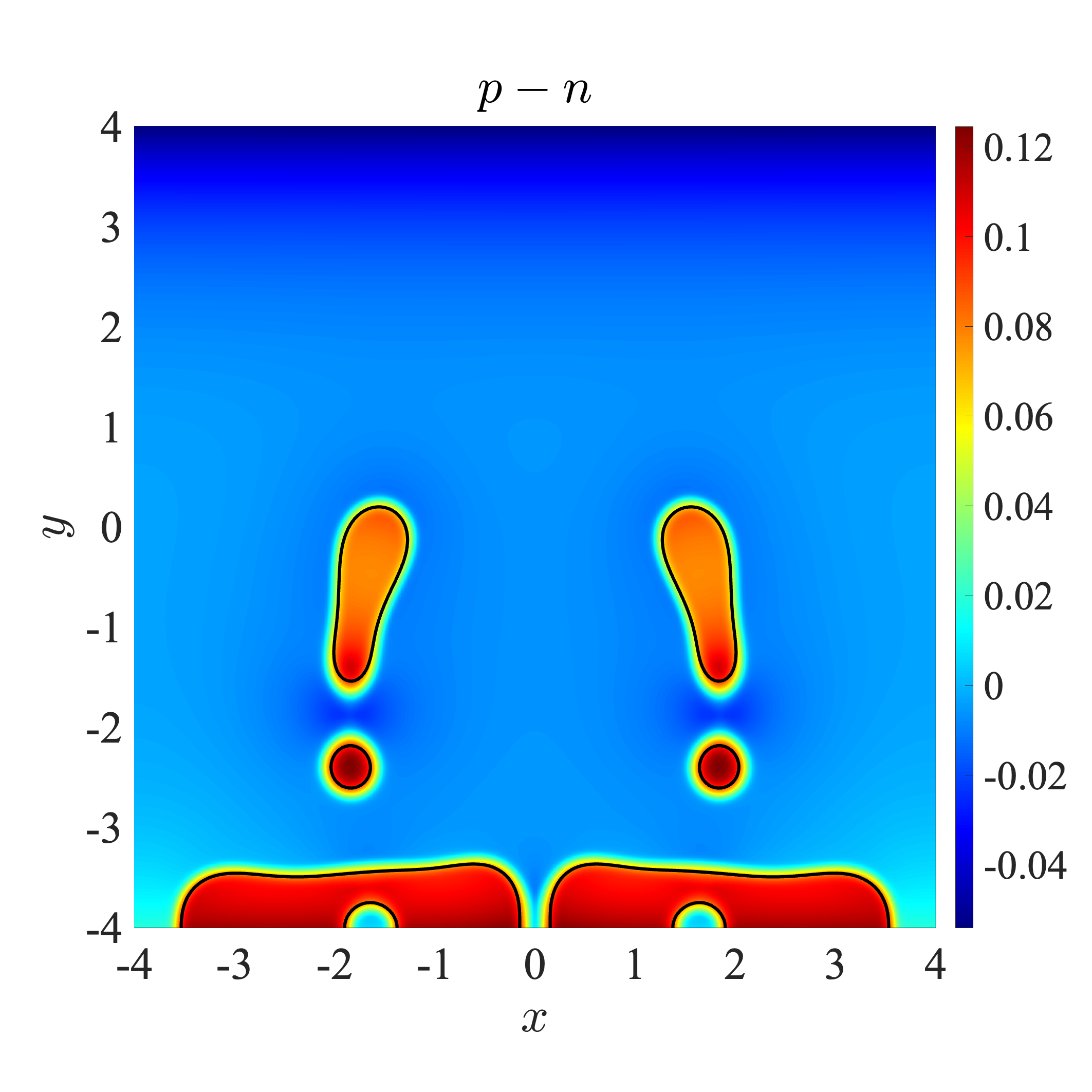}
        \label{subfig:2DropsD4N0Pump25Dif10}
	}
    \hskip -0.3cm
    \subfloat[$p-n(t=60)$.]{
		\centering
		\includegraphics[width=0.165\linewidth]{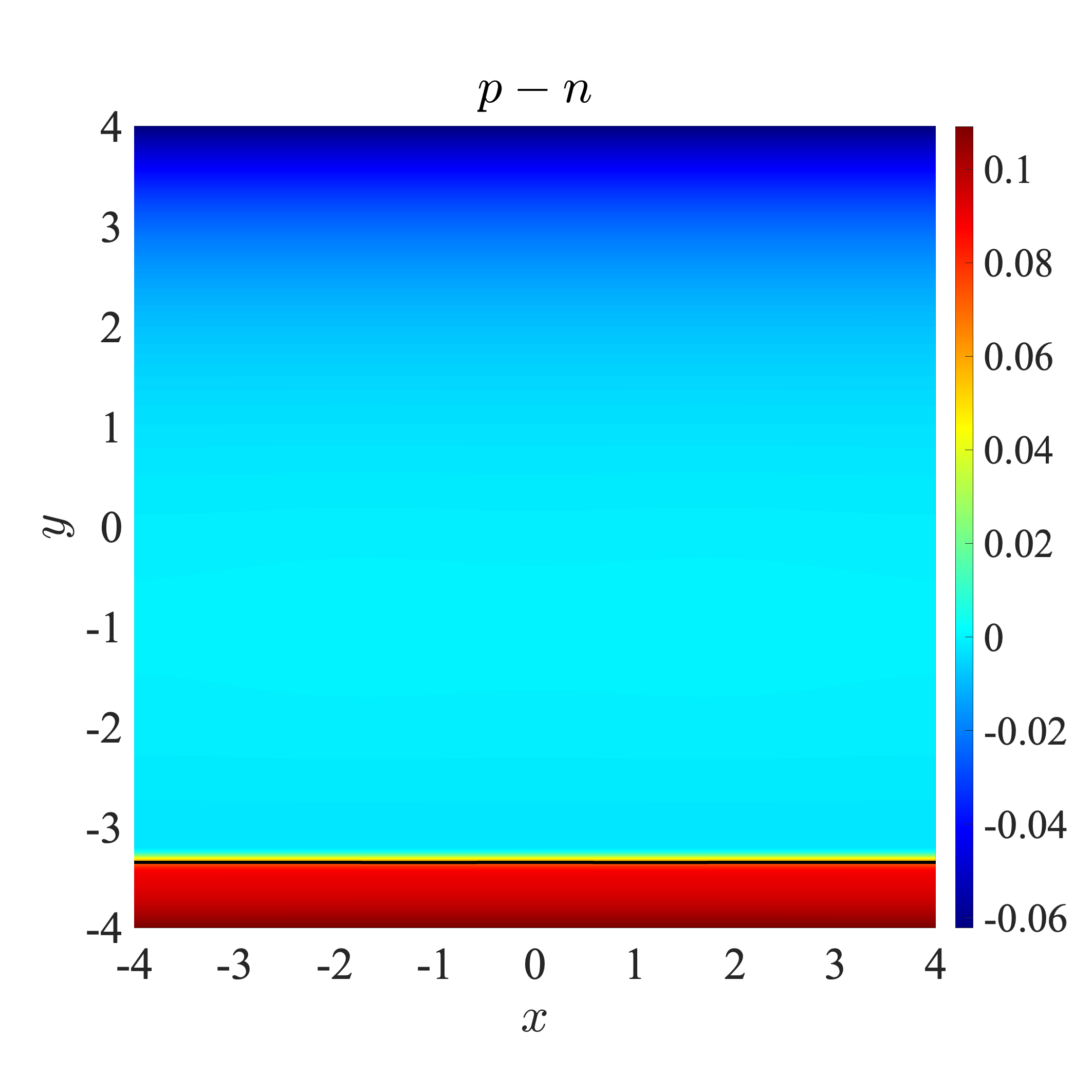}
        \label{subfig:2DropsD4N0Pump25Dif60}
	}
        \vskip -0.2cm
	\caption{The snapshots of the total charge (top) and net charge (bottom). 
    The total charge and net charge both accumulate in the droplet due to the redistribution of positive and negative ions. Here $\phi_{0b} = -4, \quad \phi_{0u} = 4$.}\label{fig:2DropsD4N0Pump25V}
\end{figure}

\section{Selective separation of droplets via active pumping and Poiseuille flow}\label{subsec:selection}

Building upon the previously observed behavior of pump-induced droplet deformation, we investigate a strategy for  selective separation  of droplets based on the presence or absence of active pumps. In this simulation, two droplets ($\psi_{L}$ and $\psi_{R}$)  are initially placed in a shear flow: one equipped with ion pumps at its interface, and the other without.
 
The computational domain is chosen as $[-8,8] \times [-4,4]$ 
with a horizontal velocity from the left to the right side, 
which is called the Poiseuille flow. 
To maintain the horizontal velocity, 
a source term $f=0.1$ is added to $u_{1}$ in the Navier-Stokes equations. 
  
\begin{figure}[!ht]
\centering
\subfloat[Initial setting of positive ion.]{
\centering
\includegraphics[width=0.33\linewidth]{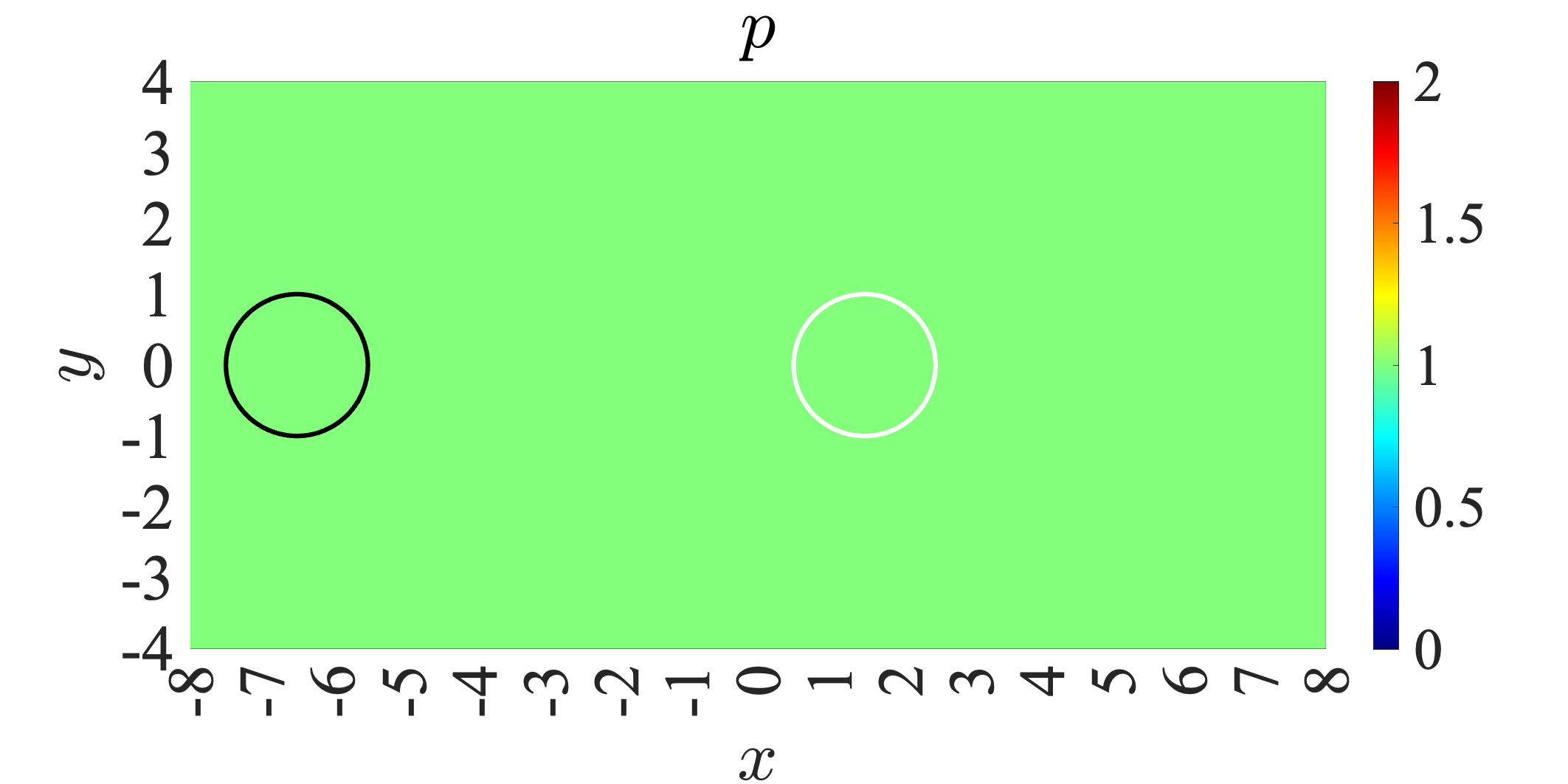}
\label{subfig:2DropsRPumpP0}
}
\hspace{-0.6cm}
\subfloat[Initial setting of negative ion.]{
\centering
\includegraphics[width=0.33\linewidth]{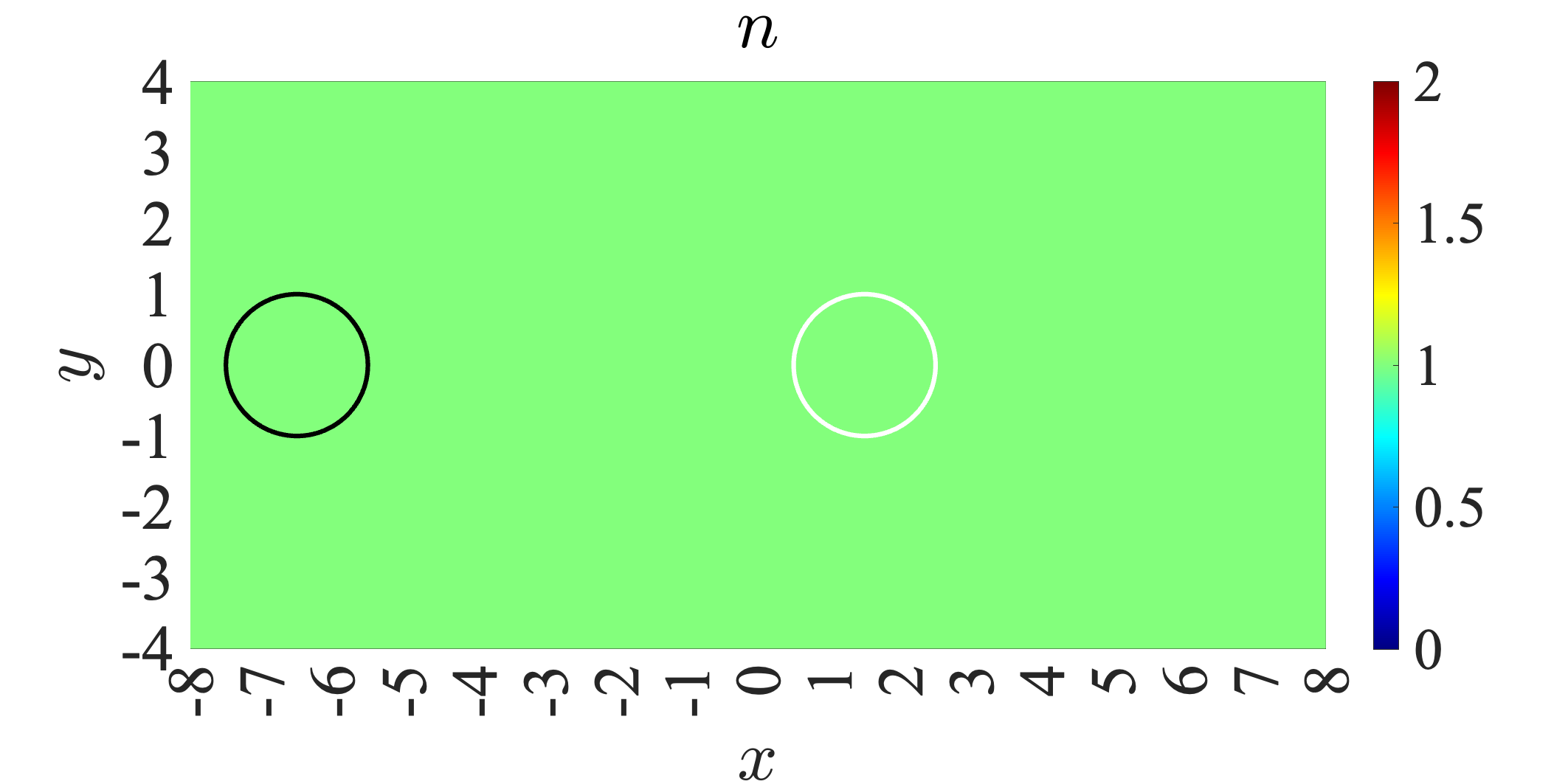}
\label{subfig:2DropsRPumpN0}
}
\hspace{-0.6cm}
\subfloat[Initial setting of electric potential.]{
\centering
\includegraphics[width=0.33\linewidth]{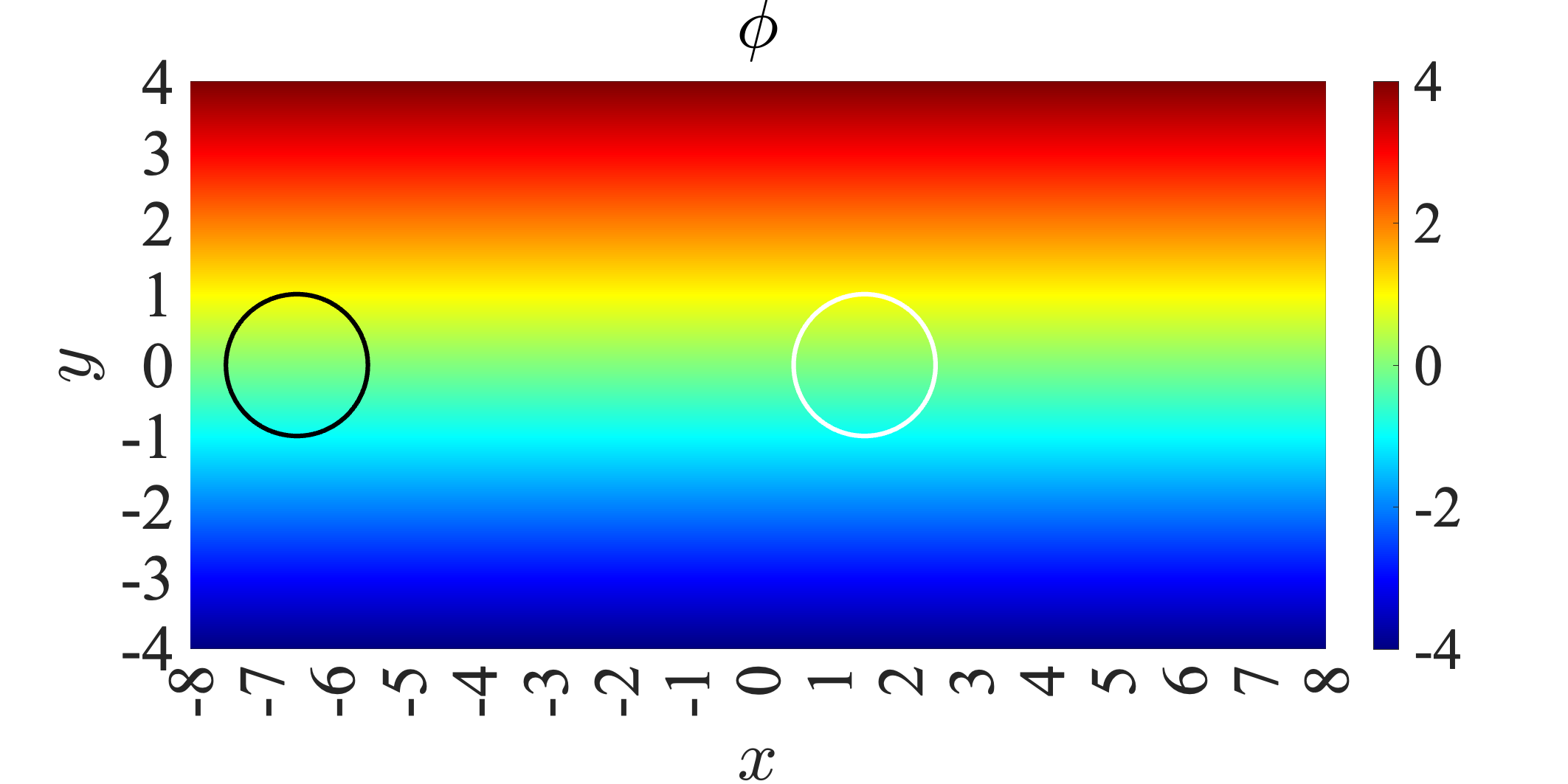}
\label{subfig:2DropsRPumpE0}
}
\\
\subfloat[Positive ion at $t = 10$.]{
\centering
\includegraphics[width=0.33\linewidth]{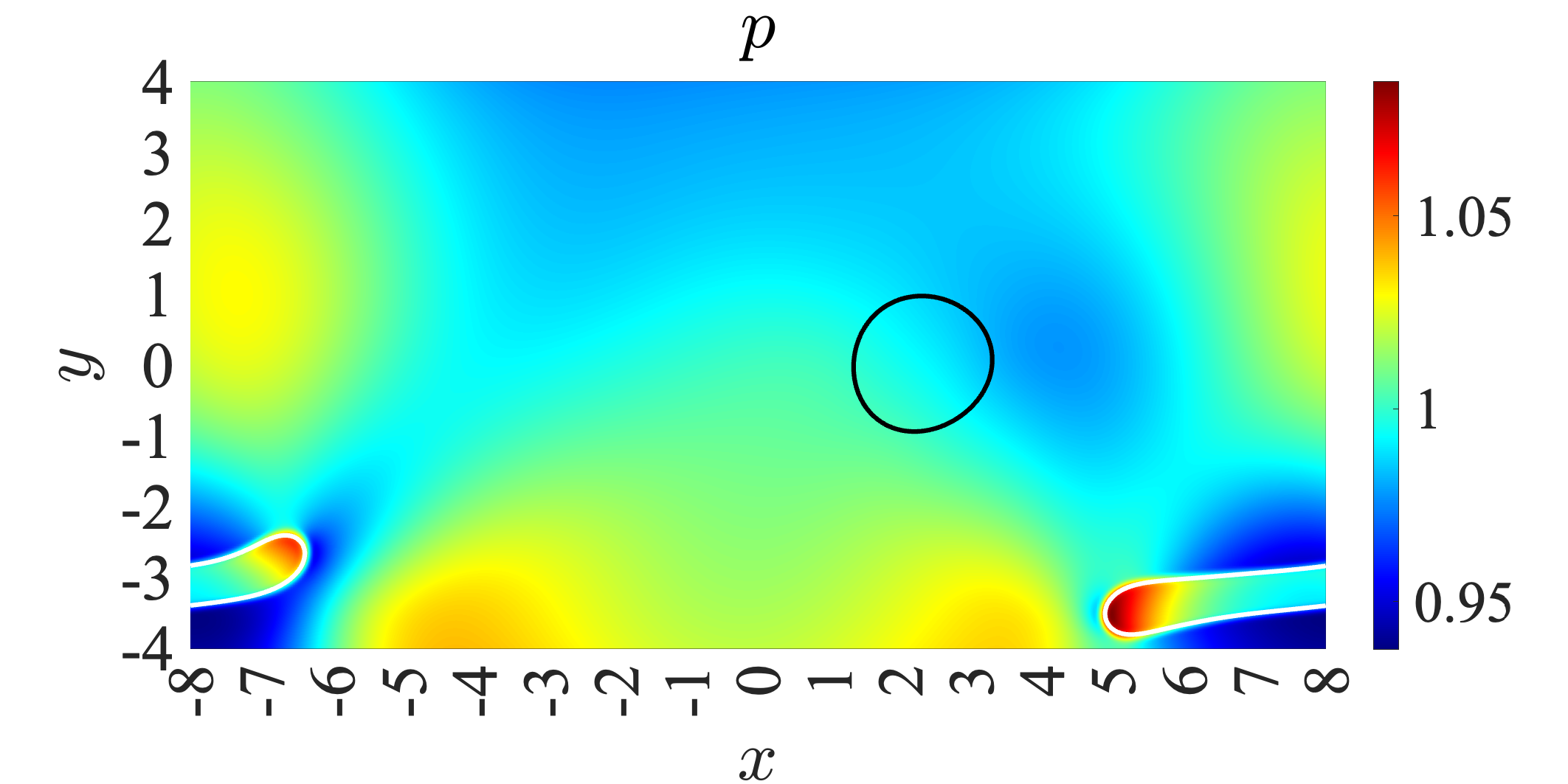}
\label{subfig:2DropsRPumpP10}
}
\hspace{-0.5cm}
\subfloat[Negative ion at $t = 10$.]{
\centering
\includegraphics[width=0.33\linewidth]{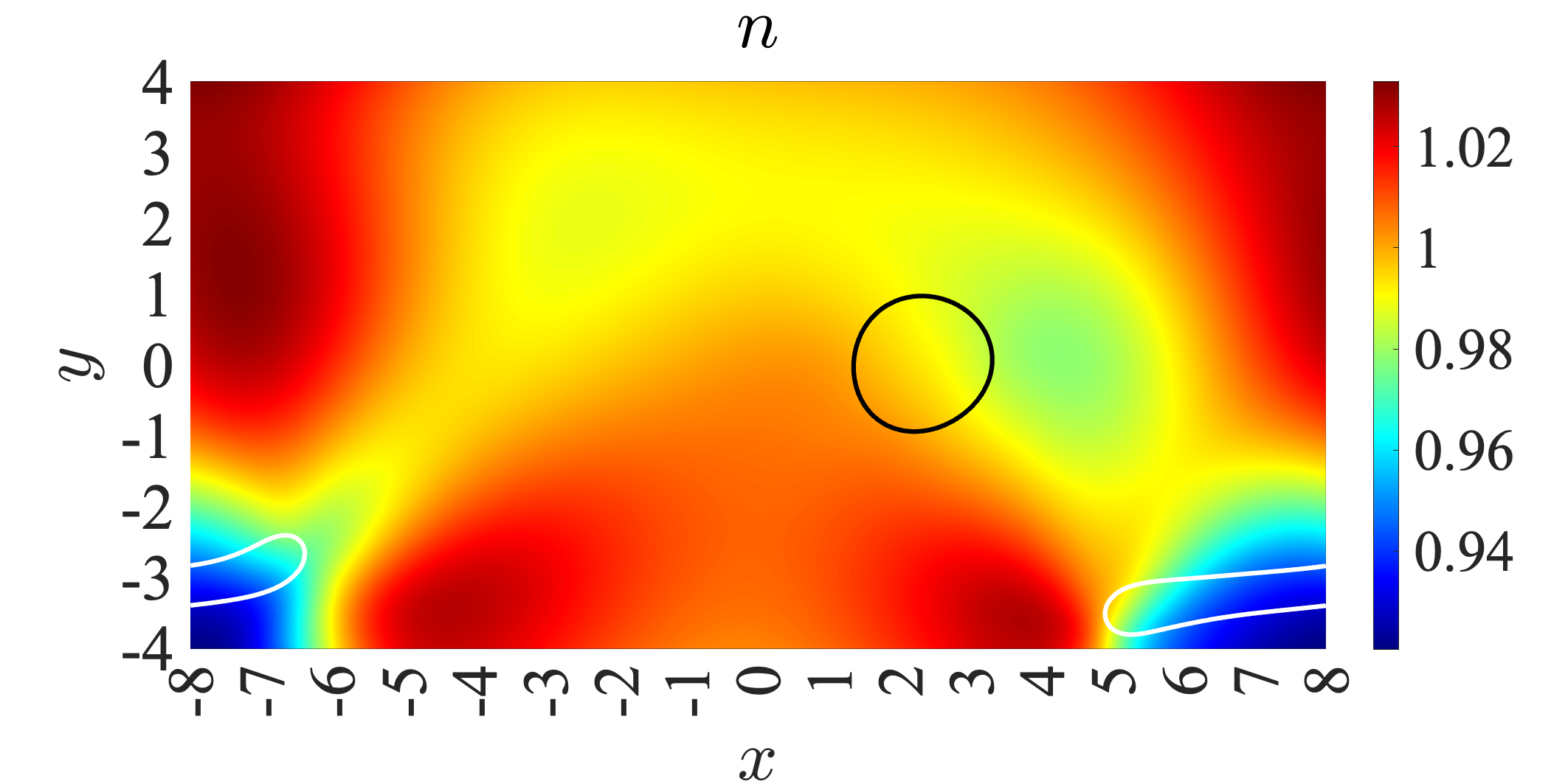}
\label{subfig:2DropsRPumpN10}
}
\hspace{-0.5cm}
\subfloat[Electric potential at $t = 10$.]{
\centering
\includegraphics[width=0.33\linewidth]{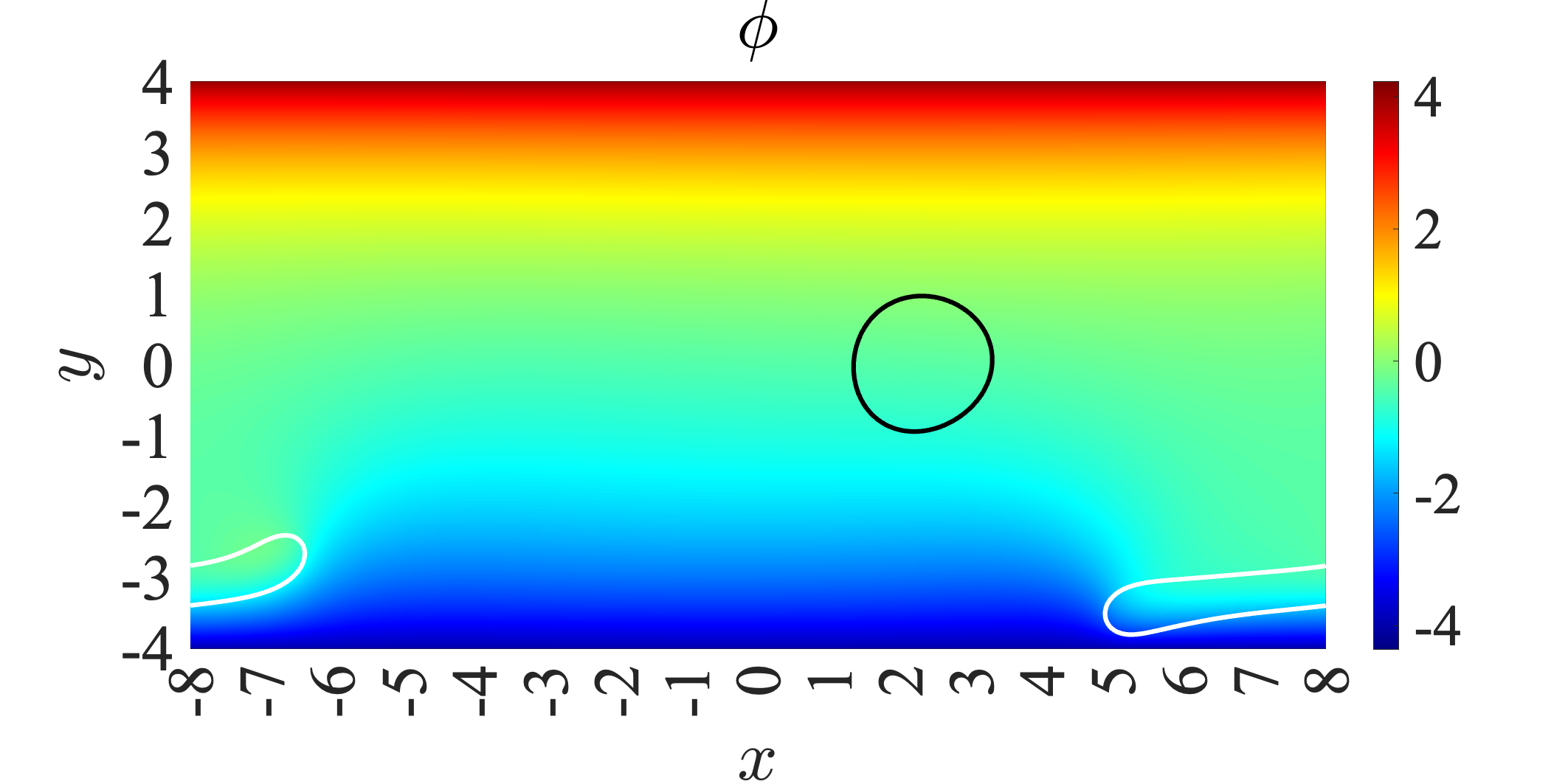}
\label{subfig:2DropsRPumpE10}
}
\\
\subfloat[Positive ion at $t = 20$.]{
\centering
\includegraphics[width=0.33\linewidth]{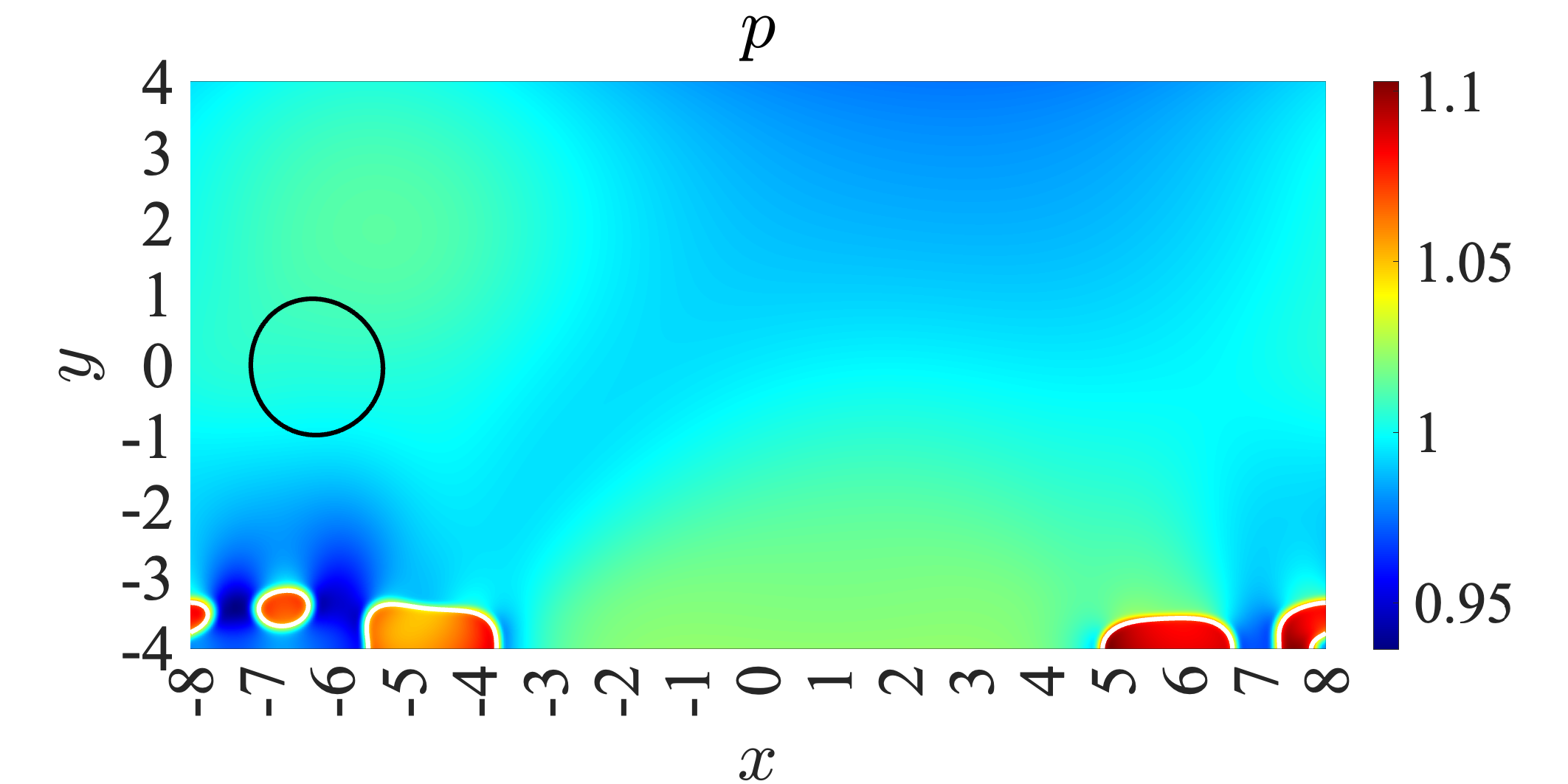}
\label{subfig:2DropsRPumpP20}
}
\hspace{-0.5cm}
\subfloat[Negative ion at $t = 20$.]{
\centering
\includegraphics[width=0.33\linewidth]{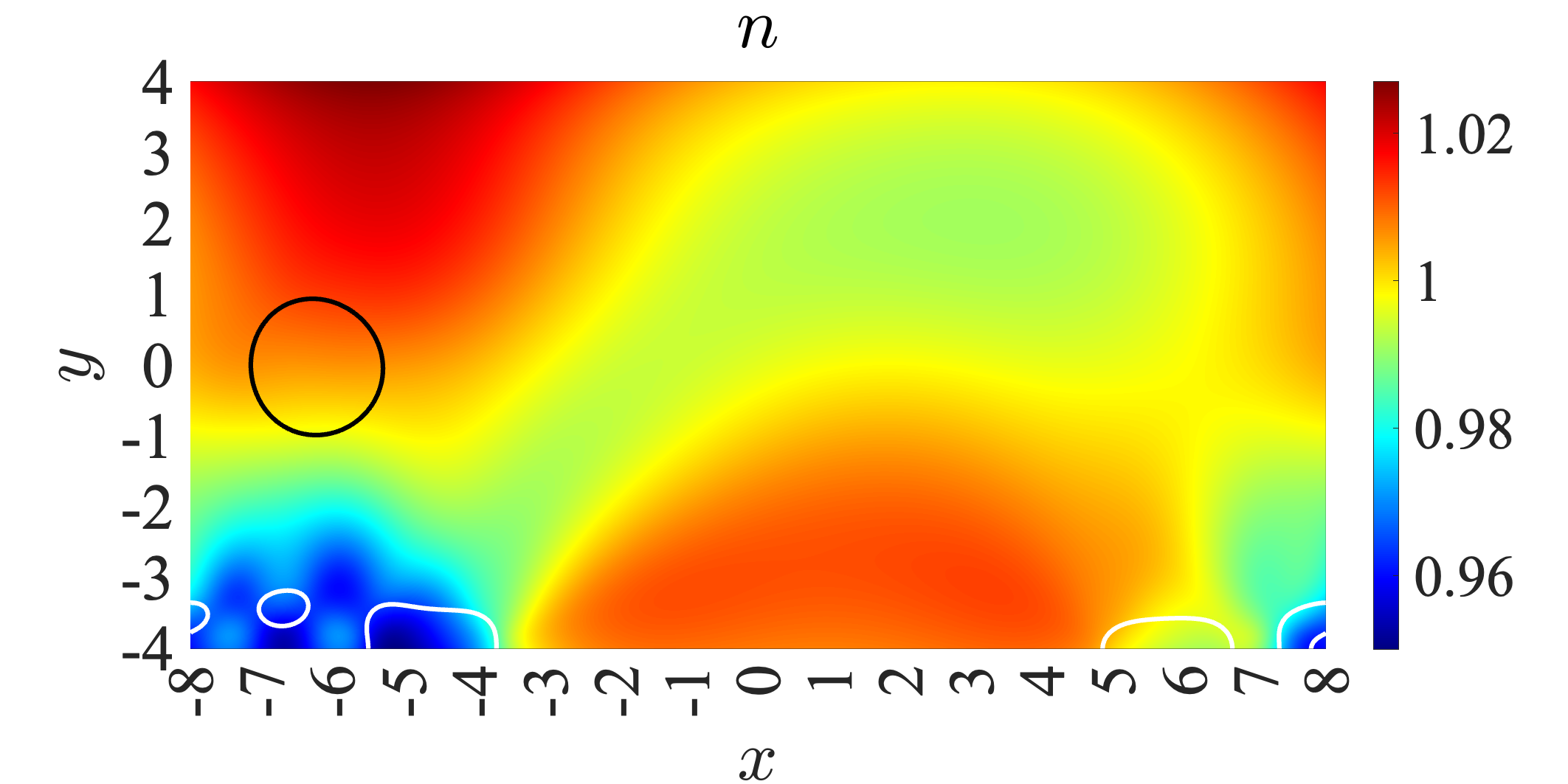}
\label{subfig:2DropsRPumpN20}
}
\hspace{-0.5cm}
\subfloat[Electric potential at $t = 20$.]{
\centering
\includegraphics[width=0.33\linewidth]{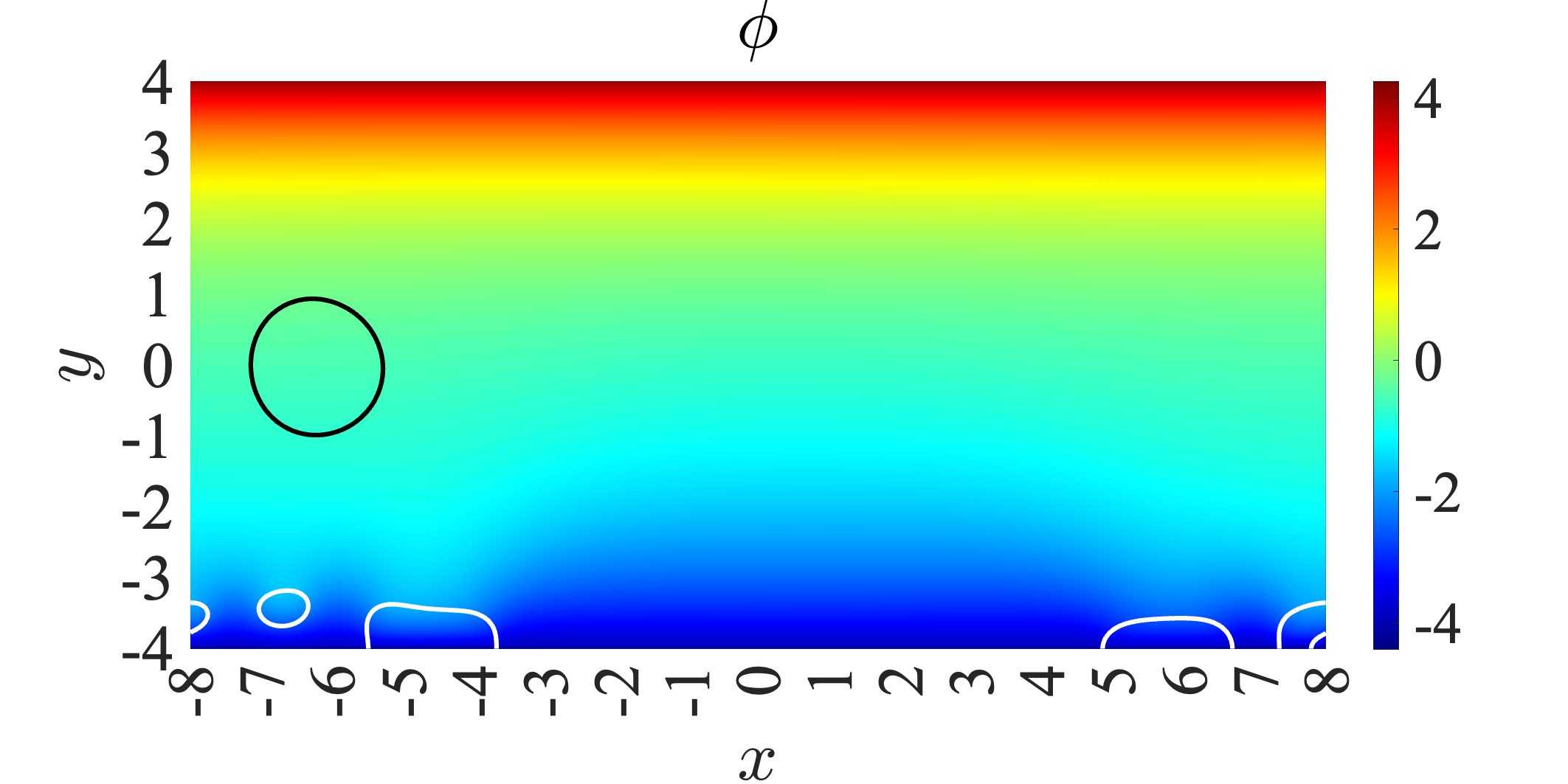}
\label{subfig:2DropsRPumpE20}
}
\\
\subfloat[Positive ion at $t = 30$.]{
\centering
\includegraphics[width=0.33\linewidth]{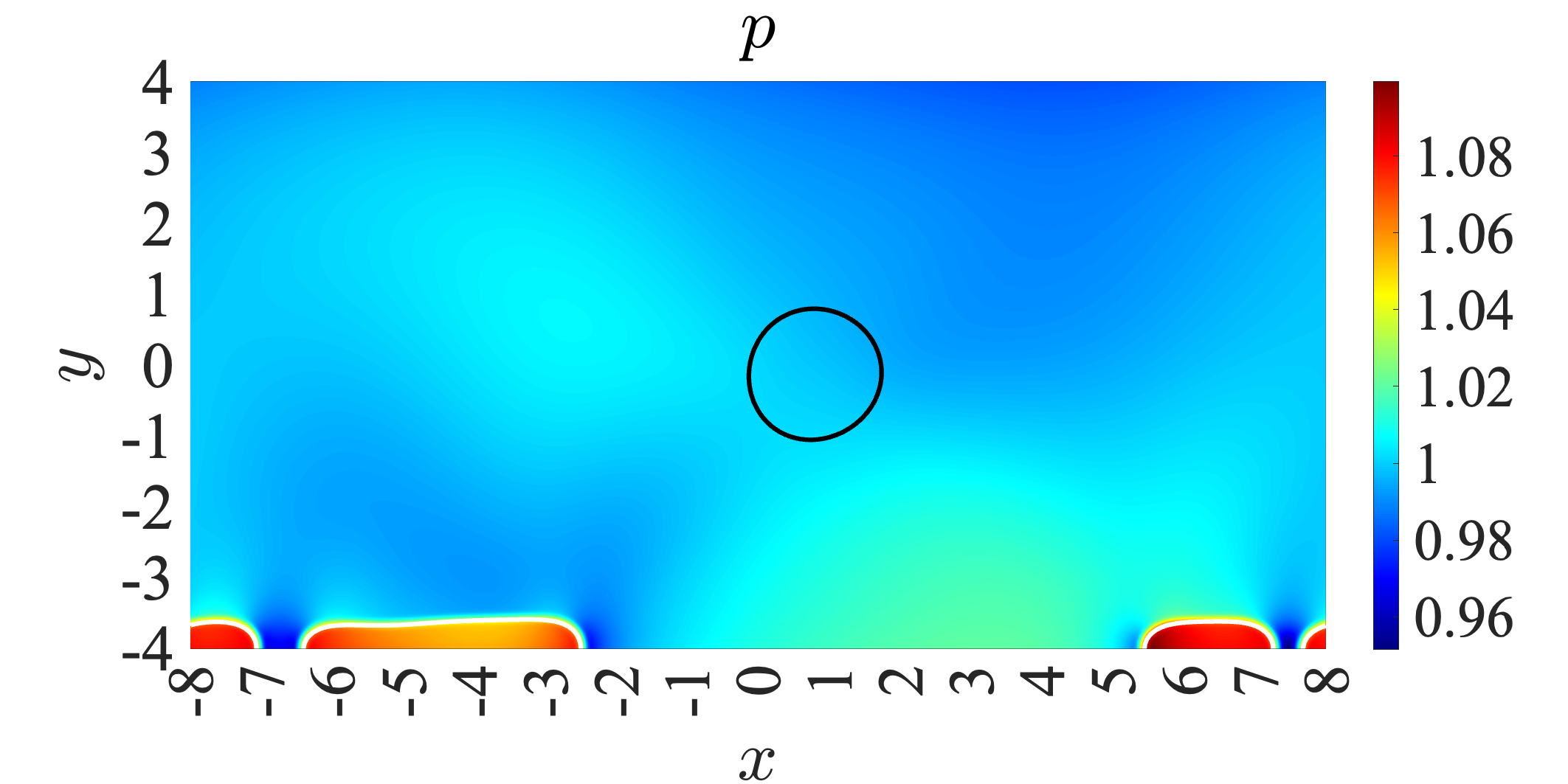}
\label{subfig:2DropsRPumpP30}
}
\hspace{-0.5cm}
\subfloat[Negative ion at $t = 30$.]{
\centering
\includegraphics[width=0.33\linewidth]{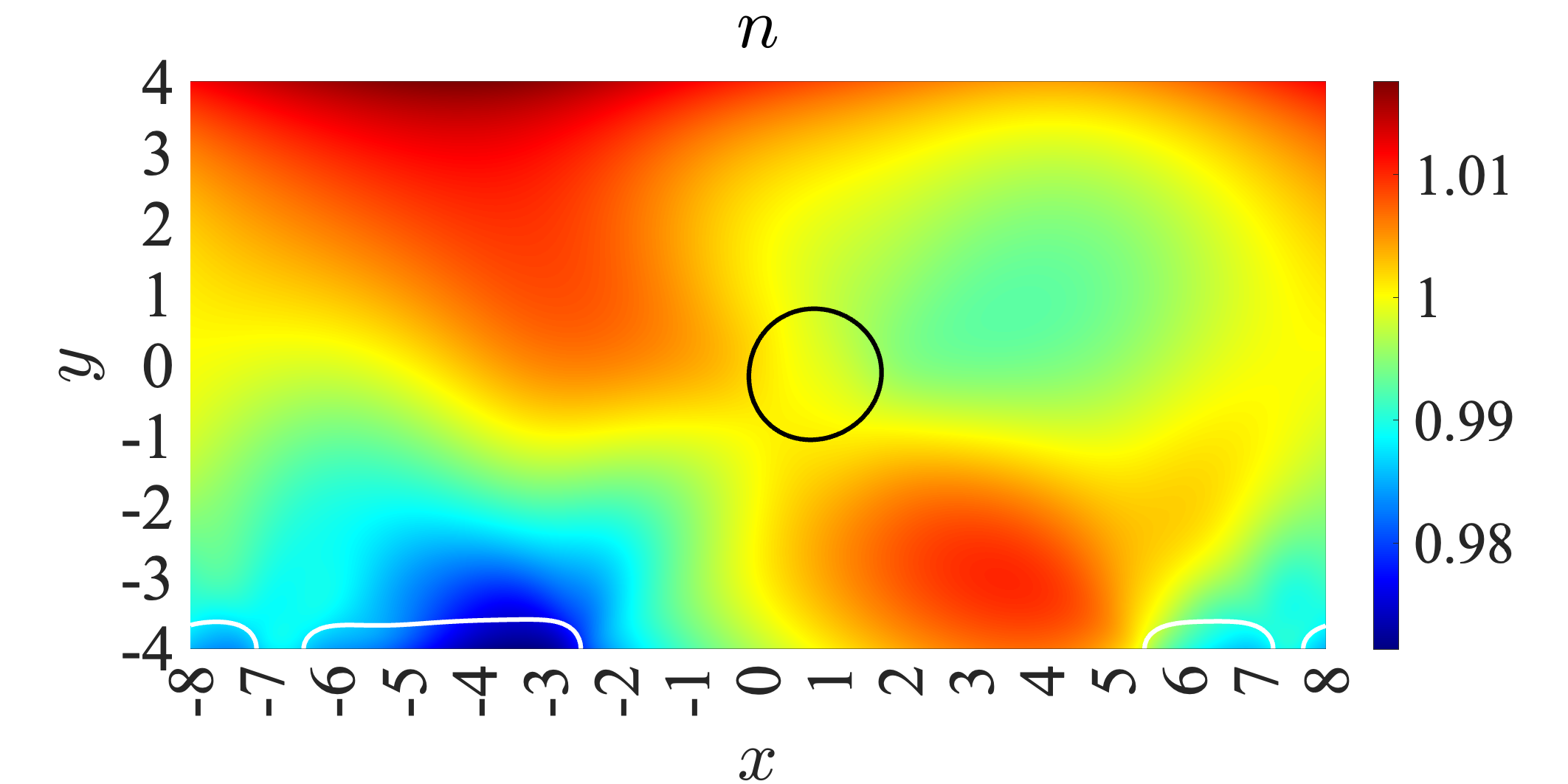}
\label{subfig:2DropsRPumpN30}
}
\hspace{-0.5cm}
\subfloat[Electric potential at $t = 30$.]{
\centering
\includegraphics[width=0.33\linewidth]{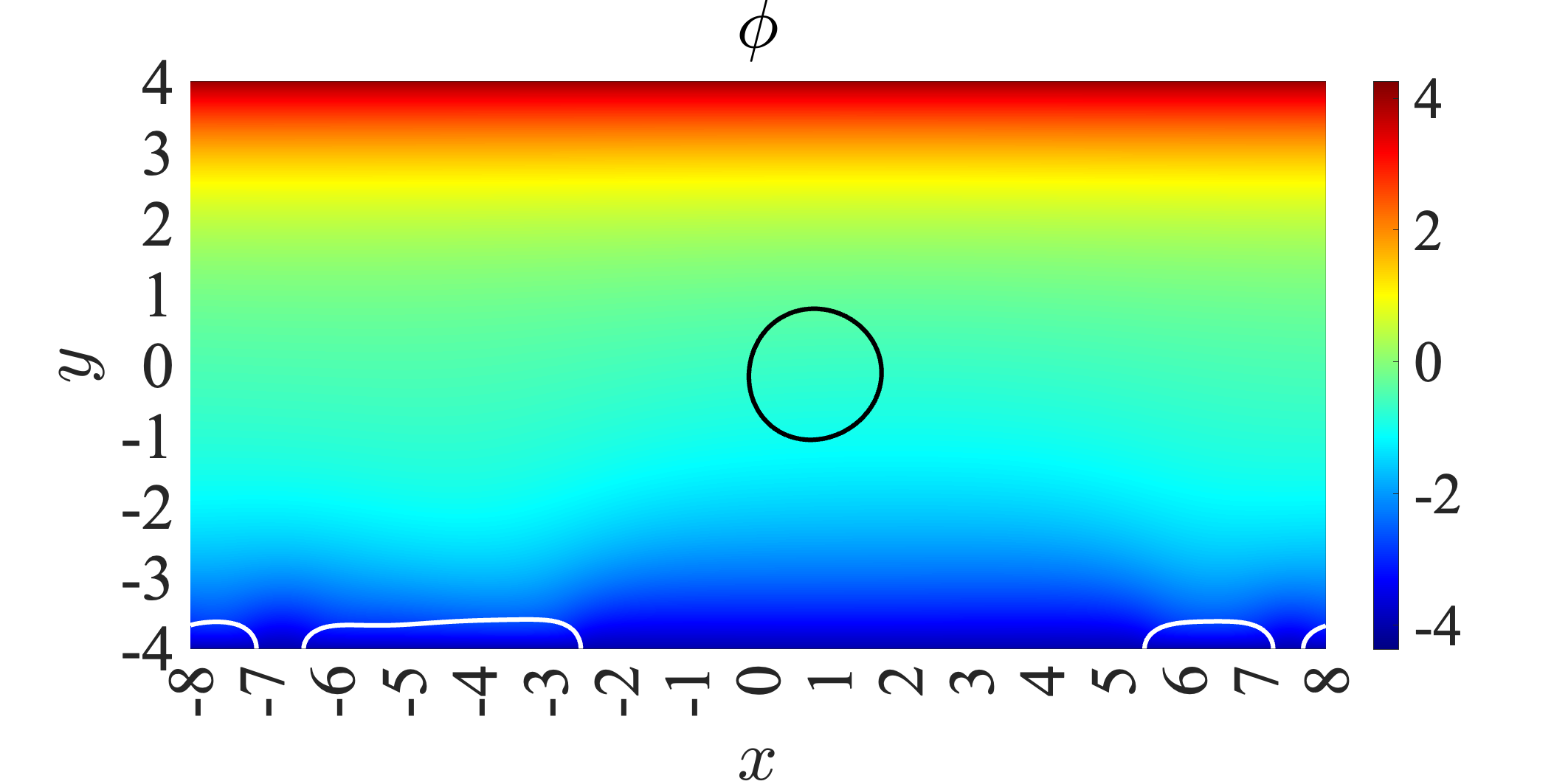}
\label{subfig:2DropsRPumpE30}
}
\caption{The snapshots for the drop filter at $t=0, 10, 20, 30$. 
The black and white solid circles represent the location of the left and right drop, 
respectively, which is denoted by the level set $\psi=0$. 
The positive ion pump is added to the right bubble which is denoted as white drop. 
The uniform distribution of positive ion and negative ion are chosen. 
However the electric potential is set as a linear distribution along $y$-axis.
}\label{fig:RightPump25}
\end{figure}

We investigate the behavior of these two bubbles under an applied electric field vertically. 
Specifically, we take the following initial conditions. 
\begin{subequations}
\begin{align}
& \psi_{L}(x,y,0) 
= \tanh\frac{\sqrt{1-((x+1.5)^{2}+y^{2})}}{\sqrt{2}\delta}, 
\\
& \psi_{R}(x,y,0) 
= \tanh\frac{\sqrt{1-((x+9.5)^{2}+y^{2})}}{\sqrt{2}\delta}, 
\\
& \bm{u}(x,y,0)=-\frac{1}{64}y^{2}+1, 
\\ 
& p(x,y,0) = 1, \quad n(x,y,0) = 1, \quad \phi(x,y,0) = y, 
\end{align}
\end{subequations}
which is shown in Fig. \ref{fig:RightPump25}, 
where the left and right drops are described as the black and white solid circles, respectively. 
We add a positive ion pump ion the right bubble. 
The uniform distribution of positive ion and negative ion are chosen. 
However the electric potential is set as a linear distribution along $y$-axis.
We take the periodic boundary conditions for all variables in the horizontal direction. 
As for $y$ direction, we take the following boundary conditions: 
\begin{equation}
\begin{aligned}
& \left. \nabla \psi_{L} \cdot \bm{n} \right|_{y = \pm 4} =
\left. \nabla \psi_{R} \cdot \bm{n} \right|_{y = \pm 4} = 0, 
\quad 
\left. \nabla \mu_{\psi_{L}} \cdot \bm{n} \right|_{y = \pm 4} = 
\left. \nabla \mu_{\psi_{R}} \cdot \bm{n} \right|_{y = \pm 4} = 0, 
\\
&
\left. \bm{u} \right|_{y = \pm 4} = \bm{0}, 
\quad 
\left. \nabla p \cdot \bm{n} \right|_{y = \pm 4} = 0, 
\quad 
\left. \nabla n \cdot \bm{n} \right|_{y = \pm 4} = 0, 
\quad 
\left. \phi \right|_{y= \pm 4} = \pm 4. 
\end{aligned}
\end{equation}
The parameters are chosen as 
\begin{equation}
\begin{aligned}
& Re = 5, \quad Ca_{E} = 0.5, \quad \zeta = 1, \quad Pe = 1, \quad Pe_{E} = 1, 
\\ 
& \alpha_{1} = 0.5, \quad \alpha_{2} = -0.5, 
\quad \beta = 2, \quad K_{0} = 3.5, 
\\ 
& \delta = 0.1, \quad M = \delta^{2}, \quad \epsilon_{r} = 1, \quad D_{i}^{r} = 1, \quad s = 2.
\end{aligned}
\end{equation}

As shown in Fig.~\ref{fig:RightPump25}, the droplet with active pumps accumulates positive ions, resulting in elevated electric potential and a net positive charge. Under the influence of the externally applied electric field, this droplet experiences a vertical Lorentz force that pulls it downward toward the bottom boundary. The resulting deformation, in conjunction with the background shear flow, leads to the droplet being elongated and eventually undergoing pinch-off and break-up near the bottom wall.

In contrast, the droplet without pumps maintains electrical neutrality and experiences no significant Lorentz force. It remains undeformed and is passively advected along with the background shear flow.

This differential response enables a clear spatial separation between the two droplets: the pumped droplet is trapped and broken near the bottom surface, while the unpumped droplet continues to move laterally. This demonstrates the potential of combining electrohydrodynamic actuation with flow control to achieve \textit{pump-mediated droplet sorting or separation}.

\section{Conclusion}
In this study, we have developed a thermodynamically consistent phase-field model to investigate how active ionic pumping across droplet interfaces modulates electrohydrodynamic behaviour. Our model systematically couples interfacial ion fluxes, electric potential governed by the Poisson equation, ionic transport via Nernst–Planck dynamics, and incompressible fluid motion through the Navier–Stokes equations. Directional pumping is implemented through prescribed fluxes on the interface, enabling the controlled accumulation of charge within the droplet.

By comparing simulations with and without active pumping, we have highlighted the critical role of surface pumps in dictating droplet dynamics. In the absence of pumps, electric fields merely polarise the droplets, resulting in symmetric charge distributions and negligible deformation. In contrast, when active pumping is present, the resulting charge asymmetry within the droplet interacts with the external electric field to generate Lorentz forces. These forces drive a rich array of behaviors, including vertical elongation, lateral displacement, crescent-shaped bending, pinch-off, and recoalescence, and the formation of star-like interfacial structures. In multi-droplet settings, electrostatic repulsion leads to increased separation and anisotropic deformation, further demonstrating the capacity for pump-driven shape modulation.
We also demonstrate that asymmetric pumping under shear flow can immobilise and rupture a droplet, while an unpumped droplet is advected downstream. This effect presents a promising mechanism for active sorting in droplet-based microfluidics.

While directional ionic transport is a natural feature of biological membranes, realised by embedded molecular pumps such as the \(\rm{Na^+/K^+}\)-ATPase, implementing similar functionality in artificial droplet systems remains a substantial challenge. In this work, we do not claim a specific physical realisation of surface-localised pumps; rather, we treat the droplet as a conceptual minimal model to probe the coupling between active ion transport, electrostatics, and interfacial hydrodynamics in a systematic and physically consistent manner. This abstraction allows us to explore the fundamental principles governing active electrohydrodynamics. It lays the groundwork for future theoretical and experimental studies in soft matter systems with heterogeneous and functionalized interfaces.
\section*{Funding}
The research of Yuzhe Qin was partially supported by the National Natural Science Foundation of China (Grant No. 12201369). 
The research of Huaxiong Huang was partially supported by the National Natural Science Foundation of China (Grant No. 12231004).

\section*{Declaration of interest} 
The authors report no conflict of interest.

\bibliographystyle{plain}
\bibliography{references} 
\appendix 
\section{Appendix}

\subsection{Non-dimensionlization}\label{sec: dimensionlization}
To solve the above physical model \eqref{main_eqn:pump} numerically, 
normalization is necessary to be introduced to improve convenience. 
We choose the characteristic length scale as $L = R$, 
where $R$ is the radius of the initial drop. 
The characteristic surface tension is defined as $\sigma_{s} = \lambda \delta$. 
The characteristic velocity scale is defined as $\tilde{u}=\sqrt{\frac{\sigma_{s}}{\rho R}}$, 
and the related characteristic time scale is $\tilde{t}=\sqrt{\frac{\rho R^{3}}{\sigma_{s}}}$. 
We choose the outer dielectric constant $\epsilon^{-}$ as the characteristic 
dielectric coefficient, and suppose there is an applied electric field $E_{\infty}$, 
then the momentum equation can be written as 
\begin{equation}\label{eqn: non-d-1}
\begin{aligned}
    &\frac{\rho \tilde{u}}{\tilde{t}} \left( \frac{\partial \bm{u}^{\prime}}{\partial t^{\prime}} 
    + \left( \bm{u}^{\prime} \cdot \nabla \right) \bm{u}^{\prime} \right) 
    +\frac{\tilde{P}}{L}\nabla P^{\prime} \\
    & = \frac{\eta \tilde{u}}{L^{2}}\nabla^{2}\bm{u}^{\prime} 
    - \frac{\lambda \delta^{2}}{L^{3}} \nabla \cdot \left( \nabla \psi \otimes \nabla \psi\right) 
    + \frac{\epsilon^{-}\left|E_{\infty}\right|^{2}}{L}
    \nabla \cdot \epsilon_{\mathrm{eff}}^{\prime} \left( \bm{E}^{\prime} \otimes \bm{E}^{\prime}
    - \frac{1}{2} \left| \bm{E}^{\prime} \right|^{2} \textbf{I} \right), 
\end{aligned}
\end{equation}
where we choose the character pressure as 
\begin{equation} 
    \tilde{P} = \frac{\sigma_{s}}{R}. 
\end{equation}
In addition, an electrical capillary number $Ca_{E}$, 
defined as $Ca_{E} = \epsilon^{-}R \left|E_{\infty}\right|^{2}/\sigma_{s}$, 
is introduced here to measure the strength of the electric field relative 
to the surface tension force, 
which means equation \eqref{eqn: non-d-1} will be transformed into 
\begin{equation}\label{eqn: non-d-2}
    \frac{\partial \bm{u}^{\prime}}{\partial t^{\prime}} 
    + \left( \bm{u}^{\prime} \cdot \nabla \right) \bm{u}^{\prime}  
    + \nabla P^{\prime} 
    = \frac{1}{Re}\nabla^{2}\bm{u}^{\prime} 
    - \delta^{\prime} \nabla \cdot \left( \nabla \psi \otimes \nabla \psi\right) 
    + Ca_{E} \nabla \cdot \epsilon_{\mathrm{eff}}^{\prime} \left( \bm{E}^{\prime} \otimes \bm{E}^{\prime} 
    - \frac{1}{2} \left| \bm{E}^{\prime} \right|^{2} \textbf{I} \right),
\end{equation} 
where the Reynold number, the dimensionless interface thickness 
and dielectric coefficient are represented as 
\begin{equation}
	Re = \frac{\sqrt{\rho R \sigma_{s}}}{\eta}, \quad 
	\delta^{\prime} = \frac{\delta}{R} \quad \mbox{and} 
	\quad \epsilon^{\prime}_{\mathrm{eff}} = \frac{1}{\frac{\left(1-\psi^{2}\right)^{2}}{\delta^{\prime} C_{m}^{\prime}}
    + \frac{1-\psi}{2}
    + \frac{1+\psi}{2\epsilon_{r}}}			
	\left(\epsilon_{r}=\frac{\epsilon^{+}}{\epsilon^{-}} \mbox{ and }
	C_{m}^{\prime}=\frac{C_{m}R}{\epsilon^{-}}	\right).  
\end{equation}
We choose the characteristic concentration as the reference concentration $\tilde{c}$, and similarly, the outer diffuse constant $D^{-}$ is set the characteristic diffuse coefficient, 
 the Nernst-Planck equation can be written as follows 
\begin{equation}
    \frac{\tilde{c}}{\tilde{t}}\frac{\partial c_{i}^{\prime}}{\partial t^{\prime}} 
    + \frac{\tilde{u} \tilde{c}}{L}\nabla \cdot \left( \bm{u}^{\prime} c_{i}^{\prime} \right) 
    = \frac{\tilde{D}\tilde{c}}{L^{2}}\nabla \cdot D_{i}^{\prime}\nabla c_{i}^{\prime} 
	+\frac{\tilde{D}\tilde{c}e \tilde{\phi}z_{i}}{k_{B}TL^{2}}\nabla \cdot D_{i}^{\prime} 
	\left( c_{i}^{\prime}  \nabla \phi^{\prime} \right)
    - \frac{I_{max}}{L} \nabla \cdot \bm{I}_{\mathrm{pump}}^{\prime}, 
\end{equation}
when we define the conductivity $\displaystyle\sigma = \tilde{D}\sum\limits_{i=1}^{N}z_{i}^{2}\tilde{c}e^{2}/k_{B}T$ 
and characteristic current density $I_{max} = \tilde{c}\tilde{u}$, 
the following equivalent form of the Nernst-Planck equation is obtained, 
\begin{equation}
	\frac{\partial c_{i}^{\prime}}{\partial t^{\prime}}
	+ \nabla \cdot \left( \bm{u}^{\prime} c_{i}^{\prime} \right)
        = \frac{1}{Pe}\nabla \cdot \left(D_{i}^{\prime} \nabla c_{i}^{\prime} \right)
+  \alpha_{i} \frac{\zeta^{2}}{Pe_{E}}\nabla \cdot\left( D_{i}^{\prime}c_{i}^{\prime} \nabla \phi^{\prime} \right)
- \nabla \cdot \bm{I}_{\mathrm{pump}}^{\prime}, 
\end{equation}
where electric relaxation time $\tilde{t}_{E}$, diffusion time $\tilde{t}_{D}$ the modified valence and the relative Peclet number $Pe$ and electric Peclet number $Pe_{E}$ are defined as follows 
\begin{equation}
\tilde{t}_{E} = \frac{\epsilon^{-}}{\sigma}, \quad 
\tilde{t}_{D} = \frac{L^{2}}{\tilde{D}}, \quad 
\alpha_{i} = \frac{z_{i}}{\sum\limits_{i=1}^{N}z_{i}^{2}}, \quad 
Pe = \frac{\tilde{t}_{D}}{\tilde{t}}, \quad 
Pe_{E} = \frac{\tilde{t}_{E}}{\tilde{t}}, 
\end{equation}  
and the effective diffusion coefficient is 
\begin{equation}
    D_{i}^{\prime} = \frac{1}{\frac{\left(1-\psi^{2}\right)^{2}}
    {\delta^{\prime} g\left(c_{i}^{\prime}\right)}
    +\frac{1-\psi}{2}
    +\frac{1+\psi}{2D_{i}^{r}}} 
    \left(\mbox{where }D_{i}^{r}=\frac{D_{i}^{+}}{D_{i}^{-}} \mbox{ and } g\left(c_{i}^{\prime}\right) = \frac{\frac{g_{i}}{\left(z_{i}e\right)^{2}}\frac{\partial\mu_{i}}{\partial c_{i}}R}{D_{i}^{-}}
    = \frac{g_{i}k_{B}TR}{\left(z_{i}e\right)^{2}D_{i}^{-}\tilde{c}}\frac{1}{c_{i}^{\prime}}
    = \frac{g_{i}^{\prime}}{c_{i}^{\prime}}\right). 
\end{equation}
For the Poisson equation, we have 
\begin{equation}
    - \frac{\epsilon^{-}\tilde{\phi}}{L^{2}} 
    \nabla \cdot \left( \epsilon_{\mathrm{eff}}^{\prime} \nabla \phi^{\prime} \right) 
    = \tilde{c} e \sum_{i=1}^{N} z_{i} c_{i}^{\prime}, 
\end{equation}
i.e. 
\begin{equation}
    - \zeta^{2} 
    \nabla \cdot \left( \epsilon_{\mathrm{eff}}^{\prime} \nabla \phi^{\prime} \right) 
    = \sum_{i=1}^{N} z_{i} c_{i}^{\prime}, 
\end{equation}
with the Debye length 
$\zeta = \sqrt{\epsilon_{0}\tilde{\phi}/\left(\tilde{c} e L^{2}\right)}$. 

For the Cahn-Hilliard equation, we have 
\begin{align}
	&\frac{1}{\tilde{t}}\frac{\partial \psi}{\partial t^{\prime}} 
	+ \frac{\tilde{u}}{L}\nabla \cdot \left( \bm{u}^{\prime} \psi \right) 
        = \frac{\mathcal{M}\tilde{\mu}}{L^{2}} \nabla^{2} \mu_{\psi}^{\prime}, \\
        &\tilde{\mu} \mu_{\psi}^{\prime} = \lambda \left( - \frac{\delta^{2}}{L^{2}} \nabla^{2} \psi 
        + F^{\prime}(\psi) \right) 
        - \frac{\epsilon^{-} \tilde{\phi}^{2}}{L^{2}}\frac{1}{2} \frac{\partial \epsilon_{\mathrm{eff}}}{\partial \psi} |\nabla\phi|^{2}, 
\end{align}
when we take $\tilde{\mu}=\sigma_{s}/L$ as the characteristic chemical potential, we have 
\begin{align}
	\mu_{\psi}^{\prime} = - \delta^{\prime} \nabla^{2} \psi + \frac{1}{\delta^{\prime}}F^{\prime}(\psi) 
	- \frac{Ca_{E}}{2} \frac{\partial \epsilon_{\mathrm{eff}}}{\partial \psi} |\nabla\phi|^{2}. 
\end{align}
By defining $M^{\prime} = \mathcal{M}\tilde{\mu}/L \tilde{u}$, we have the Cahn-Hilliard equations as follows, 
\begin{align}
	&\frac{\partial \psi}{\partial t^{\prime}} 
	+ \nabla \cdot \left( \bm{u}^{\prime} \psi \right) 
	= M^{\prime} \nabla^{2} \mu_{\psi}^{\prime}, \\
	&\mu_{\psi}^{\prime} = - \delta^{\prime} \nabla^{2} \psi 
	+ \frac{1}{\delta^{\prime}}F^{\prime}(\psi) 
	- \frac{Ca_{E}}{2} \frac{\partial \epsilon_{\mathrm{eff}}}{\partial \psi} |\nabla\phi|^{2}. 
\end{align}
So the system can be written as follows, 
\begin{subequations}
	\begin{align}
		& \frac{\partial \bm{u}^{\prime}}{\partial t^{\prime}} 
        + \left( \bm{u}^{\prime} \cdot \nabla \right) \bm{u}^{\prime} 
        + \nabla P^{\prime} 
        = \frac{1}{Re}\nabla^{2}\bm{u}^{\prime} 
        - \delta^{\prime} \nabla \cdot \left( \nabla \psi \otimes \nabla \psi\right) 
        + Ca_{E} \nabla \cdot \epsilon_{\mathrm{eff}}^{\prime} \left( \bm{E}^{\prime} \otimes \bm{E}^{\prime} 
        - \frac{1}{2} \left| \bm{E}^{\prime} \right|^{2} \textbf{I} \right),
        \\
        & \nabla \cdot \bm{u}^{\prime} = 0, 
        \\ 
		&\frac{\partial c_{i}^{\prime}}{\partial t^{\prime}} 
		+ \nabla \cdot \left( \bm{u}^{\prime} c_{i}^{\prime} \right) 
		= \frac{1}{Pe}\nabla \cdot \left( D_{i}^{\prime}\nabla c_{i}^{\prime} \right)
		+ \alpha_{i}\frac{\zeta^{2}}{Pe_{E}}\nabla \cdot \left( D_{i}^{\prime}c_{i}^{\prime}\nabla  \phi^{\prime}\right)
        -\nabla \cdot \bm{I}_{\mathrm{pump}}^{\prime}, 
		\\ 
        & \bm{I}_{\mathrm{pump}}^{\prime} 
		= I_{0} \left(\frac{\frac{1-\psi}{2}c_{i}}{K_{0} 
		+ \frac{1-\psi}{2}c_{i}}\right)^{\beta}\bm{n}, 
        \\ 
        &- \zeta^{2} \nabla \cdot ( \epsilon_{\it eff}^{\prime} \nabla \phi^{\prime} ) 
        = \sum_{i=1}^{N} z_{i} c_{i}^{\prime}, 
        \\ 
		&\frac{\partial \psi}{\partial t^{\prime}} 
		+ \nabla \cdot \left( \bm{u}^{\prime} \psi \right) 
		= M^{\prime} \nabla^{2} \mu_{\psi}^{\prime}, 
		\\
		&\mu_{\psi}^{\prime} = - \delta^{\prime} \nabla^{2} \psi 
        + \frac{1}{\delta^{\prime}}F^{\prime}(\psi) 
        - \frac{Ca_{E}}{2} \frac{\partial \epsilon_{\mathrm{eff}}}{\partial \psi} |\nabla\phi|^{2}. 
	\end{align}
\end{subequations}
If we rewrite the new pressure 
$\hat{P}^{\prime} = P^{\prime} + \frac{1}{\delta^{\prime}} F\left(\psi\right) 
+\frac{\delta^{\prime}}{2} \left|\nabla \psi \right|^{2}$
and omit all the superscript ${}^\prime$ and $\hat{}$ for simplicity, 
the above system can be written as follows, 
\begin{subequations}\label{Aeqn:simplified_system}
	\begin{align}
		& \frac{\partial \bm{u}}{\partial t} 
        + \left( \bm{u} \cdot \nabla \right) \bm{u} 
        + \nabla P 
        = \frac{1}{Re}\nabla^{2}\bm{u} 
        + \mu_{\psi} \nabla \psi 
        + \frac{Ca_{E}}{\zeta^{2}} \sum_{i=1}^{N} z_{i} c_{i} \nabla \phi,
        \label{Aeqn:u}
        \\ 
        & \nabla \cdot \bm{u} = 0, 
        \label{Aeqn:nabla_u}
        \\
		&\frac{\partial c_{i}}{\partial t} 
		+ \nabla \cdot \left( \bm{u} c_{i} \right) 
		= \frac{1}{Pe} 
		\nabla \cdot \left( D_{i} \nabla c_{i} \right) 
        + \alpha_{i} \frac{\zeta^{2}}{Pe_{E}} \nabla \cdot \left( D_{i} c_{i} \nabla \phi \right) - \nabla \cdot \bm{I}_{\mathrm{pump}},
		\label{Aeqn:ci}
        \\ 
        & \bm{I}_{\mathrm{pump}} 
		= I_{0} \left(\frac{\frac{1-\psi}{2}c_{i}}{K_{0} 
		+ \frac{1-\psi}{2}c_{i}}\right)^{\beta}\bm{n}, 
        \label{Aeqn:pump}
		\\
        &- \zeta^{2} 
        \nabla \cdot \left( \epsilon_{\mathrm{eff}} \nabla \phi \right) 
        = \sum_{i=1}^{N} z_{i} c_{i}, 
        \\ 
		&\frac{\partial \psi}{\partial t} 
		+ \nabla \cdot \left( \bm{u} \psi \right) 
		= M \nabla^{2} \mu_{\psi}, 
		\label{Aeqn:psi}
        \\
		&\mu_{\psi} = - \delta \nabla^{2} \psi 
        + \frac{1}{\delta} F^{\prime} \left( \psi \right)
        - \frac{Ca_{E}}{2} 
        \frac{\partial \epsilon_{\mathrm{eff}}}{\partial \psi} 
        \left|\nabla\phi\right|^{2}, 
        \label{Aeqn:mu_psi}\\
        &\epsilon_{\mathrm{eff}}^{-1} 
        = \frac{1-\psi}{2} 
        + \frac{1+\psi}{2\epsilon_{r}}, 
        \label{Aeqn:eps}\\
        & D_{i}^{-1} 
        = \frac{1-\psi}{2}
        + \frac{1+\psi}{2D_{i}^{r}}
        + \frac{\left(1-\psi^{2}\right)^{2}c_{i}}{\delta g_{i}}. 
        \label{Aeqn:Deff}
	\end{align}
\end{subequations}

\subsection{Pump flux derivation from mass action law}

We consider the classical active pump process as the following reaction:

\begin{equation}
E +  S_{o}+ATP \overset{k_1}{\underset{k_{-1}}{\rightleftharpoons}} ES-ATP \overset{k_2}{\underset{k_{-2}}{\rightleftharpoons}} E + \beta S_i+Pi
\end{equation}
 Here:
\begin{itemize}
  \item $E$: free pump protein
  \item $S_o$: substrate of extracellular
  \item $S_i$: substrate of intracellular
  \item $ES-ATP$: occupied pump
  \item $ATP$: required co-substrate
  \item $P_i$: phosphate (product of ATP hydrolysis)
  \item $k_1$: rate constant for complex formation
  \item $k_{-1}$: rate constant for complex dissociation
  \item $k_2$: rate constant for product formation
  \item $k_{-2}$:  rate constant for product react
\end{itemize}
Then we have 
\begin{align*}
    \frac{\mathrm{d}[E]}{\mathrm{d}t} &=(k_{-1}+k_2)[ES-ATP]-k_1[S_o] [ATP][E]-k_{-2}[S_i] [P_i][E], \\
   \frac{\mathrm{d}[S_o]}{\mathrm{d}t} &=k_{-1}[ES-ATP]-k_1[S_o] [ATP][E],\\
    \frac{\mathrm{d}[ATP]}{\mathrm{d}t} & = k_{-1}[ES-ATP]-k_1[S_o]^\beta[ATP][E],\\
    \frac{\mathrm{d}[ES-ATP]}{\mathrm{d}t} & = -(k_{-1}+k_2)[ES-ATP] +k1[S_o] [ATP][E] +k_{-2}[S_i] [P_i][E],\\
    \frac{\mathrm{d}[P_i]}{\mathrm{d}t} &= k_2[ES-ATP]-k_{-2}[S_i] [P_i][E],\\
     \frac{\mathrm{d}[S_i]}{\mathrm{d}t} &= k_2[ES-ATP]-k_{-2}[S_i]^\beta[P_i][E],
\end{align*}
We assume that the intermediate complex [ES-ATP] reaches the equilibrium state quickly:
\begin{equation}
(k_{-1}+k_2)[ES-ATP] =\left(k_1[S_o] [ATP] +k_{-2}[S_i] [P_i]\right)[E]. 
\end{equation}
The total pump number is conserved 
\begin{equation}
[E]+[ES-ATP]=[E]_0.
\end{equation}
Then 
\begin{align}
[ES-ATP] = & \frac{(k_1[S_o] [ATP] +K_{-2}[S_i] [P_i])[E]_0}{(k_{-1}+k_2)+(k1[S_o] [ATP] +K_{-2}[S_i] [P_i])} \nonumber \\ 
= & [E]_0\frac{[S_o] [ATP] +K_{P}[S_i] [P_i]}{K_M +([S_o] [ATP] +K_{P}[S_i] [P_i]) },
\end{align}
where 
\begin{equation}
K_M =\frac{k_{-1}+k2}{k_1}, K_P =\frac{k_{-2}}{k_1}.
\end{equation}
Then the net pump rate is 
\begin{align*}
\frac{\mathrm{d}S_i}{\mathrm{d}t} &=  k_2[ES-ATP]-K_{-2}[S_i] [P_i]([E]_0-[ES-ATP]) \\
&= (k_2+k_{-2}[S_i] [P_i]) [E]_0\frac{[S_o] [ATP] +K_{P}[S_i] [P_i]}{K_M +([S_o] [ATP] +K_{P}[S_i] [P_i]) }-k_{-2}[S_i] [P_i][E]_0. 
\end{align*}

So if $k_{-2}<<1$, then it is approximated by 

\begin{align*}
\frac{\mathrm{d}S_i}{\mathrm{d}t} &=  k_2[E]_0\frac{[ATP][S_o] }{K_M +[S_o] [ATP]}\\
&= J_{max}\frac{[ATP][S_o] }{K_M +[S_o]^\beta[ATP] }. 
\end{align*}
where $J_{max} = k_2[E]_0.$

\subsection{Numerical implementation}\label{sec:scheme}
Due to the nonlinearity and advection in our model, 
we need to pay more attention to the instability in computation. 
Firstly, we rewrite Eqn. \eqref{eqn:ci} as 
\begin{equation}\label{eqn: ci with pump reformulation}
\frac{\partial c_{i}}{\partial t} 
+ \nabla \cdot \left(\bm{u} c_{i} \right) 
- \nabla \cdot \left( \frac{D_{i} \alpha_{i} \zeta^{2}}{Pe_{E}} \nabla \phi c_{i}\right) 
+ \nabla \cdot \left( I_{0} \frac{\left(\frac{1-\psi}{2}\right)^{\beta}c_{i}^{\beta - 1}}
{\left( K_{0} + \frac{1-\psi}{2}c_{i}\right)^{\beta}} \bm{n} c_{i} \right) 
= \nabla \cdot \left( \frac{D_{i}}{Pe} \nabla c_{i} \right). 
\end{equation}
If we define a new velocity as 
\begin{equation}
\bm{v} = \bm{u} 
- \frac{D_{i} \alpha_{i} \zeta^{2}}{Pe_{E}} \nabla \phi 
+ I_{0} \frac{\left(\frac{1-\psi}{2}\right)^{\beta}c_{i}^{\beta-1}}
{\left( K_{0} + \frac{1-\psi}{2}c_{i}\right)^{\beta}} \bm{n}, 
\end{equation}
then equation \eqref{eqn: ci with pump reformulation} can be expressed as 
\begin{equation}
\frac{\partial c_{i}}{\partial t} 
+ \nabla \cdot \left( \bm{v} c_{i} \right) 
= \nabla \cdot \left( \frac{D_{i}}{Pe} \nabla c_{i} \right), 
\end{equation}
which is a convection-diffusion equation. 
We define the uniform time step as $\tau = t^{n+1} - t^{n}$, 
and $h_{x}, h_{y}$ as the space step along $x$ and $y$ direction. 

\subsubsection{Time Discretization}

In this section, we present the time discretization scheme used in our numerical simulations. An implicit-explicit (IMEX) approach is adopted: linear terms are treated implicitly to ensure stability, while nonlinear terms are handled explicitly for computational efficiency. To improve robustness, we apply a stabilization method for the Cahn--Hilliard equation and use a projection (pressure correction) method for the Navier--Stokes equations.

The full time discretization of the coupled system reads:
\begin{subequations}\label{scheme:pump_system}
\begin{align}
&\frac{\psi^{n+1}-\psi^{n}}{\tau} 
+ \nabla \cdot \left( \bm{u}^{n} \psi^{n} \right) 
= M \nabla^{2} \mu_{\psi}^{n+1}, 
\\
&\mu_{\psi}^{n+1} = - \delta \nabla^{2} \psi^{n+1} 
+ \frac{s}{\delta} \left( \psi^{n+1} - \psi^{n} \right)
+ \frac{1}{\delta} F^{\prime}(\psi^{n})
- \frac{Ca_E}{2} 
\frac{\partial \epsilon_{\it eff}^{n}}{\partial \psi} 
\left|\nabla \phi^{n} \right|^{2}, 
\\
&\frac{c_{i}^{n+1}-c_{i}^{n}}{\tau} 
+ \nabla \cdot \left( \bm{v}^{n} c_{i}^{n} \right) 
= - \nabla \cdot \left( \frac{D_i(\psi^{n+1}, c_i^n)}{Pe} \nabla c_i^{n+1} \right),
\label{scheme:ci}
\\
&- \zeta^2 
\nabla \cdot \left( \epsilon_{\it eff}(\psi^{n+1}) \nabla \phi^{n+1} \right) 
= \sum_{i=1}^N z_i c_i^{n+1}, 
\label{scheme:phi}
\\
&\frac{\bm{u}^{\star} - \bm{u}^{n}}{\tau} 
+ \left( \bm{u}^{\star} \cdot \nabla \right) \bm{u}^{n} 
+ \nabla P^{n} 
= \frac{1}{Re} \nabla^{2} \bm{u}^{\star} 
+ \mu_{\psi}^{n+1} \nabla \psi^{n} 
- \frac{Ca_E}{\zeta^2} \sum_{i=1}^{N} z_i c_i^n \nabla \phi^{n},
\\
&\frac{\bm{u}^{n+1} - \bm{u}^{\star}}{\tau} 
+ \nabla \left( P^{n+1} - P^{n} \right) = 0,
\\
&\nabla \cdot \bm{u}^{n+1} = 0.
\end{align}
\end{subequations}

Here, $s$ is a stabilization parameter used to suppress nonphysical oscillations in the phase-field evolution, as commonly employed in numerical solutions of Cahn--Hilliard-type equations.

\begin{remark}
In this paper, our primary objective is to develop and demonstrate a mathematical model 
that captures active ion transport across an interface and maintains concentration asymmetry. 
While the continuous system satisfies the energy law and preserves non-negativity of ion concentrations, 
we do not focus on designing structure-preserving numerical schemes in this work. 
We instead use sufficiently small time steps to ensure numerical accuracy and stability.

We acknowledge that it is possible to construct energy and positivity preserving schemes \cite{Qin2025PNPNS} using established techniques such as the stabilization method~\cite{Qin2022JCP,Qin2025IGA}, 
convex splitting~\cite{Qin2021IJNAM,Qin2025PNPNS}, 
the invariant energy quadratization (IEQ) method~\cite{Yang2016IEQ,Yang2017IEQ,Yang2018IEQ,Zhao2017IEQ}, 
and the scalar auxiliary variable (SAV) method~\cite{Qin2020SAV,Qin2022M2AS,Shen2018SAV,Shen2019SAV}. Developing such schemes and analyzing their error properties for the proposed model will be addressed in future work.
\end{remark}

\subsubsection{Spatial Discretization}

To maintain second-order spatial accuracy while suppressing numerical instabilities induced by convection, we adopt a modified upwind finite difference method~\cite{Axelsson1979Modified} in conjunction with a staggered mesh arrangement. This approach enhances stability and avoids spurious oscillations, particularly in regions with sharp gradients.

Here, we mainly illustrate the spatial discretization of the convection term. Since the Cahn--Hilliard equation involves only a single convection term associated with velocity, it does not require special treatment beyond the standard form.

For the ion concentration,  let $f$ denote a generic scalar variable of concentration ($c_i$), and define the vector quantity $\bm{w} = \bm{u} f$ or $\bm{w} = \bm{v} f$, depending on context. The convection term then takes the form $\nabla \cdot \bm{w}$. In our staggered mesh configuration, scalar variables ($f$) are located at the centers of computational cells;
Velocity components are defined on the cell edges—$v^y$ on vertical edges and $v^x$ on horizontal edges.
This mesh structure is illustrated schematically in the following figure. It ensures proper alignment of scalar and vector fields, facilitating accurate and conservative flux computations.
The modified upwind scheme is applied to discretize $\nabla \cdot \bm{w}$, with directional bias determined by the sign of the local velocity. This formulation achieves a balance between upwind stability and second-order spatial accuracy.

\begin{center}
\begin{tikzpicture}

\draw[thick] (0,0) rectangle (4,4);

\fill[black] (2.0,2.0) circle (3pt);
\node at (2.0,1.5) {$f_{j+\frac{1}{2},k+\frac{1}{2}}$};

\fill[red] (0.1,2) -- (-0.1,2.1) -- (-0.1,1.9) -- cycle; 
\node[red] at (-0.6,2) {$v^{x}_{j,k+\frac{1}{2}}$};

\fill[red] (4.1,2) -- (3.9,2.1) -- (3.9,1.9) -- cycle; 
\node[red] at (4.8,2) {$v^{x}_{j+1,k+\frac{1}{2}}$};

\fill[blue] (2,4.1) -- (1.9,3.9) -- (2.1,3.9) -- cycle; 
\node[blue] at (2,4.4) {$v^{y}_{j+\frac{1}{2},k+1}$};

\fill[blue] (2,0.1) -- (1.9,-0.1) -- (2.1,-0.1) -- cycle; 
\node[blue] at (2,-0.3) {$v^{y}_{j+\frac{1}{2},k}$};

\end{tikzpicture}

\end{center}
For the scalar variable, we have the following Taylor expansion: 
\begin{subequations}
\begin{align}
& f_{j,k+\frac{1}{2}} 
= f_{j+\frac{1}{2},k+\frac{1}{2}} 
- \frac{h_{x}}{2} \frac{\partial f_{j,k+\frac{1}{2}}}{\partial x} 
- \frac{h^{2}_{x}}{8} \frac{\partial^{2} f}{\partial x^{2}} \left(\gamma_{1},y_{k+\frac{1}{2}}\right), 
\quad 
\gamma_{1} \in \left(x_{j},x_{j+\frac{1}{2}}\right), 
\\
& f_{j,k+\frac{1}{2}} 
= f_{j-\frac{1}{2},k+\frac{1}{2}} + \frac{h_{x}}{2} \frac{\partial f_{j,k+\frac{1}{2}}}{\partial x} 
+ \frac{h^{2}_{x}}{8} \frac{\partial^{2} f}{\partial x^{2}} \left(\gamma_{2},y_{k+\frac{1}{2}}\right), 
\quad 
\gamma_{2} \in \left(x_{j-\frac{1}{2},x_{j}}\right). 
\end{align}
\end{subequations} 
To treat the convection term with the modified upwind finite difference method, 
we introduce the index function $\chi\left(v\right)$ firstly, which is 
\begin{equation}
\chi\left(v\right) = \left\{
\begin{array}{rcl}
1, & & v\geq 0, \\
0, & & v  <  0. 
\end{array}
\right.
\end{equation}
Then the convection term can be written as 
\begin{equation}
\begin{aligned}
& w^{x} 
= v^{x}\left(\chi\left(v^{x}\right)f 
+ \left(1-\chi\left(v^{x}\right)\right)f\right), \\
& w^{y} 
= v^{y}\left(\chi\left(v^{y}\right)f 
+ \left(1-\chi\left(v^{y}\right)\right)f\right). 
\end{aligned}
\end{equation}
Therefore we obtain 
\begin{equation}
\begin{aligned}
& w^{x}_{j,k+\frac{1}{2}} 
= v^{x}_{j,k+\frac{1}{2}}\left[\chi\left(v^{x}_{j,k+\frac{1}{2}}\right)f_{j,k+\frac{1}{2}} + \left(1-\chi\left( v^{x}_{j,k+\frac{1}{2}}\right)\right)f_{j,k+\frac{1}{2}}\right]
\\ 
= & v^{x}_{j,k+\frac{1}{2}} \left[\chi\left(v^{x}_{j,k+\frac{1}{2}}\right)\left(f_{j-\frac{1}{2},k+\frac{1}{2}} + \frac{h_{x}}{2} \frac{\partial f_{j,k+\frac{1}{2}}}{\partial x} 
+ \frac{h^{2}_{x}}{8} \frac{\partial^{2} f}{\partial x^{2}} \left(\gamma_{2},y_{k+\frac{1}{2}}\right)\right) \right.
\\
& \qquad \qquad \left. + \left(1-\chi\left( v^{x}_{j,k+\frac{1}{2}}\right)\right)\left(f_{j+\frac{1}{2},k+\frac{1}{2}} - \frac{h_{x}}{2} \frac{\partial f_{j,k+\frac{1}{2}}}{\partial x} 
- \frac{h^{2}_{x}}{8} \frac{\partial^{2} f}{\partial x^{2}} \left(\gamma_{1},y_{k+\frac{1}{2}}\right)\right)\right] 
\\ 
= & v^{x}_{j,k+\frac{1}{2}} \left( \chi\left(v^{x}_{j,k+\frac{1}{2}}\right)f_{j-\frac{1}{2},k+\frac{1}{2}} 
+ \left(1-\chi\left( v^{x}_{j,k+\frac{1}{2}}\right)\right) f_{j+\frac{1}{2},k+\frac{1}{2}}\right)
+ \frac{\left| v^{x}_{j,k+\frac{1}{2}} \right| h_{x}}{2} \frac{\partial f_{j,k+\frac{1}{2}}}{\partial x} 
+ O\left(h_{x}^{2}\right). 
\end{aligned}
\end{equation}
Similarly, we have 
\begin{equation}
\begin{aligned}
w^{y}_{j+\frac{1}{2},k} 
= & v^{y}_{j+\frac{1}{2},k} \left( \chi\left(v^{y}_{j+\frac{1}{2},k}\right)f_{j+\frac{1}{2},k-\frac{1}{2}} 
+ \left(1-\chi\left( v^{y}_{j+\frac{1}{2},k}\right)\right) f_{j+\frac{1}{2},k+\frac{1}{2}}\right) \\ 
& + \frac{\left| v^{y}_{j+\frac{1}{2},k} \right| h_{y}}{2} \frac{\partial f_{j+\frac{1}{2},k}}{\partial x} 
+ O\left(h_{y}^{2}\right). 
\end{aligned}
\end{equation}
The second order accuracy can be obtained through the above modification.

\subsection{Convergence Study} \label{Ap:constudy}
In this section, we conduct the convergence study to validate our numerical code.
The computational domain is set as $\Omega=[-4,4]\times[-4,4]$ which is the same as we use in the 2D experiments. 
The following smooth enough initial condition is chosen: 
\begin{subequations}
    \begin{align}
        & \psi (x, y, 0) = \sin(2\pi x)\sin(2\pi y), \\
        & u (x, y, 0) = \sin(2\pi x)\sin(\pi y), \\
        & v (x, y, 0) = \sin(\pi x)\sin(2\pi y), \\
        & P (x, y, 0) = \cos(2\pi x)\cos(2\pi y), \\
        & p (x, y, 0) = \cos(2\pi x)\cos(\pi y), \\
        & n (x, y, 0) = \cos(\pi x)\cos(2\pi y), 
    \end{align}
\end{subequations}
with the following boundary condition: 
\begin{equation}
    \nabla \psi \cdot \bm{n} |_{\partial\Omega} = 0, \quad 
    u |_{\partial\Omega} = v |_{\partial\Omega} = 0, \quad 
    \nabla P \cdot \bm{n} |_{\partial\Omega} = 0, \quad 
    \nabla p \cdot \bm{n} |_{\partial\Omega} = 0, \quad 
    \nabla n \cdot \bm{n} |_{\partial\Omega} = 0,
\end{equation}
where $u$ and $v$ are the horizontal and vertical components of velocity $\bm{u}$. 
The initial electric potential $\phi(x,y,0)$ is computed from the initial ion concentration $p$ and $n$ 
by using the Poisson equation. 
The dimensionless numbers are: 
\begin{equation}
\begin{aligned}
    & Ca_{E} = 1, \quad Pe = 1, \quad Pe_{E} = 1, \quad \zeta = 1, \quad I_{0} = 1, \quad K_{0} = 1, \\
    & M = 1, \quad \delta = 1, \quad s = 2, \quad D_{i}^{r} = 1, \quad \epsilon_{r} = 1, \quad z_{p} = 1, 
    \quad z_{n} = -1. 
\end{aligned}
\end{equation}
The model’s complexity renders an analytical solution nearly impossible. 
Hence we take the Cauchy's error to verify the convergence rate, which is used in \cite{Qin2025PNPNS}. 
Because we employ the first order scheme in time and second order scheme in space discretization, 
we take $\tau = C h^{2}$, where $\tau$ is the time step size and $h$ is the spacial step size in the uniform grids. $C=0.01$ is a constant. The final time is chosen as 0.08. 
The discrete $L^{2}$ and $L^{\infty}$ error and convergence rate are shown in the following Table
\ref{tab:L2convergencePNP} - \ref{tab:LinftyconvergenceNS}. 
The $L^{2}$ errors of all unknowns converge at almost second order, 
while the pressure is only first-order accurate in the $L^{\infty}$ norm. 
This is a well-known outcome for MAC/projection schemes due to the lower $L^{\infty}$ stability of the pressure Poisson problem. 
\begin{table}[!ht]
\centering
\caption{$L^{2}$ error and Convergence for $p$, $n$ and $\phi$}
\begin{tabular}{cccccccccc}
\hline
$N$ & $\|e(p)\|_{L^2}$ & Rate & $\|e(n)\|_{L^2}$ & Rate & $\|e(\phi)\|_{L^2}$ & Rate \\
\hline
64  & 2.2235e-02 &        & 2.2230e-02 &        & 1.1030e-03 &        \\
128 & 5.3708e-03 & 2.0499 & 5.3698e-03 & 2.0493 & 2.3133e-04 & 2.2534 \\
256 & 1.3501e-03 & 1.9922 & 1.3497e-03 & 1.9921 & 6.8167e-05 & 1.7628 \\
\hline
\end{tabular}
\label{tab:L2convergencePNP}
\end{table}

\begin{table}[!ht]
\vskip -0.2cm
\centering
\caption{$L^{2}$ error and Convergence for $u$, $v$, $P$ and $\psi$}
\begin{tabular}{cccccccccccc}
\hline
$N$ & $\|e(u)\|_{L^2}$ & Rate & $\|e(v)\|_{L^2}$ & Rate & $\|e(P)\|_{L^2}$ & Rate & $\|e(\psi)\|_{L^2}$ & Rate \\
\hline
64  & 1.0684e-02 &        & 1.0684e-02 &        & 6.8913e-02 &        & 8.7197e-03 &        \\
128 & 2.4791e-03 & 2.1076 & 2.4791e-03 & 2.1076 & 1.4876e-02 & 2.2117 & 1.9244e-03 & 2.1799 \\
256 & 6.1641e-04 & 2.0079 & 6.1641e-04 & 2.0079 & 3.9128e-03 & 1.9267 & 4.6693e-04 & 2.0431 \\
\hline
\end{tabular}
\label{tab:L2convergenceNS}
\end{table}

\begin{table}[!ht]
\vskip -0.2cm
\centering
\caption{$L^{\infty}$ error and Convergence for $p$, $n$ and $\phi$}
\begin{tabular}{cccccccc}
\hline
$N$ & $\|e(p)\|_{L^\infty}$ & Rate & $\|e(n)\|_{L^\infty}$ & Rate & $\|e(\phi)\|_{L^\infty}$ & Rate \\
\hline
64  & 4.2618e-03 &        & 4.1977e-03 &        & 3.5594e-04 &        \\
128 & 1.4360e-03 & 1.5694 & 1.4093e-03 & 1.5746 & 8.6627e-05 & 2.0388 \\
256 & 3.8618e-04 & 1.8948 & 3.7980e-04 & 1.8917 & 2.1010e-05 & 2.0437 \\
\hline
\end{tabular}
\label{tab:LinftyconvergencePNP}
\end{table}

\begin{table}[!ht]
\vskip -0.2cm
\centering
\caption{$L^{\infty}$ error and Convergence for $u$, $v$, $P$ and $\psi$}
\begin{tabular}{cccccccccc}
\hline
$N$ & $\|e(u)\|_{L^\infty}$ & Rate & $\|e(v)\|_{L^\infty}$ & Rate & $\|e(P)\|_{L^\infty}$ & Rate & $\|e(\psi)\|_{L^\infty}$ & Rate \\
\hline
64  & 4.2754e-03 &        & 4.2754e-03 &        & 4.0923e-02 &        & 5.7893e-03 &        \\
128 & 1.0290e-03 & 2.0548 & 1.0290e-03 & 2.0548 & 1.4285e-02 & 1.5185 & 1.2736e-03 & 2.1845 \\
256 & 2.5034e-04 & 2.0393 & 2.5034e-04 & 2.0393 & 5.8285e-03 & 1.2933 & 3.0895e-04 & 2.0434 \\
\hline
\end{tabular}
\label{tab:LinftyconvergenceNS}
\end{table}

\subsection{Role of active pumps in sustaining ionic asymmetry}\label{App:1d}

In this section, we investigate how active ion pumps contribute to establishing and maintaining ionic concentration asymmetry across a fixed interface in a one-dimensional domain $[0,1]$. The interval $[0,0.5)$ is designated as the outer region, and $(0.5,1]$ as the inner region, with the interface located at $x = 0.5$. To isolate the effect of the pump mechanism, we neglect fluid flow and consider only the Poisson–Nernst–Planck (PNP) equations supplemented with an interfacial pump flux.

The system is initialized with uniform concentrations of positive and negative ions:
\begin{align}
p(x,0) &= 1.5, \quad
n(x,0) = 1.5,
\end{align}
where $p(x,t)$ and $n(x,t)$ denote the concentrations of cations and anions, respectively. No-flux boundary conditions are imposed for both ion species, and homogeneous Neumann boundary conditions are applied to the electric potential.

In the absence of pumping, the system remains in static equilibrium, with uniform ion concentrations and a flat electric potential profile throughout the domain. This state serves as the baseline for evaluating the effects of active transport.

To examine the impact of the pump, we first assume that the membrane conductance is infinity, $g=\infty$ in Eq. \eqref{eqn:Deff}, i.e.
\[ D_{i}^{-1} 
        = \frac{1-\psi}{2}
        + \frac{1+\psi}{2D_{i}^{r}.}\]

  The simulation parameters are given by:
\begin{equation}
\begin{aligned}
& I_0 = 10, \quad K_0 = 0.5, \quad \beta = 2, \quad D_i^r = 1, \quad \zeta = 0.1, \quad \delta = 0.001, \\
& Pe = 1, \quad Pe_{E} = 0.005, \quad z_{p} = 1, \quad z_{n} = -1.
\end{aligned}
\end{equation}

\begin{figure}[!ht]
	\centering
	\subfloat[The profile of $\phi$.]{
		\centering
		\includegraphics[width=0.31\linewidth]{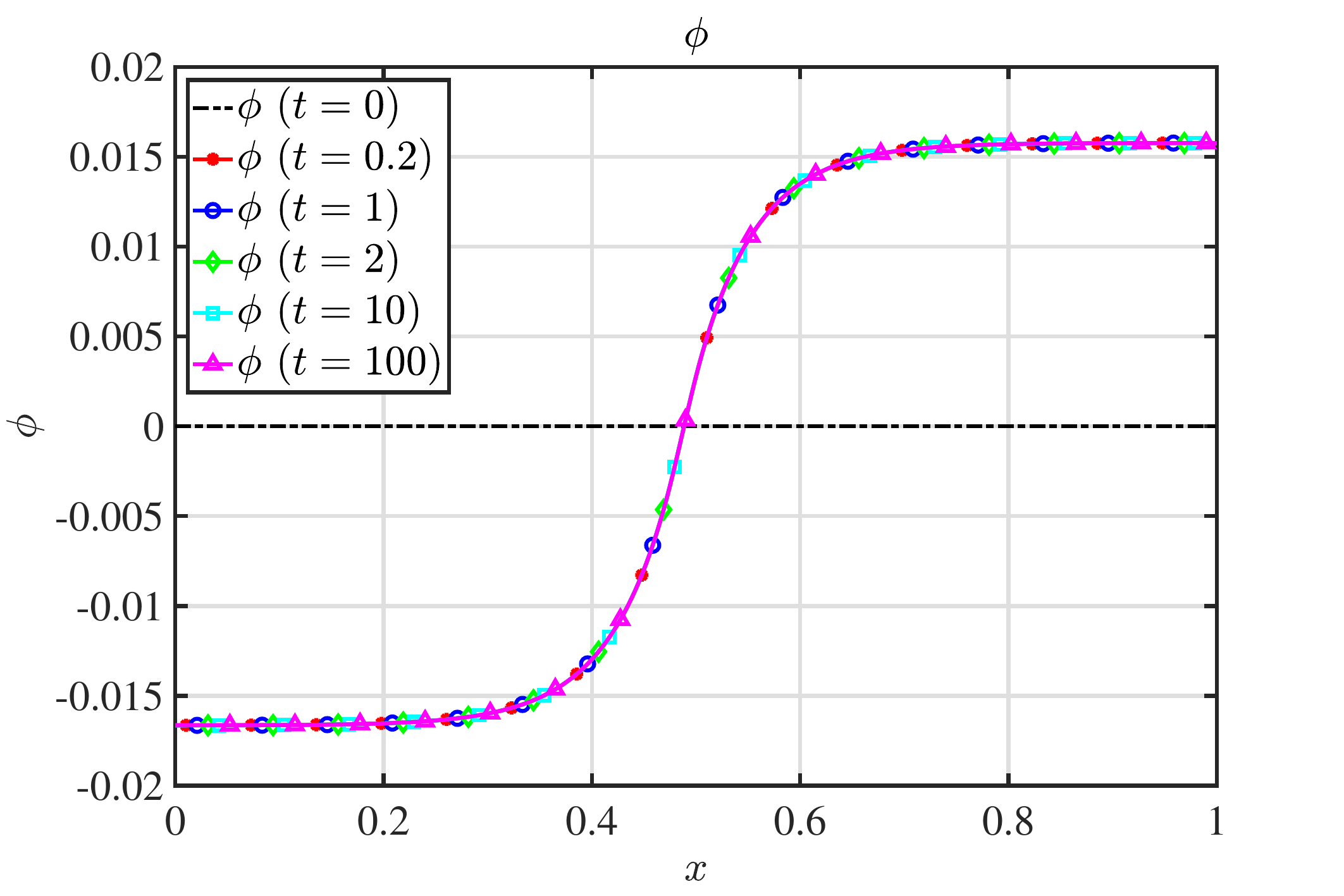}
        \label{fig:phi1D_Dirichlet0_pump}
		}
	\centering
	\subfloat[The profile of $p$.]{
		\centering
		\includegraphics[width=0.31\linewidth]{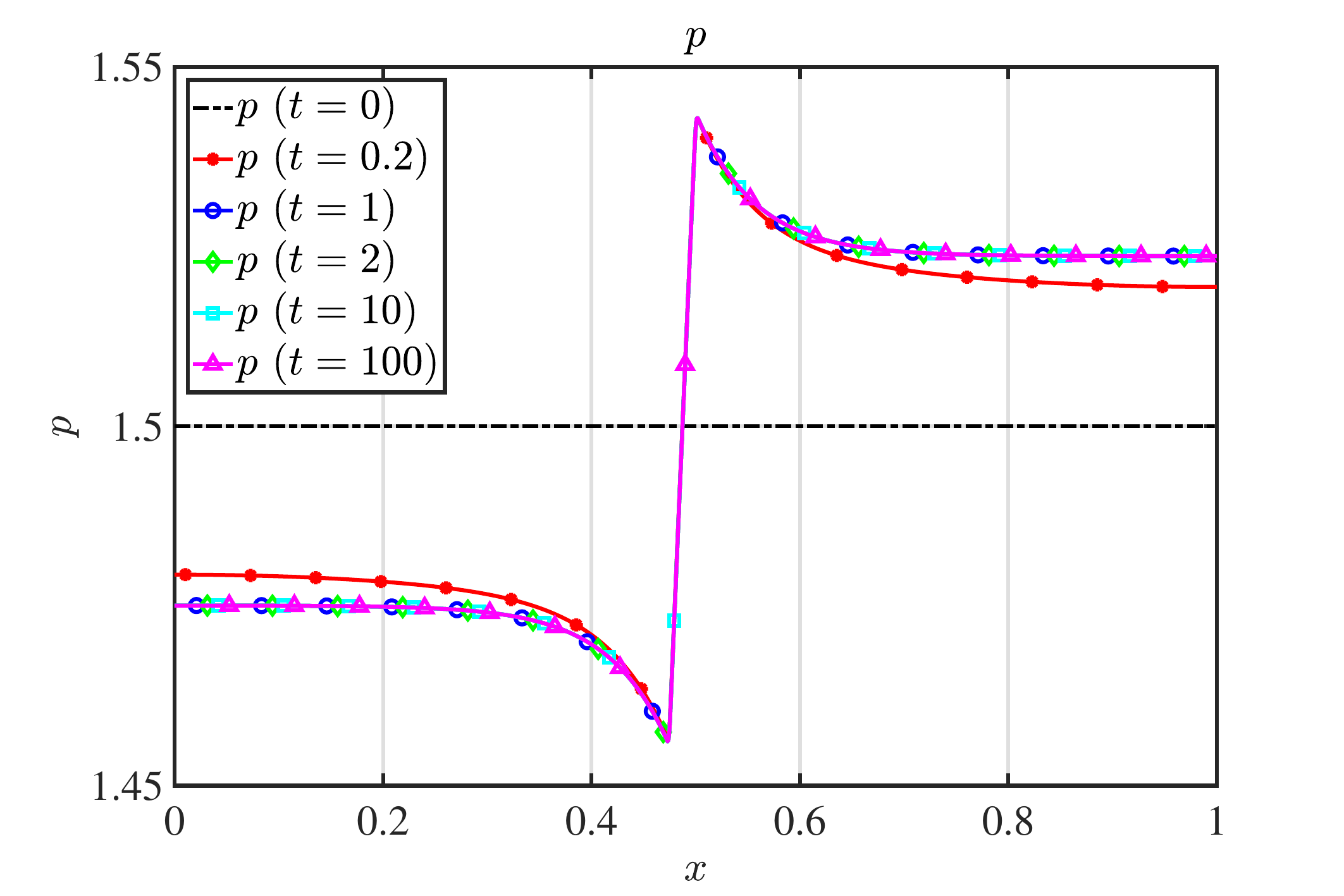}
        \label{fig:p1D_Dirichlet0_pump}
	}
	\subfloat[The profile of $n$.]{
		\centering
		\includegraphics[width=0.31\linewidth]{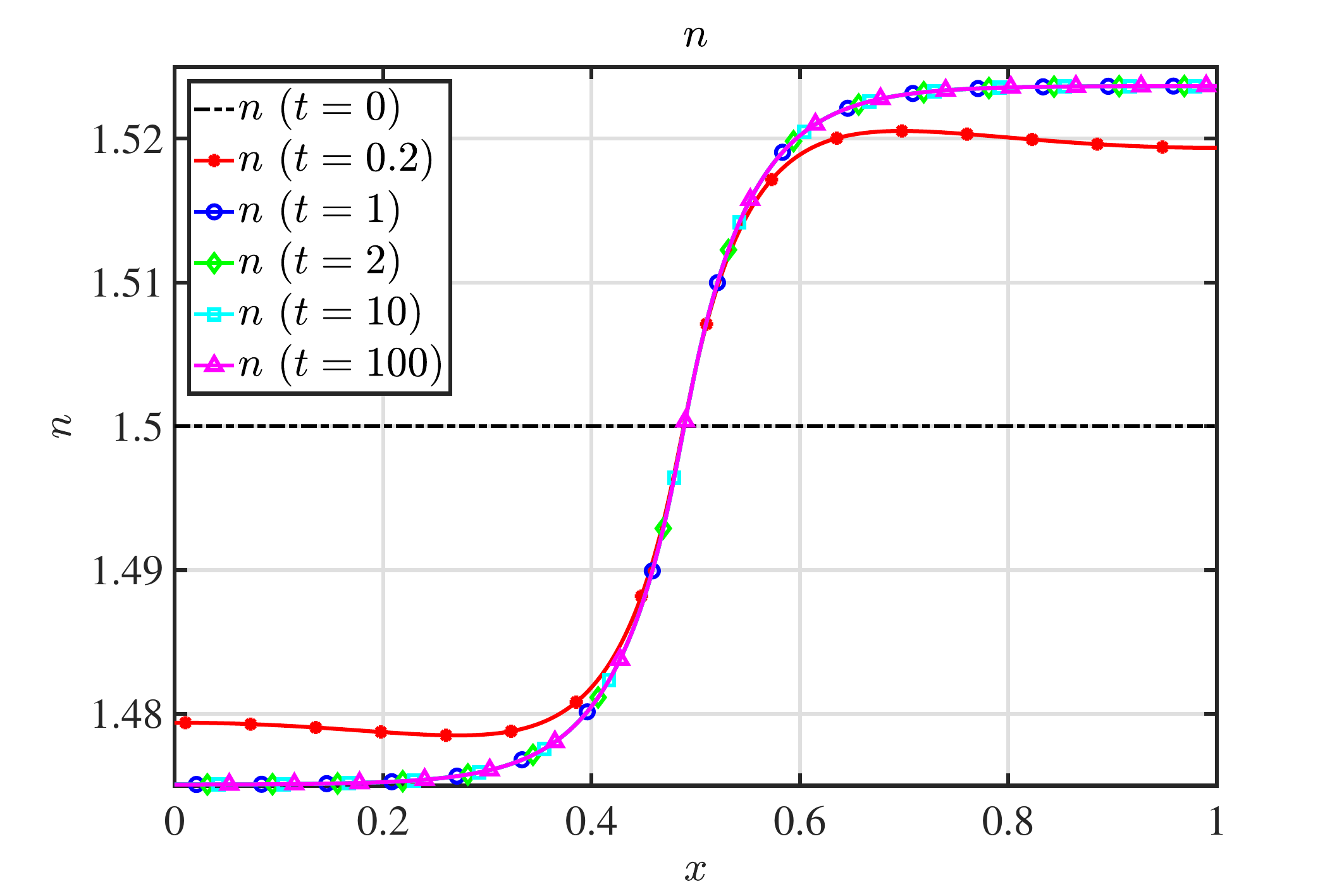}
        \label{fig:n1D_Dirichlet0_pump}
		}
    \\
    \subfloat[The profile of $p+n$.]{
		\centering
		\includegraphics[width=0.31\linewidth]{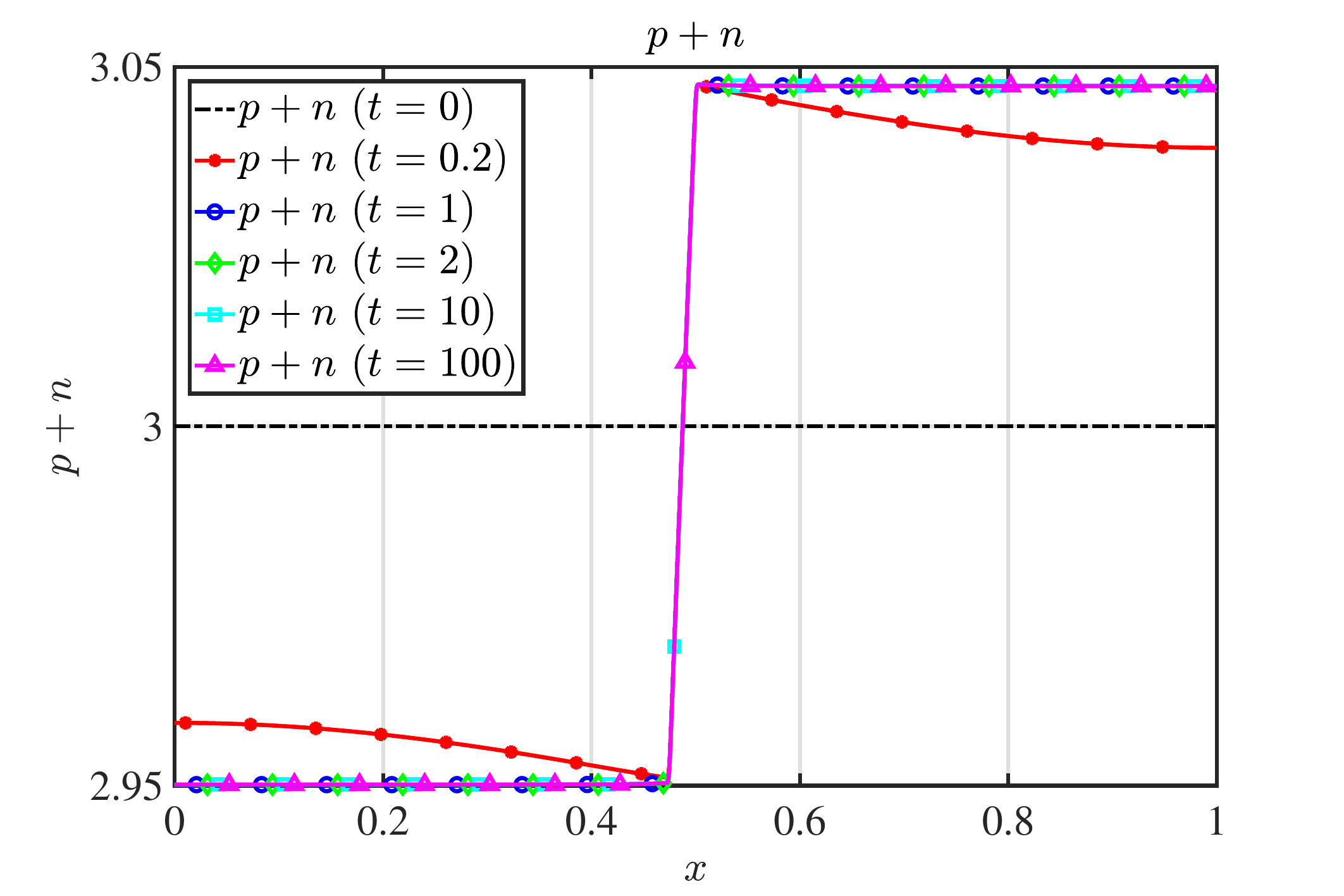}
        \label{fig:p+n1D_Dirichlet0_pump}
	}
    \subfloat[The profile of $p-n$.]{	
    \centering
		\includegraphics[width=0.31\linewidth]{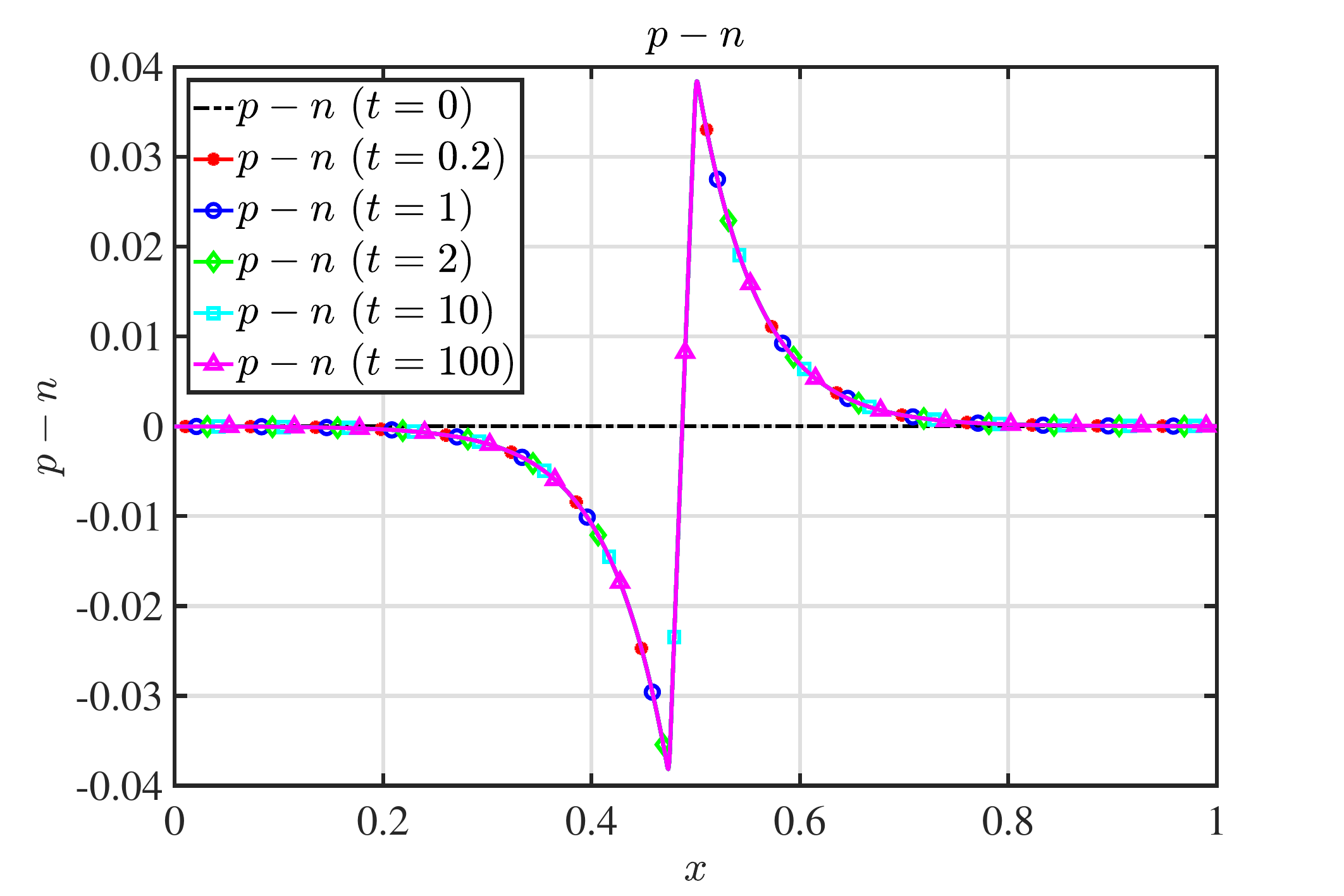}
        \label{fig:p-n1D_Dirichlet0_pump}
	}
    \subfloat[The profile of $p$ with different pump rates.]{	
    \centering
		\includegraphics[width=0.31\linewidth]{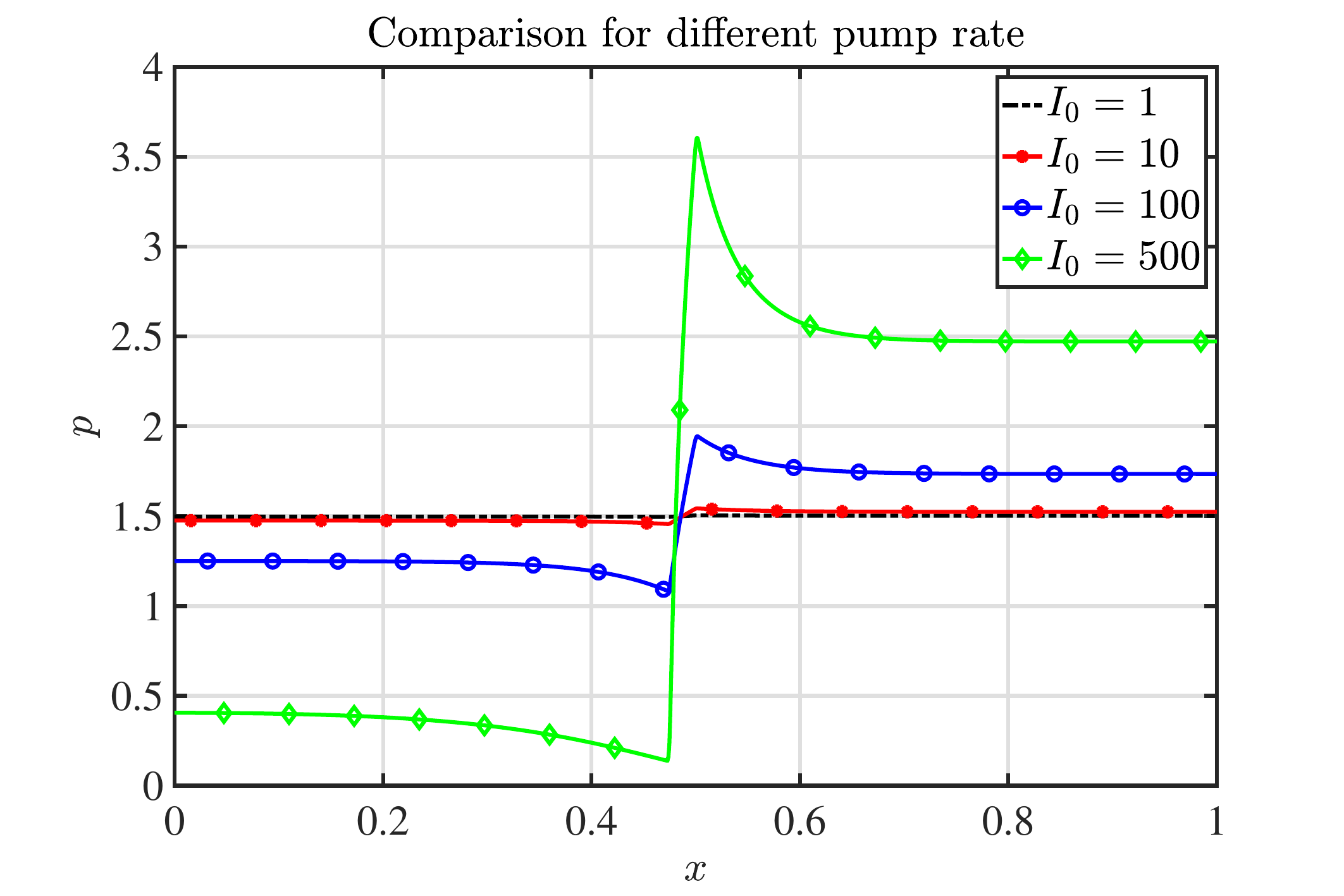}
        \label{fig:multi_pumps}
	}
    \caption{Pump effect for the 1-D situation with modified Michaelis-Menten format type pump and $\beta=2$.}
    \label{fig:1D}
\end{figure}

The resulting steady-state profiles are shown in Fig.~\ref{fig:1D} with a Michaelis–Menten (MM) type pump expression~\eqref{eq:pumpmm}. In the presence of the active pump, positive ions are directionally transported from the outer region to the inner region. Due to diffusion, the concentrations remain nearly uniform within each bulk region, away from the interface. The accumulation of positive ions near the inner side of the interface raises the local electric potential, which subsequently attracts negative ions into the inner region. Consequently, bulk electroneutrality is preserved on both sides, while a localized net charge appears near the interface due to the combined effects of the pump and the induced electric field.

Moreover, the effective ionic conductance, given by $\frac{e^2}{k_B T} D(p + n)$, remains approximately constant in the bulk regions. This observation aligns with the assumption made in~\cite{qin2023droplet}, supporting the validity of approximating bulk regions as electrically neutral and uniformly conductive in theoretical models such as the leaky dielectric framework~\cite{melcher1969electrohydrodynamics}.
The steady-state profiles of positive ion concentration under different pump strengths are shown in Fig.~\ref{fig:1D}f. In all cases, the ion concentrations remain nearly constant in the bulk regions, with sharp variations localized near the interface due to pump activity.

\begin{figure}[!ht]
\centering
        \subfloat[The profile of $p$ with different pump types.]{	
        \centering
		\includegraphics[width=0.48\linewidth]{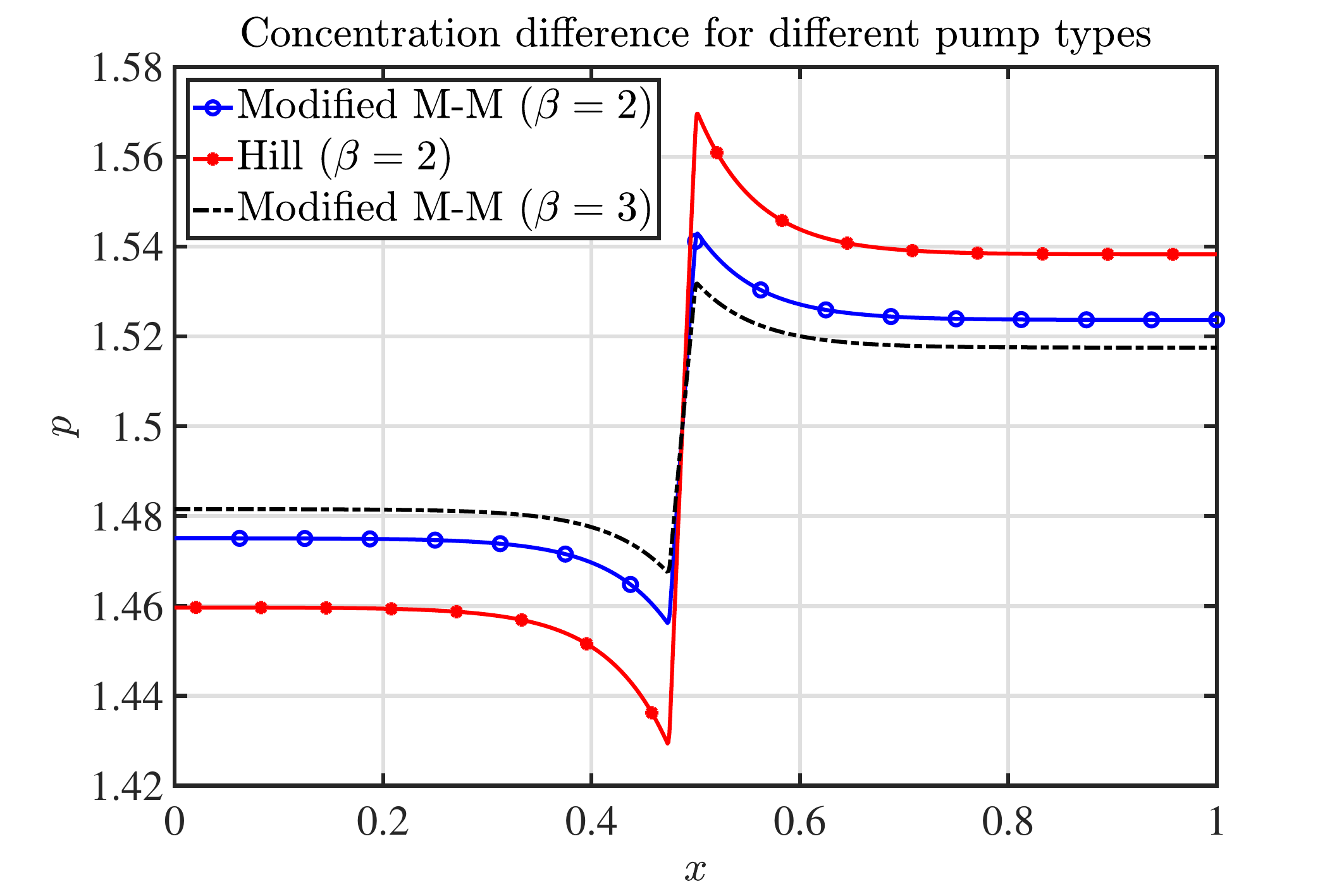}
		\label{fig:compare_p}
        }
	\subfloat[The ratio of $p$ with different pump types and rates.]{	
        \centering
		\includegraphics[width=0.48\linewidth]{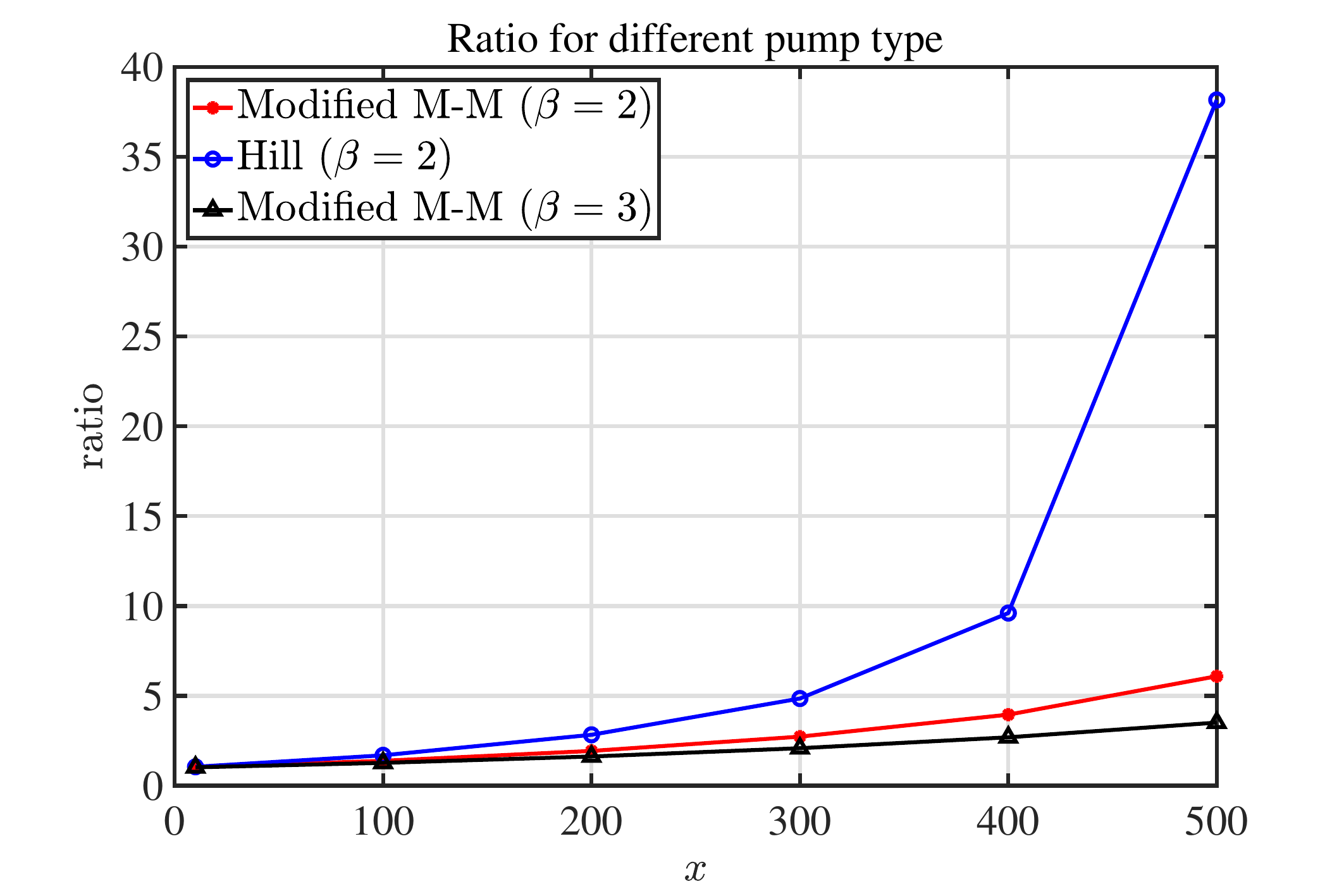}
		\label{fig:compare_beta}
        }
        \caption{Pump effect for the 1-D situation with modified Michaelis-Menten format and Hill format.}
\end{figure}

Figure~\ref{fig:compare_p} presents the equilibrium ion profiles for different values of the cooperativity parameter $\beta$. As $\beta$ increases, the concentration ratio between the inner and outer regions decreases, indicating reduced efficiency in sustaining ionic asymmetry for higher cooperativity in the MM-type (simultaneous transport) model. The figure also shows that, under the same $\beta$, the Hill-type (cooperative binding) pump is more effective than the MM-type pump at establishing ionic gradients.
Figure~\ref{fig:compare_beta}  summarizes the quantitative relationship between pump strength $I_0$ and the concentration ratio between the inner and outer bulk regions  $R:=\frac{p(x=1)}{p(x=0)}$. As $I_0$ increases, this ratio increases for both models. For $\beta = 3$, the MM model shows an approximately linear increase with $I_0$. In contrast, for $\beta = 2$, both the MM and Hill models exhibit nonlinear increases in the concentration ratio, with the Hill model showing a significantly steeper rise—highlighting the enhanced efficiency of cooperative binding at high pump strengths.

\begin{figure}[!ht]
	\centering
    \subfloat[Different membrane conductance]{
		\includegraphics[width=0.49\linewidth]{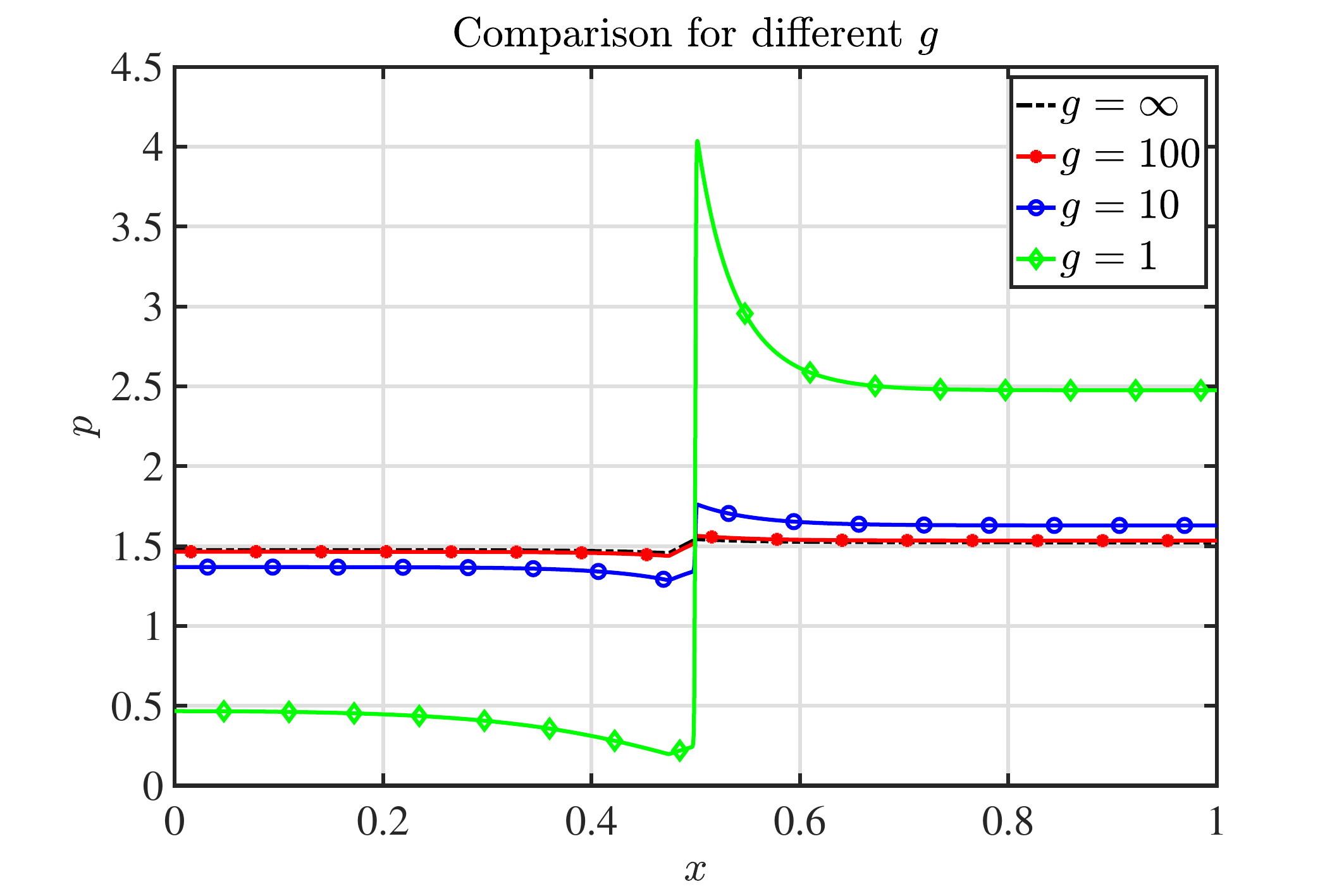}
		\label{fig:comparison_g}
        }
    \subfloat[Different dielectric constant]{
        \includegraphics[width=0.49\linewidth]{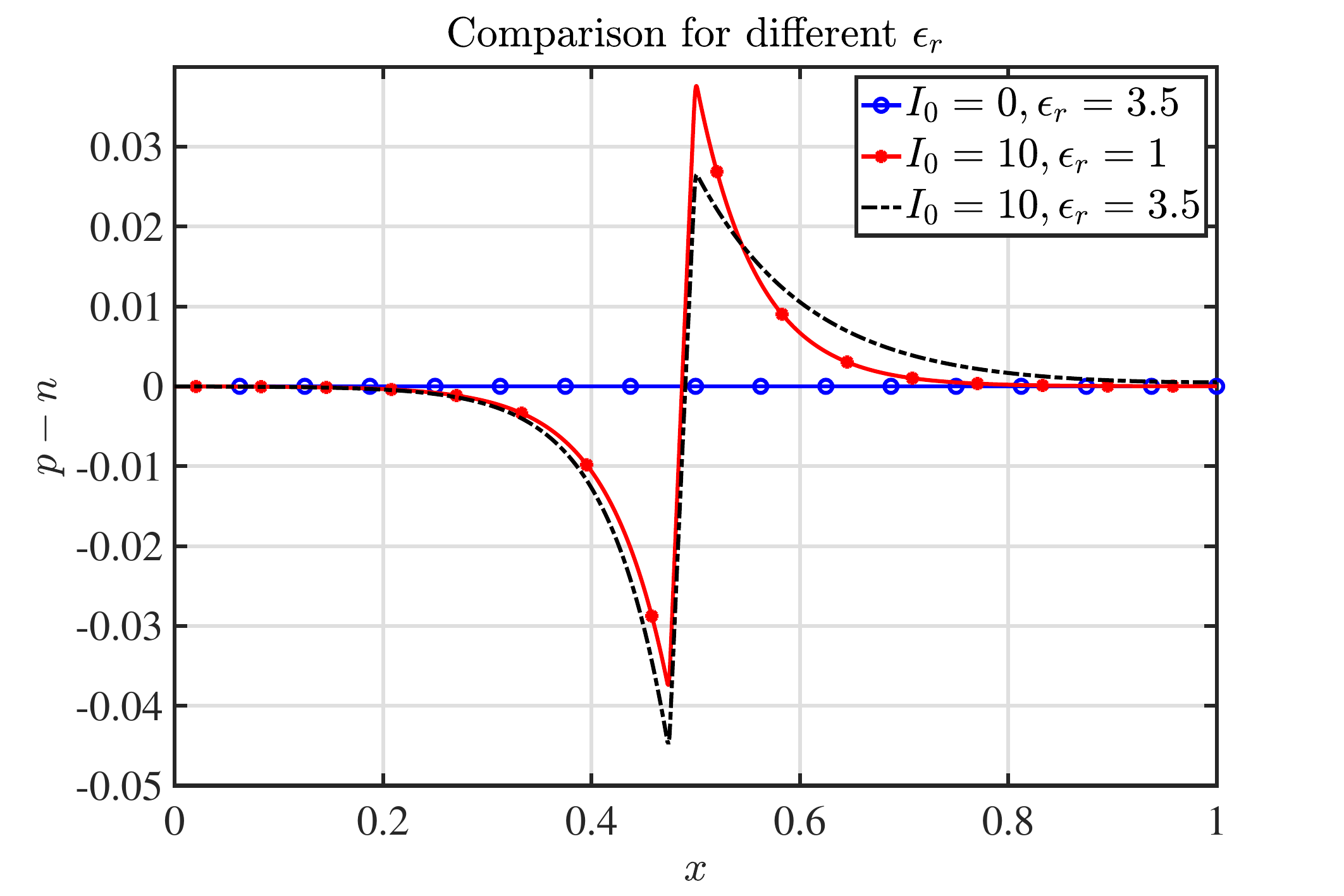}
		\label{fig:comparison_eps}
    }
    \caption{Pump effects with different $g$ and $\epsilon_{r}$. (a) Comparison of the distribution of positive ion $p$ with different $g$. The parameters are chosen as $\delta = 0.1$, $Pe = 1$, $Pe_{E}=1$, $\alpha_{p} = 0.5$, $\alpha_{n} = -0.5$, $\zeta = 0.1$, $D_{i}^{r} =1$ and 
    $\epsilon_{r} = 1$. (b) Comparison of the distribution of net charge $p-n$ with different $\epsilon_{r}$. The parameters are chosen as $\delta = 0.01$, $Pe = 1$, $Pe_{E}=1$, $\alpha_{p} = 0.5$, $\alpha_{n} = -0.5$, $\zeta = 0.1$ and $D_{i}^{r} =4.75$.}
\end{figure}

We next examine how the interfacial pump couples with (i) membrane conductance $g$ and (ii) dielectric contrast to set the steady ion partitioning. Figure~\ref{fig:comparison_g} shows that reducing $g$ suppresses passive leakage across the interface and thereby amplifies the pump-induced asymmetry: the equilibrium bulk ratio monotonically increases as $g$ decreases. In the limit of small $g$, the interior approaches a pump-limited plateau while the exterior is depleted, consistent with the formation of a narrow interfacial layer and nearly electroneutral bulks away from the interface.

We next vary the dielectric-constant ratio between the droplet interior and exterior, $\epsilon_r := \epsilon_{\mathrm{in}}/\epsilon_{\mathrm{out}}$. In the absence of pumps, the solution remains (nearly) electroneutral and the electric potential is flat in the bulk, independent of $\epsilon_r$ (Fig.~\ref{fig:comparison_eps}). With pumps active, a net space–charge layer forms at the interface, with the interior positively charged and the exterior negatively charged, reflecting the directional transport. Increasing $\epsilon_r$ enlarges the interior Debye length and thus widens the non-neutral region. Moreover, by Poisson’s equation, a larger permittivity reduces the potential and field generated per unit space charge; consequently, a greater net charge must accumulate to produce the counter-field that balances the pump flux. Consistently, on the exterior side of the interface (left in our 1D setup), the cation concentration is further depleted, i.e. the local positive charge is lower when $\epsilon_r$ is larger.

\begin{figure}[!ht]
\centering
        \subfloat[The profile of $p$  ]{	
        \centering
		\includegraphics[width=0.48\linewidth]{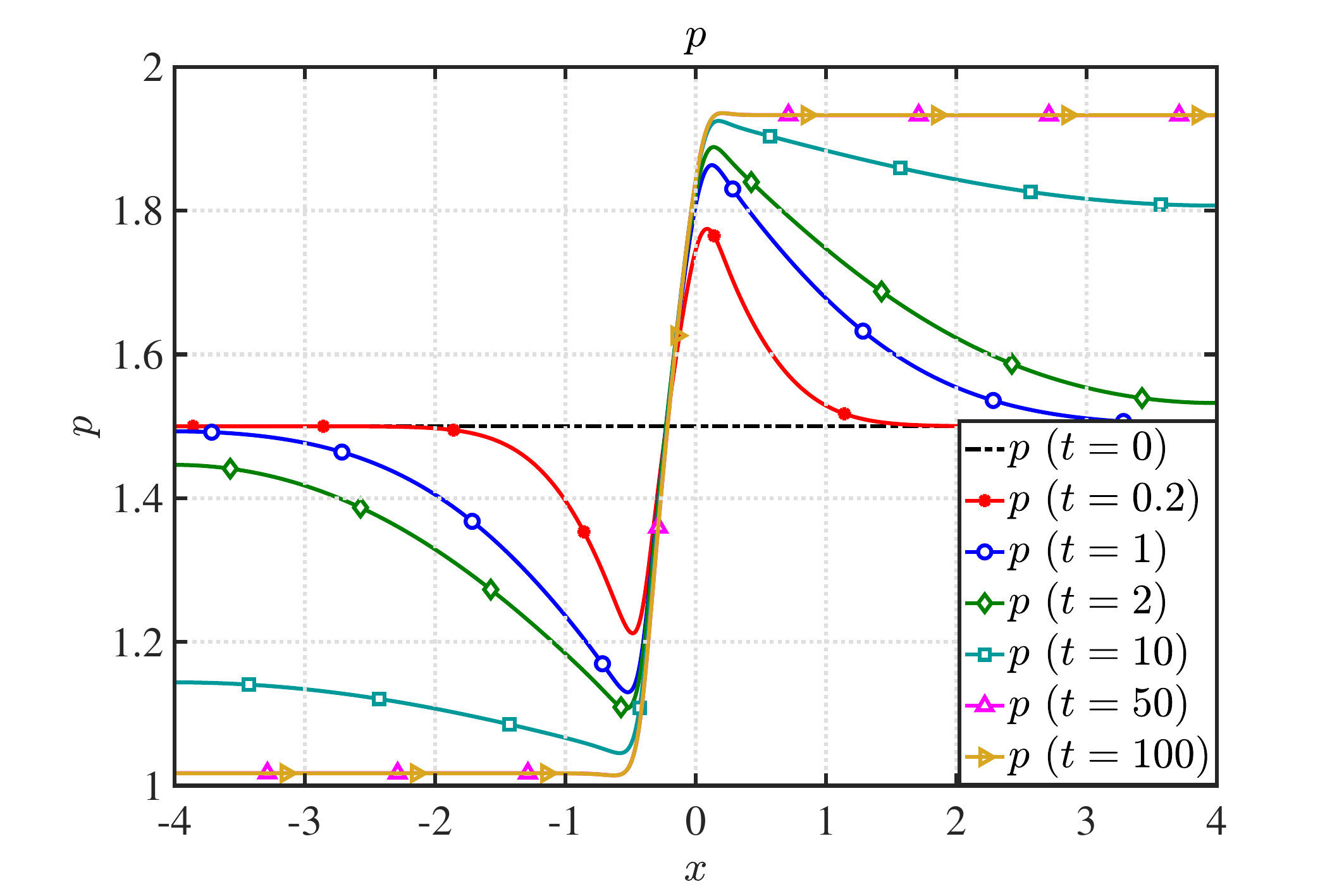}
		\label{fig:test_p}
        }
	\subfloat[The profile of $\phi$]{	
        \centering
		\includegraphics[width=0.48\linewidth]{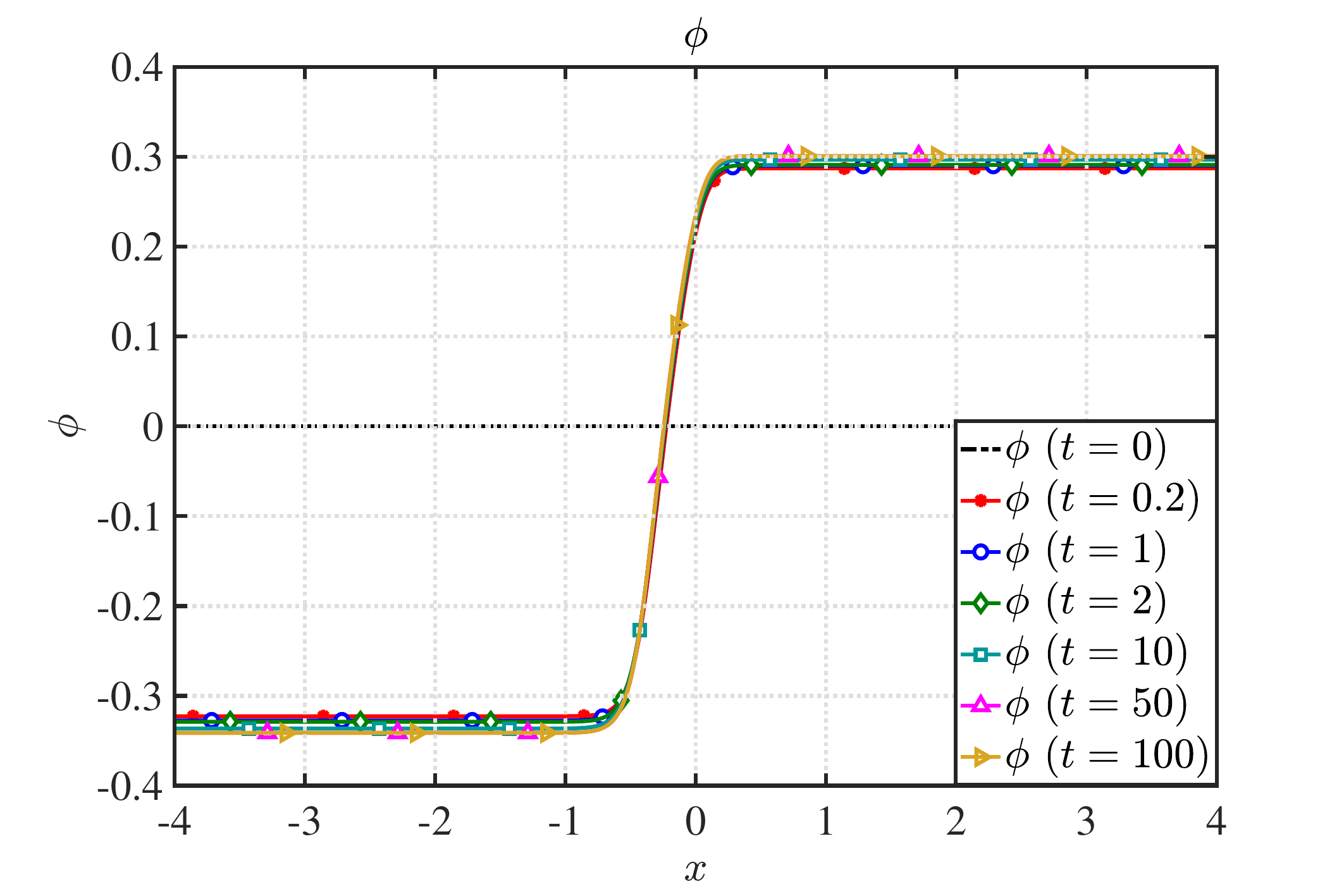}
		\label{fig:test_phi}
        }
        \caption{Ion concentration and electric potential profile with large interface thickness $\delta=0.1$. Here $Pe = 1$, $Pe_{E}=1$, $\alpha_{p} = 0.5$, $\alpha_{n} = -0.5$ and $\zeta = 0.1$.}
\end{figure}

Finally, we investigate the effect of the interface thickness $\delta$ in the phase-field label function. Figure~\ref{fig:test_phi} presents simulation results with a relatively thick interface $\delta = 0.1$. To mitigate potential boundary effects, the computational domain is extended to $[-4, 4]$. The results show that the profiles of ion concentration and electric potential remain qualitatively similar to previous cases: the inner and outer bulk regions maintain distinct, nearly constant ion concentrations, sustained by the active pump.

Based on this observation, we adopt $\delta = 0.1$ for the remaining two-dimensional simulations to reduce computational cost while preserving essential physical behavior.

\subsection{Supplementary figures}\label{subsec:figs}

\begin{figure}[!ht]
\centering 
    \vskip -0.4cm
        \subfloat[$p~(y=0,t=10)$]{
		\includegraphics[width=0.33\linewidth]{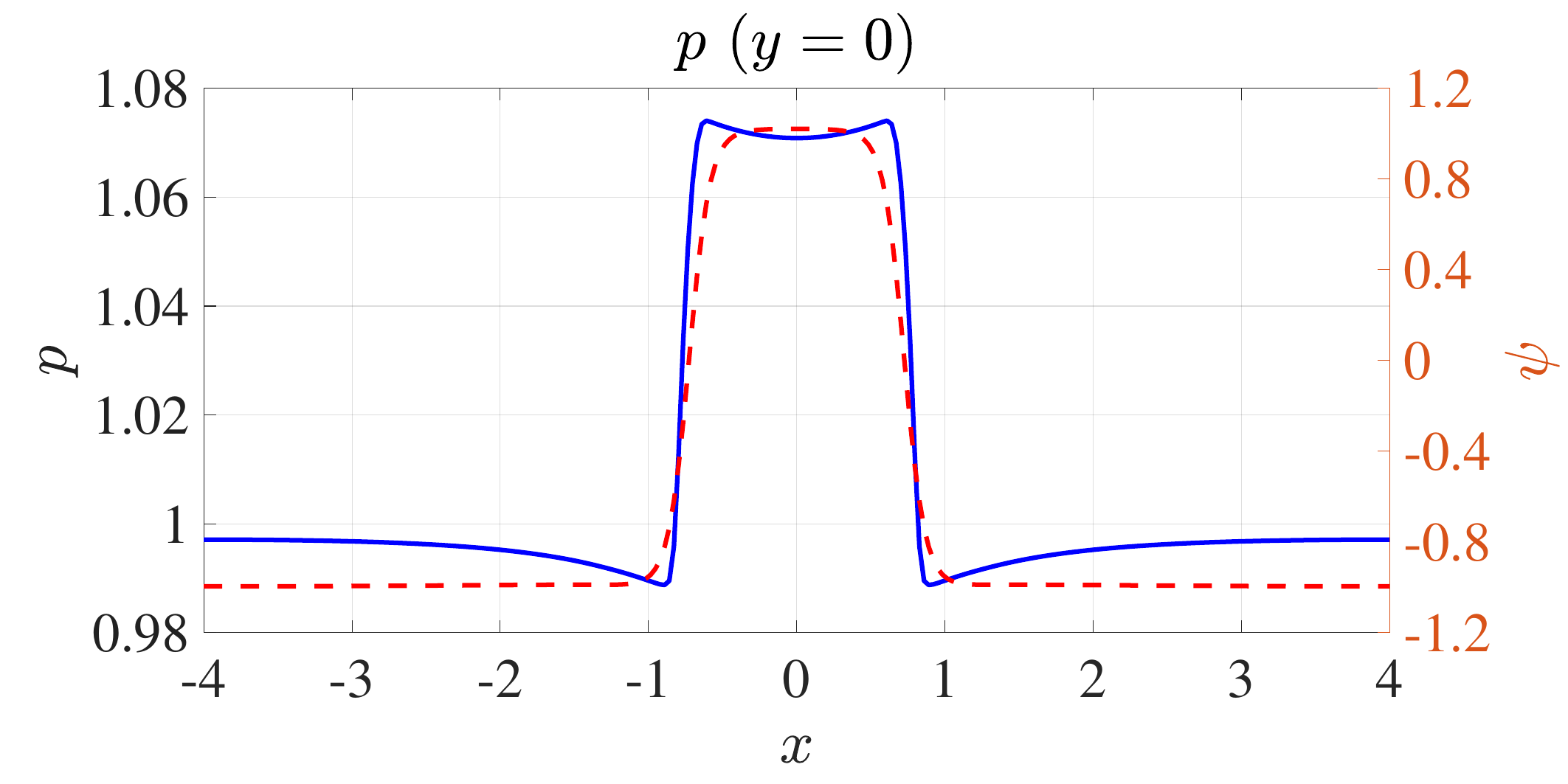}
        \label{subfig:1DropD0N0bdPump25P10y0}
		}
        \hskip -0.4cm
	\subfloat[$n~(y=0,t=10)$]{
		\includegraphics[width=0.33\linewidth]{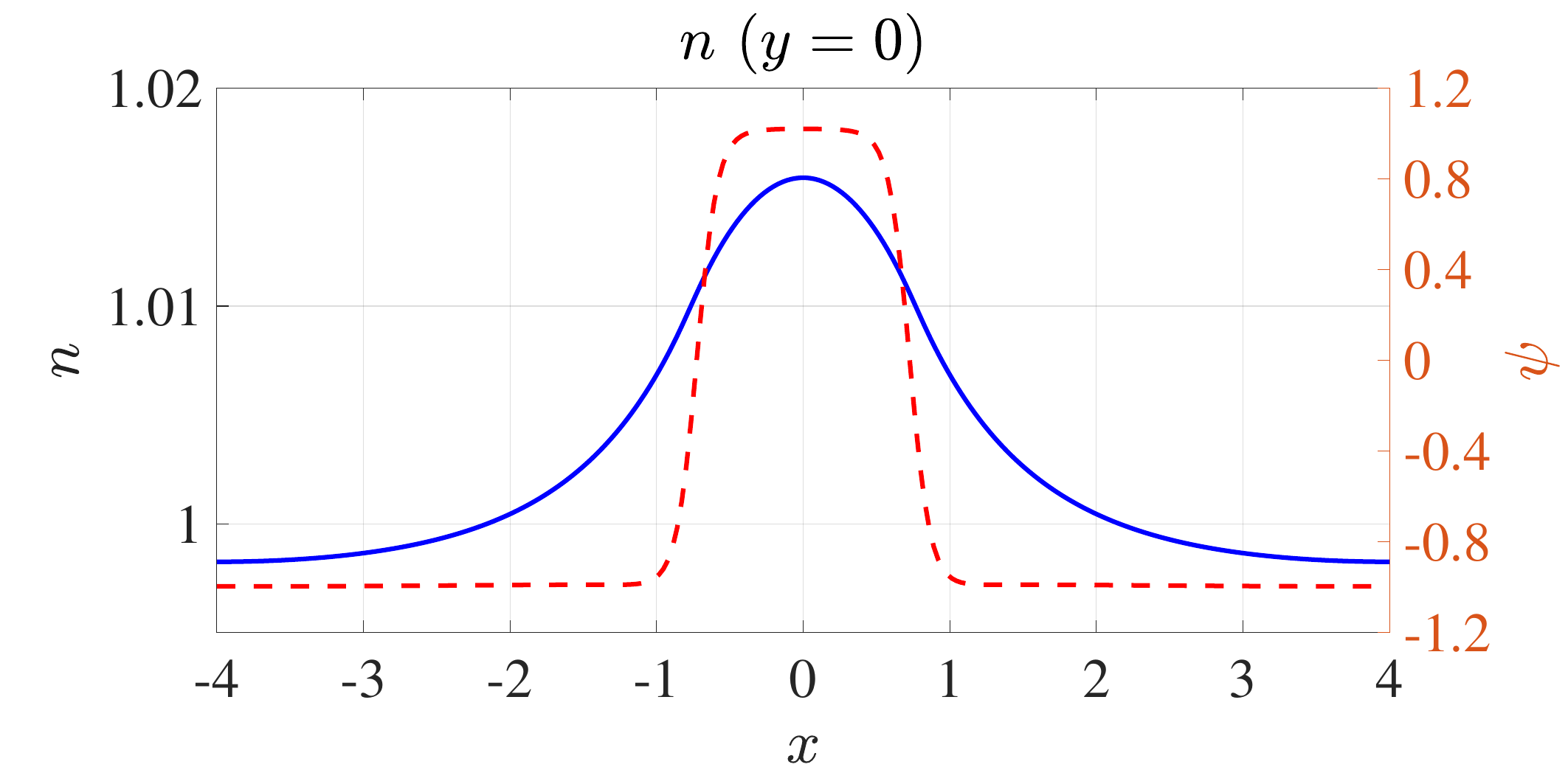}
        \label{subfig:1DropD0N0bdPump25N10y0}
		}
        \hskip -0.4cm
	\subfloat[$\phi~(y=0,t=10)$]{
		\includegraphics[width=0.33\linewidth]{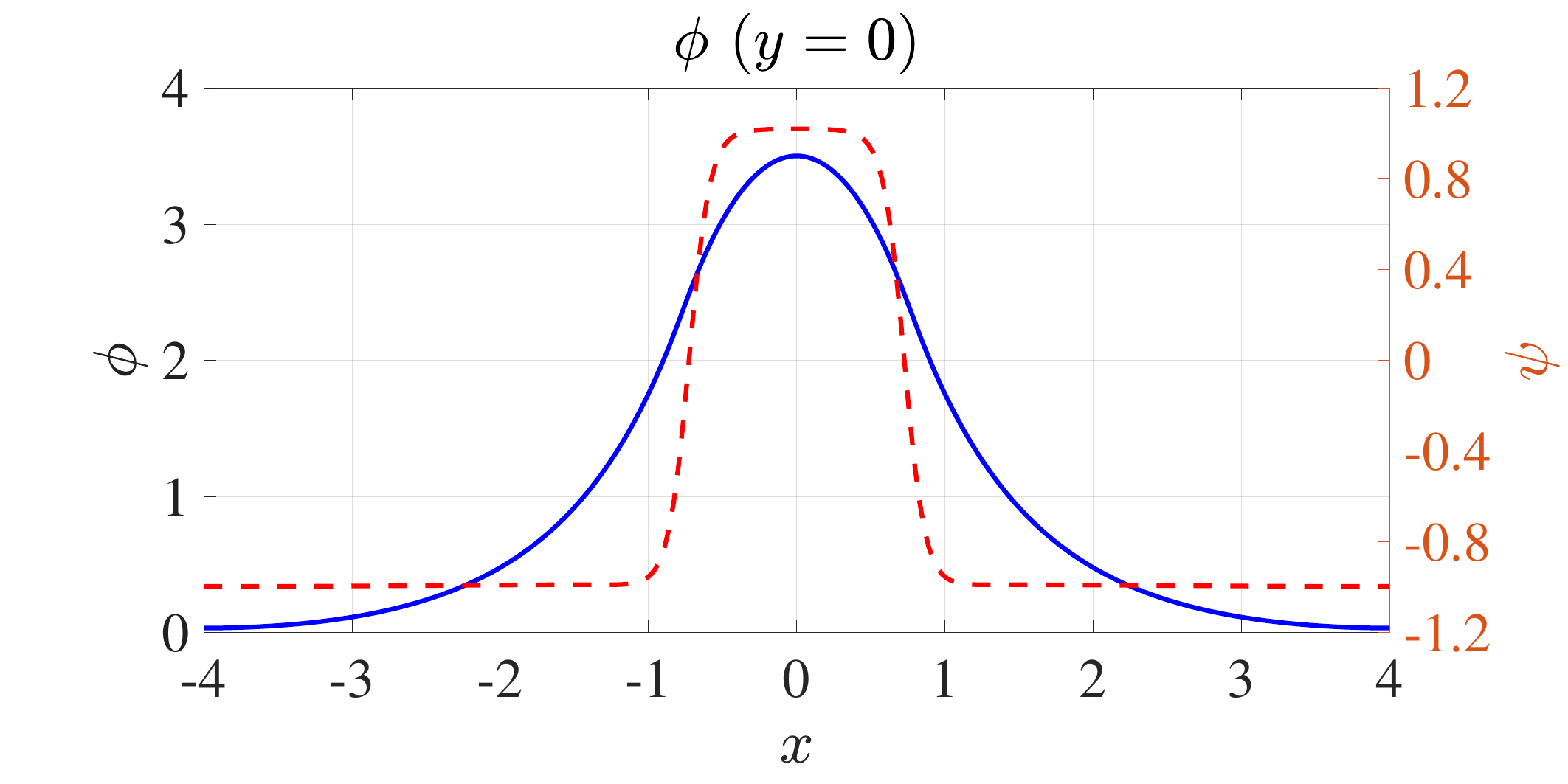}
        \label{subfig:1DropD0N0bdPump25Phi10y0}
		}
        \\
        \vskip -0.0cm
    \subfloat[$p~(x=0,t=10)$]{
		\includegraphics[width=0.33\linewidth]{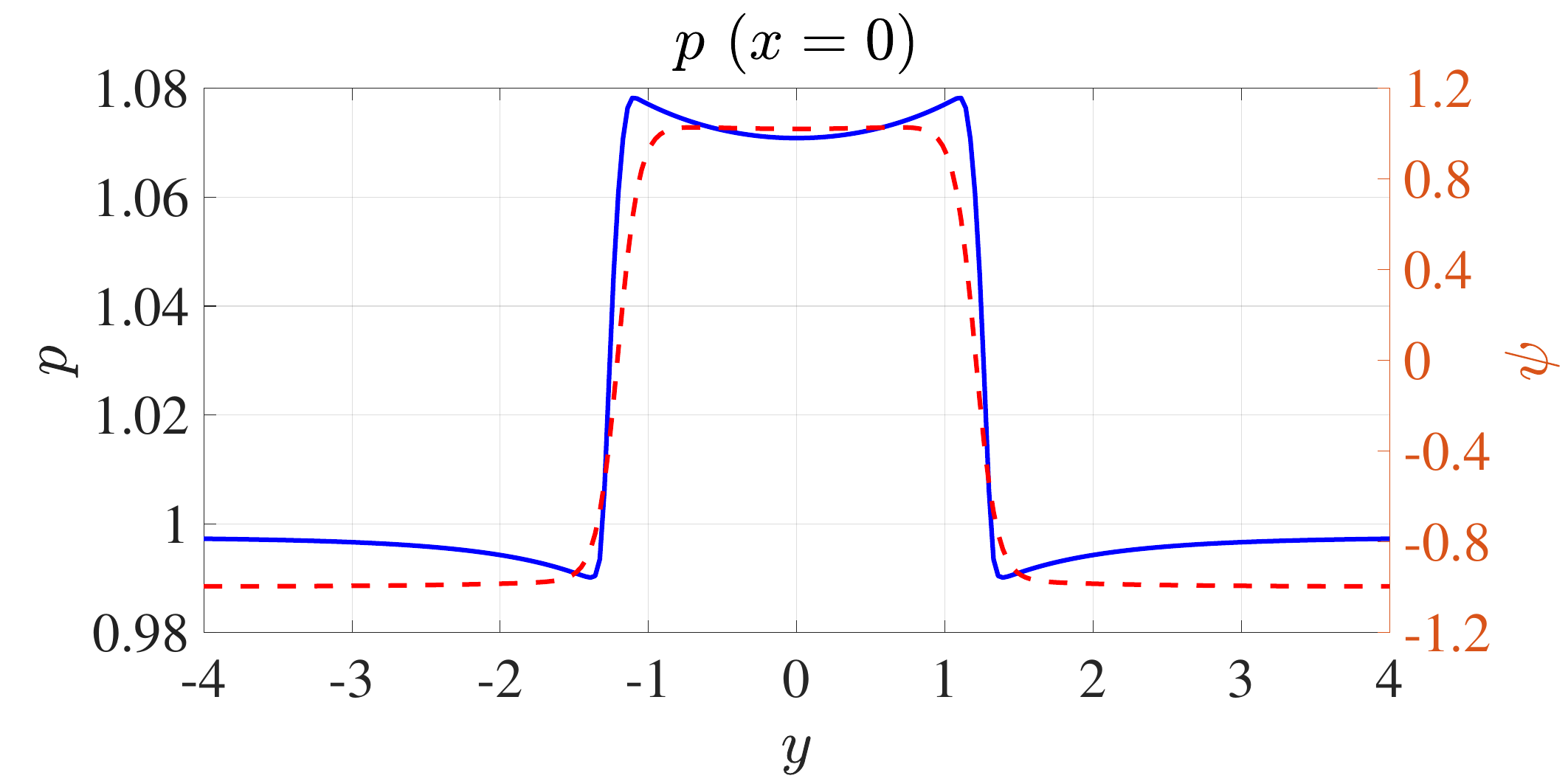}
        \label{subfig:1DropD0N0bdPump25P10x0}
		}
        \hskip -0.4cm
	\subfloat[$n~(x=0,t=10)$]{
		\includegraphics[width=0.33\linewidth]{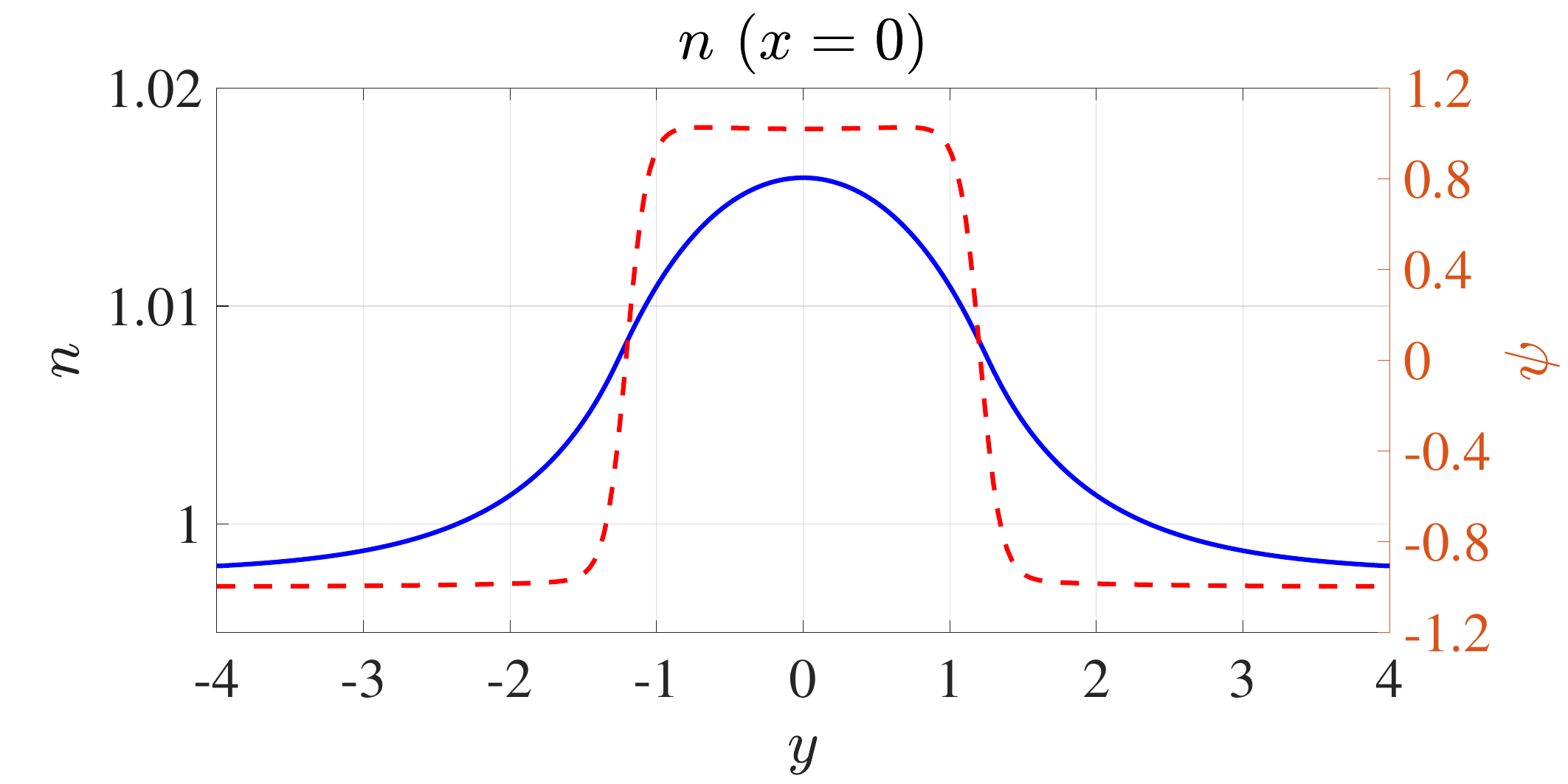}
        \label{subfig:1DropD0N0bdPump25N10x0}
		}
        \hskip -0.4cm
	\subfloat[$\phi~(x=0,t=10)$]{
		\includegraphics[width=0.33\linewidth]{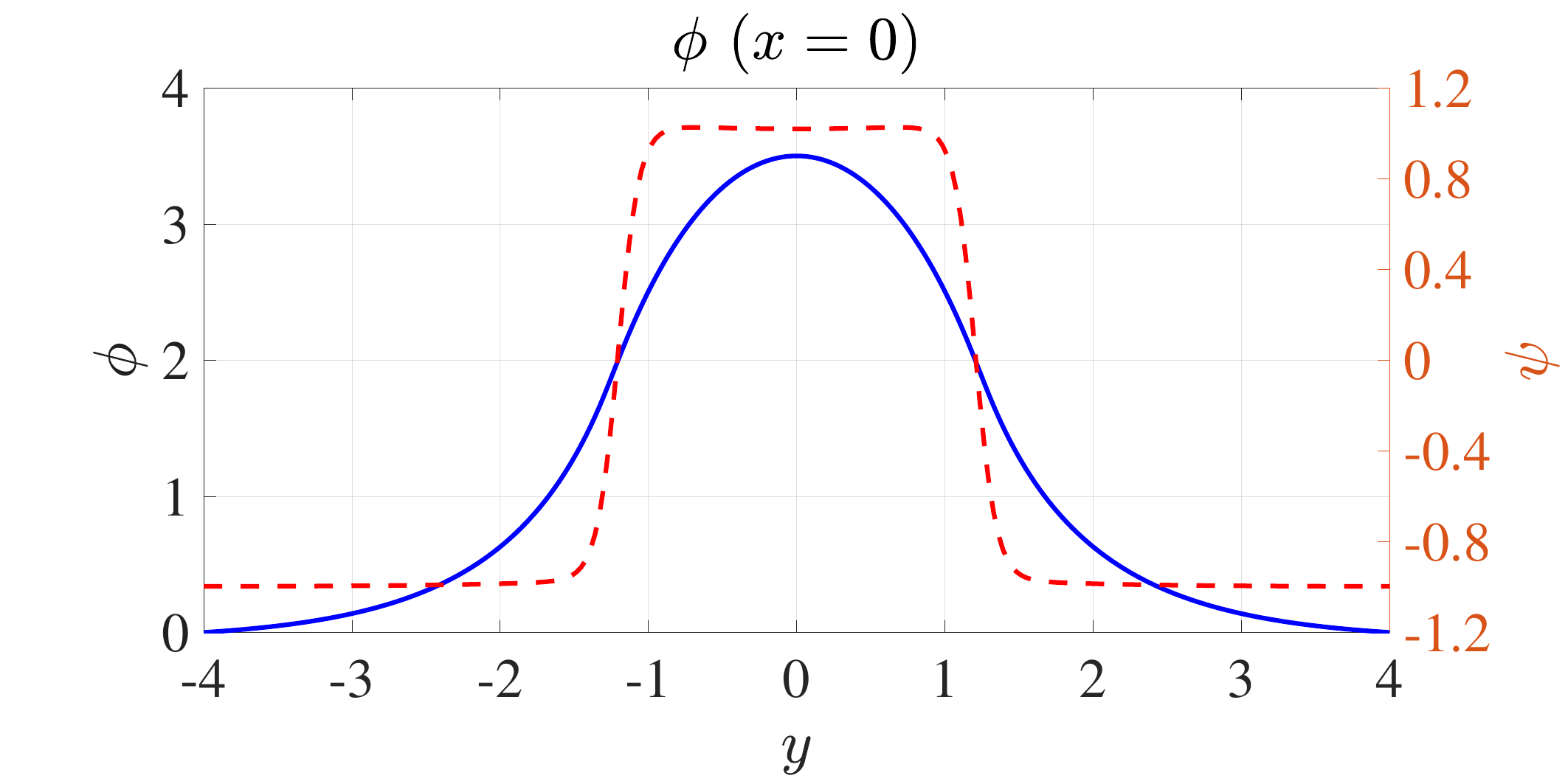}
        \label{subfig:1DropD0N0bdPump25Phi10x0}
		}
        \vskip -0.2cm
	\caption{The positive and negative ion and the electric potential distribution along $x$-axis (top) and $y$-axis (bottom) for the case in Figure \ref{fig:1DropD0N0Pump25}, respectively. The blue solid lines show the ions concentration and electric potential and the red dash lines show the label function distribution with the diffusion interface. 
    }\label{fig:1DropD0N0bdPump25Section}
\end{figure}

\begin{figure}[!ht]
    \vskip -0.2cm
		\centering
	\subfloat[$-\frac{Ca_{E}}{\zeta^{2}}\rho\nabla\phi~(y=0,t=10)$]{
		\includegraphics[width=0.48\linewidth]{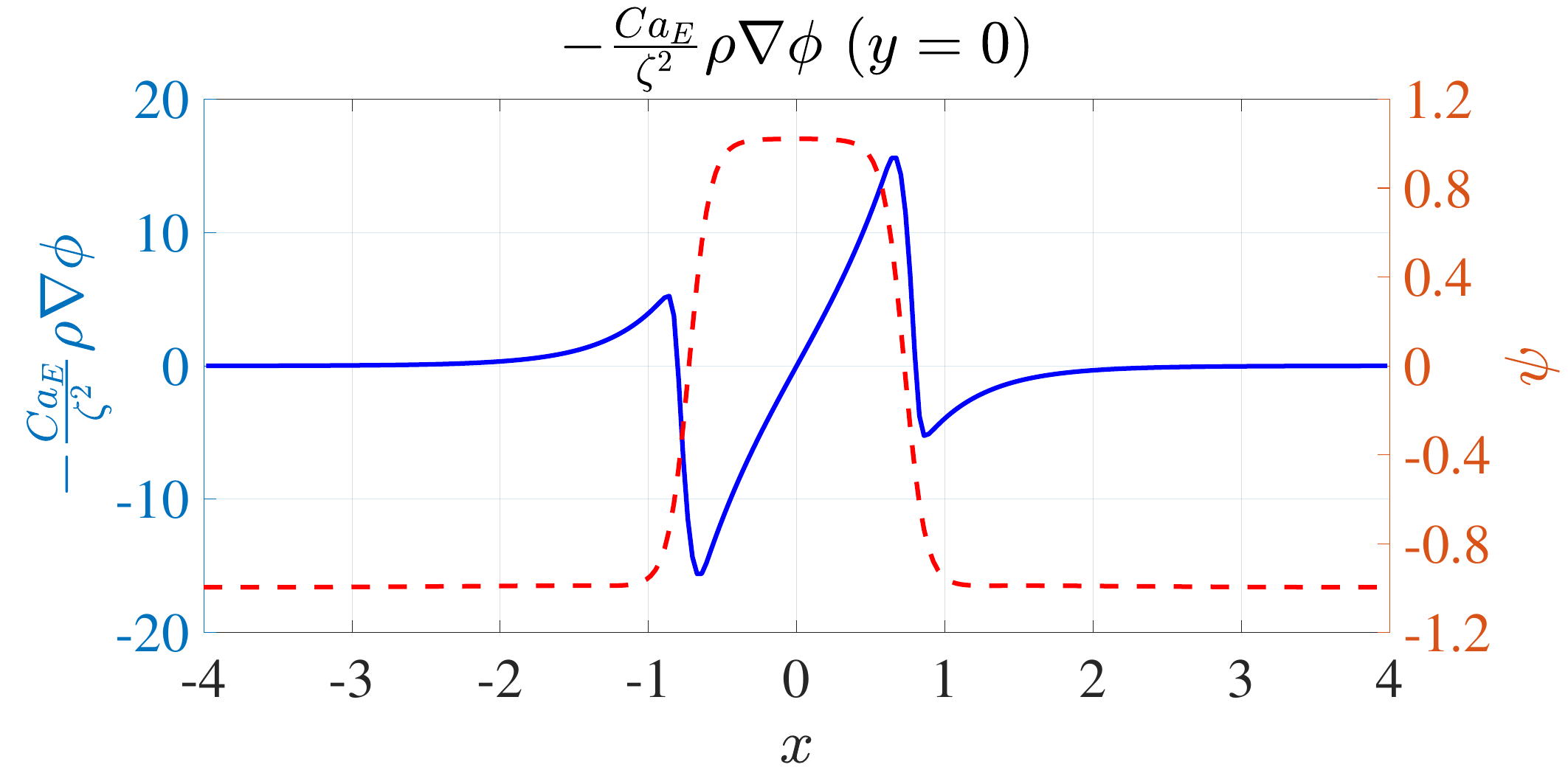}
        \label{subfig:1DropD0N0bdPump25Eforcey0}
	}
	\subfloat[$-\frac{Ca_{E}}{\zeta^{2}}\rho\nabla\phi~(x=0,t=10)$]{
		\includegraphics[width=0.48\linewidth]{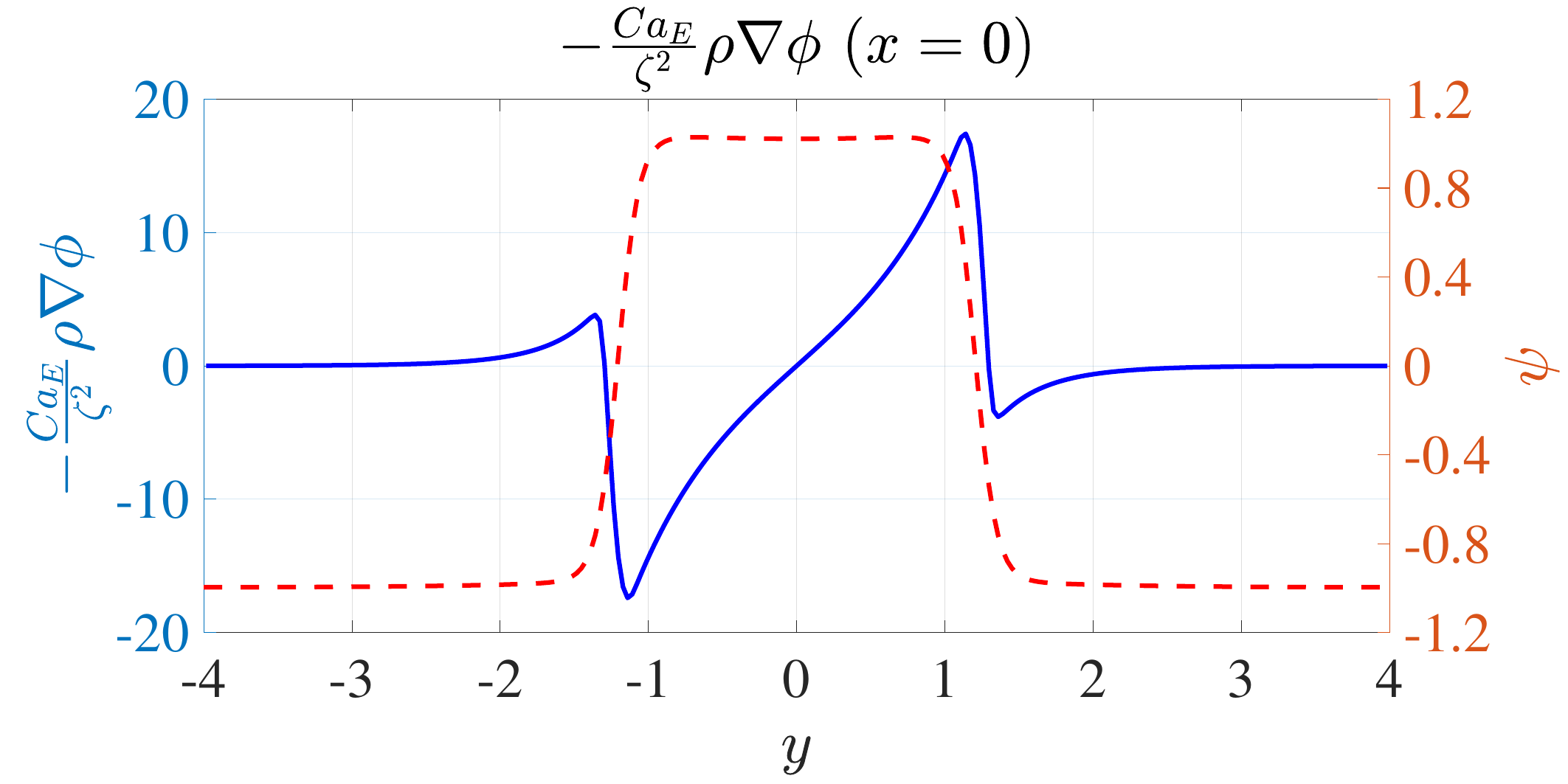}
        \label{subfig:1DropD0N0bdPump25Eforcex0}
		} 
    \vskip -0.2cm
	\caption{The electric force along $x$-axis (left) and $y$-axis (right) induced by the distribution of ions and electric potential for the case in Figure \ref{fig:1DropD0N0Pump25}, where the blue solid lines and red dash lines show the Lorentz force and label function, respectively.}\label{fig:1DropD0N0bdPump25Eforce}
\end{figure}

\begin{figure}[!ht] 
\centering 
\vskip -0.4cm
    \subfloat[$p~(y=0,t=10)$]{
		\includegraphics[width=0.33\linewidth]{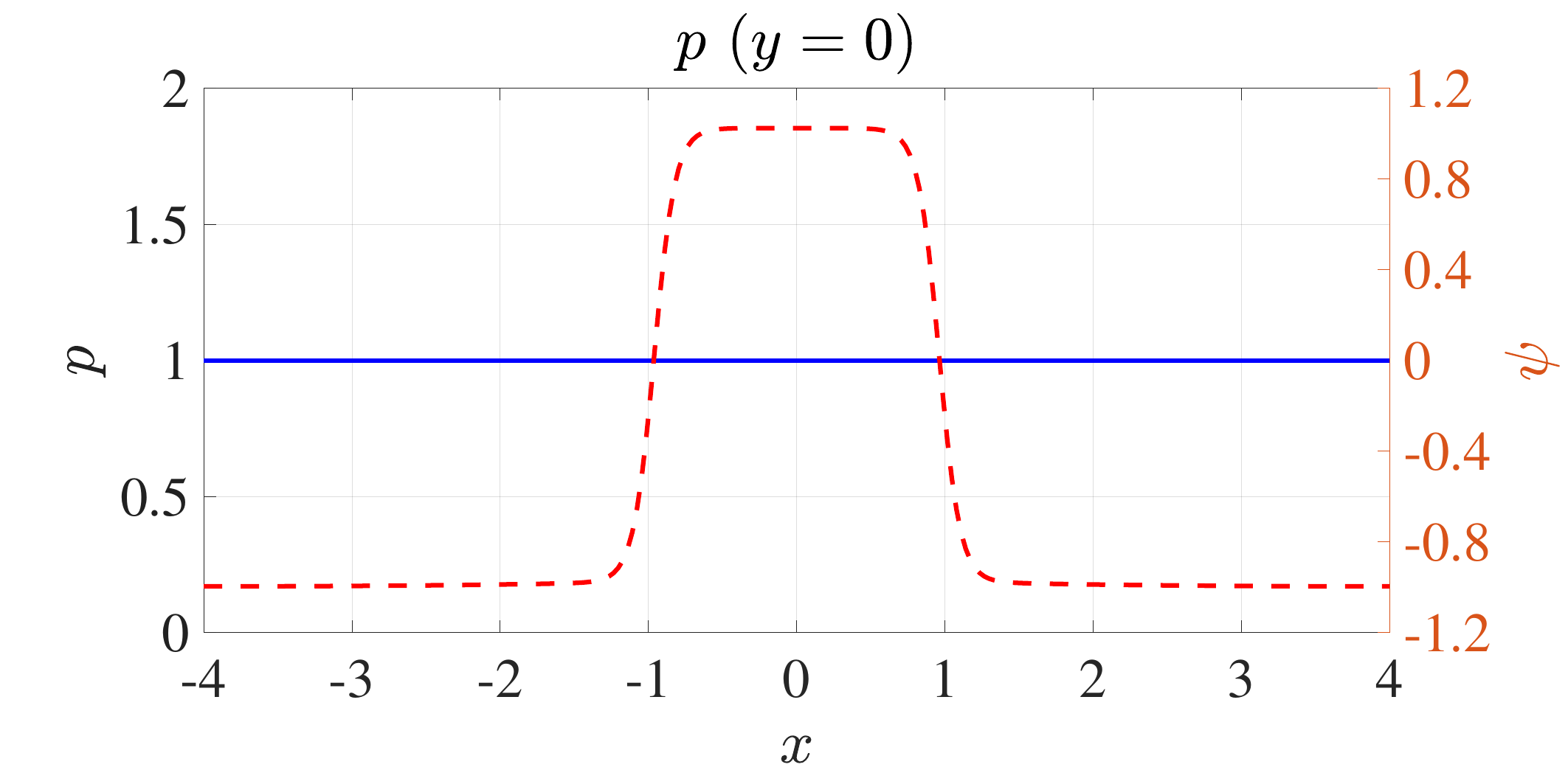}
        \label{subfig:1DropNoPumpP10y0}
		}
    \hskip -0.4cm
	\subfloat[$n~(y=0,t=10)$]{
		\includegraphics[width=0.33\linewidth]{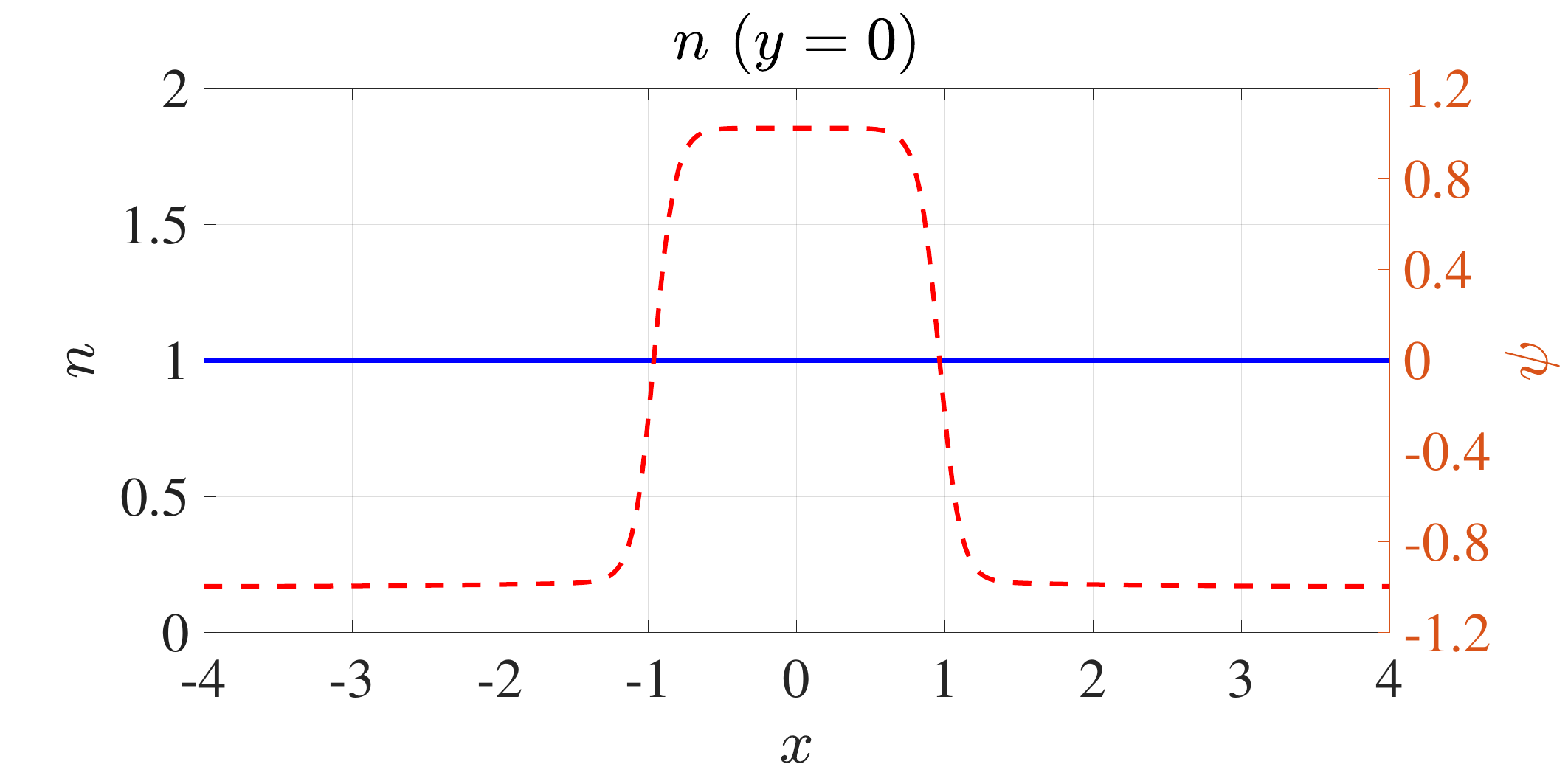}
        \label{subfig:1DropNoPumpN10y0}
		}
    \hskip -0.4cm
	\subfloat[$\phi~(y=0,t=10)$]{
		\includegraphics[width=0.33\linewidth]{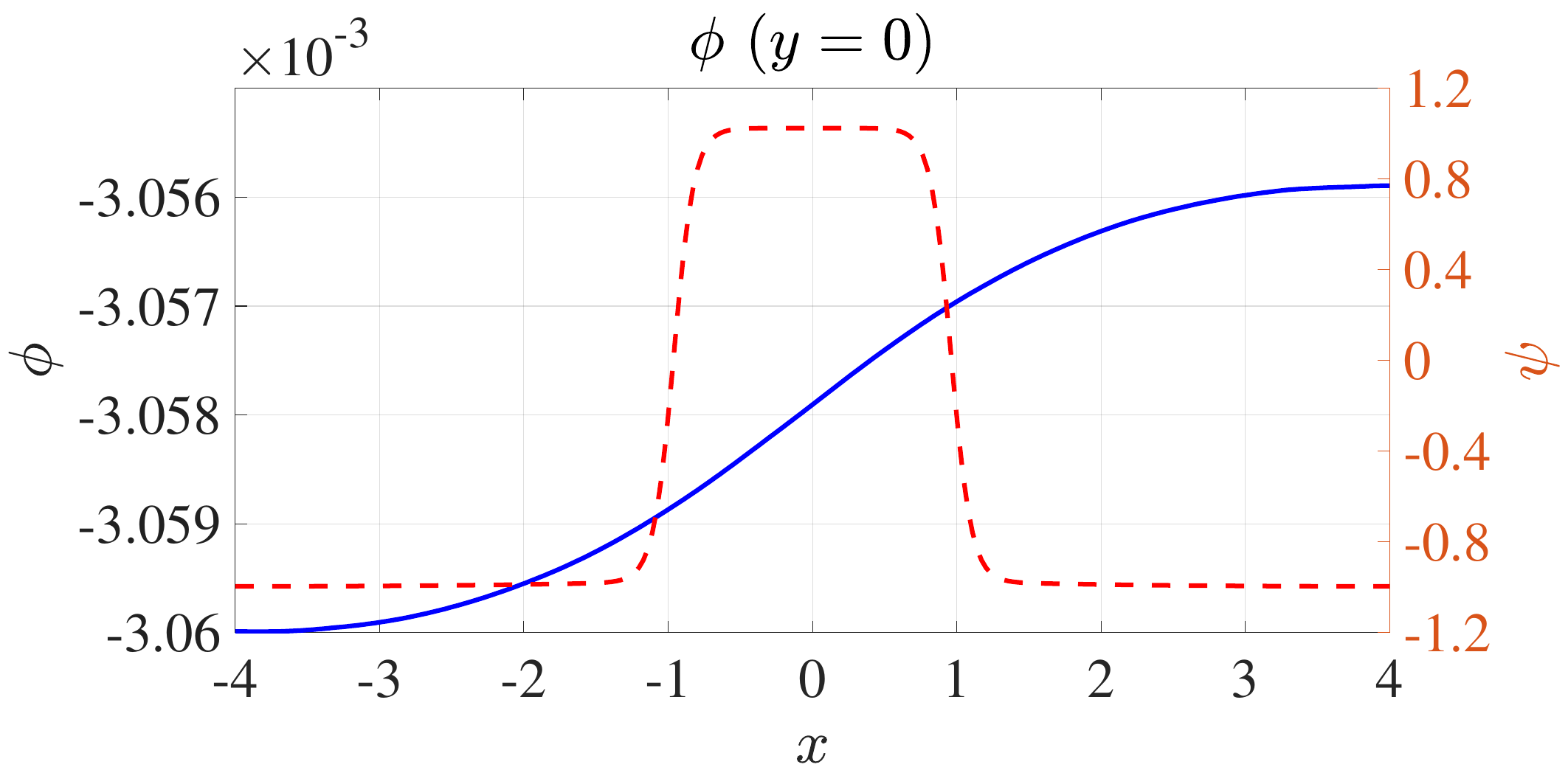}
        \label{subfig:1DropNoPumpPhi10y0}
		}
        \\
        \vskip -0.3cm
	\subfloat[$p~(x=0,t=10)$]{
		\includegraphics[width=0.33\linewidth]{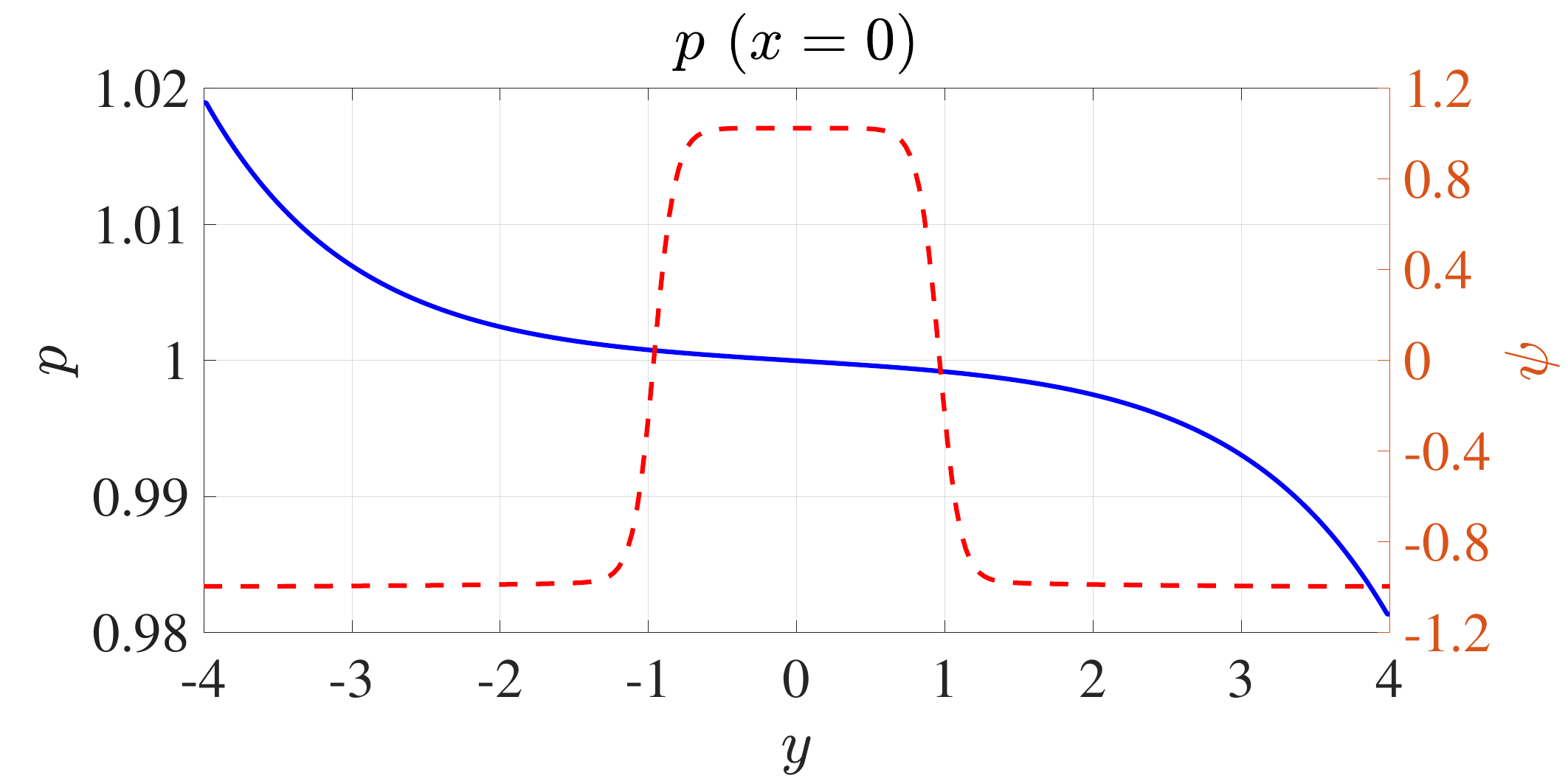}
        \label{subfig:1DropNoPumpP10x0}
		}
    \hskip -0.4cm
	\subfloat[$n~(x=0,t=10)$]{
		\includegraphics[width=0.33\linewidth]{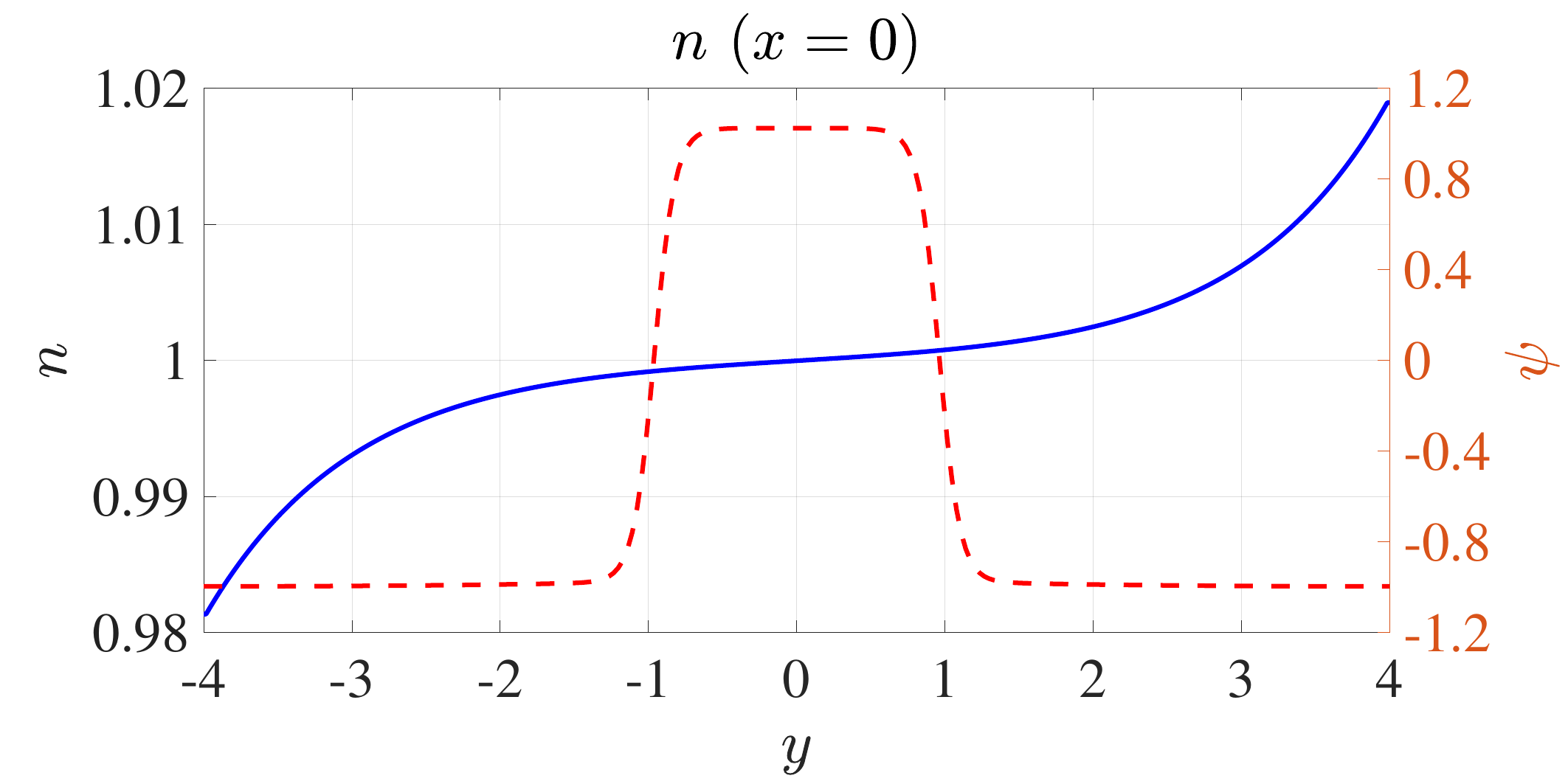}
        \label{subfig:1DropNoPumpN10x0}
		}
    \hskip -0.4cm
	\subfloat[$\phi~(x=0,t=10)$]{
		\includegraphics[width=0.33\linewidth]{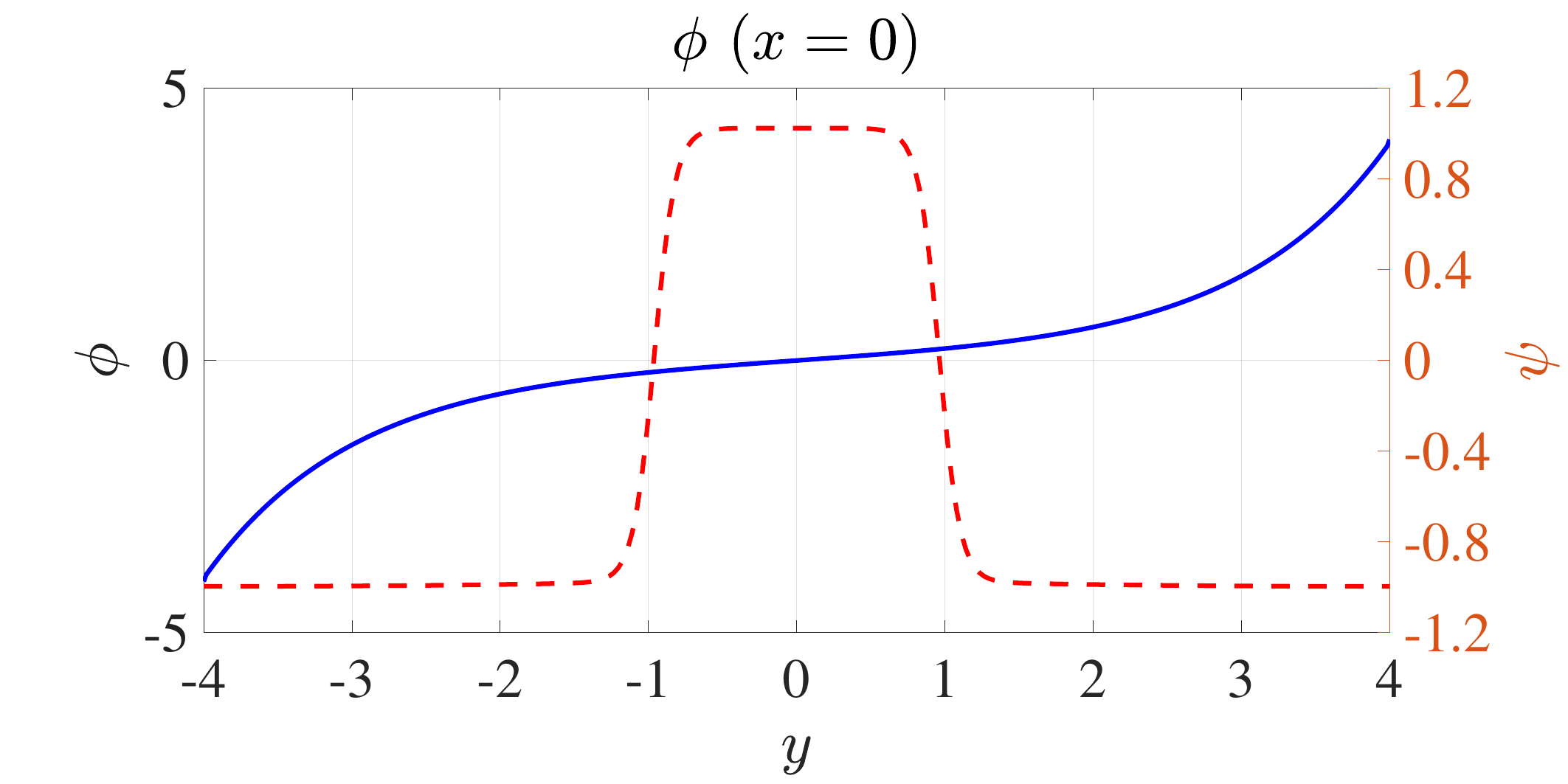}
        \label{subfig:1DropNoPumpPhi10x0}
		}
        \vskip -0.2cm
	\caption{The positive and negative ion and the electric potential distribution along $x$- (top) and $y$-axis (bottom) for the case in Figure \ref{fig:1DropNoPump}, respectively. 
    The blue solid lines show the ions concentration and electric potential and the red dash lines show the label function distribution with the diffusion interface.
    }\label{fig:1DropNoPumpSection}
\end{figure}

\begin{figure}[!ht]
\vskip -0.4cm
\centering
	\subfloat[$-\frac{Ca_{E}}{\zeta^{2}}\rho\nabla\phi~(y=0,t=10)$]{
		\includegraphics[width=0.48\linewidth]{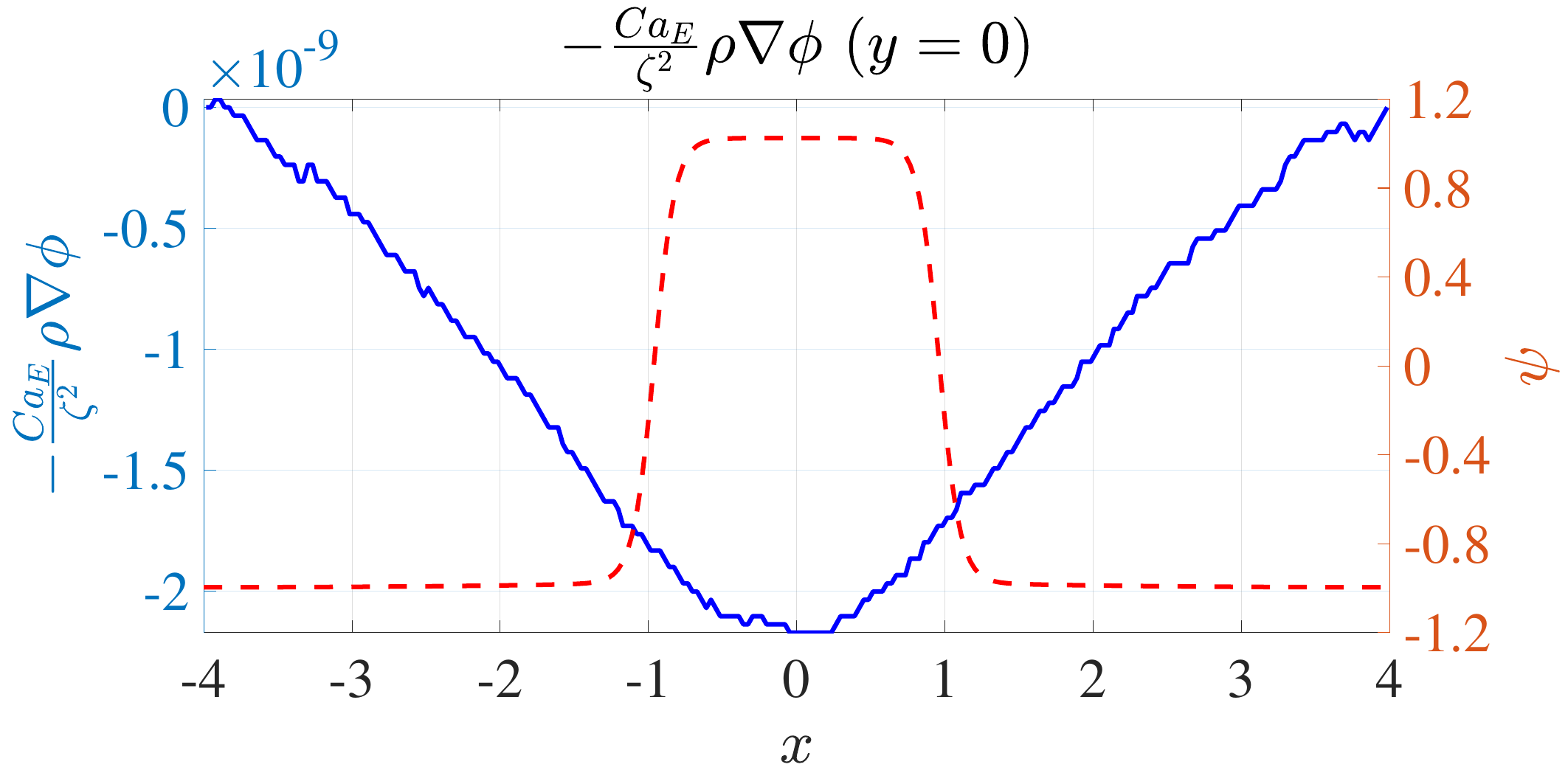}
        \label{subfig:1DropNoPumpEforcex}
		} 
	\subfloat[$-\frac{Ca_{E}}{\zeta^{2}}\rho\nabla\phi~(x=0,t=10)$]{
		\includegraphics[width=0.48\linewidth]{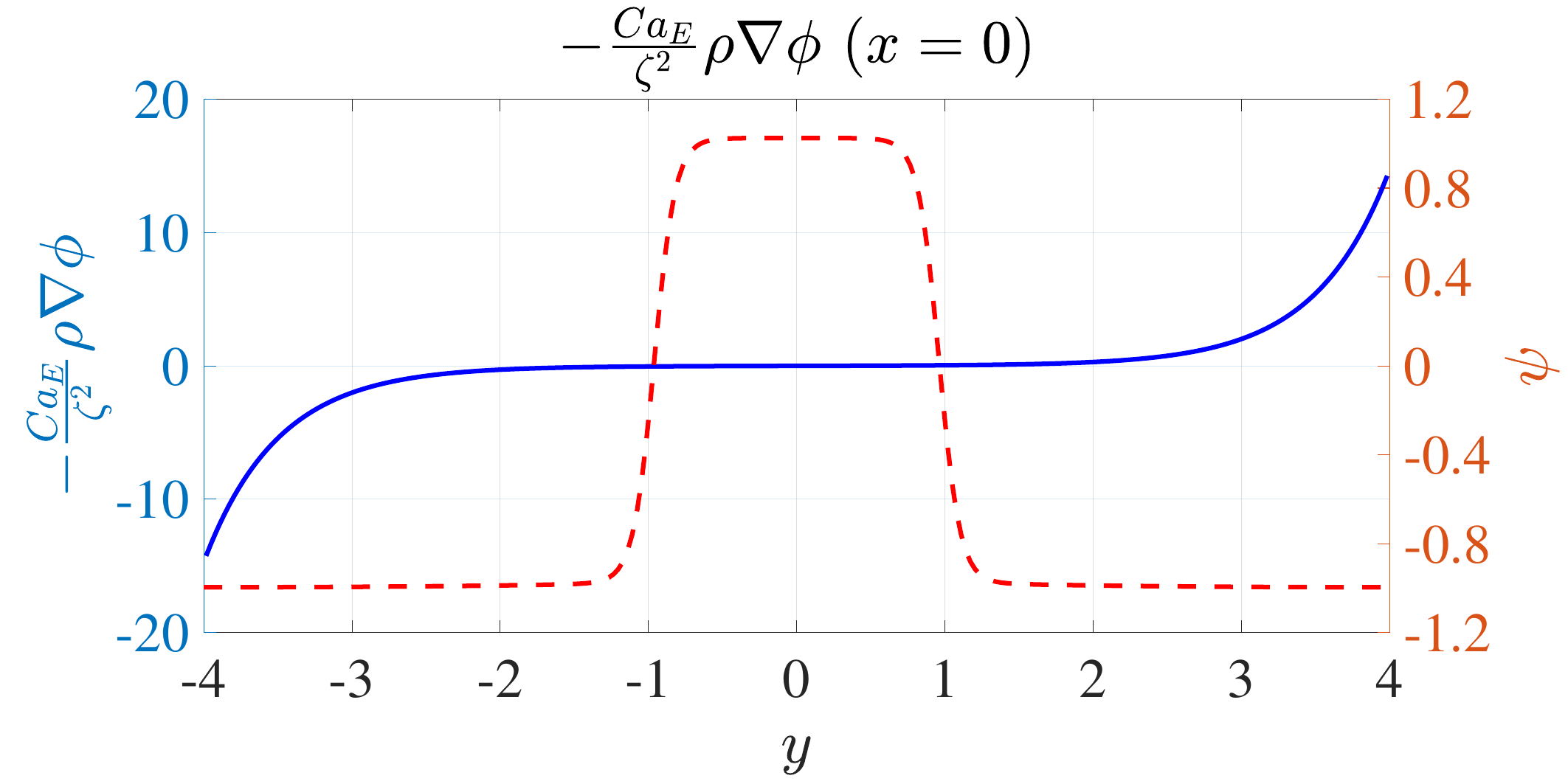}
        \label{subfig:1DropNoPumpEforcey}
	}
    \vskip -0.1cm
	\caption{The Lorentz force along $x$- (left) and $y$-axis (right) for the case in Figure \ref{fig:1DropNoPump}, respectively. The blue solid lines show the Lorentz force and the red dash lines show the label function defining the diffusion interface. }\label{fig:1DropNoPumpEforce}
\end{figure}

\begin{figure}[!ht]
	\centering
	\subfloat[$p~(x=0,t=10)$]{
		\includegraphics[width=0.33\linewidth]{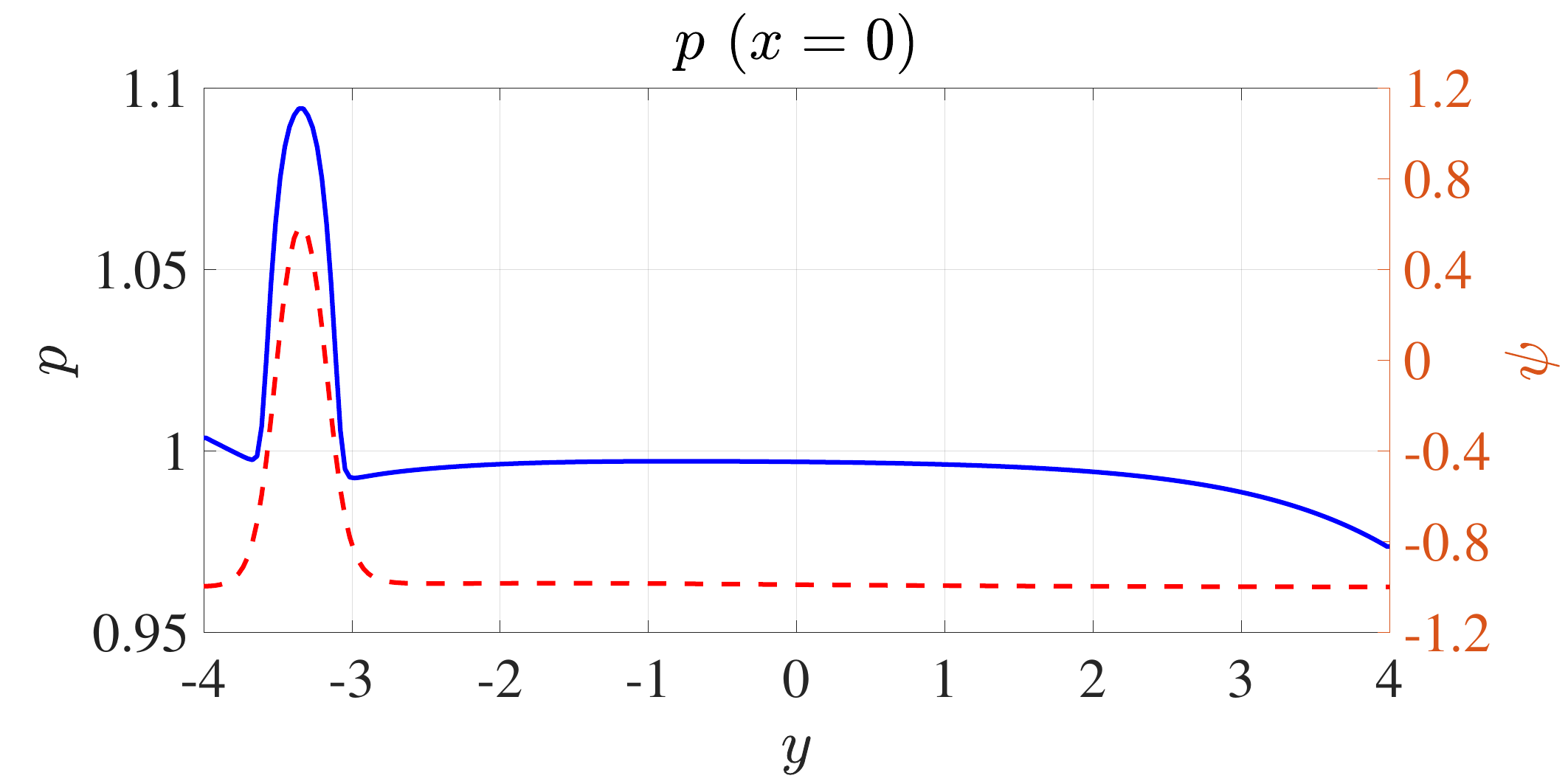}
        \label{subfig:1DropD4N0Pump25P10x0}
		}
        \hskip -0.4cm
	\subfloat[$n~(x=0,t=10)$]{
		\includegraphics[width=0.33\linewidth]{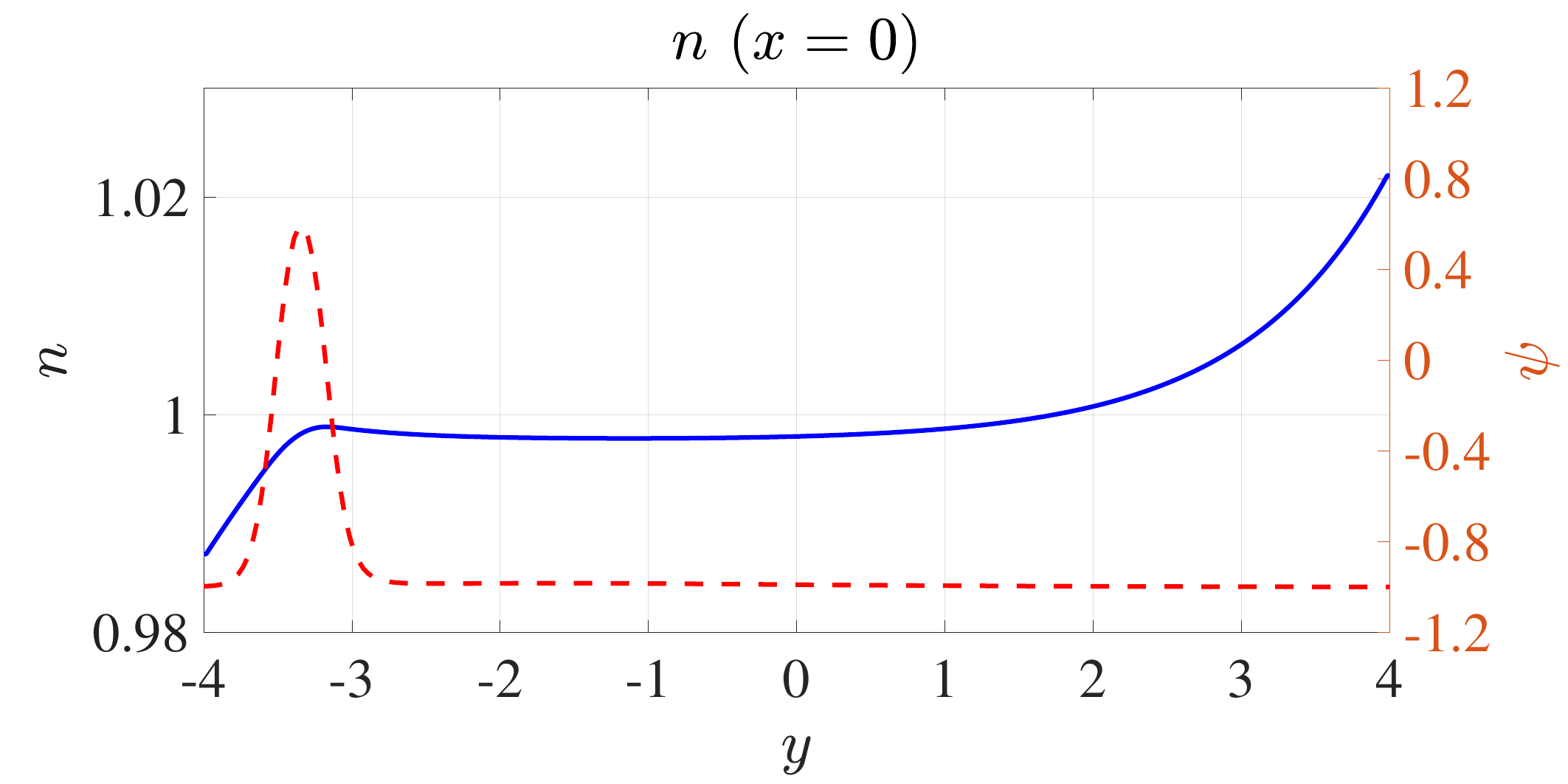}
        \label{subfig:1DropD4N0Pump25N10x0}
		}
        \hskip -0.4cm
	\subfloat[$\phi~(x=0,t=10)$]{
		\includegraphics[width=0.33\linewidth]{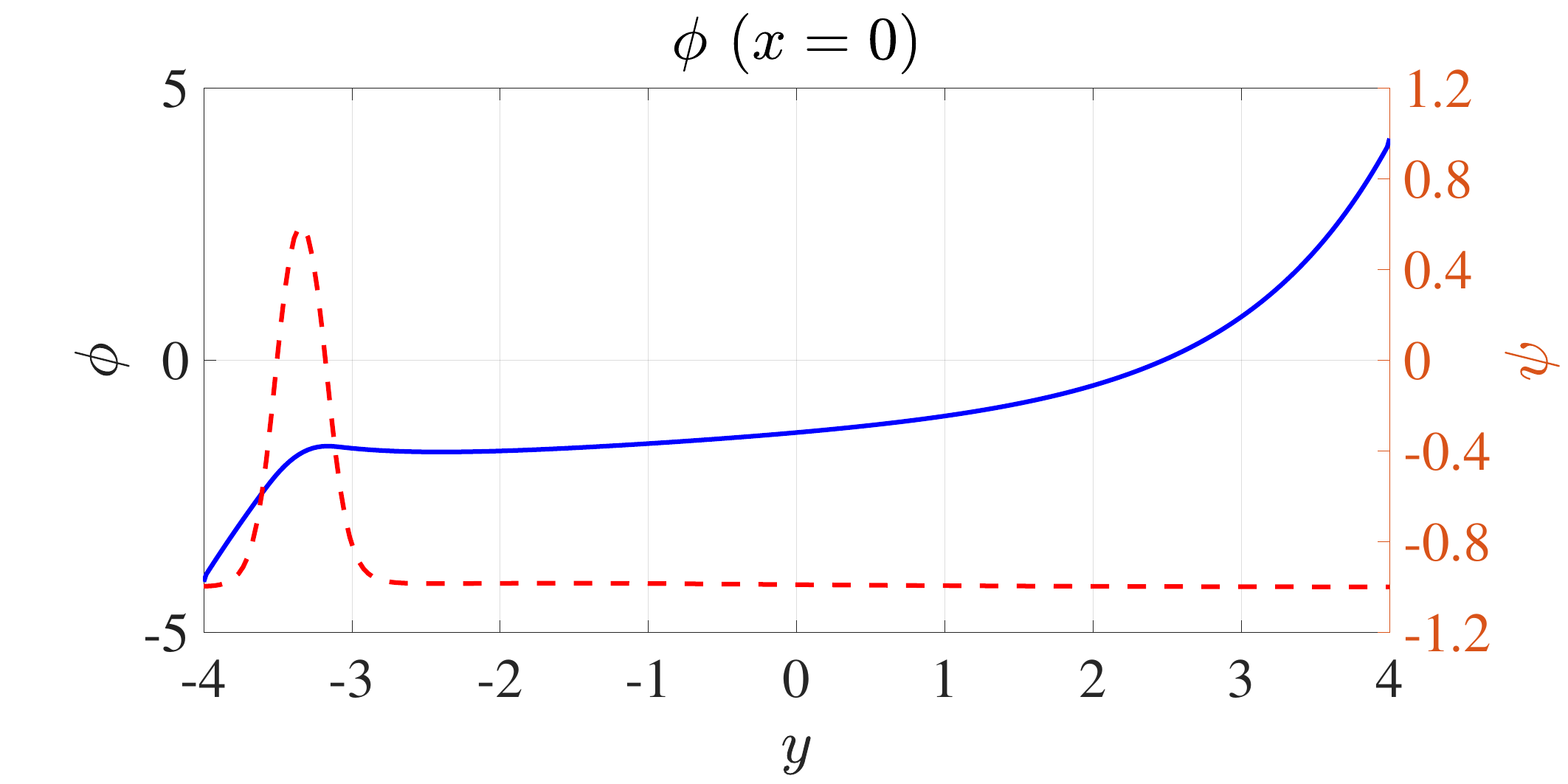}
        \label{subfig:1DropD4N0Pump25Phi10x0}
		}
	\caption{The ion distribution and electric potential distribution along $y$-axis with positive ion pump when the vertical electric field is added.
    The blue solid lines show the ion and the electric potential distribution. 
    The red dash lines show the diffuse interface. 
    }\label{fig:1DropD4N0Pump25Section}
\end{figure}

\begin{figure}[!ht]
	\centering
	\subfloat[$-\nabla P~(x=-1.2,t=5)$]{
		\includegraphics[width=0.33\linewidth]{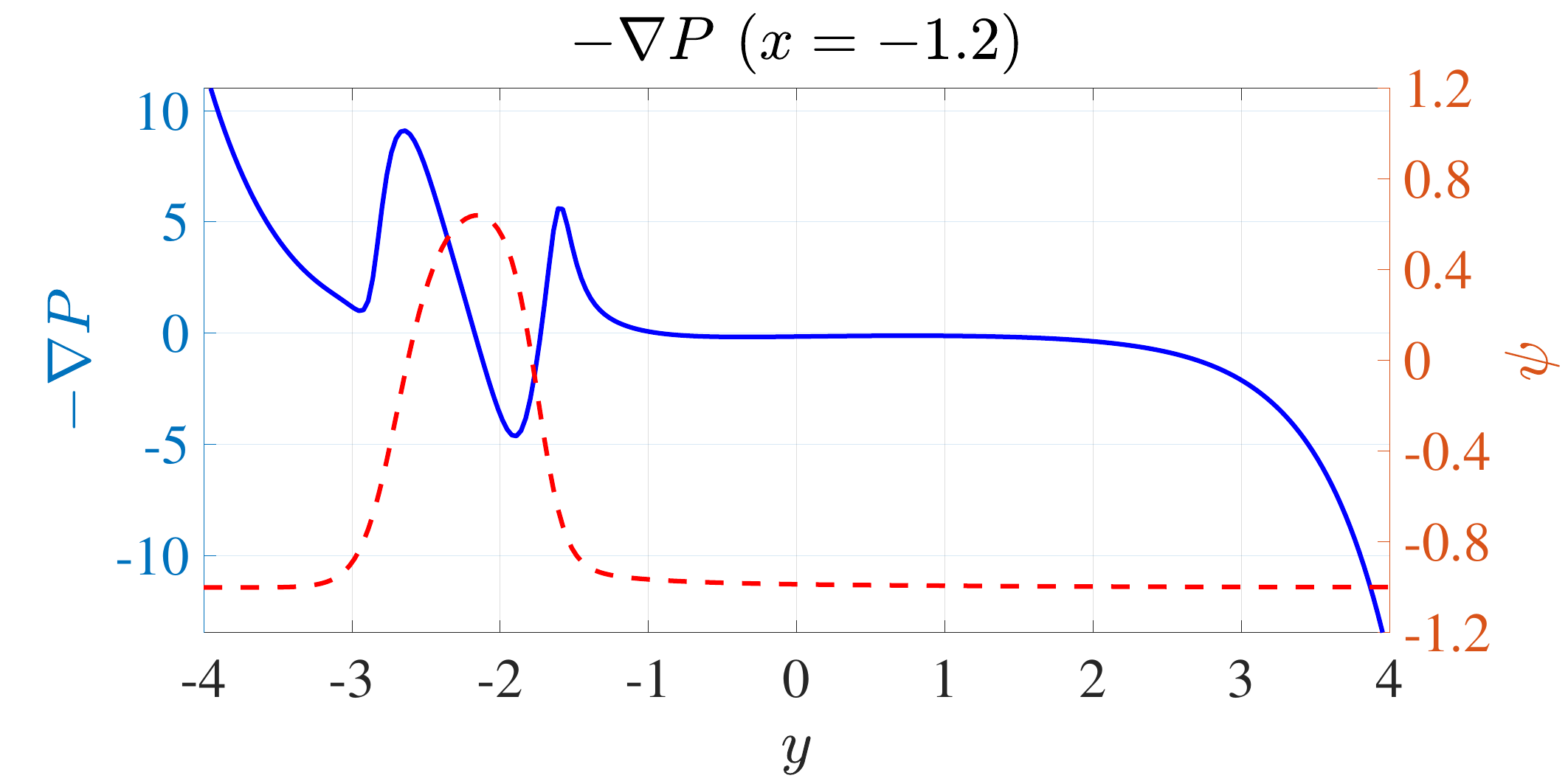}
        \label{subfig:1DropD4N0Pump25dpre5x-1d2}
		}
	\subfloat[$-\nabla P~(x=0,t=5)$]{
		\includegraphics[width=0.33\linewidth]{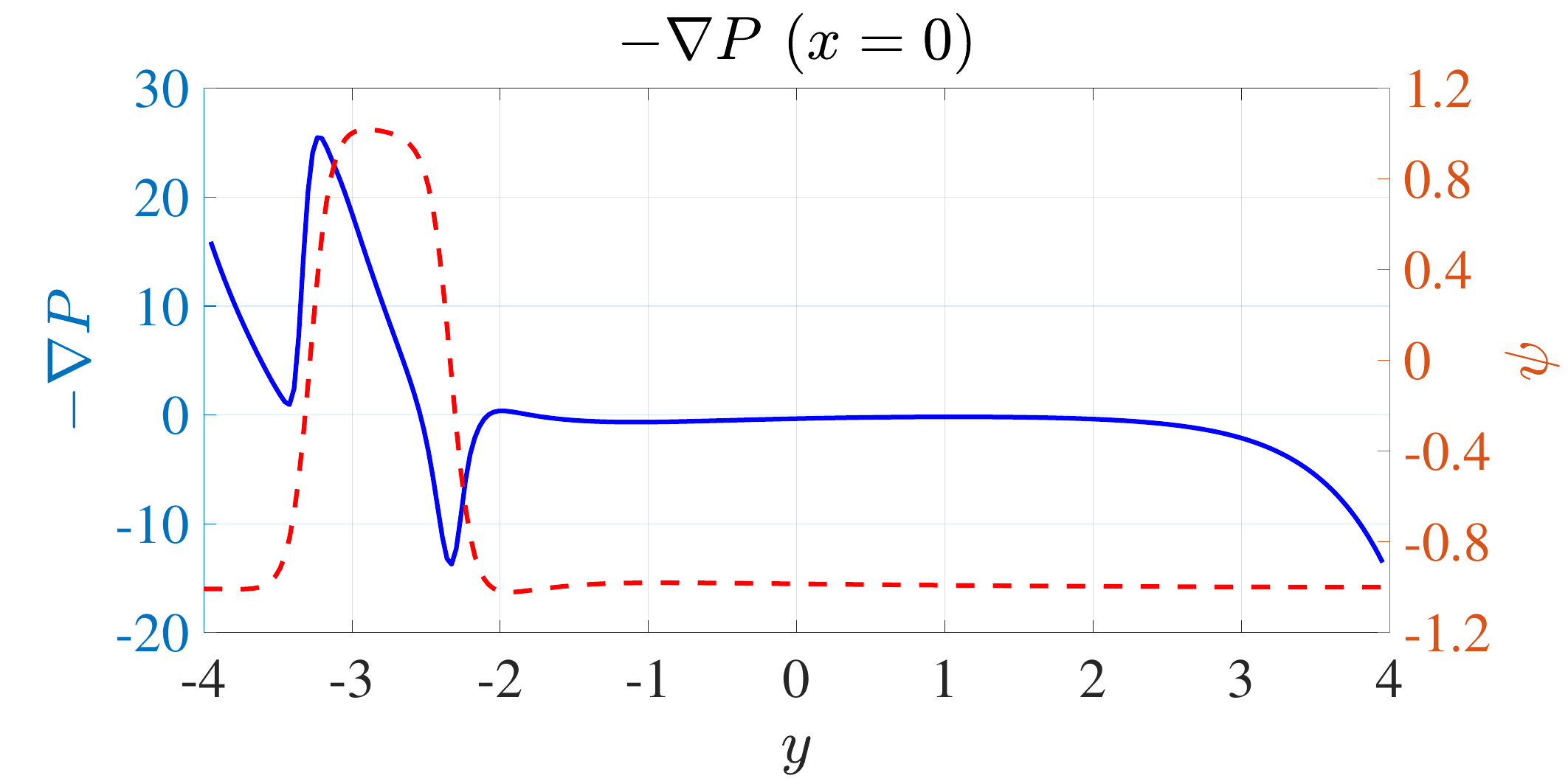}
        \label{subfig:1DropD4N0Pump25dpre5x0}
		}
	\subfloat[$-\nabla P~(x=1.2,t=5)$]{
		\includegraphics[width=0.33\linewidth]{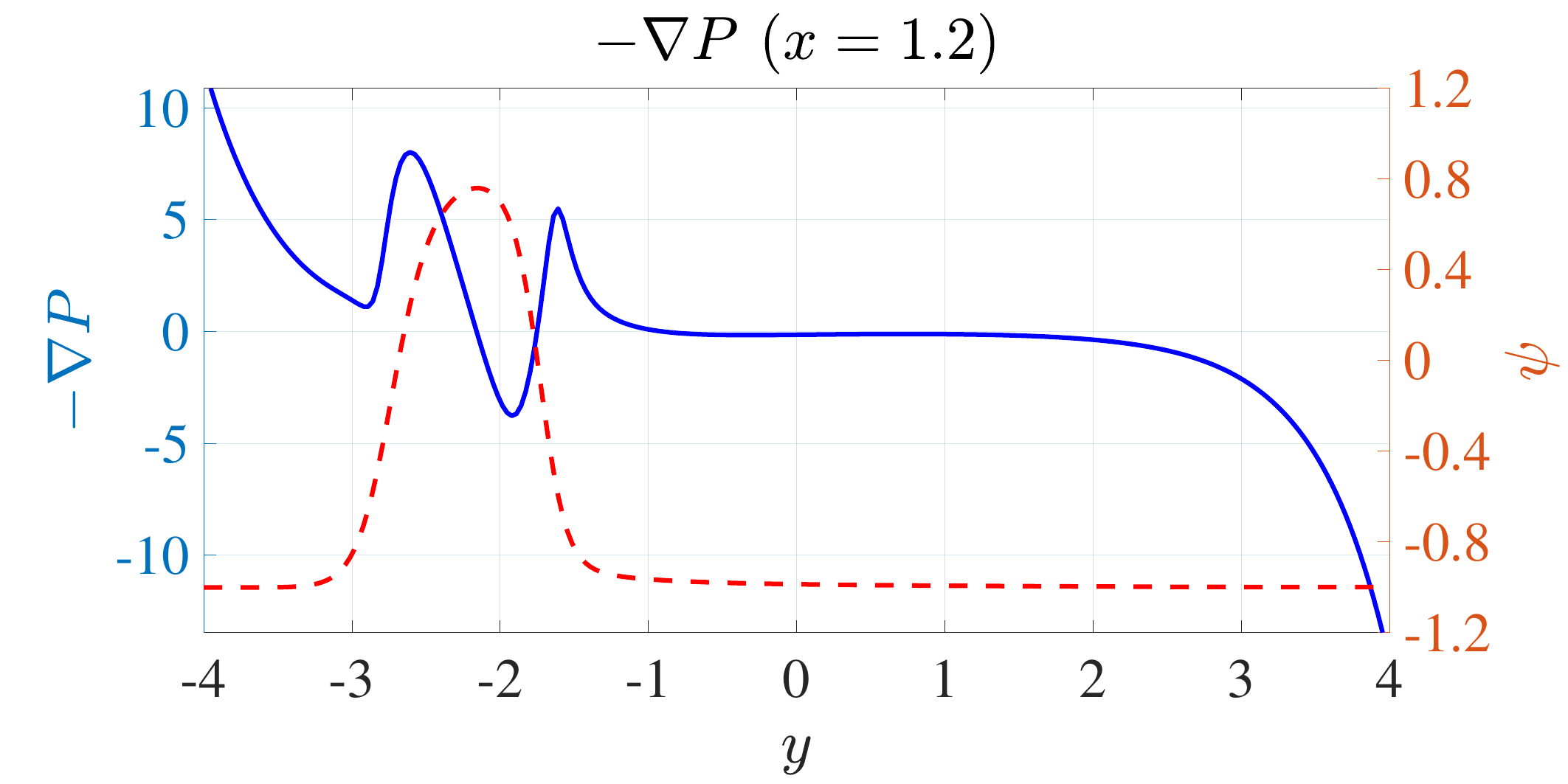}
        \label{subfig:1DropD4N0Pump25dpre5x1d2}
		}
		\\
	\subfloat[$-\nabla P~(x=-2.4,t=8)$]{
		\includegraphics[width=0.33\linewidth]{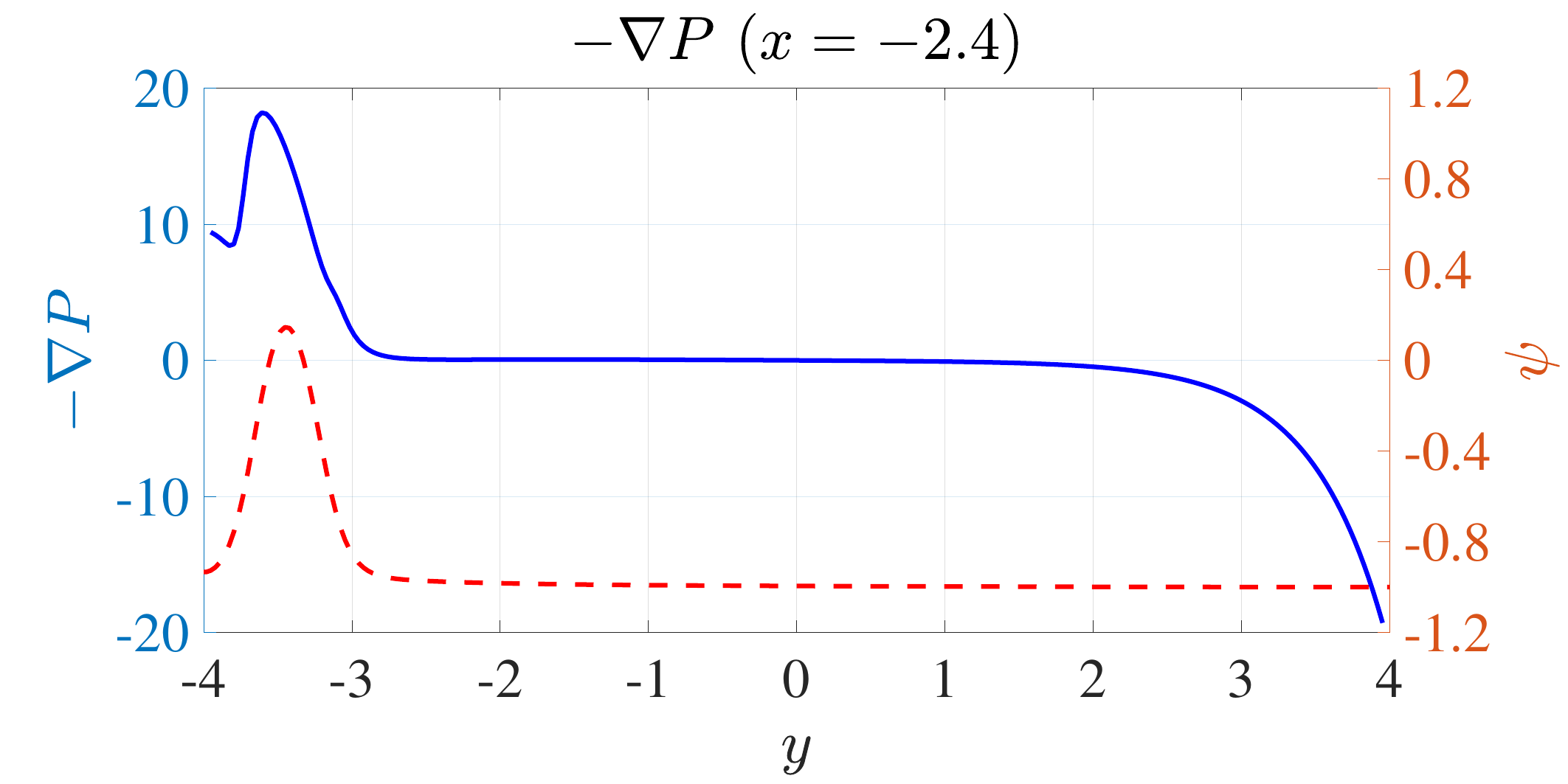}
        \label{subfig:1DropD4N0Pump25dpre8x-2d4}
		}
	\subfloat[$-\nabla P~(x=0.4,t=8)$]{
		\includegraphics[width=0.33\linewidth]{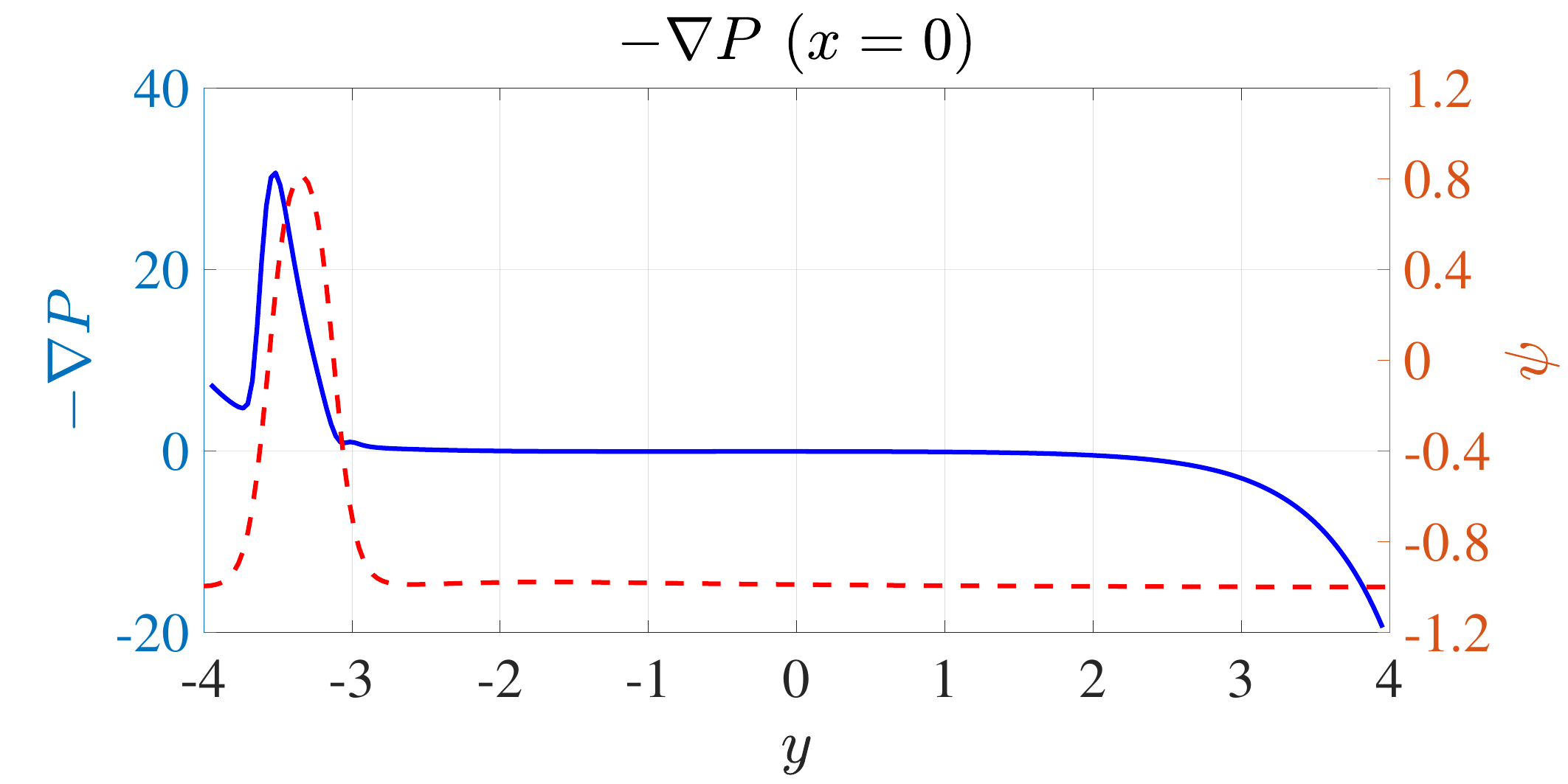}
        \label{subfig:1DropD4N0Pump25dpre8x0}
		}
	\subfloat[$-\nabla P~(x=2.4,t=8)$]{
		\includegraphics[width=0.33\linewidth]{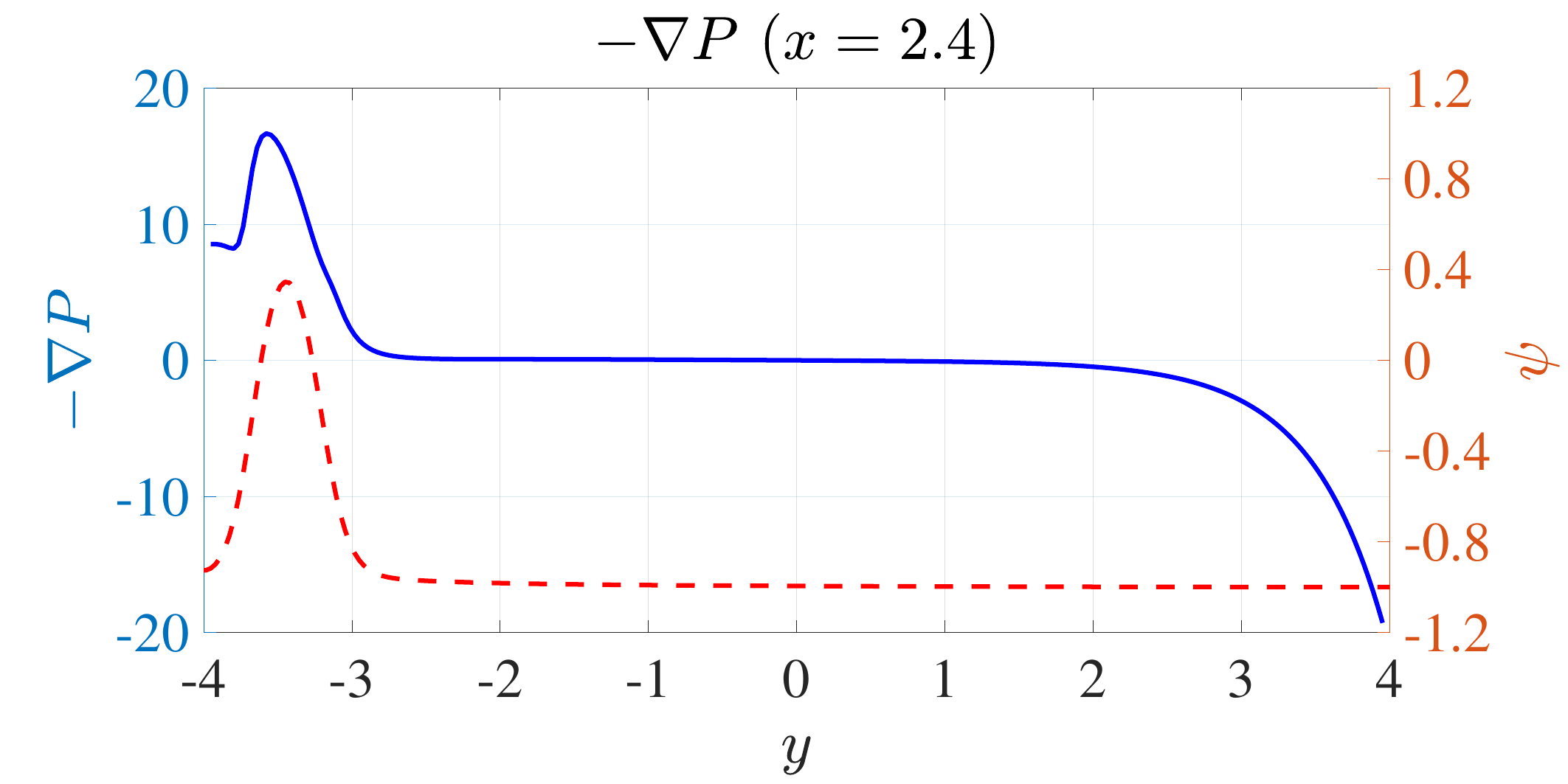}
        \label{subfig:1DropD4N0Pump25dpre8x2d4}
		}
	\caption{The snapshots of $-\nabla P$ with positive ion pump at $t=5$ (top) and $t=8$ (bottom) when the vertical electric field is added.
    The blue solid lines show $-\nabla P$ distribution at left end (left), middle (middle) and right end (right) of the drop in the horizontal direction. 
    }\label{fig:1DropD4N0Pump25dpreSection}
\end{figure}

\begin{figure}[!ht]
\vskip -0.4cm
	\centering
	\subfloat[$-\frac{Ca_{E}}{\zeta^{2}}\rho\nabla\phi~(x=-1.2,t=5)$]{
		\includegraphics[width=0.33\linewidth]{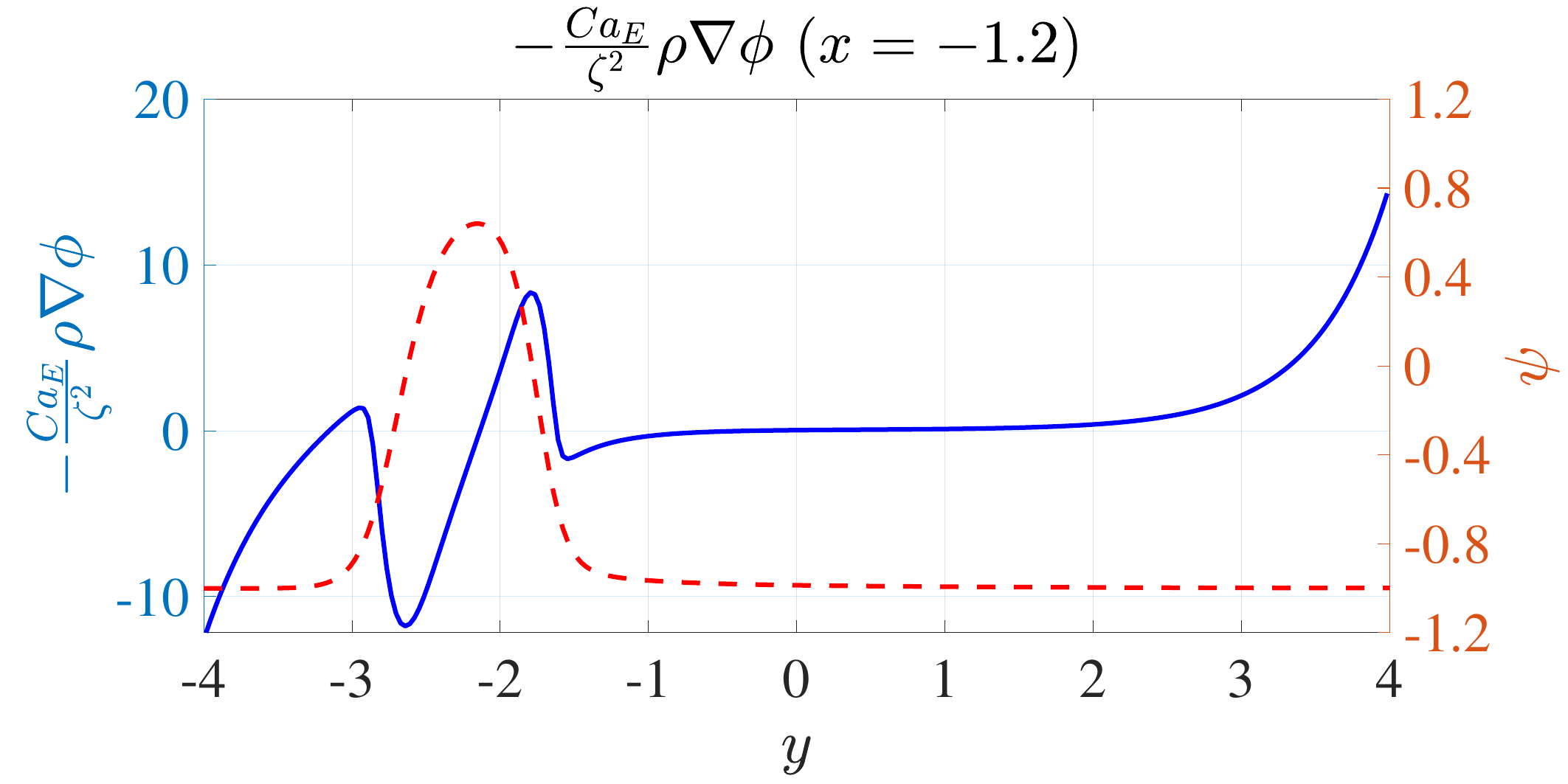}
        \label{subfig:1DropD4N0Pump25eforce5x-1d2}
		} 
	\subfloat[$-\frac{Ca_{E}}{\zeta^{2}}\rho\nabla\phi~(x=0,t=5)$]{
		\includegraphics[width=0.33\linewidth]{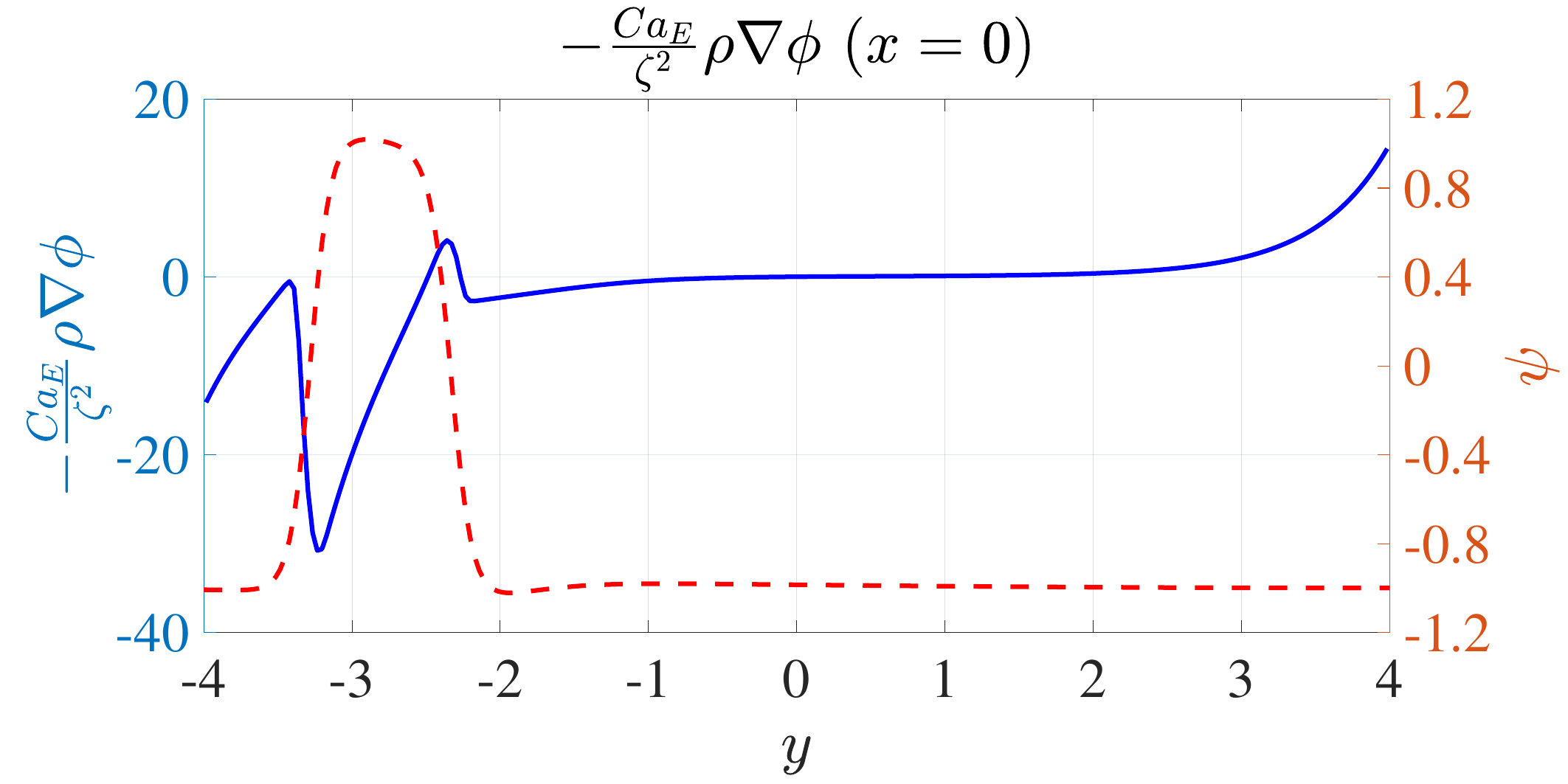}
        \label{subfig:1DropD4N0Pump25eforce5x0}
		} 
	\subfloat[$-\frac{Ca_{E}}{\zeta^{2}}\rho\nabla\phi~(x=1.2,t=5)$]{
		\includegraphics[width=0.33\linewidth]{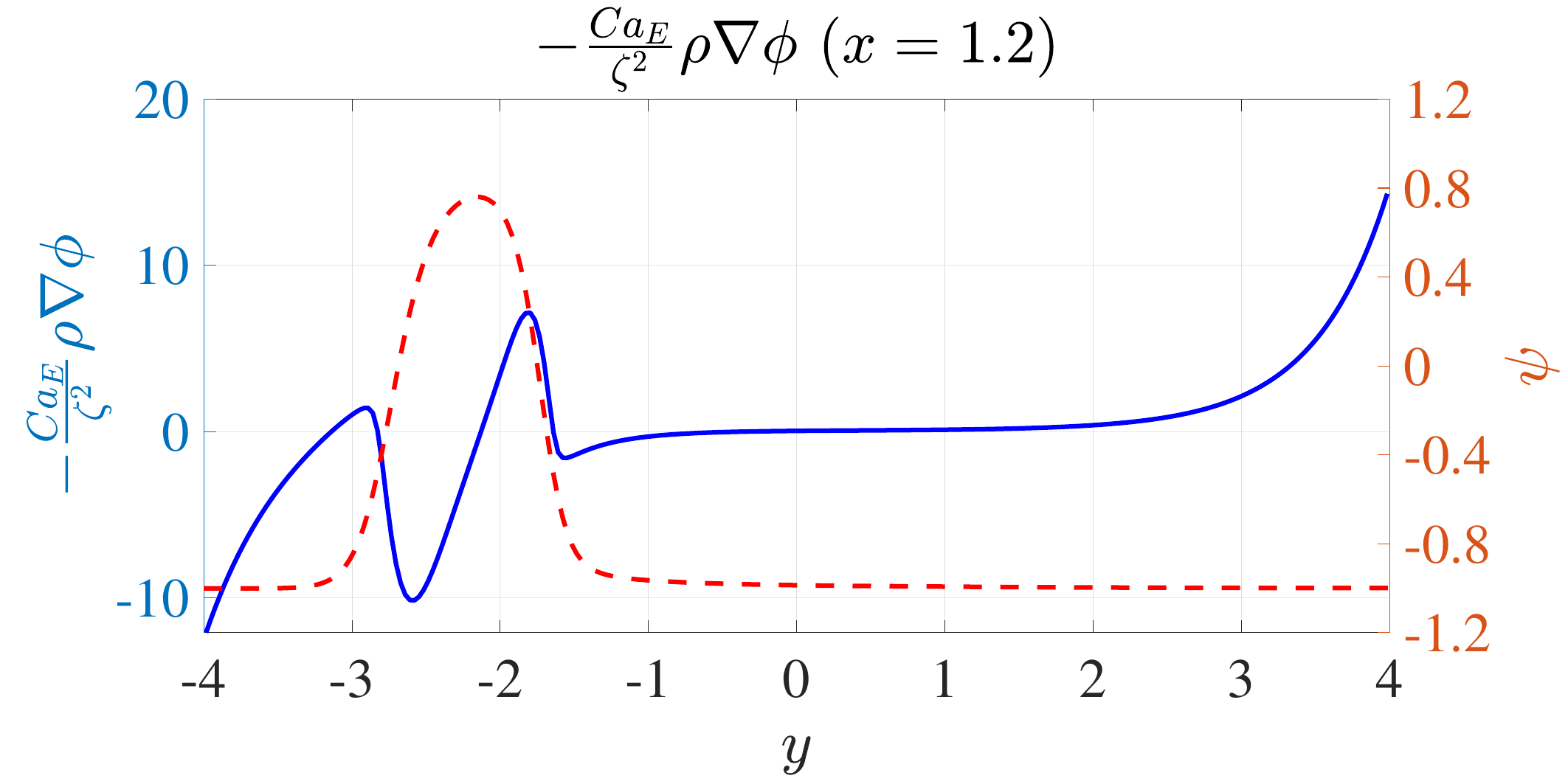}
        \label{subfig:1DropD4N0Pump25eforce5x1d2}
		}
		\\
	\subfloat[$-\frac{Ca_{E}}{\zeta^{2}}\rho\nabla\phi~(x=-2.4,t=8)$]{
		\includegraphics[width=0.33\linewidth]{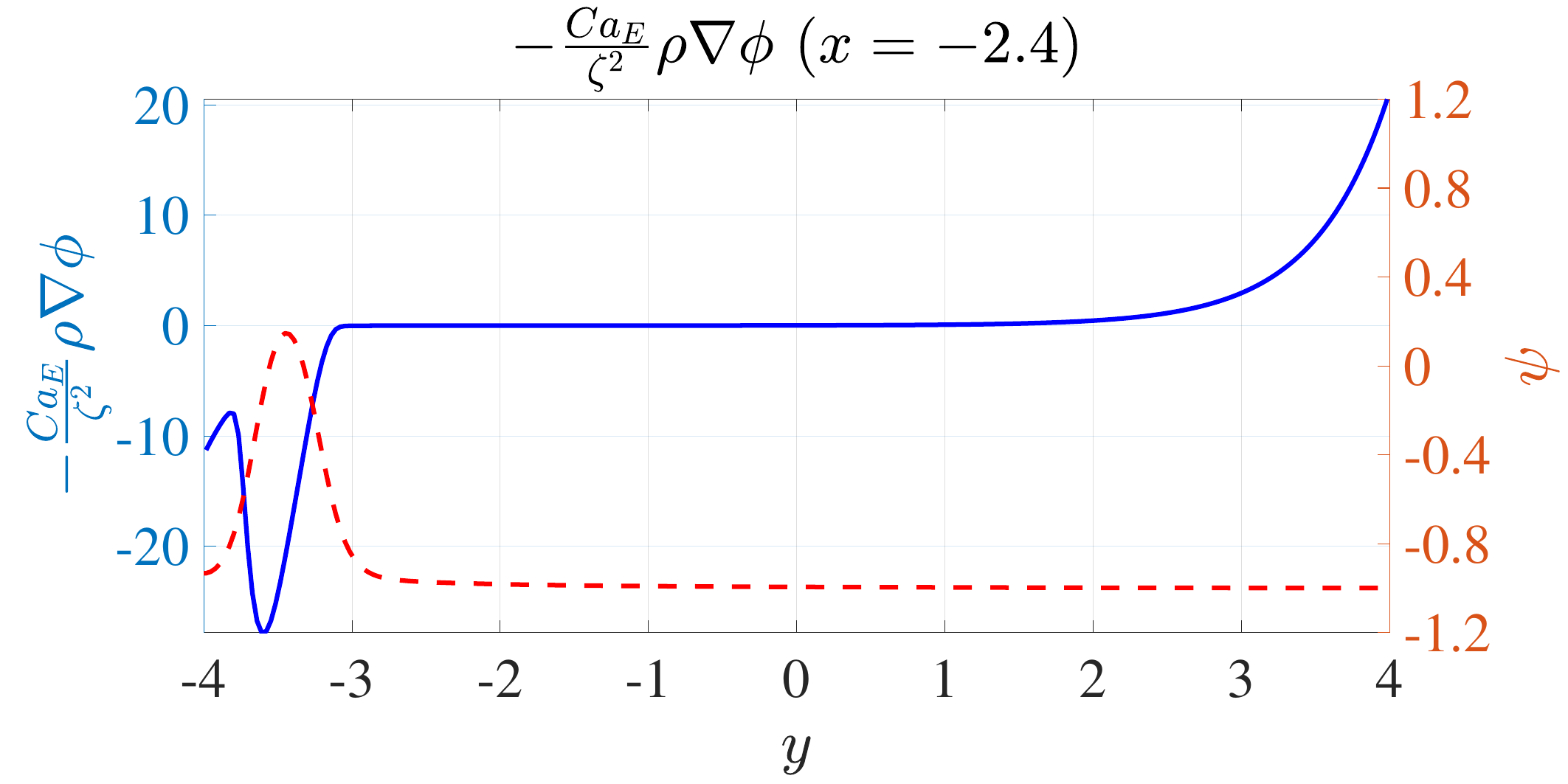}
        \label{subfig:1DropD4N0Pump25eforce8x-2d4}
		} 
	\subfloat[$-\frac{Ca_{E}}{\zeta^{2}}\rho\nabla\phi~(x=0,t=8)$]{
		\includegraphics[width=0.33\linewidth]{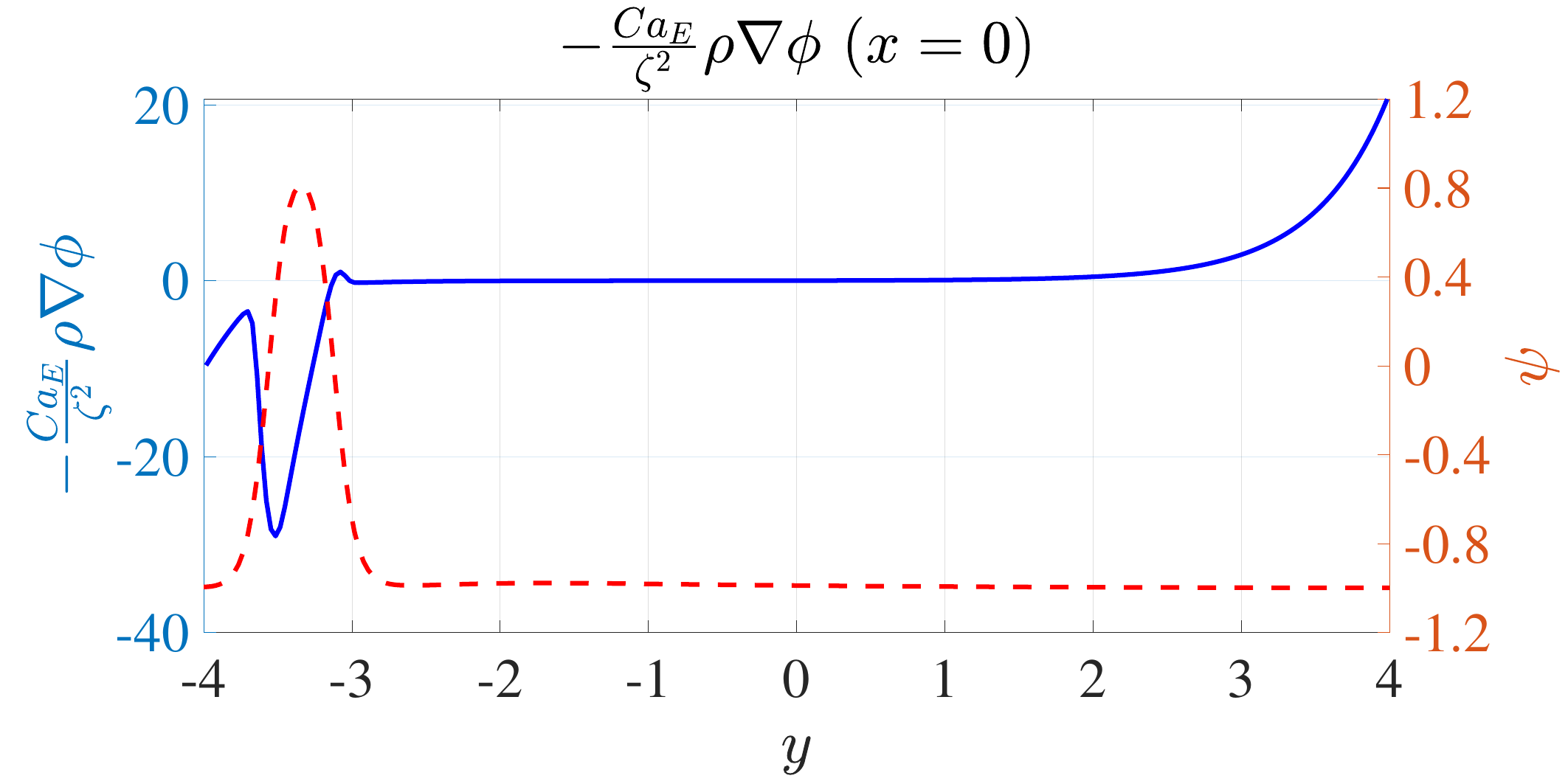}
        \label{subfig:1DropD4N0Pump25eforce8x0}
		} 
	\subfloat[$-\frac{Ca_{E}}{\zeta^{2}}\rho\nabla\phi~(x=2.4,t=8)$]{
		\includegraphics[width=0.33\linewidth]{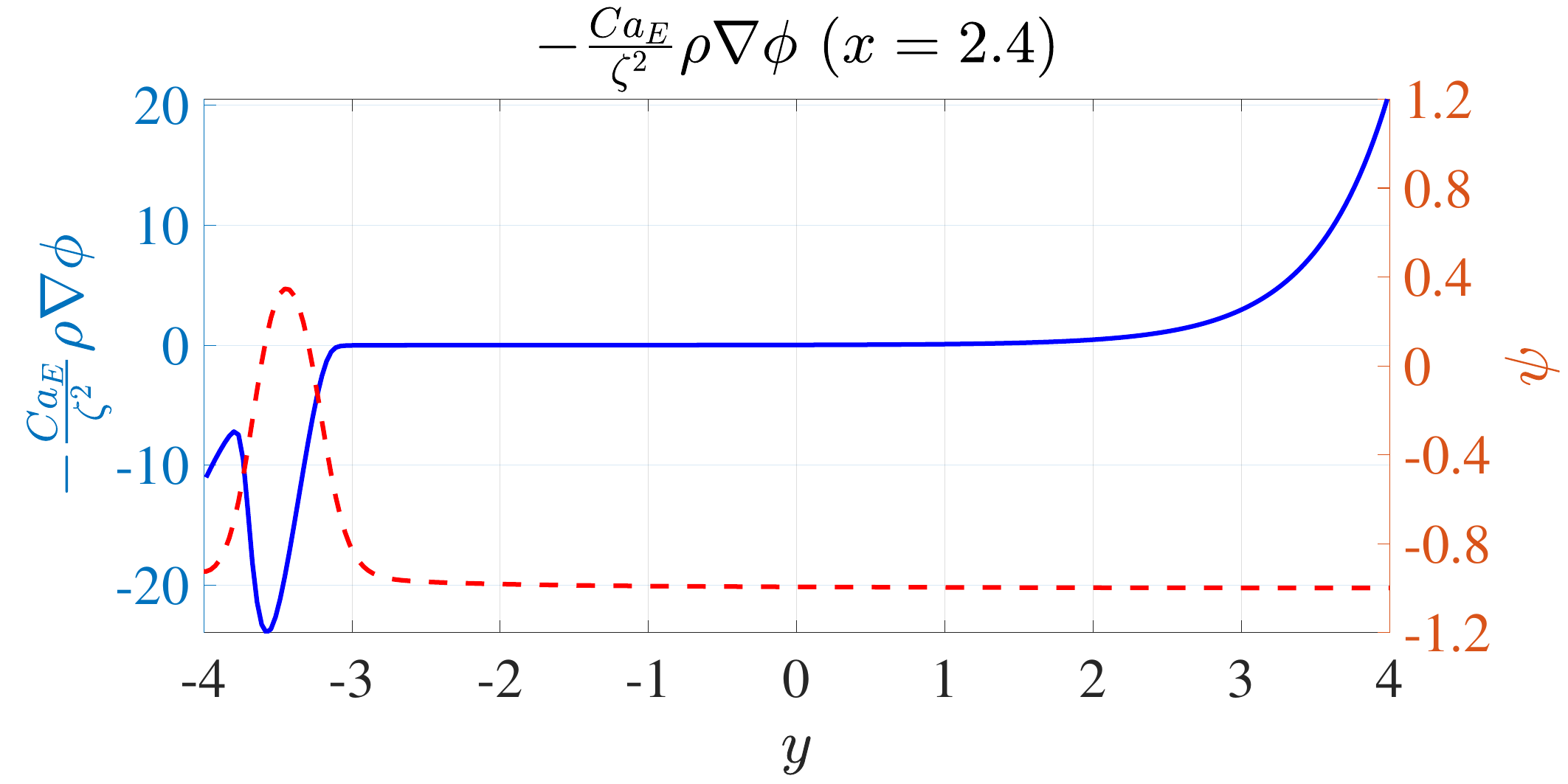}
        \label{subfig:1DropD4N0Pump25eforce8x2d4}
		}
	\caption{The snapshots of Lorentz force with positive ion pump at $t=5$ (top) and $t=8$ (bottom) when the vertical electric field is added.
    The blue solid lines show the Lorentz force distribution at left end (left), middle (middle) and right end (right) of the drop in the horizontal direction. 
    }\label{fig:1DropD4N0Pump25eforceSection}
\end{figure}

\begin{figure}[!ht]
\vskip -0.4cm
	\centering
	\subfloat[$-\frac{Ca_{E}}{\zeta^{2}}\rho\nabla\phi-\nabla P~(x=-1.2,t=5)$]{
		\includegraphics[width=0.33\linewidth]{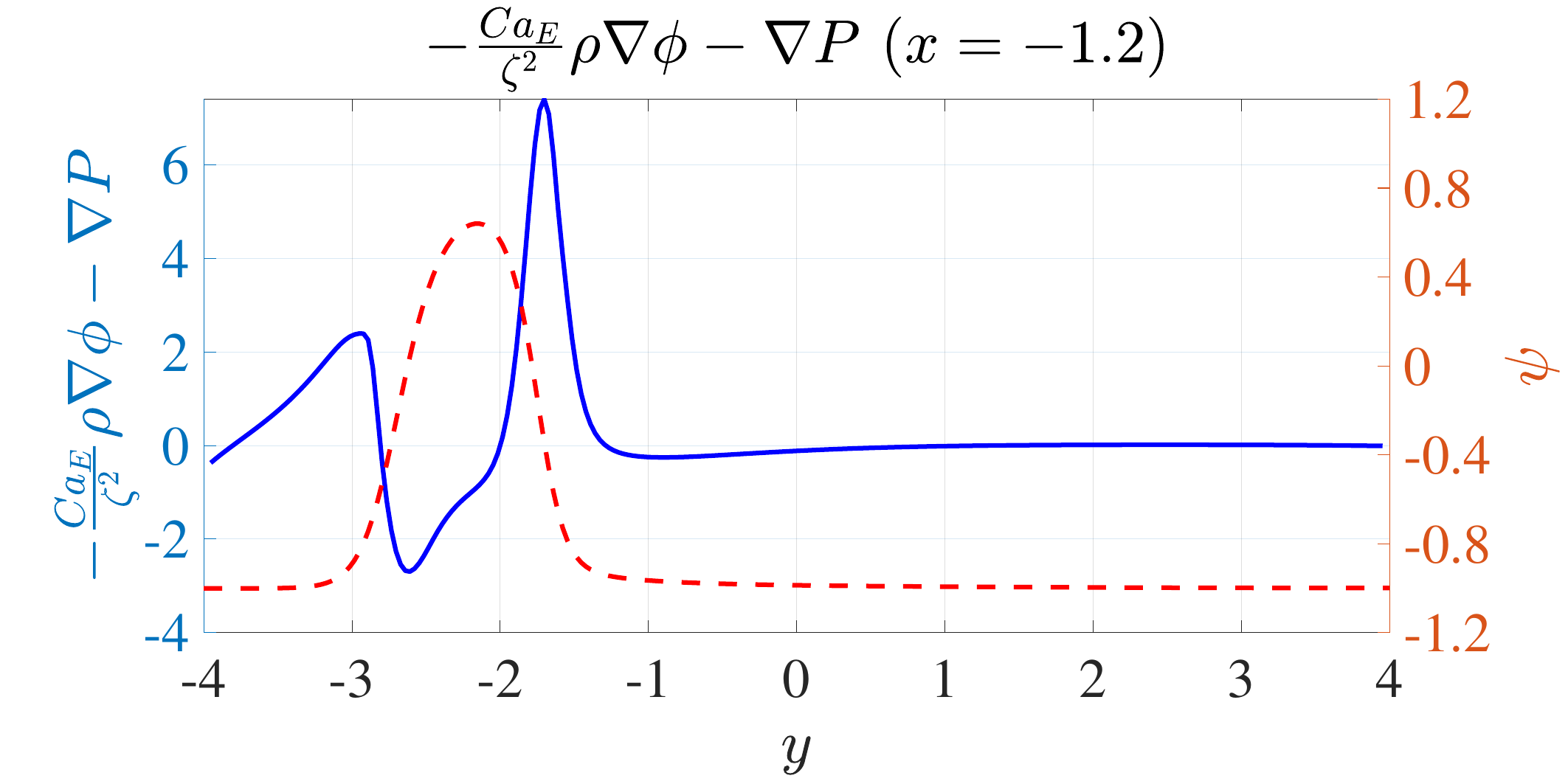}
        \label{subfig:1DropD4N0Pump25ep5x-1d2}
		} 
	\subfloat[$-\frac{Ca_{E}}{\zeta^{2}}\rho\nabla\phi-\nabla P~(x=0,t=5)$]{
		\includegraphics[width=0.33\linewidth]{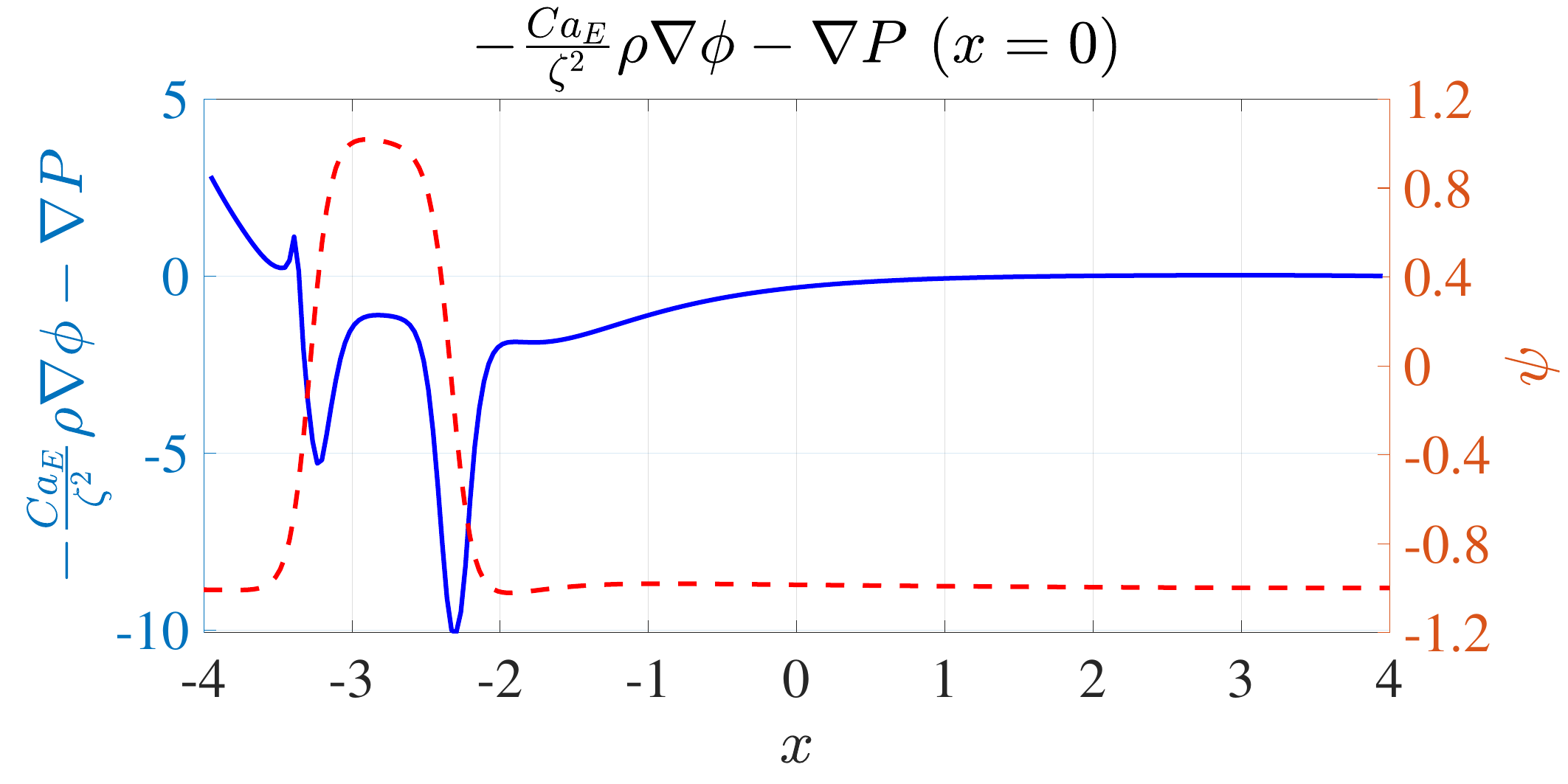}
        \label{subfig:1DropD4N0Pump25ep5x0}
		} 
	\subfloat[$-\frac{Ca_{E}}{\zeta^{2}}\rho\nabla\phi-\nabla P~(x=1.2,t=5)$]{
		\includegraphics[width=0.33\linewidth]{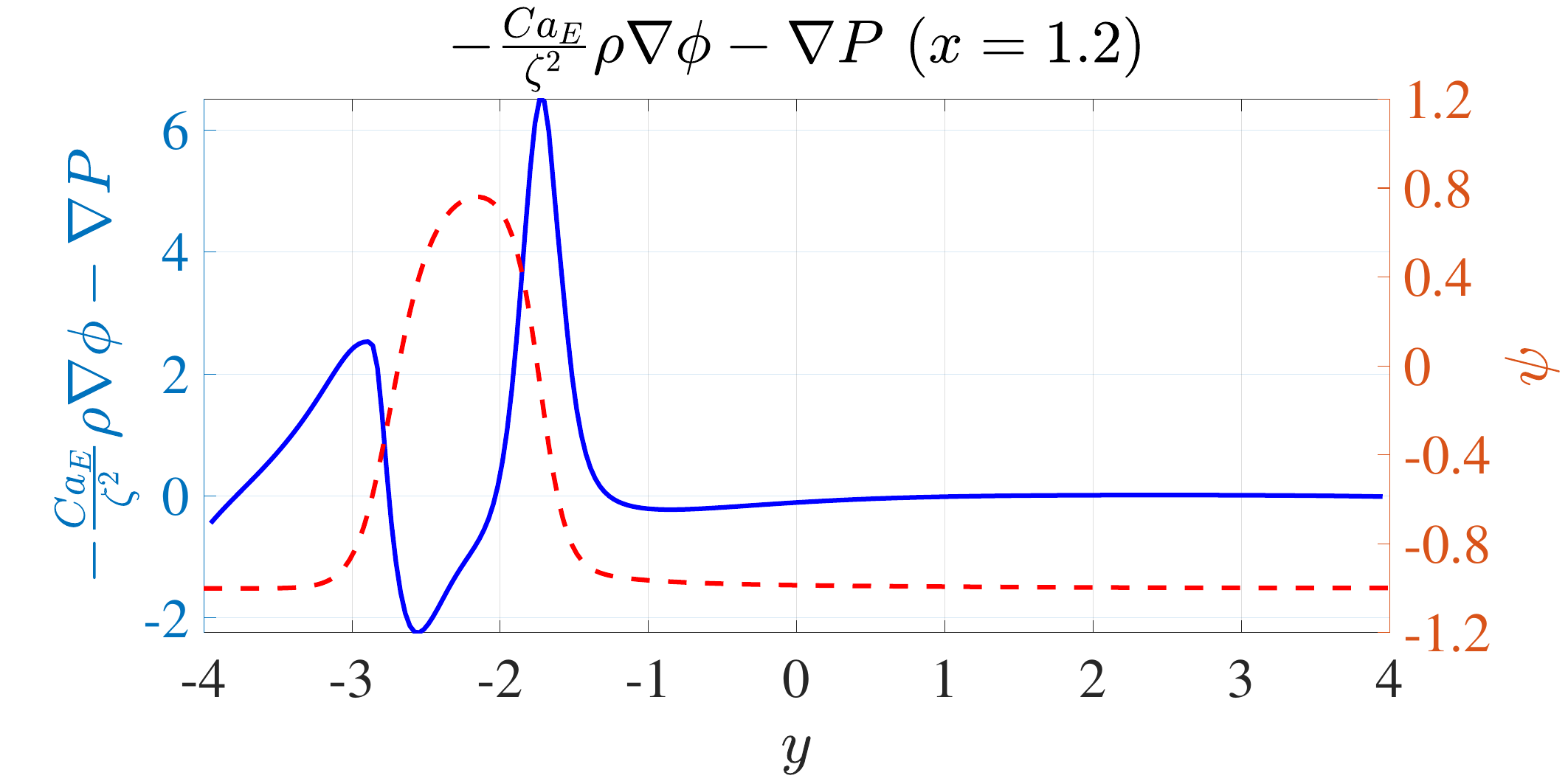}
        \label{subfig:1DropD4N0Pump25ep5x1d2}
		}
		\\
	\subfloat[$-\frac{Ca_{E}}{\zeta^{2}}\rho\nabla\phi-\nabla P~(x=-2.4,t=8)$]{
		\includegraphics[width=0.33\linewidth]{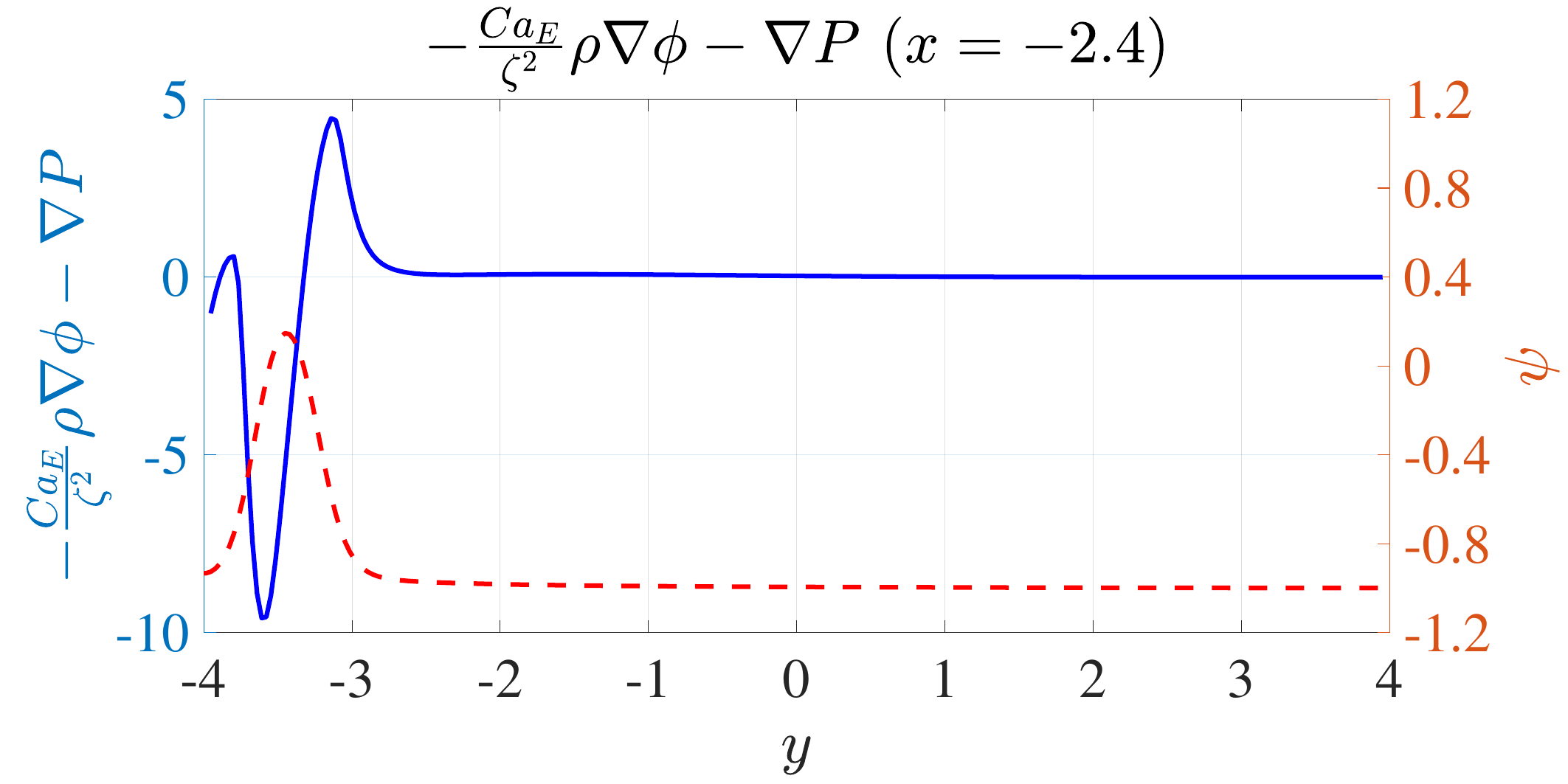}
        \label{subfig:1DropD4N0Pump25ep8x-2d4}
		} 
	\subfloat[$-\frac{Ca_{E}}{\zeta^{2}}\rho\nabla\phi-\nabla P~(x=0,t=8)$]{
		\includegraphics[width=0.33\linewidth]{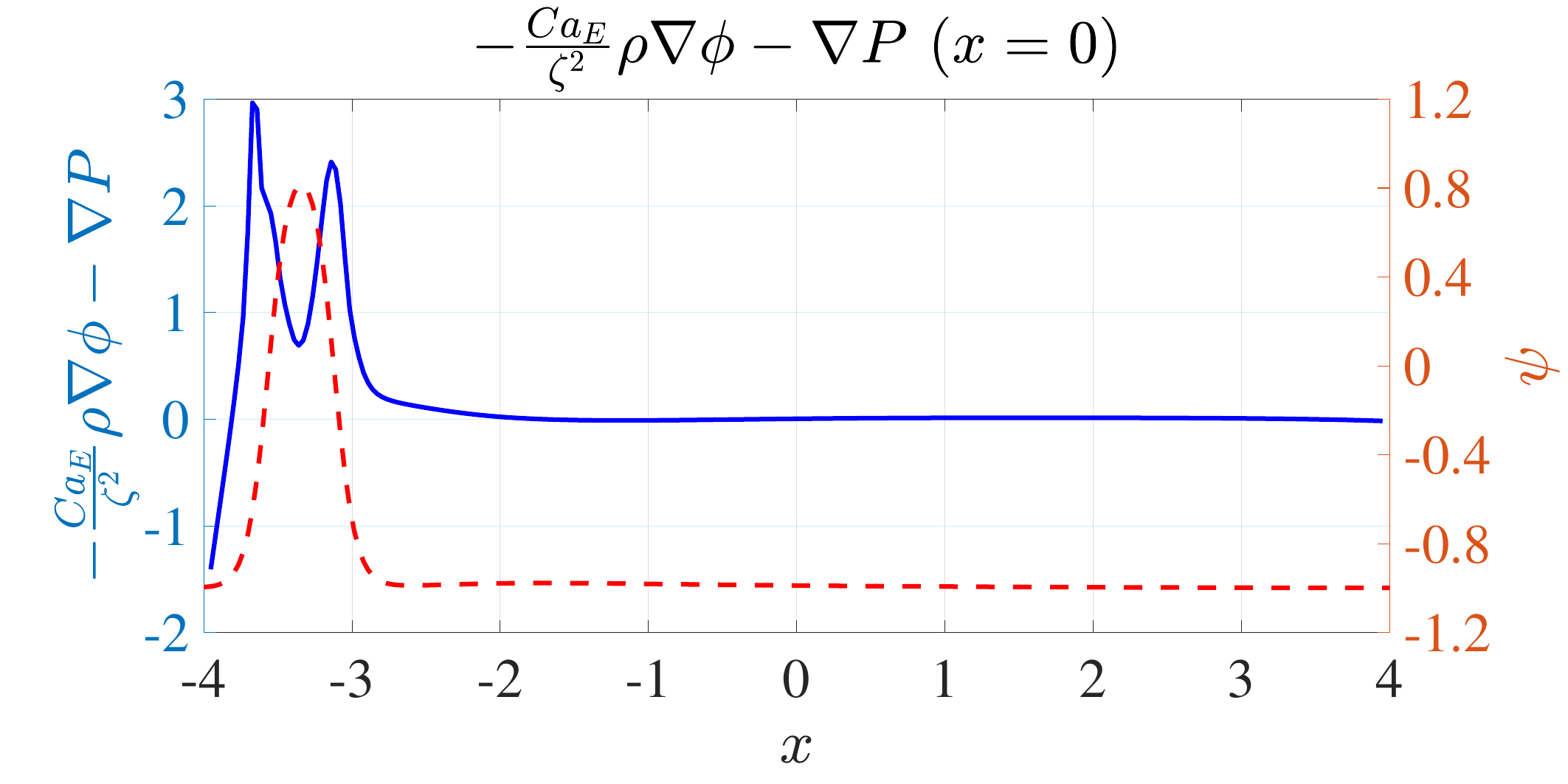}
        \label{subfig:1DropD4N0Pump25ep8x0}
		} 
	\subfloat[$-\frac{Ca_{E}}{\zeta^{2}}\rho\nabla\phi-\nabla P~(x=2.4,t=8)$]{
		\includegraphics[width=0.33\linewidth]{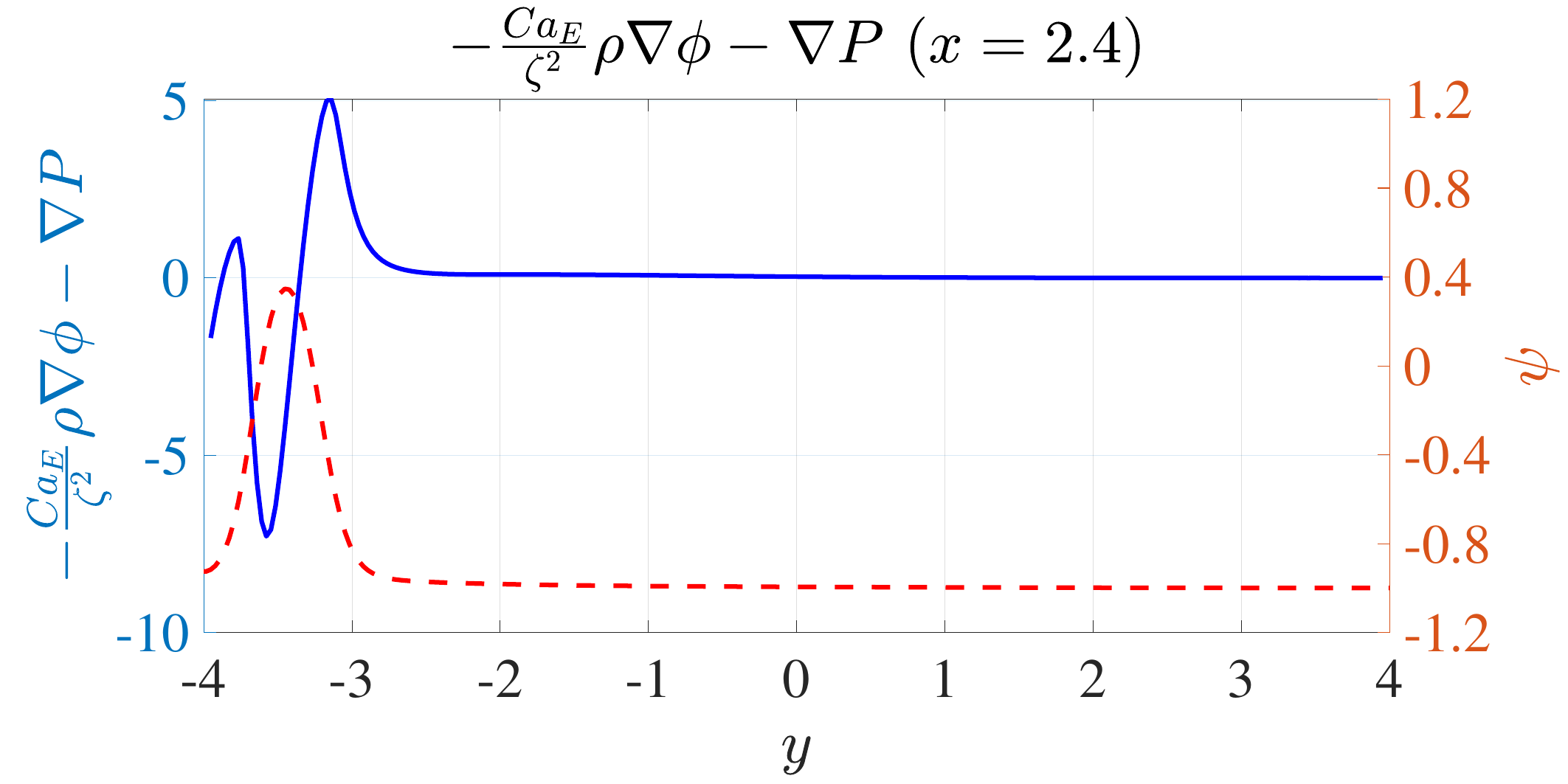}
        \label{subfig:1DropD4N0Pump25ep8x2d4}
		}
	\caption{The snapshots of $-\frac{Ca_{E}}{\zeta^{2}}\rho\nabla\phi-\nabla P$ with positive ion pump at $t=5$ (top) and $t=8$ (bottom) when the vertical electric field is added.
    The blue solid lines show $-\frac{Ca_{E}}{\zeta^{2}}\rho\nabla\phi-\nabla P$ distribution at left end (left), middle (middle) and right end (right) of the drop in the horizontal direction. 
    }\label{fig:1DropD4N0Pump25ep8Section}
\end{figure}

\begin{figure}[!ht]
	\vskip -0.4cm
    \centering 
    \subfloat[$p~(y=0,t=10)$]{
		\includegraphics[width=0.33\linewidth]{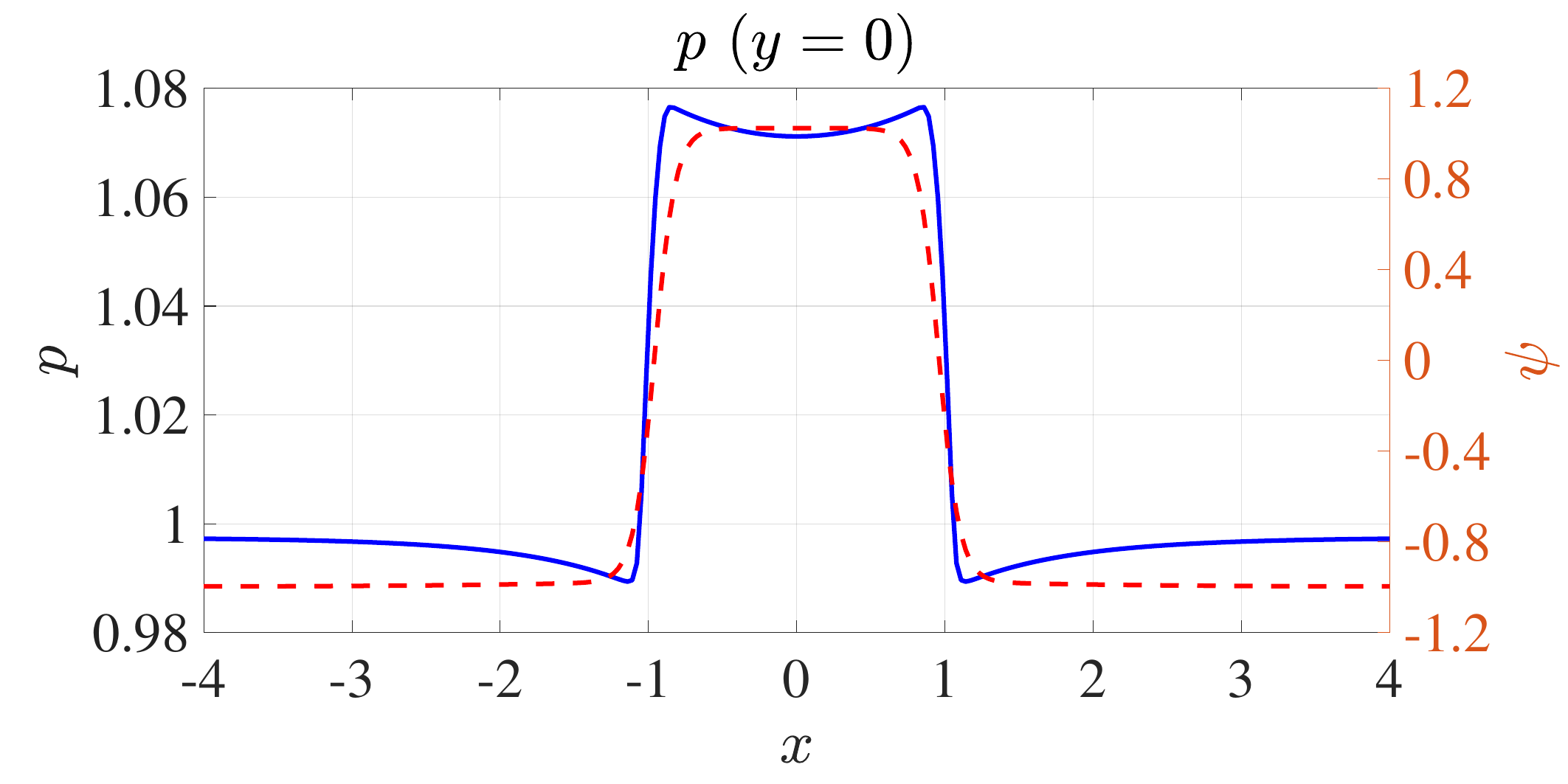}
        \label{subfig:1Drop0bdPump25P10y0}
		}
	\subfloat[$n~(y=0,t=10)$]{
		\includegraphics[width=0.33\linewidth]{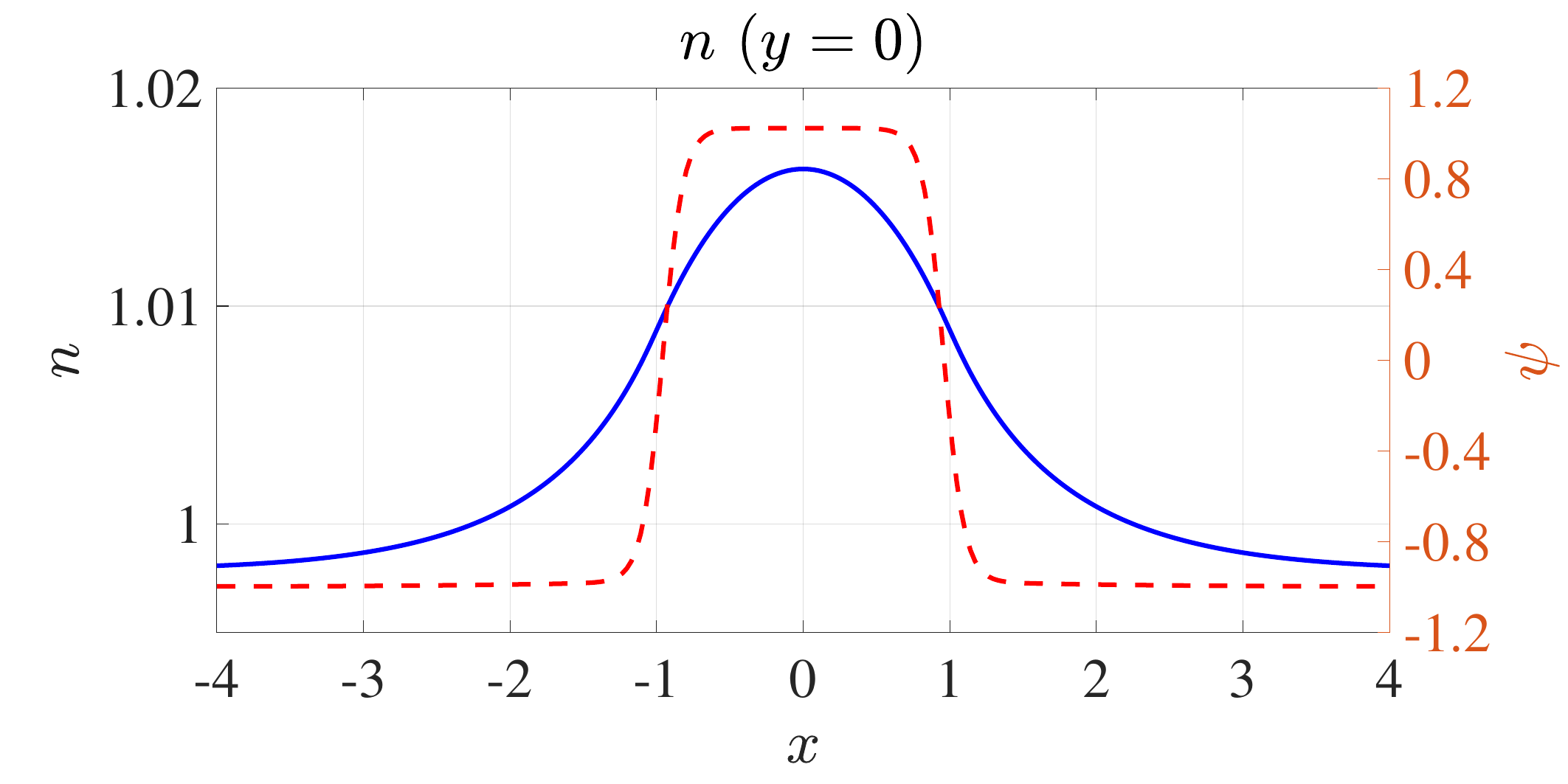}
        \label{subfig:1Drop0bdPump25N10y0}
		}
	\subfloat[$\phi~(y=0,t=10)$]{
		\includegraphics[width=0.33\linewidth]{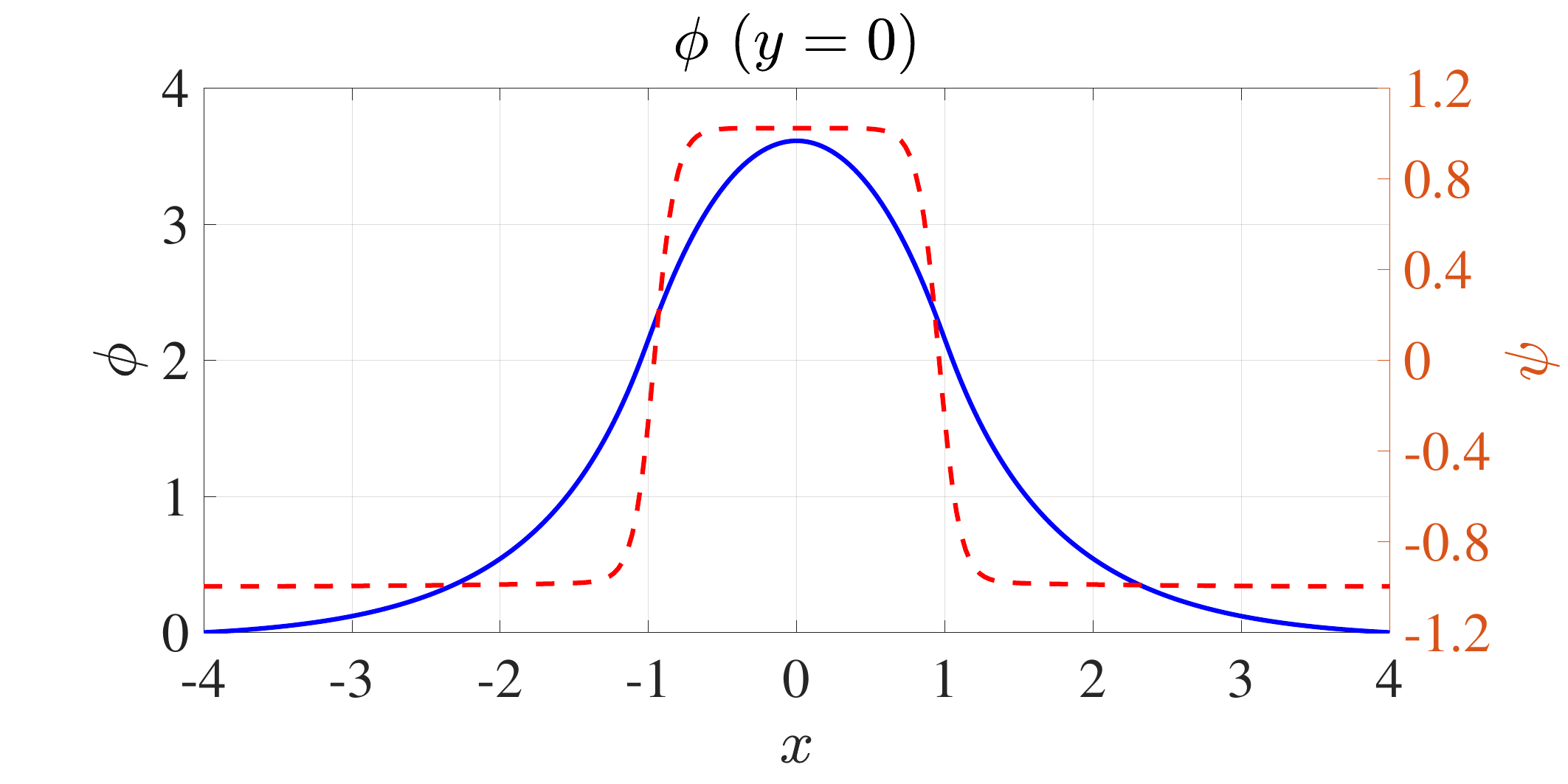}
        \label{subfig:1Drop0bdPump25Phi10y0}
		}
        \\
	\subfloat[$p~(x=0,t=10)$]{
		\includegraphics[width=0.33\linewidth]{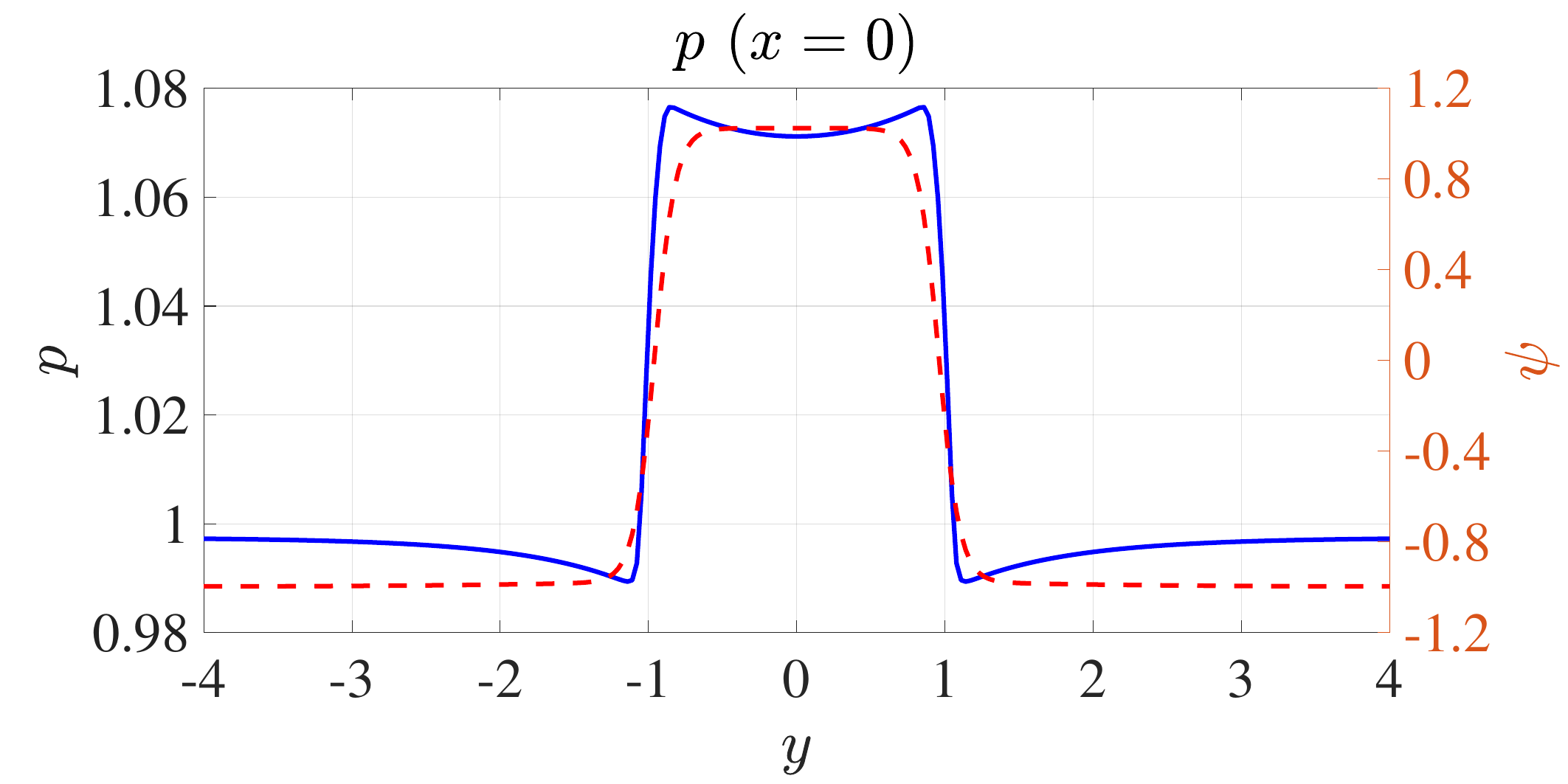}
        \label{subfig:1Drop0bdPump25P10x0}
		}
	\subfloat[$n~(x=0,t=10)$]{
		\includegraphics[width=0.33\linewidth]{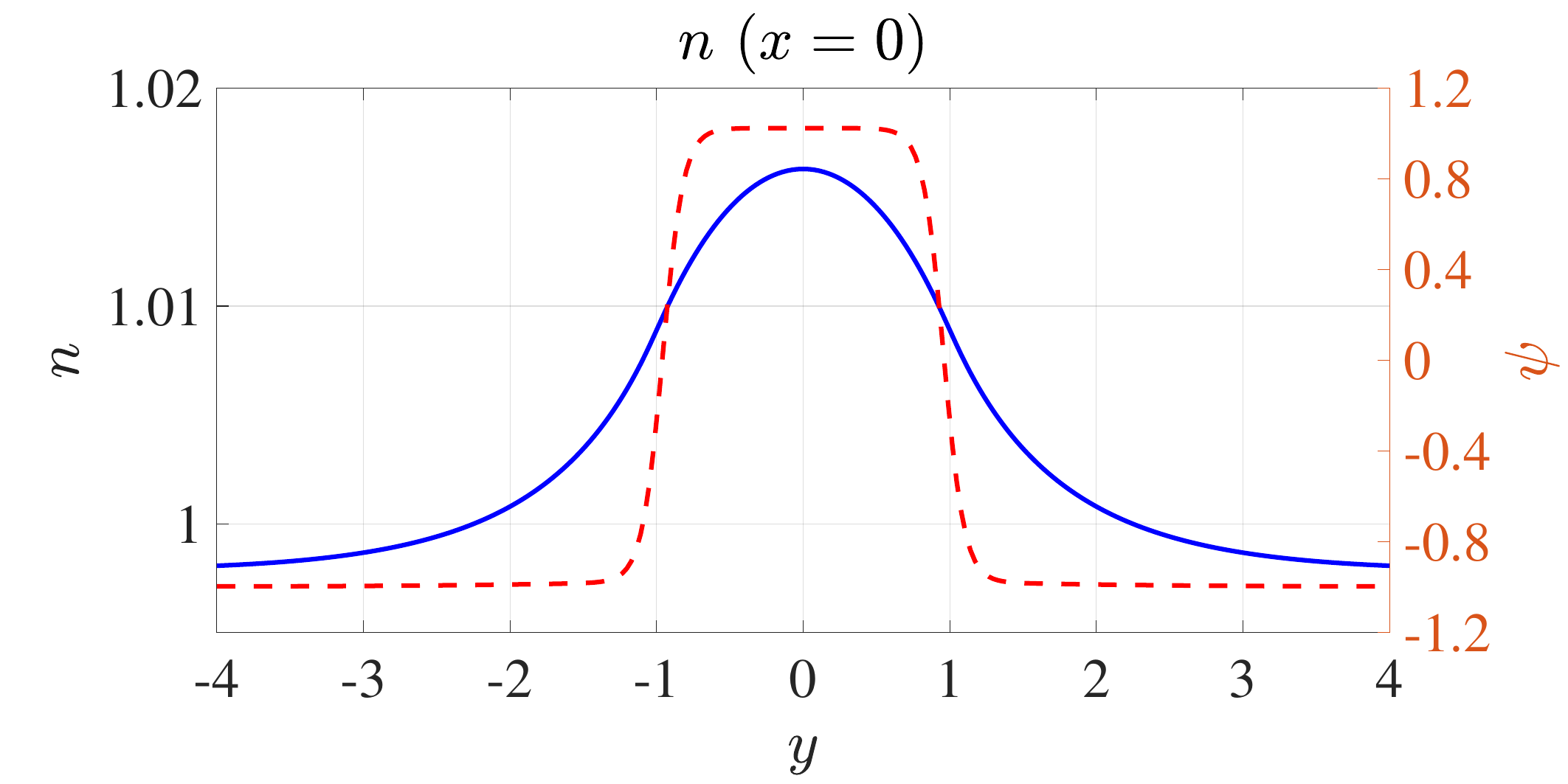}
        \label{subfig:1Drop0bdPump25N10x0}
		}
	\subfloat[$\phi~(x=0,t=10)$]{
		\includegraphics[width=0.33\linewidth]{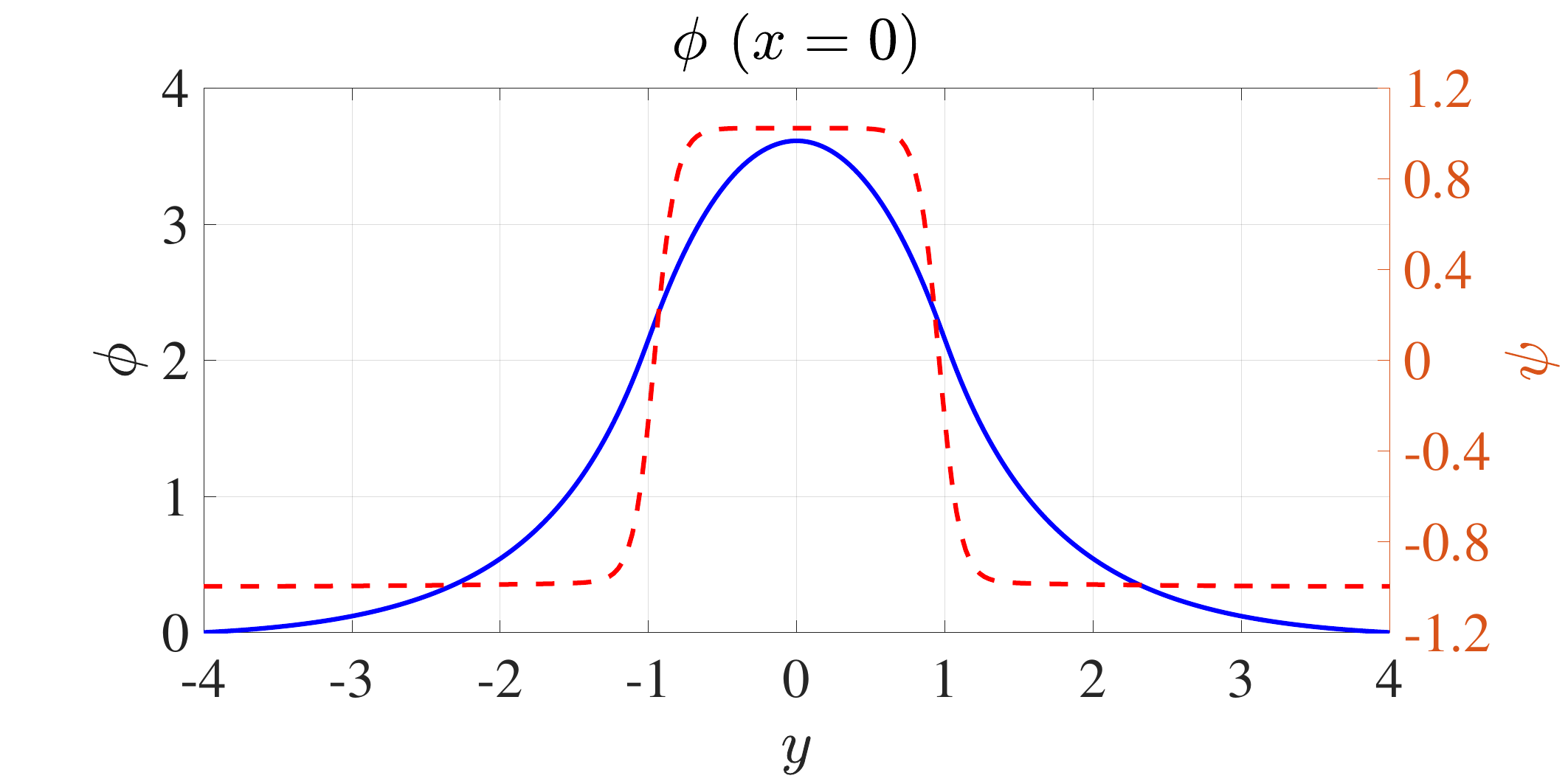}
        \label{subfig:1Drop0bdPump25Phi10x0}
		}
	\caption{The positive and negative ion and the electric potential distribution along $x$- (top) and $y$-axis (bottom) for the uniform pump case shown in Figure \ref{fig:1Drop0bdPump25} at $t=10$, respectively. The blue solid lines show the ion concentration and electric potential and the red dash lines show the label function distribution with the diffusion interface. 
    }\label{fig:1Drop0bdPump25Section}
\end{figure}

\begin{figure}[!ht]
\centering
	\subfloat[$-\frac{Ca_{E}}{\zeta^{2}}\rho\nabla \phi~(y=0,t=10)$]{
		\centering
		\includegraphics[width=0.49\linewidth]{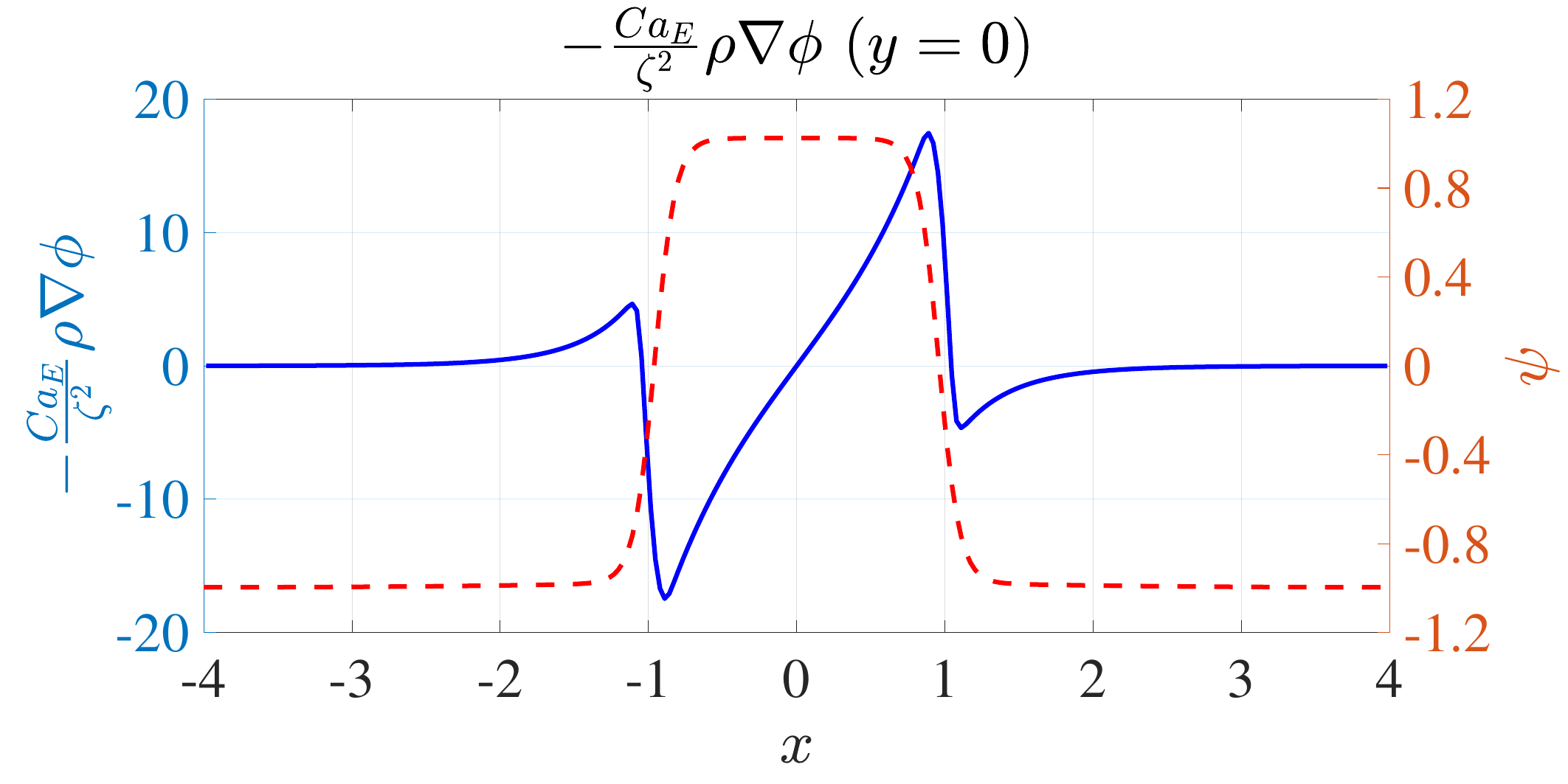}
        \label{subfig:1Drop0bdPump25Eforcey}
	}
	\subfloat[$-\frac{Ca_{E}}{\zeta^{2}}\rho\nabla \phi~(x=0,t=10)$]{
		\centering
		\includegraphics[width=0.49\linewidth]{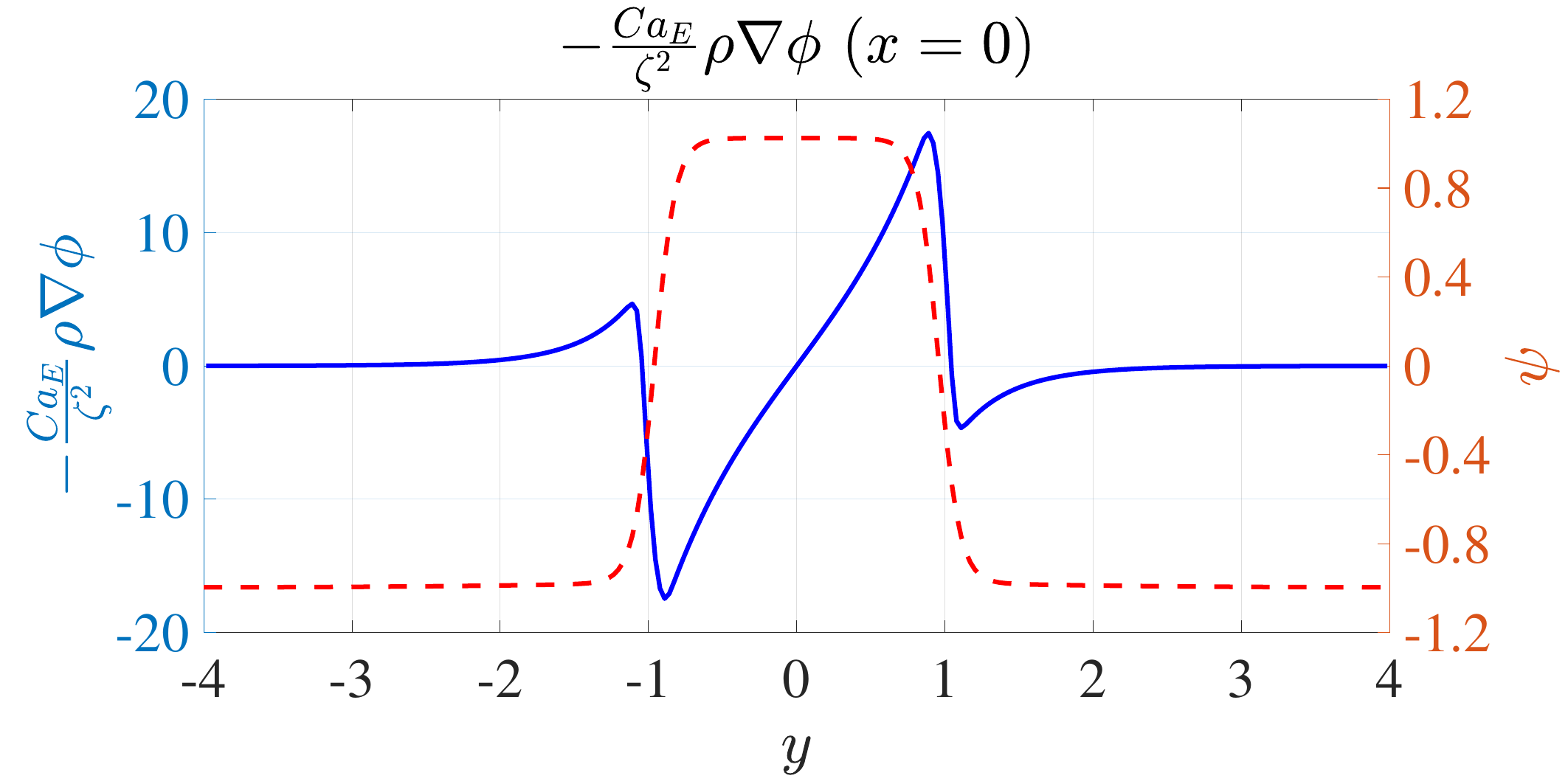}
        \label{subfig:1Drop0bdPump25Eforcex}
		} 
	\caption{The Lorentz force along $x$-axis (left) and $y$-axis (right) induced by the distribution of ions and electric potential for the droplet with uniformly distributed pumps at $t=10$.}\label{fig:1Drop0bdPump25Eforce}
\end{figure}

\begin{figure}[!ht]
	\subfloat[$-\frac{Ca_{E}}{\zeta^{2}}\rho\nabla \phi~(y=0,t=10)$]{
		\centering
		\includegraphics[width=0.49\linewidth]{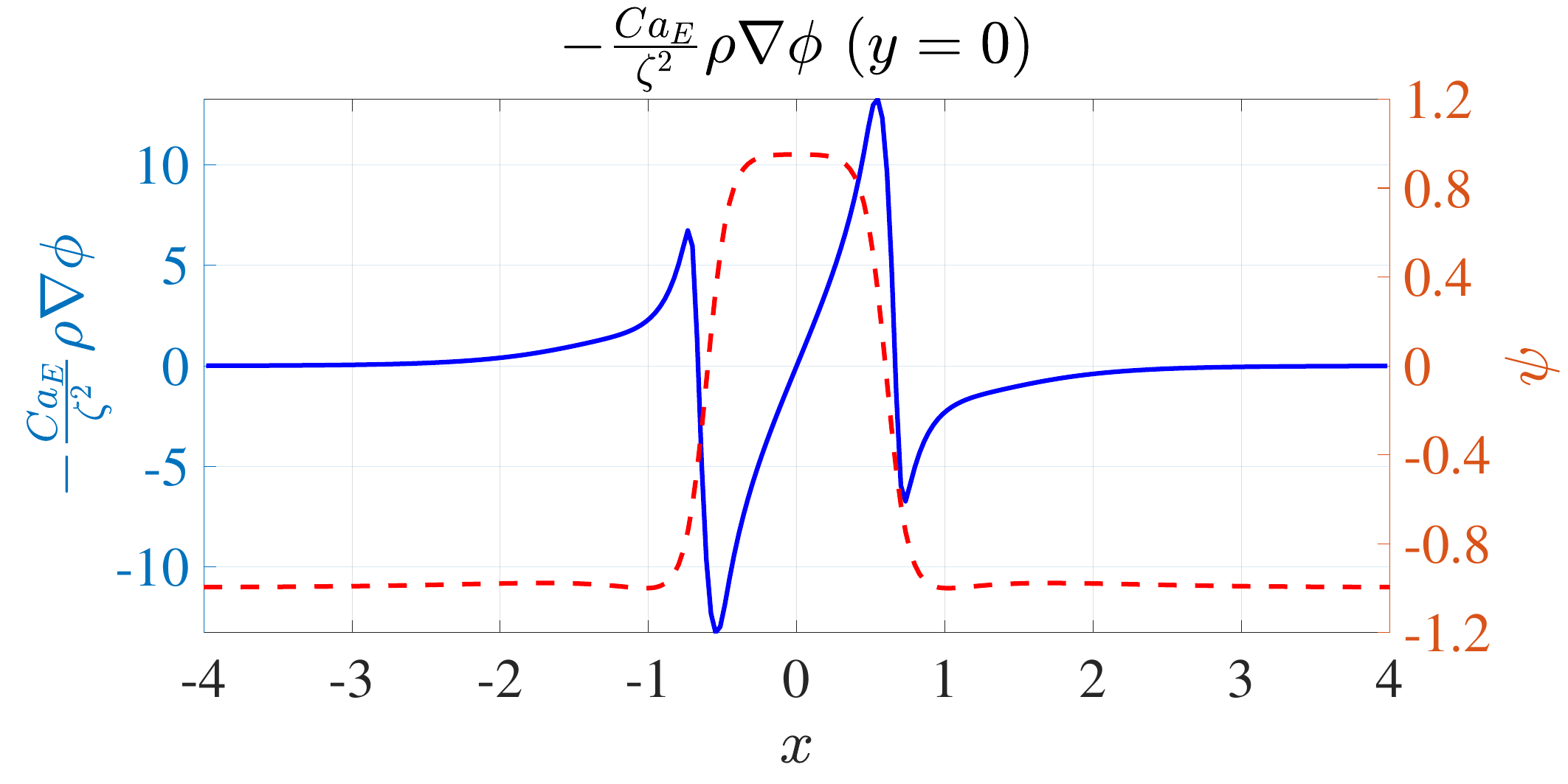}
        \label{subfig:1Drop0bdNonuPump25Eforcey}
	}
	\subfloat[$-\frac{Ca_{E}}{\zeta^{2}}\rho\nabla \phi~(x=0,t=10)$]{
		\centering
		\includegraphics[width=0.49\linewidth]{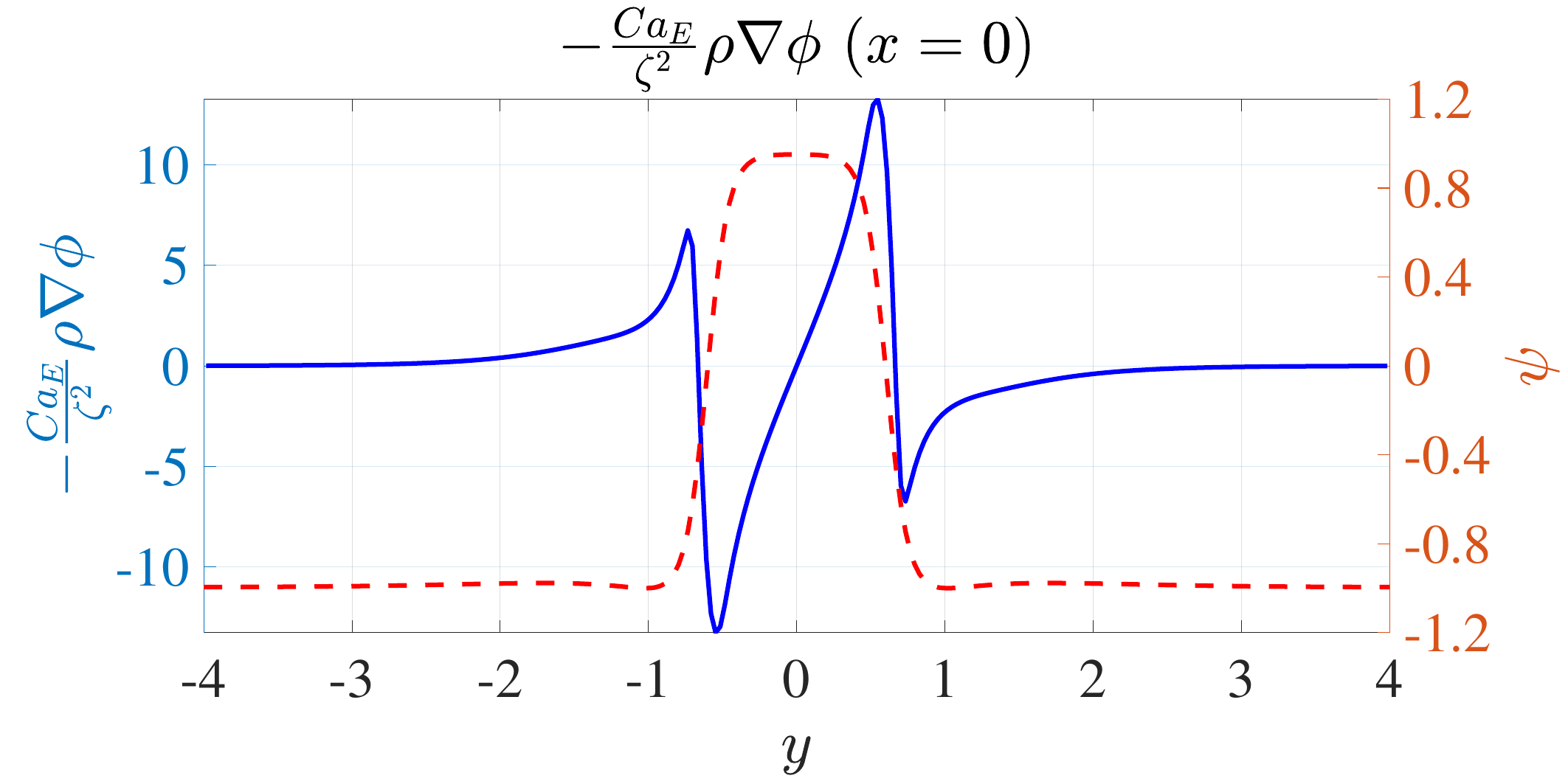}
        \label{subfig:1Drop0bdNonuPump25Eforcex}
		} 
	\caption{The Lorentz force along $x$-axis (left) and $y$-axis (right) induced by the distribution of ions and electric potential for the droplet with nonuniformly distributed pumps at $t=10$.}\label{fig:1Drop0bdNonuPump25Eforce}
\end{figure}

\begin{figure}
\hskip -0.5cm
	\begin{minipage}[b]{0.6\columnwidth}
		\subfloat[Total force at $t=10$.]{\includegraphics[width=0.9\linewidth,height=0.9\linewidth]{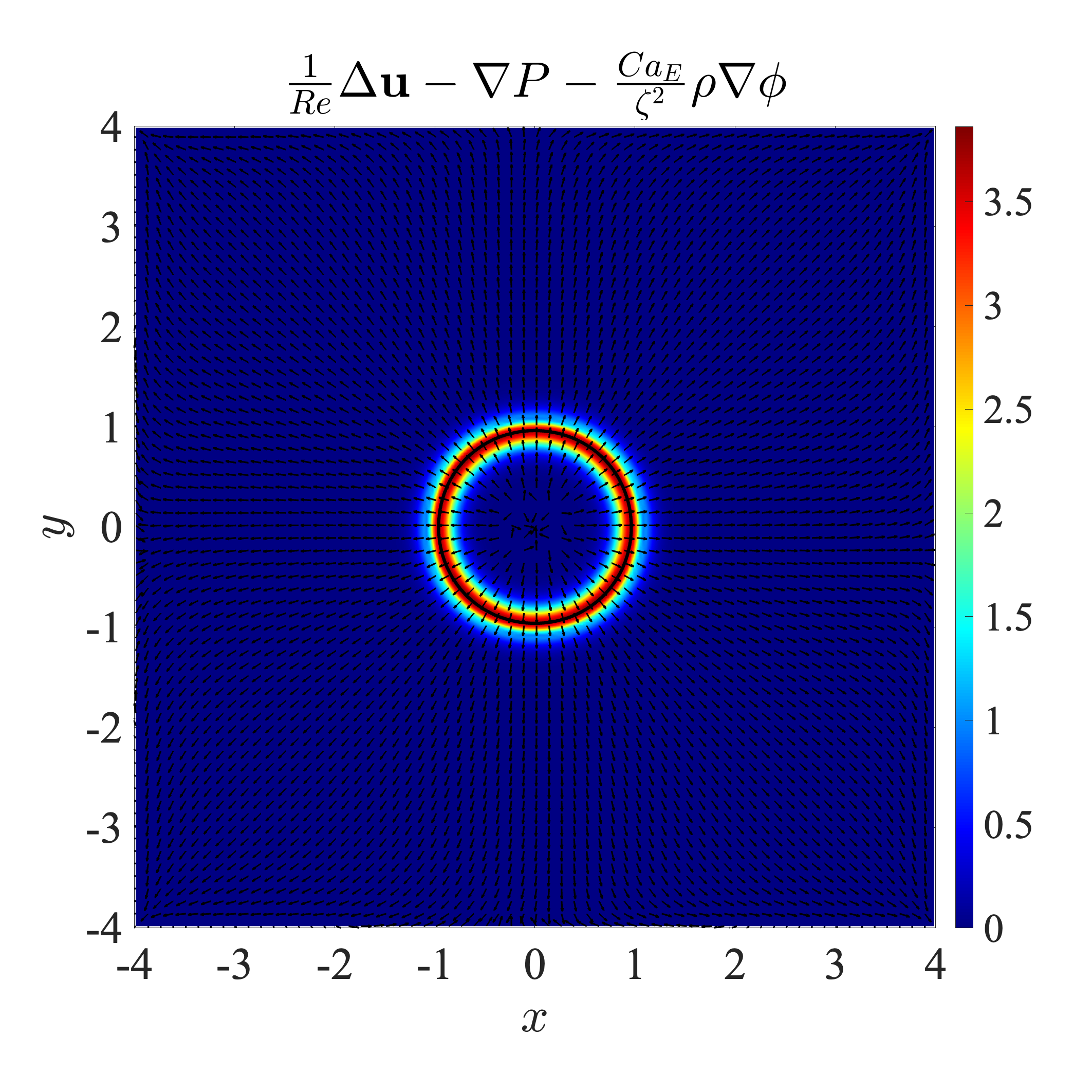}}\label{fig:1Drop0bdPump25Force}
	\end{minipage}
    \hskip -1cm
	\begin{minipage}[b]{0.4\columnwidth}
		\subfloat[Total force along $x$-axis at $t=10$.]{\includegraphics[width=1.2\linewidth,height=0.6\linewidth]{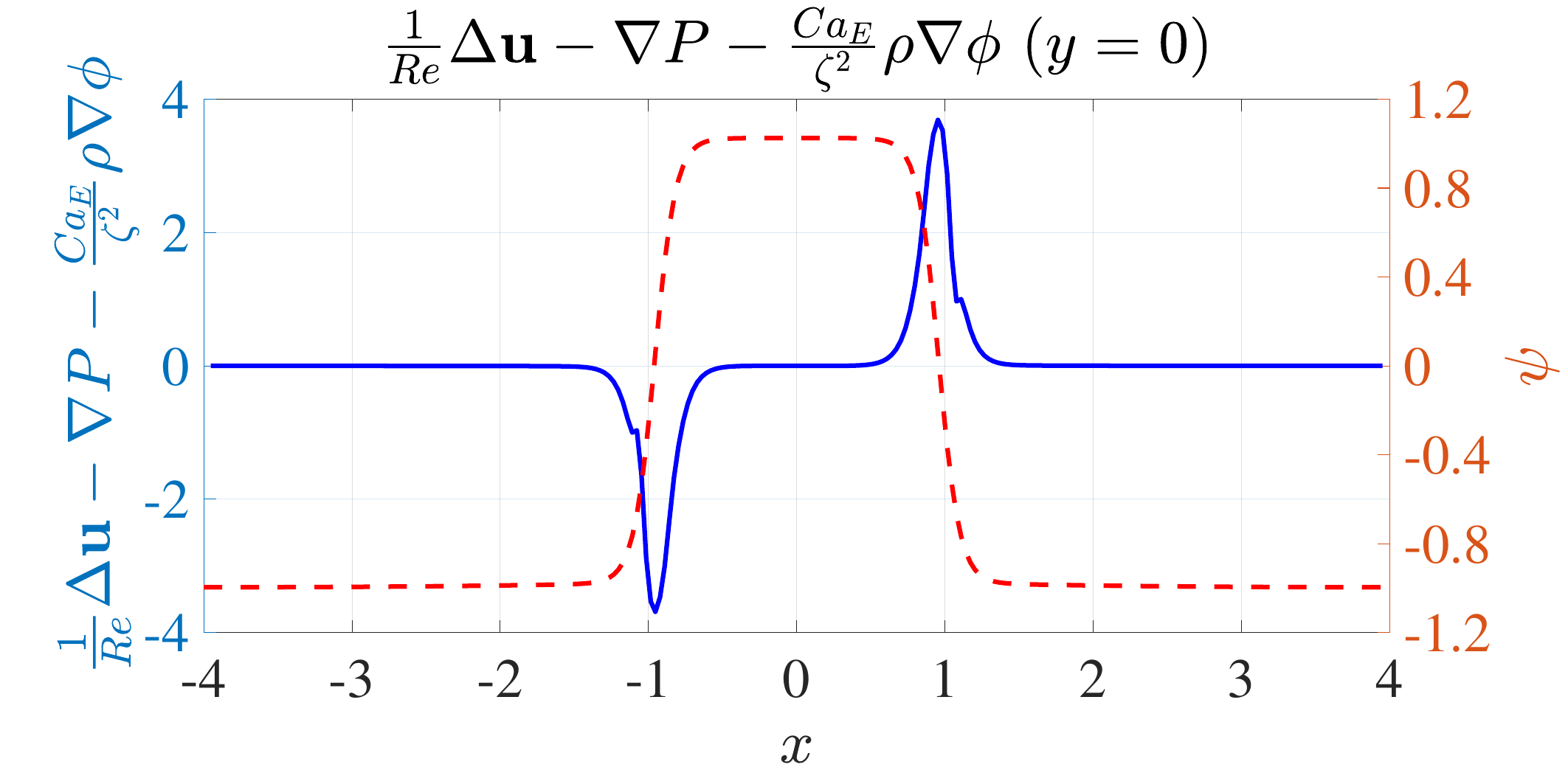}}
        \label{fig:1Drop0bdPump25Forcey=0}
		\\
		\subfloat[Total force along $y$-axis at $t=10$.]{\includegraphics[width=1.2\linewidth,height=0.6\linewidth]{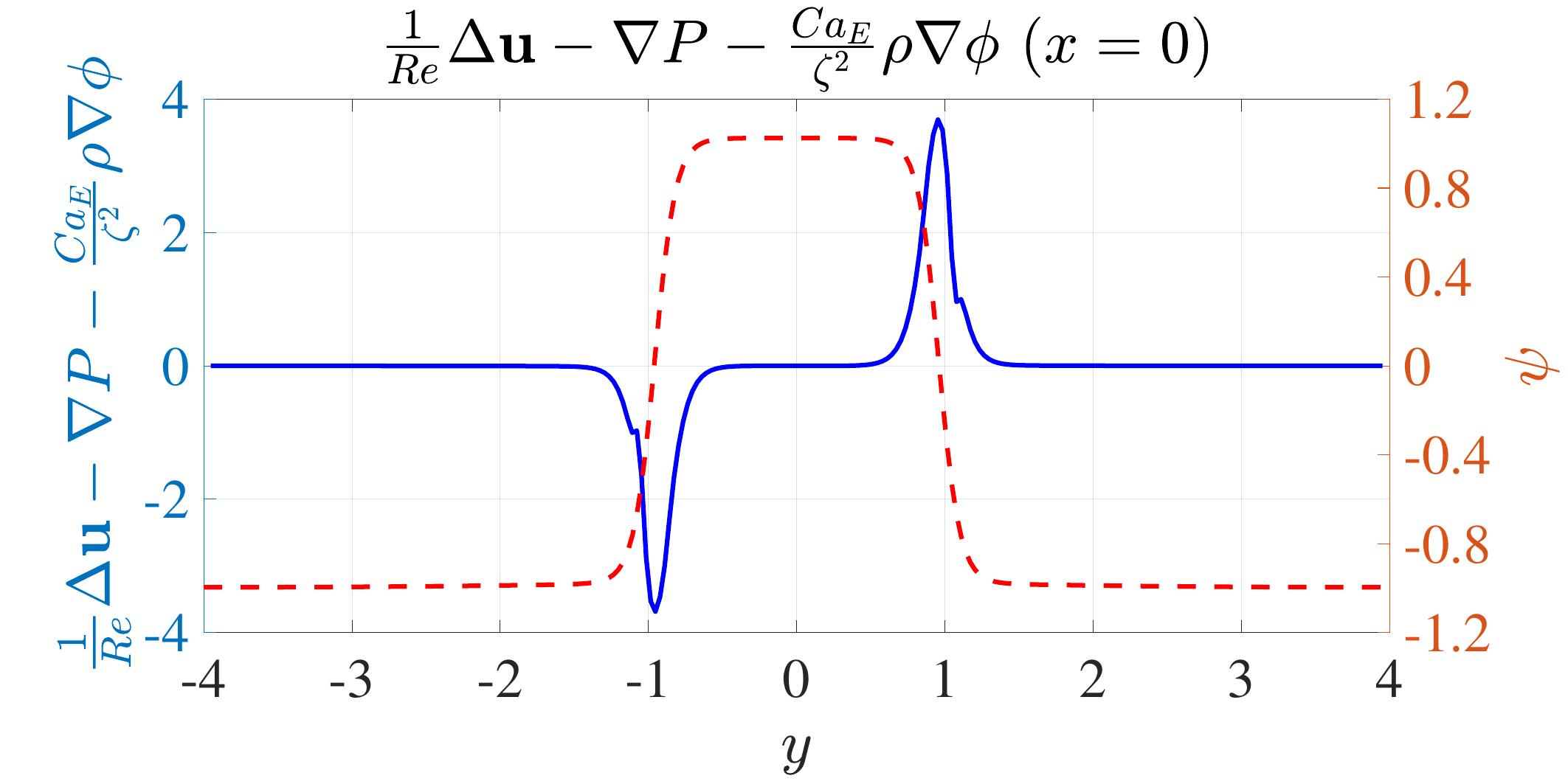}}
        \label{fig:1Drop0bdPump25Forcex=0}
	\end{minipage}
	\caption{Total Viscous stress induced force and Lorentz force for the droplet with uniformly distributed pumps. (a) 2D map; (b) Distribution along  $y=0$ line; (c) Distribution along  $x=0$ line.}\label{fig:1Drop0bdPumpstess}
\end{figure}

\begin{figure}
\hskip -0.5cm
	\begin{minipage}[b]{0.6\columnwidth}
		\subfloat[Total force at $t=10$.]{\includegraphics[width=0.9\linewidth,height=0.9\linewidth]{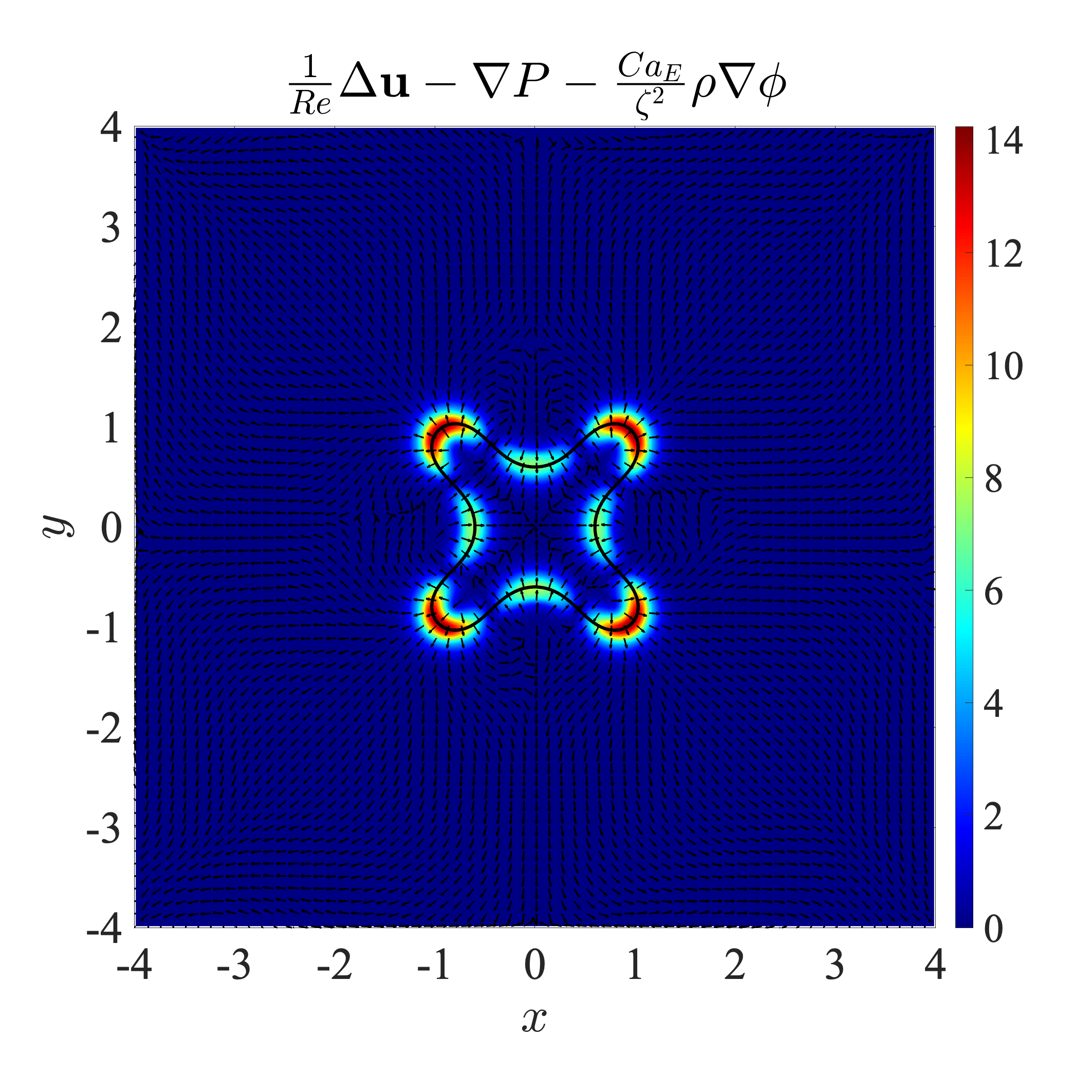}}\label{fig:1Drop0bdNonuPump25Force}
	\end{minipage}
    \hskip -1cm
	\begin{minipage}[b]{0.4\columnwidth}
		\subfloat[Total force along $x$-axis at $t=10$.]{\includegraphics[width=1.2\linewidth,height=0.6\linewidth]{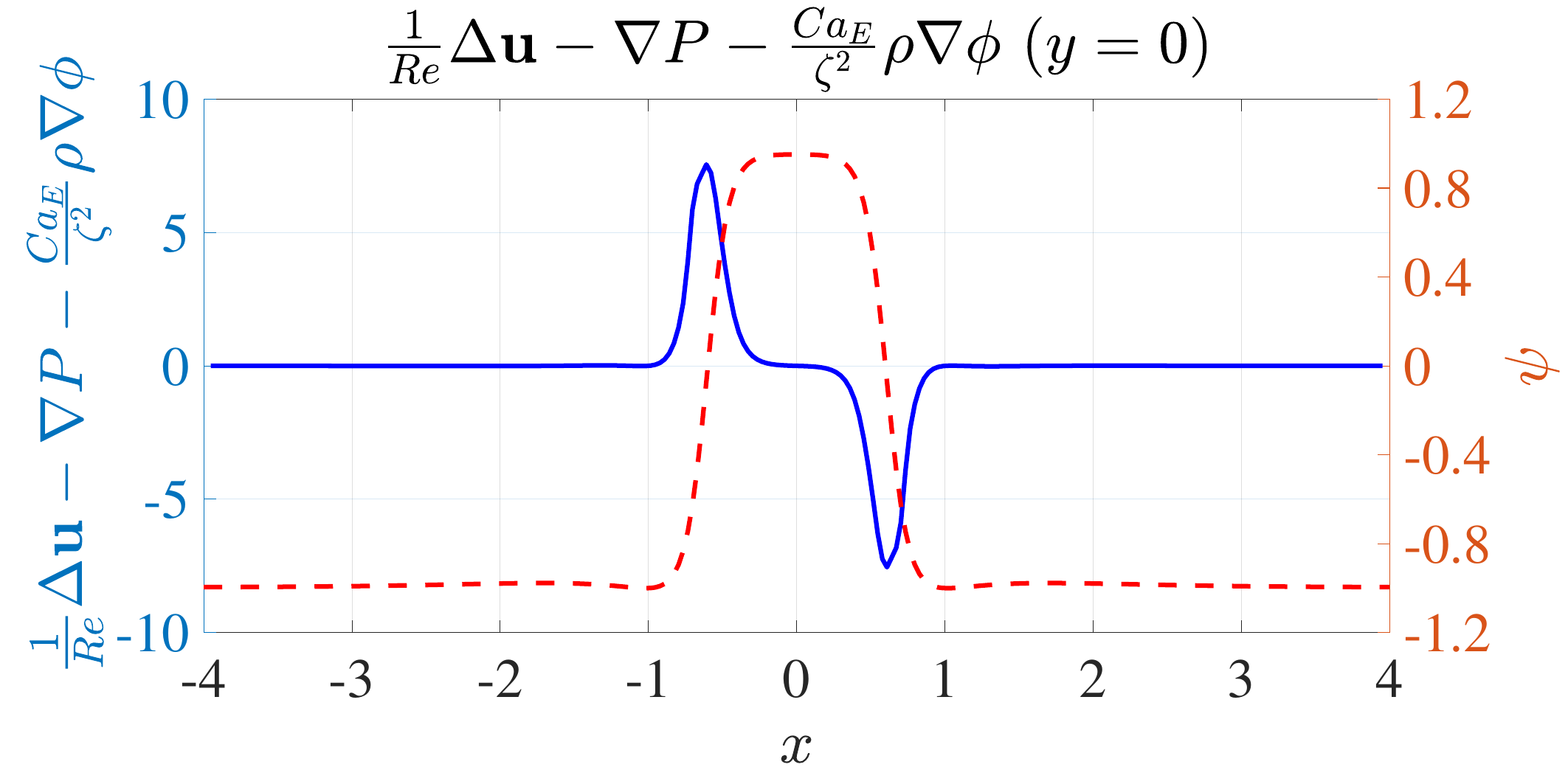}}
        \label{fig:1Drop0bdNonuPump25Forcey=0}
		\\
		\subfloat[Total force along $y$-axis at $t=10$.]{\includegraphics[width=1.2\linewidth,height=0.6\linewidth]{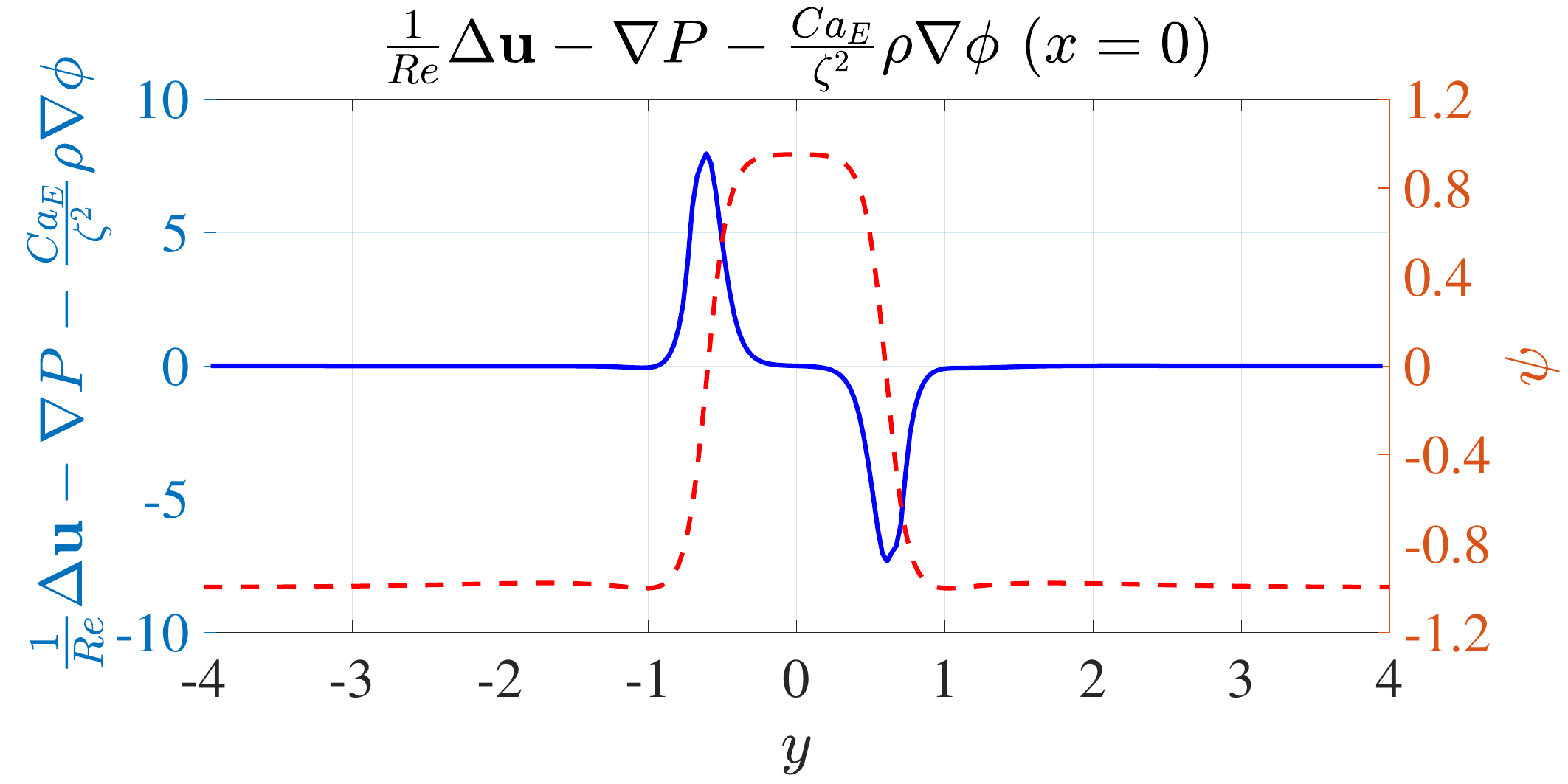}}
        \label{fig:1Drop0bdNonuPump25Forcex=0}
	\end{minipage}
	\caption{Total Viscous stress induced force and Lorentz force for the droplet with nonuniformly distributed pumps \eqref{case:nonuniformpump}. (a) 2D map; (b) Distribution along  $y=0$ line; (c) Distribution along  $x=0$ line. $I_0=25$}\label{fig:1Drop0bdNonuPumpstess}
\end{figure}

\begin{figure}[!ht]
\vskip -0.6cm
\centering 
    \subfloat[$p~(t=3)$]{
		\includegraphics[width=0.25\linewidth]{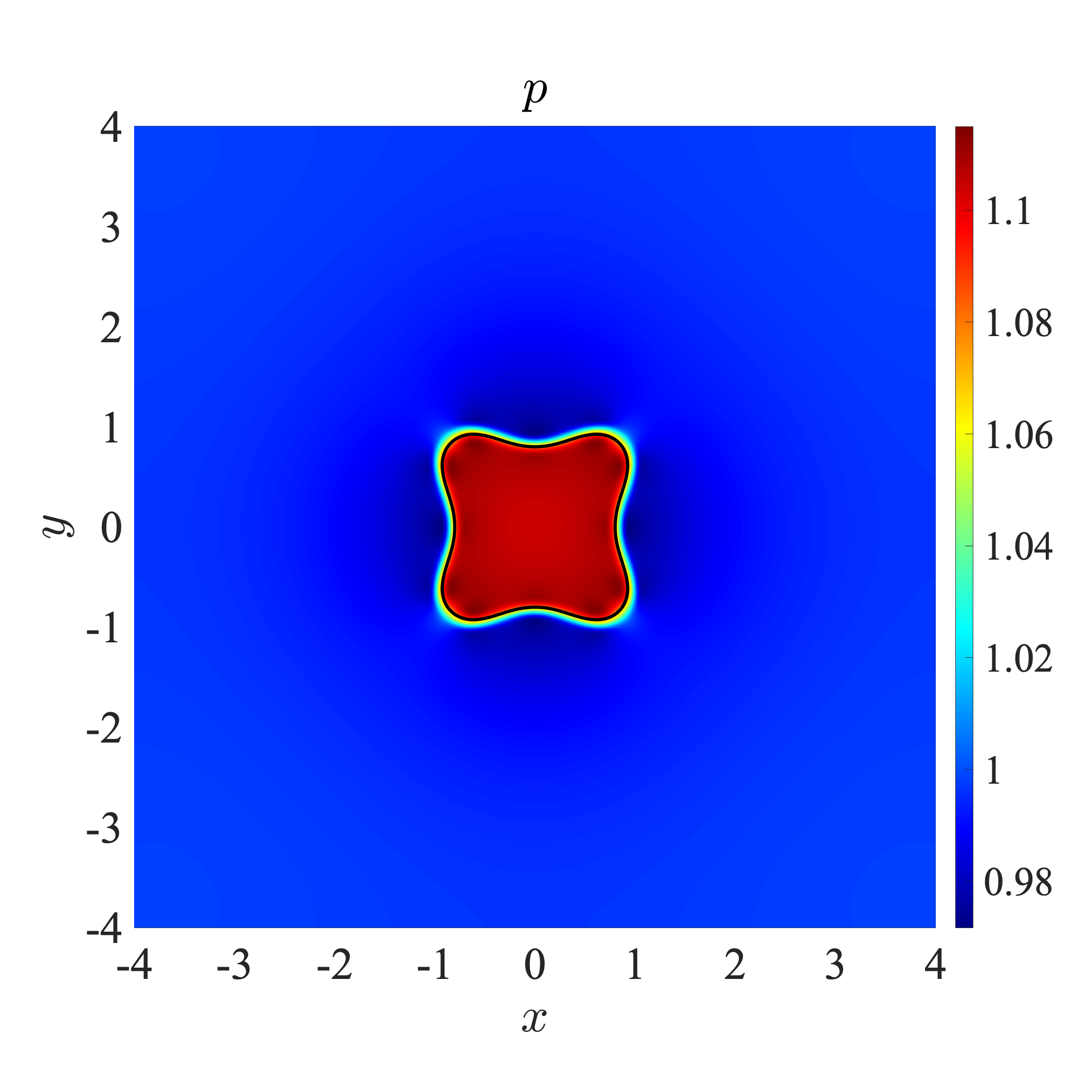}
        \label{subfig:1Drop0bdNonuPumpP3}
		}
    \subfloat[$p~(t=5)$]{
		\includegraphics[width=0.25\linewidth]{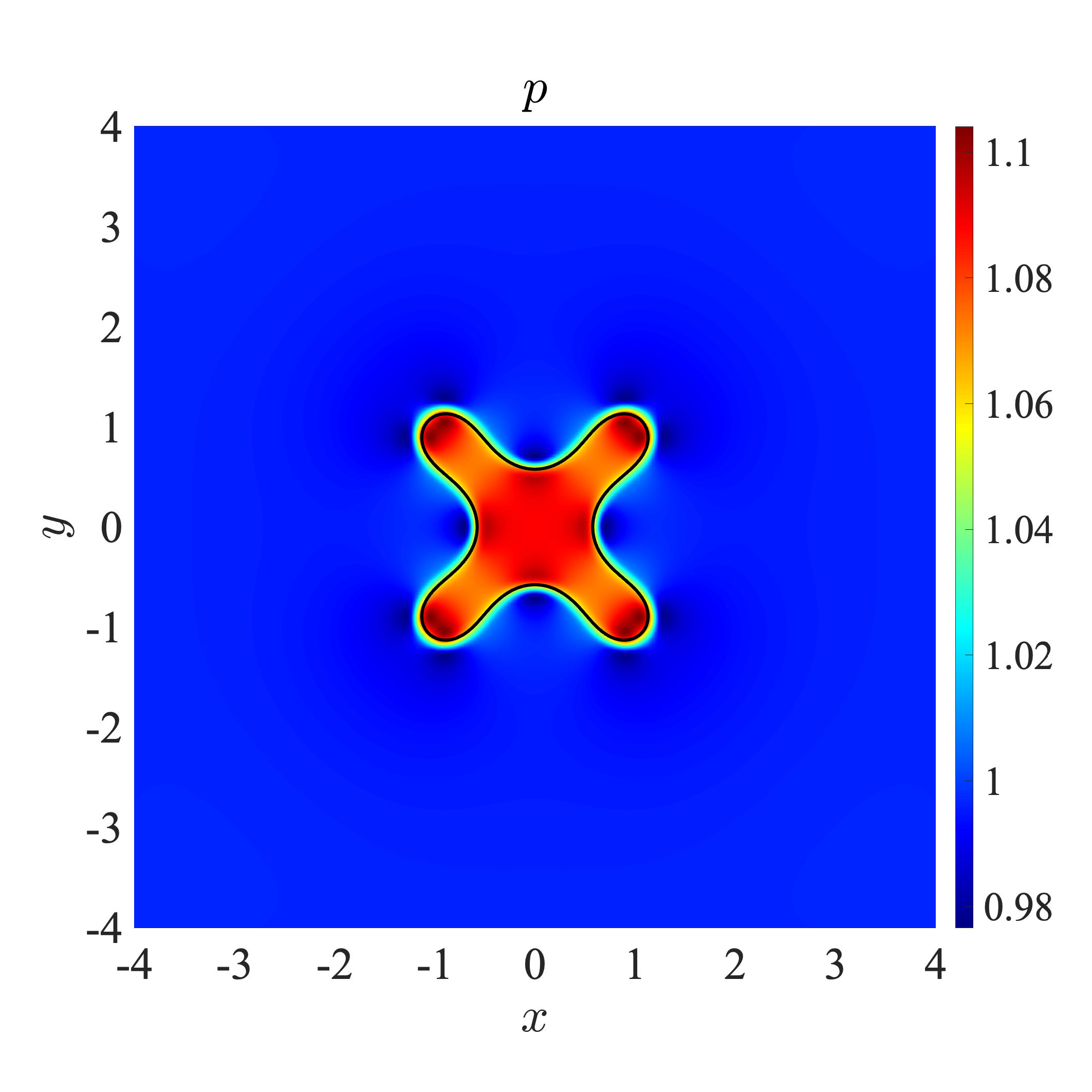}
        \label{subfig:1Drop0bdNonuPumpP5}
		}
    \subfloat[$p~(t=7.4)$]{
		\includegraphics[width=0.25\linewidth]{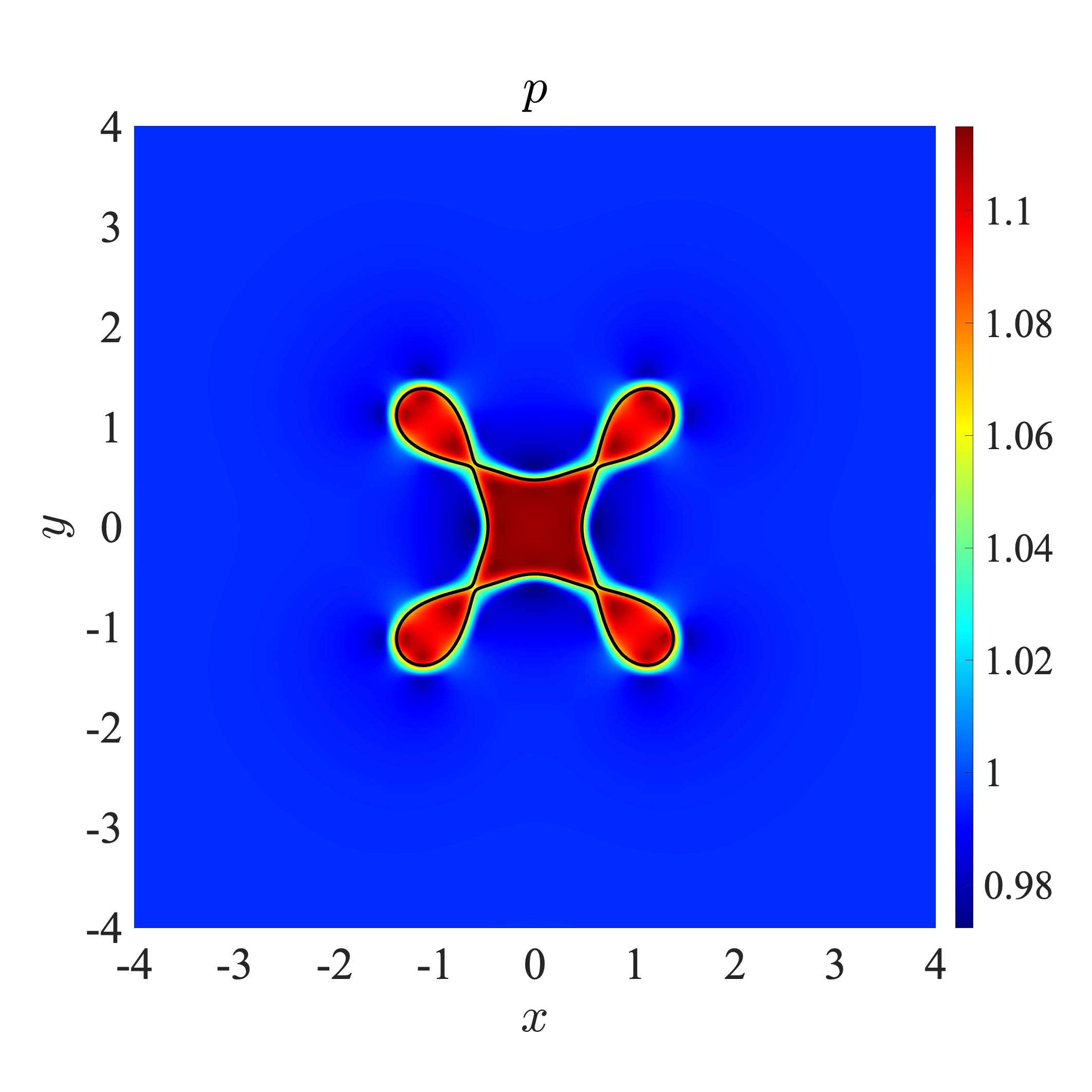}
        \label{subfig:1Drop0bdNonuPumpP7d4}
		}
	\subfloat[$p~(t=10)$]{
		\includegraphics[width=0.25\linewidth]{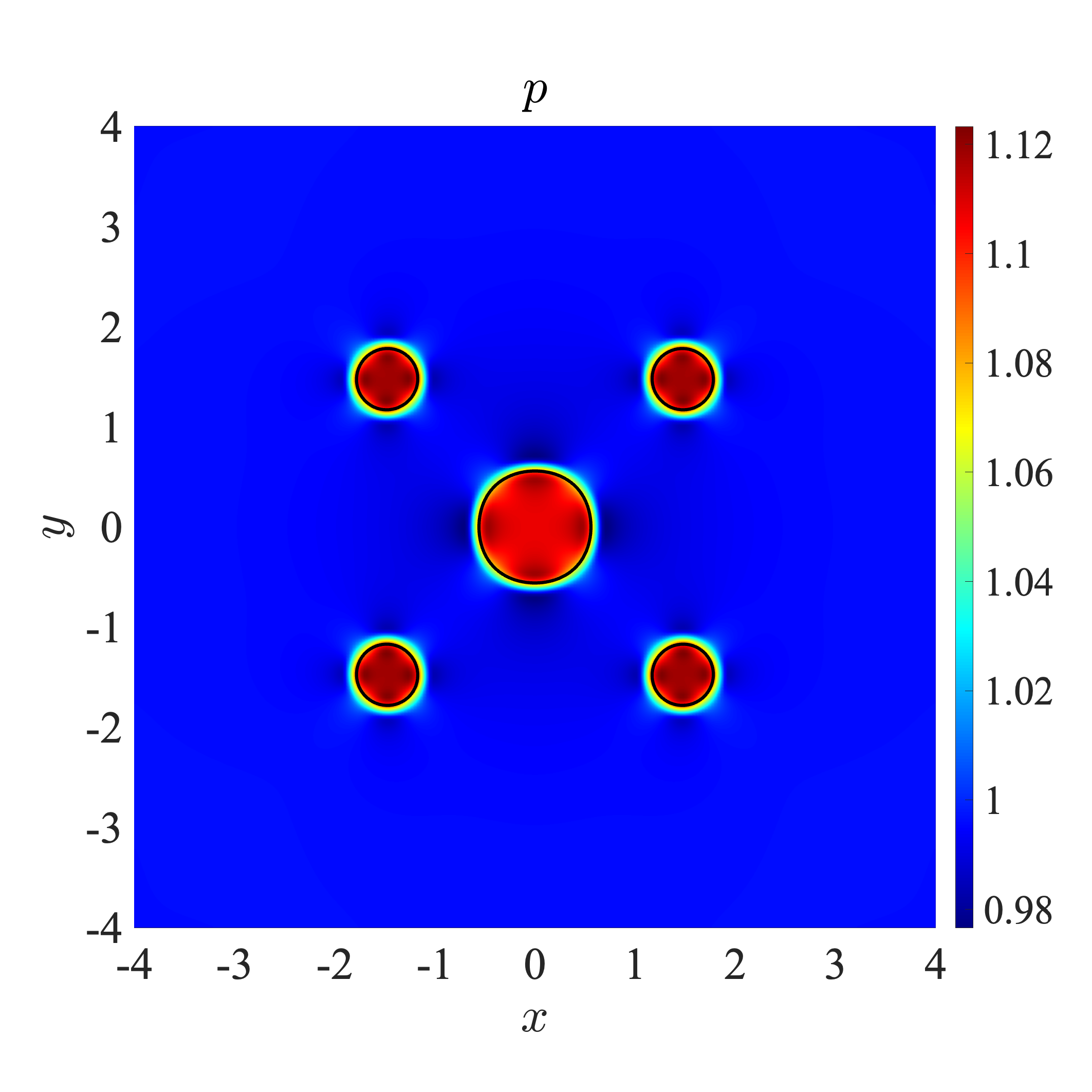}
        \label{subfig:1Drop0bdNonuPumpP10}
		}
    \\
	\subfloat[$n~(t=3)$]{
		\includegraphics[width=0.25\linewidth]{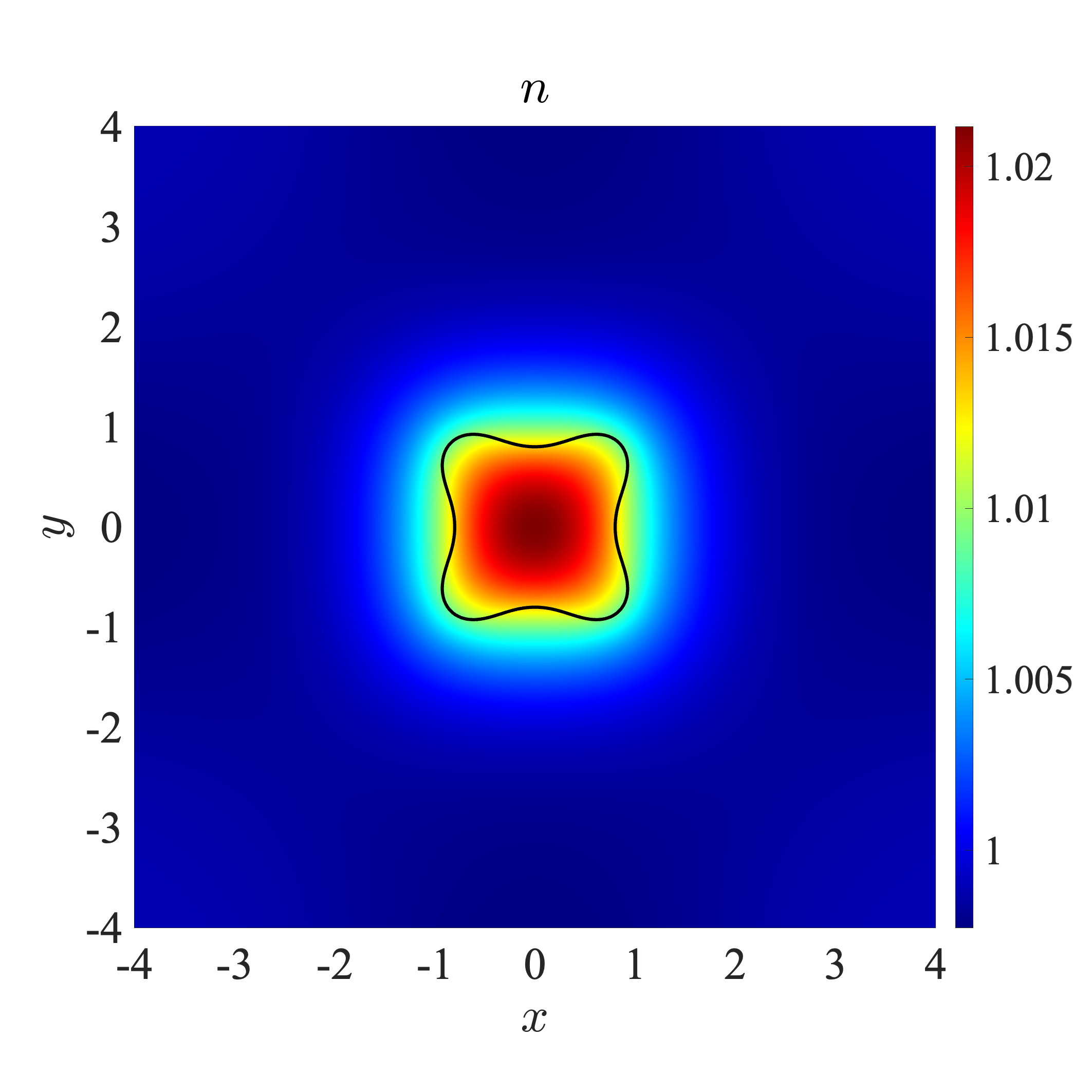}
        \label{subfig:1Drop0bdNonuPumpN3}
		}
	\subfloat[$n~(t=5)$]{
		\includegraphics[width=0.25\linewidth]{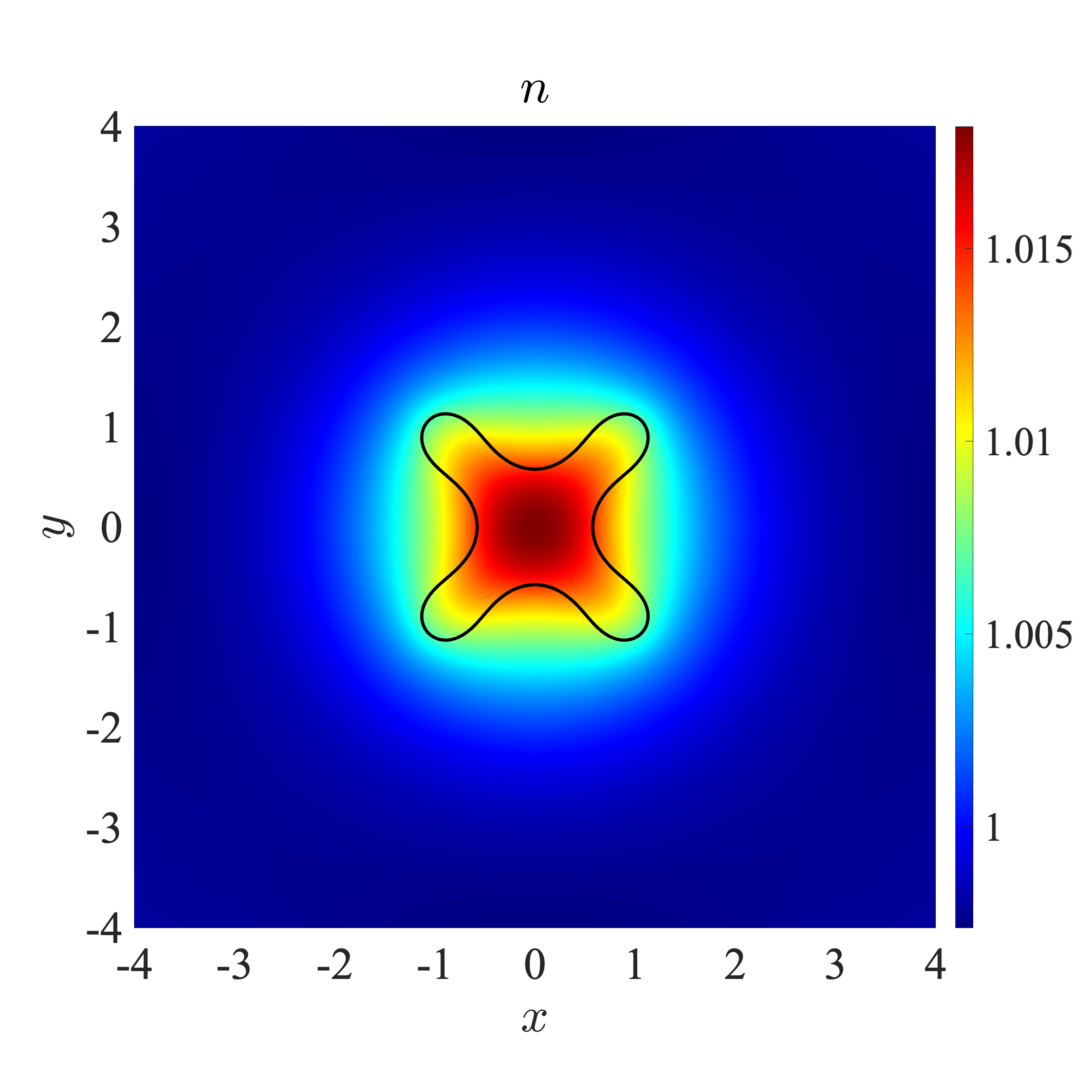}
        \label{subfig:1Drop0bdNonuPumpN5}
		}
	\subfloat[$n~(t=7.4)$]{
		\includegraphics[width=0.25\linewidth]{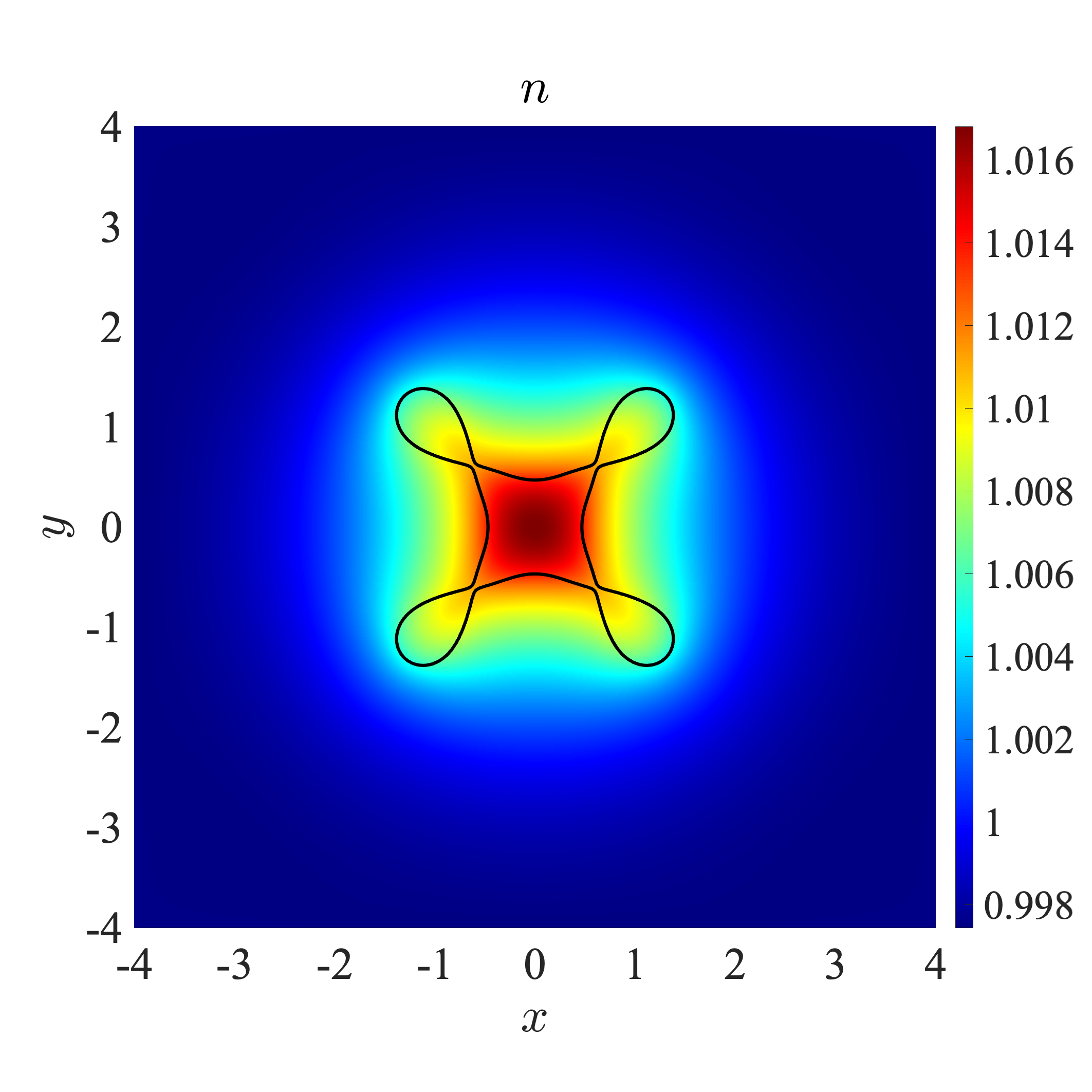}
        \label{subfig:1Drop0bdNonuPumpN7d4}
		}
	\subfloat[$n~(t=10)$]{
		\includegraphics[width=0.25\linewidth]{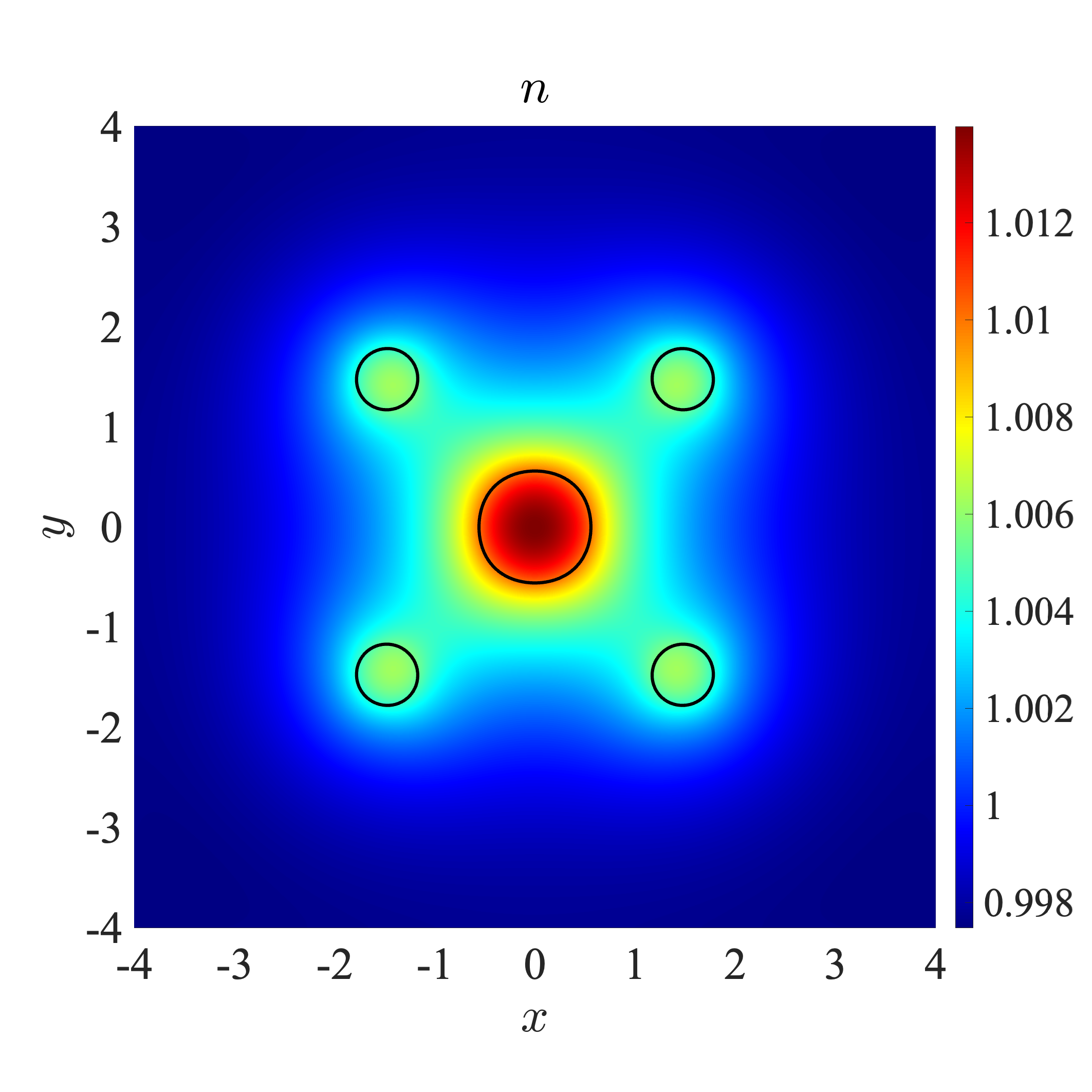}
        \label{subfig:1Drop0bdNonuPumpN10}
		}
    \\
	\subfloat[$\phi~(t=3)$]{
		\includegraphics[width=0.25\linewidth]{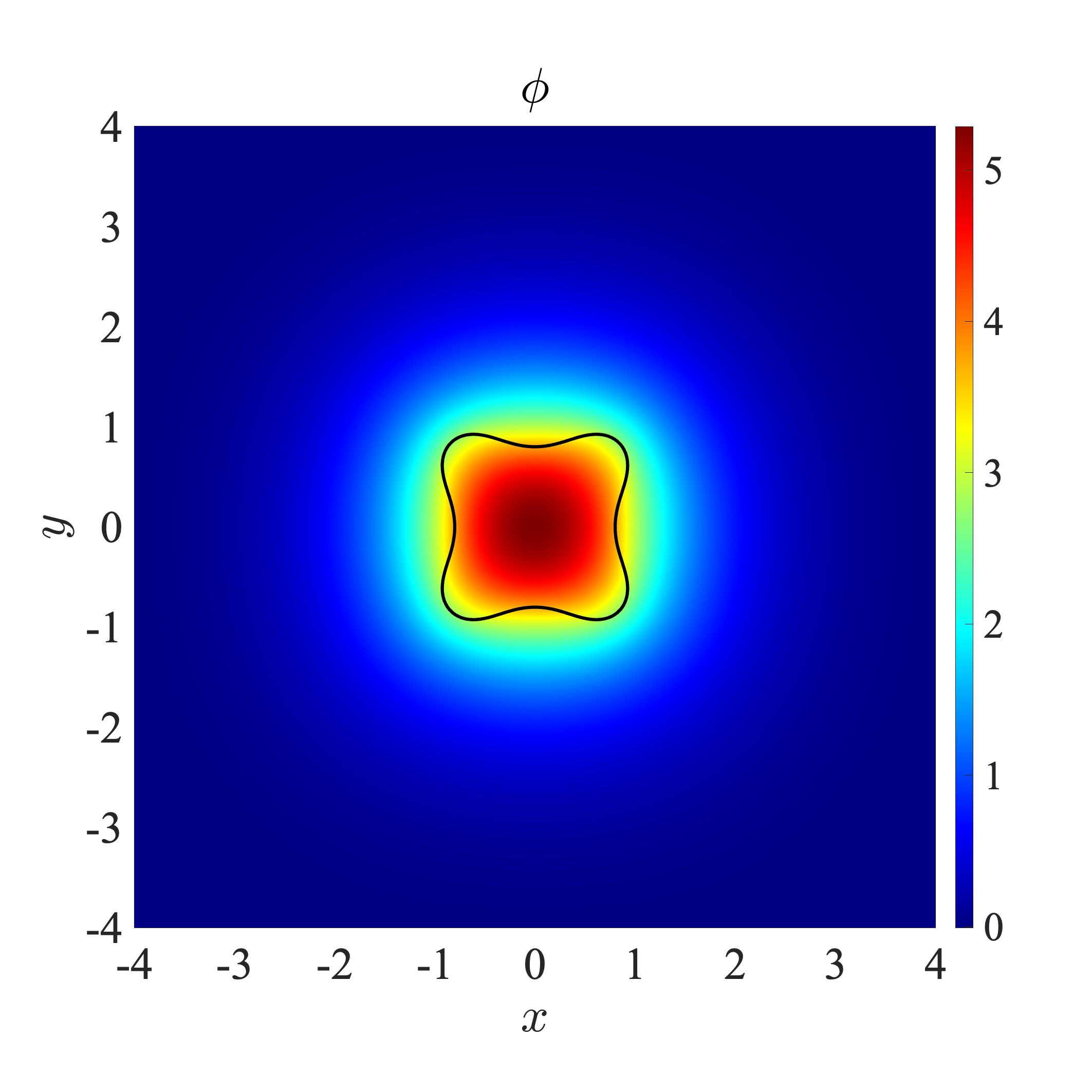}
        \label{subfig:1Drop0bdNonuPumpPhi3}
		}
	\subfloat[$\phi~(t=5)$]{
		\includegraphics[width=0.25\linewidth]{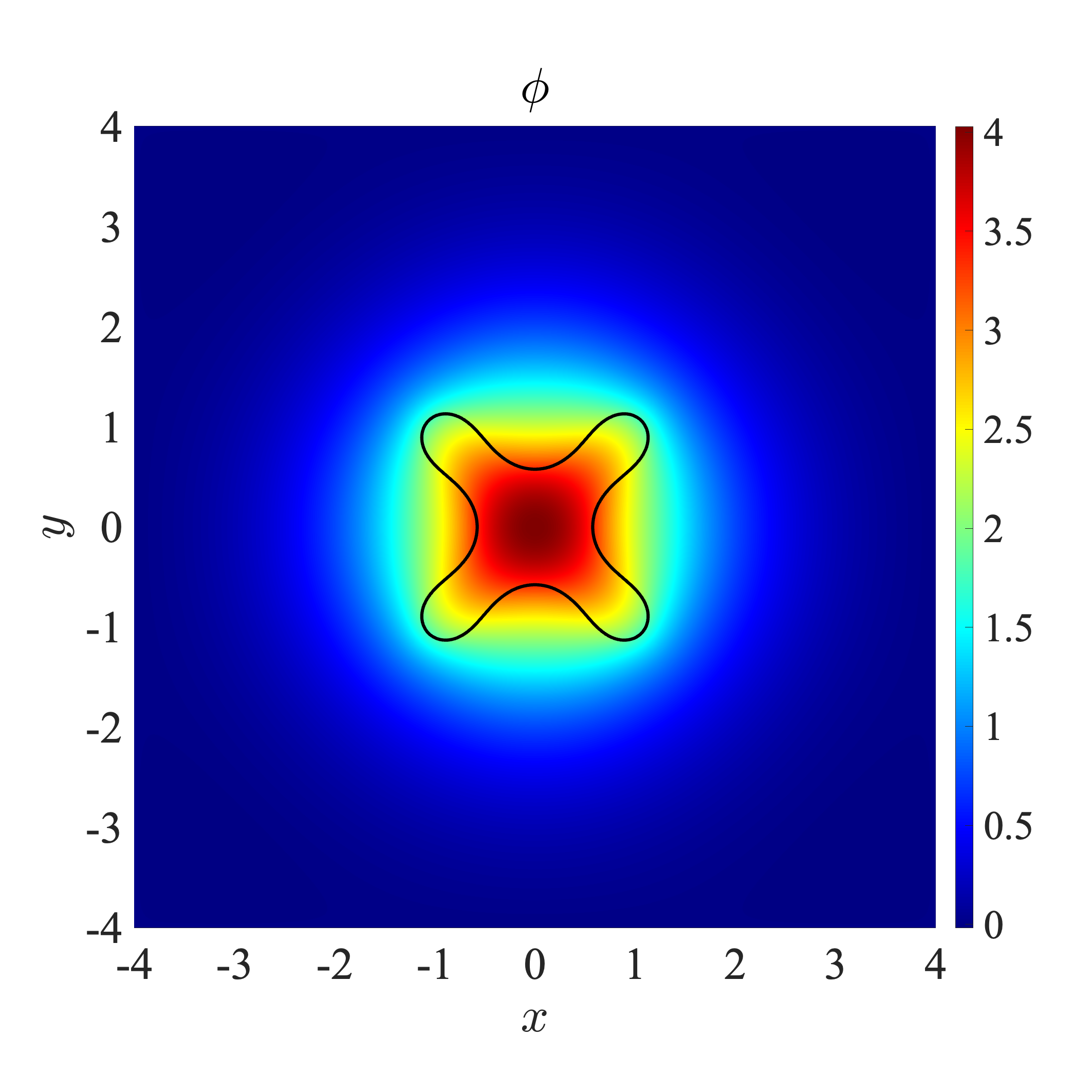}
        \label{subfig:1Drop0bdNonuPumpPhi5}
		}
	\subfloat[$\phi~(t=7.4)$]{
		\includegraphics[width=0.25\linewidth]{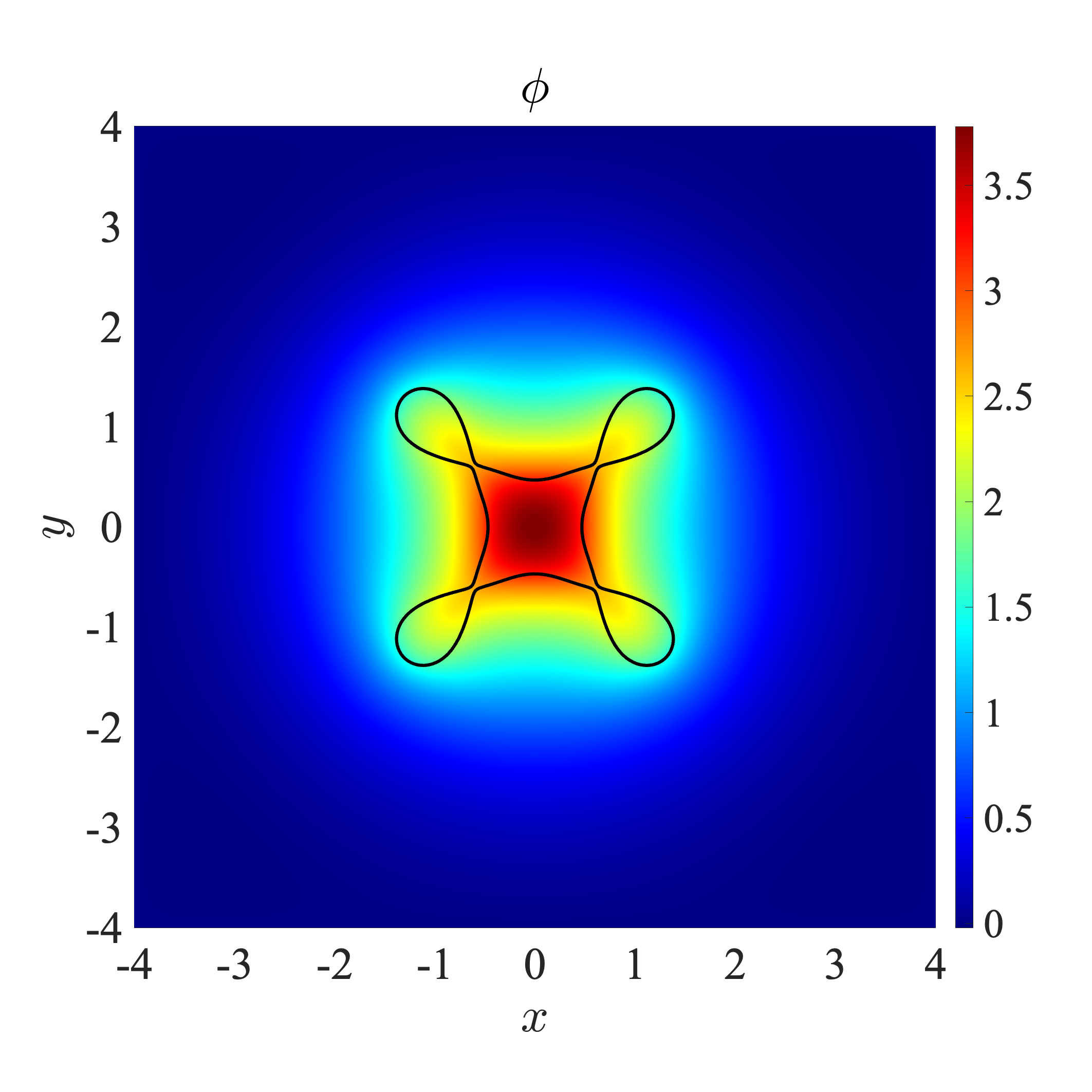}
        \label{subfig:1Drop0bdNonuPumpPhi7d4}
		}
	\subfloat[$\phi~(t=10)$]{
		\includegraphics[width=0.25\linewidth]{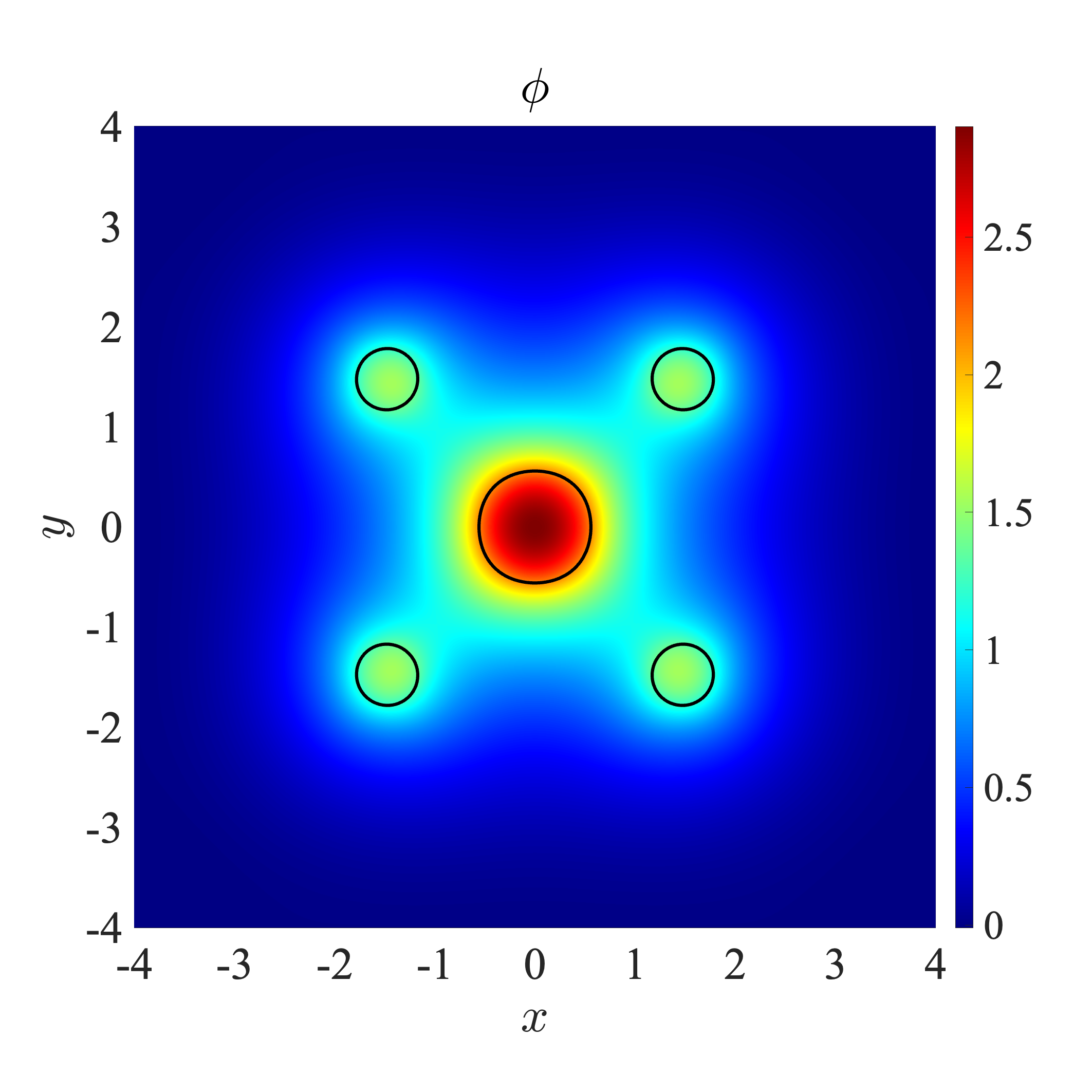}
        \label{subfig:1Drop0bdNonuPumpPhi10}
		}
	\caption{The snapshots for the drop deformation with non-uniform positive ion pump \eqref{case:nonuniformpump}. 
    The black solid circle (denoted by the level set $\psi=0$) represents the location of the drop. 
    The concentration and electric potential distribution are shown on the color map.  $I_0=30$.
    }\label{fig:1Drop0bdNonuPump}
\end{figure}

\begin{figure}[!ht]
\centering
	\subfloat[$p+n~(t = 10)$]{
		\centering
		\includegraphics[width=0.33\linewidth]{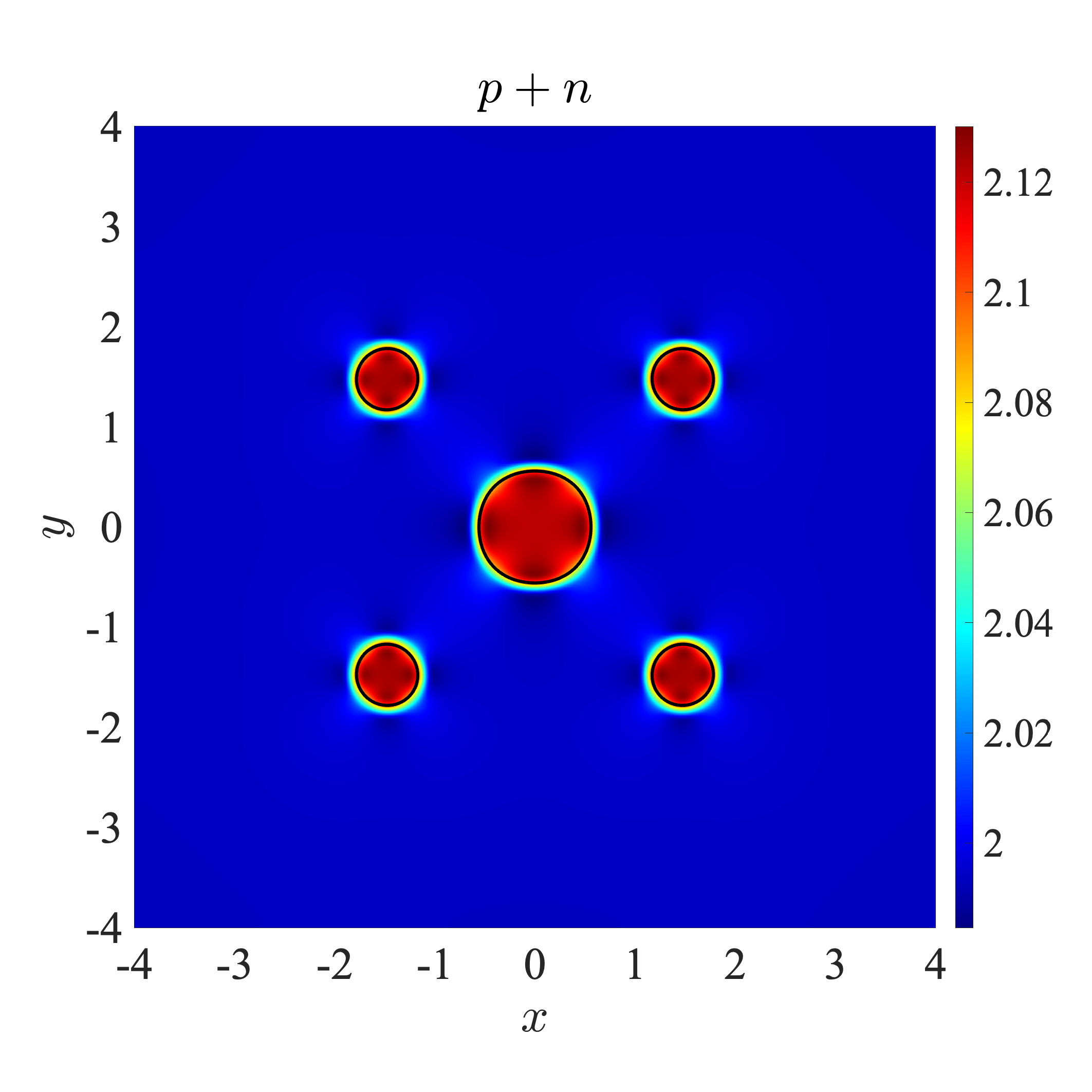}
        \label{subfig:1Drop0bdNounPumpSum10}
		} 
	\subfloat[$p-n~(t = 10)$]{
		\centering
		\includegraphics[width=0.33\linewidth]{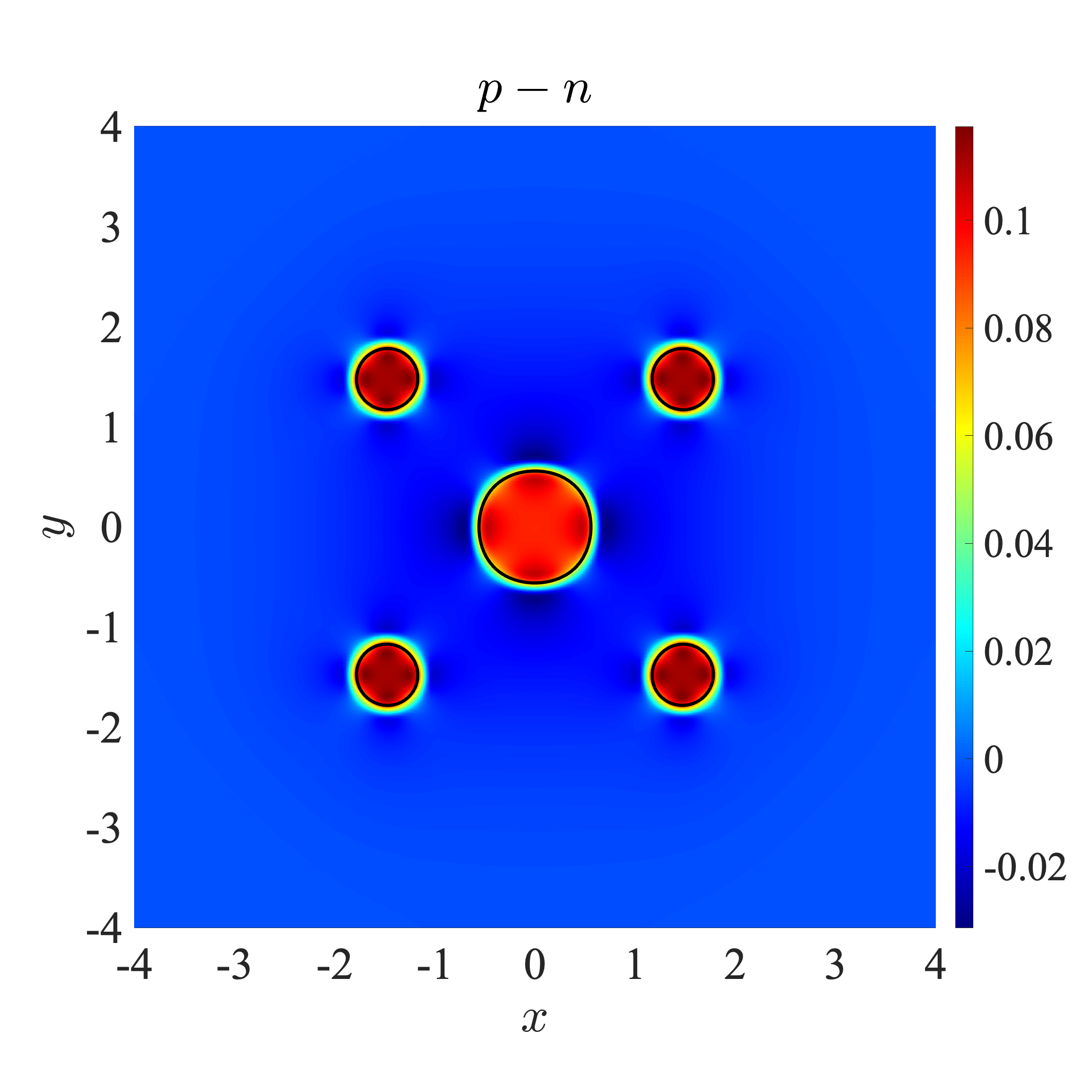}
        \label{subfig:1Drop0bdNonuPumpDif10}
	}
	    \subfloat[$\bm{u}~(t=10)$]{
		\centering
		\includegraphics[width=0.33\linewidth]{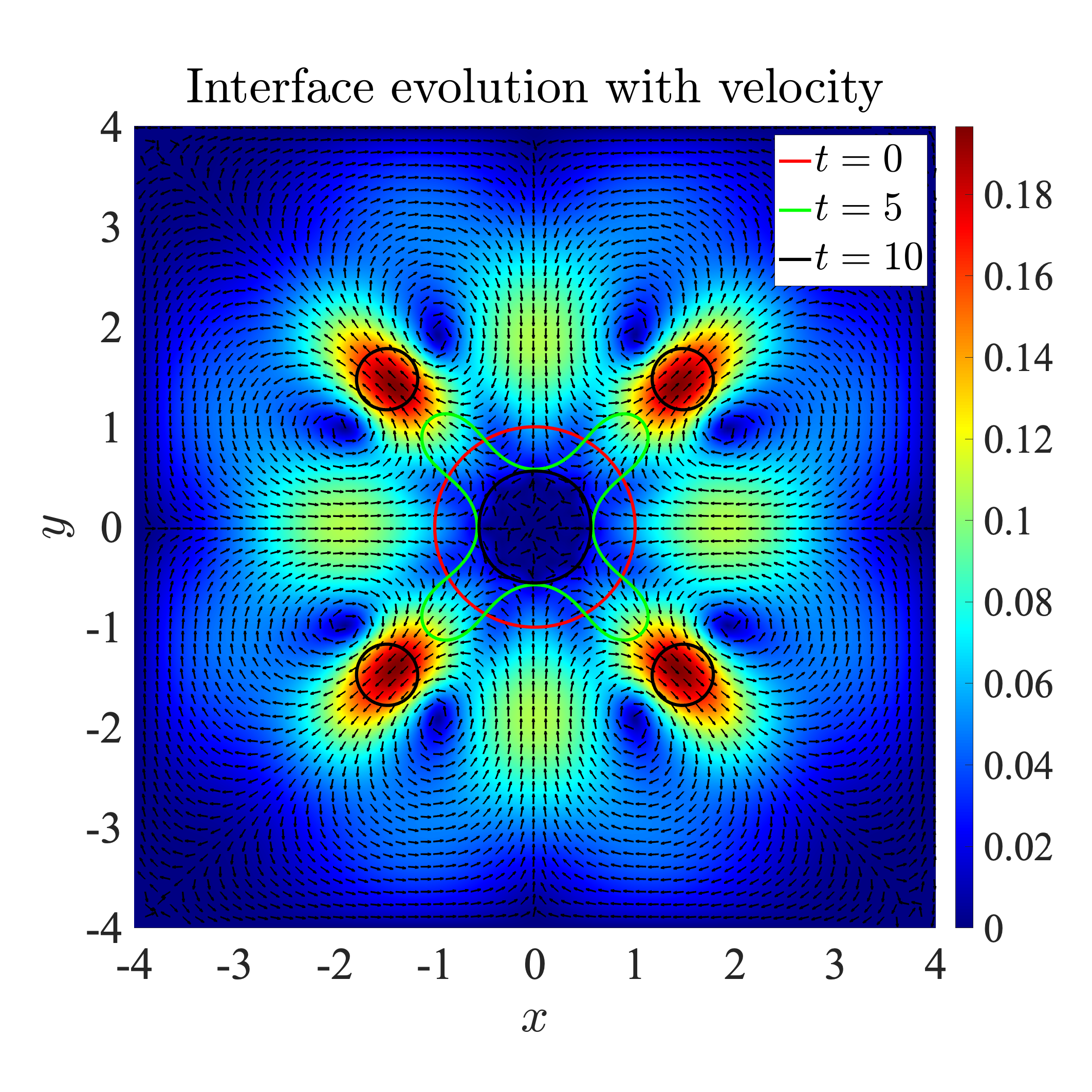}
        \label{subfig:1Drop0bdNonuPumpU10}
		}
	\caption{The snapshots of the total charge (left), net charge (middle) and the velocity (right) for case \eqref{case:nonuniformpump}. 
    The total charge and net charge both accumulate into the droplet due to the redistribution of positive and negative ions. $I_0=30$. }\label{fig:1Drop0bdNonuPumpV}
\end{figure}

\begin{figure}[!ht]
\vskip -0.5cm
    \subfloat[$-\frac{Ca_{E}}{\zeta^{2}}\nabla\phi~(x=-1.5,t=1)$]{
		\includegraphics[width=0.33\linewidth]{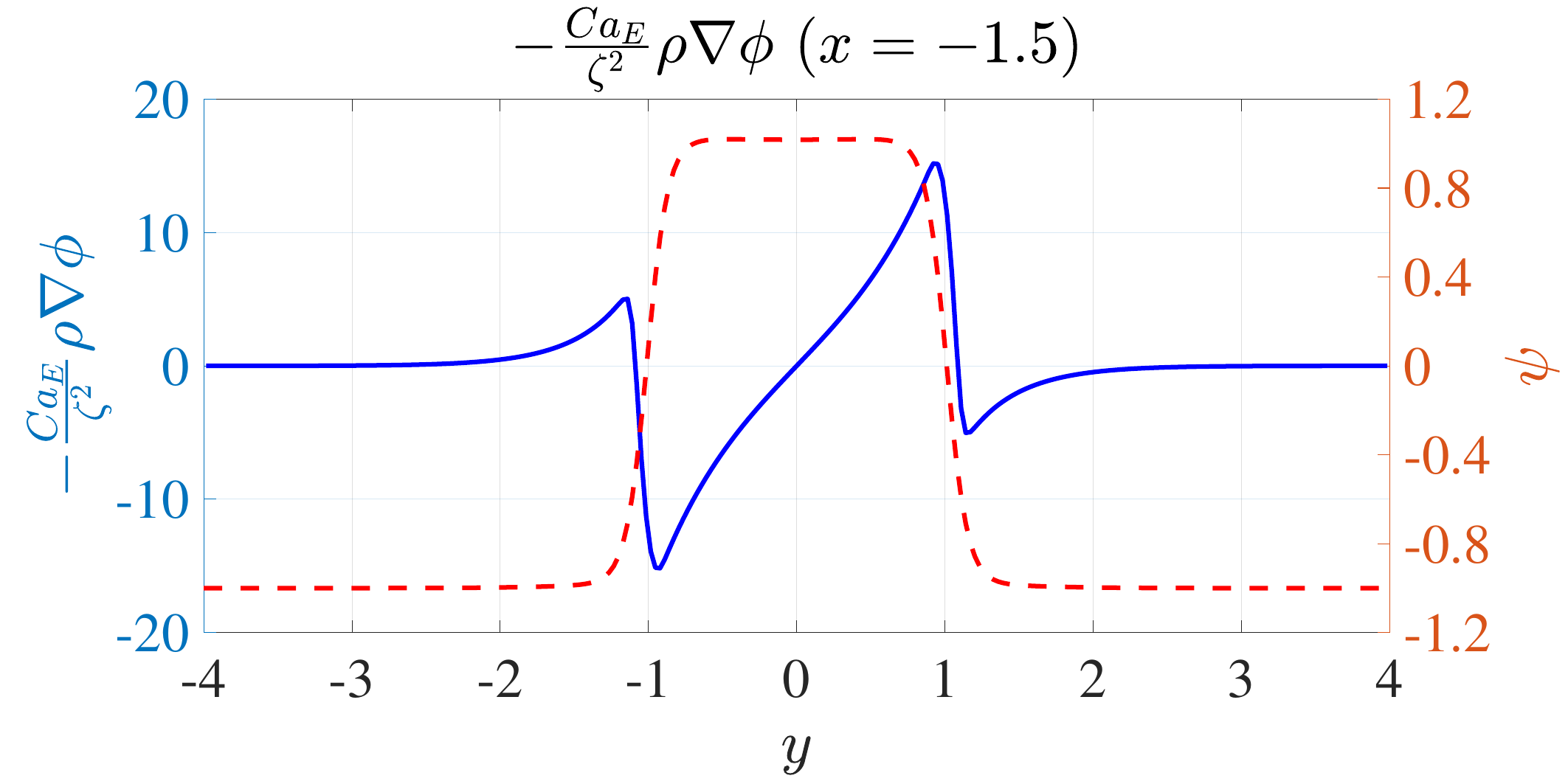}
        \label{subfig:2Drop0bdPump25Eforce1x-1d5}
		} 
	\subfloat[$-\frac{Ca_{E}}{\zeta^{2}}\nabla\phi~(x=1.5,t=1)$]{
		\centering
		\includegraphics[width=0.33\linewidth]{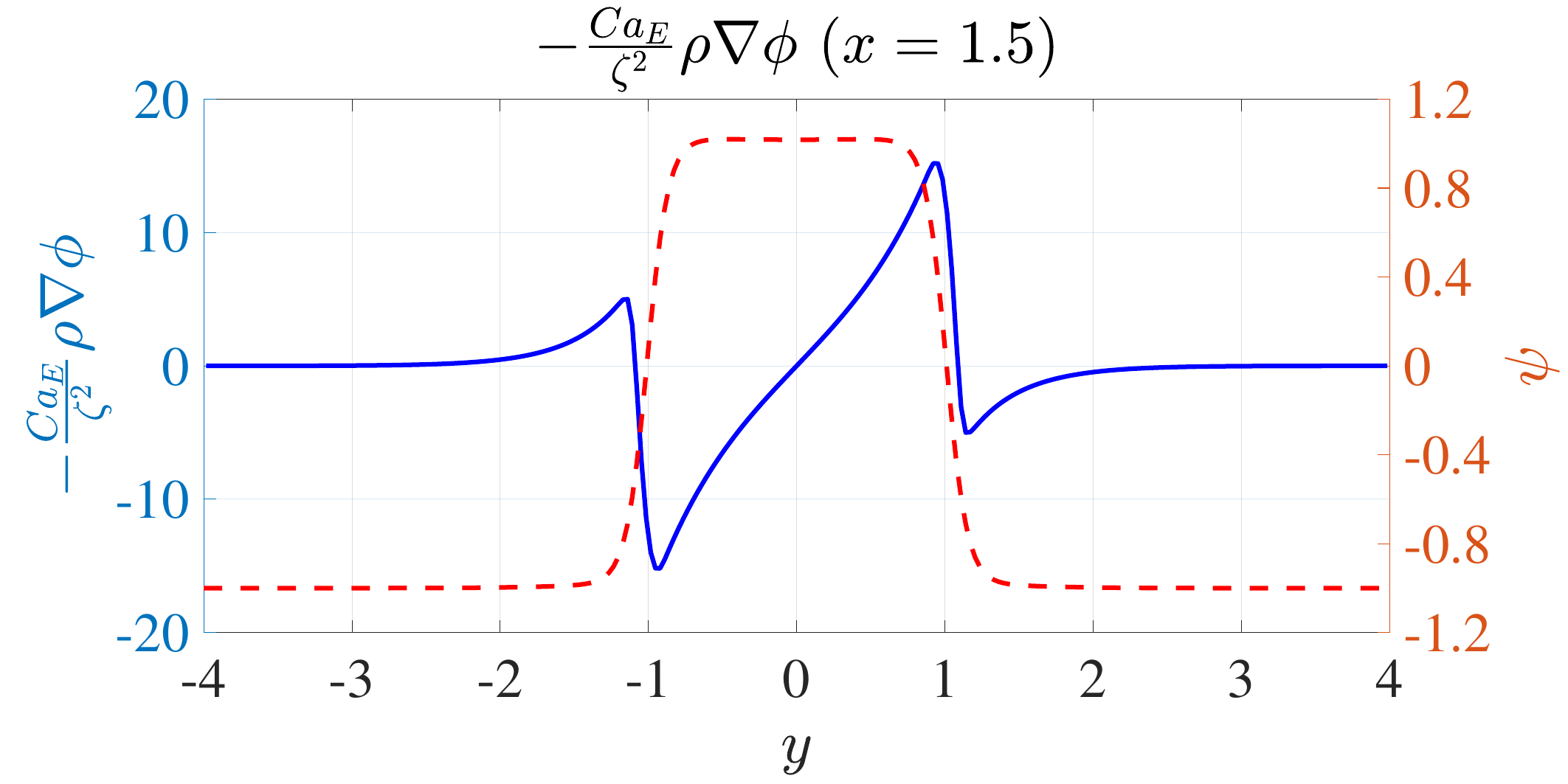}
        \label{subfig:2Drop0bdPump25Eforce1x1d5}
	}
	\subfloat[$-\frac{Ca_{E}}{\zeta^{2}}\nabla\phi~(y=0,t=1)$]{
		\centering
		\includegraphics[width=0.33\linewidth]{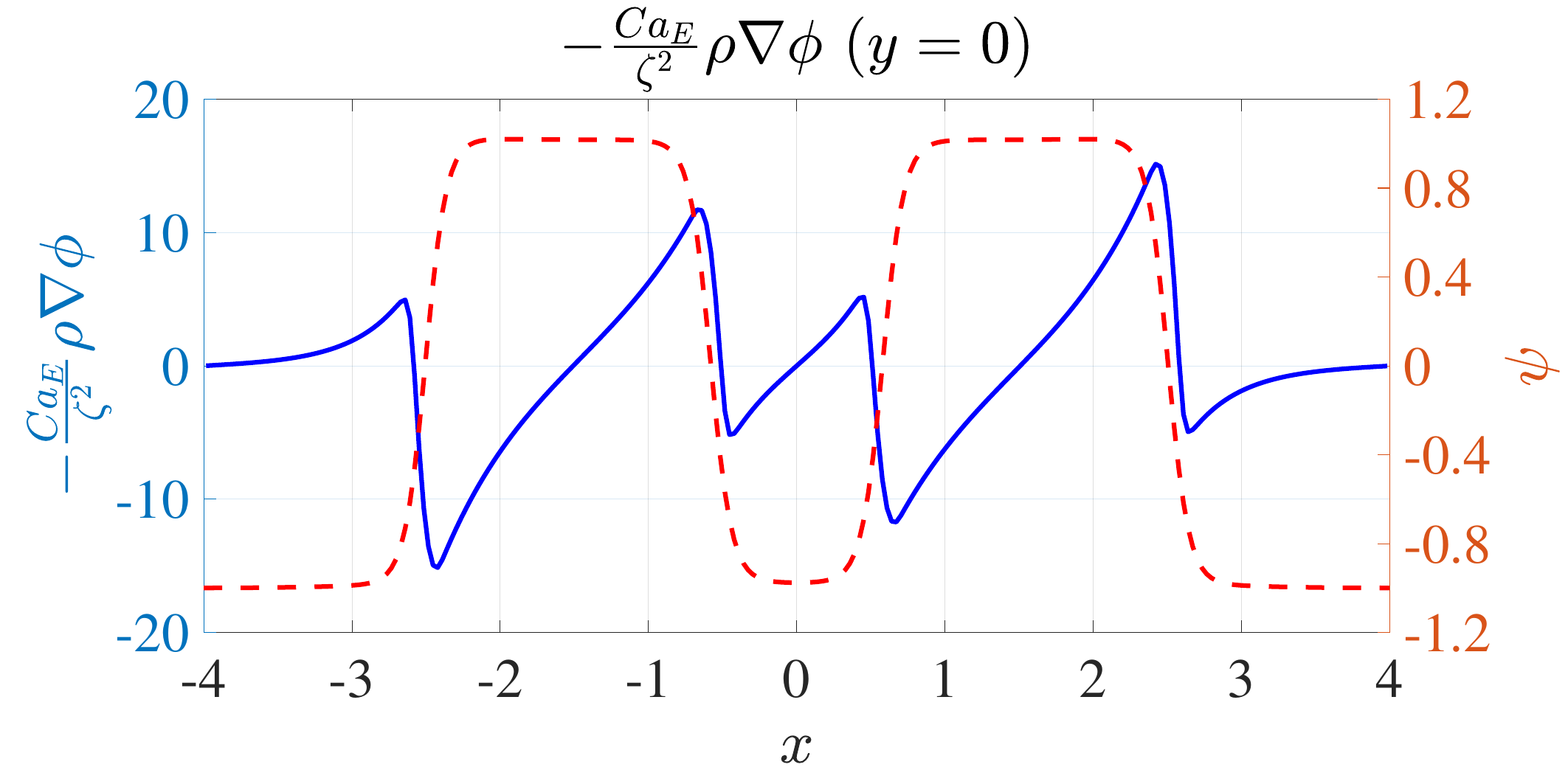}
        \label{subfig:2Drop0bdPump25Eforce1y0}
	}
	\caption{The Lorentz force along $x=-1.5$ (left), $x=1.5$ (middle) and $y=0$ (right) at $t=1$
    induced by the distribution of ions and electric potential for case \eqref{bd:2drops_b0d0} in Section \ref{subsec:2dropsnoelectric}.}\label{fig:2Drop0bdPump25Eforce}
\end{figure}

\begin{figure}
\hskip -0.5cm
	\begin{minipage}[b]{0.6\columnwidth}
		\subfloat[Total force at $t=5$.]{\includegraphics[width=0.9\linewidth,height=0.9\linewidth]{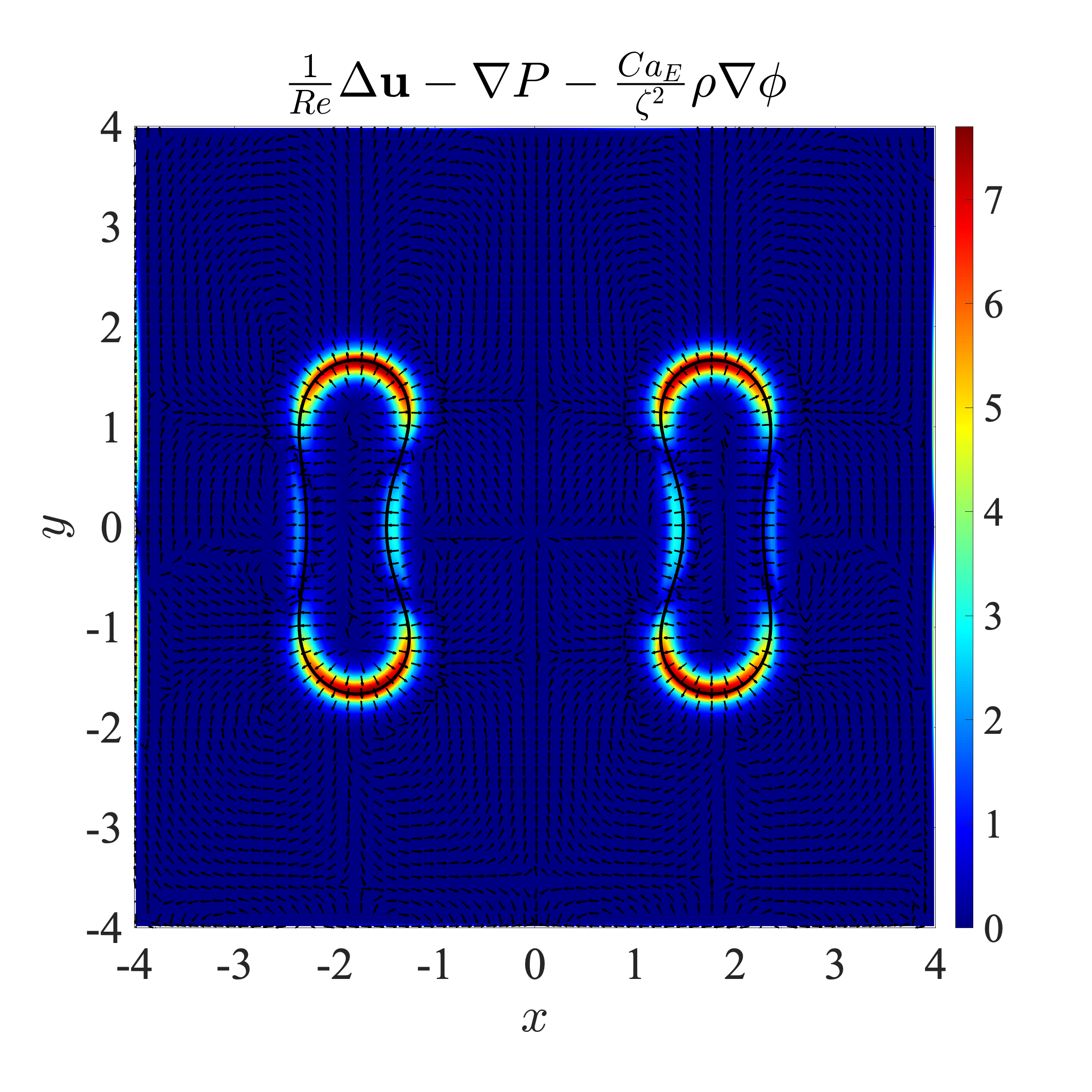}}\label{fig:2DropD0N00bdPump25Force5}
	\end{minipage}
    \hskip -1cm
	\begin{minipage}[b]{0.4\columnwidth}
		\subfloat[Total force along $x$-axis at $t=5$.]{\includegraphics[width=1.2\linewidth,height=0.6\linewidth]{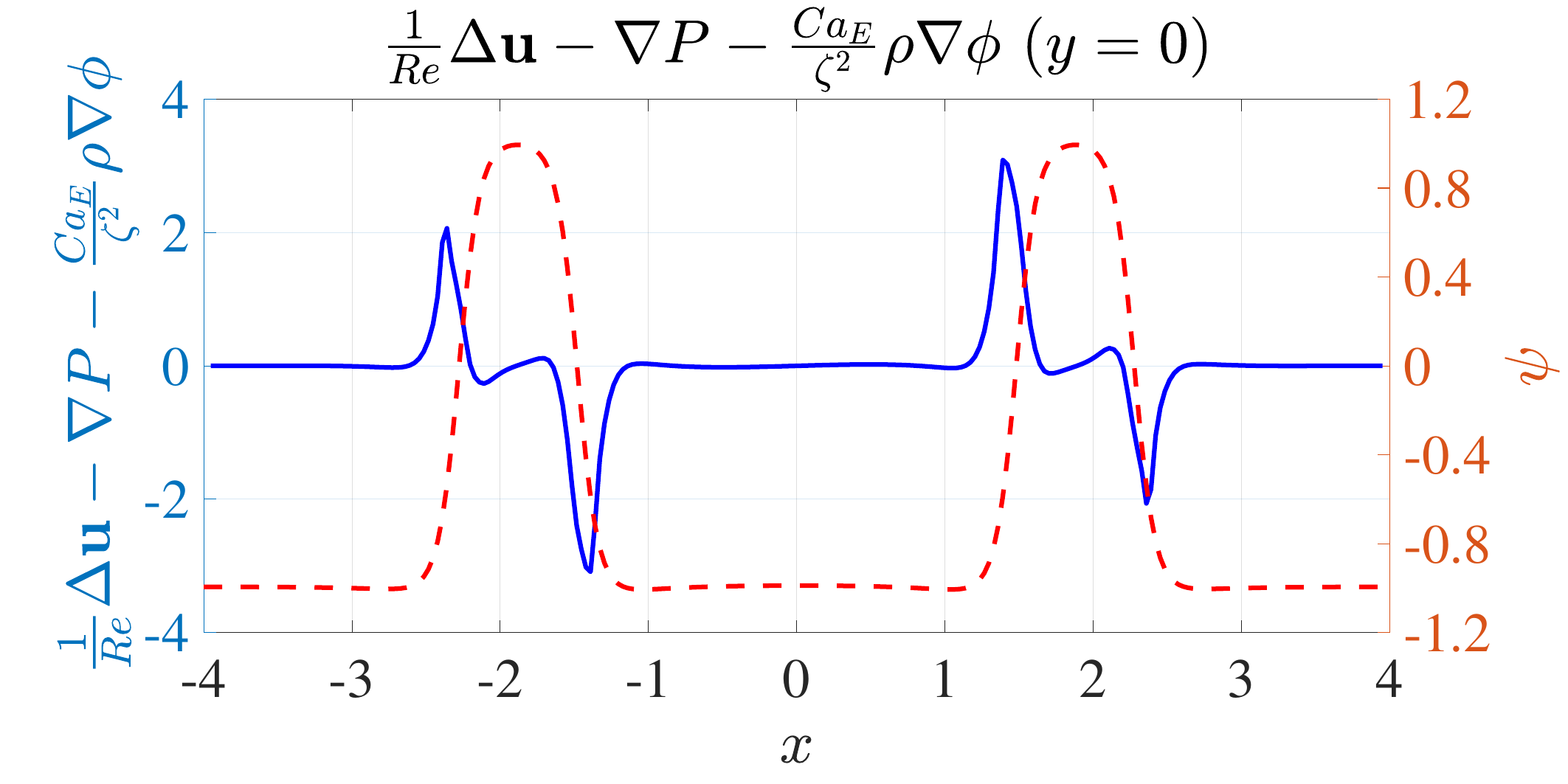}}
        \label{fig:2DropD0N0bdPump25Force5y=0}
		\\
		\subfloat[Total force along $y$-axis at $t=5$.]{\includegraphics[width=1.2\linewidth,height=0.6\linewidth]{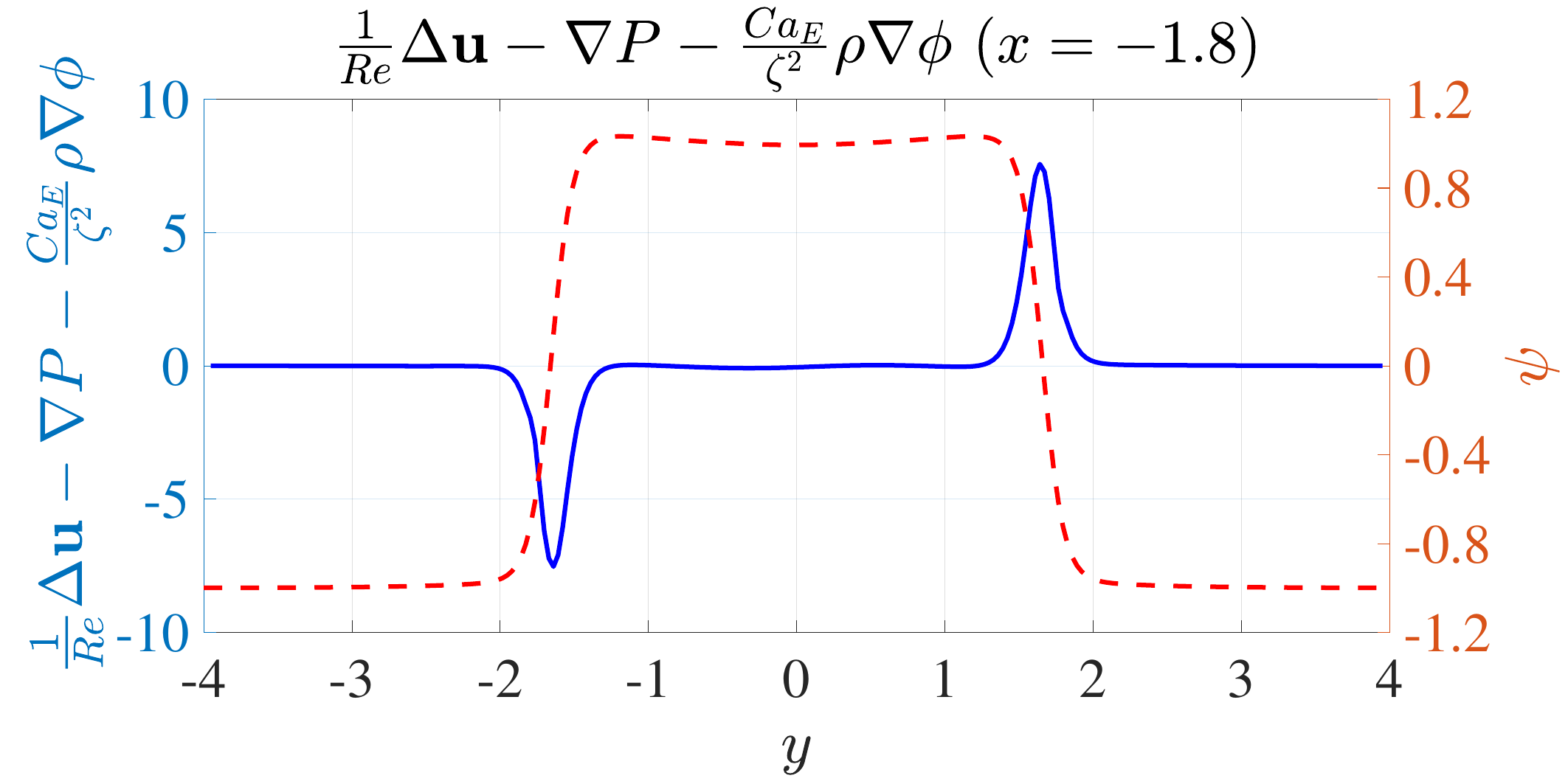}}
        \label{fig:2DropD0N0bdPump25Force5x=-1d8}
	\end{minipage}
	\caption{Total Viscous stress induced force and Lorentz force for the droplet with pump. (a) 2D map; (b) Distribution along  $y=0$ line; (c) Distribution along  $x=-1.8$ line.}\label{fig:2DropD0N0bdPump25Force}
\end{figure}

\begin{figure}[!ht]
    \vskip -0.4cm
    \centering
	\subfloat[$p~(t=1)$]{
		\includegraphics[width=0.165\linewidth]{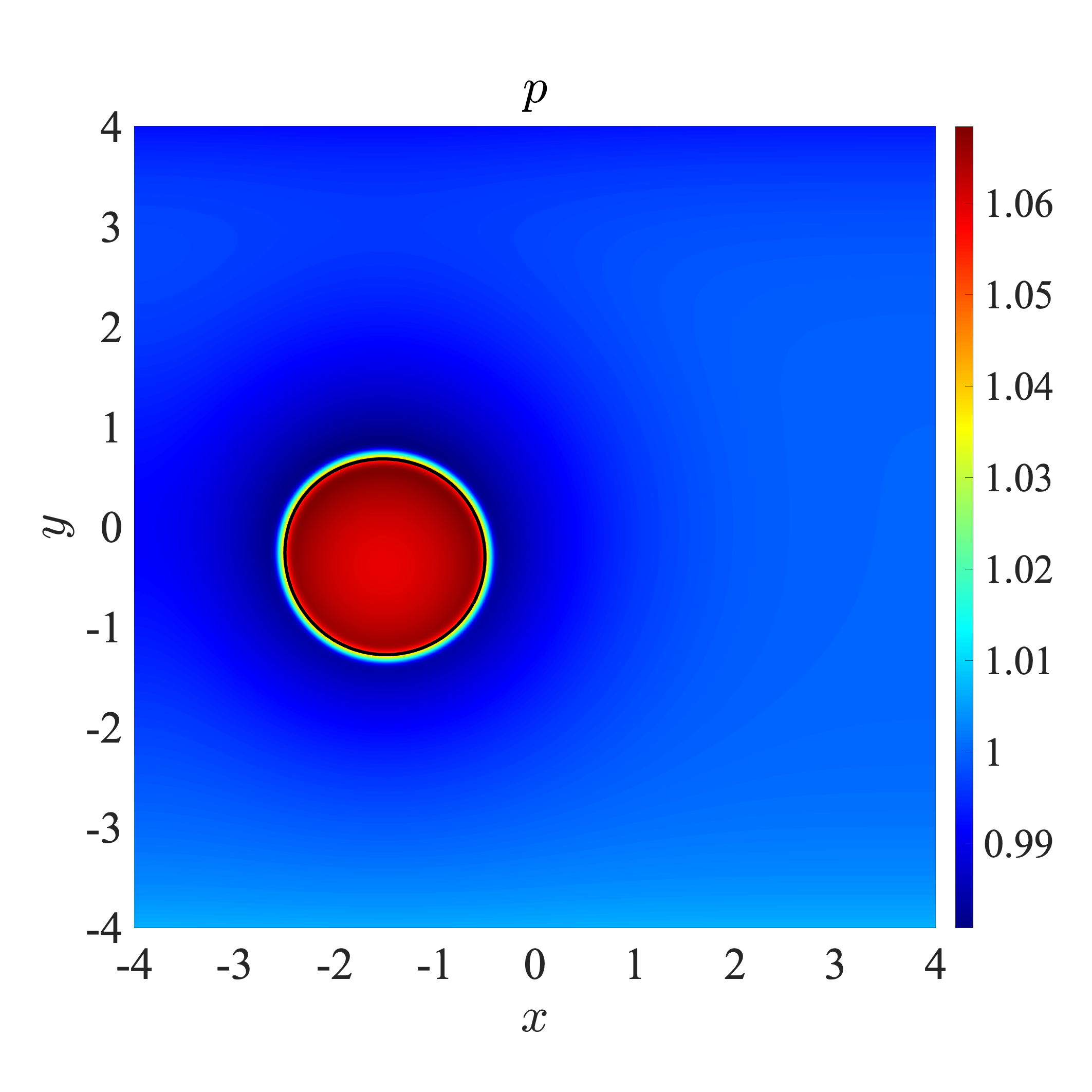}
        \label{subfig:1LeftDropD4N0Pump25P1}
		}
    \hskip -0.3cm
    \subfloat[$p~(t=3)$]{
		\includegraphics[width=0.165\linewidth]{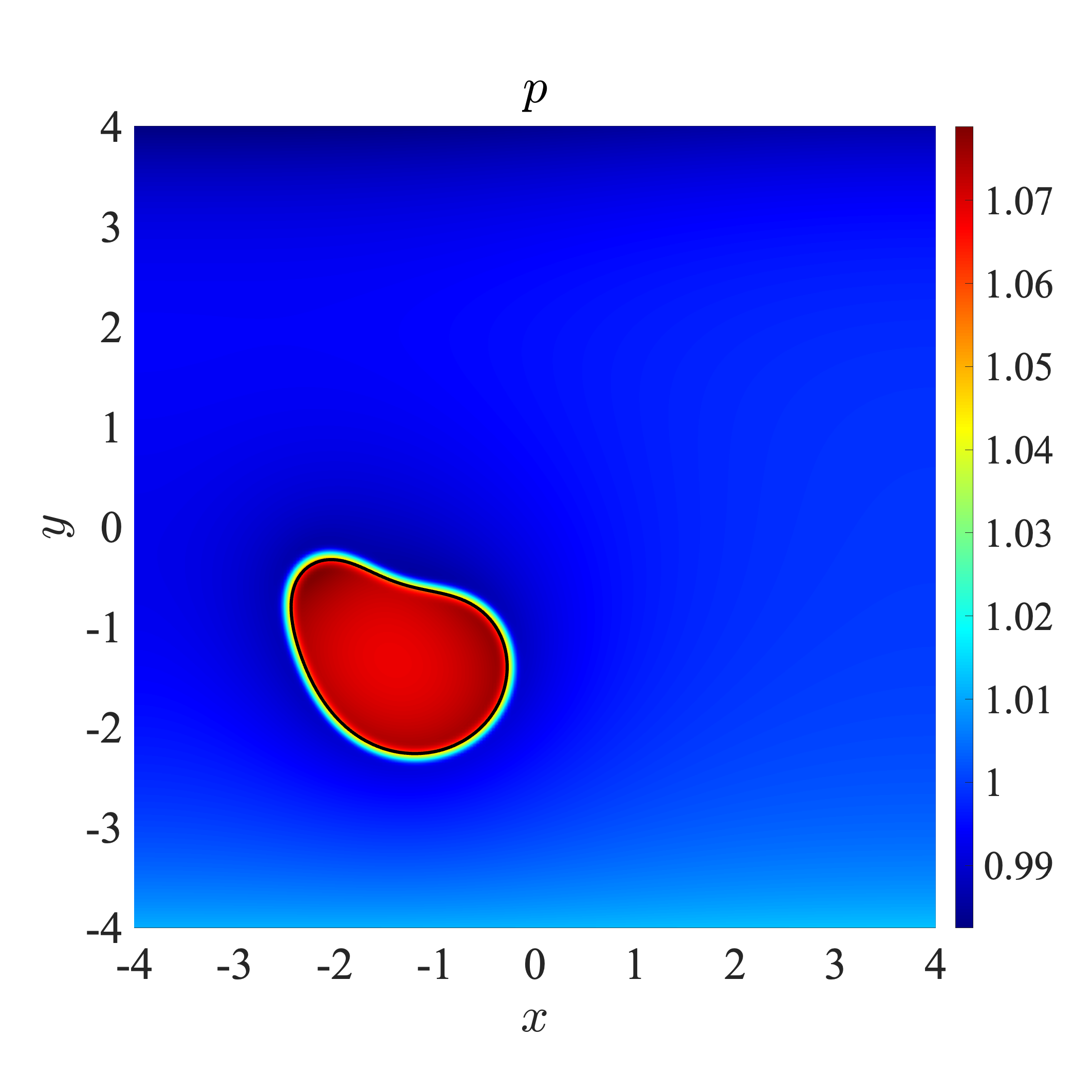}
        \label{subfig:1LeftDropD4N0Pump25P3}
		}
    \hskip -0.3cm
	\subfloat[$p~(t=5)$]{
		\includegraphics[width=0.165\linewidth]{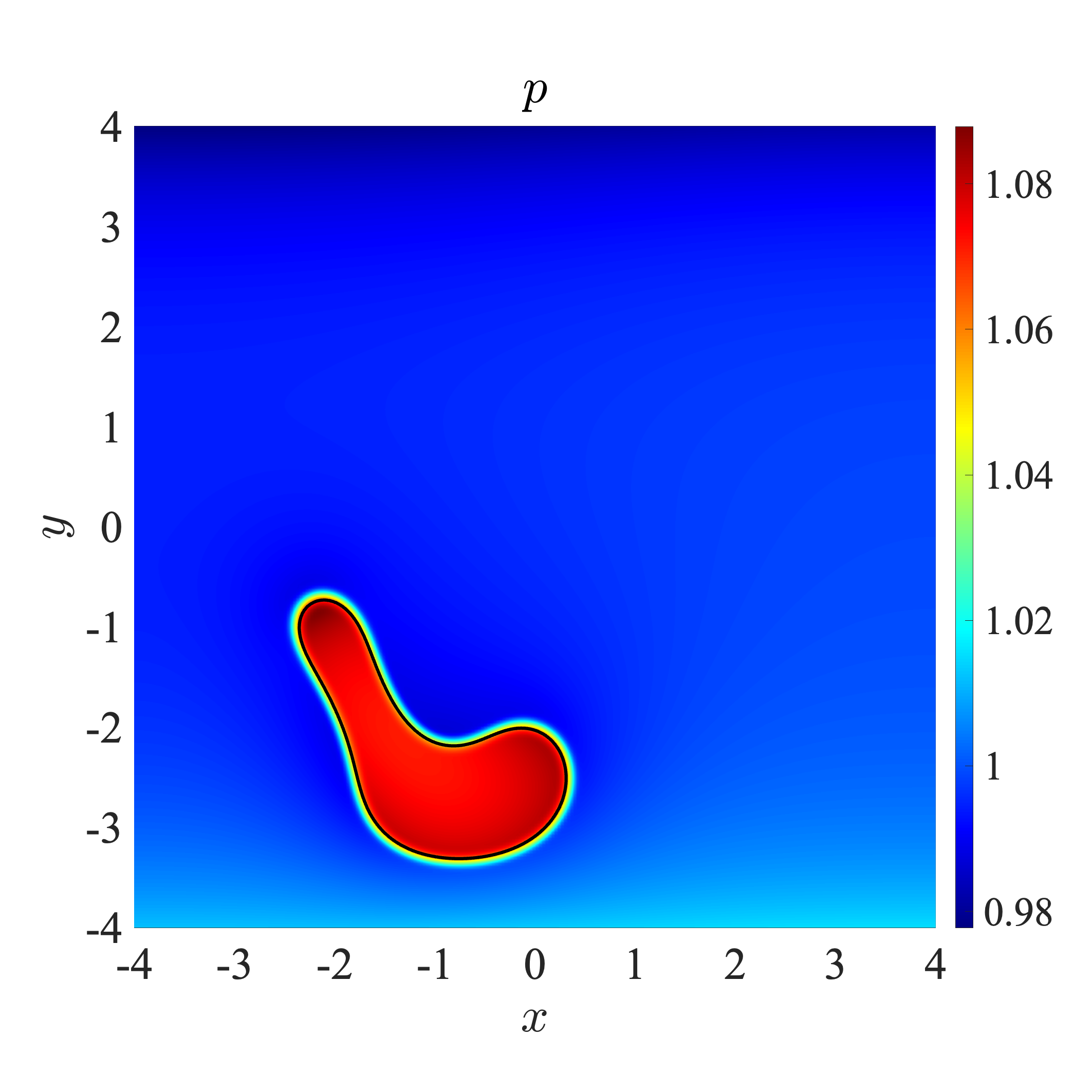}
        \label{subfig:1LeftDropD4N0Pump25P5}
		}
    \hskip -0.3cm
        \subfloat[$p~(t=8)$]{
		\includegraphics[width=0.165\linewidth]{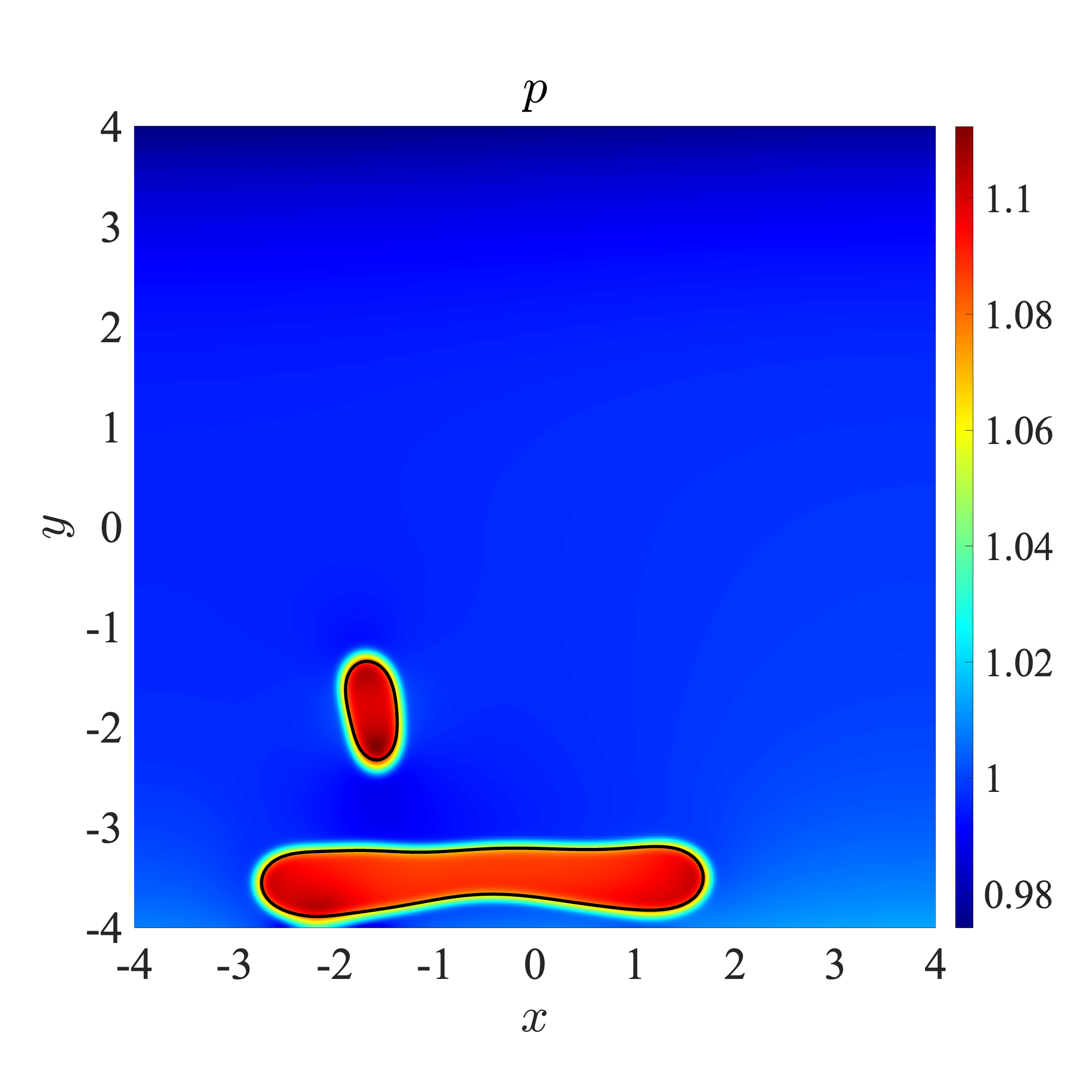}
        \label{subfig:1LeftDropD4N0Pump25P8}
		}
    \hskip -0.3cm
        \subfloat[$p~(t=16)$]{
		\includegraphics[width=0.165\linewidth]{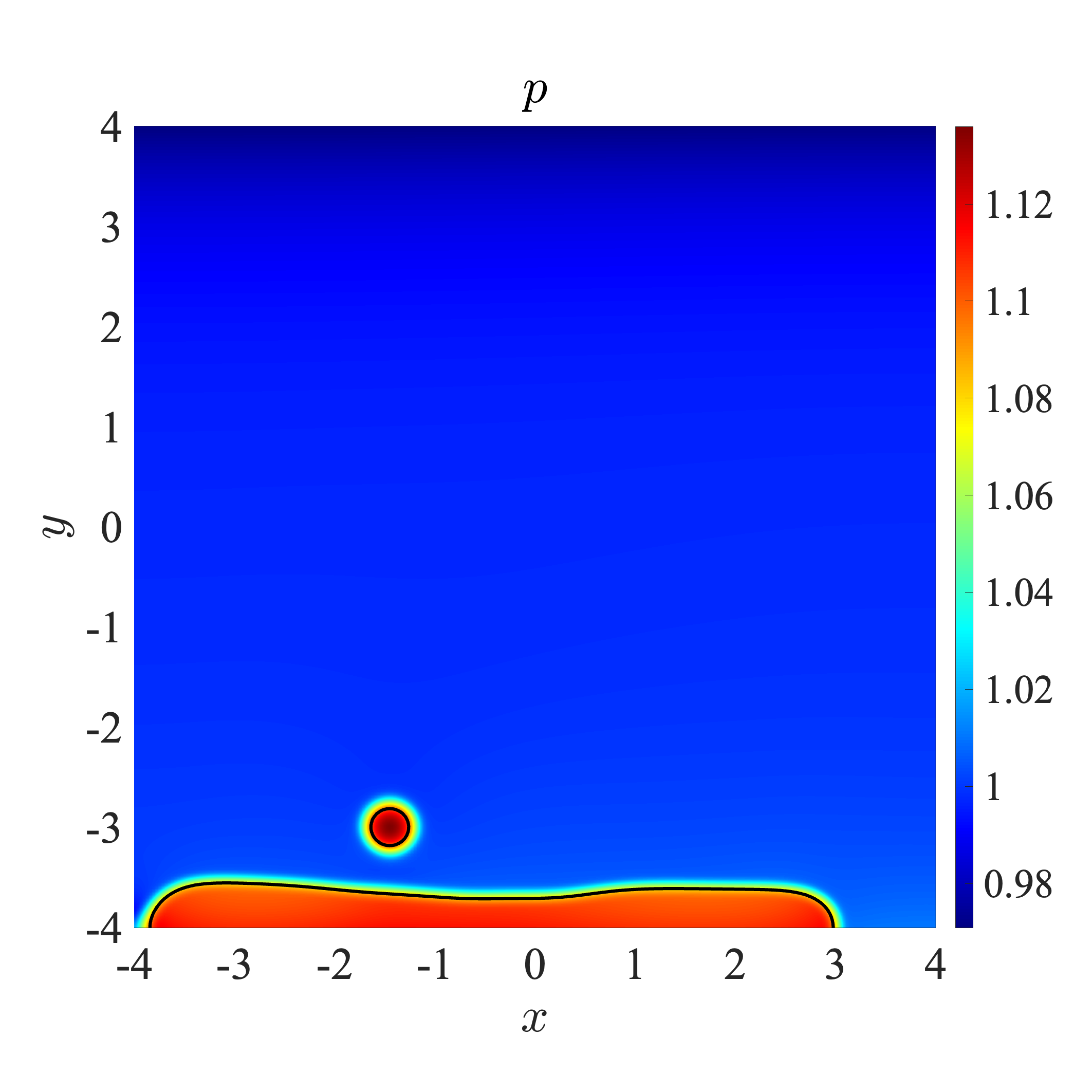}
        \label{subfig:1LeftDropD4N0Pump25P16}
		}
    \hskip -0.3cm
        \subfloat[$p~(t=80)$]{
		\includegraphics[width=0.165\linewidth]{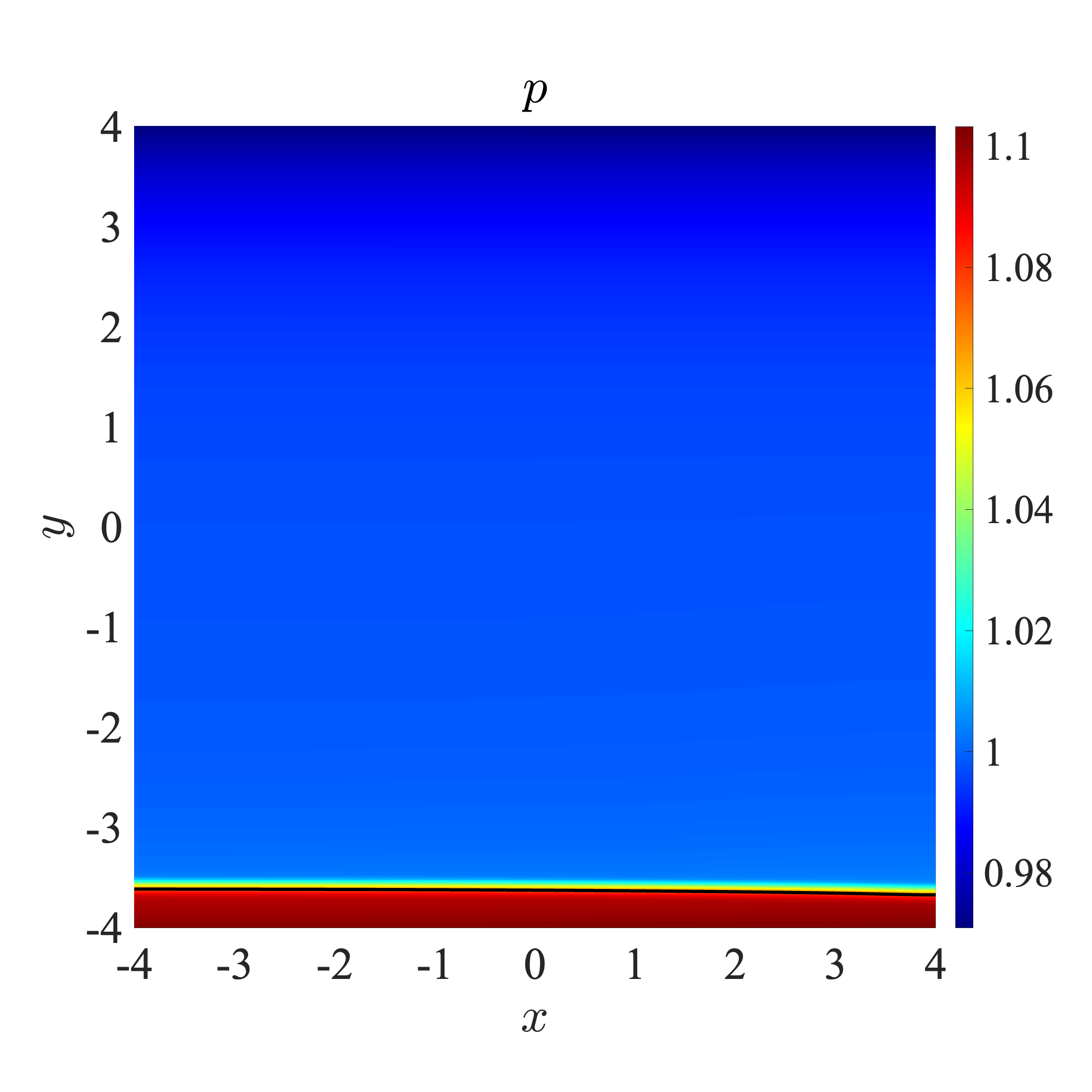}
        \label{subfig:1LeftDropD4N0Pump25P80}
		}
        \\
        \vskip -0.3cm
	\subfloat[$n~(t=1)$]{
		\includegraphics[width=0.165\linewidth]{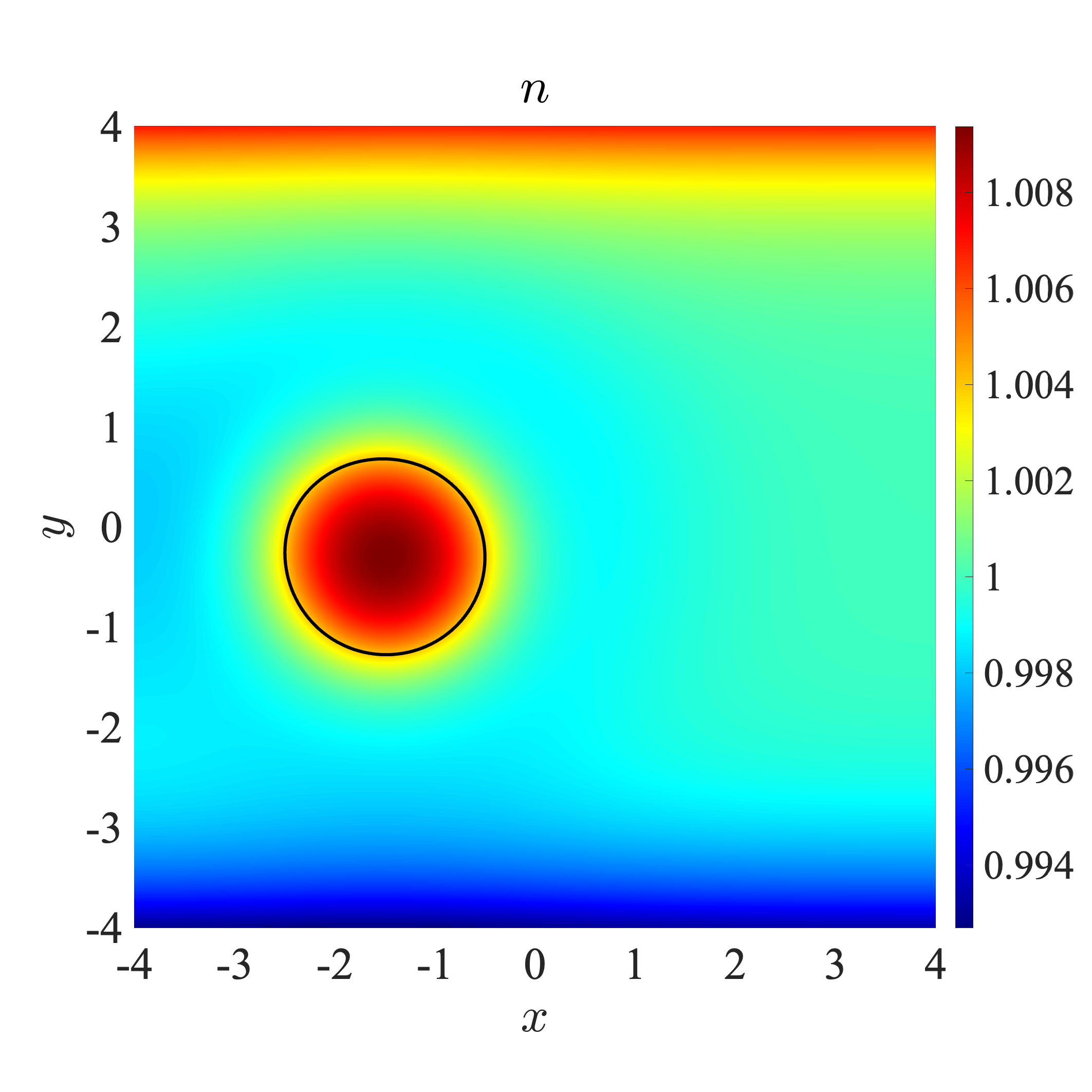}
        \label{subfig:1LeftDropD4N0Pump25N1}
		}
    \hskip -0.3cm
    \subfloat[$n~(t=3)$]{
		\includegraphics[width=0.165\linewidth]{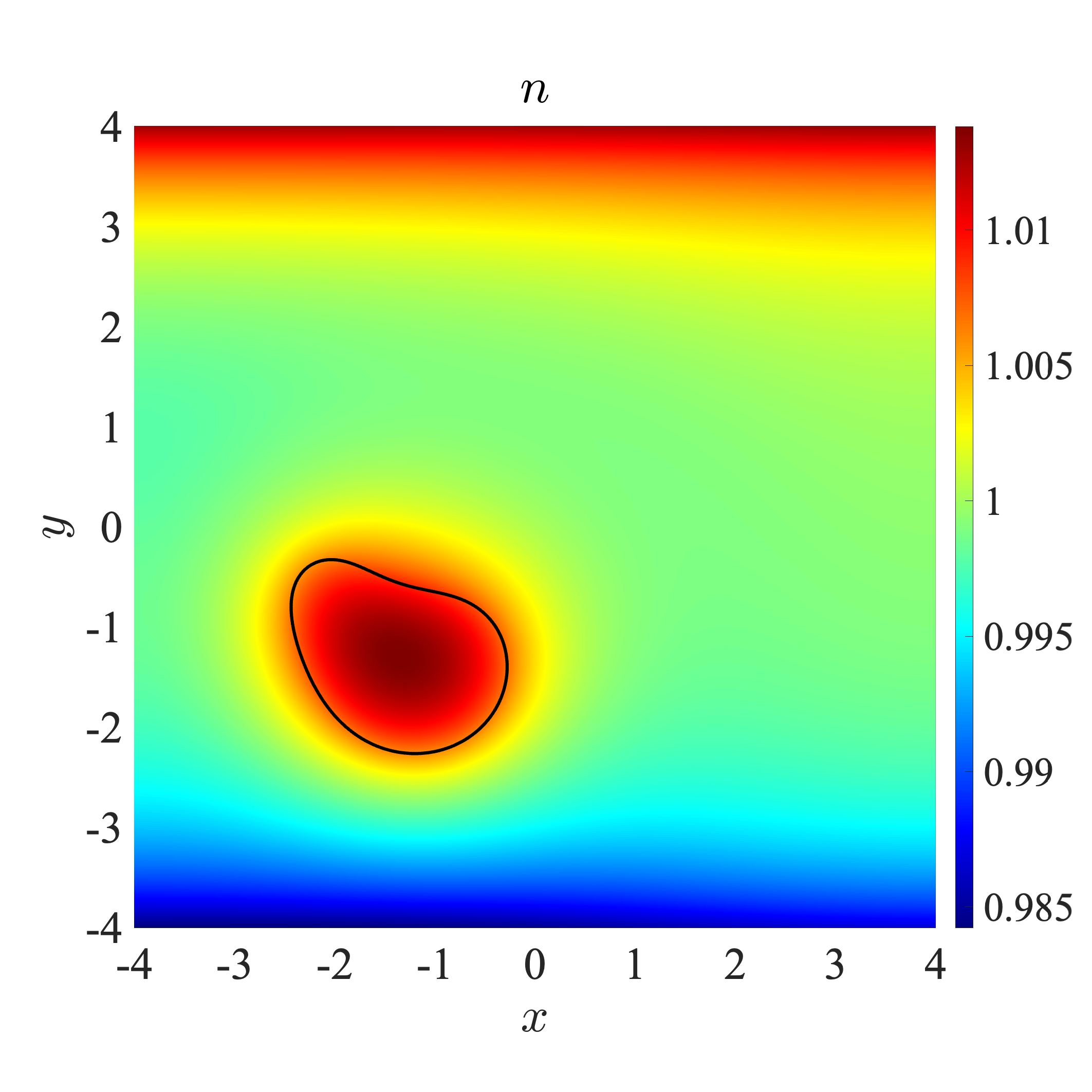}
        \label{subfig:1LeftDropD4N0Pump25N3}
		}
    \hskip -0.3cm
	\subfloat[$n~(t=5)$]{
		\includegraphics[width=0.165\linewidth]{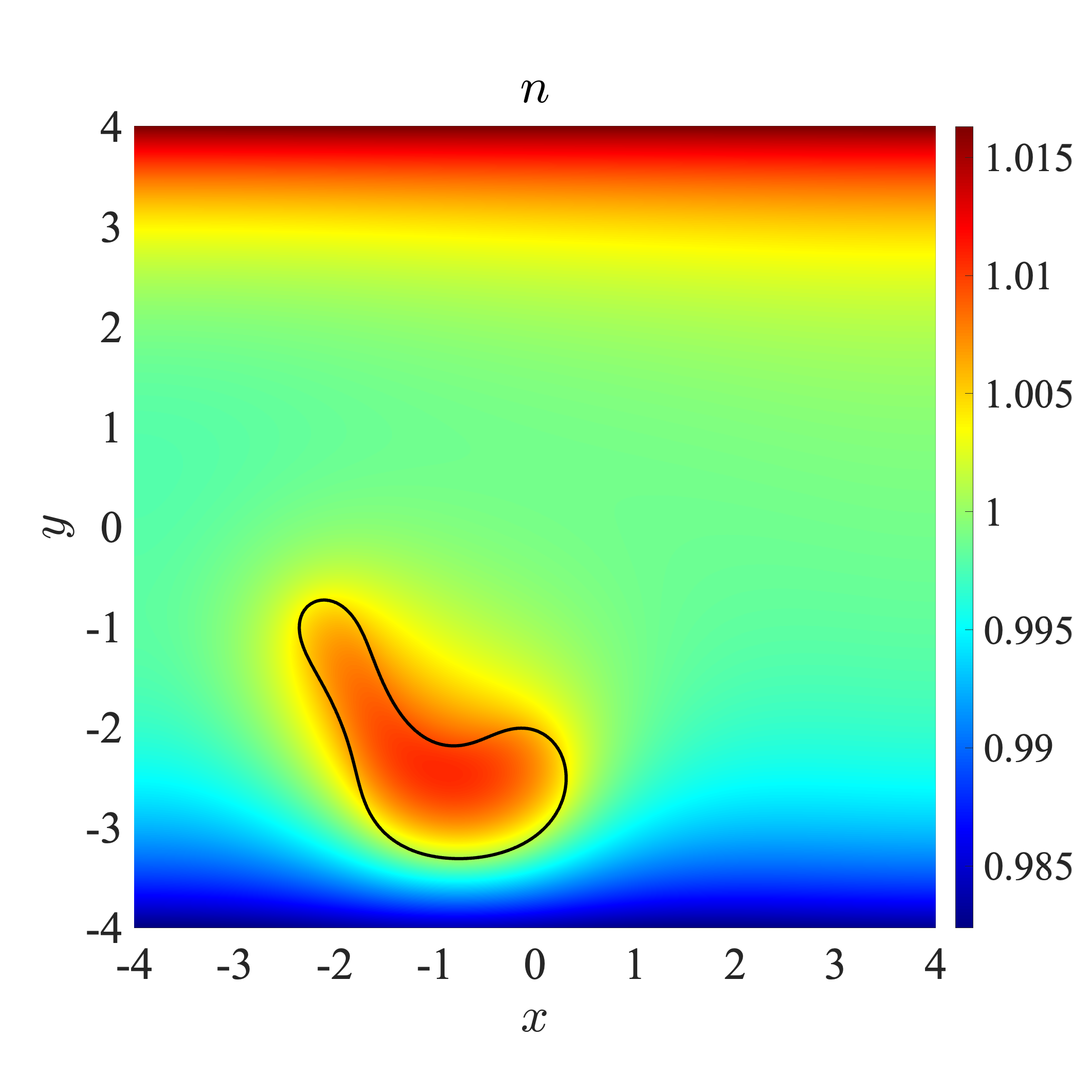}
        \label{subfig:1LeftDropD4N0Pump25N5}
		}
    \hskip -0.3cm
	\subfloat[$n~(t=8)$]{
		\includegraphics[width=0.165\linewidth]{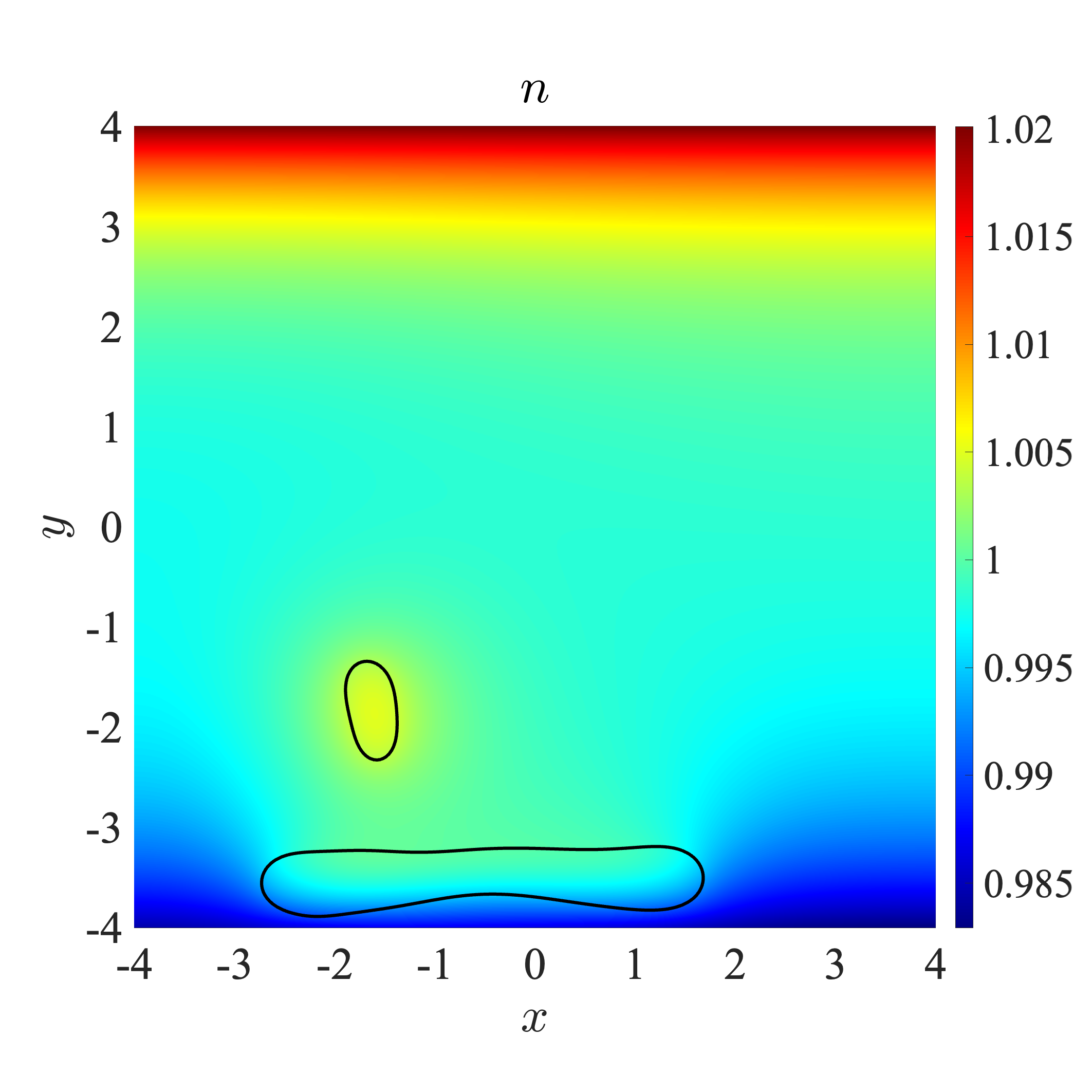}
        \label{subfig:1LeftDropD4N0Pump25N8}
		}
    \hskip -0.3cm
	\subfloat[$n~(t=16)$]{
		\includegraphics[width=0.165\linewidth]{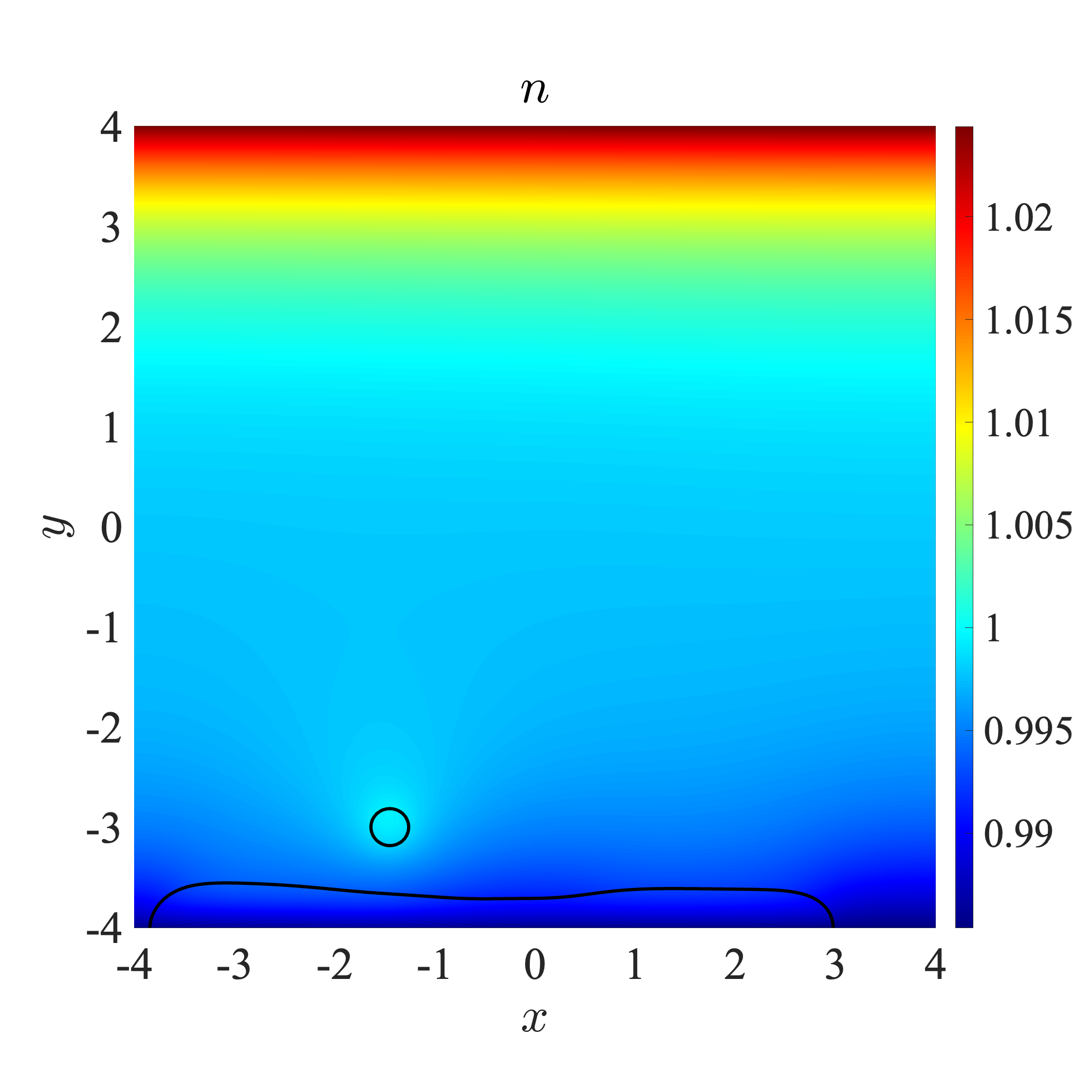}
        \label{subfig:1LeftDropD4N0Pump25N16}
		}
    \hskip -0.3cm
	\subfloat[$n~(t=80)$]{
		\includegraphics[width=0.165\linewidth]{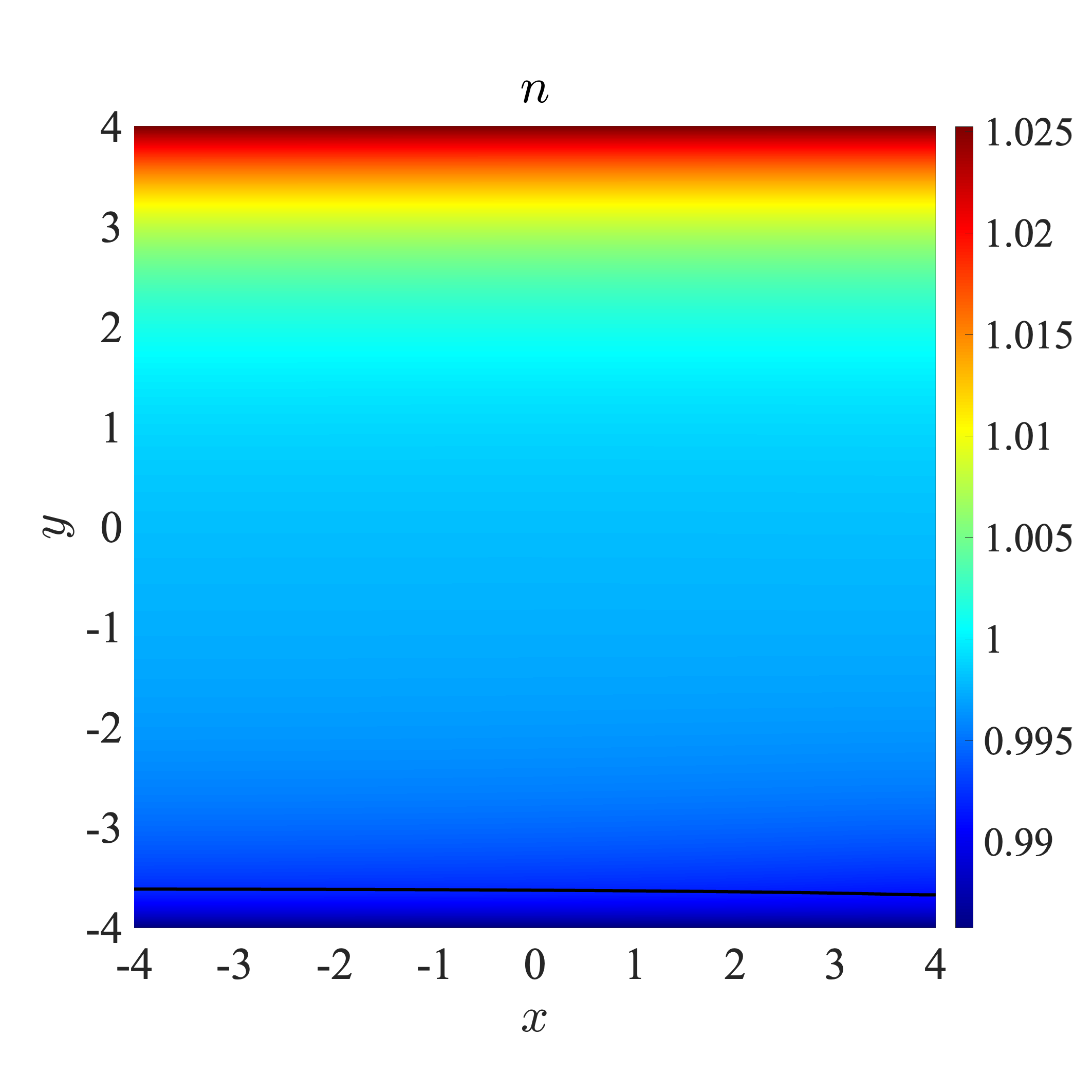}
        \label{subfig:1LeftDropD4N0Pump25N80}
		}
        \\ 
        \vskip -0.3cm
	\subfloat[$\phi~(t=1)$]{
		\includegraphics[width=0.165\linewidth]{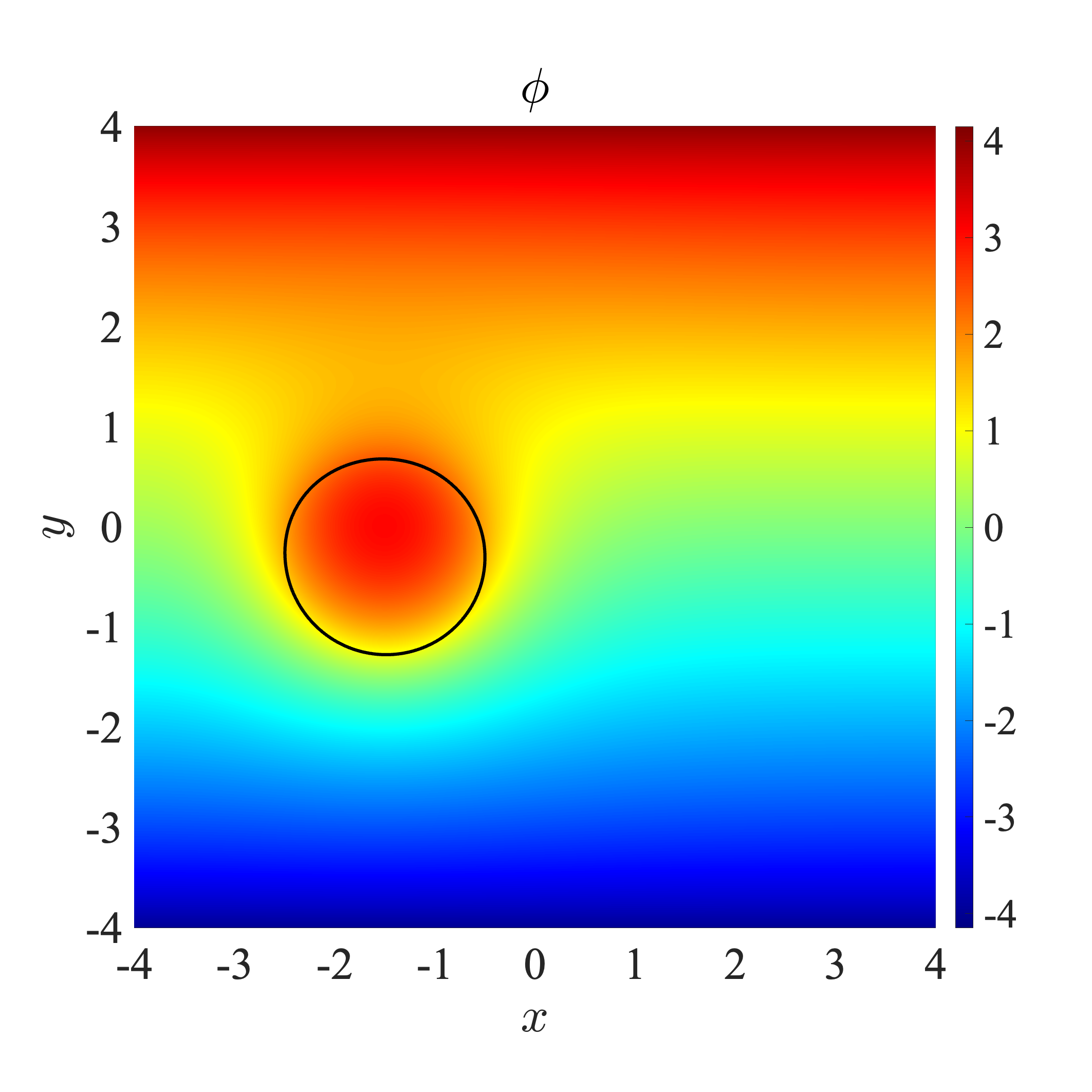}
        \label{subfig:1LeftDropD4N0Pump25Phi1}
		}
    \hskip -0.3cm
    \subfloat[$\phi~(t=3)$]{
		\includegraphics[width=0.165\linewidth]{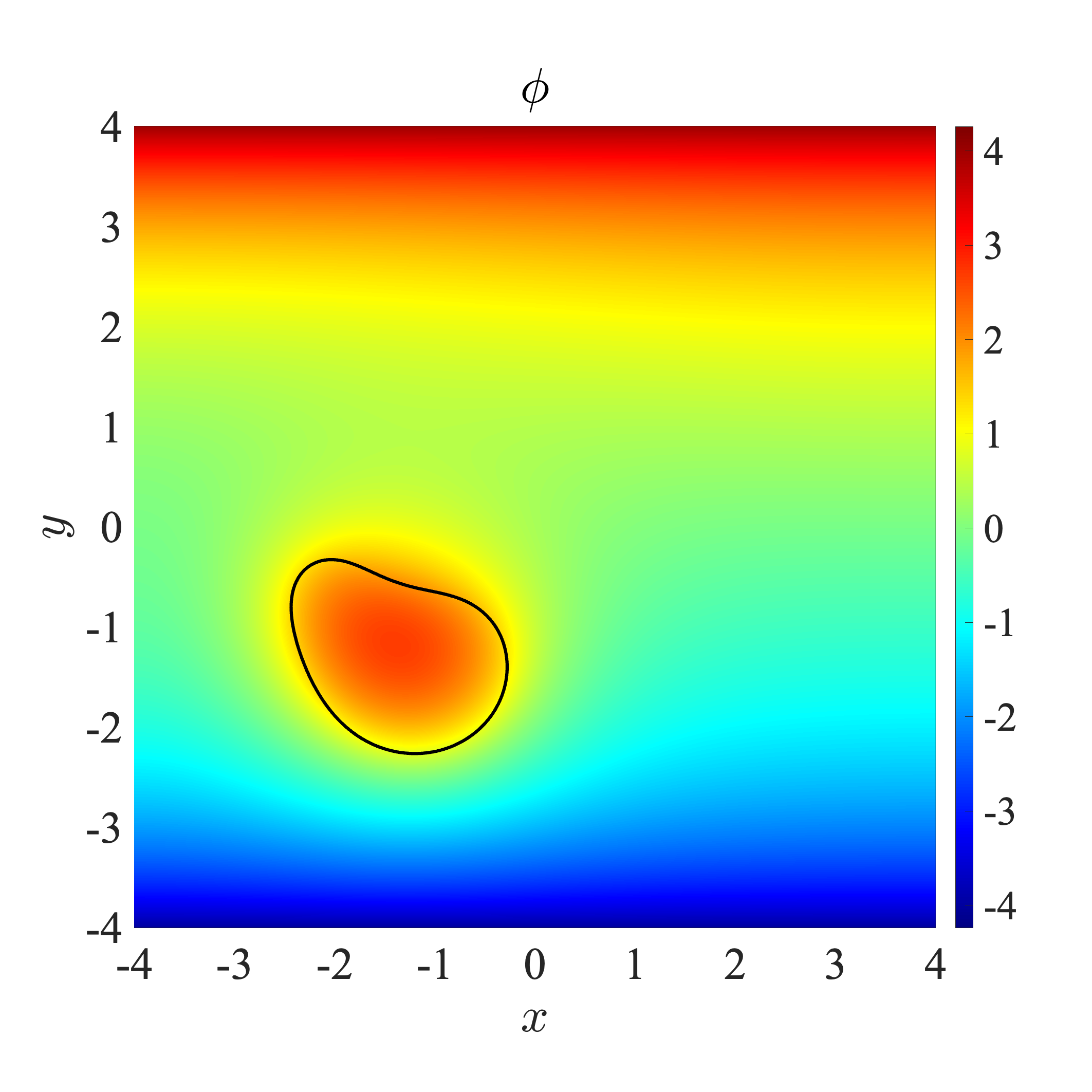}
        \label{subfig:1LeftDropD4N0Pump25Phi3}
		}
    \hskip -0.3cm
	\subfloat[$\phi~(t=5)$]{
		\includegraphics[width=0.165\linewidth]{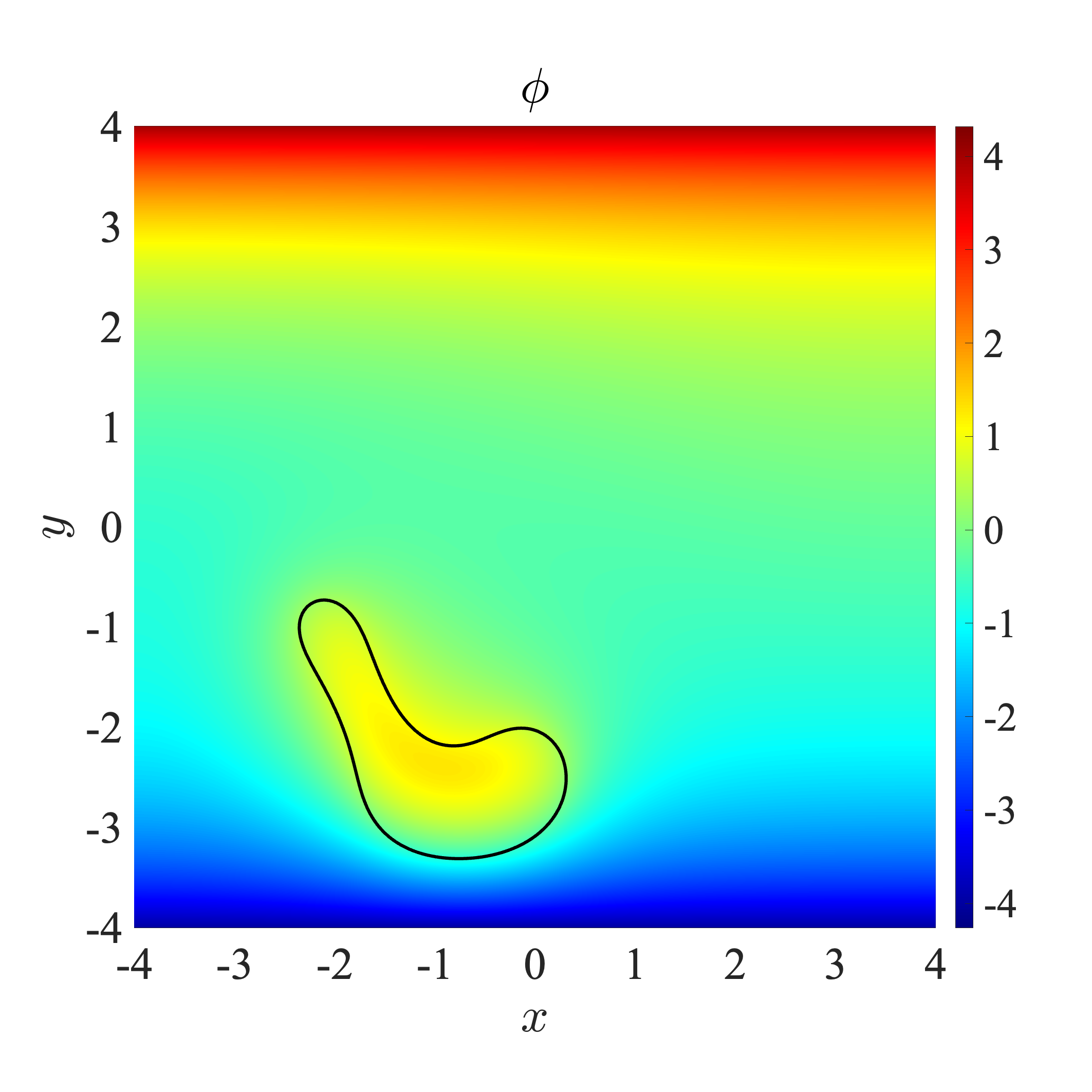}
        \label{subfig:1LeftDropD4N0Pump25Phi5}
		}
    \hskip -0.3cm
	\subfloat[$\phi~(t=8)$]{
		\includegraphics[width=0.165\linewidth]{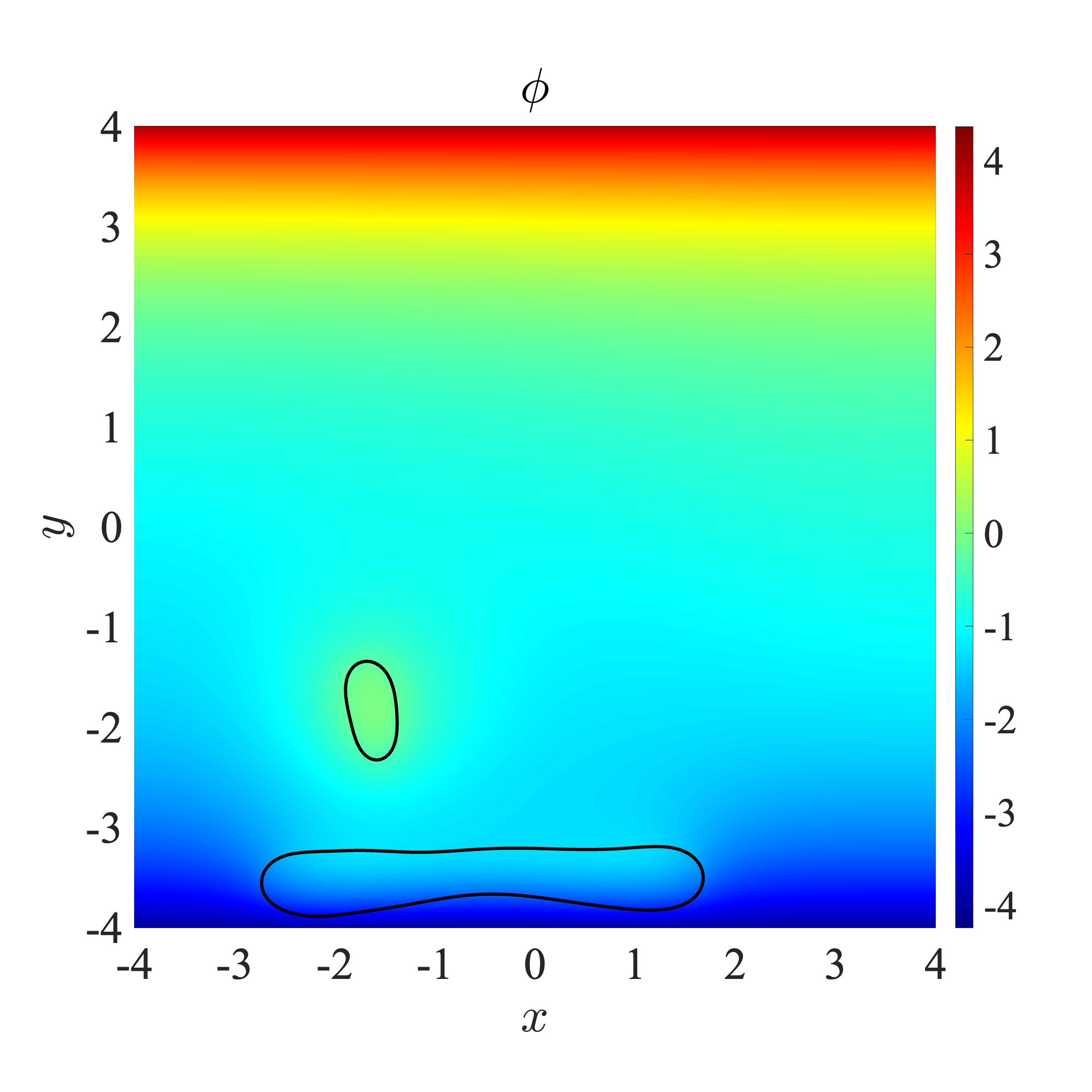}
        \label{subfig:1LeftDropD4N0Pump25Phi8}
		}
    \hskip -0.3cm
    \subfloat[$\phi~(t=16)$]{
		\includegraphics[width=0.165\linewidth]{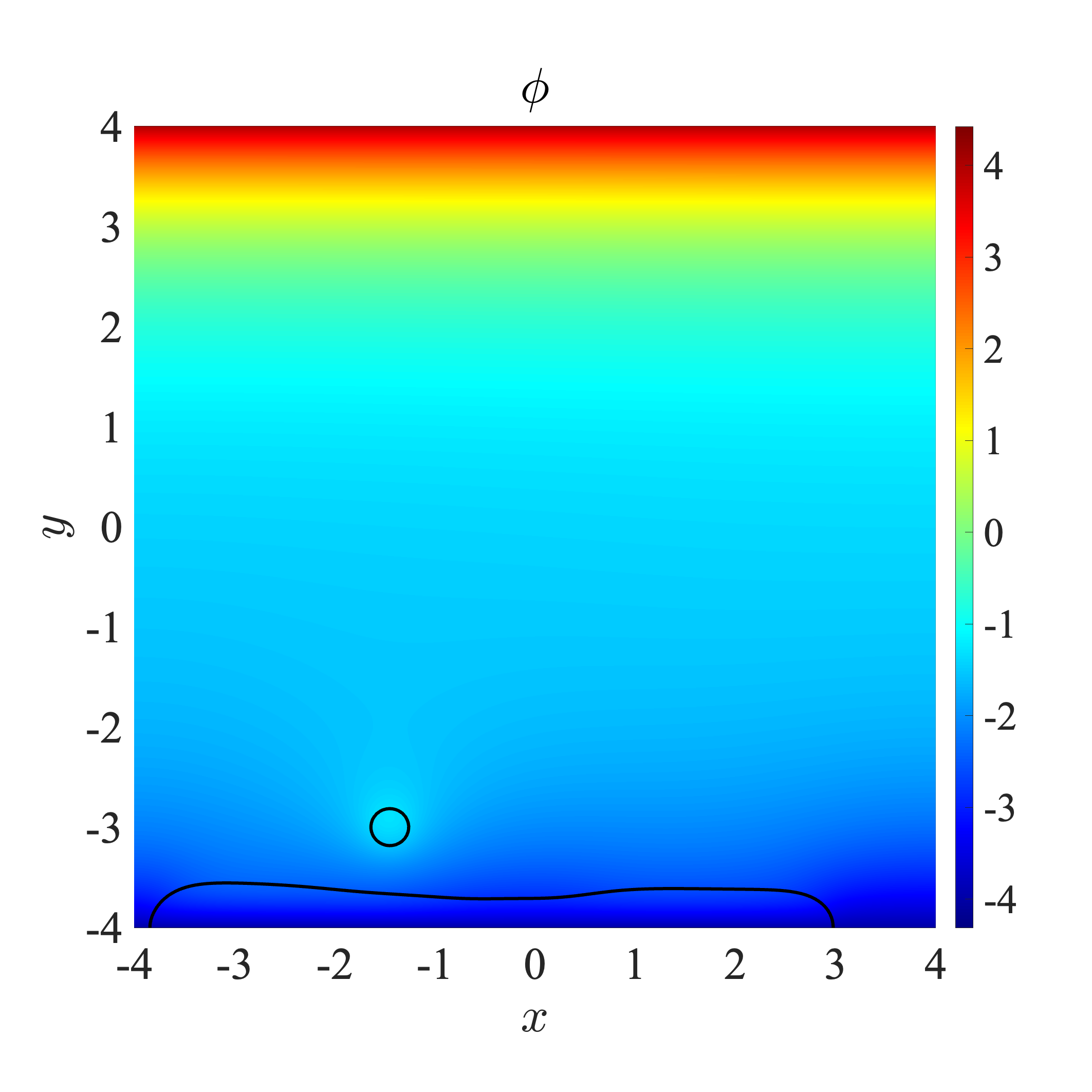}
        \label{subfig:1LeftDropD4N0Pump25Phi16}
		}
    \hskip -0.3cm
    \subfloat[$\phi~(t=80)$]{
		\includegraphics[width=0.165\linewidth]{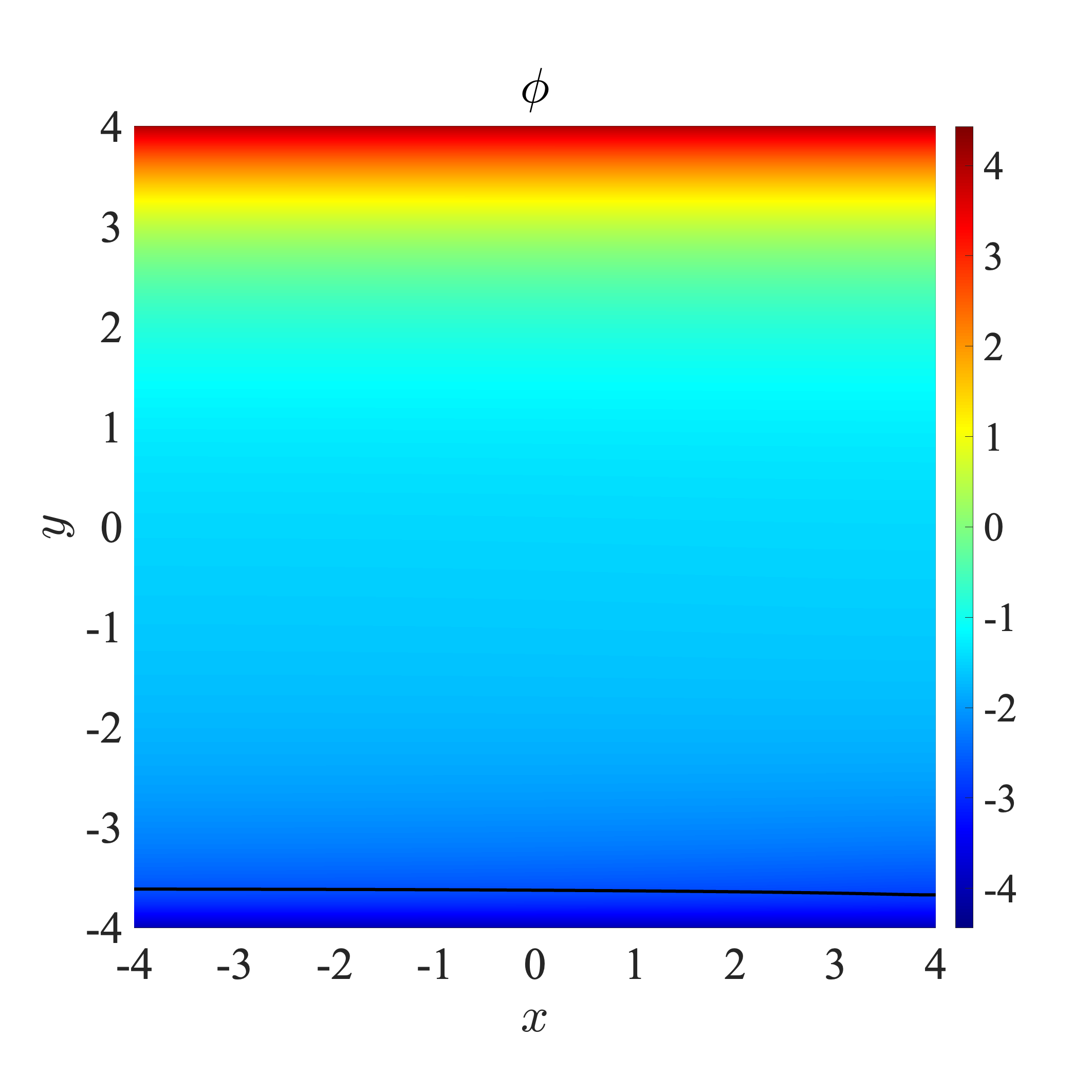}
        \label{subfig:1LeftDropD4N0Pump25Phi80}
		}
        \vskip -0.2cm
	\caption{The snapshots for the deformation and motion of drop located in the left part of the computational domain with positive ion pump when the vertical electric field is added. 
    The black solid circle represents the location of the drop, 
    which is denoted by the level set $\psi=0$. 
    The concentration and electric potential distribution are shown on the color map. 
    We choose the final time as $t = 80$.  Here \(\phi_{0b} = -4, \quad \phi_{0u} = 4.\)
    }\label{fig:1LeftDropD4N0Pump25}
\end{figure}
\end{document}